\catcode`\@=11
\newif\if@fewtab\@fewtabtrue
{\count255=\time\divide\count255 by 60
\xdef\hourmin{\number\count255}
\multiply\count255 by-60\advance\count255 by\time
\xdef\hourmin{\hourmin:\ifnum\count255<10 0\fi\the\count255}}
\def\ps@draft{\let\@mkboth\@gobbletwo
    \def\@oddfoot{\hbox to 7 cm{\tiny \versionno
       \hfil}\hskip -7cm\hfil\rm\thepage \hfil {\tiny\draftdate}}
    \def\@oddhead{}
    \def\@evenhead{}\let\@evenfoot\@oddfoot}
\def\draftdate{\number\month/\number\day/\number\year\ \ \ \hourmin }

\def\citen#1{\if@filesw \immediate\write \@auxout {\string\citation{#1}}\fi%
\@tempcntb\m@ne \let\@h@ld\relax \def\@citea{}%
\@for \@citeb:=#1\do {\@ifundefined {b@\@citeb}%
    {\@h@ld\@citea\@tempcntb\m@ne{\bf ?}%
    \@warning {Citation `\@citeb ' on page \thepage \space undefined}}%
    {\@tempcnta\@tempcntb \advance\@tempcnta\@ne
    \setbox\z@\hbox\bgroup\ifcat0\csname b@\@citeb \endcsname \relax
    \egroup \@tempcntb\number\csname b@\@citeb \endcsname \relax
    \else \egroup \@tempcntb\m@ne \fi \ifnum\@tempcnta=\@tempcntb
    \ifx\@h@ld\relax \edef \@h@ld{\@citea\csname b@\@citeb\endcsname}%
    \else \edef\@h@ld{\hbox{--}\penalty\@highpenalty
    \csname b@\@citeb\endcsname}\fi
    \else \@h@ld\@citea\csname b@\@citeb \endcsname \let\@h@ld\relax \fi}%
\def\@citea{,\penalty\@highpenalty\hskip.13em plus.13em minus.13em}}\@h@ld}
\def\@citex[#1]#2{\@cite{\citen{#2}}{#1}}%
\def\@cite#1#2{\leavevmode\unskip\ifnum\lastpenalty=\z@\penalty\@highpenalty\fi%
  \ [{\multiply\@highpenalty 3 #1%
  \if@tempswa,\penalty\@highpenalty\ #2\fi}]}   %
\makeatother % end of CITE.STY
\catcode`\@=12

\newcounter{defthm}

\def\AA            {\mbox{$A$}-\mbox{$\!A$}}
\def\AB            {\mbox{$A$}-\mbox{$B$}}

\newcommand\aai[2] {\jmath^{#1\bar{#2}}_{A}}
\def\aff           {affine Lie algebra}
\newcommand\Ai[1]  {\langle #1{,}A\rangle}

\def\alg           {algebra}
\newcommand\Ann[2] {{{\rm A}_{#1}}^{\!\!#2}}
\newcommand\AnnA[3]{{{\rm A}(#1)_{#2}}^{\!\!#3}}
\newcommand\AnnB[2]{{{{\rm A}(B)}_{#1}}^{\!\!\!\!\!#2}}
\newcommand\AnntB[2]{{{{\rm A}(\tilde B)}_{#1}}^{\!\!\!\!\!#2}}
\def\Atop          {\mbox{$A_{\rm top}$}}
\def\ATop          {{A_{\rm top}}}
\def\Av            {{A^\vee}}
\newcommand\Bar[1] {\bar{#1}} 

\newcommand\Barray[4]{\mbox{\large$[$}\!\!{\scs\begin{array}{lr}{}\\[-1.8em]
                   \scs #1\!\!&\!\!\scs #2\\[-.39em]\scs #4\!\!%
                   &\!\!\scs #3\\[-.19em]\end{array}}\!\!\mbox{\large$]$}}
\def\bc            {boundary condition}
\def\Bc            {Boundary condition}
\def\be            {\begin{equation}}
\def\bea           {\begin{equation}\begin{array}l}
\def\bearl         {\begin{array}{l}}
\def\bearll        {\begin{array}{ll}}
\newcommand\bigoplusJ[2]{\bigoplus_{#2\in\JJ}}
\def\CA            {\mbox{$\mathfrak V$}}
\def\calc          {\mbox{$\mathcal C$}}
\def\calca         {\mbox{${\mathcal C}_{\!A}$}}
\def\calcaa        {\mbox{${}_A\!{\,}{\mathcal C}_{\!A}$}}
\def\cald          {{\mathcal D}}
\def\calh          {\mbox{$\mathcal H$}}
\def\calH          {{\mathcal H}}
\def\Calh          {\mbox{$U$}}
\def\Calha         {U_{\!a}}
\def\calhT         {\mbox{$\mathcal H(\emptyset;\torus)$}}

\def\calm          {{\mathcal M}}

\def\calv          {{\mathcal V}}
\def\cats          {categories}
\def\centlA        {{\rm C}_\ell(A)}
\def\centre        {{\rm cent}}
\def\centreA       {{\rm cent}_A}
\def\centrA        {{\rm C}_r(A)}
\def\cft           {conformal field theory}
\def\Cft           {Conformal field theory}
\def\cfts          {conformal field theories}
\def\chii          {\raisebox{.15em}{$\chi$}}
\def\cir           {\,{\circ}\,}
\def\class         {classification}
\def\Class         {Classification }
\def\cocon         {coset construction}  
\def\complex       {\mbox{$\mathbb C$}}
\def\con           {conformal }
\def\Con           {Conformal }
\def\corfu         {correlation function}

\def\cvo           {chiral vertex operator}

\def\dimA          {{\rm dim}(A)}

\def\dim           {{\rm dim}}
\def\dsty          {\displaystyle }
\def\E             {\mbox{E$_7$} }

\def\eE            {{\rm e}}
\def\ee            {\end{equation}}
\def\eear          {\end{array}}
\newcommand\eev[1] {{{}^{\vee\!}}{#1}} 
\newcommand\Eev[1] {{{}^{\vee\!}}\!{#1}} 
\def\End           {{\rm End}}
\newcommand\epicture[2] {\end{picture}\\{}\\[#1.#2em]\end{array}}
\def\eps           {\varepsilon}
\def\epsnat        {\eps_{\ell}}
\def\epsnat        {\eps_{\natural}}
\def\epsnatt       {\eps_{r}}
\def\epsnatt       {\eps_{\rho}}
\def\epsnattop     {\eps_{\natural,{\rm top}}}
\def\eq            {\,{=}\,}
\def\equi          {equivalent}
\newcommand\erf[1] {(\ref{#1})}
\newcommand\Erf[2] {(\ref{#1#2})}
\newcommand\F[9]   {{\sf F}_{\!{\sss#6}#4{\sss#7},{\sss#8}#5{\sss#9}}
                   ^{\,({#1}\,{#2})\,{#3}}}
\newcommand\Fbox[1]{\shadowbox{#1}}
\def\FF            {{\sf F}}
\def\findim        {fini\-te-di\-men\-si\-o\-nal}
\newcommand\foodnode[1] {\,\footnote{~#1}}
\newcommand\Frac[2]{\mbox{\large$\frac{#1}{#2}$}}
\newcommand\Fs[6]  {{\sf F}_{\,{#5}\,{#6}}^{\,({#1}\,{#2}\,{#3})\,{#4}}}
\def\fsi           {Fro\-be\-ni\-us\hy Schur indicator}
\def\ft            {field theory}
\def\fts           {field theories}
\def\furu          {fusion rule}
\newcommand\goodnode[1] { }
\newcommand\G[9]   {{\sf G}_{\,{\sss#6}#4{\sss#7},{\sss#8}#5{\sss#9}}
                   ^{\,({#1}\,{#2})\,{#3}}}
\def\GG            {{\sf G}}
\newcommand\Gs[6]  {{\sf G}_{\,{#5}\,{#6}}^{\,({#1}\,{#2}\,{#3})\,{#4}}}
\def\Hom           {{\rm Hom}}
\def\HomA          {{\rm Hom}_{\!A}}
\def\hopf          {Hopf algebra}
\newcommand\hsp[1] {\mbox{\hspace{#1 em}}}
\def\hy            {$\mbox{-\hspace{-.66 mm}-}$}
\def\I             {\mbox{$\II$}}
\newcommand\iaa[2] {\imath_{#1#2}^{\,A}}
\def\id            {\mbox{\sl id}}
\def\ide           {identification}  
\def\idM           {\id_{\dot M}}
\def\idN           {\id_{\dot N}}
\def\ii            {{\rm i}}
\def\II            {\mathcal I}

\def\iN            {\,{\in}\,}
\def\In            {\prec}
\def\IN            {\,{\In}\,}
\def\inda          {{\rm Ind}_{A\!}}
\def\indar         {\widetilde{{\rm Ind}}_{A\!}}
\def\indaz         {{\rm Ind}_{A_{2r}\!}}

\def\Infdim        {Infinite-dimensional }
\def\Intro         {Introduction }

\def\J             {\mbox{$\JJ$}}

\def\JJ            {\mathcal J}
\def\kma           {Kac\hy Moody algebra}

\newcommand\labl[1]{\label{#1}\ee}
\newcommand\Labl[2]{\label{#1#2}\ee}
\def\lhs           {left hand side}
\def\lie           {Lie algebra}
 
\def\llb           {\mbox{\large[}}

\def\Loc           {{\rm Hom}_{\rm loc}}
\newcommand\lr[2]  {\langle #1{,}#2\rangle}
\newcommand\lra[2] {\langle #1{,}#2\rangle_{\!A}^{}}
\def\lrb           {\mbox{\large]}}

\newcommand\m[7]   {m_{#1#2,#3#4}^{\;#5#6;\,#7}}
\def\M             {{\dot M}}
\newcommand\mC[7]  {C_{#1#2,#3#4}^{\;#5#6;\,#7}}
\newcommand\mCb[8] {C_{#1#2,#3#4}^{(#8)#5#6;\,#7}}

\def\mimo          {minimal model}
\def\mtc           {modular tensor category}
\def\mtcs          {modular tensor categories}
\def\Modinv        {Modular invarian}
\def\modinv        {modular invarian}
\def\n             {{\dot N}}
\newcommand\N[3]   {{N_{#1#2}}^{\!\!#3}}
\def\nxt           {\raisebox{.08em}{\rule{.44em}{.44em}}\hsp{.4}}
\def\Nxt           {\raisebox{.08em}{\rule{.44em}{.44em}}}
\def\Nu            {{\mathcal V}}
\def\oa            {operator algebra}
\def\obj           {{\mathcal O}bj}
\def\objc          {{\mathcal O}bj(\calc)}
\def\objca         {{\mathcal O}bj(\calca)}
\def\one           {{\bf1}}
\def\One           {0}
\def\opl           {\,{\oplus}\,}
\def\Opl           {\!\!\!\!{\oplus}\!\!\!\!}
\def\oT            {^{\rm t}}

\def\oti           {\,{\otimes}\,}
\def\Oti           {{\otimes}}
\def\parfu         {partition function}

\def\q             {quantum }
\def\Q             {Quantum }
\def\qed           {\hfill\checkmark\medskip}
\def\QFT           {Quantum Field Theory}        
\def\qfts          {quantum field theories}
\long\def\query#1{\hskip 0pt{\vadjust{\everypar={}\small\vtop to 0pt{\hbox{}%
     \vskip -13pt\rlap{\hbox to 49.0pc{\hfil{\vtop{\hsize=8pc\tolerance=6000%
     \hfuzz=.5pc\rightskip=0pt plus 5.5em\noindent#1}}}}\vss}}}}%
\def\r             {{\rho}}
\newcommand\R[5]   {{\sf R}^{(#1\,#2)#3}_{#4\,#5}}
\newcommand\rhobas[8]{\rho_{(#1#2)(#3#4)}^{#8\,(#5#6);#7}}
\newcommand\Rm[5]  {{\sf R}^{-\,(#1\,#2)#3}_{\;#4\,#5}}
\def\rmr           {{\rm r}}
\newcommand\Rp[5]  {{\sf R}^{+\,(#1\,#2)#3}_{\;#4\,#5}}

\newcommand\rbas[4]{\rho_{#2\,,\,#3}^{#1\, #4}}
\def\reals         {\mbox{$\mathbb R$}}
\def\rep           {representation}
\def\Rep           {Representation}
\def\Repth         {Representation theory}
\def\resp          {respectively}
\def\rhs           {right hand side}
\def\rmX           {{\rm X}}
\def\RR            {{\sf R}} 
\newcommand\Rs[4]  {{\sf R}^{#1\,(#2\,#3)#4}}        

\newcommand\Rss[3] {{\sf R}^{(#1\,#2)#3}}
\def\scs           {\scriptstyle} 
\newcommand\sect[1]{\section{#1}\setcounter{equation}0\setcounter{defthm}0} 
\def\slz           {\mbox{$\mathfrak{sl}(2)$}}
\def\smallone      {{\bf1}}
\def\sss           {\scriptscriptstyle}
\def\Star          {\,{\star}\,}
\def\stc           {statistic}

\def\su            {\mathfrak{su}}
\newcommand\sumI[2]{\sum_{#2\in\II}}
\newcommand\sumJ[2]{\sum_{#2\in\JJ}}
\def\sym           {symmetry} 
\def\syms          {sym\-me\-tries} 
\newcommand\tbas[2]{|\chii_{#1};{#2}\torus\rangle}
\newcommand\tdbas[2]{\langle \chii_{#1};{#2}\torus|}
\def\tc            {tensor category}
\def\tcs           {tensor categories}
\def\tft           {topological field theory}
\def\Tft           {Topological field theory}
\def\tfts          {topological field theories}
\def\theequation   {\thesection.\arabic{equation}}    
\def\torus         {{\rm T}}
\def\top           {_{\rm top}}
\def\tparfu        {torus partition function}
\def\tr            {{\rm tr}\,}
\def\Tr            {{\rm tr}}
\def\twodim        {two-di\-men\-si\-o\-nal}
\def\U             {\,} 
\def\V             {\mbox{$\mathcal V$}} 
\def\Vect          {{\rm Vect}}
\def\Vee           {^{\vee}}

\def\voa           {vertex operator algebra}     
\newcommand\void[1]{}
\def\wrt           {with respect to }
\def\wzwm          {WZW model}
\newcommand\x[4]   {\lambda_{(#1,#2) #3}^{\,#4}}
\newcommand\xx[4]  {\lambda_{(#1,#2) #3}^{}}
\newcommand\y[4]   {\Upsilon^{(#1,#2) #3}_{\bar{#4}}}
\newcommand\yy[4]  {\Upsilon^{(#1,#2) #3}_{}}
\newcommand\Zdfct[4]{Z^{#1|#2}_{#3 #4}}
\def\zet           {\mbox{$\mathbb Z$}}

\documentclass[12pt]{article}

\renewcommand\theequation{\thesection.\arabic{equation}} 
\usepackage{latexsym,amssymb,amsfonts,epsf,color,colordvi,fancybox}
\usepackage[dvips]{graphics} %
\newtheorem{defthm}{\whattheorem}[section]
\newcommand\dt[1]  {\noindent\def\whattheorem{#1}\pagebreak[0]\begin{defthm}{}%
                   \samepage{$\!\!${\rm:}\nopagebreak\\[-1.91em]{}}\end{defthm}}
\newcommand\dtl[2] {\noindent\def\whattheorem{#1}\pagebreak[0]\begin{defthm}{}%
                   \samepage{$\!\!${\rm:}\label{#2}\nopagebreak\\[-1.91em]{}}%
                   \end{defthm}}

\begin{document}

\begin{flushright}  {~} \\[-12mm]
{\sf hep-th/0204148}\\[1mm]{\sf PAR-LPTHE 02-25}\\[1mm]
{\sf April 2002} \end{flushright}

\begin{center} \vskip 14mm
{\Large\bf TFT CONSTRUCTION OF RCFT CORRELATORS}\\[4mm]
{\Large\bf I: PARTITION FUNCTIONS}\\[20mm] 
{\large 
J\"urgen Fuchs$\;^1$ \ \ \ Ingo Runkel$\;^2$ \ \ \ Christoph Schweigert$\;^2$}
\\[8mm]
$^1\;$ Institutionen f\"or fysik~~~~{}\\
Universitetsgatan 5\\ S\,--\,651\,88\, Karlstad\\[5mm]
$^2\;$ LPTHE, Universit\'e Paris VI~~~{}\\
4 place Jussieu\\ F\,--\,75\,252\, Paris\, Cedex 05
\end{center}
\vskip 20mm

\begin{quote}{\bf Abstract}\\[1mm]
We formulate rational conformal field theory in terms of a symmetric 
special Frobenius algebra $A$ and its representations. $A$ is an algebra 
in the modular tensor category of Moore\hy Seiberg data of the underlying 
chiral CFT. The multiplication on $A$ corresponds to the OPE of boundary 
fields for a single boundary condition. General boundary conditions are 
$A$-modules, and (generalised) defect lines are \AA-bimodules.
\\
The relation with three-dimensional TFT is used to express CFT data, like
structure constants or torus and annulus coefficients, as invariants of 
links in three-manifolds. We compute explicitly the ordinary and twisted 
partition functions on the torus and the annulus partition functions. We  
prove that they satisfy consistency conditions, like modular invariance 
and NIM-rep properties.
\\
We suggest that our results can be interpreted in terms of non-commutative
geometry over the modular tensor category of Moore\hy Seiberg data.
\end{quote}
\vfill
\newpage

%%%%%%%%%%%%%%%%%%%%%%%%%%%%%%%%%%%%%%%%%%%%%%%%%%%%%%%%%%%%%%%%%%%%%%%%

\sect{Introduction and summary} 

%%%%%%%%%%%%%%%%%%%%%%%%%%%%%%%%%%%%%%%%%%%%%%%%%%%%%%%%%%%%%%%%%%%%%%%%

The use of \twodim\ \cft\ in
string theory, statistical mechanics and condensed matter physics has
often focussed on issues related to bulk fields on closed world sheets. 
But among the multitude of applications of CFT there are also many -- 
like the study of percolation probabilities, of defects in condensed matter 
systems and of string perturbation theory in D-brane backgrounds -- that 
require an understanding of CFT on world sheets with boundary, and in particular
of conformally invariant boundary conditions. These aspects have been
investigated intensively over the last few years. Apart from its important 
physical applications, the study of boundary conditions is also considerably 
contributing to increase our structural insight in conformal field theory.
Further progress can be expected to result from the analysis of defect 
lines, a subject that so far has attracted comparatively moderate attention.

\medskip

In the present paper we make transparent the behavior of rational 
conformal field theories on arbitrary (orientable) world sheets, 
including boundaries and defect lines. This is achieved by combining tools 
from topological field theory with concepts from non-commutative algebra,
making ample use of two basic facts:
\\[.23em]
\nxt The Moore\hy Seiberg data of a rational chiral CFT give rise to a
     topological field theory in three dimensions, and thereby to invariants
     of links in three-manifolds. 
\\[.23em]
\nxt The Moore\hy Seiberg data give rise to a modular tensor category \calc.
     One can set up algebra and representation theory in this category \calc\
     in very much in the same way as it is usually done in the categories of
     vector spaces or of super-vector spaces.

\medskip

A modular tensor category is actually nothing else than a 
basis-independent formulation of the Moore\hy Seiberg data.
An important motivation to adopt this framework is the observation that there
exist several rather different mathematical formalisations of the
physical intuition of a chiral conformal field theory, i.e.\ of the
chiral algebra, its space of physical states and of the properties of chiral
vertex operators and conformal blocks associated to these states. Two 
prominent examples of such formalisms are the one based on local algebras of
observables on the circle, and hence nets of subfactors, and the one of 
vertex algebras. 
Both frameworks involve quite intricate mathematical structures. Accordingly,
in both settings the explicit treatment of even modestly complicated models 
proves to be difficult. 

A major problem is to work out the representation theory of the vertex 
algebras, \resp\ to find the (physically relevant) representations of the 
local algebras of observables. As a 
consequence, there have been various attempts to extract the relevant part
of the information about the representation category of the chiral algebra
and to encode it in simpler structures. These attempts have been particularly 
successful for rational theories, for which the representation category is
semisimple and there are only finitely
many inequivalent irreducible representations. In the present paper we 
require the chiral algebra to be rational. However, we do not insist on
choosing the maximally extended chiral algebra. This allows us to deal also
with symmetry breaking boundary conditions, as well as theories for which
the left- and right-moving chiral algebras are different.

The attempts to formalise aspects of the representation theory of rational
conformal field theories have lead, among other results, to 
new algebraic notions, like truncated quantum groups (see e.g.\ 
\cite{masc3,vecs,szla,jf25,fugv3}), weak Hopf algebras 
\cite{szla3,bons,bosz2,etno} and double triangle algebras \cite{ocne7,ocne9}. 
A more direct approach is to formalise the properties 
of the representation category itself. This gives rise to the notion of a
modular tensor category \cite{TUra}, which we will explain in detail in 
section 2, and of module categories \cite{ostr}. Schematically:

\begin{center}
\begin{tabular}{ll}
\begin{tabular}{l}
\multicolumn1c{formalisations of}  \\[-2pt]
\multicolumn1c{the chiral algebra} \\[3pt] \hline \\[-1pt]
nets of subfactors \\[11pt]
vertex algebras \\[-1pt]{}
\end{tabular}
\mbox{\hspace{3em}}&
\begin{tabular}{l}
\multicolumn1c{structures capturing the}\\[-2pt]
\multicolumn1c{representation category} \\[3pt] \hline\\[-7pt]
truncated quantum groups\\[1pt]
weak Hopf algebras\\[1pt]
double triangle algebras\\[1pt]
modular tensor categories 
\end{tabular}
\end{tabular}
\end{center}

\noindent
By making use of double triangle algebras and weak Hopf algebras, aspects 
of rational CFT have been analyzed in \cite{bppz2,pezu6,pezu7,pezu8}.
The present paper develops an approach to rational CFT that is
based on modular categories.

\medskip

The Moore\hy Seiberg data captures the {\em chiral\/} aspects of rational 
conformal field theory. To arrive at a {\em full\/} conformal field 
theory with local correlation functions, additional input is required.  
This can already be seen in the example of a free boson. Here one can 
choose $\hat{\mathfrak u}(1)$ as the chiral algebra, which has a unitary 
irreducible highest weight representation for every charge $q\iN\mathbb{R}$. 
There are many consistent CFTs associated to these chiral data, for example 
those describing a free boson compactified on a circle of some given radius.
These models possess (modulo T-duality) in particular different modular
invariant torus partition functions. So the first structure that comes to mind 
as additional information for the construction of a full conformal field theory
is the choice of a modular invariant for the bulk theory. However, this 
proves to be too naive, as one knows of many examples of modular invariant 
bilinear combinations of characters that do not arise as the partition 
function of any consistent CFT at all \cite{vers,scya5,fusS,gann17}.

It is therefore a crucial insight \cite{fuRs} that complete information on 
how to construct a full CFT from given chiral data is contained in the 
structure of a {\em symmetric special Frobenius algebra\/} in \calc. (We will 
explain in detail below what is meant by an algebra $A$ in \calc. It is 
an object, with specific properties, in the category \calc, and 
thereby corresponds to some representation of the chiral algebra -- the
algebra object $A$ must in particular not be confused with the chiral algebra 
\CA\ itself.) Already at this point a significant advantage of tensor 
categories becomes apparent: Once one accepts the idea of doing algebra 
and representation theory in the setting of general tensor categories
rather than the category of complex vector spaces, one can directly use 
standard algebraic and
representation theoretic concepts, like, in the case at hand, the notion 
of a Frobenius algebra. Indeed, all mathematical concepts that are needed 
in the approach to CFT that is developed here
can already be found in standard textbooks on associative algebras 
\cite{CUre,FAde} and category theory \cite{MAcl,KAss}.

\medskip

We can show that every symmetric special Frobenius algebra object in the 
\mtc\ of a chiral CFT leads to 
a full CFT that is consistent on all orientable world sheets; Morita
equivalent algebras yield the same CFT. Conversely, we establish that every
unitary rational full CFT -- provided only that it possesses one 
boundary condition preserving the chiral algebra
  \foodnode{Since the chiral algebra is not required to
  be maximally extended, the boundary condition is still allowed to break
  part of the bulk symmetry. For brevity, in this paper we will sometimes
  refer to boundary conditions that preserve the chiral algebra as
  `conformal boundary conditions' or also just as `boundary conditions'.}
at all -- determines uniquely a (Morita class of)
symmetric special Frobenius algebra(s).  
The Frobenius algebra in question is actually nothing else than the algebra of
boundary fields associated to one given boundary condition of the theory.
It is associative due to the associativity of the operator product of boundary 
fields; the non-degenerate bilinear invariant form that turns it into a Frobenius 
algebra expresses the non-degeneracy of the two-point functions of boundary
fields on the disk. Our results can thus be briefly summarised by saying that 
we are able to {\em construct the correlation functions of a unitary rational 
conformal field theory starting from just one of its boundary conditions\/}.

In fact, every boundary condition of a full CFT gives rise to a symmetric 
special Frobenius algebra object, and all such algebra objects are Morita 
equivalent and hence lead to one and the same CFT.
Moreover, when a given full CFT can be constructed from a Frobenius
algebra $A$, then any of its \bc s gives back an algebra in the Morita
class of $A$. (On the other hand, we cannot, as yet, exclude the
possibility that there exists a \bc\ $M$ of some CFT $C$ such that
the CFT reconstructed from the Frobenius algebra $A$ that arises from 
$M$ does not coincide with the original CFT $C$.)

\medskip

Apart from its conceptual aspects, the formalism presented in this paper
is also of considerable practical and computational value. The main point 
is that {\em structure constants\/} -- OPE constants as well as coefficients 
of the torus and annulus \parfu s -- {\em are given as link invariants 
in three-manifolds\/}.  Computing the value of an invariant is 
straightforward once we have gathered three ingredients: the 
Moore\hy Seiberg data (i.e.\ the fusing and braiding matrices), the 
structure constants for the multiplication of the algebra object $A$, and 
the \rep\ matrices describing the action of $A$ on its irreducible modules. 

We will treat the Moore\hy Seiberg data as given. It is of course a 
non-trivial problem to obtain these data from a chiral CFT; here, however, 
we are concerned with building a full CFT given all the chiral information.
Finding an algebra and a multiplication involves solving a 
nonlinear associativity constraint. It turns out that this constraint is 
equivalent to the sewing constraint for boundary structure constants of 
a {\em single\/} boundary condition. Solving this nonlinear equation is 
not easy, but still much simpler 
than finding a solution to the full set of nonlinear constraints involving
all boundary structure constants as well as the bulk-boundary couplings 
and the bulk structure constants. Finally, finding the representations of 
$A$ is a {\em linear\/} problem. 

There is a concept that allows us to systematically construct examples 
of symmetric special Frobenius algebras: simple currents \cite{scya6}, that 
is, the simple objects of \calc\ with quantum dimension one. As we will 
explain elsewhere, for algebra objects that contain only simple currents 
as simple subobjects the associativity constraints reduce to a cohomology 
problem for abelian groups that can be solved explicitly. Algebras built 
from simple currents describe modular invariants of `D-type'. Often,
in particular for all WZW models, they provide representatives for almost 
all Morita classes of algebras. It is, however, a virtue of the formalism 
developed in this paper that it treats exceptional modular invariants,
including those of automorphism type, on the same footing as simple 
current modular invariants. (The structure of full conformal field theories
having an exceptional modular invariant is therefore not really exceptional.)

An important aspect of our construction is that it is possible to prove that 
the resulting structure constants of the CFT solve all sewing constraints. 
{}From a computational point of view, the present formalism thus allows us 
to generate a solution to the full set of sewing constraints from a solution 
to a small subset of these constraints. This way one can also check the 
consistency of boundary conditions that have been proposed in the literature.

\medskip

Modular tensor categories possess in particular a braiding which accounts for 
the braid group statistics of \twodim\ field theories. Thus there is a natural
notion of commutativity with respect to this braiding. It is therefore worth 
emphasising that the Frobenius algebras $A$ we consider are not necessarily 
(braided-) commutative. As a consequence, our construction constitutes a natural 
generalisation of {\em non-commutative\/} algebra to modular tensor categories. 
This allows us to summarise our results in the following dictionary between 
physical concepts in CFT and notions in the theory of associative algebras:

\begin{center} \fbox{ \begin{tabular}{l|l} {}&\\[-.8em]
\multicolumn1{c|}{physical concepts} & \multicolumn1c{algebraic notions}
\\&\\[-.93em] \hline &\\[-.8em]
boundary conditions        & \ $A$-modules             \\&\\[-.8em]
defect lines               & \ \AA-bimodules           \\&\\[-.8em]
boundary fields $\,\Psi^{MN}_i$& $\ {\rm Hom}_A(M\oti i,N)$ 
 \\&\\[-.8em]
bulk fields $\,\Phi_{ij}$
             & $\ {\rm Hom}_{A|A}((A\oti j)^-\!,(A\oti\bar\imath)^+)$ 
 \\&\\[-.8em]
disorder fields $\,\Phi^{B_1,B_2}_{ij}$
             & $\ {\rm Hom}_{A|A}((B_1\oti j)^-\!,(B_2\oti\bar\imath)^+)$
\\[.5em] \end{tabular} }
\end{center}

Let us explain the entries of this table in some detail. Boundary 
conditions will be shown to be in correspondence with (left) modules of $A$.
In particular, simple modules correspond to elementary boundary conditions,
while direct sums of simple modules indicate the presence of Chan\hy Paton
multiplicities. If $M$
is a left $A$-module, then for any object $i$ of \calc, $M\oti i$ is a left 
$A$-module, too. As a consequence, it makes sense to consider 
left $A$-module morphisms from $M\oti i$ to another module $N$,
i.e.\ morphisms from $M\oti i$ to $N$ that intertwine the action of $A$ 
on the two objects. For each such morphism there is a boundary field
changing the boundary condition from $M$ to $N$ and carrying the chiral
label $i$.

For non-commutative algebras, it is natural to consider not only left
(or right) modules, but also bimodules, i.e.\ objects that carry an action 
of $A$ both from the left and from the right, such that both actions commute.
We show that bimodules correspond to (generalised) defect lines. The 
trivial defect line -- i.e.\ no defect at all -- is $A$ itself, and the
tensor product (over $A$) of bimodules is `fusion' of defect lines. 

Given a bimodule $B_1$, one can endow the object $B_1\oti i$ for any object 
$i$ of \calc\ with the structure of a bimodule in two different ways: The 
left action of $A$ is just the one inherited from the left action of $A$ on
$B_1$, while a right action of $A$ can be defined by using either the
braiding of $i$ and $A$ or the inverse of this braiding. We denote the 
two resulting bimodules by $(B_1\oti i)^+$ and $(B_1\oti i)^-$.

A particular bimodule is $A$ itself. The degeneracy of a bulk field
with chiral labels $i$ and $j$ is again given by a space of morphisms -- 
the space ${\rm Hom}_{A|A}((A\oti j)^-\!,(A\oti\bar\imath)^+)$ of bimodule 
morphisms. This suggests the following re-interpretation of bulk fields: 
they ``change'' the trivial defect $A$ to itself. It is therefore natural 
to generalise bulk fields and consider fields with chiral
labels $i,j$ that change the defect line of type corresponding to the 
bimodule $B_1$ to a defect line of some other type $B_2$. The degeneracy 
of these disorder fields is described by the space ${\rm Hom}_{A|A}
((B_1\oti j)^-\!,(B_2\oti\bar\imath)^+)$ of bimodule morphisms.

\medskip

For each such type of fields there are partition functions that count the
corresponding states. For boundary fields these are linear, for bulk fields
bilinear combinations of characters with non-negative integral coefficients.
This way, every full rational conformal field theory gives rise to a collection
of combinatorial data -- essentially the dimensions of the morphism spaces
introduced in the table above. Clearly, these data must satisfy various
consistency constraints, both among each other and with the underlying 
category \calc, in particular with the fusion rules of \calc. 
Concrete instances of such consistency conditions have been obtained
from a variety of arguments, see e.g.\
\cite{dore3,prss3,sasT2,boev3,boek2,bppz2,pezu4,pezu5,scst,fuSc16,gann17}.
In particular, the annulus
partition functions provide non-negative integral matrix representations
(NIM-reps) of the fusion rules of \calc, while the partition functions
of defect line changing operators give rise to NIM-reps of the double
fusion rules. 

One important result of the present paper is a rigorous proof of these 
relations. As a word of warning, let us point out that, by themselves,
the problems of classifying modular invariants or NIM-reps are {\em not
physical\/} problems. (Still, the classification of such combinatorial 
data can be a useful auxiliary task.) Indeed, as already mentioned,
they tend to possess solutions 
that do not describe the partition functions of any conformal field theory 
(see e.g.\ \cite{vers,scya5,fusS,gann17}). In contrast, in our approach, 
the partition functions arise as special cases of \corfu s. Indeed,
in a forthcoming publication our approach will be extended to general 
amplitudes (compare \cite{fuRs}), and it will be shown that the system
of amplitudes, with arbitrary insertions and on arbitrary world sheets, 
satisfies all factorisation and locality constraints, and that they are 
invariant under the relevant mapping class groups. This result 
guarantees that only physical solutions occur in our approach.

\medskip

A brief outline of the paper is as follows. We start in section 2 with 
a review of some facts about modular tensor categories and topological
field theory. In sections 3 and 4 we investigate symmetric special 
Frobenius algebra objects and their representation theory, respectively,
and show how these structures arise in \cft. In the remainder of the paper 
these tools are employed to deduce various properties of torus and annulus 
partition functions and to study defect lines. As an illustration how 
our approach works in practice, two examples accompany our development 
of the general theory: The free boson compactified at a radius of 
rational square, and the \E modular invariant of the $\su(2)$ WZW model.

Some of our results have been announced in \cite{fuRs,fuRs2}.

%%%%%%%%%%%%%%%%%%%%%%%%%%%%%%%%%%%%%%%%%%%%%%%%%%%%%%%%%%%%%%%%%%%%%%%%
\newpage 
{\small \tableofcontents} 
%%%%%%%%%%%%%%%%%%%%%%%%%%%%%%%%%%%%%%%%%%%%%%%%%%%%%%%%%%%%%%%%%%%%%%%%
\newpage

\sect{Modular tensor categories and chiral CFT}\label{sec:mod-cat-cCFT} 

%%%%%%%%%%%%%%%%%%%%%%%%%%%%%%%%%%%%%%%%%%%%%%%%%%%%%%%%%%%%%%%%%%%%%%%%

\subsection{Modular tensor categories} \label{sec:mod-cat}

As already pointed out in the introduction, the framework we are going 
to use is the one of modular tensor categories \cite{TUra,KAss,KArt,BAki}. 
Let us explain in some detail what these structures are and why it is 
natural and appropriate to work in this setting.

In \cft, modular tensor categories arise in the form of \rep\ categories
of rational \voa s \cite{FRlm,frhl,KAc4},%
 \foodnode{It is still a conjecture that the \rep\ category of every
 rational VOA is modular. There is no general proof, but the property has
 been established for several important classes of VOAs, compare e.g.\
 \cite{hule5,kalu3-6,fink}, and it is expected that possible exceptions
 should better be accounted for by an appropriate refinement of the
 qualification `rational'.}
which in turn constitute a concrete mathematical realisation of the 
physical concepts of a chiral \alg\ and its primary fields. It has been 
demonstrated by Moore and Seiberg \cite{mose3,Mose} that the basic 
properties of a rational chiral \cft\ can be encoded in a small collection 
of data -- like braiding and fusing matrices and the modular $S$-matrix -- 
and relations among them -- like the pentagon and hexagon identities.
One must be aware, however, of the fact that the usual presentation of
those data implicitly involves various non-canonical basis choices.
As a consequence, the fusing matrices, for instance, enjoy a large gauge 
freedom, whereas only their gauge-invariant part has a physical meaning.
Posing the Moore\hy Seiberg data in a basis-free setting leads rather
directly to the concept of a modular tensor category.
As an additional benefit, this formulation supplies us with
a powerful graphical calculus.

In the sequel we start out by reviewing details of the mathematical machinery 
that is required to understand \mtcs; only afterwards we return to the origin 
of these structures in rational \cft. A {\em category\/} \calc\ consists 
of two types of data: A class $\objc$ of {\em objects\/} and a 
family of {\em morphism\/} sets $\Hom(U,V)$, one for each pair $U,V\iN\objc$.
In the situation of our interest, the objects are the \rep s of the chiral 
\alg\ \CA\ of the CFT (a rational \voa), and the morphisms are the 
intertwiners between \CA-\rep s.

Morphisms can be composed when the relevant objects match, i.e.\ the
composition $g\cir f$ of $f\iN\Hom(U,V)$ and $g\iN\Hom(Y,Z)$ exists if
$Y{=}\,V$. This operation of composition is associative, and for every object
$U$ the endomorphism space $\End(U)\,{\equiv}\,\Hom(U,U)$ contains
a distinguished element, the identity morphism $\id_U$, satisfying
$g\cir\id_U\eq g$ for all $g\iN\Hom(U,V)$ and $\id_U\cir f\eq f$ for all 
$f\iN\Hom(Y,U)$.  The categories \calc\ of our interest are complete \wrt 
direct sums (this can always be assumed without loss of generality) and 
come enriched with quite a bit of additional structure; we introduce 
this structure in three steps.

\smallskip\noindent\nxt~%
First, \calc\ is a semisimple abelian strict tensor category 
with the complex numbers as ground ring.

Let us explain the various qualifications appearing in this statement.
{\em Abelianness\/} \cite[chapter\,VIII]{MAcl} means that there is a zero 
object 0 and the morphisms possess various natural properties 
familiar from vector spaces. Concretely, every morphism set is an abelian
group, and composition of morphisms is bilinear;
every morphism has a kernel and a cokernel (they are defined by a universal
property); every monomorphism is the kernel of its cokernel, and every
epimorphism is the cokernel of its kernel; finally, every morphism $f$ can 
be written as the composition $f\eq h\cir g$ of a monomorphism $h$ and an 
epimorphism $g$. 

\noindent
In a {\em tensor\/} category (see e.g.\ \cite[chapter\,VII]{MAcl} or
\cite[chapter\,XI]{KAss}) there is a tensor product $\otimes$,
both of objects and of morphisms. The tensor product on objects has a unit 
element, which is denoted by $\one$; for $f\iN\Hom(U,Y)$ and $g\iN\Hom(V,Z)$,
the tensor product morphism $f\oti g$ is an element of $\Hom(U\Oti V,Y\Oti Z)$.
The endomorphisms $\End(\one)$ of the tensor unit $\one$ form a commutative ring 
$k$, called the {\em ground ring\/}, and every morphism set is a $k$-module. 
The operations of composition and of forming the tensor product of morphisms 
are bilinear and compatible in an obvious manner. In the present context we 
require that the ground ring is the field of complex numbers, $k\eq\complex$; 
then the morphism sets are complex vector spaces.

\noindent
In any \tc\ there is a family of isomorphisms between $U\oti(Y\Oti Z)$ 
and $(U\Oti Y)\oti Z$, with $U,Y,Z$ any triple of objects, and families of
isomorphisms between $U\oti\one$ and $U$ as well as between $\one\oti U$ and 
$U$, for any object $U$. They are called associativity and unit constraints, 
\resp, and are subject to the so-called pentagon identity (assuring
that any two possibilities of bracketing multiple tensor products are related
via the associativity constraints) and triangle identities (compatibility
between associativity and unit constraints). A tensor category is called 
{\em strict\/} if all these isomorphisms are identities, so that the tensor 
product of objects is associative and $\one\oti U\eq U\eq U\oti\one$. By the
coherence theorems \cite[chapter\,VII.2]{MAcl}, there is no loss of generality
in imposing this strictness property. On the other hand, when dealing 
with certain other structures below, we will often have to be careful 
not to mix up equality and isomorphy of objects.

\noindent
Finally, the meaning of semisimplicity is analogous as for other algebraic 
structures. A {\em simple\/} (or irreducible) object $U$ of an abelian 
tensor category is an object whose endomorphisms are given by the ground 
ring, $\End(U)\eq k\,\id_U$, i.e.\
$\End(U)\eq\complex\,\id_U$ for the \cats\ considered here;%
  \foodnode{In a general category, this property rather characterises an
  {\em absolutely simple\/} object, while simplicity of an object means that
  it does not possess a non-trivial proper subobject. In semisimple
  categories, absolutely simple implies simple, and in any abelian category
  over an algebraically closed ground field the two notions are equivalent.}
in particular, the tensor unit $\one$ is automatically simple. 
A {\em semisimple\/} category is then characterised by the property that 
every object is the direct sum of finitely many simple objects.

\medskip

Semisimplicity of a \tc\ \calc\ implies in particular {\em dominance\/} of 
\calc, which means that there exists a family $\{U_i\}^{}_{i\in\II}$ of
simple objects with the following property: for any $V,W\iN\objc$ every 
morphism $f\iN\Hom(V,W)$ can be decomposed into a finite sum 
  \be  f = \sum_r g_r \cir h_r  \ee
with
  \be  h_r \in \Hom(V,U_i) \qquad{\rm and}\qquad g_r \in \Hom(U_i,W)  \labl{1dom}
for suitable members $U_i\eq U_i(r)$ (possibly with repetitions) of this
family.

Since in the categories we are considering, the morphism sets are 
vector spaces (over \complex), it is convenient to introduce a shorthand 
notation for their dimension:
  \be \dim\,\Hom(X,Y) =: \lr XY  \Labl lr
for $X,Y\iN\objc$. As a consequence of semisimplicity we have 
$\lr XY\eq\lr YX$.

A convenient way to visualise morphisms in an abelian \tc\ is via graphs in 
which lines stand for identity morphisms; thus $\id_U$ and $f\iN\Hom(U,V)$ 
are depicted as
  \bea  \begin{picture}(130,44)(0,29)
  \put(42,0)  {\begin{picture}(0,0)(0,0)
              \scalebox{.38}{\includegraphics{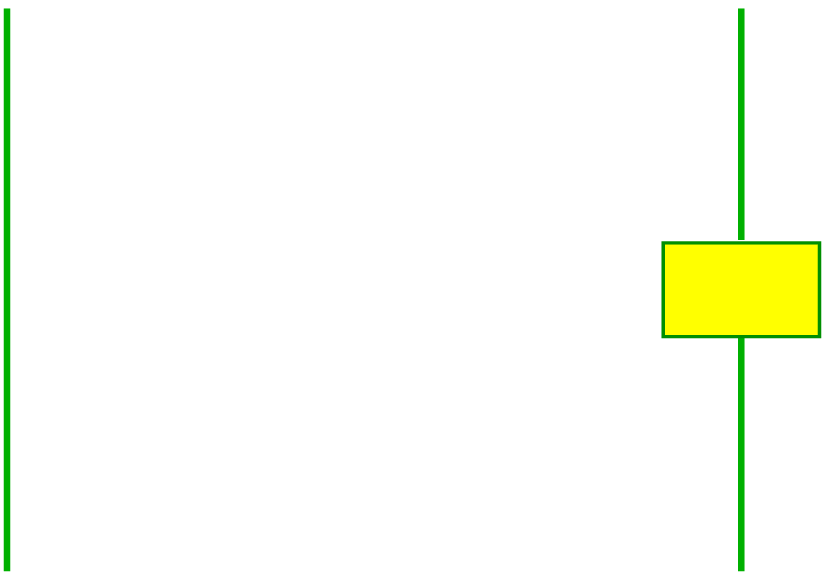}} \end{picture}}
  \put(-2,28)     {$\id_U^{}\;=$} 
  \put(39.3,-6.6) {\scriptsize$U$} 
  \put(39.8,65.2) {\scriptsize$U$} 
  \put(84,27.8)   {$f =$} 
  \put(121,29.8)  {\scriptsize$f$} 
  \put(120.5,-6.6){\scriptsize$U$} 
  \put(120.8,65.2){\scriptsize$V$} 
  \epicture17 \labl{morph}
We follow the convention that such pictures are read from bottom to top. 
Because of $\End(\one)\eq\complex$ we have $\id_\smallone\eq1\iN\complex$;
accordingly, lines labelled by the tensor unit $\one$ can and will be omitted,
so that in the pictorial description morphisms in $\Hom(\one,U)$ or 
$\Hom(U,\one)$ emerge from and disappear into `nothing', \resp.
Composition of morphisms amounts to concatenation of lines, while the 
tensor product corresponds to juxtaposition:
  \bea  \begin{picture}(292,76)(0,29)
  \put(0,0)     {\begin{picture}(0,0)(0,0)
                \scalebox{.38}{\includegraphics{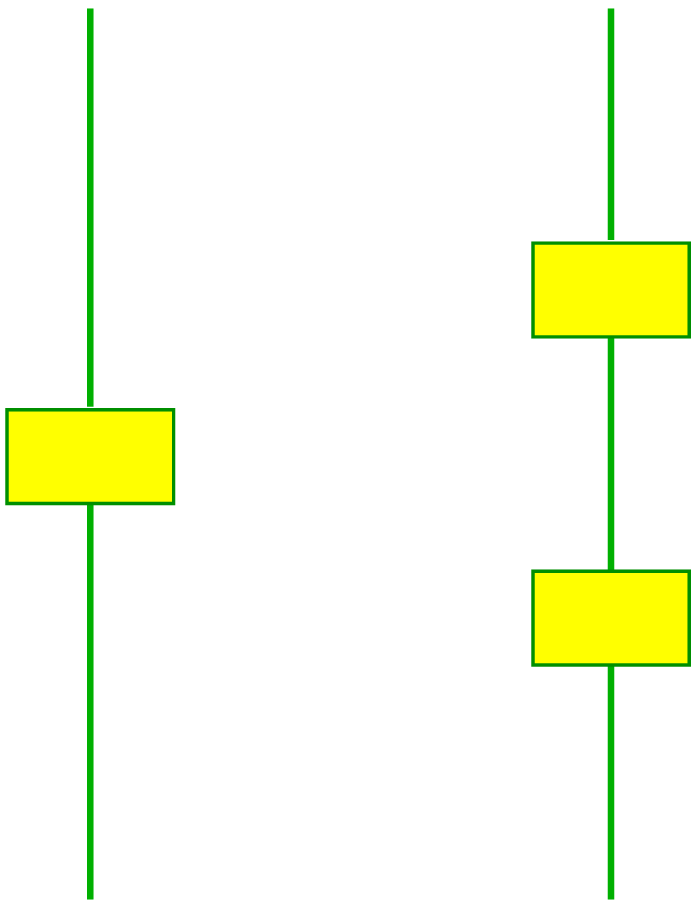}} \end{picture}}
  \put(1.6,47.7)  {\scriptsize$g\cir f$} 
  \put(6.1,-7.2)  {\scriptsize$U$} 
  \put(6.2,102.1) {\scriptsize$W$} 
  \put(35,47)     {$=$} 
  \put(63.4,-7.2) {\scriptsize$U$} 
  \put(64.2,29.8) {\scriptsize$f$} 
  \put(69,47.6)   {\scriptsize$V$} 
  \put(64.5,66.8) {\scriptsize$g$} 
  \put(63.4,102.1){\scriptsize$W$} 
  \put(121,47)    {and}
  \put(182,17)  {\begin{picture}(0,0)(0,0)
                \scalebox{.38}{\includegraphics{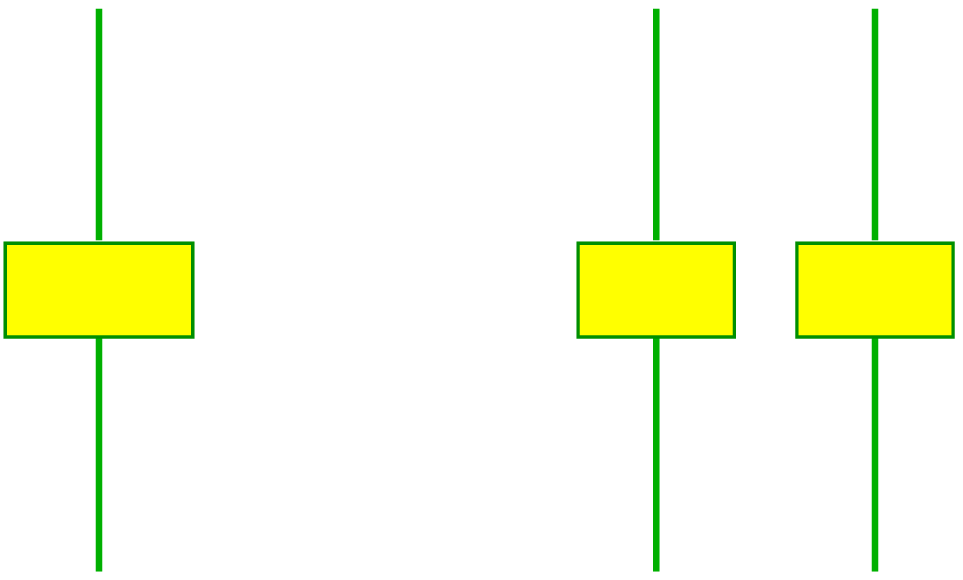}} \end{picture}}
  \put(183.3,10)  {\scriptsize$U\Oti V$} 
  \put(183.7,47)  {\scriptsize$f\oti g$} 
  \put(182.5,83.3){\scriptsize$Y\Oti Z$} 
  \put(223,47)    {$=$} 
  \put(250.9,9.5) {\scriptsize$U$} 
  \put(251.7,47)  {\scriptsize$f$} 
  \put(251.3,83.3){\scriptsize$Y$} 
  \put(275.5,9.5) {\scriptsize$V$} 
  \put(276.7,48.7){\scriptsize$g$} 
  \put(275.5,83.3){\scriptsize$Z$} 
  \epicture15 \labl{jux}

\noindent\nxt~% 
Second, \calc\ is a {\em ribbon category\/}, that is 
\cite[chapter\,XIV.3]{KAss}, a strict \tc\ supplemented with three additional 
ingredients: a duality, a braiding, and a twist.
A (right) {\em duality\/} on a tensor%
 \foodnode{Often the existence of a duality is included in the definition of
 the term `tensor category'. What we refer to as a tensor category here is
 then called a {\em monoidal\/} category.}
category \calc\ associates to 
every $U\iN\obj(\calc)$ another object $U\Vee{\in}\,\obj(\calc)$ and morphisms
  \be  b_U \in\Hom(\one,U\Oti U\Vee)\,, \qquad
  d_U \in \Hom(U\Vee\Oti U,\one) \,,  \Labl1d
and to every morphism $f\iN\Hom(U,Y)$ the morphism
  \be  f\Vee := (d_Y\oti\id_{U\Vee}) \circ (\id_{Y\Vee}\oti f\oti\id_{U\Vee})
  \circ (\id_{Y\Vee}\oti b_U) \ \in \Hom(Y\Vee\!{,}\,U\Vee) \,.  \ee
$U\Vee$ is called the object (right-)\,dual to $U$,
and $f\Vee$ the morphism (right-)\,dual to $f$; the duality morphisms $d_U$ 
and $b_U$ are also known as the evaluation and co-evaluation morphisms, \resp.
A {\em braiding\/} on a tensor category \calc\ allows one to `exchange' objects;
it consists of a family of isomorphisms $c_{U,V}\iN\Hom(U\Oti V,V\Oti U)$,
one for each pair $U,V\iN\objc$. Finally, a {\em twist\/} is a family of 
isomorphisms $\theta_U$, one for each $U\iN\objc$.
Graphically, the braiding and twist and their inverses and the duality will be 
depicted as follows:
  \bea \begin{picture}(400,155)(25,0)
              \put(0,96){\begin{picture}(0,0)(0,0)
  \put(0,0)     {\begin{picture}(0,0)(0,0)
                \scalebox{.38}{\includegraphics{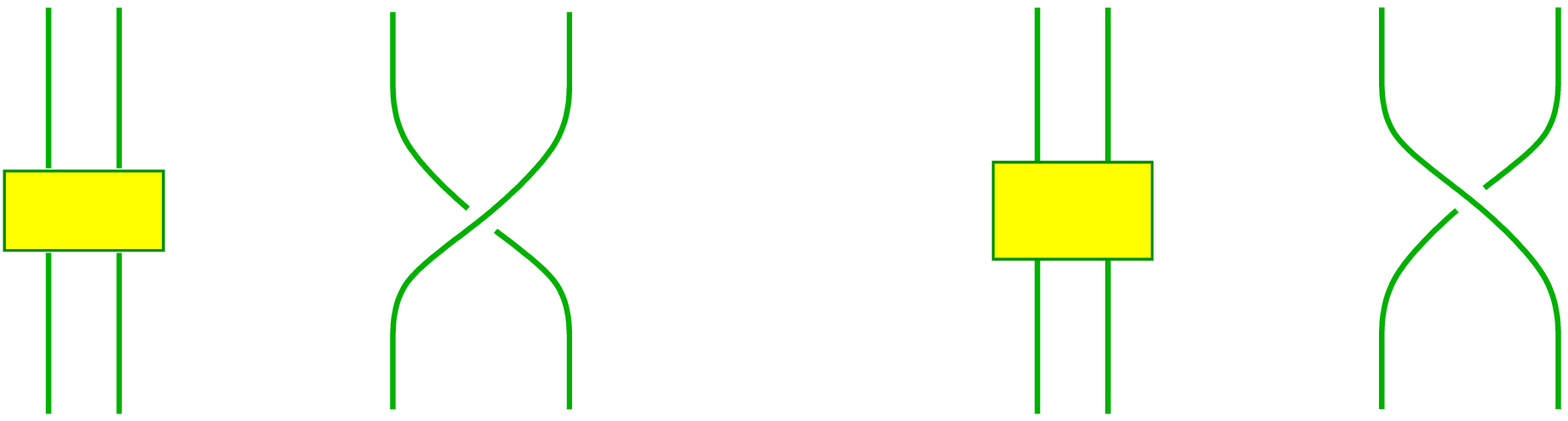}} \end{picture}}
  \put(2.2,-7)    {\scriptsize$U$} 
  \put(2.9,55.9)  {\scriptsize$V$} 
  \put(3,26.7)    {\scriptsize$c_{U,V}$}
  \put(12.7,-7)   {\scriptsize$V$} 
  \put(13.2,55.9) {\scriptsize$U$} 
  \put(32,24.5)   {$=$} 
  \put(48,-7)     {\scriptsize$U$} 
  \put(48,55.9)   {\scriptsize$V$} 
  \put(71.3,-7)   {\scriptsize$V$} 
  \put(71.3,55.9) {\scriptsize$U$} 
  \put(131.5,-7)  {\scriptsize$U$} 
  \put(140.9,-7)  {\scriptsize$V$} 
  \put(131.4,25.2){\scriptsize$c_{V,U}^{-1}$}
  \put(132.2,55.9){\scriptsize$V$} 
  \put(140.9,55.9){\scriptsize$U$} 
  \put(159,24.5)  {$=$} 
  \put(175.7,-7)  {\scriptsize$U$} 
  \put(175.7,55.9){\scriptsize$V$} 
  \put(198.6,-7)  {\scriptsize$V$} 
  \put(199.3,55.9){\scriptsize$U$} 
              \end{picture}}
              \put(265,96){\begin{picture}(0,0)(0,0)
  \put(0,0)     {\begin{picture}(0,0)(0,0)
                \scalebox{.38}{\includegraphics{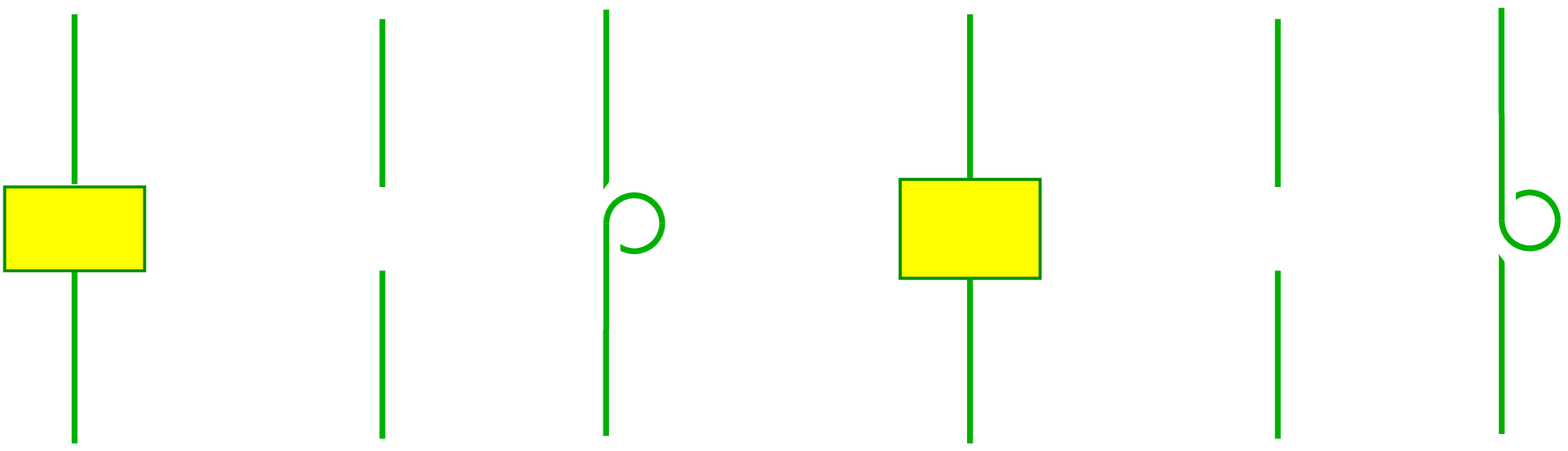}} \end{picture}}
  \put(5.5,-7)    {\scriptsize$U$} 
  \put(5.5,25.5)  {\scriptsize$\theta_U$}
  \put(5.5,55.9)  {\scriptsize$U$} 
  \put(26,24.6)   {$=$} 
  \put(45,24.6)   {\scriptsize$\theta$} 
  \put(44,-7)     {\scriptsize$U$} 
  \put(44,55.9)   {\scriptsize$U$} 
  \put(58,24.6)   {$=$} 
  \put(72,-7)     {\scriptsize$U$} 
  \put(72,55.9)   {\scriptsize$U$} 
  \put(112,25.6)  {\scriptsize$\theta_U^{-1}$}
  \put(116,-7)    {\scriptsize$U$} 
  \put(116,55.9)  {\scriptsize$U$} 
  \put(136,24.6)  {$=$} 
  \put(153,24.6)  {\scriptsize$\theta^{-1}$} 
  \put(154,-7)    {\scriptsize$U$} 
  \put(154,55.9)  {\scriptsize$U$} 
  \put(170,24.6)  {$=$} 
  \put(180,-7)    {\scriptsize$U$} 
  \put(180,55.9)  {\scriptsize$U$} 
              \end{picture}}
              \put(0,0){\begin{picture}(0,0)(0,0)
  \put(0,0)     {\begin{picture}(0,0)(0,0)
                \scalebox{.38}{\includegraphics{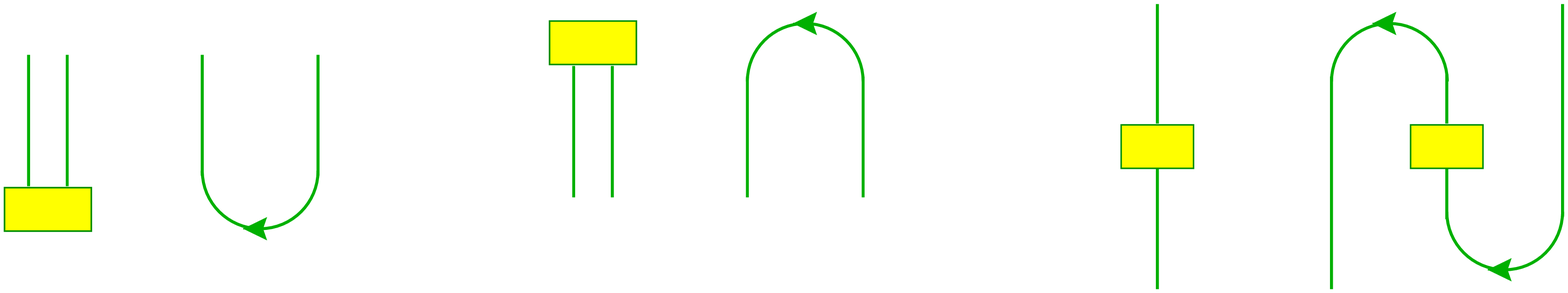}} \end{picture}}
  \put(6.2,17.5)  {\scriptsize$b_U$}
  \put(3.8,59)    {\scriptsize$U$} 
  \put(12.5,59)   {\scriptsize$U^\vee$} 
  \put(29,34)     {$=$} 
  \put(46.8,59)   {\scriptsize$U$} 
  \put(71,59)     {\scriptsize$U^\vee$} 
  \put(128,12)    {\scriptsize$U^\vee$} 
  \put(135.8,57.3){\scriptsize$d_U$}
  \put(142.2,12)  {\scriptsize$U$} 
  \put(158,40)    {$=$} 
  \put(171.5,12)  {\scriptsize$U^\vee$} 
  \put(201,12)    {\scriptsize$U$} 
  \put(270,-7.6)  {\scriptsize$V^\vee$} 
  \put(268.9,32.0){\scriptsize$f^\vee$}
  \put(270,70.3)  {\scriptsize$U^\vee$} 
  \put(294,32)    {$=$} 
  \put(310,-7.6)  {\scriptsize$V^\vee$} 
  \put(339.6,32.2){\scriptsize$f$}
  \put(364,70.3)  {\scriptsize$U^\vee$} 
              \end{picture}}
  \epicture{-1}3 \labl{limos}
Actually we should think of these morphisms as ribbons rather than lines
-- this is the reason for the terminology. For example, the twist $\theta$, 
braiding $c$ and duality morphism $b$ are drawn as
  \bea \begin{picture}(280,56)(0,27)
  \put(0,0)     {\begin{picture}(0,0)(0,0)
                \scalebox{.38}{\includegraphics{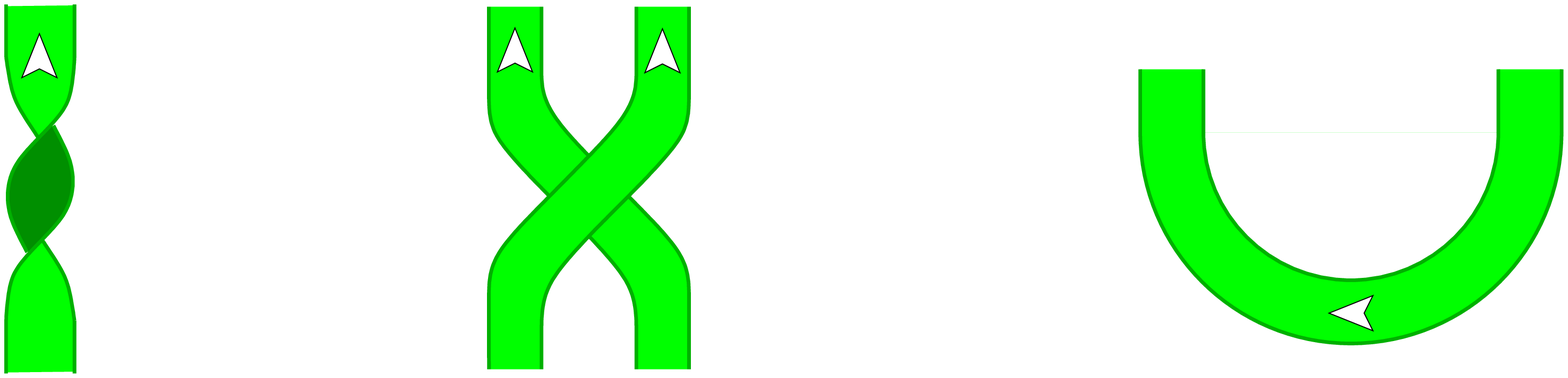}} \end{picture}}
  \put(-39,46)   {$\theta_U\,=$} 
  \put(5,20)     {\scriptsize$U$} 
  \put(53,46)    {$c_{U,V}^{}\,=$} 
  \put(102.5,20) {\scriptsize$U$} 
  \put(133,20)   {\scriptsize$V$} 
  \put(193,46)   {$b_U\,=$} 
  \put(237,60)   {\scriptsize$U$} 
  \epicture-5 \labl{ribbon}
In the sequel the interpretation of graphs with lines as ribbon graphs
will be implicit.
The duality, braiding and twist are subject to a number of consistency
conditions, which precisely guarantee that the visualisation via ribbons 
is appropriate, so that in particular the graphs obtained by their 
composition share the properties of the correspondingly glued ribbons.
More concretely, one has to impose duality identities, functoriality
and tensoriality of the braiding, functoriality of the twist, and 
compatibility of the twist with duality and with braiding.
In the notation of \erf{limos} these properties look as follows:
  \begin{eqnarray}
  &&  \begin{picture}(400,85)(8,0)
              \put(0,0){\begin{picture}(0,0)(0,0)
  \put(0,0)     {\begin{picture}(0,0)(0,0)
                \scalebox{.38}{\includegraphics{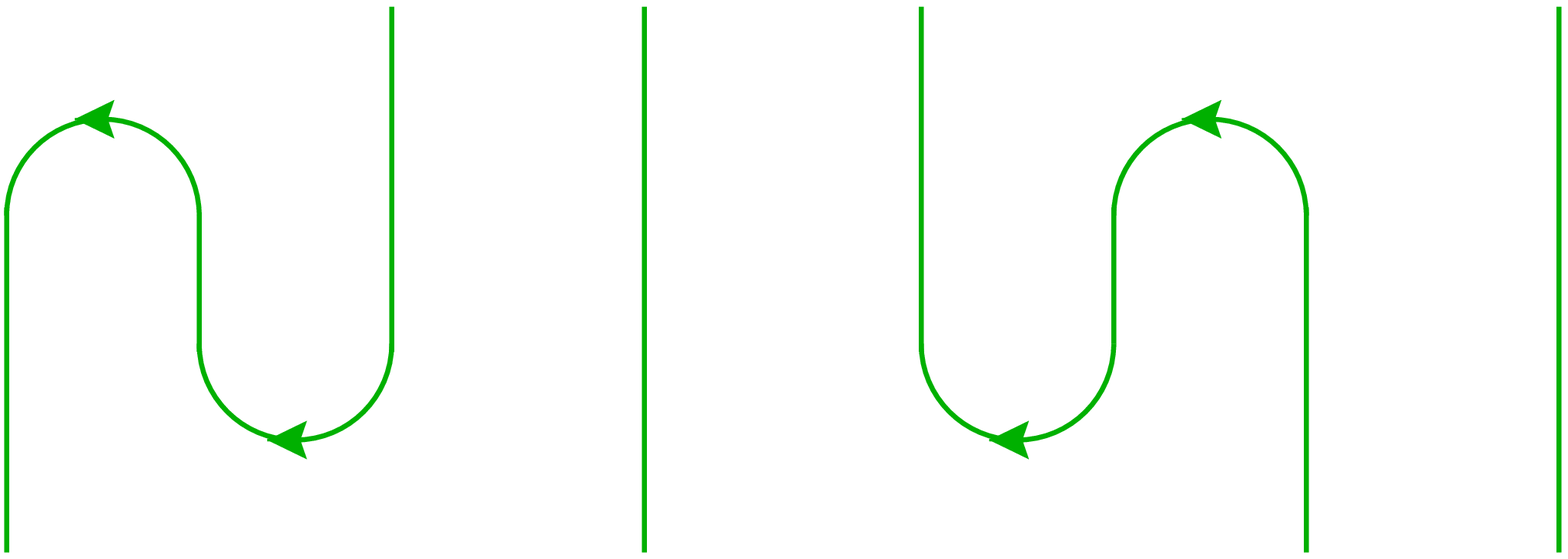}} \end{picture}}
  \put(-3.2,-7)  {\scriptsize$U\Vee$} 
  \put(51,82.8)  {\scriptsize$U\Vee$} 
  \put(69,39)    {$=$} 
  \put(88,-7)    {\scriptsize$U\Vee$} 
  \put(88,82.8)  {\scriptsize$U\Vee$} 
  \put(129,82.8) {\scriptsize$U$} 
  \put(182.4,-7) {\scriptsize$U$} 
  \put(200,39)   {$=$} 
  \put(218.5,-7) {\scriptsize$U$} 
  \put(218.5,82.8){\scriptsize$U$} 
  \put(58,-34)   {\footnotesize\Fbox{axioms for (right-) duality}}
              \end{picture}}
              \put(290,0){\begin{picture}(0,0)(0,0)
  \put(0,10)    {\begin{picture}(0,0)(0,0)
                \scalebox{.38}{\includegraphics{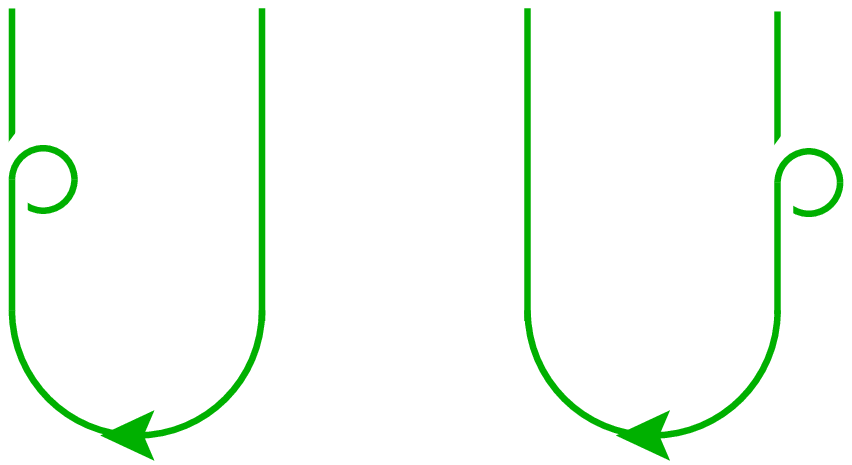}} \end{picture}}
  \put(.9,62.8)  {\scriptsize$U$} 
  \put(26,62.8)   {\scriptsize$U\Vee$} 
  \put(41,36)     {$=$} 
  \put(57.4,62.8) {\scriptsize$U$} 
  \put(82.5,62.8) {\scriptsize$U\Vee$} 
  \put(3,-34)    {\footnotesize\Fbox{\begin{tabular}c{$\!\!$duality and twist:$\!\!$}
                                \\$\theta_{U\Vee}=(\theta_U)\Vee$\end{tabular}}}
              \end{picture}}
  \end{picture}
  \nonumber\\
  &&  \begin{picture}(400,120)(8,0)
              \put(0,0){\begin{picture}(0,0)(0,0) 
  \put(0,0)     {\begin{picture}(0,0)(0,0)
                \scalebox{.38}{\includegraphics{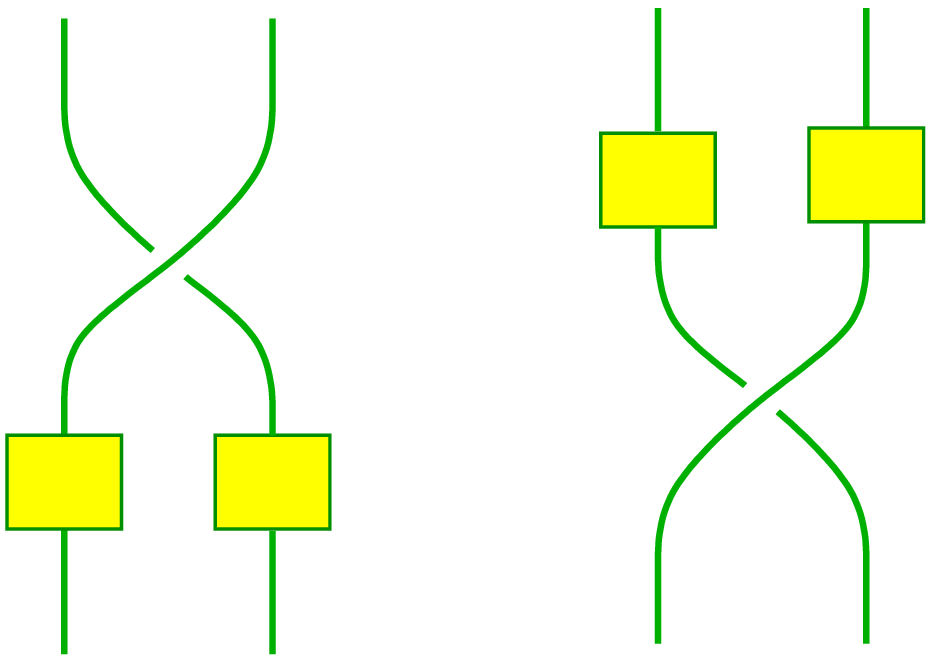}} \end{picture}}
  \put(2.9,-7)   {\scriptsize$U$} 
  \put(3.9,17.8) {\scriptsize$f$} 
  \put(3.3,73)   {\scriptsize$X$} 
  \put(-3.1,44)  {\scriptsize$c_{Y,X}^{}$} 
  \put(26.8,-7)  {\scriptsize$V$} 
  \put(27.4,18.7){\scriptsize$g$} 
  \put(27.3,73)  {\scriptsize$Y$} 
  \put(49,31)    {$=$} 
  \put(69.4,-7)  {\scriptsize$U$} 
  \put(70.2,51.7){\scriptsize$g$} 
  \put(69.4,74)  {\scriptsize$X$} 
  \put(89,27.3)  {\scriptsize$c_{U,V}^{}$} 
  \put(91.8,-7)  {\scriptsize$V$} 
  \put(92.6,51.4){\scriptsize$f$} 
  \put(93.3,74)  {\scriptsize$Y$} 
  \put(-8,-34)   {\footnotesize\Fbox{functoriality of braiding}}
              \end{picture}}
              \put(200,0){\begin{picture}(0,0)(0,0)
  \put(0,0)     {\begin{picture}(0,0)(0,0)
                \scalebox{.38}{\includegraphics{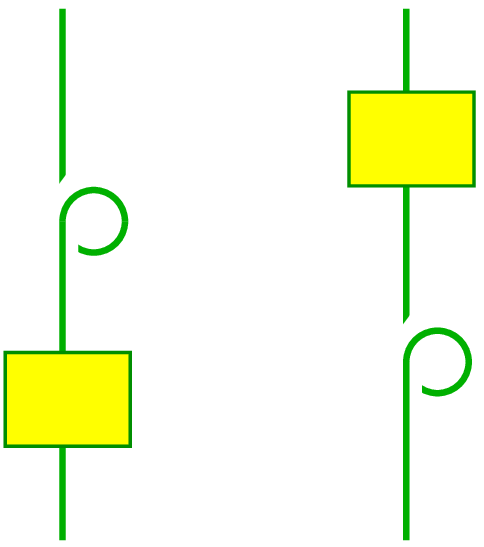}} \end{picture}}
  \put(4.4,-7)   {\scriptsize$U$} 
  \put(4.9,14.2) {\scriptsize$f$} 
  \put(3.8,63)   {\scriptsize$V$} 
  \put(22,26)    {$=$} 
  \put(42.3,-7)  {\scriptsize$U$} 
  \put(43.0,42.7){\scriptsize$f$} 
  \put(41.8,63)  {\scriptsize$V$} 
  \put(-16,-34)  {\footnotesize\Fbox{functoriality of twist}}
              \end{picture}}
  \end{picture} %%\nonumber \\[-4.6em]{}
  \label{ribax1} %% \\[3.9em]{}\nonumber
  \end{eqnarray}
  \bea  \begin{picture}(360,95)(0,25)
              \put(0,0){\begin{picture}(0,0)(0,0)
  \put(0,5)     {\begin{picture}(0,0)(0,0)
                \scalebox{.38}{\includegraphics{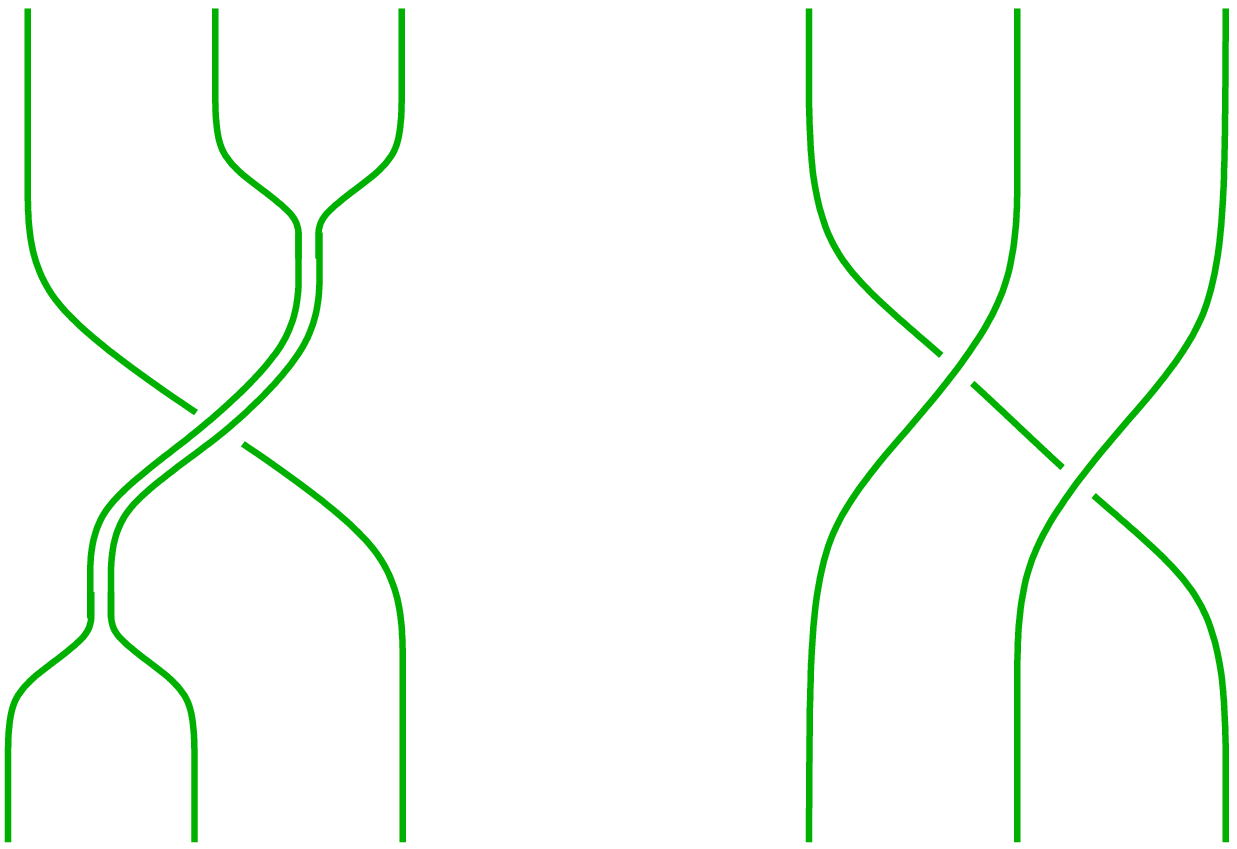}} \end{picture}}
  \put(-8.3,50.5) {\scriptsize$c_{U\otimes V,Y}^{\phantom x}$} 
  \put(-2.3,-2.5) {\scriptsize$U$} 
  \put(1.2,100.4) {\scriptsize$Y$} 
  \put(18.1,-2.5) {\scriptsize$V$} 
  \put(21,100.4)  {\scriptsize$U$} 
  \put(41.1,-2.5) {\scriptsize$Y$} 
  \put(42,100.4)  {\scriptsize$V$} 
  \put(60,48)     {$=$} 
  \put(83.0,56)   {\scriptsize$c_{U,Y}^{\phantom x}$} 
  \put(85.5,-2.5) {\scriptsize$U$} 
  \put(86.5,100.4){\scriptsize$Y$} 
  \put(108.8,-2.5){\scriptsize$V$} 
  \put(108.8,100.4){\scriptsize$U$} 
  \put(132.2,-2.5){\scriptsize$Y$} 
  \put(132.2,100.4){\scriptsize$V$} 
  \put(123.1,44)  {\scriptsize$c_{V,Y}^{\phantom x}$} 
  \put(11,-34)    {\footnotesize\Fbox{tensoriality of braiding}}
              \end{picture}}
              \put(220,0){\begin{picture}(0,0)(0,0)
  \put(0,0)     {\begin{picture}(0,0)(0,0)
                \scalebox{.38}{\includegraphics{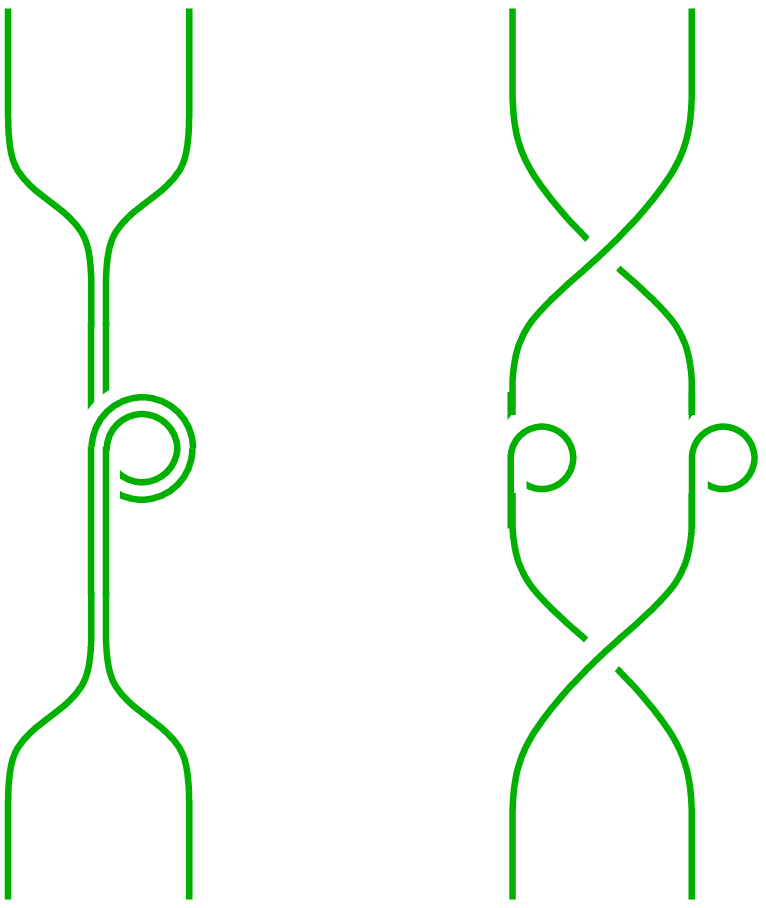}} \end{picture}}
  \put(-13.8,48) {\scriptsize$\theta_{U\otimes V}$} 
  \put(-2.5,-7)  {\scriptsize$U$} 
  \put(-2.2,101.3){\scriptsize$U$} 
  \put(17.4,-7)  {\scriptsize$V$} 
  \put(18.0,101.3){\scriptsize$V$} 
  \put(30,46)    {$=$} 
  \put(46.8,45)  {\scriptsize$\theta_{\!V}^{\phantom x}$} 
  \put(53.1,-7)  {\scriptsize$U$} 
  \put(53.8,101.3){\scriptsize$U$} 
  \put(72.2,-7)  {\scriptsize$V$} 
  \put(72.8,101.3){\scriptsize$V$} 
  \put(83.3,45)  {\scriptsize$\theta_{\!U}^{\phantom x}$} 
  \put(-5.4,-34) {\footnotesize\Fbox{braiding and twist}}
              \end{picture}}
  \epicture41 \labl{ribax2}

In a ribbon category there is automatically also a left duality; it is defined 
on objects by $\eev U\,{:=}\,U\Vee$ and left duality morphisms $\tilde b_U$ and 
$\tilde d_U$; the latter, as well as left-dual morphisms $\Eev f$, are given by
  \bea  \begin{picture}(410,54)(0,25)
  \put(0,0)   {\begin{picture}(0,0)(0,0)
              \scalebox{.38}{\includegraphics{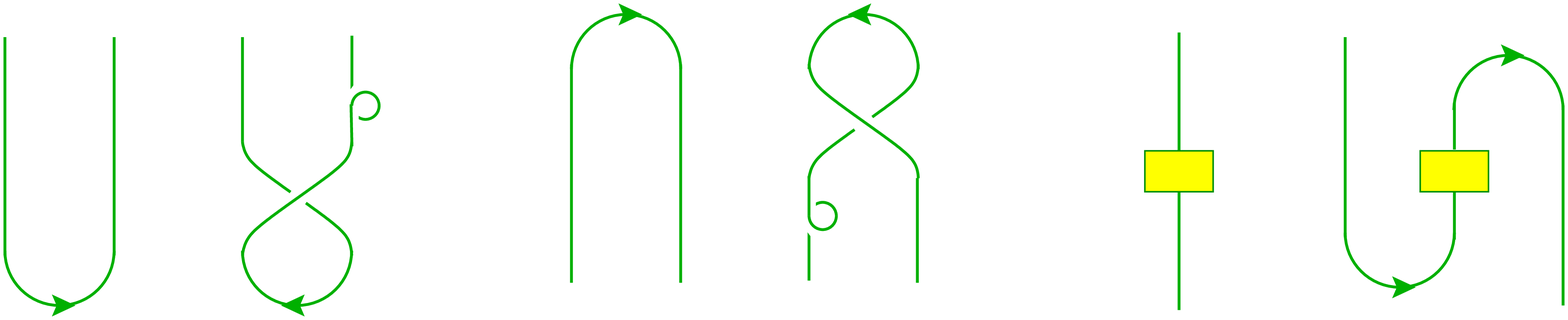}} \end{picture}}
  \put(-3.5,72)   {\scriptsize$\Eev U$} 
  \put(25.5,72)   {\scriptsize$U$} 
  \put(41.3,31)   {$=$} 
  \put(55.5,72)   {\scriptsize$\Eev U$} 
  \put(84.9,72)   {\scriptsize$U$} 
  \put(139.9,-1)  {\scriptsize$U$} 
  \put(165.5,-1)  {\scriptsize$\Eev U$} 
  \put(182.1,31)  {$=$} 
  \put(199.8,-1)  {\scriptsize$U$} 
  \put(225.5,-1)  {\scriptsize$\Eev U$} 
  \put(291.2,-8)  {\scriptsize$\Eev V$} 
  \put(291.2,73.2){\scriptsize$\Eev U$} 
  \put(291.9,33.5){\scriptsize$\Eev f$} 
  \put(316,33)    {$=$} 
  \put(335.2,72.9){\scriptsize$\Eev U$} 
  \put(362.9,34.2){\scriptsize$f$} 
  \put(388.6,-8)  {\scriptsize$\Eev V$} 
  \epicture13 \labl{left-dual}
One can check that this left duality coincides with the right duality not 
only on objects, but also on morphisms, i.e.\ $\Eev f\eq f\Vee$; categories
with a coinciding left and right duality are called {\em sovereign\/}.
It follows e.g.\ that the double dual $(U\Vee)^{\!\vee}_{}$ of an object 
$U$ is isomorphic (though in general not equal) to $U$.
In fact, a natural isomorphism between $(U\Vee)^{\!\vee}_{}$ and $U$ can be
obtained with the help of the twist (see e.g.\ \cite[Chapter\,2.2]{BAki}):
For any object $U$ we have a morphism $\delta_U\,{:=}\,\psi_U^{-1}\cir
\theta_U\iN\Hom(U,(U\Vee)^{\!\vee}_{})$, with
  \be  \psi_U := (\id_U \oti d_{U^\vee}) \cir (\id_U \oti c_{(U\Vee)
  ^{\!\vee}_{},U\Vee}^{-1}) \cir (b_U \oti \id_{(U\Vee)^{\!\vee}_{}})
  \,\in \Hom((U\Vee)^{\!\vee}_{},U) \,.  \ee
The properties of the twist $\theta_U$ are precisely such that these
morphisms are tensorial, i.e.\ satisfy $\delta_{V\otimes W}\eq\delta_V
\oti\delta_W$,
preserve the unit, $\delta_\one\eq\id_\one$, and are compatible with the 
duality in the sense that $\delta_{U^\vee}\eq{((\delta_U)^\vee)}^{-1}$.

Further, once we have two dualities, we can also define left and right traces of 
endomorphisms, via
  \bea \begin{picture}(107,37)(0,22)
  \put(0,0)   {\begin{picture}(0,0)(0,0)
              \scalebox{.38}{\includegraphics{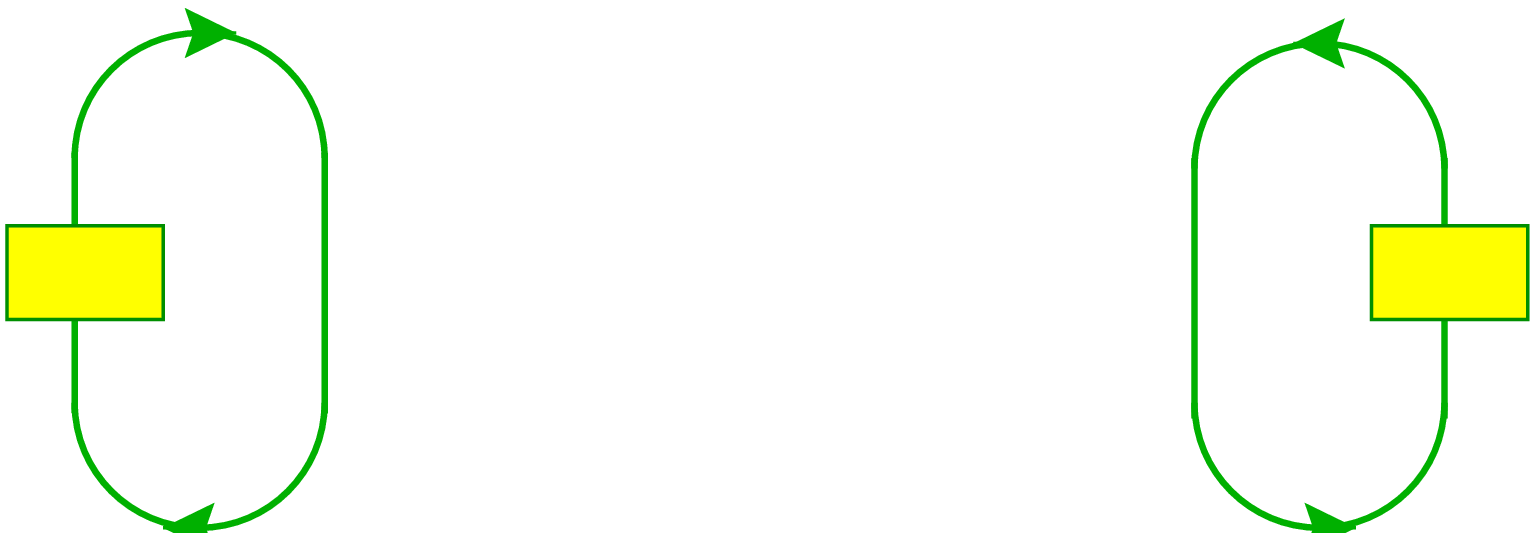}} \end{picture}}
  \put(-53.8,26) {${\rm tr}_{\rm r}(f)\;=$} 
  \put(7.4,26.7) {\scriptsize$f$} 
  \put(76.8,26)  {${\rm tr}_{\rm l}(f)\;=$} 
  \put(155.4,26.7){\scriptsize$f$} 
  \epicture07 \labl{trace}
Both traces are cyclic, 
  \be  {\rm tr}_{\rm l,r}(g\cir f) = {\rm tr}_{\rm l,r}(f\cir g)  \ee
and obey
  \be  {\rm tr}_{\rm l,r}(f\oti g)
  = {\rm tr}_{\rm l,r}(f)\,{\rm tr}_{\rm l,r}(g) \,.  \ee
In the case at hand, where the left duality is constructed from the right 
duality by \erf{left-dual}, the two notions of trace coincide; thus the 
category \calc\ is {\em spherical\/} \cite{bawe2}. The trace of the 
identity morphism is known as the {\em quantum dimension\/} of an object,
  \be  \dim(U) := {\rm tr}(\id_U) \,.  \ee
The quantum dimension
is additive under direct sums and multiplicative under tensor products.

For any self-dual object $U$ of a sovereign \tc\ and any isomorphism $f$ in
the space $\Hom(U,U^\vee)$ one introduces the endomorphism $\Nu_U\iN\Hom(U\Vee
\!{,}\,U\Vee)$ as (\hsp{-.34}\cite{fuSc16}, see also \cite{bant5,fgsv,fffs3})
  \bea \begin{picture}(29,70)(0,27)
  \put(0,0)   {\begin{picture}(0,0)(0,0)
              \scalebox{.38}{\includegraphics{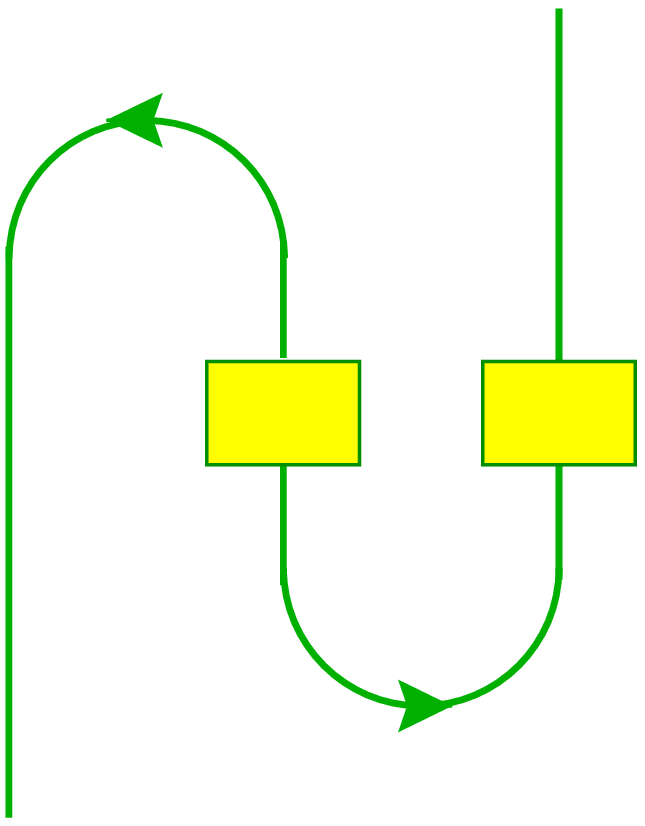}} \end{picture}}
  \put(-43.8,36)  {$\Nu_U^{}\, =$}
  \put(-3.4,-7)   {\scriptsize$U^\vee$} 
  \put(24.4,42.9) {\scriptsize$f^{-1}$} 
  \put(58.6,43.6) {\scriptsize$f$} 
  \put(56.4,93)   {\scriptsize$U^\vee$} 
  \epicture13 \labl{fsi}

One can show \cite{fuSc16} that $\Nu_U$ is in fact an automorphism, and that for 
simple self-dual $U$ it does not depend on $f$ and satisfies
  \be  \Nu_U = \nu_U^{}\, \id_{U\Vee}  \labl{fs}
with
  \be  \nu_U^{}\in\{\pm1\}\,. \labl{fs2}
The sign $\nu_U$ is called the {\em \fsi\/} of the object $U$.
In agreement with the terminology in the \rep\ theory of groups and Lie \alg s, 
self-dual objects $U$ with $\nu_U\eq1$ are called real (or orthogonal), while
those with $\nu_U\eq{-}1$ are called pseudo-real (or symplectic, or 
quaternionic).

\smallskip\noindent\nxt~% 
Third, \calc\ is a {\em modular tensor category\/} (or, briefly, {\em modular
category\/}), that is, a semisimple abelian ribbon category with ground field% 
  \foodnode{In the original definition of a \mtc\ \cite{TUra}, semisimplicity
  is replaced by dominance and instead of abelianness also only a weaker  
  property is imposed, nor is it required that the ground ring is \complex.
  But in the context of rational \cft\ \mtcs\ appear naturally in this more
  restricted form.}
\complex\ that has only a finite number of isomorphism classes of simple 
objects and a non-degenerate $s$-matrix.
\\
To explain the latter property, first note that in a modular tensor category
the family of simple objects that appears in the definition of dominance 
(see formula \erf{1dom}) can be chosen such that it contains precisely one object
out of each isomorphism class of simple objects and hence in particular is 
finite, $|\I|\,{<}\,\infty$. Also, the dual $U^\vee$ of any member $U$ of 
the family is isomorphic to another member $\Bar U$, and as representative 
of the class of the tensor unit $\one$ we choose $\one$ itself. Note that 
$\Bar U\eq U$ iff $U$ is self-dual; in particular, $\Bar{\one}\eq\one$.
{}From now on we denote the elements of the family by $U_i$ with 
$i\iN\{0,1,2,...\,, |\I|{-}1\}$ and set $U_0\eq\one$. 

One then defines an 
$|\I|\,{\times}\,|\I|$\,-matrix $s\eq(s_{i,j})^{}_{i,j\in\II}$ by
  \be  s_{i,j} := {\rm tr}(c_{U_i,U_j}^{}c_{U_j,U_i}^{}) \,,  \ee
or pictorially,
  \bea \begin{picture}(104,27)(0,27)
  \put(42,0)  {\begin{picture}(0,0)(0,0)
              \scalebox{.38}{\includegraphics{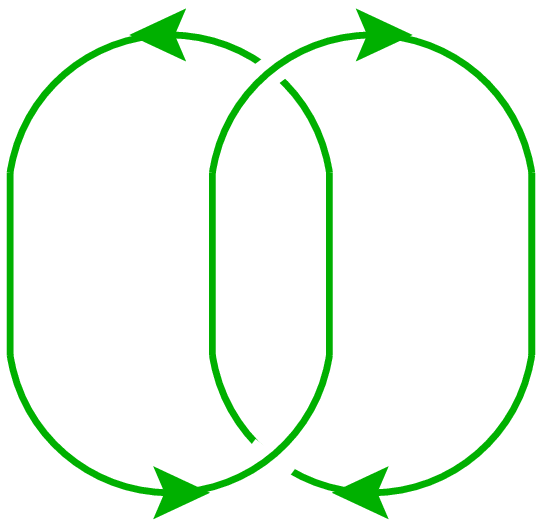}} \end{picture}}
  \put(0,25.3)  {$s_{i,j}\;=$} 
  \put(59.4,22.8){\scriptsize$i$} 
  \put(80.3,22.8){\scriptsize$j$}
  \epicture11 \labl{smat}
(Here and in the sequel we often simplify notation by writing the label $i$
in place of $U_i$.)
The final requirement to be imposed on \calc\ in order that it is a modular
category is that this square matrix $s$ is non-degenerate.
This latter property provides in fact an explanation of the qualification
`modular': When combined with the other axioms, it implies that the matrices
$s$ and $t\eq{\rm diag}(\theta_{U_i})$ (when multiplied with overall constants
that are expressible through the $\theta_{U_j}$ and the \q dimensions 
$\dim(U_j)$) generate a projective \rep\ of the modular group SL$(2,\zet)$.

The tensor product of objects induces on the set 
of isomorphism classes of \calc\ the structure of a commutative and
associative ring over the integers, called the Grothendieck ring $K_0(\calc)$
of \calc. (Conversely, the \tc\ \calc\ may be thought of as a categorification 
of the ring $K_0(\calc)$ \cite{beRn}.) A distinguished
basis of this ring is given by the isomorphism classes of the objects 
$U_i$ with $i\iN\I$. In this basis, the structure constants are the 
non-negative integers $\,\dim\,\Hom(U_i\oti U_j,U_k)$.  

\medskip

The mapping $U_k\,{\mapsto}\,\Bar U_k$ is an involution on the finite set 
\I, which induces an involution $k\,{\mapsto}\,\bar k$ on the set of labels by
$U_{\bar k}\eq\Bar U_k$. With this convention, $U_{\bar k}^\vee$ is isomorphic
to $U_k$, for every $k\iN\I$; let us then fix an isomorphism
  \be  \pi_k \in \Hom(U_k,U_{\Bar k}^\vee)  \labl{def-pik}
for each $k\iN\I$, and for each pair $k,\bar k$ with $k\,{\ne}\,\bar k$ 
perform the choice in such a way that the number $p_k$ defined by
  \bea \begin{picture}(174,50)(0,34)
  \put(0,0)   {\begin{picture}(0,0)(0,0)
              \scalebox{.38}{\includegraphics{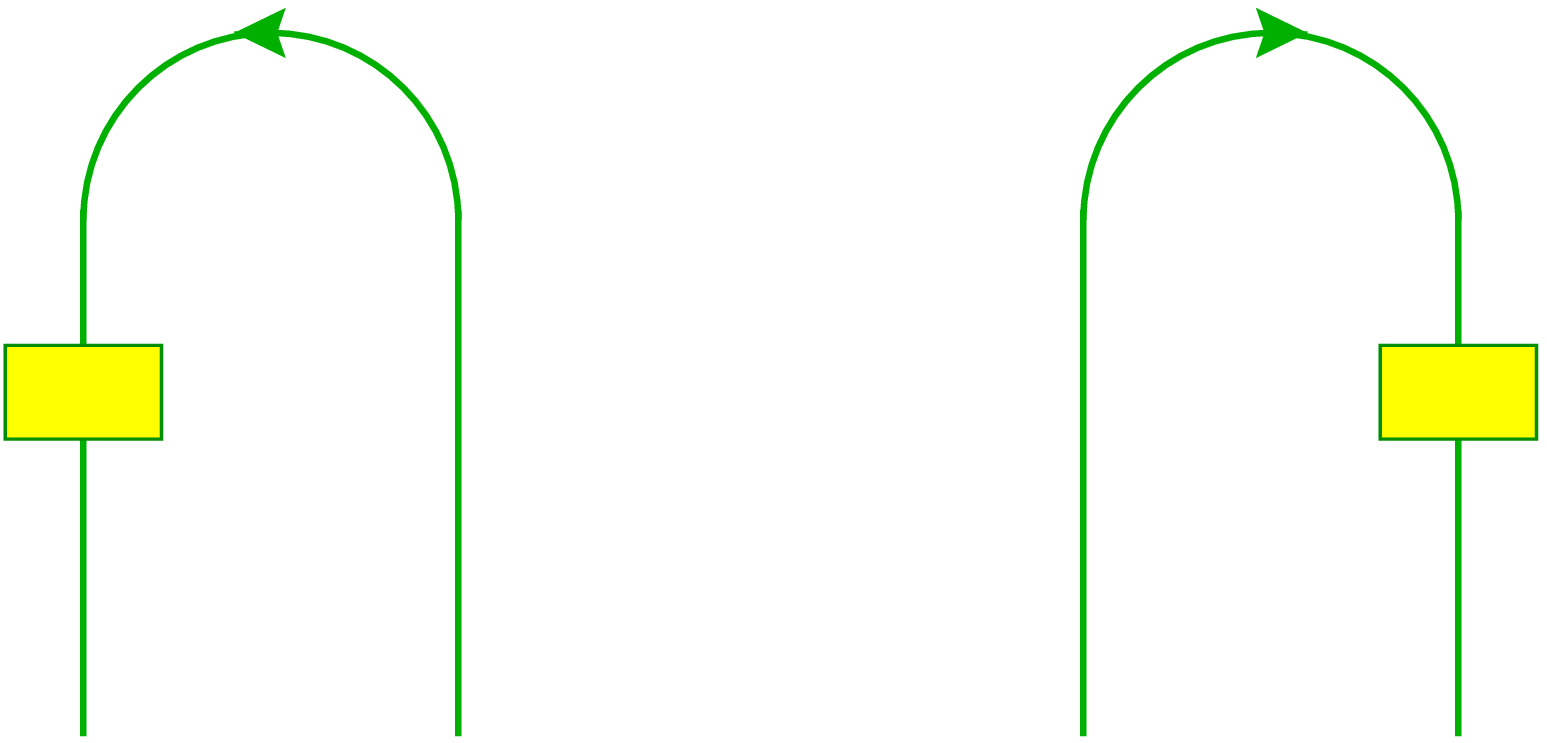}} \end{picture}}
  \put(6.5,-8.2){\scriptsize$k$}
% \put(-1,60)   {\scriptsize$\bar k^\vee$}
  \put(48,-8.2) {\scriptsize$\bar k$}
  \put(4.6,37.6){\scriptsize$\pi_k$}
  \put(73,36)   {$=\ \ p_k$}
  \put(155.6,37.6){\scriptsize$\pi_{\bar k}$}
  \put(116.5,-8.2){\scriptsize$k$}
% \put(162,60)  {\scriptsize$k^\vee$}
  \put(157,-8.2){\scriptsize$\bar k$}
  \epicture25 \labl{def-pi}
is equal to 1. Using sovereignty of $\calc$,
one can then show that for these values of $k$ we have 
consistently $p_{\bar k}\eq1$, too. In contrast, for the self-dual objects 
$U_k\eq U_{\bar k}$ in the family, we are not free to choose the number 
$p_k$; rather, it follows directly from its definition in \erf{def-pi} 
that $p_k$ coincides with the \fsi\ \erf{fs2} of $U_k$:
  \be  p_k= \nu_k  \qquad {\rm for}\ \ U_k\,{\cong\,}U_k^\vee\,. \labl{eq:p=nu}

A particularly simple example of a modular 
tensor category, which can serve as a guide to the general theory, is the
category \V\ of \findim\ vector spaces over the complex numbers. 
The category \V\ has a single isomorphism class of simple objects -- the 
class of the one-dimensional vector space \complex\ -- and has trivial twist
and braiding. In \cft, this category arises for meromorphic models, i.e.\ 
models with a single primary field, such as the $E_8$ WZW theory at level 1.  

\subsection{Fusing and braiding matrices}

Let us now explain the meaning of the various properties of a modular category 
in the \cft\ context. The (simple) objects of \calc\ are the (irreducible) \rep s
of the chiral \alg\ \CA, and the morphisms of \calc\ are \CA-intertwiners. 
The tensor product is the (fusion) tensor product of \CA-\rep s, with the
tensor unit given by the vacuum \rep\ (identity field and its descendants).
Thus the isomorphism classes of simple objects correspond to the primary \cvo s,
and the Grothendieck ring of \calc\ is the fusion ring of the \cft.

The duality in \calc\ encodes the existence of conjugate \CA-representations,
and the twist is determined by the fractional part of the conformal weight: 
  \be  \theta_U = \exp(-2\pi\ii\Delta_U)\,\id_U   \ee
for simple objects $U$. The braiding of \calc\ accounts for the presence of 
braid group statistics (see e.g.\ \cite{Froh,Schr,frrs,long3}) in two 
dimensions, and the matrix $s$ coincides, up to normalisation, with the 
modular S-matrix of the CFT that implements the modular transformation 
$\tau\,{\mapsto}\,{-}\frac1\tau$ on the characters of primary fields:
  \be  s_{i,j} = S_{i,j} / S_{0,0} \,.  \ee
(Conversely, $S_{0,0}$ and thereby $S$ is recovered from the data of the \mtc\
by requiring $S\eq S_{0,0}\,s$ to be unitary.)
In terms of $s$, the quantum dimensions are
  \be  \dim(U_i) = s_{i,0} = S_{i,0} / S_{0,0} \,.  \ee

All the axioms of \calc\ can be understood as formalisations of properties of 
primary \cvo s in rational CFT. Often such properties are presented in a
form where explicit basis choices in the three-point coupling spaces 
have been made. To make contact with such a formulation we fix%
 \foodnode{Our strategy is to keep these bases as general as possible. 
 More specific basis choices can be interesting for purposes different from
 ours, e.g.\ for an efficient numerical encoding of the defining data
 of a modular category as discussed in \cite{bobt,boqu}.} 
once and for all bases $\{\x ijk\alpha\}$ in the coupling spaces 
$\Hom(U_i\oti U_j,U_k)$, as well as dual bases $\{\y ijk\alpha\}$ in 
$\Hom(U_k,U_i\oti U_j)$. We depict the basis morphisms as follows:
  \bea  \begin{picture}(215,44)(0,32)
  \put(0,0)   {\begin{picture}(0,0)(0,0)
              \scalebox{.38}{\includegraphics{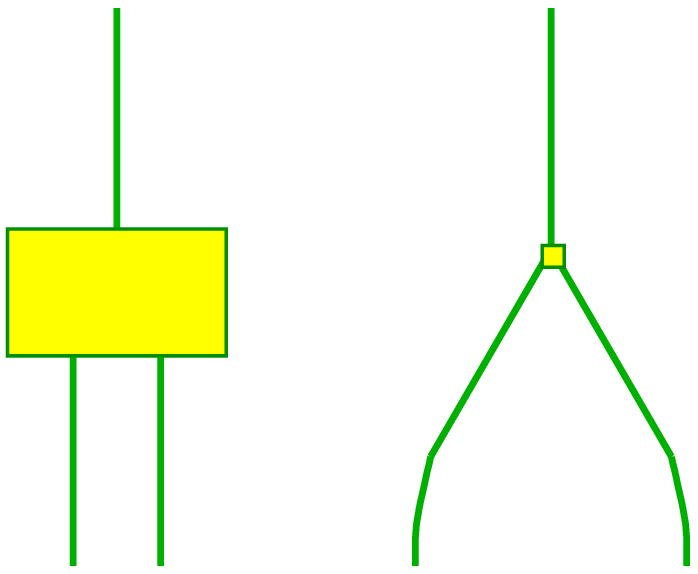}} \end{picture}}
  \put(1.1,31.0) {\scriptsize$\x ijk\alpha$}
  \put(5.8,-7)   {\scriptsize$i$}
  \put(10.7,65.3){\scriptsize$k$}
  \put(15.1,-7)  {\scriptsize$j$}
  \put(35.2,28)  {$=$}
  \put(43.8,-7)  {\scriptsize$i$}
  \put(58.7,65.3){\scriptsize$k$}
  \put(64.2,33.6){\tiny$\alpha$}
  \put(73.1,-7)  {\scriptsize$j$}
  \put(140,0) {\begin{picture}(0,0)(0,0)
              \scalebox{.38}{\includegraphics{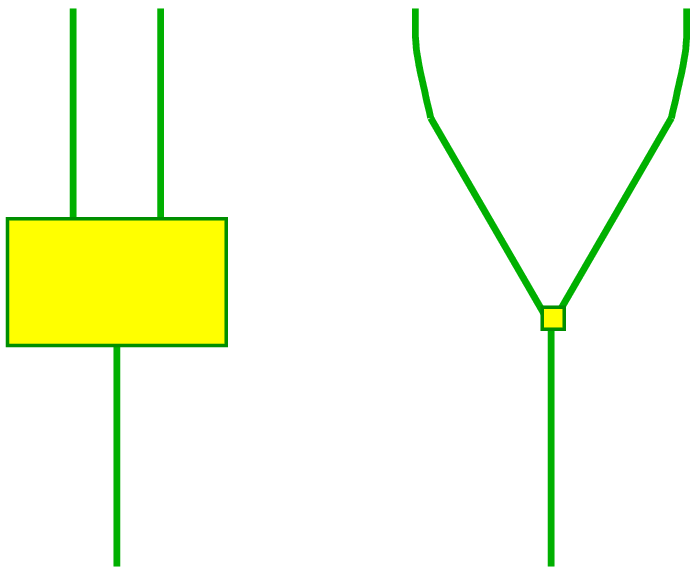}} \end{picture}}
  \put(140.9,29.2){\scriptsize$\y ijk\alpha$}
  \put(146.2,65.3){\scriptsize$i$}
  \put(150.3,-7) {\scriptsize$k$}
  \put(156.6,66.2){\scriptsize$j$}
  \put(176.2,30) {$=$}
  \put(183.3,65.3){\scriptsize$i$}
  \put(198.2,-7) {\scriptsize$k$}
  \put(203.4,27.3){\tiny$\bar\alpha$}
  \put(214.6,66.2){\scriptsize$j$}
  \epicture24 \labl{xijk,yijk}
Duality of the bases means that
  \bea  \begin{picture}(89,63)(0,40)
  \put(0,0)   {\begin{picture}(0,0)(0,0)
              \scalebox{.38}{\includegraphics{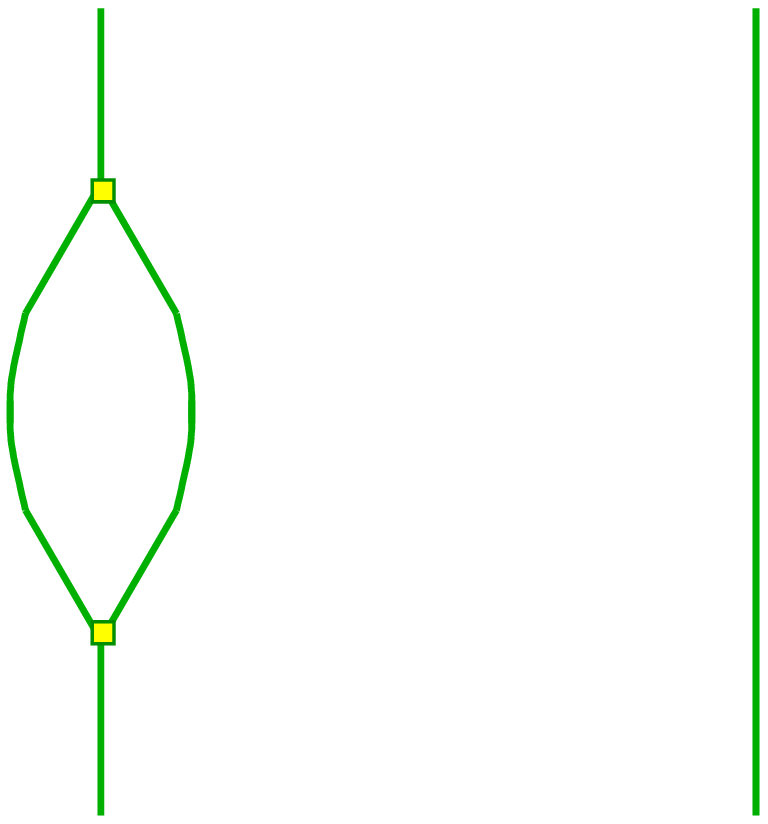}} \end{picture}}
  \put(-4.4,44)  {\scriptsize$i$}
  \put(4.4,20)   {\tiny$\alpha$}
  \put(4.4,67.8) {\tiny$\bar\beta$}
  \put(9.3,-7)   {\scriptsize$k$}
  \put(9.6,92.8) {\scriptsize$k$}
  \put(23.1,44)  {\scriptsize$j$}
  \put(38.2,44)  {$=\ \delta_{\alpha,\beta}$}
  \put(81.2,-7)  {\scriptsize$k$}
  \put(81.6,92.8){\scriptsize$k$}
  \epicture29 \labl{xy-dual}
By the dominance property of \calc\ we also have the completeness relation
  \bea \begin{picture}(150,69)(0,39)
  \put(0,0)   {\begin{picture}(0,0)(0,0)
              \scalebox{.38}{\includegraphics{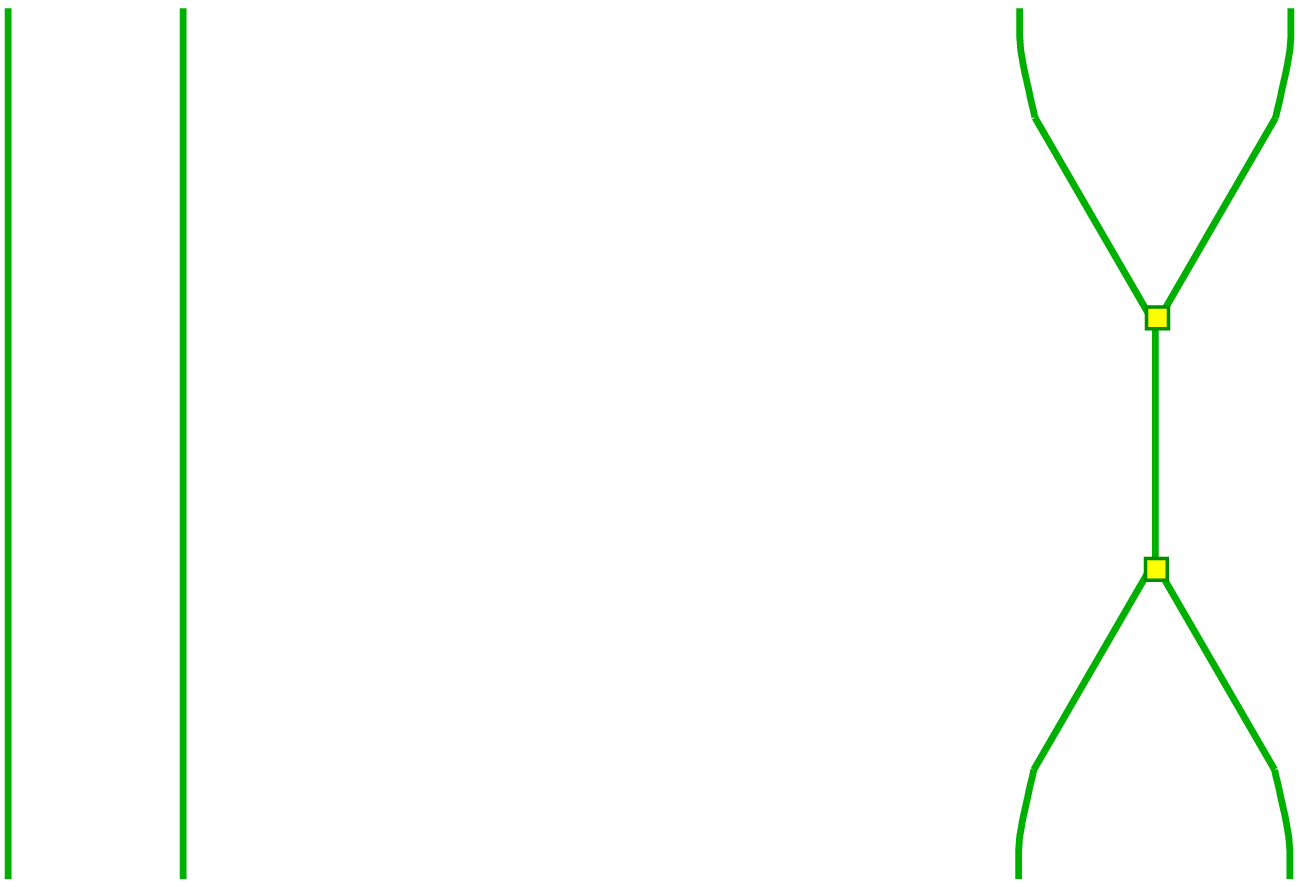}} \end{picture}}
  \put(-0.9,-7)  {\scriptsize$i$}
  \put(-0.6,100) {\scriptsize$i$}
  \put(17.5,-7)  {\scriptsize$j$}
  \put(17.8,101) {\scriptsize$j$}
  \put(41.2,47.2){$=\ \dsty\sum_{k\in\mathcal I}\sum_{\gamma}$}
  \put(110.8,-7) {\scriptsize$i$}
  \put(110.8,100){\scriptsize$i$}
  \put(120.1,45.7){\scriptsize$k$}
  \put(129.2,33.5){\tiny$\gamma$}
  \put(129.2,61.6){\tiny$\bar\gamma$}
  \put(139.1,-7) {\scriptsize$j$}
  \put(139.4,101){\scriptsize$j$}
  \epicture28 \labl{xy-comp}
When all three labels $i,j,k$ are generic, no basis in $\Hom(U_k,U_i\oti U_j)$
or $\Hom(U_i\oti U_j,U_k)$ is distinguished. In contrast, when one of the
labels equals 0, i.e.\ when one of the simple objects involved is the tensor 
unit $\one$, then it is natural to make the choice
  \be  \x i0i\circ = \x 0ii\circ = \id_{U_i} = \y i0i\circ = \y 0ii\circ
  \,, \labl{llyy}
(which is possible due to strictness of the \tc\ \calc). Here we have 
used the symbol $\circ$ in order to indicate that the coupling label can
take only a single value. In the sequel, for notational simplicity we suppress
such unique labels, both in the formulas and in the pictures. Thus e.g.\ the
pictorial form of the relation \erf{llyy} is
  \bea \begin{picture}(265,45)(0,28)
  \put(0,0)   {\begin{picture}(0,0)(0,0)
              \scalebox{.38}{\includegraphics{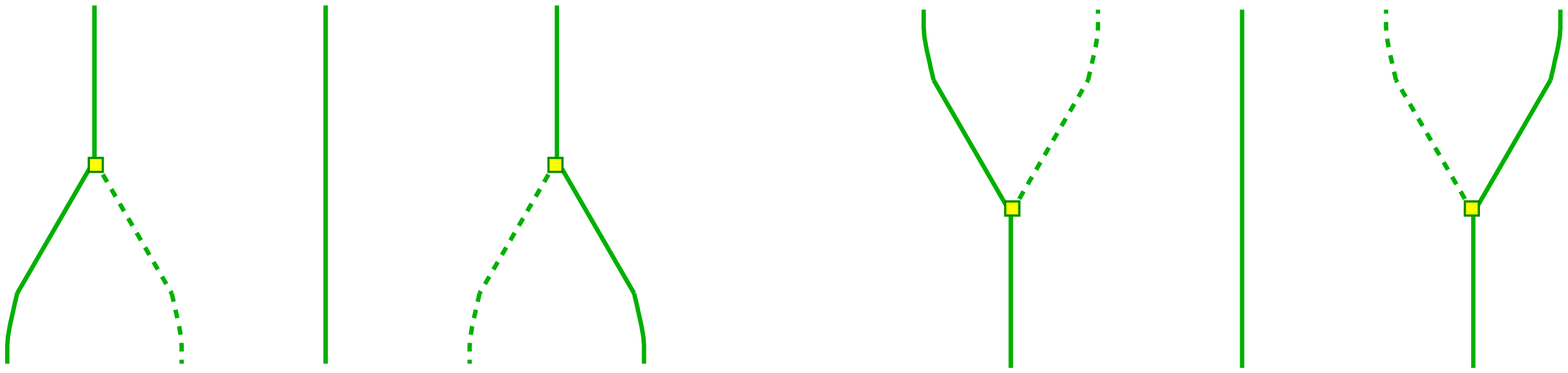}} \end{picture}}
  \put(-1.1,-7)  {\scriptsize$i$}
  \put(14.1,66)  {\scriptsize$i$}
  \put(28.9,-7)  {\scriptsize$0$}
  \put(34.2,30)  {$=$}
  \put(54.2,-7)  {\scriptsize$i$}
  \put(54.2,66)  {\scriptsize$i$}
  \put(67.2,30)  {$=$}
  \put(77.8,-7)  {\scriptsize$0$}
  \put(93.6,66)  {\scriptsize$i$}
  \put(107.7,-7) {\scriptsize$i$}
  \put(156.1,66) {\scriptsize$i$}
  \put(170.9,-7) {\scriptsize$i$}
  \put(184.9,66) {\scriptsize$0$}
  \put(191.2,30) {$=$}
  \put(210.2,-7) {\scriptsize$i$}
  \put(210.2,66) {\scriptsize$i$}
  \put(224.2,30) {$=$}
  \put(233.8,66) {\scriptsize$0$}
  \put(249.6,-7) {\scriptsize$i$}
  \put(264.9,66) {\scriptsize$i$}
  \epicture20 \labl{kokok}
Since the spaces $\Hom(\Bar U_k \oti U_k,\one)$ are one-dimensional,
the morphisms $\xx {\bar k}k00$ and $\yy 0{\bar k}k0$ are proportional to the 
respective combinations of dualities and $\pi$s (as defined in \erf{def-pik}),
i.e.\ there are numbers $\lambda_k$ and $\tilde\lambda_k$ such that
  \bea \begin{picture}(285,44)(0,25)
  \put(0,0)   {\begin{picture}(0,0)(0,0)
              \scalebox{.38}{\includegraphics{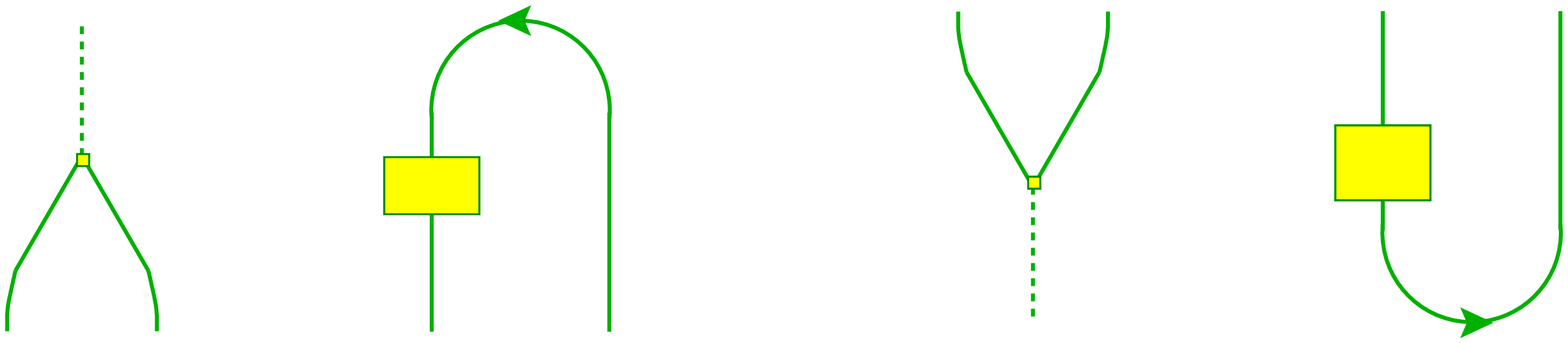}} \end{picture}}
  \put(-1.5,-8)  {\scriptsize$\bar k$}
  \put(12.5,59)  {\scriptsize$0$}
  \put(25.1,-8)  {\scriptsize$k$}
  \put(36.2,24)  {$=\, \lambda_k$}
  \put(72.7,26.7){\scriptsize$\pi_{\bar k}$}
  \put(74.9,-8)  {\scriptsize$\bar k$}
  \put(107.1,-8) {\scriptsize$k$}
  \put(170.8,61.3) {\scriptsize$\bar k$}
  \put(184.2,-5) {\scriptsize$0$}
  \put(197.7,61.3) {\scriptsize$k$}
  \put(205.5,26) {$=\, \tilde\lambda_k$}
  \put(242.3,29.7){\scriptsize$\pi_{\bar k}^{-1}$}
  \put(247.1,61.3) {\scriptsize$\bar k$}
  \put(279.1,61.3) {\scriptsize$k$}
  \epicture20 \labl{xy-pi}
The normalisation condition \erf{xy-dual} implies that the constants of 
proportionality are related by
  \be  \lambda_k\,\tilde\lambda_k = (\dim\, U_k)^{-1}_{} \,.  \labl{xy-pi-norm}

Recall now that \calc\ is a strict \tc, i.e.\ that the tensor product of objects
is strictly associative. Nevertheless, once we have chosen bases as above,
there are two distinct distinguished bases for the morphism space
$\Hom(U_i\Oti U_j\Oti U_k,U_l)$, corresponding to its two decompositions
$\bigoplus_q\!\Hom(U_i\Oti U_j,U_q)\oti\Hom(U_q\Oti U_k,U_l)$ and
$\bigoplus_p\!\Hom(U_j\Oti U_k,U_p)\oti\Hom(U_i\Oti U_p,U_l)$, \resp.
The coefficients of the basis transformation between the two are known
as the {\em fusing\/} matrices, \FF-matrices, or 6j-symbols of \calc. 
We denote them as follows:
  \bea \begin{picture}(240,60)(0,38)
  \put(0,0)   {\begin{picture}(0,0)(0,0) 
              \scalebox{.38}{\includegraphics{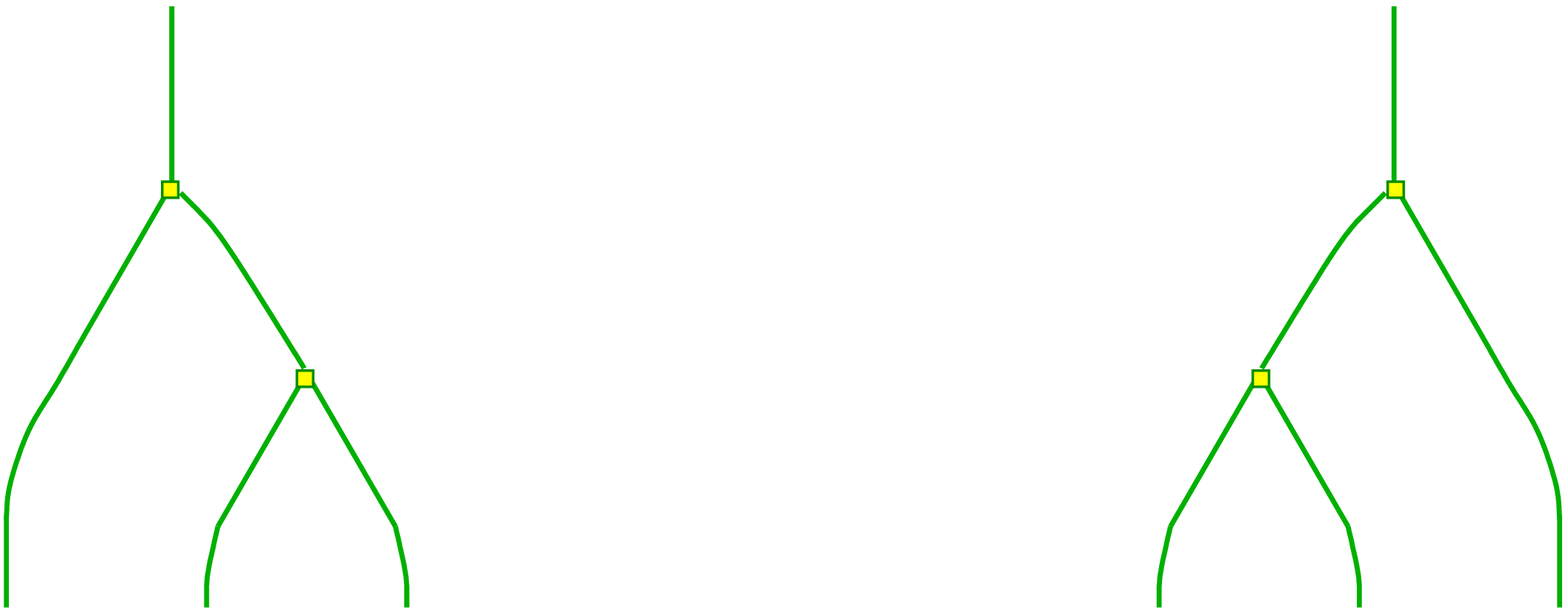}} \end{picture}}
  \put(-0.7,-8)  {\scriptsize$i$}
  \put(18.2,61.7){\tiny$\alpha$}
  \put(24.8,93.5){\scriptsize$l$}
  \put(28.1,-8)  {\scriptsize$j$}
  \put(42.1,48)  {\scriptsize$p$}
  \put(38.6,33.1){\tiny$\beta$}
  \put(58.1,-8)  {\scriptsize$k$}
  \put(76,42)    {$=\ \dsty\sum_q\sum_{\gamma,\delta}\;
                 \F{i\U j}klpq\alpha\beta\gamma\delta$}
  \put(170.7,-8) {\scriptsize$i$}
  \put(181.1,34.5){\tiny$\gamma$}
  \put(190.1,50) {\scriptsize$q$}
  \put(199.5,-8) {\scriptsize$j$}
  \put(210.4,61.6){\tiny$\delta$}
  \put(206.6,93.5){\scriptsize$l$}
  \put(229.5,-8) {\scriptsize$k$}
  \epicture26 \labl{def-fmat}
By composing with the morphism dual to the one on the \rhs, we arrive at the
formula
  \bea \begin{picture}(160,81)(0,36)
  \put(0,0)   {\begin{picture}(0,0)(0,0)
              \scalebox{.38}{\includegraphics{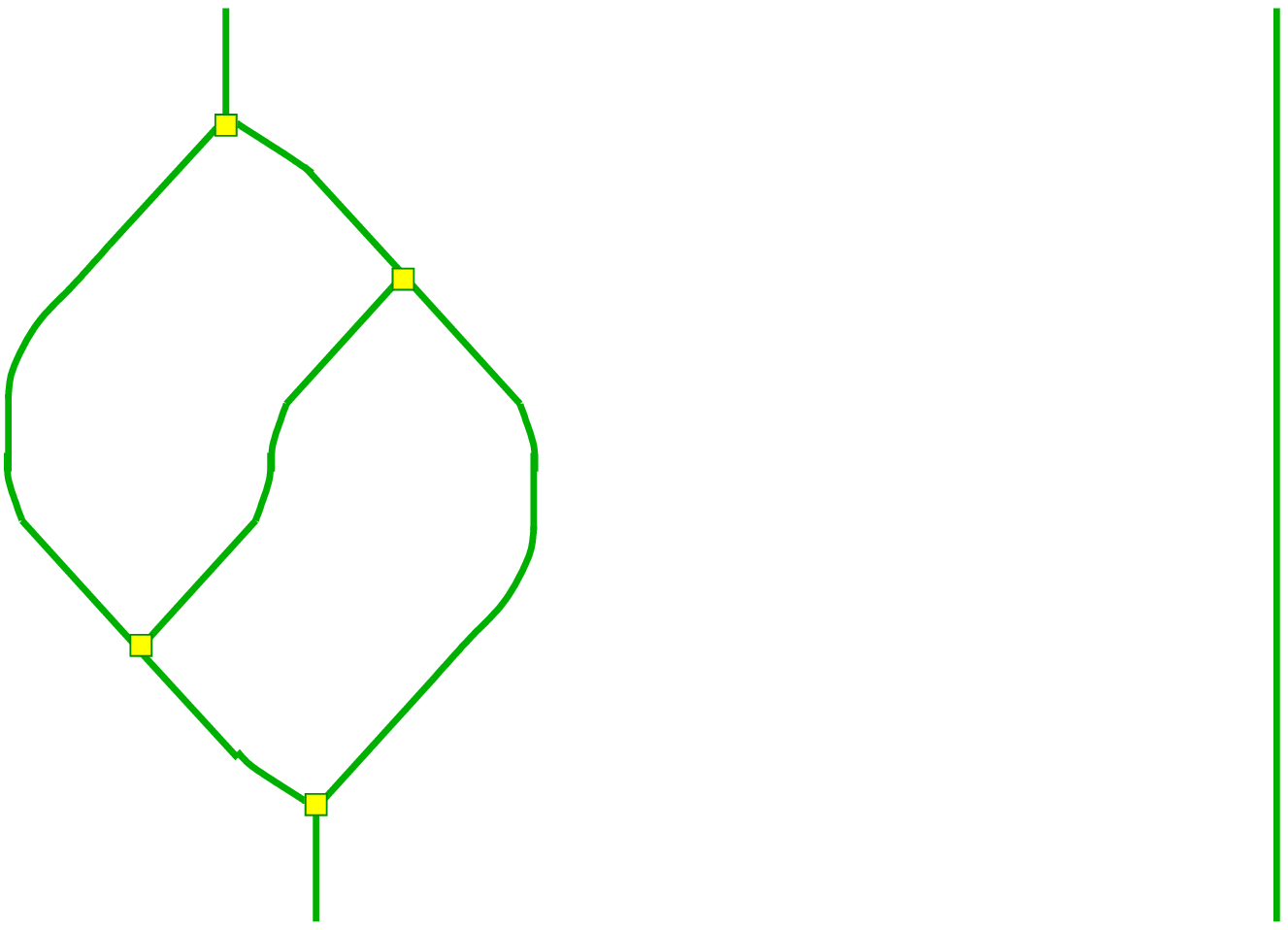}} \end{picture}}
  \put(-4.9,58)  {\scriptsize$i$}
  \put(11,26)    {\tiny$\overline\gamma$}
  \put(24.1,108) {\scriptsize$l$}
  \put(24.0,86)  {\tiny$\alpha$}
  \put(21.5,16)  {\scriptsize$q$}
  \put(25.5,52)  {\scriptsize$j$}
  \put(39.5,84.4){\scriptsize$p$}
  \put(34.5,-8)  {\scriptsize$l$}
  \put(34.3,17)  {\tiny$\bar\delta$}
  \put(42.7,67)  {\tiny$\beta$}
  \put(53.7,46)  {\scriptsize$k$}
  \put(78,48)    {$=\ \F{i\U j}klpq\alpha\beta\gamma\delta$}
  \put(142.9,-8) {\scriptsize$l$}
  \put(142.9,108){\scriptsize$l$}
  \epicture27 \labl{fmat}
for the \FF-matrices.
When any of the labels $i,j,k$ equals $\One$, then the \lhs\ of formula 
\erf{fmat} degenerates (if non-zero) to the \lhs\ of \erf{xy-dual}, leading to
  \be  \F{\One\U j}kllj{}\beta{}\delta = \delta_{\beta,\delta} \,, \qquad
  \F{i\U \One}klki\alpha{}{}\delta = \delta_{\alpha,\delta} \,, \qquad
  \F{i\U j}\One ljl\alpha{}\gamma{} = \delta_{\alpha,\gamma} \,.  \labl{eq:fnorm}
For the morphisms dual to those appearing in \erf{def-fmat}, there is an
analogous relation
  \bea \begin{picture}(240,59)(0,40)
  \put(0,0)   {\begin{picture}(0,0)(0,0)
              \scalebox{.38}{\includegraphics{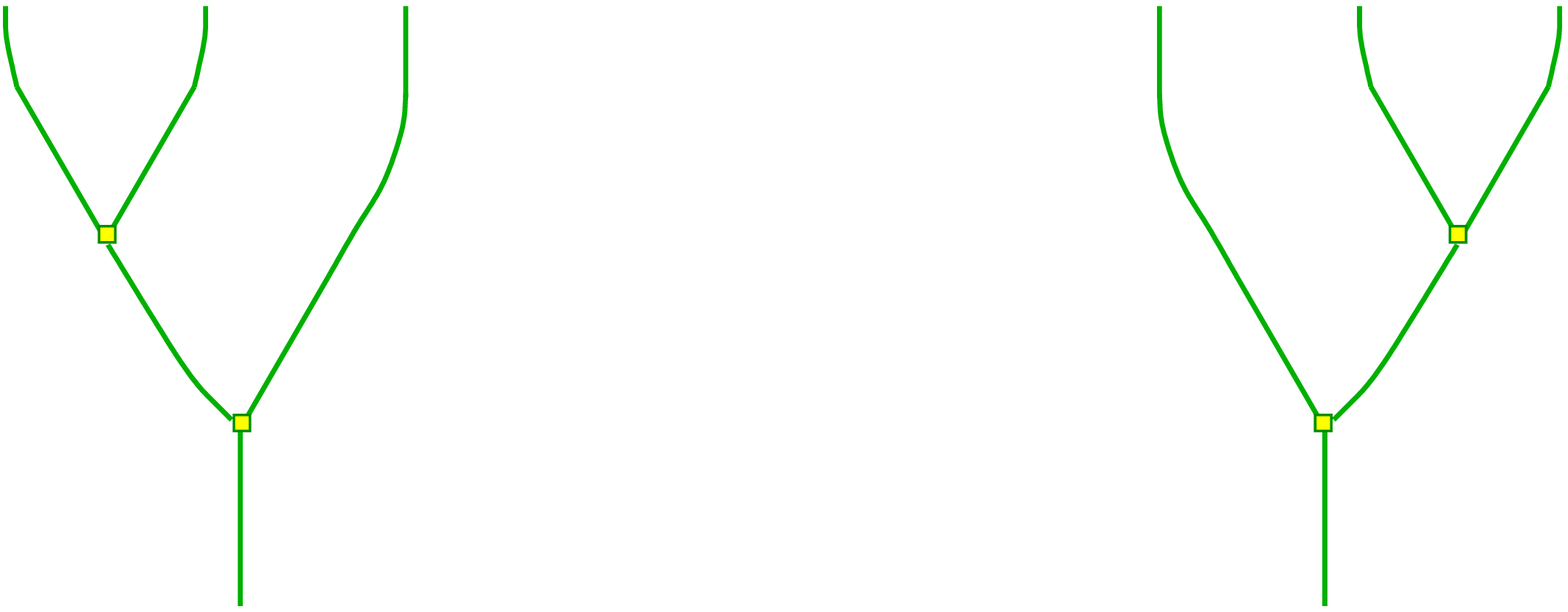}} \end{picture}}
  \put(-0.5,93.5){\scriptsize$i$}
  \put(18.6,54.9){\tiny$\bar\gamma$}
  \put(34.5,-8)  {\scriptsize$l$}
  \put(18.5,38.9){\scriptsize$q$}
  \put(29.4,94.9){\scriptsize$j$}
  \put(38.8,25.7){\tiny$\bar\delta$}
  \put(58.3,93.5){\scriptsize$k$}
  \put(76,42)    {$= \ \dsty\sum_p\sum_{\alpha,\beta}\;
                 \F{i\U j}klpq\alpha\beta\gamma\delta$}
  \put(171.7,93.5){\scriptsize$i$}
  \put(195.9,-8) {\scriptsize$l$}
  \put(200.1,25.7){\tiny$\bar\alpha$}
  \put(201.2,94.9){\scriptsize$j$}
  \put(210.5,38.2){\scriptsize$p$}
  \put(220.2,54.6){\tiny$\bar\beta$}
  \put(230.7,93.5){\scriptsize$k$}
  \epicture22 \labl{fmat-dual}
It is convenient to introduce a separate symbol $\GG$
for the inverse of $\FF$. It is defined as
  \bea \begin{picture}(420,65)(0,45)
  \put(0,0)   {\begin{picture}(0,0)(0,0) 
              \scalebox{.38}{\includegraphics{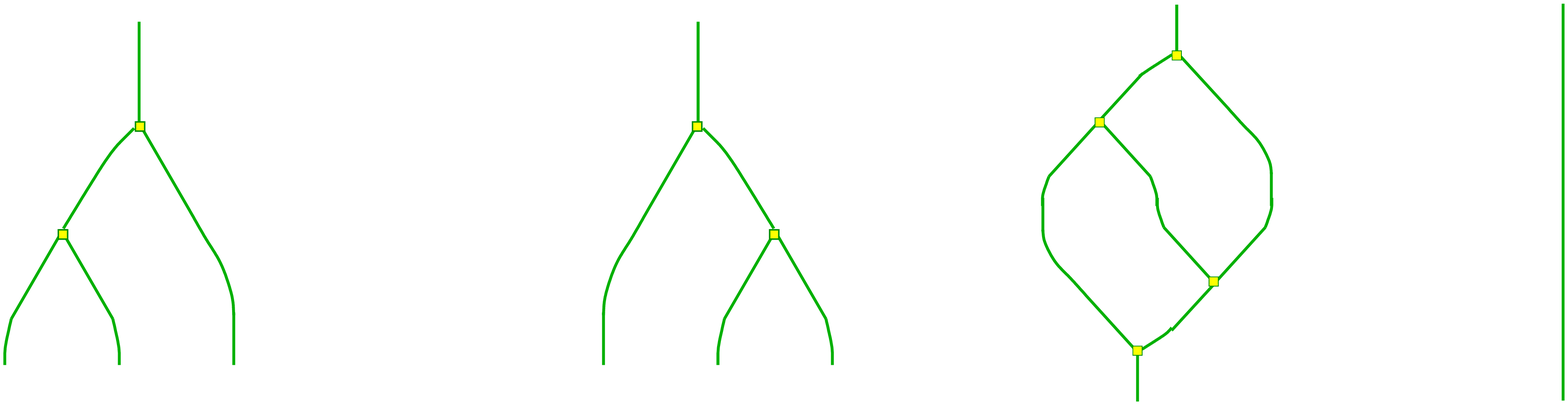}} \end{picture}}
   \put(34,103) {\scriptsize$l$}
   \put(27,72) {\scriptsize$\beta$}
   \put(17,58) {\scriptsize$p$}
   \put(8,44) {\scriptsize$\alpha$}
   \put(0,2) {\scriptsize$i$}
   \put(30,2) {\scriptsize$j$}
   \put(58,2) {\scriptsize$k$}
  \put(70,42)    {$=\ \dsty\sum_{q,\gamma,\delta}\;
                 \G{i\U j}klpq\alpha\beta\gamma\delta$}
   \put(179,103) {\scriptsize$l$}
   \put(173,70) {\scriptsize$\gamma$}
   \put(194,62) {\scriptsize$q$}
   \put(194,42) {\scriptsize$\delta$}
   \put(156,2) {\scriptsize$i$}
   \put(185,2) {\scriptsize$j$}
   \put(214,2) {\scriptsize$k$}
  \put(230,42)    {\rm i.e.}
   \put(304,108) {\scriptsize$l$}
   \put(290,86) {\scriptsize$p$}
   \put(265,48) {\scriptsize$i$}
   \put(294,48) {\scriptsize$j$}
   \put(332,48) {\scriptsize$k$}
   \put(306,16) {\scriptsize$q$}
   \put(295,-8) {\scriptsize$l$}
   \put(283,65) {\scriptsize$\alpha$}
   \put(303,83) {\scriptsize$\beta$}
   \put(293,20) {\scriptsize$\bar\gamma$}
   \put(314,37) {\scriptsize$\bar\delta$}
   \put(406,108) {\scriptsize$l$}
   \put(406,-8) {\scriptsize$l$}
  \put(345,42)   {$=\ \G{i\U j}klpq\alpha\beta\gamma\delta$}
  \epicture28 \labl{gmat}
Combining the braiding morphisms with the basis choice \erf{xijk,yijk}
provides the {\em braiding\/} matrices \RR:
  \bea \begin{picture}(150,62)(0,31)
  \put(0,0)   {\begin{picture}(0,0)(0,0)
              \scalebox{.38}{\includegraphics{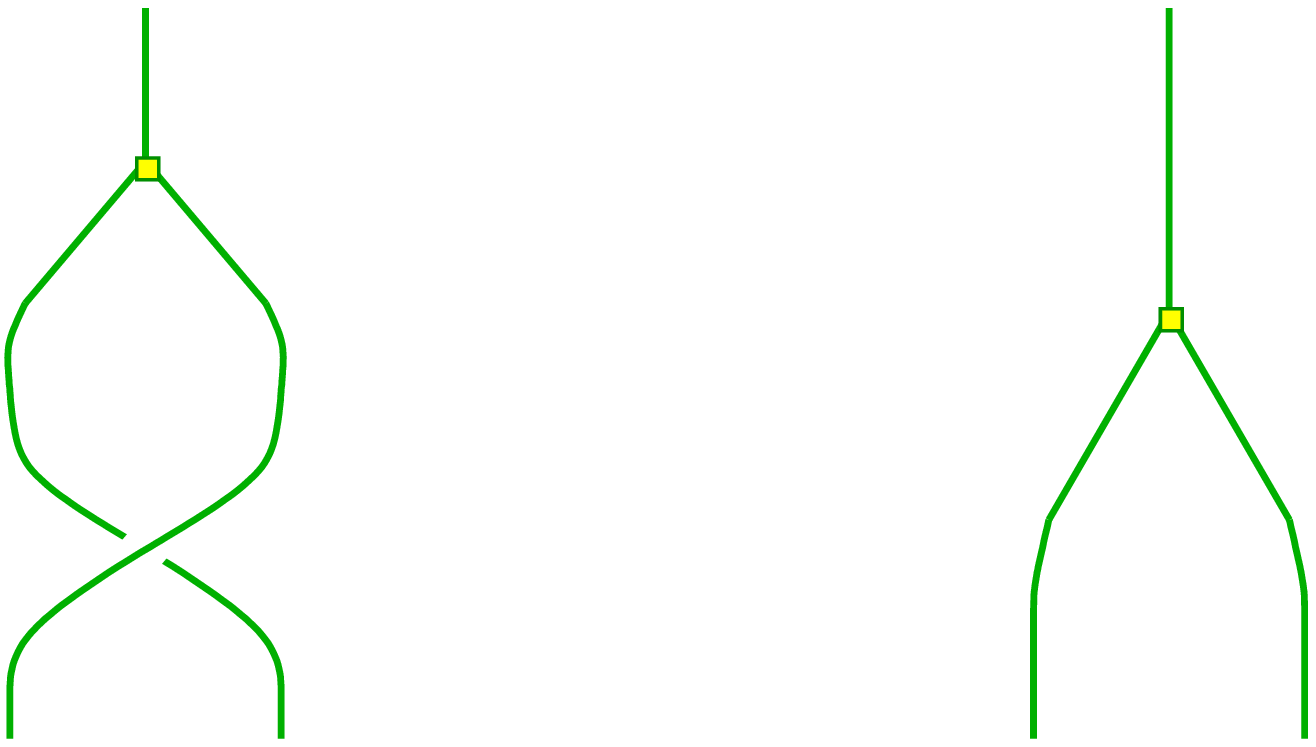}} \end{picture}}
  \put(45,36)    {$=:\ \dsty\sum_{\beta}\R ijk\alpha\beta$}
  \put(-0.7,-7)  {\scriptsize$i$}
  \put(13.9,84)  {\scriptsize$k$}
  \put(17,66)    {\tiny$\alpha$}
  \put(28.7,-7)  {\scriptsize$j$}
  \put(112.3,-7) {\scriptsize$i$}
  \put(126.6,84) {\scriptsize$k$}
  \put(129,50.7) {\tiny$\beta$}
  \put(140.7,-7) {\scriptsize$j$}
  \epicture21 \labl{rmat}
The number that is obtained when the braiding $c_{i,j}$ (`over-braiding') 
is replaced by $c_{j,i}^{-1}$ (`under-braiding') is denoted by 
$\Rm ijk\alpha\beta$, and instead of $\R ijk\alpha\beta$ one also often
writes $\Rp ijk\alpha\beta$. One easily checks that
  \be  \sum_\beta \R ijk\alpha\beta \, \Rm jik\beta\gamma
  = \delta_{\alpha,\gamma}^{}  \ee
and, using the compatibility of twist and braiding and functoriality of the
twist (see \erf{ribax2} and \erf{ribax1}), that
  \be  \sum_\beta \R ijk\alpha\beta \, \R jik\beta\gamma
  = \Frac{\theta_k}{\theta_i\,\theta_j}\,\delta_{\alpha,\gamma}^{} \,. 
  \labl{eq:RR=th3}

Here the complex number $\theta_k$ is defined by $\theta_{U_k}\,{=:}\,
\theta_k\,\id_{U_k}$, i.e.\ specifies the twist of the simple object $U_k$.
We can express $\theta_k$ in terms of special matrix elements of \FF\ and \RR.
To this end one rewrites the twist morphism of the object $U_k$ as
  \bea \begin{picture}(218,61)(0,38)
  \put(0,0)   {\begin{picture}(0,0)(0,0)
              \scalebox{.38}{\includegraphics{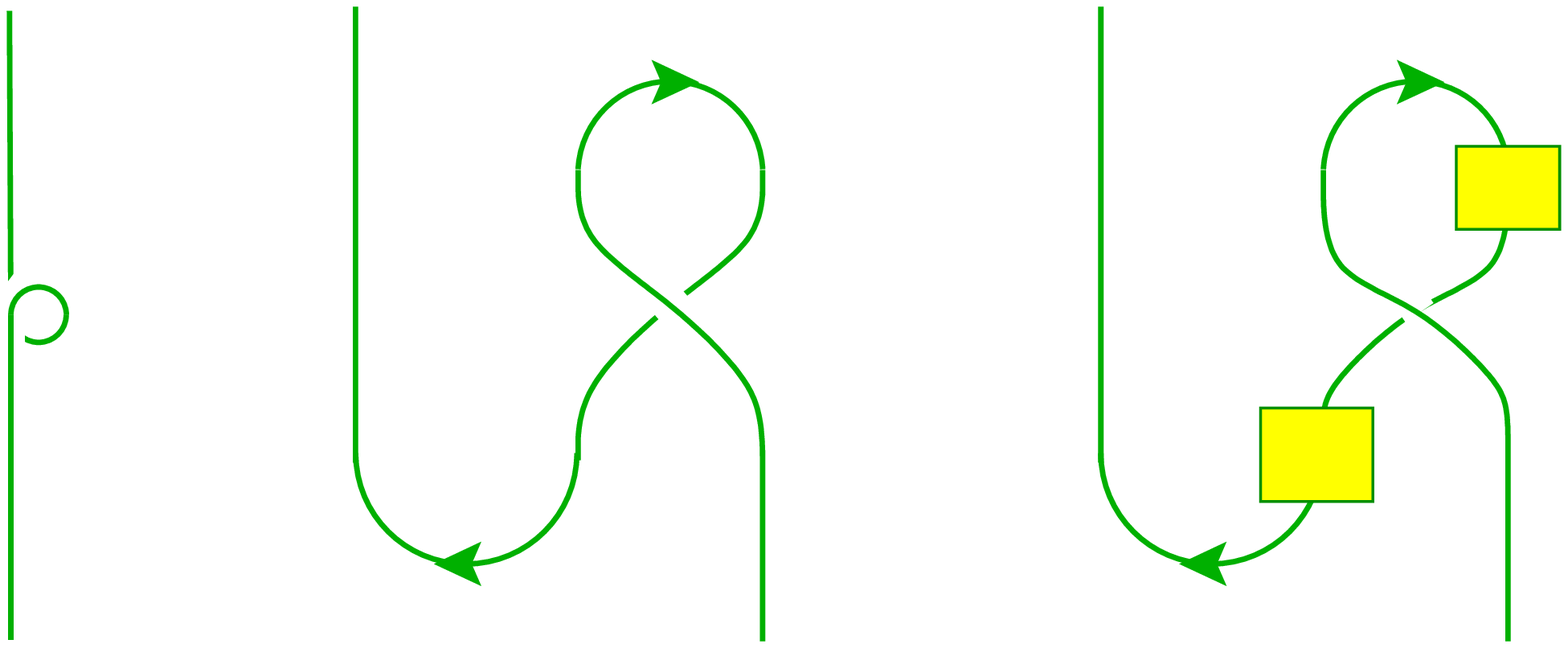}} \end{picture}}
  \put(25,46) {$=$}
  \put(125,46){$=$}
  \put(2,93)      {\scriptsize$k$}
  \put(48,93)     {\scriptsize$k$}
  \put(150,93)    {\scriptsize$k$}
  \put(2,-8)      {\scriptsize$k$}
  \put(103,-8)    {\scriptsize$k$}
  \put(205,-8)    {\scriptsize$k$}
  \put(181,39)    {\scriptsize$\bar k$}
  \put(204.3,62)  {\scriptsize$\pi_{\bar k}$}
  \put(175.5,24.5){\scriptsize$\pi_{\bar k}^{-1}$}
  \epicture27 \labl{twist-f-r}
and then computes the constant by which the \rhs\ differs from $\id_{U_k}$
by first using the identity \erf{xy-pi}, then \erf{rmat} and finally \erf{fmat}
(with $p\eq q\eq0$, $i\eq k$ and $j\eq\bar k$). The result is
  \be  \theta_k = \dim(U_k)\, \Fs k{\bar k}kk\One\One \,
  \Rs-{\bar k}k{\,\One} \,.  \labl{eq:theta-from-F}
The \fsi\ of a self-dual simple object $U_k$ is encoded in the $\FF$ matrix as 
well. To see this take formula \erf{fmat} with $p\eq q\eq0$, $i\eq k\eq l$
and $j\eq\bar k$ for a not necessarily self-dual $U_k$. Then apply \erf{xy-pi}
and \erf{xy-pi-norm} to the \lhs\ and use \erf{def-pi} to cancel the morphisms 
$\pi_k$.  The resulting relation reads
  \be  \dim(U_k)\, \Fs k{\bar k}kk\One\One 
  = p_k \, \lambda_k\, / \lambda_{\bar k} \,. \labl{eq:dimF=pll}
Finally specialise to $k=\bar k$ and employ \erf{eq:p=nu} to arrive at
  \be  \nu_k = \dim(U_k)\,\Fs kkkk\One\One \,.  \labl{eq:nu=F}
Also note that combining \erf{eq:theta-from-F} (or rather, the analogous 
equation obtained when using the inverse braiding) and \erf{eq:dimF=pll} 
one has, for any simple $U_k$,
  \be \Rs{}{\bar k}k{\;\One} = p_k\, \theta_k^{-1}\, \lambda_{\bar k}\,
  / \lambda_k \,.  \labl{eq:Rkk0=pllt} 

\smallskip 

For comparison with the literature, we note that our convention for the
\FF- and \RR-matrices is related to the one of \cite{Mose} by
  \be
  \Fs jklipq  \ \hat= \ \FF_{\!p,q}^{\phantom x}{\Barray jkli}
  \qquad{\rm and}\qquad \Rs{}ijk \ \hat= \ \Omega^k_{ji} \ee
(compare formula~(2.4) and example~2.8 in~\cite{Mose}). Note that often 
also the composite quantities ${\sf B} \,{\sim}\,\Omega^- \FF \, \Omega^+$ 
are used, see e.g.\ formula (3.3) in \cite{Mose}.

\subsection{Three-manifolds and ribbon graphs}

To determine a \corfu\ of a rational CFT, we specify it as a particular element
in the relevant space of conformal blocks. A very convenient characterisation
of conformal blocks is via ribbon graphs in three-manifolds. In this formulation,
the coefficients in the expansion of a CFT correlator in terms of a chosen
basis of conformal blocks are obtained as invariants of closed three-manifolds
with embedded ribbon graphs. To explain this construction we need to introduce 
the concepts of a ribbon graph and of a three-dimensional topological field theory.
(For more details see e.g.\ 
\cite{witt27,frki,joSt5,Bbrt,walk,dujn,kasc,chfs,TUra,KAss,BAki,KOsu};
this quick introduction follows section 2 of \cite{fffs3}.)

A {\em ribbon graph\/} consists of an oriented three-manifold $M$, possibly 
with boundaries, together with embedded ribbons and coupons. A {\em ribbon\/}
is an oriented rectangle, say $[-1/10,1/10]\,{\times}\,[0,1]$, together with 
an orientation for its {\em core\/} $\{0\}\,{\times}\,[0,1]$. The two subsets
$[-1/10,1/10]\,{\times}\,\{0\}$ and $[-1/10,1/10]\,{\times}\,\{1\}$
are the {\em ends\/} of the ribbon. A {\em coupon\/} is an
oriented rectangle with two preferred opposite edges, called top and bottom. 
The embeddings of ribbons and coupons into $M$ are demanded to be injective.
A ribbon minus its ends does not
intersect any other ribbon, nor any coupon, nor the boundary of $M$.
A coupon does not intersect any other coupon nor the boundary of $M$.
The ends of a ribbon must either lie on one of the preferred edges of some
coupon or on $\partial M$. For ribbons ending on a coupon, the orientation of
the ribbon and of the coupon must agree.  

Choosing an orientation for the ribbons and coupons is equivalent
to choosing a preferred side -- henceforth called the `white' side --
which in the drawings will usually face
the reader. The opposite (`black') side is drawn in a darker
shade, as has already been done in figure \erf{ribbon} above. 
We use open arrows to indicate the orientation of the ribbon's core.

Each constituent ribbon of a ribbon graph is labelled (sometimes called 
`colored') by a
(not necessarily simple) object of the modular category \calc, and each 
coupon is labelled by a morphism of \calc. Which space the morphism belongs 
to depends on the ribbons ending on the coupon.  If a ribbon labelled by 
$U$ is `incoming', the relevant object is $U$ when the ribbon is attached
to the bottom of the coupon, and $U^\vee$ when it is attached to its top.
If it is `outgoing', then the convention for the object is the other way round.
As an illustration, the coupon in
  \bea \begin{picture}(40,38)(0,32)
  \put(0,0)  {\begin{picture}(0,0)(0,0)
              \scalebox{.38}{\includegraphics{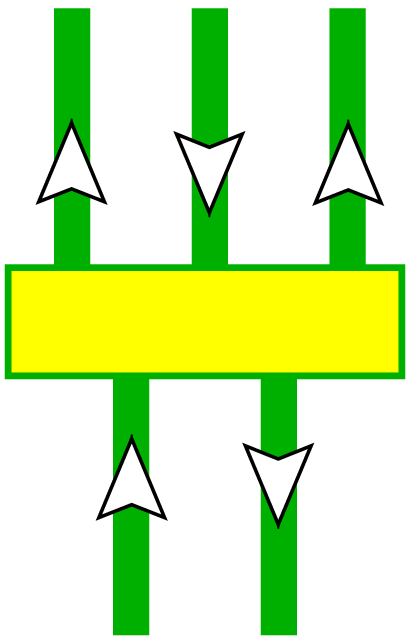}} \end{picture}}
  \put(-3.2,60) {\scriptsize$X$}
  \put(14,60)   {\scriptsize$Y$}
  \put(42,60)   {\scriptsize$Z$}
  \put(20,33.3) {\scriptsize$\varphi$}
  \put(4.3,3)   {\scriptsize$U$}
  \put(34,3)    {\scriptsize$V$}
  \epicture13 \labl{coupon}
is labelled by an element $\varphi\iN\Hom(U\oti V^\vee{,}\,
X\oti Y^\vee\Oti\,Z)$.

Consider now a ribbon graph in $S^3$. 
We can assign an element in $\Hom(\one,\one)$, i.e.\ a complex
number, to it as follows: Regard $S^3$ as $\reals^3 \,{\cup}\, \{\infty\}$,
with $\reals^3$ parametrised by Cartesian coordinates $(x,y,z)$. Deform 
the ribbon graph such that all its coupons are rectangles in the 
$x$-$y$-plane, with white sides facing upwards, and such that the 
top and bottom edges are parallel to the $x$-axis, with the bottom edge 
having a smaller $y$-coordinate than the top edge. Deform the
ribbons to lie in a small neighbourhood of the $x$-$y$-plane. 
In short, the ribbon graph must be arranged in such a manner that the
bends, twists and crossings of all ribbons can be expressed as dualities,
twists and braidings as appearing in \erf{limos} and \erf{ribbon}. 
The element in $\Hom(\one,\one)$ is then obtained by reading the graph from 
$y\eq{-}\infty$ to $y\eq{+}\infty$ and interpreting it 
as a concatenation of morphisms in \calc.
Let us display an example:
  \begin{eqnarray} \begin{picture}(450,88)(10,0)
  \put(25,0)  {\begin{picture}(0,0)(0,0)
              \scalebox{.38}{\includegraphics{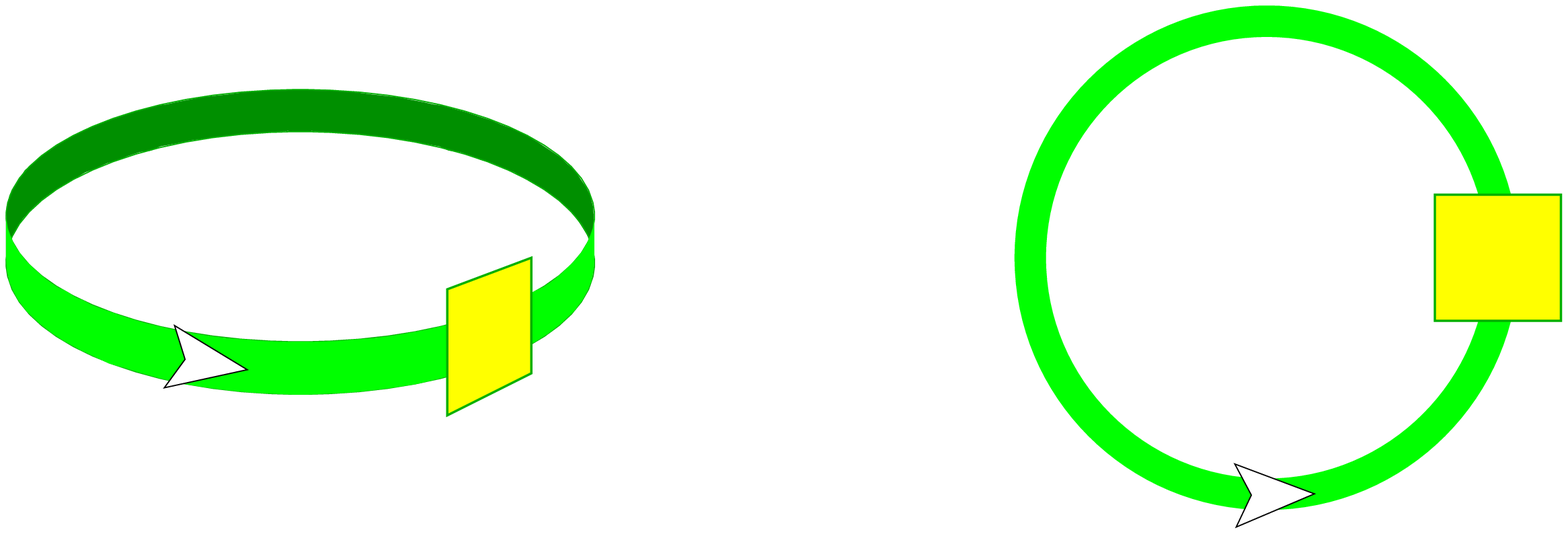}} \end{picture}}
  \put(138,40){$\stackrel{{\rm deform}
     }{-\!\!\!-\!\!\!-\!\!\!-\!\!\!\!\longrightarrow}$}
  \put(295,40){$\stackrel{\;\rm interpret\ }{\stackrel{{\rm as~morph.}
     }{-\!\!\!-\!\!\!-\!\!\!-\!\!\!\!\longrightarrow}}
     \ \ \,d_U \cir (id_{U^\vee} \oti f) \cir \tilde b_U $}
  \put(32,21)  {\scriptsize$U$}
  \put(102,30) {\scriptsize$f$}
  \put(256,80) {\scriptsize$U$}
  \put(267,43) {\scriptsize$f$}
  \end{picture} \nonumber \\[-2.9em]{} \label{eq:ribboon-example}
  \\[-.4em]{}\nonumber \end{eqnarray}
One of the non-trivial results following from the defining relations of 
a modular category is that different ways to translate a ribbon graph into
a morphism give rise to one and the same value for the ribbon graph.
In other words, the value for the ribbon graph is invariant under the
various local moves that transform those different descriptions into
each other.

All pictures in sections \ref{sec:mod-case-N=1}--\ref{sec:N=1-repn} below
directly stand for morphisms in \calc. The first genuine
ribbon graphs will occur in section \ref{sec:con-mf-rib}.
We will usually simplify the pictures involving ribbon
graphs by replacing all ribbons by lines. When doing so, it is understood
that the ribbon lies in the plane of the paper, with the white
side facing up (this convention is known as `blackboard framing').
Also note that, strictly speaking, the definition of a ribbon
graph given above forbids annulus-shaped ribbons. Whenever such a ribbon 
occurs below it is understood to be of the form displayed in 
\erf{eq:ribboon-example}, with $f$ the identity morphism in $\Hom(U,U)$.

\subsection{Topological field theory} \label{sec:TFT}

So far we have used the modular category \calc\ only to assign numbers to 
ribbon graphs in $S^3$; for this purpose it is actually sufficient that
\calc\ is a ribbon category. That \calc\ is even modular implies
the highly non-trivial result that it gives rise to a full
three-dimensional topological field theory. By definition, a
three-dimensional TFT is a pair $(Z,\calh)$ of assignments that
associate algebraic structures to geometric data -- extended surfaces
and cobordisms -- and satisfy various properties, to be be outlined below.

By an {\em extended surface\/} we mean an oriented
closed compact two-manifold $\rmX$ with a finite number of disjoint
oriented arcs (the remnants, at the topological level, of the local
coordinates that one must choose around the insertion points of the 
world sheet) labelled by pairs $(U,\eps)$ with $U\iN\objc$ and 
$\eps\iN\{\pm1\}$, and with a lagrangian subspace $L(\rmX)$ of the
first homology group $H_1(\rmX,\reals)$. The {\em opposite\/} $-\rmX$ of 
an extended surface is obtained from $\rmX$ by reversing the orientation 
of $\rmX$, reversing the orientation of all arcs and replacing $\eps$ by 
$-\eps$. To make explicit the insertions $U,V,\cdots$ of an extended 
surface we will sometimes write $(U,V,\cdots ; \rmX)$ instead of $\rmX$.
This notation assumes that all signs $\eps$ are $+1$, but does not
encode the positions and orientations of the arcs; when the sign $\eps$ 
for some object is $-1$, then we write $(U,-)$ instead of $U$.

The first datum, $\calh$, of the TFT assigns to each extended surface 
$\rmX$ a finite-dimensional vector space $\calh(\rmX)$. This space is
constructed from appropriate tensor products of morphism spaces of \calc;
one has 
  \be  \calh(\emptyset) = \complex \,.  \ee
The CFT interpretation of $\calh(\rmX)$ is as the space of conformal 
blocks on $\rmX$. Thus we have, for example, $\dim\, \calh(k;S^2)
\eq\delta_{k,0}$, in agreement with the fact that
the space of one-point blocks on the sphere is zero-dimensional except
for the insertion of a vacuum. Similarly,
$\dim\, \calh(k,(\ell,-);S^2)$\linebreak[0]$\eq \delta_{k,\ell}$, 
i.e.\ the two-point
blocks on the sphere vanish unless the two fields are dual to
each other. More generally, 
  \be  \dim\, \calh(i,j,(k,-);S^2) = \N ijk  \ee 
(recall that $\N ijk\eq\dim\,\Hom(U_i\oti U_j,U_k)\eq\dim\,\Hom(U_i\oti
U_j\oti U_k^\vee{,}\,\one)$).
Each homeomorphism $f{:}\ X\,{\to}\, Y$ of extended surfaces
induces a linear map $f_\natural{:}\ \calh(X)\,{\to}\,\calh(Y)$.

Consider now a manifold $M$ with ribbon graph whose boundary can be
written as a disjoint union $\partial M\eq\partial_1 M\,{\sqcup}\,\partial
_2 M$. Turn $\partial_{1} M$ and $\partial_{2} M$ into extended surfaces by 
taking as arcs the ends of ribbons, with orientation induced by the ribbons. 
In addition fix a Lagrangian subspace (this is part of the defining data
of an extended surface it is not determined by $M$ and the embedded 
ribbon graph). When a ribbon ending on $\partial M$ is labelled by $U$, 
then the corresponding arc is labelled by $(U,+1)$ if the core of the 
ribbon points away from the surface, and by $(U,-1)$ otherwise. Denote
the extended surface $\partial_2 M$ by $\partial_+ M$
and $-\partial_1 M$ by $\partial_- M$.
Then the triple $(M, \partial_-M, \partial_+M)$ is called a
{\em cobordism\/} from $\partial_-M$ to $\partial_+M$.

The second datum, $Z$, of the TFT assigns a linear map 
  \be  Z(M, \partial_-M, \partial_+M) :\quad \calh(\partial_-M)
  \rightarrow \calh(\partial_+M) \ee
to every cobordism. Let us mention two of its properties. The first concerns 
the normalisation of $Z$.  Let $\rmX$ be an extended surface and 
$M$ the manifold $\rmX\,{ \times}\, [0,1]$ with embedded ribbon graph
given by straight ribbons connecting the arcs
in $\rmX \,{\times}\, \{0\}$ to the arcs in $\rmX\,{\times}\,\{1\}$, 
with cores oriented from $0$ to $1$. Then 
  \be  Z(M,\rmX,\rmX) = \id_{\calH(\rmX)} \,.  \labl{eq:TFT-Zid}
The second property is functoriality. Let $M_1$ and $M_2$ be two
three-manifolds with ribbon graphs and let 
$f{:}\ \partial_+M_1 \,{\to}\, \partial_-M_2$ be a 
homeomorphism of extended surfaces, and let $M$ be the manifold 
obtained from glueing $M_1$ to $M_2$ using $f$. Then
  \be  Z(M,\partial_-M_1,\partial_+M_2) =
  \kappa^m\, Z(M_2,\partial_-M_2,\partial_+M_2) \cir f_\natural \cir
  Z(M_1,\partial_-M_1,\partial_+M_1) \,,  \labl{eq:Zfunct}
where $m$ is an integer (see the review in \cite{fffs3} for details) and
$\kappa\eq S_{0,0} \sumI_j \theta_j^{-1} \dim(U_j)^2$, which is
called the {\em charge\/} of the modular category \calc.

In our application we will not need the properties of $Z$ in their most 
general form.  But we will use the following special cases and 
consequences: For the manifolds $S^2\,{\times}\,S^1$ and $S^3$ without
ribbon graph we have 
  \be  Z(S^2{\times}S^1 ;\emptyset;\emptyset) = 1 \qquad {\rm and}
  \qquad Z(S^3 ;\emptyset;\emptyset) = S_{0,0} \,.  \labl{norm-Z}
More generally, for $S^3$ 
with embedded ribbon graph, the number assigned by $Z$ is 
$S_{0,0}$ times the number obtained by translating the ribbon graph
to a morphism of \calc\ as described in the previous section.

Functoriality of $Z$ implies that the invariant of a ribbon graph
in any closed three-manifold can be related to an invariant in $S^3$ by
the use of surgery along links. This will be used in section
\ref{sec:toruszero} and \ref{sec:one-point-blocks} to relate 
invariants of $S^2\,{\times}\,S^1$ and $S^3$.

We will also make intensive use of a trace formula that is obtained
as follows: For $\rmX$ an extended surface, consider the three-manifold
$N\eq\rmX\,{\times}\,[0,1]$ with embedded ribbon graph such that
$\partial_- N\eq\partial_+ N\eq\rmX$ as extended surfaces.
Let $M$ be the closed three-manifold with ribbon graph obtained
from $N$ by identifying $(x,0)$ with $(x,1)$ for all $x\iN\rmX$. Then
  \be  Z(M,\emptyset,\emptyset) = {\rm tr}_{\calH(\rmX)} \,
  Z(N,\rmX,\rmX) \,.  \labl{eq:TFT-Ztrace}
This trace formula, too, is a consequence of the functoriality of $Z$.

\subsection{The case ${N_{ij}}^k\iN\{0,1\}$}\label{sec:mod-case-N=1}

In this section we specialise the general treatment above to the case 
that the dimensions ${N_{ij}}^k$ of all coupling spaces $\Hom(U_i{\otimes}
U_j,U_k)$ are either 0 or 1. This greatly simplifies both notation and
calculation. It may be thought of as a `meta-example'; in particular, it
encompasses both concrete examples that we study in this paper.

The main notational simplification is that the multiplicity indices
disappear, so that the fusing and braiding matrices take the form
  \be
  \Fs ijk\ell pq  \qquad {\rm and } \qquad \Rs{}ijk \,, \ee
\resp. (But we must still choose bases in the morphism spaces, and the 
form of the fusing and braiding matrices does depend on this choice.)
The quantum dimensions $\dim(U_k)$ (which in a modular category are positive 
real numbers) can be obtained from $\FF$ and $\RR$,
by taking the absolute value of \erf{eq:theta-from-F} 
(since $\theta_k$ is a phase). The twist eigenvalues $\theta_k$ 
follow from \erf{eq:theta-from-F} as well, and finally the \fsi s are 
given by \erf{eq:nu=F}. (In practice, it is often easier to obtain
$\dim(U_k)$ and $\theta_k$ by some other means, though.)

To reconstruct the $S$-matrix from $\FF$ and $\RR$, as well as for later use, 
the following identities prove to be useful:
  \bea \begin{picture}(400,58)(0,34)
  \put(0,0)   {\begin{picture}(0,0)(0,0)
              \scalebox{.38}{\includegraphics{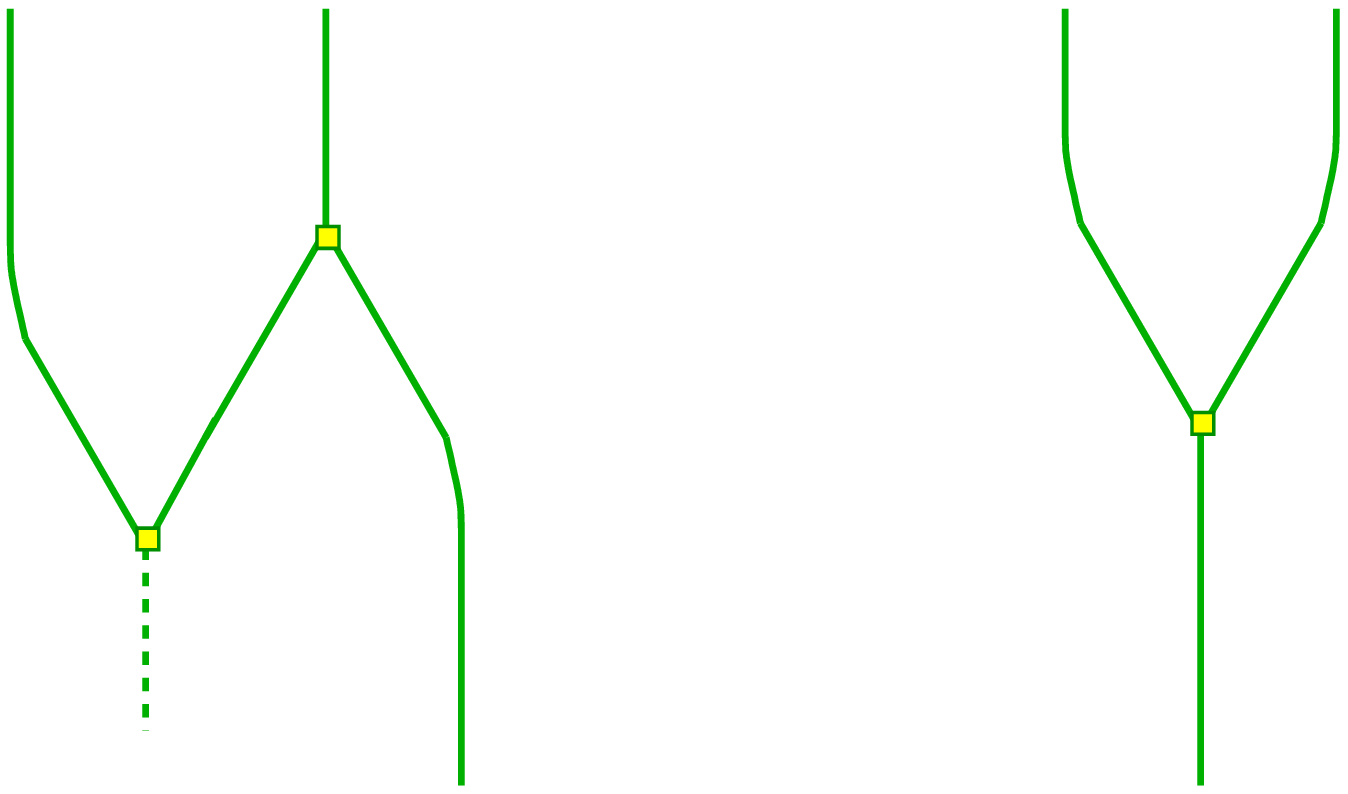}} \end{picture}}
  \put(-.3,89.1)   {\scriptsize$k$}
  \put(20.0,45.9)  {\scriptsize$\bar k$}
  \put(34.5,89.1)  {\scriptsize$i$}
  \put(48.3,-7.6)  {\scriptsize$j$}
  \put(62,40)      {$=\; \Fs k{\bar k}jji\One$}
  \put(115.5,89.1) {\scriptsize$k$}
  \put(130.3,-7.6) {\scriptsize$j$}
  \put(144.9,89.1) {\scriptsize$i$}
  \put(186,40)     {and}
  \put(240,0) {\begin{picture}(0,0)(0,0)
              \scalebox{.38}{\includegraphics{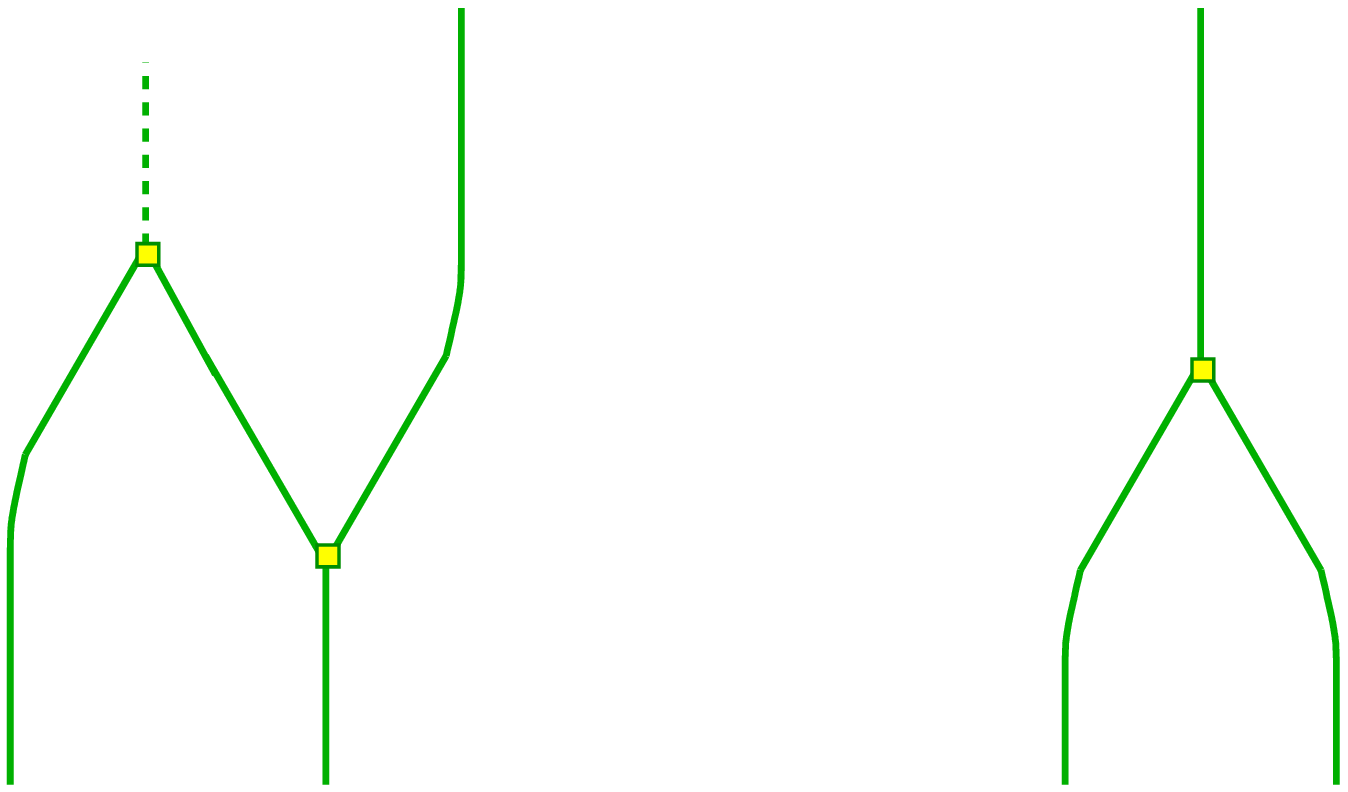}} \end{picture}}
  \put(238.7,-7.6) {\scriptsize$k$}
  \put(267.7,42.4) {\scriptsize$\bar k$}
  \put(273.8,-7.6) {\scriptsize$i$}
  \put(289.1,89.1) {\scriptsize$j$}
  \put(301,40)     {$= \,\Gs k{\bar k}jj\One i$}
  \put(354.4,-7.6) {\scriptsize$k$}
  \put(371.1,89.1) {\scriptsize$j$}
  \put(384.1,-7.6) {\scriptsize$i$}
  \epicture19 \labl{eq:rel-3pt-cpl}
as well as
  \be \Gs ijk\ell pq = \frac{\Rs{}jkq\, \Rs{}iq\ell}{\Rs{}ijp\, \Rs{}pk\ell}\,
  \Fs kji\ell pq \;.  \ee
The first equation, for instance, follows by composing with the three-point 
coupling dual to the \rhs\ and noticing that the \lhs\ thereby becomes a 
special case ($q\eq\One$) of the graph on the \lhs\ of \erf{fmat} that 
gives the general \FF-matrix element. The last equation relating the 
\FF-matrix to its inverse can be seen by following the sequence of
moves indicated below (this is nothing but the hexagon identity):
  \begin{eqnarray}  \begin{picture}(400,69)(26,0)
  \put(0,0)   {\begin{picture}(0,0)(0,0)
              \scalebox{.38}{\includegraphics{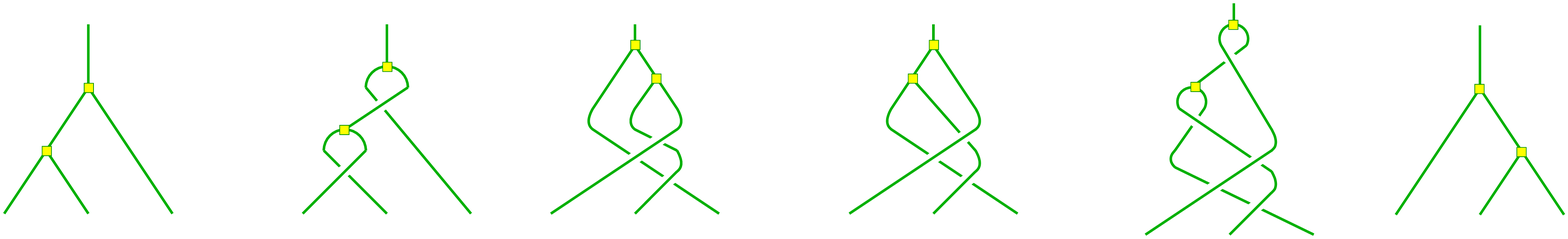}} \end{picture}}
   \put(49,30) {$\stackrel{\RR^- \,\cdot\, \RR^-}{{-}\!\!{-}\!\!\!\longrightarrow}$}
   \put(143,30) {$=$}
   \put(224,30) {$\stackrel{\FF}{\longrightarrow}$}
   \put(300,30) {$\stackrel{\RR^+ \,\cdot\, \RR^+}{{-}\!\!{-}\!\!\!\longrightarrow}$}
   \put(392,30) {$=$}
   \put(0,-1) {\scriptsize$i$}
   \put(24,-1) {\scriptsize$j$}
   \put(49,-1) {\scriptsize$k$}
   \put(87,-1) {\scriptsize$i$}
   \put(112,-1) {\scriptsize$j$}
   \put(137,-1) {\scriptsize$k$}
   \put(160,-1) {\scriptsize$i$}
   \put(184,-1) {\scriptsize$j$}
   \put(209,-1) {\scriptsize$k$}
   \put(247,-1) {\scriptsize$i$}
   \put(272,-1) {\scriptsize$j$}
   \put(297,-1) {\scriptsize$k$}
   \put(408,-1) {\scriptsize$i$}
   \put(431,-1) {\scriptsize$j$}
   \put(457,-1) {\scriptsize$k$}
   \put(334,-6) {\scriptsize$i$}
   \put(358,-6) {\scriptsize$j$}
   \put(384,-6) {\scriptsize$k$}
   \put(25,65) {\scriptsize$l$}
   \put(112,65) {\scriptsize$l$}
   \put(184,65) {\scriptsize$l$}
   \put(272,65) {\scriptsize$l$}
   \put(360,72) {\scriptsize$l$}
   \put(432,65) {\scriptsize$l$}
   \put(15.3,36.2) {\scriptsize$p$}
   \put(121.1,42) {\scriptsize$p$}
   \put(191.1,53.4) {\scriptsize$p$}
   \put(266,54) {\scriptsize$q$}
   \put(368.2,57) {\scriptsize$q$}
   \put(442,36) {\scriptsize$q$}
  \end{picture} \nonumber\\[.23em]{} \end{eqnarray}

The $S$-matrix can then be expressed through the \FF- and \RR-matrices by 
the moves in the following figure: 
  \begin{eqnarray}&&  \begin{picture}(444,80)(26,0)
  \put(100,0)   {\begin{picture}(0,0)(0,0)
              \scalebox{.38}{\includegraphics{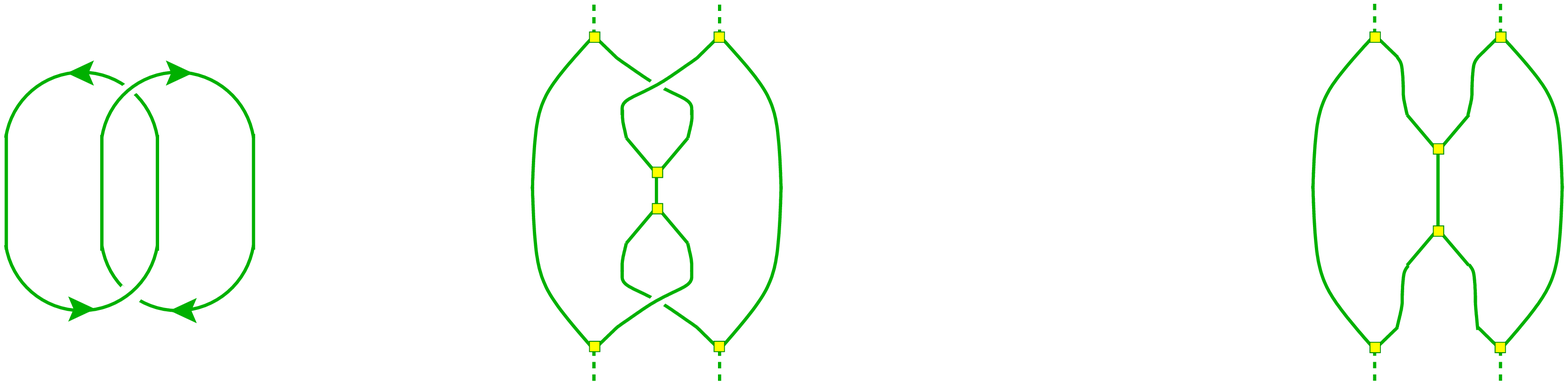}} \end{picture}}
  \put(2.8,43)     {$\dsty\frac1{\dim(U_i)\,\dim(U_j)}$}
  \put(94.7,32.6)  {\scriptsize$j$}
  \put(161.1,32.6)  {\scriptsize$i$}
  \put(175,43)     {$=\ \dsty\sumI_k$}
   \put(218,26) {\scriptsize$\bar\jmath$}
   \put(281,26) {\scriptsize$\bar\imath$}
   \put(244,43) {\scriptsize$k$}
   \put(239,59) {\scriptsize$i$}
   \put(260,59) {\scriptsize$j$}
   \put(239,29) {\scriptsize$i$}
   \put(260,29) {\scriptsize$j$}
  \put(297,43)     {$=\;\ \dsty\sumI_k\Rs{}ijk\,\Rs{}jik$}
   \put(398,26) {\scriptsize$\bar\jmath$}
   \put(462,26) {\scriptsize$\bar\imath$}
   \put(434,43) {\scriptsize$k$}
   \put(418.8,65) {\scriptsize$j$} 
   \put(441.5,65) {\scriptsize$i$}
   \put(418.8,21) {\scriptsize$j$}
   \put(442.1,21) {\scriptsize$i$}
  \end{picture} \nonumber\\[-.17em]{} \label{s-rgf} \end{eqnarray}
Using now the two relations in \erf{eq:rel-3pt-cpl} as well as
\erf{eq:RR=th3}, this results in 
  \be  S_{i,j} = S_{\One,\One}\,
  \dim(U_i)\, \dim(U_j)\, \sum_{k\in\II} \frac{\theta_k}{\theta_i\,\theta_j}
  \,\Gs{\bar \jmath}jii\One k\, \Fs{\bar \jmath}jiik\One \,.  \ee

Recall that we chose the basis in the spaces of three-point couplings in 
such a way that an \FF-matrix is equal to one (see \erf{eq:fnorm}) when 
one of the `ingoing' objects is the tensor unit $\one$. It is convenient 
to have a similar behavior when the `outgoing' object is $\one$. The 
following lemma implies that we can always make choices such that
$\Fs ijk\One{\bar\imath}{\bar k}{=}\, p_i p_j p_k$:
\vspace{-.9em}

\dt{Lemma} 
\mbox{$\ $}\\[-.99em]
If $\dim\,\Hom(X{\otimes}Y,Z)\iN\{0,1\}$ for all simple objects
$X,Y,Z\in\objc$, then there is a choice of basis in the spaces of
three-point couplings such that 
  \be  \Fs ijk\ell pq=1  \ee
whenever the $\FF$-matrix element is allowed to be non-zero by the 
fusion rules and one or more of the $i,j,k$ are $\One$, and that
  \be  F_{ijk} \equiv \Fs ijk\One {\bar \imath}{\bar k} = p_i\, p_j\, p_k
  \,.  \ee

\smallskip\noindent
Proof:\\ Recall the conventions \erf{kokok} and \erf{xy-pi} that we have
made already. For the sake of this proof, we will specialise
the choice of basis further such that $\lambda_k\eq\lambda_{\bar k}$.  
By definition of \FF\ we have
  \bea \begin{picture}(200,58)(0,29)
  \put(0,0)   {\begin{picture}(0,0)(0,0)
              \scalebox{.38}{\includegraphics{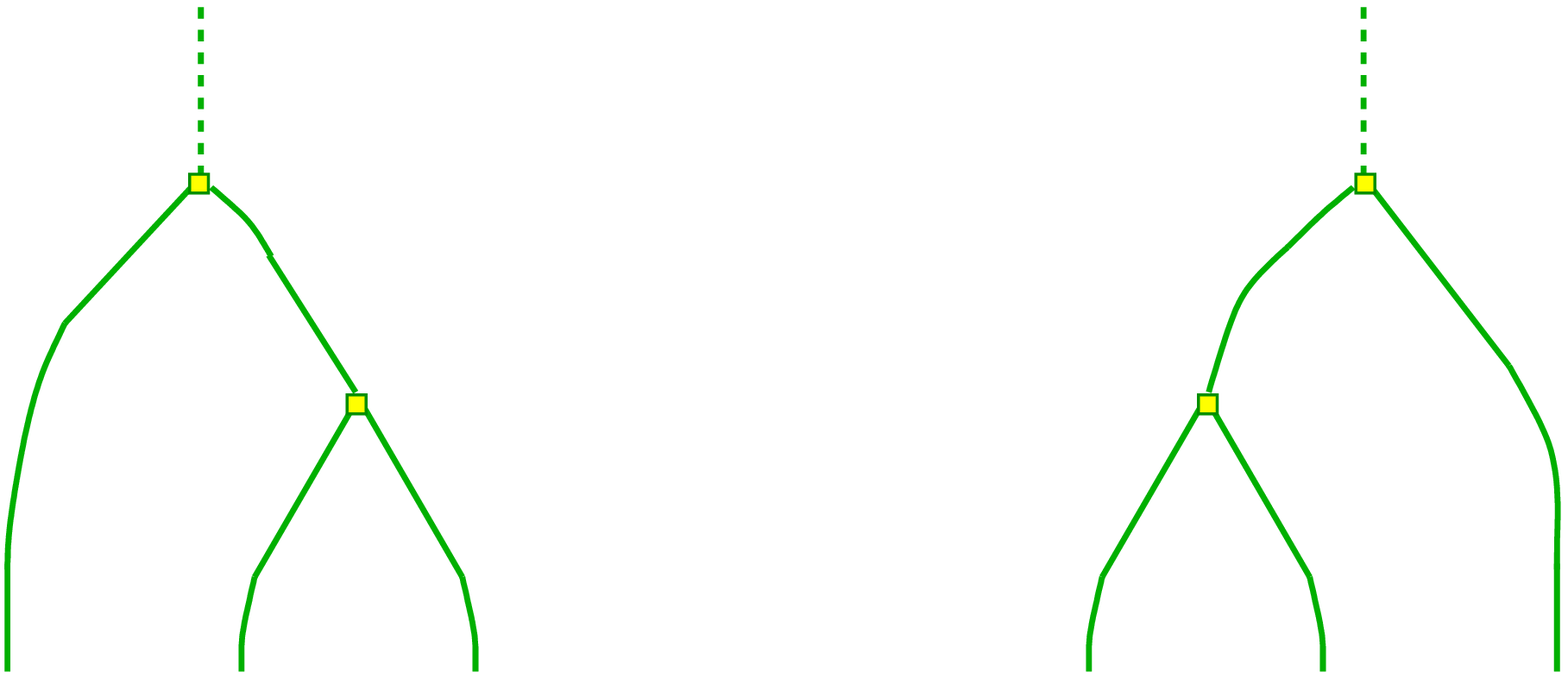}} \end{picture}}
  \put(-1.3,-7.5)  {\scriptsize$k$}
  \put(28.8,-7.5)  {\scriptsize$i$}
  \put(39.6,48.9)  {\scriptsize$\bar k$}
  \put(57.2,-7.5)  {\scriptsize$j$}
  \put(90,37)      {$=\;\ F_{kij}$}
  \put(134.8,-7.5) {\scriptsize$k$}
  \put(153.7,48.9) {\scriptsize$\bar\jmath$}
  \put(166.4,-7.5) {\scriptsize$i$}
  \put(195.5,-7.5) {\scriptsize$j$}
  \epicture21 \label{eq:Fijk-def}\labl{fijk}
Then we have the relation
  \bea \begin{picture}(195,60)(0,30)
  \put(0,0)   {\begin{picture}(0,0)(0,0)
              \scalebox{.38}{\includegraphics{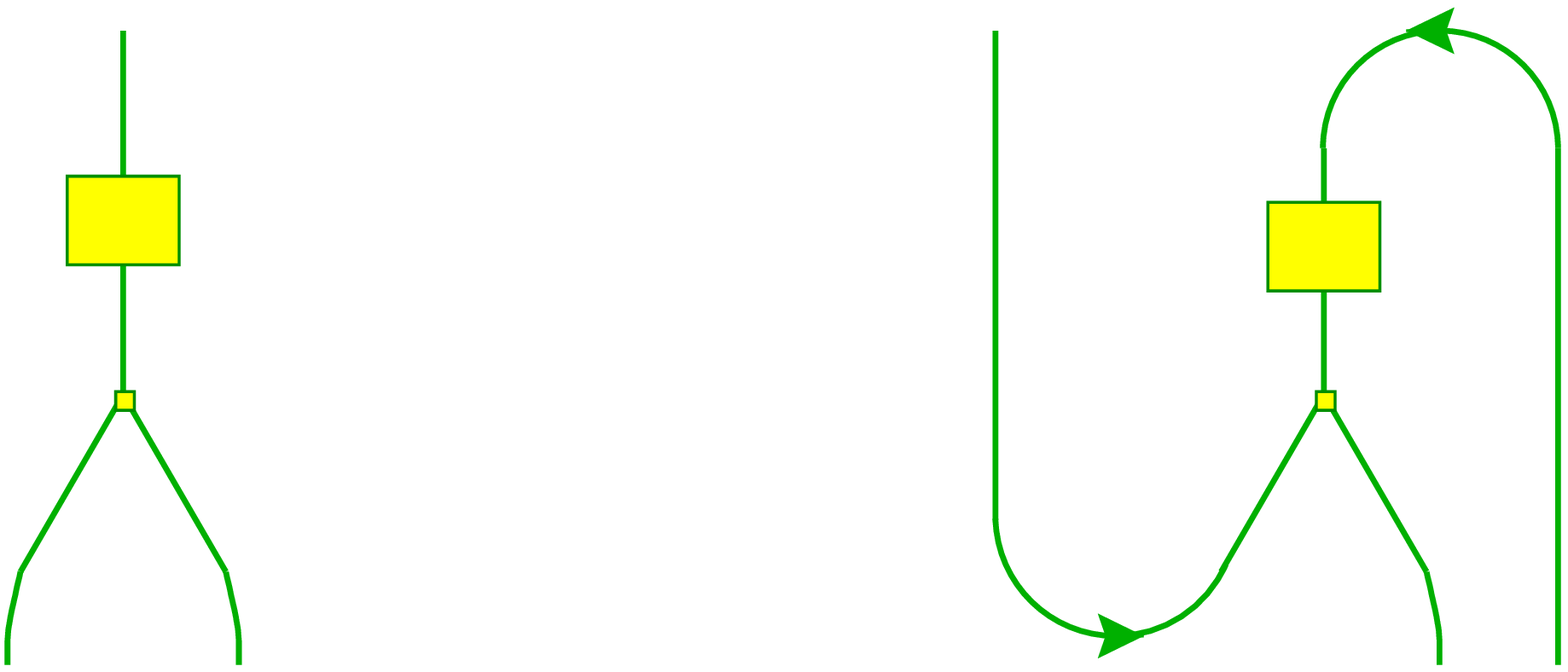}} \end{picture}}
  \put(-1.3,-7.5)  {\scriptsize$i$}
  \put(12.3,86.4)  {\scriptsize$k\Vee$}
  \put(11.3,57)    {\scriptsize$\pi_{\bar k}$}
  \put(28.8,-7.5)  {\scriptsize$j$}
  \put(50,40)      {$=\,\ p_k\,\dsty\frac{\lambda_j}{\lambda_k}\,F_{kij}$}
  \put(124.7,86.4) {\scriptsize$k\Vee$}
  \put(166.3,54)   {\scriptsize$\pi_{\bar\jmath}$}
  \put(183.5,-7.5) {\scriptsize$i$}
  \put(198.5,-7.5) {\scriptsize$j$}
  \epicture21 \labl{fijk-rel} 
which when iterated results in
  \be   F_{kij}\,F_{jki}\,F_{ijk} = p_i\, p_j\, p_k \,.
  \labl{eq:Fijk-rel}
When the three coupling spaces that form an orbit upon iteration,
i.e.\ $\Hom(i{\otimes}j,{\bar k})$, $\Hom(k{\otimes}i,{\bar \jmath})$ and
$\Hom(j{\otimes}k,{\bar \imath})$, are mutually distinct, then we
can link the choice of basis in the latter two to the choice for the
first in such a manner that $F_{kij}\eq F_{jki}\eq p_i p_j p_k$. Relation
\erf{eq:Fijk-rel} then implies that $F_{ijk}\eq p_i p_j p_k$ as well.
\\
That the three coupling spaces are not mutually distinct can happen
only if $i\eq j\eq k$. Let us abbreviate from here on $F_{iii}$ by $F_i$. 
Define the two numbers $B_i$ and $R_i$ via
  \bea \begin{picture}(370,60)(0,31)
  \put(0,0)   {\begin{picture}(0,0)(0,0)
              \scalebox{.38}{\includegraphics{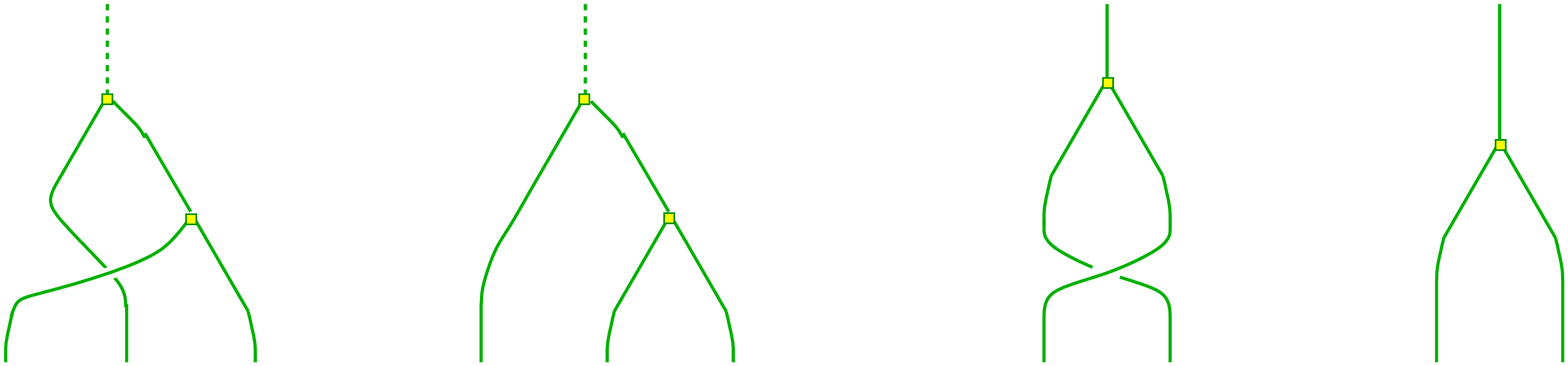}} \end{picture}}
  \put(-1.0,-7.5)  {\scriptsize$i$}
  \put(27.5,-7.5)  {\scriptsize$i$}
  \put(40.3,48.5)  {\scriptsize$\bar\imath$}
  \put(57.7,-7.5)  {\scriptsize$i$}
  \put(74,35)      {$=\,\ B_i$}
  \put(111.3,-7.5) {\scriptsize$i$}
  \put(141.3,-7.5) {\scriptsize$i$}
  \put(145.3,48.5) {\scriptsize$\bar\imath$}
  \put(171.3,-7.5) {\scriptsize$i$}
  \put(244.6,-7.5) {\scriptsize$i$}
  \put(259.8,88.5) {\scriptsize$\bar\imath$}
  \put(273.8,-7.5) {\scriptsize$i$}
  \put(296,35)     {$=\,\ R_i$}
  \put(337.9,-7.5) {\scriptsize$i$}
  \put(352.5,88.5) {\scriptsize$\bar\imath$}
  \put(366.6,-7.5) {\scriptsize$i$}
  \epicture21 \labl{def-b-r} 
The number $B_i$ exists because the space $\Hom(U_i\oti U_i \oti U_i, \one)$ has
dimension one and $R_i$ is just $\Rs{}ii{\bar\imath}$, compare formula \erf{rmat}.

Next note the relations
  \be F_i F_i F_i = p_i \,, \qquad
  F_i = p_i \theta_i B_i R_i \,, \qquad
  B_i B_i = 1/\theta_i \,, \qquad
  R_i R_i = 1/\theta_i \,.  \labl{eq:FRB-rel}
The second relation, for instance, is obtained by the moves
(we omit the obvious labels)
  \bea  \begin{picture}(403,47)(0,39)
  \put(0,0)   {\begin{picture}(0,0)(0,0)
              \scalebox{.38}{\includegraphics{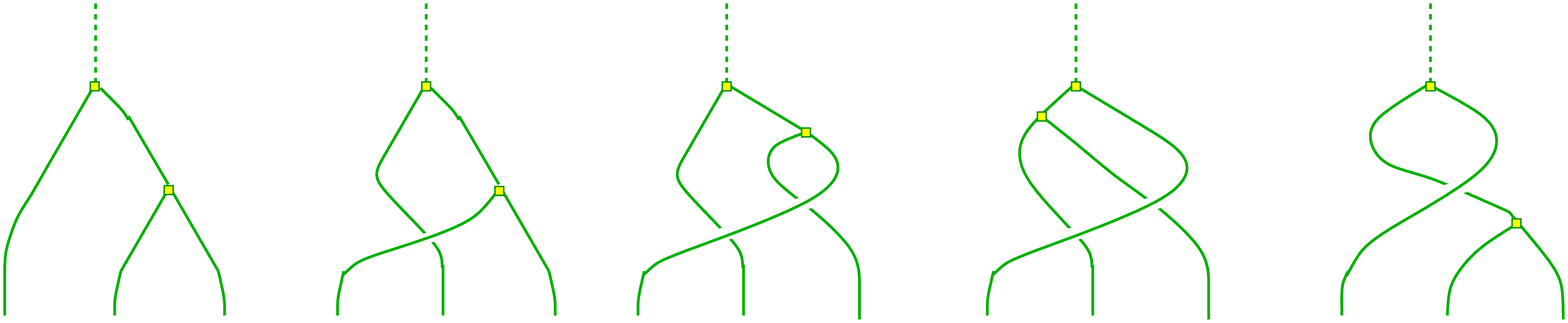}} \end{picture}}
  \put(-24,38)     {$R_iB_i$}
  \put(64,38)      {$=\; R_i$}
  \put(153,38)     {$=$}
  \put(238,38)     {$=\; F_i$}
  \put(331,38)     {$=\; F_i$}
  \epicture18 \labl{f-br}
together with formula \erf{eq:Rkk0=pllt} for $\Rs{}i{\bar\imath}\One$.
Using the relations \erf{eq:FRB-rel}, it follows that
  \be
  F_i F_i = (p_i \theta_i B_i R_i)^2 = 1 \,,
  \ee
and thus $p_i\eq F_i F_i F_i\eq F_i$, which proves the lemma. 
\qed

\subsubsection{Example: Free boson}

The first example is the modular category associated to the chiral data 
of a free boson field compactified on a circle of rational radius squared. 
The generic chiral algebra of this theory is $\hat{\mathfrak u}(1)$.
If the compactification radius $R$ fulfills (in the normalisation we 
choose) $R^2\eq p/q$ with $p,q$ coprime, then $\hat{\mathfrak u}(1)$ can 
be extended by a pair of Virasoro primary fields of weight $N\eq pq$.

The modular category for the Moore\hy Seiberg data of this theory can
be presented as follows (see e.g.\ \cite{FRke}). There is a simple object
$U_k$ for each integer $k$. The simple objects $U_k$ and $U_{k+2N}$ are
isomorphic, so that the category has $2N$ isomorphism classes 
of simple objects; we label them as $\I\eq\{0,1,\dots,2N{-}1\}$. 
(Note that we should not ignore these isomorphisms and pretend that 
$U_k$ and $U_{k+2N}$ are equal. This is, for instance, in line with the 
fact that the operator product of two fields of ${\mathfrak u}(1)$ charge
$j/\sqrt{2N}$ and $k/\sqrt{2N}$ in the standard range, i.e.\ $j,k\iN\II$,
does not contain any field with charge in the standard range when 
$j{+}k\,{\ge}\,2N$.)
The conformal weight of the primary field labelled by $k\iN\I$, whose 
non-integral part determines the twist of the object $U_k$, is
  \be
  \Delta(k) = \Bigg\{ \begin{array}{ll}
  \textstyle{\frac{1}{4N}}\, k^2  & {\rm for}\ k\,{\le}\,N\,, \\{}\\[-.8em]
  \textstyle{\frac{1}{4N}}\, (2N{-}k)^2  & {\rm for}\ k\,{>}\,N\,. \end{array}
  \labl{rq:Z2N-Delta}

The representation $k$ of the extended chiral algebra contains 
$\hat{{\mathfrak u}}(1)$-representations of charges 
$q(k)\eq (k{+}2Nm)/\sqrt{2N}$ for every integer $m$ (in the convention 
$\Delta(k)\eq (q(k))^2/2$).  {}From the conformal weights we read off that
  \be
  \theta_k = \eE^{-2\pi \ii \Delta(k)} = \eE^{-\pi \ii k^2 /2N}\,.  \ee
To give the braiding and fusing matrices, it is useful to 
introduce the function $[\,\cdot\,]{:}\ \mathbb{Z}\,{\rightarrow}\,\I$ that
associates to $n$ the element of \I\ with which it coincides modulo $2N$,
i.e.\ $[n]\eq n-2N\,\sigma(n)$ with $\sigma(n)\,{\equiv}\,(n{-}[n])/2N$ the
unique integer such that $n\,{-}\,2N\sigma(n)$ is in \I. Then the 
fusion rules, which furnish the abelian group $\mathbb{Z}_{2N}$, read
  \be [i] \star [j] = [i{+}j] \,, \ee
and with our conventions for the coupling spaces
the non-zero braiding and fusing matrices are given by
  \be \begin{array}{l}
  \Fs ijk{[i{+}j{+}k]}{[j{+}k]}{[i{+}j]} = 
  (-1)^{(i{+}k{+}1)\{j\sigma(i{+}j{+}k)\,+\,(j+k)(\sigma(i+j)+\sigma(j+k))\}}\,, 
  \\{}\\[-.5em]
  \Rs{}k\ell{[k{+}\ell]} = (-1)^{(k+\ell)\sigma(k+\ell)} 
  \, \eE^{- \pi \ii k \ell /(2N) } \,.  \end{array} \labl{eq:Z2N-F}
Two points should be noted: First, had we chosen another way to represent the
isomorphism classes of simple objects, these formulas would look different. 
In particular, they depend on our specific choice for the set \I, and
indeed are not invariant under the shift $n\,{\mapsto}\,n{+}2N$.  
Second, by redefining the bases of the coupling spaces 
$\Hom(j{\otimes}k,[j{+}k])$ by a factor
$(-1)^{j(k+1)\sigma(j+k)}$, the \FF's could be made to look 
much simpler, namely $\Fs ijk{[i{+}j{+}k]}{[j{+}k]}{[i{+}j]}\eq
(-1)^{\sigma(j{+}k)\,i}$, which is the form used in \cite{brSc2}.
This gauge does not, however, fulfill $\Fs ijk\One{\bar\imath}{\bar k}
\eq p_i p_j p_k$ (which we use in simplifying the explicit expressions 
in the examples treated), whereas \erf{eq:Z2N-F} does.

\subsubsection{Example: \slz\ WZW models} \label{sec:modcat-su2}

The chiral algebra of the \slz\ WZW model at level $k$ is the affine Lie 
algebra $\slz_k$. It has $k{+}1$ isomorphism classes of integrable highest
weight modules, which we label by their highest weights (twice the spin):
$(0),\,(1),...\,,(k)$. The corresponding \con weights read 
  \be \Delta(n) = \frac{n\,(n+2)}{4\,(k+2)} \,,  \ee
and the quantum dimensions are
  \be \dim\,(n) = S_{0,n} \,/\, S_{0,0} = 
  {\textstyle \sin\big(\Frac{\pi(n+1)}{k+2}\big) \,/\, 
  \sin\big(\Frac{\pi}{k+2}\big) } \,.  \ee
The fusion rules read
  \be (m) \star (n) = \sum_{p=|m-n|}^{\min(m+n,2k-m-n)} (p) \,, \ee
where $p$ is increased in steps of 2.

The concrete expressions for the braiding and fusing matrices depend on
the chosen normalisation of the chiral vertex operators.
We will use the form in which they naturally arise as quantum group 6j-symbols,
as given e.g.\ in \cite{kire6,hswy}:
  \be
  \Rss rst = (-1)^{(r+s-t)/2} \, \eE^{-\ii\pi( \Delta(t)-\Delta(r)-\Delta(s) )} 
  \,, \qquad
  \Fs rstupq = \Big\{ \begin{array}{ccc} \!\!\scs t/2 \!\!&\! \scs s/2
  \!\!&\! \scs p/2 \!\!\!\\[-.2em]
  \!\!\scs r/2 \!\!&\! \scs u/2 \!\!&\! \scs q/2 \!\!\!\end{array} \Big\}_{\!q}
  \labl{eq:su2-RFmat}
with
  \be\begin{array}{rr}
  \Big\{ \begin{array}{ccc} \!\!\scs a \!\!&\! \scs b \!\!&\! \scs e \!\!\!
  \\[-.2em]
  \!\!\scs d \!\!&\! \scs c \!\!&\! \scs f \!\!\!\end{array} \Big\}_{\!q} 
  \!\!\!& := (-1)^{a+b-c-d-2e}
  \big( [2e{+}1]\,[2f{+}1] \big)^{1/2}
  \Delta(a,b,e)\,\Delta(a,c,f)\,\Delta(c,e,d)\,\Delta(d,b,f)\; \\
  & \times 
  \sum_{z} (-1)^z \,[z{+}1]!\; \Big( \,
  [z{-}a{-}b{-}e]! \, [z{-}a{-}c{-}f]! \, [z{-}b{-}d{-}f]! \, [z{-}d{-}c{-}e]! \\
  & \times\,
  [a{+}b{+}c{+}d{-}z]! \, [a{+}d{+}e{+}f{-}z]! \, [b{+}c{+}e{+}f{-}z]!
  \, \Big)^{\!-1} \eear \ee
and
  \be  \Delta(a,b,c) :=  \sqrt{
  [-a{+}b{+}c]! \, [a{-}b{+}c]! \, [a{+}b{-}c]! \, / \,
  [a{+}b{+}c{+}1]! } \,. \ee
The symbols $[n]$ and $[n]!$ stand for $q$-numbers and $q$-factorials,
\resp, i.e.
  \be
  [n] = \frac{ \sin\big( {\textstyle\frac{\pi n}{k+2}} \big) }
  { \sin\big( {\textstyle\frac\pi{k+2}} \big) }\, , \qquad
  [n]! = \prod_{m=1}^n [m] \, , \qquad [0]! = 1 \,.  \ee 
(The fusion rules ensure that the numbers that appear as arguments of
$[\,\cdot\,]$ are always integral.) The
range of the summation is such that the arguments are non-negative, 
i.e.\ $z$ runs over all integers (in steps of 1) from
$\max(a{+}b{+}e,a{+}c{+}f,b{+}d{+}f,c{+}d{+}e)$ to 
$\min(a{+}b{+}c{+}d,a{+}d{+}e{+}f,$\linebreak[0]$b{+}c{+}e{+}f)$.  

It is worth pointing out that, while these expressions solve the
pentagon and hexagon identities, to find the normalisation of WZW 
conformal blocks such that their transformation behaviour under 
fusion and braiding is described by the matrices \erf{eq:su2-RFmat} 
is a non-trivial task. In the sequel, however,
we will mainly be interested in partition functions; these are 
independent of the normalisation of the chiral vertex operators.  

%%%%%%%%%%%%%%%%%%%%%%%%%%%%%%%%%%%%%%%%%%%%%%%%%%%%%%%%%%%%%%%%%%%%%%%%
\newpage

\sect{Frobenius objects and algebras of open string states} 

%%%%%%%%%%%%%%%%%%%%%%%%%%%%%%%%%%%%%%%%%%%%%%%%%%%%%%%%%%%%%%%%%%%%%%%%

\subsection{Algebra objects} \label{sec:alg-obj}

In the construction of correlation functions a central role will 
be played by objects with additional structure in the modular category 
of the chiral CFT, namely so-called algebra objects. 
It is quite natural to study such extra structures for objects of a \tc;
for some category theoretic background, see \cite[chapter VI]{MAcl}
and \cite[chapters\,2,3]{pare28}.
In short, an \alg\ in a \tc\ \calc\ is an object $A\iN\objc$ 
together with a product, i.e.\ a morphism between $A\oti A$ and $A$, 
that is associative and has a unit. More concretely:
\vspace{-.9em}

\dtl{Definition}{def:alg-obj}
An {\em algebra object\/}, or simply an {\em algebra\/}, in a \tc\
\calc\ is a triple $(A,m,\eta)$, where $A$ is an object of \calc, 
$m\iN\Hom(A\Oti A,A)$ and $\eta\iN\Hom(\one,A)$, such that the
multiplication morphism $m$ and the unit morphism $\eta$ fulfill 
  \be
  m\circ (m\oti\id_A) = m \circ (\id_A\oti m) \qquad {\rm and} \qquad
  m \circ(\eta\oti \id_A) = \id_A = m \circ (\id_A\oti \eta) \,.
  \labl{eq:alg}

\medskip\noindent
Our pictorial notation for $m$ and $\eta$ and the resulting pictures 
expressing formulas \erf{eq:alg} are as follows:
  \begin{eqnarray} \begin{picture}(397,60)(0,0)
  \put(0,0)   {\begin{picture}(0,0)(0,0)
              \scalebox{.38}{\includegraphics{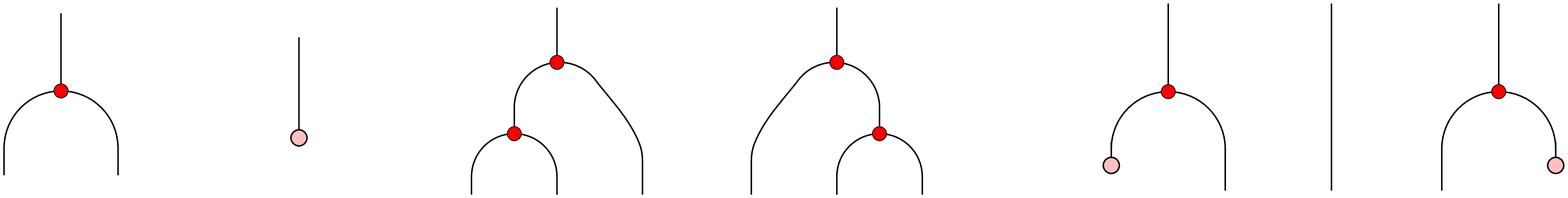}} \end{picture}}
  \put(-30,25)    {$m\,=$}
  \put(-3.5,-4.1) {\scriptsize$A$}
  \put(12.8,52.5) {\scriptsize$A$}
  \put(26.5,-4.1) {\scriptsize$A$}
  \put(50.6,25)   {$\eta\,=$}
  \put(76.7,3.7)  {\scriptsize$\one$}
  \put(76.1,45.7) {\scriptsize$A$}
  \put(121.1,-8.4){\scriptsize$A$}
  \put(143.8,-8.4){\scriptsize$A$}
  \put(144.8,53.8){\scriptsize$A$}
  \put(166.0,-8.4){\scriptsize$A$}
  \put(180.2,25)  {$=$}
  \put(194.6,-8.4){\scriptsize$A$}
  \put(217.5,-8.4){\scriptsize$A$}
  \put(218.5,53.8){\scriptsize$A$}
  \put(240.5,-8.4){\scriptsize$A$}
  \put(307.2,54.8){\scriptsize$A$}
  \put(320.6,-7.7){\scriptsize$A$}
  \put(333.2,25)  {$=$}
  \put(349.6,-7.7){\scriptsize$A$}
  \put(350.3,54.8){\scriptsize$A$}
  \put(365.4,25)  {$=$}
  \put(378.7,-7.7){\scriptsize$A$}
  \put(394.3,54.8){\scriptsize$A$}
  \end{picture} \nonumber\\[-.87em]{} \label{m-eta} \end{eqnarray} 
Later on, we will often suppress the label $A$ on the algebra lines
when this is unambiguous, e.g.\ when the presence of morphisms depicted by
small circles -- products or units -- indicates that we are
dealing with morphisms involving only $A$ (and possibly $\one$).

To call $A$ an algebra is appropriate because in the particular
case that \calc\ is the category of vector spaces over \complex\ (or some
other field), the prescription reduces to the conventional notion of an \alg.
In every \tc\ the tensor unit $\one$ provides a trivial example of an \alg;
its product is $m\eq\id_\smallone\,{\equiv}\,\id_{\smallone\otimes\smallone}
\eq\id_\smallone\oti\id_\smallone$, and its unit is $\eta\eq\id_\smallone$. 
A class of less trivial examples, present in any \tc\ with duality, is given 
by objects of the form $A\eq U\oti U\Vee$; in these cases one can take 
  \be  m = \id_U\oti d_U\oti\id_{U\Vee} \qquad{\rm and}\qquad 
  \eta = b_U \,.  \labl{mUU}
However, it turns out that many interesting \alg s are not 
of this special form.

Consider now an \alg\ $A$ in a semisimple \tc\ \calc. The object $A$ is
a finite direct sum of simple subobjects $U_i$. 
We fix once and for all bases $\{\iaa i\alpha\}$ in the embedding spaces 
$\Hom(U_i,A)$, as well as dual bases $\{\aai i\alpha\}$ in $\Hom(A,U_i)$.
We draw these morphisms as
  \bea \begin{picture}(200,46)(0,23)
  \put(0,0)   {\begin{picture}(0,0)(0,0)
              \scalebox{.38}{\includegraphics{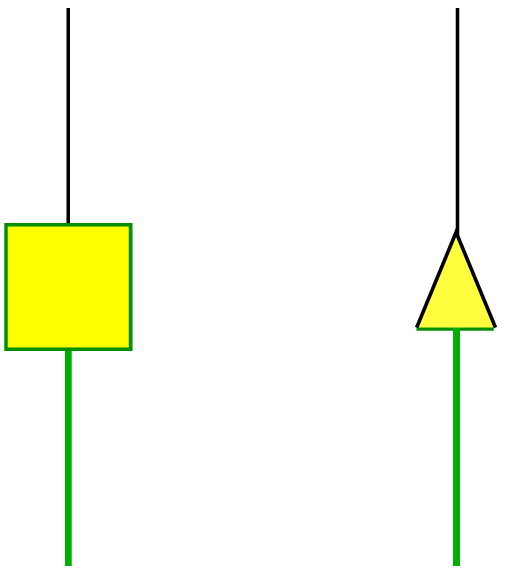}} \end{picture}}
  \put(140,0) {\begin{picture}(0,0)(0,0)
              \scalebox{.38}{\includegraphics{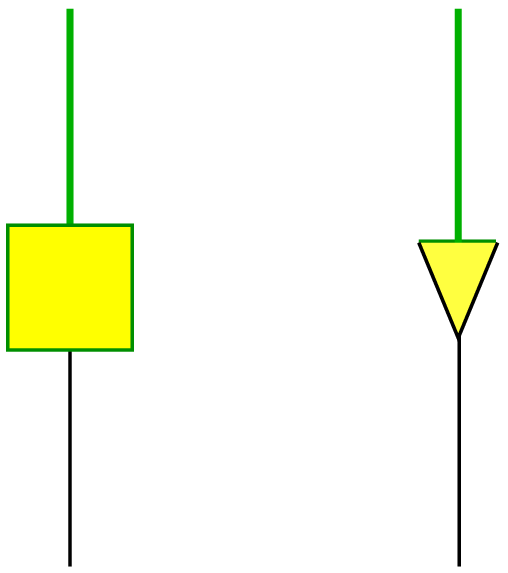}} \end{picture}}
  \put(2.2,29.5)  {\scriptsize$\iaa i\alpha$}
  \put(4.3,65.3)  {\scriptsize$A$}
  \put(5.3,-7)    {\scriptsize$i$}
  \put(25.2,28.5) {$=:$}
  \put(48.6,-7)   {\scriptsize$i$}
  \put(47.6,65.3) {\scriptsize$A$}
  \put(48.1,28.4) {\tiny$\alpha$}
  \put(90.2,28.5) {and}
  \put(142.3,29.2){\scriptsize$\aai i\alpha$}
  \put(143.6,-7.8){\scriptsize$A$}
  \put(145.7,65.3){\scriptsize$i$}
  \put(164.2,28.5){$=:$}
  \put(188.4,65.3){\scriptsize$i$}
  \put(186.1,-7.8){\scriptsize$A$}
  \put(188.2,32.2){\tiny$\bar\alpha$}
  \epicture13 \labl{iaa,aai}
In accordance with the notation \Erf lr,
the dimensions of these coupling spaces will be abbreviated by
  \be  \Ai i:=\dim\,\Hom(U_i,A)=\dim\,\Hom(A,U_i)\,. \labl{Ai}
Thus as an object we have $A \,{\cong}\, \bigoplus_{k\in\II}\!
U_k^{\oplus \Ai k}$. The unit morphism $\eta$ is non-zero, and hence 
  \be  \Ai\One = \dim\,\Hom(\one,A) \ge 1 \,.  \ee

We may express the product $m$ \wrt the bases $\{\iaa i\alpha\}$ of
$\Hom(U_i,A)$ and $\{\aai i\alpha\}$ of $\Hom(A,U_i)$, for $i\eq a,b,c$,
as compared to the basis $\{\x abc\delta\}$ of $\Hom(U_a\oti U_b,U_c)$ that
we introduced in \erf{xijk,yijk}. This way we characterise $m$ by the 
collection of numbers $\m a\alpha b\beta c\gamma\delta$ that are defined as
  \bea \begin{picture}(170,54)(0,27)
  \put(0,0)   {\begin{picture}(0,0)(0,0)
              \scalebox{.38}{\includegraphics{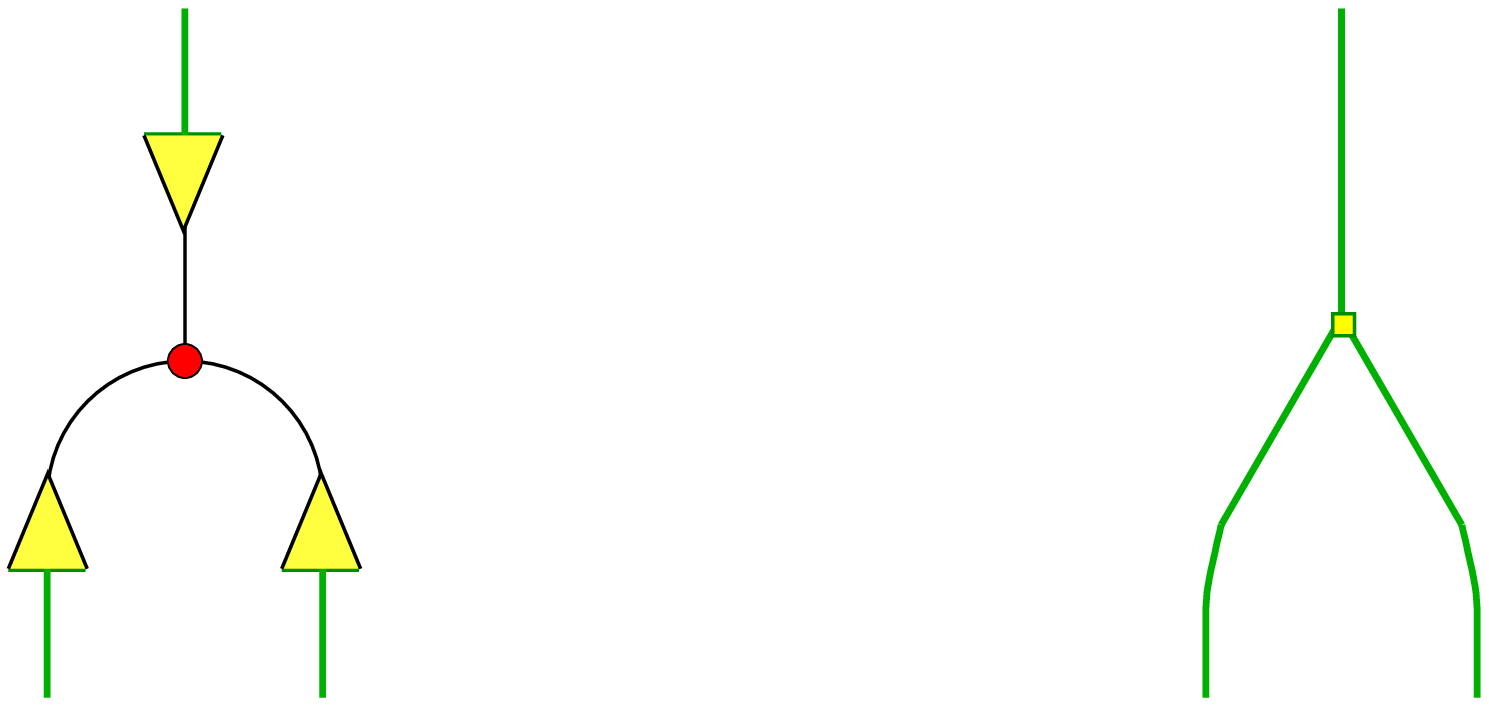}} \end{picture}}
  \put(2.8,-7)    {\scriptsize$a$}
  \put(2.9,16.7)  {\tiny$\alpha$}
  \put(18.9,81.3) {\scriptsize$c$}
  \put(18.1,58.7) {\tiny$\bar\gamma$}
  \put(33.8,-8)   {\scriptsize$b$}
  \put(32.8,16.7) {\tiny$\beta$}
  \put(56.2,32)   {$=\,\ \dsty\sum_{\delta=1}^{\N ab{\,c}}
                   \m a\alpha b\beta c\gamma\delta$}
  \put(130.4,-7)  {\scriptsize$a$}
  \put(146.6,81.3){\scriptsize$c$}
  \put(151.1,40.2){\tiny$\delta$}
  \put(160.7,-8)  {\scriptsize$b$}
  \epicture15 \label{eq:m-in-basis} \labl{def-mabc}
Upon use of formula \erf{def-fmat} one finds that
in terms of these numbers $\m a\alpha b\beta c\gamma\delta$,
the associativity property of $m$ is expressed as
  \be
  %% \sum_f \sum_{\rho=1}^{\N abf}\sum_{\sigma=1}^{\N fcd}
  \sum_{\varphi=1}^{\Ai f} \m a\alpha b\beta f\varphi\rho\,
  \m f\varphi c\gamma d\delta\sigma
  = \sum_{e\in\II} \sum_{\varepsilon=1}^{\Ai e}\sum_{\varrho=1}^{\N bce}
    \sum_{\tau=1}^{\N aed}
  \m b\beta c\gamma e\varepsilon\varrho\, \m a\alpha e\varepsilon d\delta\tau\,
  \F{a\U b}cdef\tau\varrho\rho\sigma \,.  \labl{eq:general-assoc-m-basis} 
This shows that the concept of associativity we are dealing with is most
natural indeed: When expressing the associativity property in terms of bases,
it involves precisely the \FF-matrices, a behavior that is familiar from
the \rep\ theory of \alg s over $\complex$, like e.g.\ universal enveloping
\alg s of simple Lie \alg s.

\subsection{From boundary conditions to algebras} \label{sec:alg2bCFT}

We will now argue that each boundary condition of a rational CFT 
determines an algebra in the modular category for that CFT. To explain 
this relationship, let us start on the CFT side, i.e.\ with some 
boundary condition $M$.

For definiteness take the world sheet geometry to be the upper half
plane. Then the boundary is the real line $\reals$.
Let $M$ be a boundary condition that preserves the chiral 
algebra \CA. We do not assume, however, that \CA\ is maximally
extended, so $M$ may break some of the symmetries of the model;
but we do assume that the theory is still rational \wrt \CA.

The space \calh\ of states that corresponds to boundary fields living 
on a boundary segment with boundary condition $M$ -- first introduced  
in \cite{card9} -- is often referred to as the `state space
on the upper half plane with boundary condition $M$' or, in string
terminology, as the space of open string states for open strings
stretching from one D-brane to itself. \calh\ is
organised in modules of one copy of the chiral \alg\ \CA, rather than in
modules of $\CA\oti\CA$ as for the bulk:
  \be  \calh = \bigoplus_{a\in\II} n^a\, \Calha \,.  \labl{eq:bnd-H}
The sum runs over (representatives for the isomorphism classes of) 
irreducible highest weight modules
$\Calha$ of \CA, and $n^a\iN\zet_{\ge0}$ specifies their multiplicities.
In the CFTs we consider, the representation theory of \CA\ gives rise
to a modular category \calc\ whose simple objects are the irreducible 
\CA-modules. The notation $\Calha$ for these modules is chosen so as
to conform with our convention for the representatives for isomorphism 
classes of simple objects in \calc, see section \ref{sec:mod-cat} above.

There is a one-to-one correspondence between states in $\calh$ and
fields that live on the boundary $M$. Let us denote primary boundary
fields (i.e.\ those that correspond to highest weight states) on $M$ as 
$\Psi_{a\alpha}(x)$. Here $a$ labels the \CA-representation, while 
$\alpha$ is a multiplicity index that runs from 1 to $n^a$. Since $\Psi$ 
is a boundary field, its argument $x$ takes values in $\reals$ only. 

In simple cases, for instance in minimal models, the operator product
expansion of boundary fields takes the form
  \be  \Psi_{a\alpha}(x)\,\Psi_{b\beta}(y) =
  \sum_{c,\gamma} C_{a\alpha,b\beta}^{\;\;c\gamma}\,
  (x{-}y)^{\Delta_c-\Delta_a-\Delta_b}\, \llb\, \Psi_{c\gamma}(y) + 
  {\rm terms\ with\ descendants} \,\lrb\, .  \labl{eq:bope-minmod}
This defines the boundary operator product coefficients
(or structure constants)
$C_{a\alpha,b\beta}^{\;\;c\gamma}$ of primary boundary fields on $M$. 

In the general case two related additional features must be taken into 
account: First, there may be more than one independent way for 
representations $a$ and $b$ to fuse to $c\,$. Second, for a given coupling 
of $a$ and $b$ to $c$, the primary field of $c$ does not necessarily appear%
  \foodnode{In fact, with a suitable basis choice the primary field appears 
  for at most one of the couplings.}
on the \rhs\ of \erf{eq:bope-minmod}.
The dimension of the space of couplings is the fusion rule $\N abc$.

To obtain a formulation of the OPE that covers the generic situation we 
regard a coupling of $a$ and $b$ to $c$ as a prescription on how to associate
to every vector $v$ in the highest weight module $\Calh_a$ of the chiral
\alg\ a linear map from $\Calh_b$ to $\Calh_c$.
A basis in the space of couplings is then a collection of $\N abc$
maps ${V_{ab}}^{\!c,\delta}{:}\;\Calh_a\,{\to}\,\Hom(\Calh_b,\Calh_c)
[[z,z^{-1}]] \,{\cdot}\, z^{\Delta_c-\Delta_a-\Delta_b}$, $\delta\eq1,2,
...\,,\N abc$; the basis elements are known as chiral vertex operators.%
 \foodnode{$\Hom(\Calh_b,\Calh_c)[[z,z^{-1}]]$ denotes the space of
 Laurent series with values in the space $\Hom(\Calh_b,\Calh_c)$.  
 To make the concept of chiral vertex operators precise, one should
 work in the vertex operator algebra setting, where $z$ is a 
 formal variable, so that one is dealing with formal Laurent series.}
Further, we denote by $\{v^D_d\}_{\!D}^{}$ an orthonormal basis of
$L_0$-eigenvectors in $\Calh_d$, with $v^0_d\,{\equiv}\,v_d^{}$ a
highest weight vector, and by $\Psi_{d\alpha}^D$ the corresponding 
descendant field of the primary $\Psi_{d\alpha}\,{\equiv}\,
\Psi_{d\alpha}^0$. Then the OPE reads
  \be \bearll \Psi_{a\alpha}(x)\,\Psi_{b\beta}(y) \!\!
  &= \dsty \sumI_c\sum_{\gamma=1}^{n^c}\sum_{\delta\eq1}^{\N abc}
  \mC a\alpha b\beta c\gamma\delta \, (x{-}y)^{\Delta_c-\Delta_a-\Delta_b} 
  \\{}\\[-.7em] &\hsp{4.2}
  \dsty\sum_C \langle v^C_c | {V_{ab}}^{\!c,\delta}(v_a^{};z{=}1) | v_b^{}
  \rangle\, (x{-}y)^{\Delta(v^C_c)-\Delta_c}\, \Psi_{c\gamma}^{C}(y) \,.
  \eear \labl{eq:bope-general}
Note that here we consider only special cases of boundary operator products,
since we are interested in a single \bc\ $M$. To be fully general, we must
also deal with boundary fields $\Psi_{d\alpha}^{MM'}$ that change the
\bc, from $M$ to $M'$, and hence operator products $\Psi_{a\alpha}^{MM'}
\!(x)\,\Psi_{b\beta}^{M'M''}\!(y)$ expanded in boundary fields 
$\Psi_{c\gamma}^{MM''}\!(y)$. Correspondingly, the operator product
coefficients acquire three more labels $M,M',M''$. 
As we will discuss in section \ref{bc-to-rep}, boundary changing
fields play a natural role in the categorical setup as well; they 
occur in the CFT interpretation of representations of algebra objects.

The sewing constraint \cite{lewe3} that arises from the factorisation of
the correlator of four boundary fields on a disk with boundary condition
$M$ then looks as follows:
  \be
  \sum_{\varphi=1}^{n_f} \mC a\alpha b\beta f\varphi\rho\,
  \mC f\varphi c\gamma d\delta\sigma
  = \sum_{e\in\II} \sum_{\varepsilon=1}^{n_e}\sum_{\varrho=1}^{\N bce}
    \sum_{\tau=1}^{\N aed}
  \mC b\beta c\gamma e\varepsilon\varrho\, \mC a\alpha e\varepsilon d\delta
  \tau\, \F{a\U b}cdef\tau\varrho\rho\sigma \,.  \labl{eq:bope-sewing}

It proves to be convenient to describe the boundary structure constants 
with the help of the concepts of the {\em category\/} \calca\ {\em of 
$A$-modules\/} and of a {\em module category\/}, which will be described 
in section \ref{rep-and-mod} below.
They correspond to the generalised 6j-symbols ${}^{\sss(1)}\!{F}$ (defined
via formula \erf{gz6j}) of \calca\ regarded as a module category, and the
relation \erf{eq:bope-sewing} is nothing but the corresponding generalisation 
of the pentagon identity. The identification between the boundary 
structure constants $C$ and the quantities ${}^{\sss(1)}\!{F}$ also appears
in the formalism used in \cite{bppz2}, where it is obtained via the
relation with weak \hopf s; our formulas \erf{eq:bope-general} and 
\erf{eq:bope-sewing} correspond to (4.11) and (4.29) of \cite{bppz2}
for a single \bc. In the Cardy case the ${}^{\sss(1)}\!{F}$ 
specialise to the ordinary \FF-matrices
(this was first observed for the A-series of Virasoro minimal models in 
\cite{runk} and established in general in \cite{bppz2,fffs}); the sewing 
constraint \erf{eq:bope-sewing} is then just the pentagon identity for $\FF$. 

{}From the data provided by the decomposition \erf{eq:bnd-H} of the state 
space and by formula \erf{eq:bope-general} for the structure constants
we can obtain an algebra $A$ in \calc, which we call the \alg\ of open
string states. This is achieved as follows. As an object in \calc, the 
algebra of open string states coincides with \calh\ as given in formula 
\erf{eq:bnd-H}, i.e.
  \be  A \cong \bigoplus_{a\in\II} n^a\, U_a \,.  \labl{eq:A=H}
Now let us choose a basis in $\Hom(U_a,A)$ as described in section
\ref{sec:alg-obj}, and define a multiplication $m$ on $A$ in this basis via 
\erf{def-mabc}, by demanding that the coefficients
$\m a\alpha b\beta c\gamma\delta$ are just the structure constants:
  \be  \m a\alpha b\beta c\gamma\delta    
  :=  \mC a\alpha b\beta c\gamma\delta \,.  \labl{eq:m=c}
With this assignment the consistency condition \erf{eq:bope-sewing}
fulfilled by the structure constants of the boundary fields on $M$
coincides with the associativity condition \erf{eq:general-assoc-m-basis} 
for the algebra $A$. The unit of $A$ is the identity field on $M$.
We conclude that indeed the fields living on a boundary with \bc\
$M$ that preserves \CA\ provide us with an algebra object $A$ in \calc.

\medskip

The algebra objects arising from boundary conditions enjoy an
additional important property which derives from the non-degeneracy of
the two-point \corfu s. Non-degeneracy means that for any field
$\Psi$ on the $M$-boundary there exists at least one%
  \foodnode{We cannot formulate this condition in terms of primary fields
  alone, for the same reason that we had to adopt the more complicated
  form \erf{eq:bope-general} of the OPE. E.g.\ in WZW models generically
  it takes a (horizontal) descendant to get a non-zero two-point function
  with a primary field.}
$\Psi'$ such that 
  \be  \langle \, \Psi(x) \,\Psi'(y) \, \rangle \neq 0 \,.
  \labl{eq:2pt-nondeg}
If there were a field in the theory that had zero two-point
function with all other fields, it would decouple from the theory, 
i.e.\ every correlator involving that field would vanish.

To evaluate the two-point functions one can use the OPE
\erf{eq:bope-general} together with the one-point functions
$\langle \Psi(x)\rangle$ of boundary fields. The latter can be non-zero 
only for boundary fields of weight zero, but (as is natural in view
of the connection of our results with those in two-dimensional lattice TFT)
we allow for the possibility that there can be more of those than just the 
identity field. A superposition of two elementary boundaries, for example,
has at least two fields of weight zero. In the case at hand, the 
one-point functions are determined once we know them for the 
primary boundary fields $\Psi_{a\alpha}$ on $M$.

In the category theoretic setting, the collection of one-point functions of
boundary fields on $M$ will give rise to a morphism $\eps\iN\Hom(A,\one)$.
The non-degeneracy \erf{eq:2pt-nondeg} translates into the non-degeneracy 
of the composition $\eps \cir m$. That is, the matrix 
$G(a)_{\alpha\beta}$ defined by
  \bea \begin{picture}(145,46)(0,23)
  \put(0,0)   {\begin{picture}(0,0)(0,0)
              \scalebox{.38}{\includegraphics{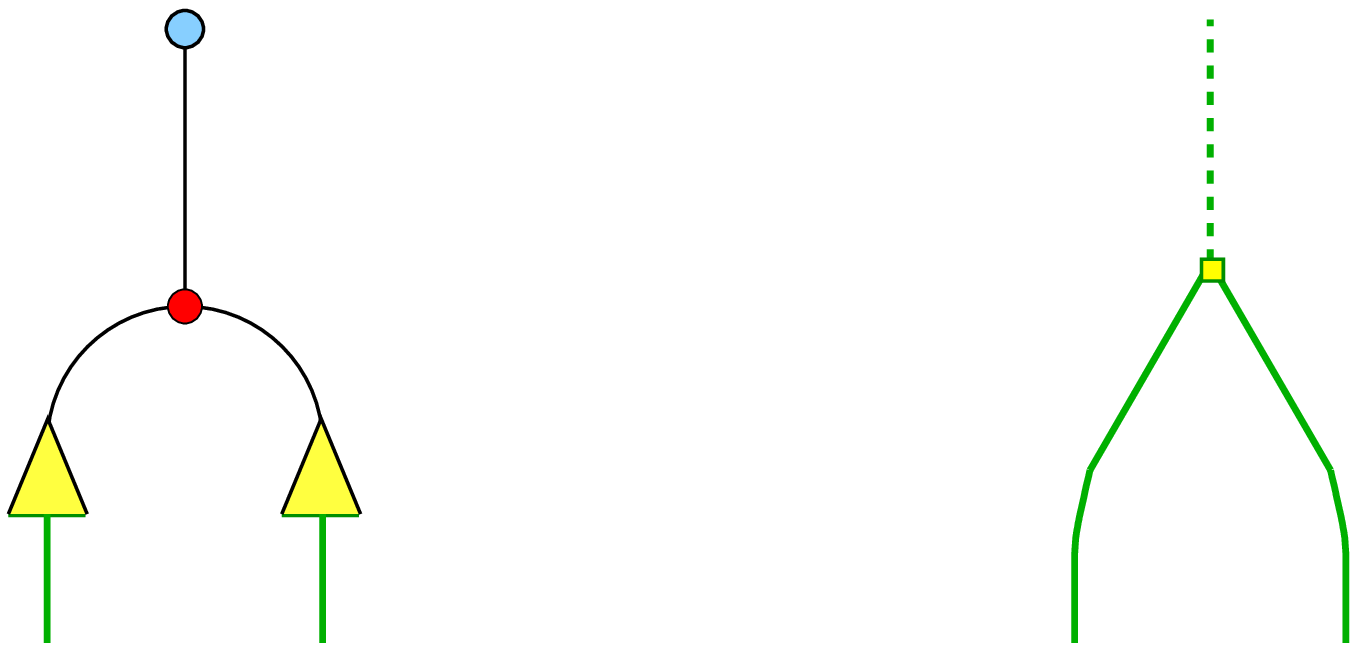}} \end{picture}}
  \put(2.8,-6.7)  {\scriptsize$a$}
  \put(2.9,16.7)  {\tiny$\alpha$}
  \put(33.5,-7.5) {\scriptsize$\bar a$}
  \put(32.8,16.7) {\tiny$\beta$}
  \put(54.4,31.2) {$=\,\ G(a)_{\alpha\beta}$}
  \put(115.6,-6.7){\scriptsize$a$}
  \put(135.7,40.4){\tiny$0$}
  \put(145.2,-7.5){\scriptsize$\bar a$}
  \epicture12 \labl{def-gaab}
is invertible as a matrix in the multiplicity labels $\alpha,\beta$.

In fact, the one-point functions $\langle \Psi_{a\alpha}(x)\rangle$ 
of primary boundary fields on the upper half plane (or equivalently, 
on the unit disk) are themselves of a special form. First of all, 
by conformal invariance they can be non-zero only for $a\eq0$.
A more interesting result follows from the sewing constraint for the 
boundary one point functions on an annulus. 

Consider a cylinder of length $L$ and circumference $T$ with boundary 
condition $M$ at both ends. On one of the boundaries insert a 
boundary field $\Psi_{0\alpha}$ of weight zero.  The cylinder can be 
conformally mapped to an annulus of inner radius $q$ and outer radius
$1$, and alternatively to a half-annulus (with the two half-circular 
boundaries identified) in the upper half plane, with the two ends 
stretching from $-1$ to $-\tilde q$ and from $\tilde q$ to $1$, where
  \be
  q := \exp( -2 \pi L / T ) \qquad {\rm and} \qquad
  \tilde q := \exp(-\pi T/L) \,.  \ee
This results in the equation
  \be
  \langle M(\Psi_{0\alpha}) |\, q^{L_0+\bar L_0 - c/12}\, | M \rangle
  = \Tr_{\calH} \big(\, \tilde q^{\,L_0-c/24}\,\Psi_{0\alpha}\,\big) \,,
  \labl{eq:1pt=1pt}
where $| M(\Psi_{0\alpha})\rangle$ denotes the boundary state for the 
$M$-boundary with an insertion of the boundary field $\Psi_{0\alpha}$ 
and $|M\rangle$ the boundary state without field insertion.
On the \rhs\ of \erf{eq:1pt=1pt}, the boundary field $\Psi_{0\alpha}$ 
is interpreted as an operator on the boundary state space $\calh$.

Let us now assume that the CFT under consideration 
does not possess any state of negative conformal weight (as is
in particular the case for all unitary theories) and that it
has a unique vacuum (i.e., state $|0\rangle$ of weight zero) in the 
{\em bulk\/}, which we take to be normalised as $\langle0|0\rangle\eq1$. 
Then in the limit of infinite length $L$ only the vacuum propagates 
on the \lhs\ of \erf{eq:1pt=1pt}, so that
  \be  %% \lim_{L\to\infty}
  \langle M(\Psi_{0\alpha}) |\, q^{L_0+\bar L_0 - c/12}\, | M \rangle
  = q^{-c/12} \,(1 + O(q^\kappa))\,
  \langle M(\Psi_{0\alpha}) |0\rangle \, \langle 0 | M \rangle
  \qquad{\rm for}\ \ L\to\infty  \labl{eq:1pt=1pt-lhs}
with $\kappa\,{>}\,0$.
Finding the $L\,{\to}\,\infty$ limit of the \rhs\ of \erf{eq:1pt=1pt}
requires slightly more work. In accordance with the notation introduced 
in \erf{eq:bope-general}, let $|\Psi_{d\delta}^D\rangle$ denote a basis
of \calh, and $\langle\Psi_{d\delta}^D|$ a dual basis, 
i.e.\ $\langle \bar\Psi_{c\gamma}^C|\Psi_{d\delta}^D\rangle
= \delta_{c,d} \delta_{\gamma,\delta} \delta_{C,D}$. Then
  \be\begin{array}{ll}
  \Tr_\calH\, \tilde q^{\,L_0-c/24}\,\Psi_{0\alpha} \!\!
  &= \dsty\sumI_d \sum_{\delta=1}^{n^d} \sum_{D\;{\rm desc.}}
  \langle \Psi_{d\delta}^D| \,\tilde q^{\,L_0-c/24}\,\Psi_{0\alpha}\,
  |\Psi_{d\delta}^D\rangle
  \\{}\\[-.8em]
  &= \sum_{d,\delta} \langle \Psi_{d\delta}| \,\Psi_{0\alpha}\,
  |\Psi_{d\delta}\rangle \, \chii_d(\tilde q)
  %\\{}\\[-.8em]
   = \sum_{d,\delta} \sumI_c C_{0\alpha,d\delta}^{\;\; d\delta}
  \,S_{d,c} \,\chii_c(q^2)
  \\{}\\[-.55em]
  & = q^{-c/12}\,(1 + O(q^\kappa)) \sum_{d,\delta} 
  C_{0\alpha,d\delta}^{\;\; d\delta} \,S_{d,0}
  \qquad{\rm for}\ \ L\to\infty \,.  \end{array}\labl{eq:1pt=1pt-rhs}
In the first step, the trace is written out in terms of the bases.
The crucial step is the second one. Since $\Psi_{0\alpha}$
is a field of weight zero, it behaves just like the identity field as far
as commutation with elements of the chiral algebra is concerned, and this
in turn implies that
  \be  \langle \Psi_{d\delta}^D| \,\tilde q^{\,L_0-c/24}\,\Psi_{0\alpha}\,
  |\Psi_{d\delta}^D\rangle
  = \tilde q^{\,\Delta(\Psi_{d\delta}^D)-c/24}\, \langle \Psi_{d\delta}
  | \,\Psi_{0\alpha}\,|\Psi_{d\delta} \rangle \,,  \ee 
so that the sum over descendants $D$ gives rise to the character $\chii_d$. 
The third step involves a modular transformation of this character,
upon which finally the limit as $L\,{\to}\,\infty$ can be taken, leaving
only the contribution of the vacuum character.%
  \foodnode{Here we assume that of the representations $U_i$ of the chiral
  algebra \CA, the vacuum \rep\ $U_0$ is the only one which contains a
  state of weight zero.}

Comparing the results \erf{eq:1pt=1pt-lhs} and \erf{eq:1pt=1pt-rhs} we 
conclude that the one-point functions of weight zero boundary fields on 
the unit disk are given by
  \be
  \langle \Psi_{0\alpha} \rangle = {\rm const}
  \sumI_d \sum_{\delta=1}^{n^d} S_{d,0}\, C_{0\alpha,d\delta}^{\;\; d\delta}
  \,. \ee
Note that here the sum runs over all primary boundary fields on the 
$M$-boundary, not only over those of zero weight. When
translated back into the language of algebra objects, this relation 
reads (using the basis \erf{iaa,aai})
  \be
  \eps \cir \iaa 0\alpha = {\rm const'} 
  \sum_{d\in\II} \sum_{\delta=1}^{\Ai d} m_{0\alpha,d\delta}^{\;\; d\delta} 
  \; \dim(U_d) \,.  \labl{CFT-eps}

The additional properties \erf{def-gaab} and \erf{CFT-eps} 
of the multiplication $m$ mean that 
-- provided that the CFT under consideration has a unique bulk vacuum --
the algebra object $A$ can be turned into what will be called a symmetric 
special Frobenius algebra. It is the aim of the next section to explain 
these notions. Before entering that discussion, let us make some short 
remarks on the situation that $Z_{00}\,{>}\,1$ for the torus \parfu, 
taking $Z_{00}\eq2$ as an example. In this case there exist two distinct 
bulk fields of weight zero. All correlators are independent of the 
insertion points of these weight zero fields, thus they form a topological 
subsector of the theory. In particular, their operator product
algebra is commutative. As a consequence there is a projector basis
$\{P_1, P_2\}$ for the weight zero fields, in which their operator 
products take the form $P_1 P_1\eq P_1$, $P_2 P_2\eq P_2$ and 
$P_1 P_2\eq 0$. The identity field is then given by 
$\one\eq P_1 \,{+}\, P_2$. The projector fields $P_1, P_2$ act on the 
space of fields via the OPE, and thereby decompose the space of fields 
into eigenspaces. Let us denote by $\phi_1(z,\bar z)$ bulk fields 
for which $P_1 \phi_1(z,\bar z)\eq\phi_1(z,\bar z)$, and analogously 
for $\phi_2$. Mixed correlators then vanish:
  \be  \bearll
  \langle\, \phi_1(z,\bar z)\, \phi_2(w,\bar w)\, \cdots\, \rangle \!\!
  &= \langle\, (P_1\phi_1(z,\bar z))\, \phi_2(w,\bar w)\, \cdots\, \rangle
  \\[.6em]
  &= \langle\, \phi_1(z,\bar z)\, (P_1\phi_2(w,\bar w))\, \cdots\, \rangle
  = 0 \,.  \eear \ee
We conclude that a CFT with $Z_{00}\eq2$ should be interpreted
as a superposition of two CFTs each of which has $Z_{00}\eq1$.
By a {\em superposition\/} of two \cfts, CFT$_1$ and CFT$_2$,
with the same central charge we mean the theory in which fields
are pairs $\Phi\eq(\phi_1,\phi_2)$ of fields in the two constituent
CFTs and \corfu s are sums
  \be  \langle\, \Phi(z,\bar z)\, \Phi(w,\bar w)\, \cdots\, \rangle
  = {\langle\, \phi_1(z,\bar z)\, \phi_1(w,\bar w)\, \cdots\, \rangle}_1
  + {\langle\, \phi_2(z,\bar z)\, \phi_2(w,\bar w)\, \cdots\, \rangle}_2
  \,.  \labl{cft12}
Requiring that a CFT cannot be written as a superposition of this type
has a counterpart in the underlying algebra object, which in this case
must be indecomposable in the sense of definition 9(ii) of \cite{ostr}.

\subsection{Frobenius algebras}

The construction of correlation functions that will be discussed in
sections \ref{sec:torus-pf} and \ref{sec:annulus-pf} (and further in 
forthcoming papers) uses a symmetric special Frobenius algebra object as 
an input.  In the present section we explain the qualifiers special, 
symmetric, and Frobenius, and we show that algebras with these
properties are natural from the CFT point of view.
This is the content of theorem \ref{thm:alg-from-bc} below.

{}From a computational point of view, the construction of such an algebra 
object requires the solution of the boundary factorisation constraint 
\erf{eq:bope-sewing} for boundary fields living on {\em one single\/} 
boundary -- rather than a
simultaneous solution of all constraints involving all boundary
conditions as well as boundary changing fields, which form a much larger
system. It turns out that the construction of the algebra is the 
{\em only\/} place where a {\em non-linear\/} constraint ever must
be solved.  Finding the other boundary conditions and structure constants 
then amounts to solving systems of linear equations only. 

The notions of co-algebra (which is the notion dual to that of an 
algebra), Frobenius algebra, special and symmetric given below are 
straightforward extensions to the category setting of the corresponding 
concepts in algebras over $\complex$, see e.g.~\cite{CUre}.

\dtl{Definition}{def:coalg}
A {\em co-algebra\/} $A$ in a \tc\ \calc\ is an object with a coassociative 
coproduct 
$\Delta$ and a counit $\eps$, i.e.\ morphisms $\Delta\iN\Hom(A,A\Oti A)$ and 
$\eps\iN\Hom(A,\one)$ that satisfy
  \be  (\Delta\oti\id_A)\circ\Delta = (\id_A\oti\Delta)\circ\Delta
  \qquad{\rm and}\qquad
  (\eps\oti\id_A) \circ\Delta = \id_A = (\id_A\oti\eps) \circ \Delta \,. \ee

\medskip\noindent
The pictures for these morphisms and their properties are obtained by
turning upside-down those for an \alg:
  \bea \begin{picture}(377,43)(0,14)
  \put(0,0)   {\begin{picture}(0,0)(0,0)
              \scalebox{.38}{\includegraphics{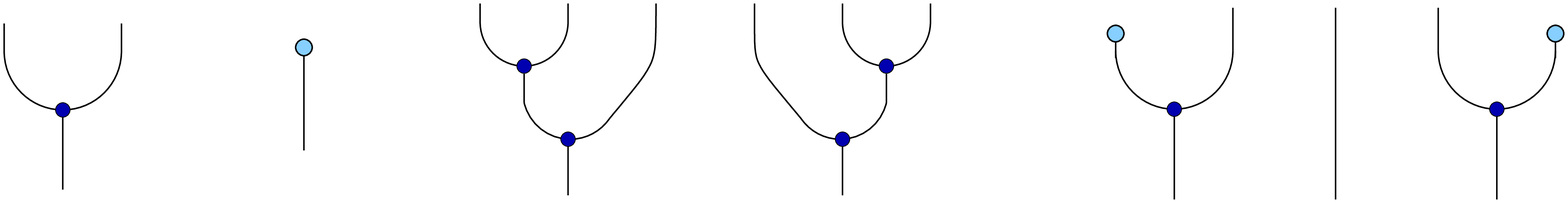}} \end{picture}}
  \put(-30,25)    {$\Delta\,=$}
  \put(-2.8,51.9) {\scriptsize$A$}
  \put(11.9,-6.2) {\scriptsize$A$}
  \put(28.1,51.9) {\scriptsize$A$}
  \put(48.8,25)   {$\eps\,=$}
  \put(75.5,46.8) {\scriptsize$\one$}
  \put(73.9,3.5)  {\scriptsize$A$}
  \put(119.9,54.7){\scriptsize$A$}
  \put(142.5,54.7){\scriptsize$A$}
  \put(142.4,-7.7){\scriptsize$A$}
  \put(165.5,54.7){\scriptsize$A$}
  \put(177.8,25)  {$=$}
  \put(191.5,54.7){\scriptsize$A$}
  \put(212.7,-7.7){\scriptsize$A$}
  \put(213.5,54.7){\scriptsize$A$}
  \put(236.5,54.7){\scriptsize$A$}
  \put(298.7,-7.7){\scriptsize$A$}
  \put(314.8,54.7){\scriptsize$A$}
  \put(325.9,25)  {$=$}
  \put(340.0,-7.7){\scriptsize$A$}
  \put(340.6,54.7){\scriptsize$A$}
  \put(356.1,25)  {$=$}
  \put(367.1,54.7){\scriptsize$A$}
  \put(381.9,-7.7){\scriptsize$A$}
  \epicture06 \labl{Delta-eps}

We will be interested in objects that possess both an \alg\ and a co-\alg\
structure and for which these structures are interrelated in a special way.
This is encoded in the definition of a Frobenius algebra (see 
\cite{stri2,muge8}
and references therein). Note that provided the \tc\ has a braiding
(a condition that is not needed in the Frobenius case) one may combine the
\alg\ and co-\alg\ structure also into the one of a bialgebra; this amounts
to a different compatibility condition between product and coproduct.
Bi- or Hopf algebras in \tcs\ do not occur in our discussion of \bc s, 
though they play an important role in other contexts (see e.g.\ 
\cite{pare11,maji25,lyub3,pare26,KElu}), and
we refrain from describing those structures here.

\dtl{Definition}{def:frob}
\mbox{$\ $}\\[-1.07em]
A {\em Frobenius \alg\/} in a \tc\ \calc\ is an object that is both
an \alg\ and a co-\alg\ and for which the product and coproduct are
related by
  \be  (\id_A\oti m) \circ (\Delta\oti\id_A)
  = \Delta \circ m = (m\oti\id_A) \circ (\id_A\oti\Delta) \,.  \labl{1f}

\medskip\noindent
In pictures,
  \bea \begin{picture}(225,60)(0,26)
  \put(0,0)   {\begin{picture}(0,0)(0,0)
              \scalebox{.38}{\includegraphics{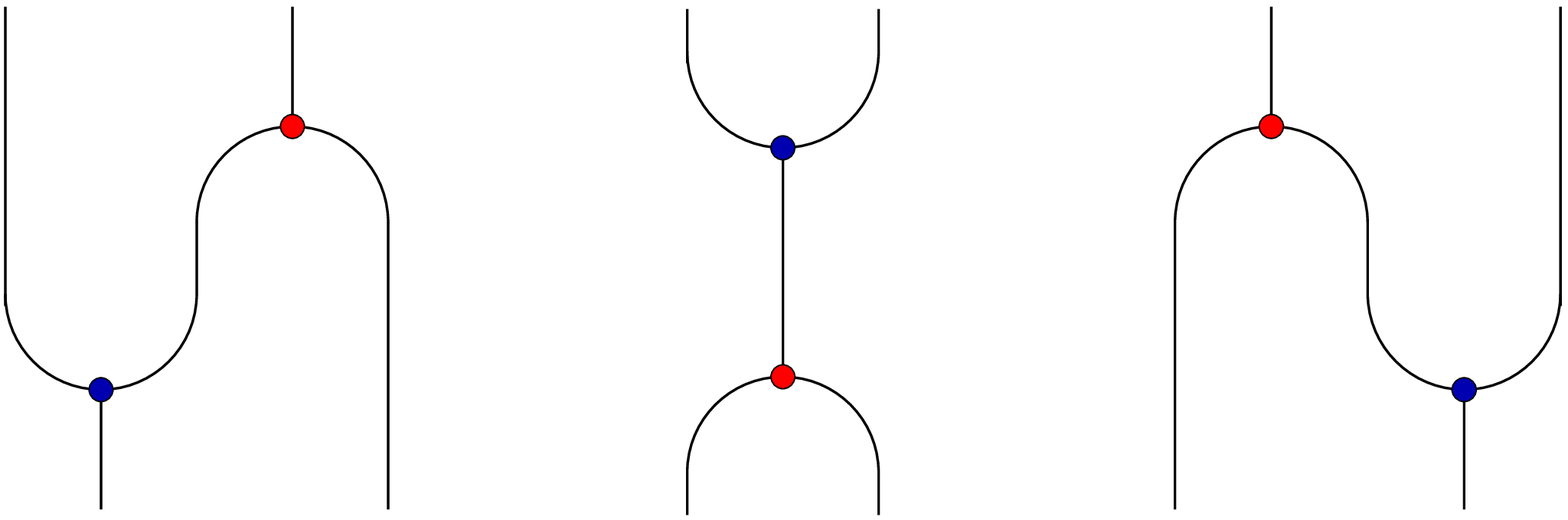}} \end{picture}}
  \put(-2.5,77)   {\scriptsize$A$}
  \put(10.0,-7.9) {\scriptsize$A$}
  \put(38.6,77)   {\scriptsize$A$}
  \put(51.5,-7.9) {\scriptsize$A$}
  \put(74,33)     {$=$}
  \put(94.4,-7.9) {\scriptsize$A$}
  \put(95.5,77)   {\scriptsize$A$}
  \put(122.2,-7.9){\scriptsize$A$}
  \put(123.0,77)  {\scriptsize$A$}
  \put(140,33)    {$=$}
  \put(164.3,-7.9){\scriptsize$A$}
  \put(179.5,77)  {\scriptsize$A$}
  \put(205.9,-7.9){\scriptsize$A$}
  \put(220.7,77)  {\scriptsize$A$}
  \epicture25 \labl{frob}

To be able to discuss additional properties of the algebras that come
from boundary conditions, let us further introduce the following notions.
\vspace{-.9em}

\dtl{Definition}{df:spec-etc}
\mbox{$\ $}\\[-1.07em]
(i) A {\em special} \alg\/ in a \tc\ \calc\ is an object that is both
an \alg\ and a co-\alg\ such that
  \be
  \eps\cir\eta = \beta_\smallone\, \id_\one 
  \qquad {\rm and}\qquad
  m\cir\Delta = \beta_{\!A}\, \id_A \ee
for non-zero numbers $\beta_\smallone$ and $\beta_{\!A}$. In pictures,
  \bea \begin{picture}(235,44)(0,35)
  \put(0,0)   {\begin{picture}(0,0)(0,0)
              \scalebox{.38}{\includegraphics{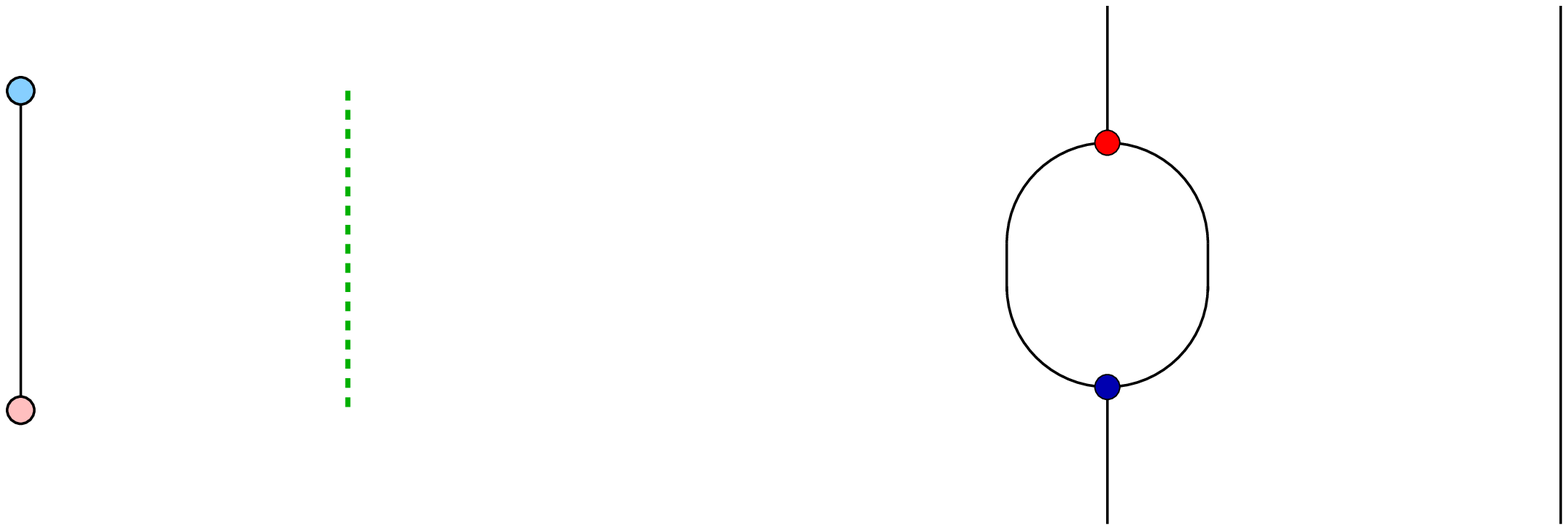}} \end{picture}}
  \put(16,38)     {$=\ \beta_\smallone$}
  \put(92,38)     {and}
  \put(162.0,-7.9){\scriptsize$A$}
  \put(162.9,81.9){\scriptsize$A$}
  \put(195,38)    {$=\ \beta_{\!A}$}
  \put(230.2,-7.9){\scriptsize$A$}
  \put(231.2,81.9){\scriptsize$A$}
  \epicture19 \labl{spec2}
(ii) A {\em symmetric} \alg\/ in a sovereign \tc\ \calc\ is an \alg\ object 
$(A,m,\eta)$ together with a morphism $\eps\iN\Hom(A,\one)$ such that the
two morphisms $\Phi_1, \Phi_2 \iN \Hom(A,A^\vee)$ defined as
  \be  
  \Phi_1 := [(\eps\cir m)\oti \id_{A^\vee}] \circ (\id_A \otimes b_A)
  \qquad {\rm and} \qquad
  \Phi_2 :=  [\id_{A^\vee}\oti (\eps\cir m)] \circ (\tilde b_A \oti \id_A)
  \labl{eq:Phi-def}
are equal. 
In pictorial notation the morphisms $\Phi_1$ and $\Phi_2$ are given by
  \bea \begin{picture}(257,61)(0,29)
  \put(42,0)  {\begin{picture}(0,0)(0,0)
              \scalebox{.38}{\includegraphics{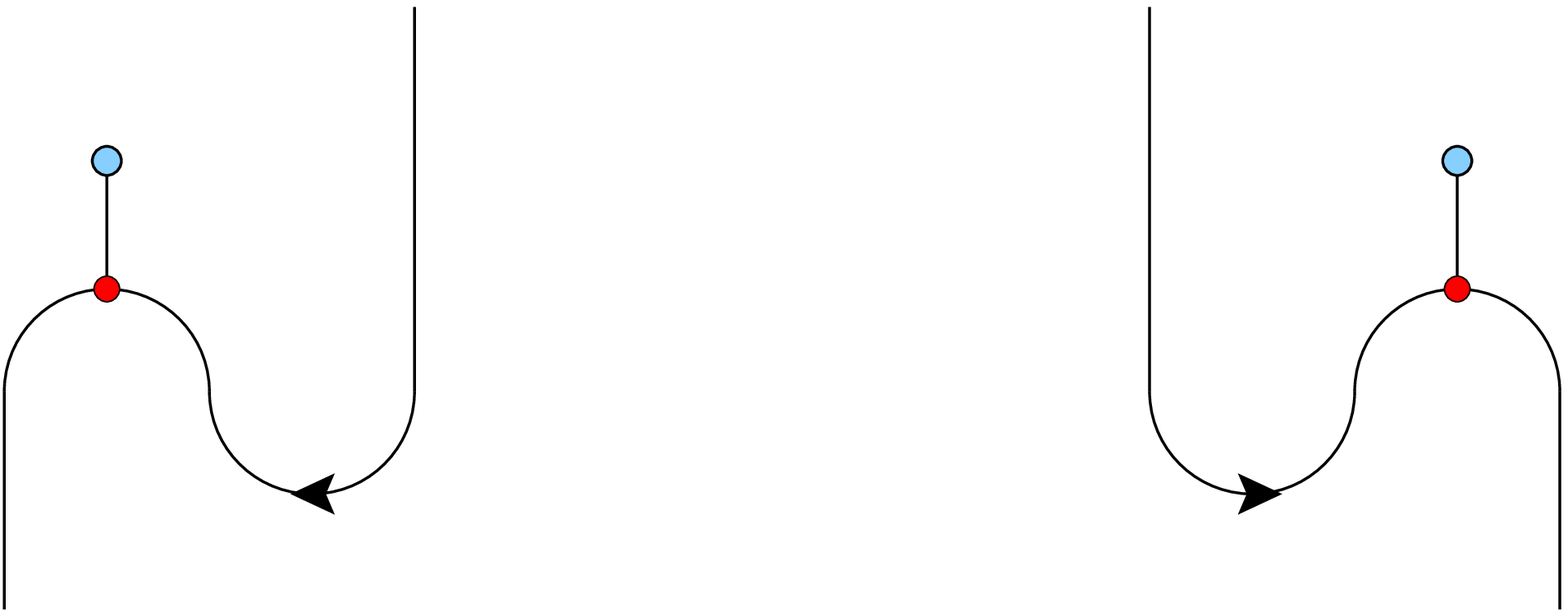}} \end{picture}}
  \put(0,31)      {$\Phi_1\ =$}
  \put(38.7,-8.5) {\scriptsize$A$}
  \put(92.2,84.2) {\scriptsize$A\Vee$}
  \put(152.5,31)  {$\Phi_2\ =$}
  \put(190.2,84.2){\scriptsize$A\Vee$}
  \put(246.3,-8.5){\scriptsize$A$}
  \epicture20 \labl{Phi12}
(iii) A {\em commutative\/} algebra in a braided tensor category \calc\ is 
an algebra object $(A,m,\eta)$ for which $m\cir c_{A,A}\eq m$.
\\[.26em]
(iv)  A {\em haploid\/} algebra in a tensor category is an
algebra object $A$ for which $\dim\,\Hom(\one,A)\eq1$.
\smallskip

\dt{Remark}
\begin{itemize}
\item[(i)]
In the study of algebras over $\complex$ \cite{CUre}, symmetric algebras 
constitute an interesting subclass of Frobenius algebras. They contain for 
instance all group algebras. Also, the property of being symmetric is considerably
weaker than being commutative (abelian).
For algebras in \mtcs, commutativity (with respect to the braiding $c$)
and triviality of the twist (i.e.\ $\theta_{\!A}\eq\id_A$) together
imply that $A$ is symmetric.
\\[-1.7em]
\item[(ii)]
Algebras possessing the same \rep\ theory are called Morita equivalent 
(see e.g.\ \cite{pare13,pare14,fuRs,ostr,muge8}). It turns out that 
neither haploidity nor commutativity are invariant under Morita equivalence. 
As will be discussed in more detail elsewhere, the algebra objects
$U \Oti U^\vee$ described below definition \ref{def:alg-obj} turn out to
be Morita equivalent for different (not necessarily simple) $U$. 
But while for $U\eq\one$ the resulting algebra is certainly 
haploid and commutative, for a general object $U$ this need not be the case. 
\\[-1.7em]
\item[(iii)]
In practice, haploidity is a quite useful property, because a haploid algebra 
is automatically symmetric, as follows from corollary \ref{lem:hap-sym} below. 
Furthermore it can be shown \cite[section\,3.3]{ostr}
  % un-numbered corollary
that every Morita class of algebras that cannot be decomposed non-trivially 
in a direct sum  has at least one haploid representative. (For the notion of 
direct sum of \alg s see proposition \ref{prop:alg-add} below.)
\\[-1.7em]
\item[(iv)]
The algebras $A$ of our interest are not necessarily (braided-) commutative. 
However, commutativity is still an interesting property, because in CFT 
terms it immediately allows for an interpretation of $A$ as an extension of 
the chiral algebra (see section \ref{sec:left-right-chiral} below). Note, 
however, that according to point (ii) above, a non-commutative algebra can 
still correspond to an extension. Commutative algebras $A$ in a braided 
tensor category have also been studied under the name `quantum subgroups'
\cite{wass6,ocne9}. This terminology has its origin in the fact \cite{kios}
 % Theorem 2.2.
that commutative algebras in the representation category of a group $G$ are 
in one-to-one correspondence with the algebras of functions on 
homogeneous spaces $G/H$, and thereby with subgroups $H$ of $G$.
\end{itemize}

In section \ref{sec:alg2bCFT} above we have seen that
algebra objects are a natural structure to consider from the CFT point
of view, as they can be obtained from boundary conditions. In the rest of
this section we show that:
\vspace{-.9em}

\dtl{Theorem}{thm:alg-from-bc}
(i)\hsp{.6}If an algebra can be endowed with the structure of 
     a symmetric special Frobenius algebra, then this structure is unique
     up to a normalisation constant $\xi\iN\complex^\times$.
\\[.15em]
(ii) Every algebra coming from a boundary condition that preserves the 
     chiral algebra $\CA$ 
     of a (rational, unitary) CFT with unique bulk vacuum  can be endowed
     with the structure of a symmetric special Frobenius algebra.

\medskip\noindent
%add here comment:
Note that
the results of this paper rely only on the existence of a symmetric special
Frobenius algebra. Part (ii) of theorem \ref{thm:alg-from-bc} guarantees the
existence of such an algebra in a large class of theories.
It does not exclude, however, the existence of such an algebra in 
theories not belonging to this class, in particular in certain non-unitary
theories (e.g.\ non-unitary minimal models).

\medskip
The proof of the theorem will require several lemmata.

\dtl{Lemma}{lem:frob-unique}
Let $(A,m,\eta)$ be an algebra and let $\eps \iN \Hom(A,\one)$.
Then the following holds.\\[.13em]
(i)\hsp{.6}If there exists a Frobenius structure on $A$ with counit $\eps$,
then it is unique. \\[.15em]
(ii) There exists a Frobenius structure on $A$ if and only
if $\Phi_1$ as defined in \erf{eq:Phi-def} is invertible.

\medskip\noindent
Proof:\\
(i) We must show that the coproduct $\Delta$ is unique.
If there exists any
coproduct $\Delta$ satisfying the Frobenius property, then 
$\Phi_1$ as defined in \erf{eq:Phi-def} is 
invertible, with (left- and right-) inverse $\Phi_1^{-1}$ given by
  \be  \Phi_1^{-1}
  = [d_A \oti \id_A] \circ [\id_{A^\vee}\oti (\Delta\cir\eta)] \,.
  \labl{Phi1inv}
Analogously, $\Phi_2$ is invertible as well. In pictures, the inverses are
  \bea \begin{picture}(247,59)(0,26)
  \put(42,0)  {\begin{picture}(0,0)(0,0)
              \scalebox{.38}{\includegraphics{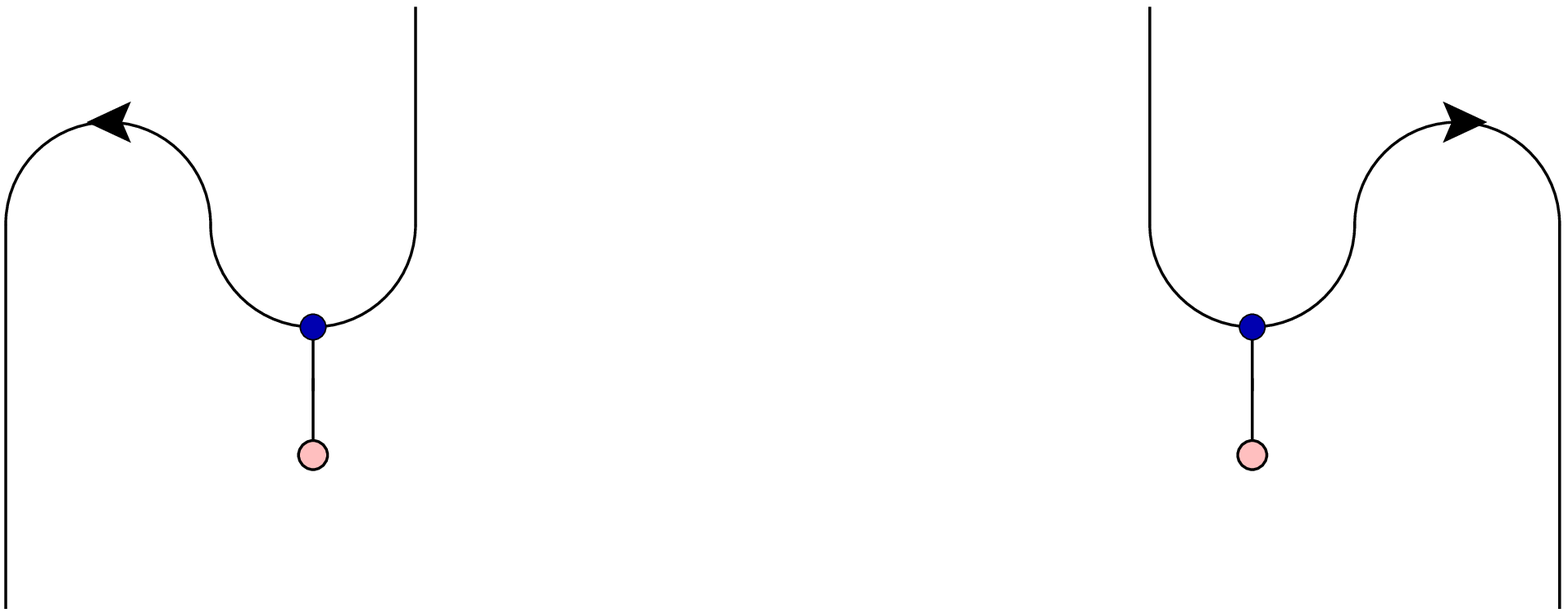}} \end{picture}}
  \put(0,33)      {$\Phi_1^{-1}=$}
  \put(37.7,-8.7) {\scriptsize$A\Vee$}
  \put(94.4,85.2) {\scriptsize$A$}
  \put(152.5,33)    {$\Phi_2^{-1}=$}
  \put(192.8,85.2){\scriptsize$A$}
  \put(245.9,-8.7){\scriptsize$A\Vee$}
  \epicture13 \label{eq:inv-Phi}\labl{Phi12-}
Moreover, application of the Frobenius property also shows that we can
express, conversely, the coproduct through the product and the morphism
\erf{Phi1inv}:
  \be  \Delta = (\id_A \oti m) \circ (\id_A \oti \Phi_1^{-1} \oti \id_A)
  \circ (b_A \oti \id_A)   \labl{Delta}
(a picture can be found below in \erf{Delta-Phi}). Now for given $m$ and 
$\eps$, $\Phi_1$ is uniquely determined by \erf{eq:Phi-def}. Since in
addition the morphism $\Phi_1^{-1}$ is uniquely determined by its property
of being inverse to $\Phi_1$, it follows that $\Delta$ is unique as well.

\medskip\noindent
(ii) We have already seen in the proof of part (i) that $A$ being Frobenius
implies that the morphism $\Phi_1$ is invertible, with inverse given by 
\erf{Phi1inv}.  Conversely, let us now assume invertibility of $\Phi_1$. 
One quickly checks that $\Phi_2$ as given in \erf{eq:Phi-def} can be 
expressed through $\Phi_1$ as%
 \foodnode{Note that $\Phi_2$ does not, in general, coincide with the
 morphism $\Phi_1^\vee$ that is dual to $\Phi_1$. Indeed, 
 $\Phi_1^\vee$ is an element of $\Hom(A^{\vee\vee}\!{,}\,A^\vee)$, and
 while $A^{\vee\vee}$ is certainly isomorphic to $A$ there is no reason why
 the two objects should be equal.}
  \be
  \Phi_2 = [\id_{A^\vee} \oti d_A] \circ 
  [\id_{A^\vee} \oti \Phi_1 \oti \id_A ] \circ
  [ \tilde b_A \oti\id_A]  \,.  \ee
Consequently it is invertible as well, with inverse given by
  \be
  \Phi_2^{-1} = [\id_A \oti \tilde d_A] \circ 
  [\id_{A} \oti \Phi_1^{-1} \oti \id_{A^\vee} ] \circ
  [ b_A \oti\id_{A^\vee}]  \,.  \labl{eq:Phi2inv-Phi1inv}
It follows that we can define a candidate coproduct $\Delta$ by formula 
\erf{Delta}, and yet another candidate $\Delta'$ by the corresponding 
formula with $\Phi_2^{-1}$, i.e.
  \be  \Delta': = (m \oti \id_A) \circ (\id_A \oti \Phi_2^{-1} \oti \id_A)
  \circ (\id_A \oti \tilde b_A) \,.  \labl{Delta'}
Note that because of the relation \erf{eq:Phi2inv-Phi1inv},
we can also write
  \be  \Delta = (\id_A \oti m) \circ (\Phi_2^{-1} \oti \id_A \oti \id_A)
  \circ (\tilde b_A \oti \id_A)  \labl{Delta=}
and
  \be  \Delta' = (m \oti \id_A) \circ (\id_A \oti \id_A \oti \Phi_1^{-1})
  \circ (\id_A \oti b_A) \,,  \labl{Delta'=}
respectively. Pictorially,
  \begin{eqnarray} \begin{picture}(340,200)(0,0)
  \put(46,0)  {\begin{picture}(0,0)(0,0)
              \scalebox{.38}{\includegraphics{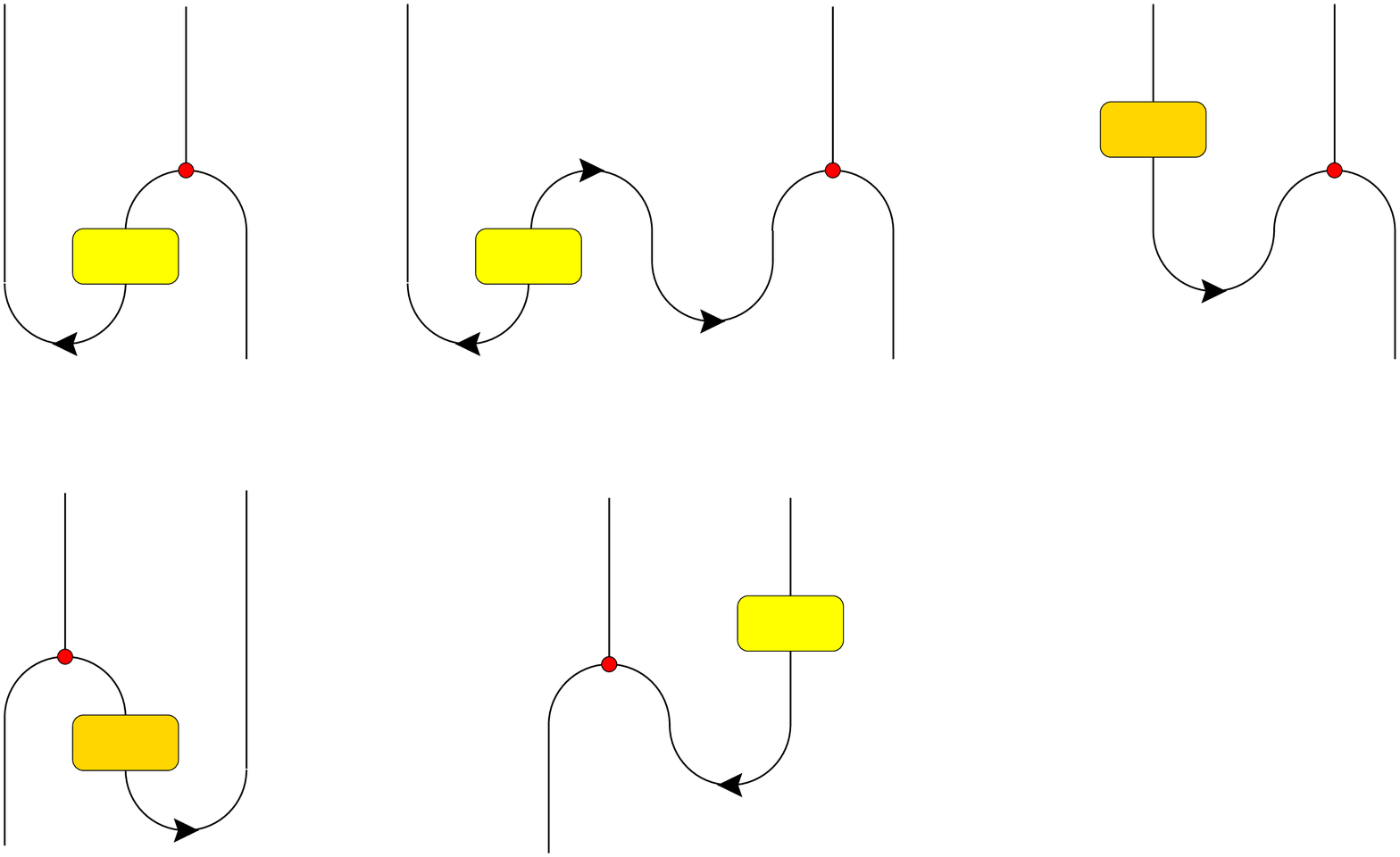}} \end{picture}}
  \put(-34,94)     {and}
  \put(0,40)       {$\Delta'\,:=$}
  \put(0,152)      {$\Delta\;:=$}
  \put(42.7,-6.9)  {\scriptsize$A$}
  \put(43.7,197.3) {\scriptsize$A$}
  \put(56.9,85.9)  {\scriptsize$A$}
  \put(67.2,23.1)  {\scriptsize$\Phi_2^{-1}$}
  \put(67.2,133.9) {\scriptsize$\Phi_1^{-1}$}
  \put(83.7,197.3) {\scriptsize$A$}
  \put(97.4,103.4) {\scriptsize$A$}
  \put(98.4,85.9)  {\scriptsize$A$}
  \put(115,152)    {$=$}
  \put(134.7,197.3){\scriptsize$A$}
  \put(135,40)     {$=$}
  \put(158.1,133.9){\scriptsize$\Phi_1^{-1}$}
  \put(166.1,-7.2) {\scriptsize$A$}
  \put(180.4,85.9) {\scriptsize$A$}
  \put(218.3,50.3) {\scriptsize$\Phi_1^{-1}$}
  \put(221.8,85.9) {\scriptsize$A$}
  \put(231.7,197.3){\scriptsize$A$}
  \put(243.4,103.4){\scriptsize$A$}
  \put(270,152)    {$=$}
  \put(300.2,162.9){\scriptsize$\Phi_2^{-1}$}
  \put(303.3,197.3){\scriptsize$A$}
  \put(345.7,197.3){\scriptsize$A$}
  \put(358.2,103.4){\scriptsize$A$}
  \end{picture} \nonumber \\[-3.5em]{}\label{Delta-Phi} \\[1.2em]\nonumber
  \end{eqnarray} 
However, the two morphisms $\Delta$ and $\Delta'$ defined this way actually
coincide. To see this, we compose the morphisms as given by \erf{Delta=} and 
\erf{Delta'=}, \resp, with $\Phi_2\oti\id_A$. In the case of $\Delta$ this yields
immediately the morphism $(\id_{A^\vee} \oti m) \circ (\tilde b_A\oti\id_A)$,
while in the case of $\Delta'$ the same result follows with the help of
associativity:
  \bea \begin{picture}(425,233)(0,33)
  \put(0,0)  {\begin{picture}(0,0)(0,0)
             \scalebox{.38}{\includegraphics{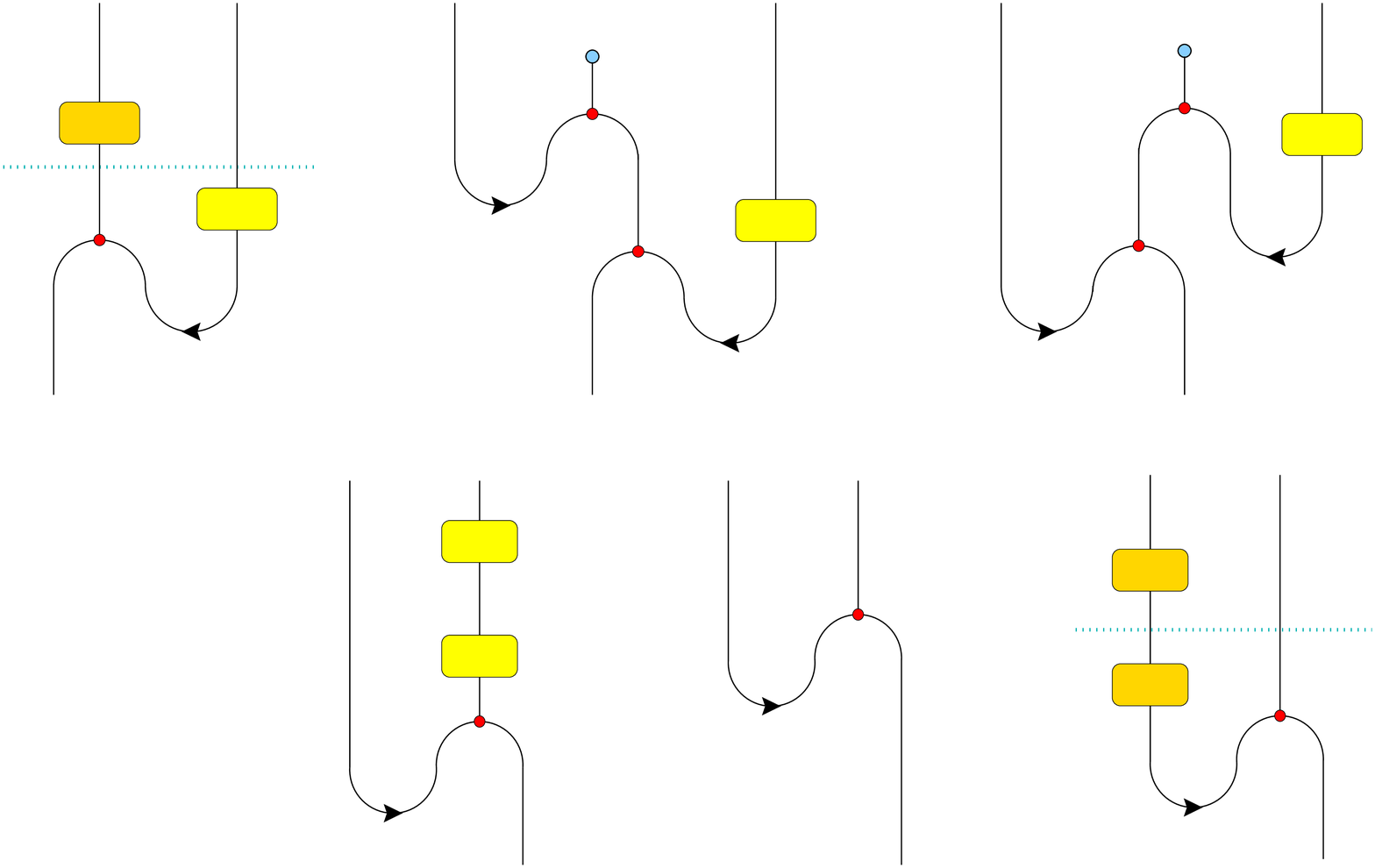}} \end{picture}}
  \put(12.9,132.3) {\scriptsize$A$}
  \put(25.2,220.1) {\scriptsize$\Phi_2$}
  \put(27.2,261.8) {\scriptsize$A$}
  \put(64.1,194.4) {\scriptsize$\Phi_1^{-1}$}
  \put(68.8,261.8) {\scriptsize$A$}
  \put(109,199)    {$\equiv$}
  \put(134.3,261.8){\scriptsize$A$}
  \put(174.7,132.3){\scriptsize$A$}
  \put(225.3,190.8){\scriptsize$\Phi_1^{-1}$}
  \put(229.5,261.8){\scriptsize$A$}
  \put(260,199)    {$\stackrel{\rm assoc.}{=}$}
  \put(296.5,261.8){\scriptsize$A$}
  \put(350.9,132.3){\scriptsize$A$}
  \put(389.1,217.1){\scriptsize$\Phi_1^{-1}$}
  \put(393.5,261.8){\scriptsize$A$}
  \put(81,57)      {$\equiv$}
  \put(101.7,118.8){\scriptsize$A$}
  \put(138,95)     {\scriptsize$\Phi_1^{-1}$}
  \put(139.1,61)   {\scriptsize$\Phi_1$}
  \put(140.2,118.8){\scriptsize$A$}
  \put(153.6,-9.2) {\scriptsize$A$}
  \put(186,57)     {$=$}
  \put(215.5,118.8){\scriptsize$A$}
  \put(254.7,118.8){\scriptsize$A$}
  \put(266.7,-9.2) {\scriptsize$A$}
  \put(298,57)     {$=$}
  \put(337.2,52)   {\scriptsize$\Phi_2^{-1}$}
  \put(339.8,86.7) {\scriptsize$\Phi_2$}
  \put(341.8,119.5){\scriptsize$A$}
  \put(379.9,119.5){\scriptsize$A$}
  \put(393.7,-9.2) {\scriptsize$A$}
  \epicture22 \labl{Delta-Delta}
The coassociativity property then follows from the fact that
associativity of $m$ implies equality of $(\Delta\oti\id_A)\cir\Delta'$ and
$(\id_A\oti\Delta')\cir\Delta$\,:
  \bea \begin{picture}(260,82)(0,41)
  \put(0,0)  {\begin{picture}(0,0)(0,0)
             \scalebox{.38}{\includegraphics{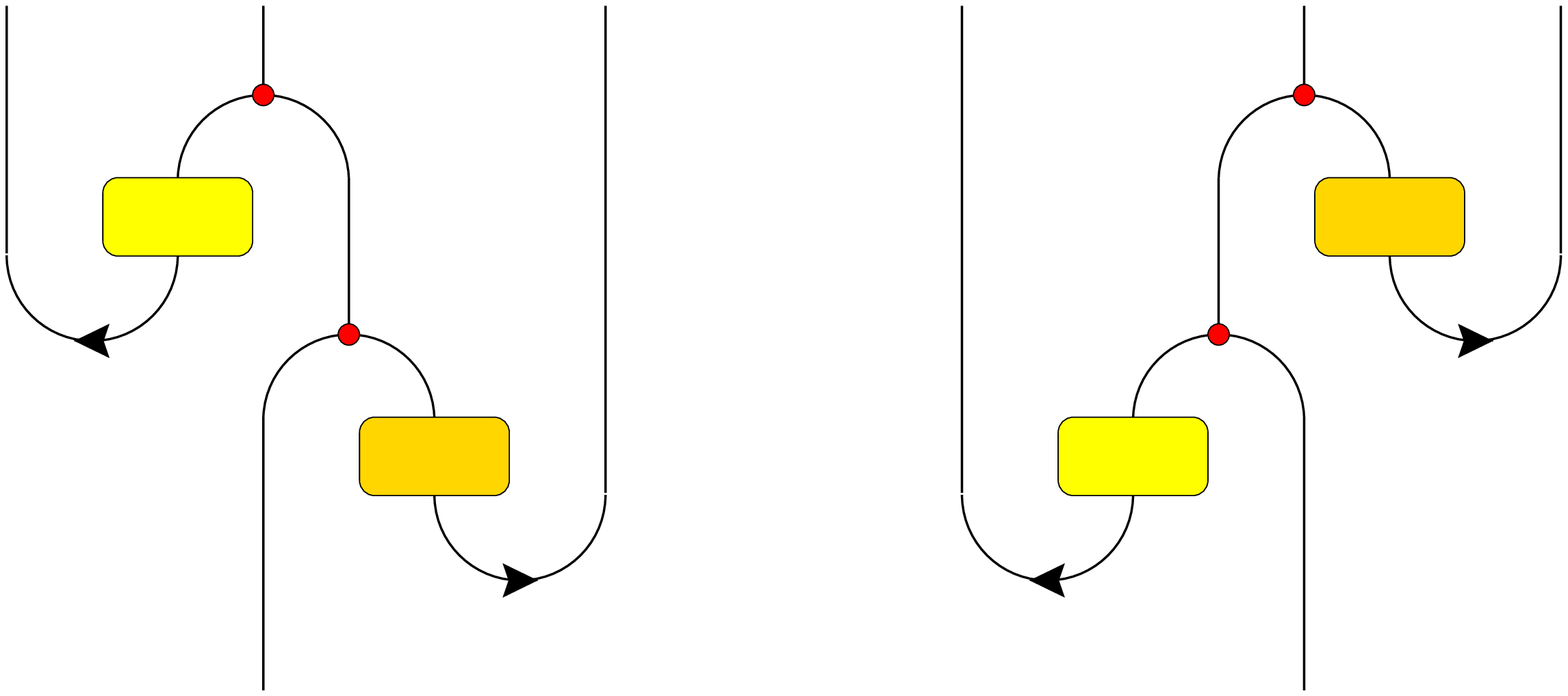}} \end{picture}}
  \put(-2.8,113.9) {\scriptsize$A$}
  \put(21.2,73.9)  {\scriptsize$\Phi_1^{-1}$}
  \put(37.5,-8.6)  {\scriptsize$A$}
  \put(38.4,113.9) {\scriptsize$A$}
  \put(63.2,35.9)  {\scriptsize$\Phi_2^{-1}$}
  \put(93.7,113.9) {\scriptsize$A$}
  \put(120,57)     {$=$}
  \put(150.3,113.9){\scriptsize$A$}
  \put(174.8,35.9) {\scriptsize$\Phi_1^{-1}$}
  \put(204.5,-8.6) {\scriptsize$A$}
  \put(205.3,113.9){\scriptsize$A$}
  \put(215.7,73.9) {\scriptsize$\Phi_2^{-1}$}
  \put(246.3,113.9){\scriptsize$A$}
  \epicture29 \labl{ddp-dpd}
That the counit properties $(\id_A\oti\eps)\cir\Delta\eq\id_A$ and
$(\eps\oti\id_A)\cir\Delta'\eq\id_A$ are satisfied follows directly
from the definition of $\Phi_1$ and $\Phi_2$. Finally, the Frobenius 
property \erf{frob} follows again easily from associativity of $m$, using 
$\Delta$ for showing the first equality, and $\Delta'$ for the second:
  \begin{eqnarray} \begin{picture}(433,109)(0,0)
  \put(0,0)  {\begin{picture}(0,0)(0,0)
             \scalebox{.38}{\includegraphics{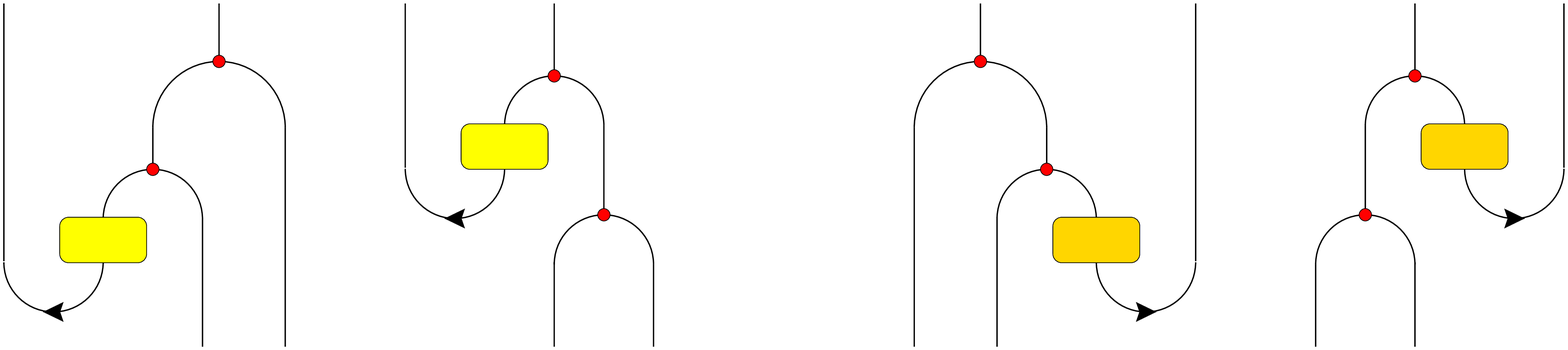}} \end{picture}}
  \put(-2.8,98.8)  {\scriptsize$A$}
  \put(20.9,27.8)  {\scriptsize$\Phi_1^{-1}$}
  \put(51.7,-8.5)  {\scriptsize$A$}
  \put(57.4,98.8)  {\scriptsize$A$}
  \put(74.7,-8.5)  {\scriptsize$A$}
  \put(91.8,44)    {$=$}
  \put(109.4,98.8) {\scriptsize$A$}
  \put(131.9,53.9) {\scriptsize$\Phi_1^{-1}$}
  \put(149.9,98.8) {\scriptsize$A$}
  \put(149.1,-8.5) {\scriptsize$A$}
  \put(177.1,-8.5) {\scriptsize$A$}
  \put(206,44)     {and}
  \put(247.7,-8.5) {\scriptsize$A$}
  \put(266.7,98.8) {\scriptsize$A$}
  \put(271.1,-8.5) {\scriptsize$A$}
  \put(296.8,27.8) {\scriptsize$\Phi_2^{-1}$}
  \put(326.3,98.8) {\scriptsize$A$}
  \put(344.2,44)   {$=$}
  \put(358.8,-8.5) {\scriptsize$A$}
  \put(386.4,-8.5) {\scriptsize$A$}
  \put(386.6,98.8) {\scriptsize$A$}
  \put(398.3,53.9) {\scriptsize$\Phi_2^{-1}$}
  \put(430.0,98.8) {\scriptsize$A$}
  \end{picture} \nonumber \\[-1.9em] {} \label{frob-dpd}
  \end{eqnarray} 
\mbox{$\ $}\qed

\dt{Definition}
For $A$ an algebra in a sovereign \tc, the morphisms $\epsnat$ and
$\epsnatt$ in $\Hom(A,\one)$ are defined as
  \bea \begin{picture}(205,52)(0,33)
  \put(0,0)  {\begin{picture}(0,0)(0,0)
             \scalebox{.38}{\includegraphics{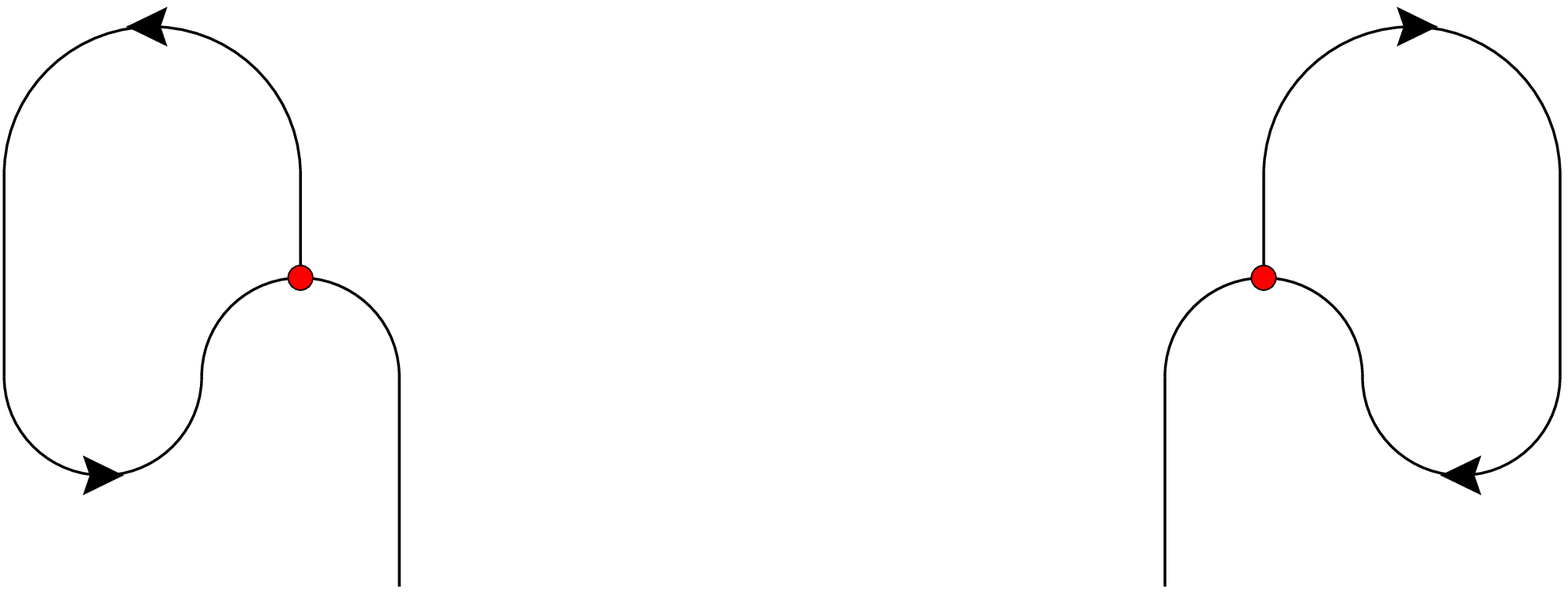}} \end{picture}}
  \put(-35.8,41.4) {$\epsnat\,:=$}
  \put(51.5,-8.5)  {\scriptsize$A$}
  \put(82,41.4)    {and}
  \put(123.8,41.4) {$\epsnatt\,:=$}
  \put(157.7,-8.5) {\scriptsize$A$}
  \epicture17 \label{eq:eps-def} \labl{epsnat}

\dtl{Lemma}{lem:epsnat-sym}
Let $A$ be an algebra in a sovereign \tc. Then we have:
\\[.13em]
(i)\hsp{.6}$A$ together with $\eps\,{:=}\,\xi\,\epsnat$ is a symmetric 
algebra, for any $\xi \iN \complex$.
The same holds with $\eps\eq\xi\epsnatt$.
\\[.13em]
(ii) If $A$ is a symmetric Frobenius algebra, then $\epsnat\eq\epsnatt$.

\medskip\noindent
Proof:\\
(i) According to definition \ref{df:spec-etc}(ii) we must show that 
$\Phi_2\eq\Phi_1$. This is verified as follows. We have
  \bea \begin{picture}(337,92)(0,37)
  \put(0,0)  {\begin{picture}(0,0)(0,0)
              \scalebox{.38}{\includegraphics{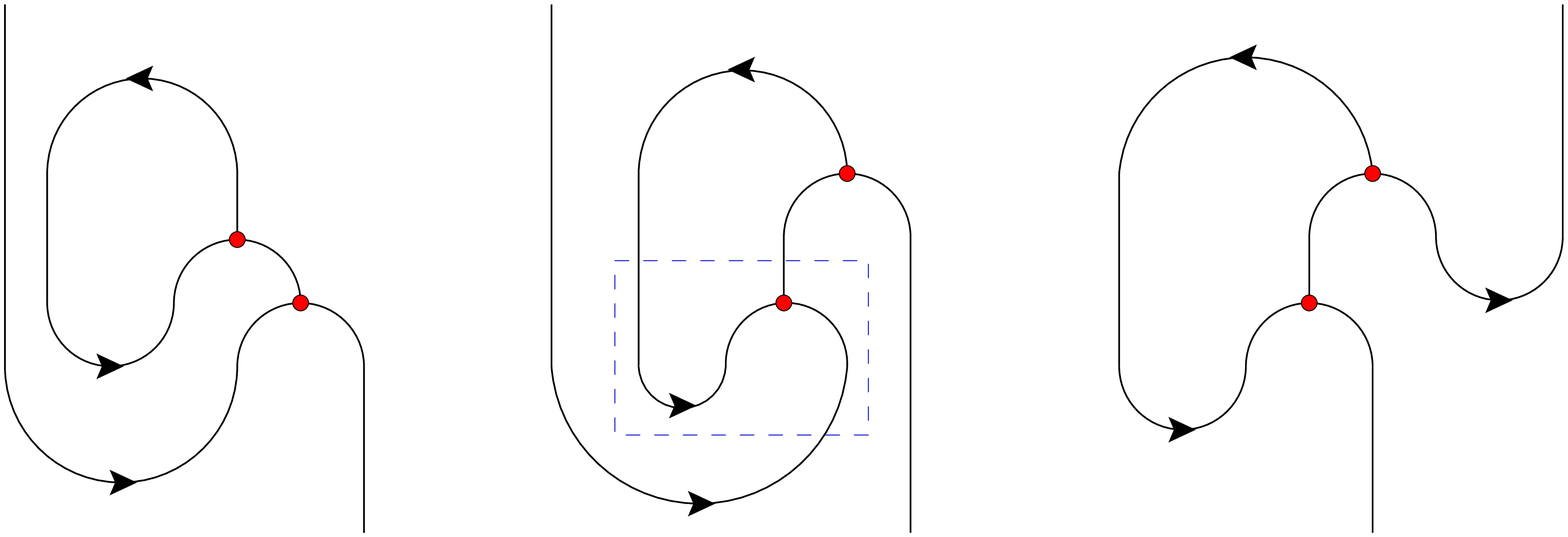}} \end{picture}}
  \put(-50,52)      {$\Phi_2\ =\; \xi$}
  \put(-4.1,118.5)  {\scriptsize$\Av$}
  \put(74.5,-8.8)   {\scriptsize$A$}
  \put(88,52)       {$=\; \xi $}
  \put(114.1,118.5) {\scriptsize$\Av$}
  \put(193.3,-8.8)  {\scriptsize$A$}
  \put(211.4,52)    {$=\; \xi $}
  \put(292.5,-8.8)  {\scriptsize$A$}
  \put(333.1,118.5) {\scriptsize$\Av$}
  \epicture26 \labl{Phi12nat} 
Here the first step is the definition, the second uses associativity, and
the third identity follows by replacing the left-dual of the morphism that 
is enclosed in a dashed box by its right-dual, which is allowed because we 
are working in a sovereign category. That the \rhs\ of \erf{Phi12nat} equals 
$\Phi_1$ follows by using once more associativity and the definition of 
$\eps\eq\xi\epsnat$.
\\[.13em]
(ii) Since $A$ is a Frobenius \alg, the morphisms $\Phi_{1,2}$ are 
invertible. And since
$A$ is symmetric, we have $\Phi_1\eq\Phi_2$. Thus in particular the
equalities $\Phi_1\cir\Phi_2^{-1}\eq\Phi_2\cir\Phi_1^{-1}\eq\id_{A^\vee}$ 
and $\Phi_1^{-1}{\circ}\,\Phi_2\,{=}\,\Phi_2^{-1}{\circ}\,\Phi_1\,{=}\,\id_A$
hold. Pictorially, we have
  \bea \begin{picture}(220,89)(0,49)
  \put(0,0)  {\begin{picture}(0,0)(0,0)
             \scalebox{.38}{\includegraphics{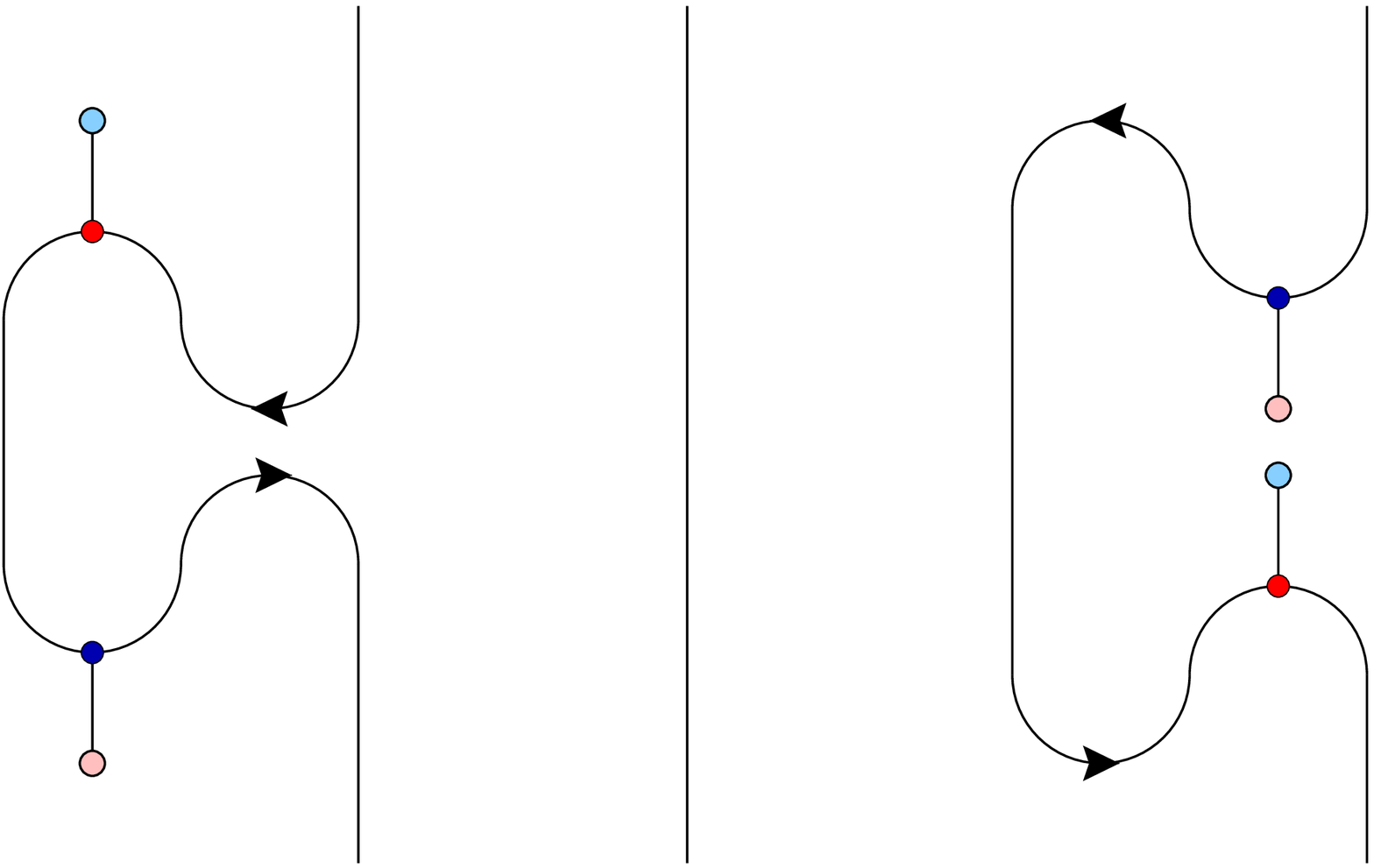}} \end{picture}}
  \put(52.5,-9.3)  {\scriptsize$A$}
  \put(52.5,137.5) {\scriptsize$A$}
  \put(78,63)      {$=$}
  \put(101.5,-9.3) {\scriptsize$A$}
  \put(102.5,137.5){\scriptsize$A$}
  \put(126.9,63)   {$=$}
  \put(207.5,-9.3) {\scriptsize$A$}
  \put(208.1,137.5){\scriptsize$A$}
  \epicture33  \label{eq:sym-prop} \labl{Phi1-Phi2inv}
as well as the mirror images of these identities.
Consider now the transformations
  \bea \begin{picture}(221,56)(0,33)
  \put(0,0)  {\begin{picture}(0,0)(0,0)
             \scalebox{.38}{\includegraphics{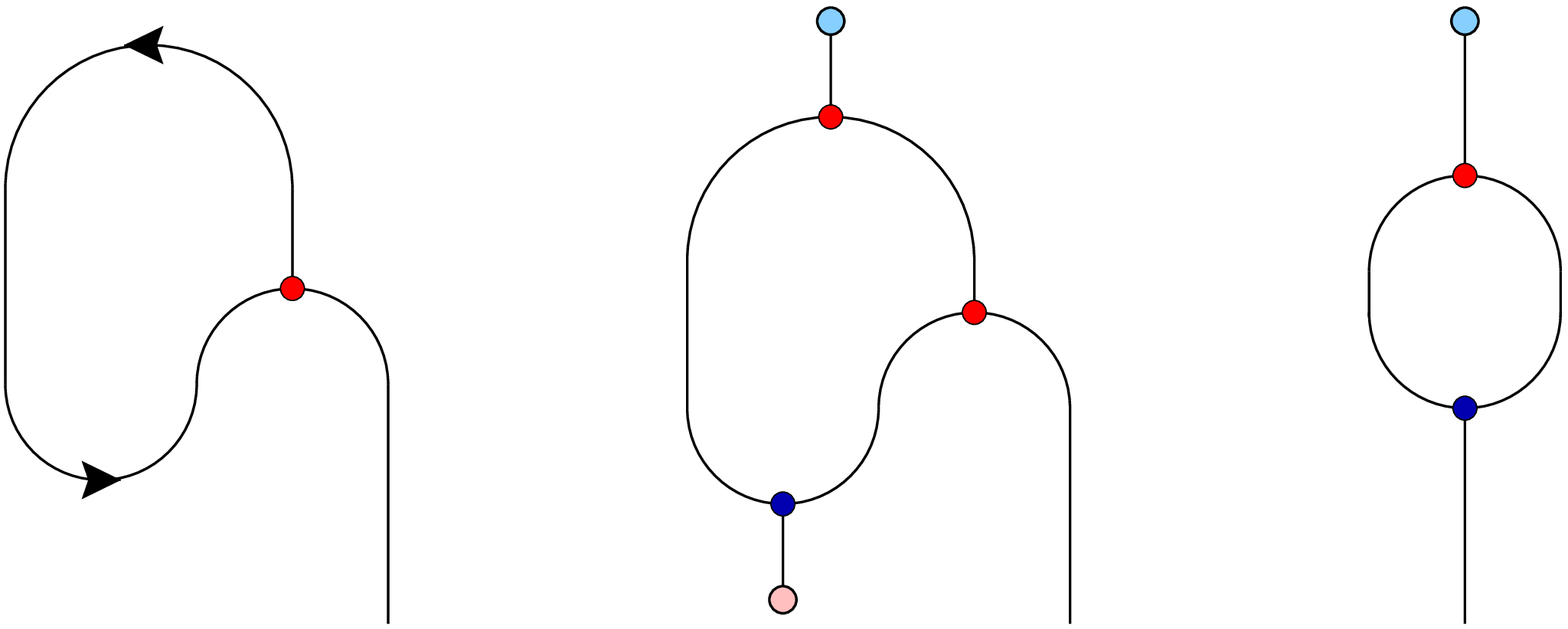}} \end{picture}}
  \put(51.5,-8.5)  {\scriptsize$A$}
  \put(74.4,44.4)  {$=$}
  \put(148.8,-8.5) {\scriptsize$A$}
  \put(171.9,44.4) {$=$}
  \put(206.5,-8.5) {\scriptsize$A$}
  \epicture21 \label{eq:eps-nat-xfer} \labl{epsnato}
where in the first step we used \erf{eq:sym-prop} to insert unit and
counit, while the second step is an application of the Frobenius and
unit properties. Analogously one shows a mirror version of
\erf{eq:eps-nat-xfer}; combining the two identities then 
establishes the lemma. 
\qed

\dtl{Corollary}{lem:hap-sym}
If $A$ is a haploid algebra, then it is symmetric for any choice of
$\eps\iN\Hom(A,\one)$.

\medskip\noindent
Proof: \\
By definition, for haploid algebras the morphism space $\Hom(A,\one)$ 
has dimension one. Furthermore, $\epsnat\,{\ne}\,0$ since $\epsnat\cir
\eta\eq\dim(A)$. Thus $\epsnat$ forms a basis of $\Hom(A,\one)$ and 
any choice of $\eps$ is proportional to $\epsnat$.
\qed

As an immediate consequence of the relation \erf{Phi1-Phi2inv}, for
every symmetric special Frobenius \alg\ the 
normalisations $\beta_\one$ and $\beta_{\!A}$ in \erf{spec2} obey
  \be  \beta_\one\,\beta_{\!A} = \dim(A) \,.  \ee 
In the present section we keep these normalisations explicitly. But 
from section \ref{sec:representations} onwards we will simplify the 
presentation by assuming (without loss of generality) that the
 coproduct is normalised such that $\beta_{\!A}\eq1$, and hence 
$\beta_\one\eq\dim(A)$.
\vspace{-.3em}

\dtl{Lemma}{lem:spec-epsnat}
For a symmetric Frobenius algebra $A$ the following statements
are equivalent:\\[.13em]
(i)\hsp{.6}$A$ is special.\\[.13em]
(ii) The counit obeys%
 \foodnode{Recall from lemma~\ref{lem:epsnat-sym} that 
 $\epsnat\eq\epsnatt$ for symmetric Frobenius algebras.}
$\eps\eq\beta\,\epsnat$ for some non-zero number $\beta$.

\medskip\noindent
Proof:\\
(i)$\,\Rightarrow$\,(ii)\,: Composing the specialness property
$m\cir\Delta\eq\beta_{\!A}\,\id_A$ (where $\beta_{\!A}$ is non-zero)
with $\eps$ yields $\eps\eq\beta_{\!A}^{-1} \eps\cir m\cir\Delta$, which is
the last expression in \erf{eq:eps-nat-xfer}. Backtracking the
steps in \erf{eq:eps-nat-xfer} thus gives $\eps\eq\beta_{\!A}^{-1}\epsnat$. 
\\[.13em]
(ii)\,$\Rightarrow$\,(i)\,: By composition with the unit we get
  \be
  \eps \circ \eta = \beta\, \epsnat\circ\eta = \beta\, \dim(A) \ne 0 \,,
  \ee
where in the last step the unit property is used to obtain
the trace of $\id_A$. Second, consider the moves
  \bea \begin{picture}(350,68)(0,41)
  \put(0,0)  {\begin{picture}(0,0)(0,0)
             \scalebox{.38}{\includegraphics{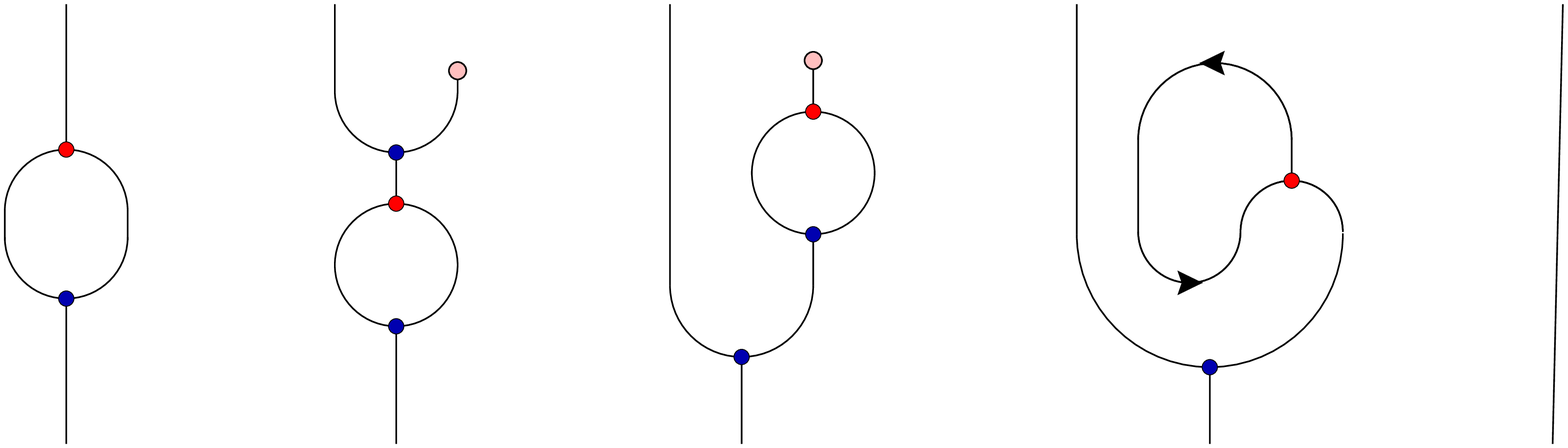}} \end{picture}}
  \put(11.5,-9.5)  {\scriptsize$A$}
  \put(12.1,102.9) {\scriptsize$A$}
  \put(46.9,48.1)  {$=$}
  \put(71.7,102.9) {\scriptsize$A$}
  \put(84.5,-9.5)  {\scriptsize$A$}
  \put(121.9,48.1) {$=$}
  \put(146.1,102.9){\scriptsize$A$}
  \put(161.1,-9.5) {\scriptsize$A$}
  \put(211.6,48.1) {$=$}
  \put(236.6,102.9){\scriptsize$A$}
  \put(265.5,-9.5) {\scriptsize$A$}
  \put(312.5,48.1) {$=\dsty\frac1\beta$}
  \put(342.5,-9.5) {\scriptsize$A$}
  \put(344.1,102.9){\scriptsize$A$}
  \epicture27 \labl{epsnatb}
In the first step a counit is inserted, the second step combines the
Frobenius property and coassociativity, the third step is again
backtracking the steps \erf{eq:eps-nat-xfer}, and finally the assumption 
$\eps\eq\beta\epsnat$ is inserted and the counit property is used.
\qed
\vspace{-.3em}

\dtl{Lemma}{lem:spec-exists}
For any algebra $(A,m,\eta)$ the following two statements
are equivalent:\\[.13em]
(i) There exist $\eps\iN\Hom(A,\one)$ and $\Delta \iN \Hom(A,A\Oti A)$ such 
that $(A,m,\eta,\Delta,\eps)$ is a symmetric special Frobenius algebra.
\\[.13em]
(ii) The morphism $\Phi_1'\iN\Hom(A,A^\vee)$, defined as in \erf{eq:Phi-def},
but with $\epsnat$ (as defined in \erf{eq:eps-def}) in
place of $\eps$, is invertible.
\\[.13em]
(The same holds when taking $\epsnatt$ instead of $\epsnat$).

\medskip\noindent
Proof:\\
(i)\,$\Rightarrow$\,(ii)\,: 
When $A$ is symmetric special Frobenius, we have
$\eps\eq\beta\epsnat$ (by lemma \ref{lem:spec-epsnat}), 
and hence $\Phi_1\eq\beta \Phi_1'$, with
a non-zero number $\beta$. Further, since $A$ is Frobenius, 
the morphism $\Phi_1$ is invertible, and hence so is $\Phi_1'$.
\\[.13em] 
(ii)\,$\Rightarrow$\,(i)\,: We must find morphisms
$\eps$ and $\Delta$ such that $(A,m,\eta,\Delta,\eps)$ is a 
symmetric special Frobenius algebra. For $\eps$ we choose
$\eps\,{:=}\,\epsnat$. {}From lemma \ref{lem:epsnat-sym}(i) it follows
that $A$ is symmetric. Since by assumption $\Phi_1$ is invertible, 
there exists a $\Delta$ that turns $A$ into a Frobenius algebra; this
morphism $\Delta$ is the one given in \erf{Delta-Phi}.
Thus $(A,m,\eta,\Delta,\eps)$ is a symmetric Frobenius algebra. 
But because of $\eps\eq\epsnat$, lemma \ref{lem:spec-epsnat} 
implies that $A$ is special, too.
\qed

\dtl{Remark}{remark:alg-c-vect}
Let us add two comments on the case
where $\calc$ is the category of complex vector spaces.
\\[-1.4em]
\begin{itemize}
\item[(i)] Suppose $A$ is an algebra over $\complex$ with basis 
$\{b_i\}$. Then the morphism $\epsnat$ is the trace of the multiplication 
operator, in the sense that, with the product given by 
$b_i\,{\times}\, b_j\eq\sum_k\! m_{ij}^{\ k}\,b_k$, it satisfies 
$\epsnat(b_i)\eq\sum_k\! m_{ki}^{\ k}$. For example, when $A$ 
is the algebra of functions over some finite set, then a basis is given 
by the delta functions, for which the product reads $b_i\,{\times}\,b_j\eq
\delta_{i,j}b_j$, and we obtain $\epsnat(b_i)\eq1$. Thus in this case 
$\epsnat$ is an integral with a measure that weighs every point evenly.
\\[-1.7em]
\item[(ii)] It is easy to construct algebras $(A,m,\eta)$ such that 
$\Phi_1$ is not invertible when setting $\eps\eq\epsnat$. Take e.g.\ 
$A\eq\complex^2$ with a basis $\{e,n\}$ such that $e$ acts as the unit 
element while $n\,{\times}\,n\eq0$. Then $\epsnat(e)\eq2$ and 
$\epsnat(n)\eq0$, and as a consequence $\Phi_1(n)\eq0$.
\end{itemize}

\medskip\noindent
Proof of theorem \ref{thm:alg-from-bc}:\\[.3em]
(i) Given an algebra object $(A,m,\eta)$,
suppose there exist $\eps$ and $\Delta$ such that 
$(A,m,\eta,\Delta,\eps)$ is a symmetric special Frobenius algebra.
By lemma \ref{lem:spec-epsnat} we have $\eps\eq\xi \epsnat$ 
for some $\xi\iN\complex^\times$. But $\epsnat$ is already
fixed in terms of the multiplication $m$ and thus
$\eps$ is fixed up to a normalisation constant $\xi$
in terms of $(A,m,\eta)$. By lemma \ref{lem:frob-unique}
the coproduct $\Delta$ is uniquely fixed in terms of 
$(A,m,\eta,\eps)$ and by lemma \ref{lem:epsnat-sym},
$(A,m,\eta,\eps)$ is symmetric.
\\[.3em]
(ii) In section \ref{sec:alg2bCFT} it was shown that $(A,m,\eta)$ 
with $A$ given in terms of the CFT data by \erf{eq:A=H} and
multiplication $m$ defined in terms of boundary structure constants
by \erf{eq:m=c} is an algebra object. The counit $\eps$ 
was defined in terms of boundary one-point functions on the 
upper half plane and has the property \erf{CFT-eps}.
Note that the calculation leading to \erf{CFT-eps} is based on
the assumption that there are no states of negative conformal
weight; this assumption is fulfilled in every unitary CFT.
The invertibility of the matrix $G(a)$ in
\erf{def-gaab} is equivalent to the statement 
that the morphism $\Phi_1$ in \erf{Phi12} is invertible.
Furthermore, comparing \erf{CFT-eps} to $\epsnatt$, as
defined in \erf{epsnat} and expressed in a basis, shows that 
$\eps\eq{\rm const}'\, \epsnatt$. The constant is non-zero since
$\Phi_1$ is invertible.
Thus lemma \ref{lem:spec-exists} is applicable, since
$\Phi_1'\eq({\rm const}')^{-1} \Phi_1$ is invertible as well.
\qed

\subsection{The associated topological algebra}

An algebra object $A$ in a tensor category $\calc$ defines 
automatically also an algebra over $\complex$, to be denoted as \Atop\
and called the {\em topological algebra associated to\/} $A$. The
complex algebra $\Atop$ is defined
as follows. As a vector space, \Atop\ is the morphism space
  \be \Atop := \Hom(\one,A) \,.  \labl{Atop}
The multiplication $m\top$ and unit $\eta\top$ (regarded as a map 
$\complex\,{\to}\,\Atop$) are given by
  \be
  m\top(\alpha\Oti\beta) := m\cir(\alpha\oti\beta) \,,\qquad
  \eta\top(1) := \eta \,, \ee
where $\alpha,\beta\iN\Atop$.

In the CFT interpretation introduced in section \ref{sec:alg2bCFT}, 
$\Atop$ is the algebra of boundary fields (on a fixed boundary condition) 
that transform in the vacuum representation of the chiral algebra. The 
fields of weight zero in this set form a topological subsector of the 
boundary theory, hence the name $\Atop$. The presence of such a subsector 
fits well with the description \cite{bape,fuhk,duJo2,abra,kamo,stri2}
of \twodim\ lattice TFTs via symmetric special Frobenius algebras in the
category of complex vector spaces.

\medskip

In the sequel we will deduce various somewhat technical results, which will
be instrumental later on. The reader not interested in the proofs of these
results may proceed directly to section \ref{sec:sum-prod-op}.

Let us choose bases for \Atop\ and its dual in such a way that
  \be  b_i \in \Hom(\one,A) \,, \qquad  b^j \in \Hom(A,\one) \,, \qquad
  b^j \cir b_i = \delta_i^{\;j} \,.  \labl{eq:Atop-bas}

\dt{Lemma} 
Let $A$ be a symmetric special Frobenius algebra in
$\calc$ and \Atop\ its associated topological algebra. Setting 
  \be 
  \Delta\top(\alpha) := \sum_{i,j} \llb(b^i\Oti b^j) \cir \Delta \cir
  \alpha \lrb \; \,b_i \Oti b_j \,, \qquad{\rm and} \qquad
  \eps\top(\alpha) := \eps\cir\alpha \labl{co-top}
for all $\alpha \iN \Atop$
turns \Atop\ into a symmetric Frobenius algebra over \complex.

\medskip\noindent
Proof:\\
The statement follows by direct computation. The calculations all boil down to
the observation that for a graph with \Atop-lines, products and coproducts 
{\em without loops\/} and with arbitrary basis elements $b_i$ and $b^j$
inserted at the external lines is equal to the same graph with 
$A$-lines, products and coproducts.
\qed

Note, however, that with $\eps\top$ defined as above, \Atop\ is not
necessarily special. Indeed, the definition of specialness contains
graphs with loops and the above argument no longer applies.  
For example, let $u$ and $v$ be two non-isomorphic simple objects in
\calc\ and let $A\eq(u{\oplus}v) \Oti (u{\oplus}v)^\vee$, with product as
defined in formula \erf{mUU}, and $\eps\eq\epsnat$ as counit. A basis of 
$\Atop$ is given by the duality morphisms $\{\tilde b_u,\tilde b_v\}$ of
simple objects $u,v$. One verifies that (with $x$ and $y$ standing for any 
of $u,v$) $m\top(\tilde b_x,\tilde b_y)\eq\delta_{x,y} \tilde b_x$ and
$\eps\top(\tilde b_x)\eq\dim(u{\oplus}v)\,\dim(x)$. However,
by the same reasoning as in remark \ref{remark:alg-c-vect} one shows that
$\epsnattop(b_x)\,{\equiv}\,1$. It follows that
for $\dim(u)\,{\ne}\,\dim(v)$, $\eps\top$ as defined by \erf{co-top}
is not proportional to
$\epsnattop$, and hence by lemma \ref{lem:spec-epsnat} \Atop\ cannot be special.

\dtl{Definition}{def:cent-Atop}
Let $A$ be an algebra in $\calc$ and \Atop\ the associated topological 
algebra. The {\em relative center of \Atop\ with respect to\/} $A$ is
the subspace
  \be  \centreA(\Atop) := \big\{\, \alpha\iN\Atop \,|\, 
  m \cir (\alpha\Oti\id_A) \eq m\cir(\id_A\Oti\alpha) \,\big\} \,.  \ee

Elements in $\centreA(\Atop)$ in particular commute with
all elements of \Atop. Thus the relative center is a subalgebra 
of the ordinary center $\centre(\Atop)$ of $\Atop$ (defined in the
usual way for algebras over $\complex$). Also note that the unit
$\eta$ always lies in $\centreA(\Atop)$. Thus we have the inclusions
  \be  \{ \xi \eta \,|\, \xi \iN \complex \} \subseteq
  \centreA(\Atop) \subseteq \centre(\Atop) \subseteq \Atop \,.  \ee
The subset $\centreA$ turns out to be useful in the description of
the torus partition function. Indeed, we will see that
$Z_{00}\eq\dim(\centreA(\Atop))$. The proof will be given in section 
\ref{sec:torus-pf}; it relies on the following lemma.
\vspace{-.9em}

\dtl{Lemma}{lem:alpha-cent}
For any symmetric special Frobenius algebra $A$ with $\beta_{\!A}\eq1$
one has
  \bea \begin{picture}(300,66)(0,33)
  \put(150,0) {\begin{picture}(0,0)(0,0)
              \scalebox{.38}{\includegraphics{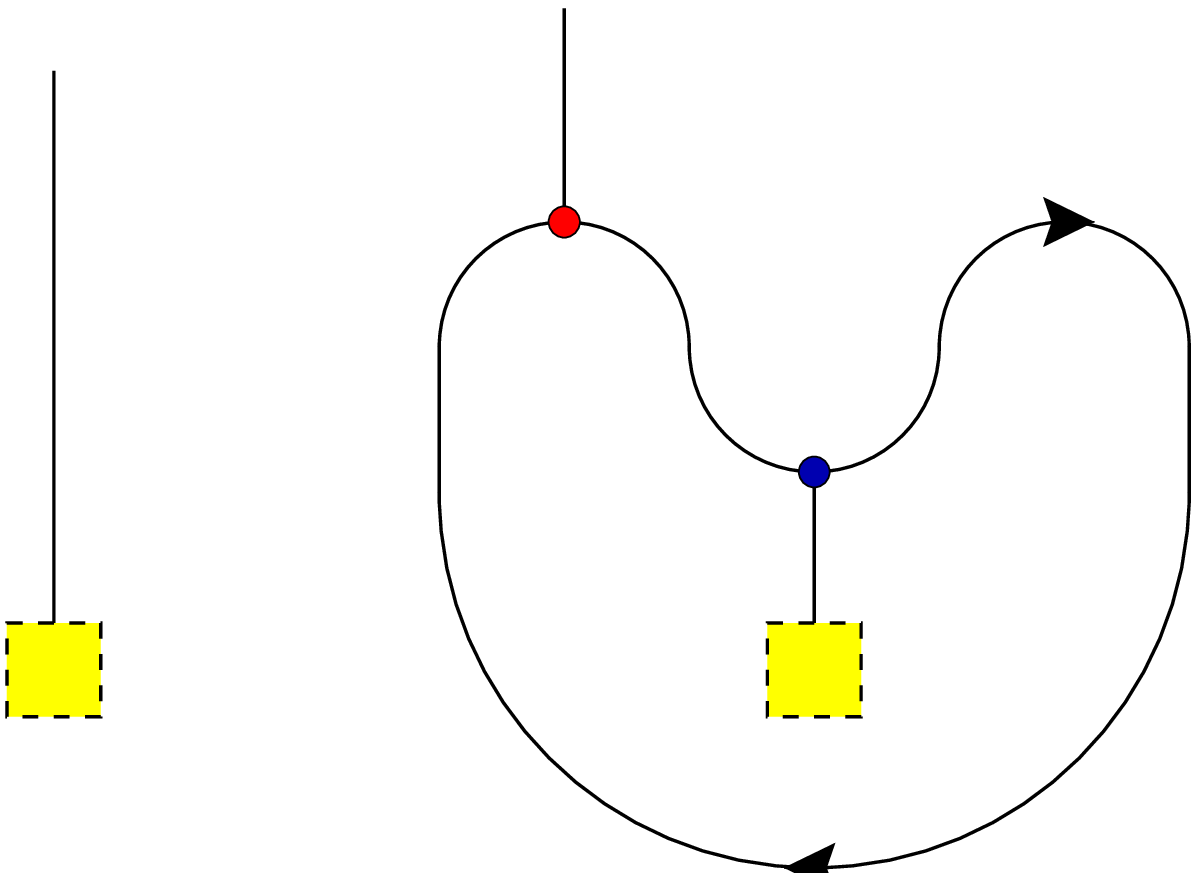}} \end{picture}}
  \put(0,49.5)      {$\alpha\in\centreA(\Atop)\quad\ \Longleftrightarrow$}
  \put(153.6,91.7)  {\scriptsize$A$}
  \put(153.3,21.3)  {\tiny$\alpha$}
  \put(172.9,49.5)  {$=$}
  \put(208.8,98.6)  {\scriptsize$A$} 
  \put(236.6,21.3)  {\tiny$\alpha$}
  \epicture12 \labl{centreA} 
\medskip\noindent
Proof:\\
We have the following equivalences:
  \begin{eqnarray} \begin{picture}(430,83)(12,0)
  \put(0,0)   {\begin{picture}(0,0)(0,0)
              \scalebox{.38}{\includegraphics{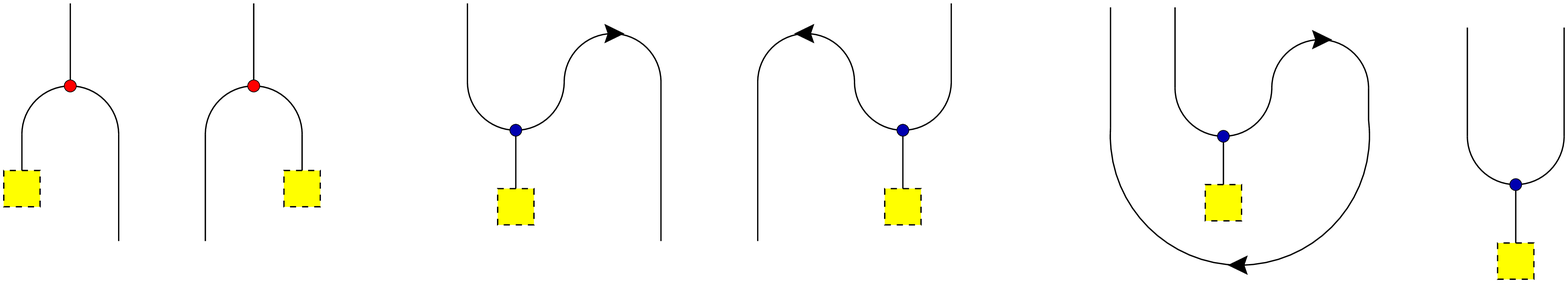}} \end{picture}}
  \put(42.9,37.5)   {$=$}
  \put(3.3,24.7)    {\tiny$\alpha$}
  \put(82.5,24.7)   {\tiny$\alpha$}
  \put(17.0,82.5)   {\scriptsize$A$}
  \put(68.0,82.5)   {\scriptsize$A$}
  \put(29.0,1.5)    {\scriptsize$A$}
  \put(54.0,1.5)    {\scriptsize$A$}
  \put(97.9,37.5)   {$\Longleftrightarrow$}
  \put(400.9,37.5)  {$=$}
  \put(143.5,20.0)  {\tiny$\alpha$}
  \put(253.3,20.0)  {\tiny$\alpha$}
  \put(129.7,82.5)  {\scriptsize$A$}
  \put(267.1,82.5)  {\scriptsize$A$}
  \put(184.3,1.5)   {\scriptsize$A^\vee$}
  \put(212.1,1.5)   {\scriptsize$A^\vee$}
  \put(277.9,39.5)  {$\Longleftrightarrow$}
  \put(195.9,39.5)  {$=$}
  \put(344.9,21.0)  {\tiny$\alpha$}
  \put(427.0,4.0)   {\tiny$\alpha$}
  \put(312.1,82.5)  {\scriptsize$A$}
  \put(330.4,82.5)  {\scriptsize$A$}
  \put(412.0,78)    {\scriptsize$A$}
  \put(440.3,78)    {\scriptsize$A$}
  \end{picture} \nonumber \\[-1.7em]{} \label{centreA-2} \end{eqnarray}
The first of these equivalences is obtained by composing the first equality 
on the left with 
$\Phi_2^{-1}$ on the left and with $\Phi_1^{-1}$ on the \rhs\ (these inverses
exist by the Frobenius property, and they are equal by the symmetry property)
and then using  the Frobenius property, while the second equivalence follows
upon composition of the middle identity with a duality morphism. The lemma 
is now established if the equality on the \rhs\ of \erf{centreA} 
can be shown to be equivalent to the one on the \rhs\ of \erf{centreA-2}.
But this is just a special case, obtained by setting $X\eq\one$, of 
lemma \ref{lem:alphaX} below.
\qed

\medskip
In the following considerations the braiding between the algebra $A$ and 
arbitrary objects $X$ plays a role.

\dtl{Lemma}{lem:alphaX}
Let $A$ be a special Frobenius algebra with $\beta_{\!A}\eq1$. For any 
object $X$ and any morphism $\alpha\iN\Hom(X,A)$ we have
  \bea \begin{picture}(340,65)(0,30)
  \put(0,0)   {\begin{picture}(0,0)(0,0)
              \scalebox{.38}{\includegraphics{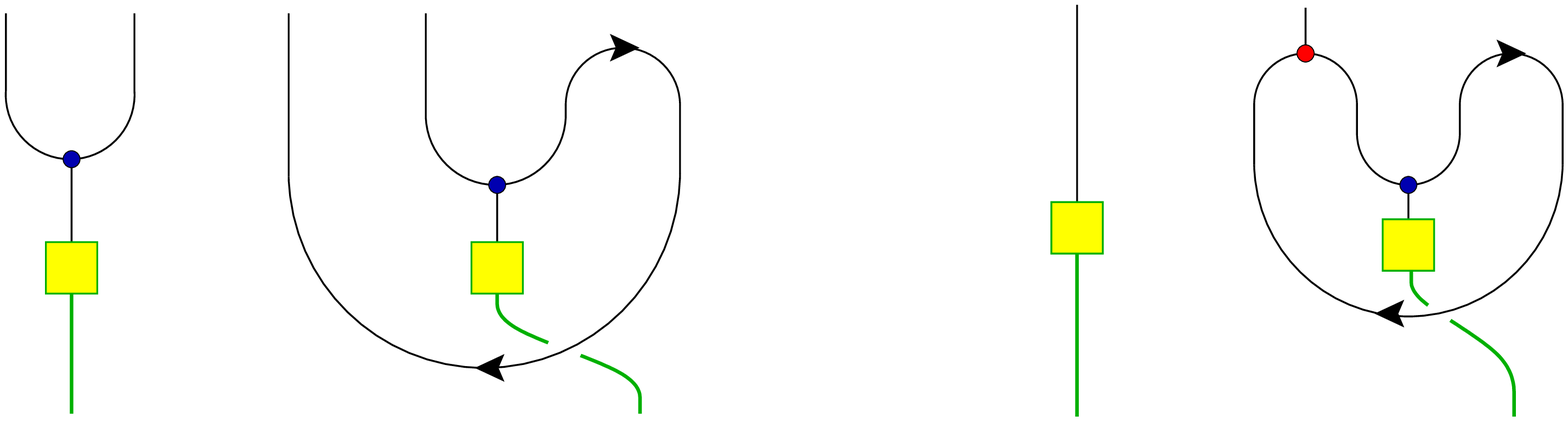}} \end{picture}}
   \put(35,39)  {$=$}
   \put(230,39) {$=$}
   \put(164,39) {$\Longleftrightarrow$}
   \put(-2,87)  {\scriptsize$A$}
   \put(24,87)  {\scriptsize$A$}
   \put(54,87)  {\scriptsize$A$}
   \put(82,87)  {\scriptsize$A$}
   \put(82,87)  {\scriptsize$A$}
   \put(212,87) {\scriptsize$A$}
   \put(257,87) {\scriptsize$A$}
   \put(9.1,-8.6){\scriptsize$X$}
   \put(124,-8.6){\scriptsize$X$}
   \put(212,-8.6){\scriptsize$X$}
   \put(298,-8.6){\scriptsize$X$}
   \put(11,29)  {\scriptsize$\alpha$}
   \put(96,29)  {\scriptsize$\alpha$}
   \put(213,37) {\scriptsize$\alpha$}
   \put(279,33) {\scriptsize$\alpha$}
  \epicture19 \labl{eq:cent-lem}
as well as the analogous equivalence in which all braidings are replaced by 
inverse braidings.

\medskip\noindent
Proof:\\
With the braidings chosen as in picture \erf{eq:cent-lem}, the two 
directions are shown as follows:
\\[.13em]
$\Rightarrow$\,: This is seen immediately by
composing both sides of the equality on the \lhs\ of \erf{eq:cent-lem} with $m$ 
and then using specialness of $A$.
\\[.13em]
$\Leftarrow$\,: Starting from the \rhs\ of \erf{eq:cent-lem}, and using the
Frobenius and (co-)associativity properties, we have
  \begin{eqnarray} \begin{picture}(430,121)(13,0)
  \put(0,0)   {\begin{picture}(0,0)(0,0)
              \scalebox{.38}{\includegraphics{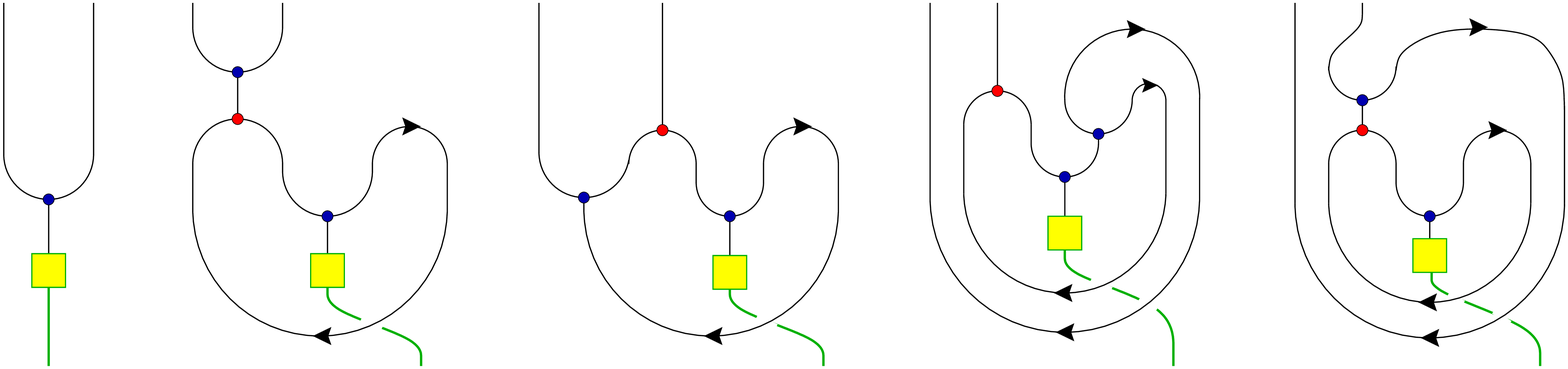}} \end{picture}}
  \put(10.0,-10)    {\scriptsize$X$}
  \put(124.9,-10)   {\scriptsize$X$}
  \put(248.1,-10)   {\scriptsize$X$}
  \put(355.4,-10)   {\scriptsize$X$}
  \put(467.2,-10)   {\scriptsize$X$}
  \put(38.9,60)     {$=$}
  \put(147.4,60)    {$=$}
  \put(265.3,60)    {$=$}
  \put(377.9,60)    {$=$}
  \put(12.1,29.0)   {\tiny$\alpha$}
  \put(97.2,29.0)   {\tiny$\alpha$}
  \put(220.5,28.5)  {\tiny$\alpha$}
  \put(323.2,40.0)  {\tiny$\alpha$}
  \put(434.6,33.0)  {\tiny$\alpha$}
  \put(-2,116)      {\scriptsize$A$}
  \put(24.5,116)    {\scriptsize$A$}
  \put(55.0,116)    {\scriptsize$A$}
  \put(82.9,116)    {\scriptsize$A$}
  \put(161.1,116)   {\scriptsize$A$}
  \put(198.5,116)   {\scriptsize$A$}
  \put(280.6,116)   {\scriptsize$A$}
  \put(301.2,116)   {\scriptsize$A$}
  \put(392.4,116)   {\scriptsize$A$}
  \put(413.4,116)   {\scriptsize$A$}
  \end{picture} \nonumber\\[.9em]{} \label{centreA-3} \end{eqnarray}
If we now substitute again the equality on the \rhs\ of \erf{eq:cent-lem} 
in the last 
expression, we arrive at the equality on the \lhs\ of \erf{eq:cent-lem}.
For the inverse braidings the equivalence is shown in the same way. 
\qed

\subsection{Sums, products, and the opposite algebra} \label{sec:sum-prod-op}

Given two (symmetric special Frobenius) algebras $A$ and $B$ in a ribbon
category, we can define algebra structures on $A \opl B$ and $A \oti B$.
It is also possible to twist the product of $A$ by the braiding, 
giving rise to an algebra $A_{\rm op}$.
In section \ref{sec:torus-pf} we will encounter a physical interpretation 
of these operations: Denoting the torus partition function obtained from 
an algebra $A$ as $Z(A)$, and defining $\tilde Z(A)_{kl}\,{:=}\,Z(A)
_{\bar k l}$, we have the matrix equations $\tilde Z(A{\oplus} B)
\eq\tilde Z(A)\,{+}\,\tilde Z(B)$, $\tilde Z(A \Oti B)\eq\tilde 
Z(A)\,\tilde Z(B)$ and $Z(A_{\rm op})\eq Z(A)\oT$.

These expressions suggest the interpretation that $A \,{\oplus}\, B$ 
describes a superposition of two CFTs -- in the sense introduced
before formula \erf{cft12} -- while $A \oti B$ defines a (in general 
non-commutative) product of two CFTs that are associated to the
same chiral data. Note that this product is different from the usual 
product ${\rm CFT}_1 \times {\rm CFT}_2$, where the new stress tensor is
given by $T_1 + T_2$ and thus the central charges add. The CFT resulting
from $A \oti B$ has the same stress tensor 
and central charge as the ones resulting from $A$ and $B$. 

In the rest of this section we will make precise the notions
of $A \opl B$, $A \oti B$ and $A_{\rm op}$ and prove some of
properties of these \alg s.
\vspace{-.8em}

\dt{Proposition} [{\em Opposite algebra\/}]
\\[.13em]
(i)\hsp{.6}$A\eq(A,m,\eta)$ is an algebra if and only if 
$A_{\rm op} \,{:=}\, (A,m \cir (c_{A,A})^{-1}, \eta)$ is an algebra.%
   \foodnode{Note that we take the inverse braiding in this definition.
   It is this choice that is needed to prove propositions
   \ref{prop:bimod-leftmod} and \ref{prop:Ztilde}. The first of these links
   the definition of $A_{\rm op}$ to that of the tensor product, while the
   second links that of the
   tensor product to that of the graph for the torus partition function.
   In this sense the conventions implicit in the graph \erf{Zij2} fix
   the convention for $A_{\rm op}$.}
\\[.15em]
(ii) $A\eq(A,m,\eta,\Delta,\eps)$ is a symmetric special 
Frobenius algebra if and only if 
  \be  A_{\rm op} := (A,m\cir(c_{A,A})^{-1}{,}\,\eta, c_{A,A} \cir \Delta, \eps) 
  \labl{Aop} 
is a symmetric special Frobenius algebra.

\medskip\noindent
Proof:\\
The statement (i) follows by a straightforward application of definition
\ref{def:alg-obj}. Similarly, for obtaining (ii) one checks easily 
from the definitions in \ref{def:coalg}, \ref{def:frob} and \ref{df:spec-etc} 
that the respective properties of $A$ and $A_{\rm op}$ follow from each other.
For example, that $A$ symmetric implies $A_{\rm op}$ symmetric is seen as
follows: 
  \bea \begin{picture}(395,69)(0,33)
  \put(0,0)   {\begin{picture}(0,0)(0,0)
              \scalebox{.38}{\includegraphics{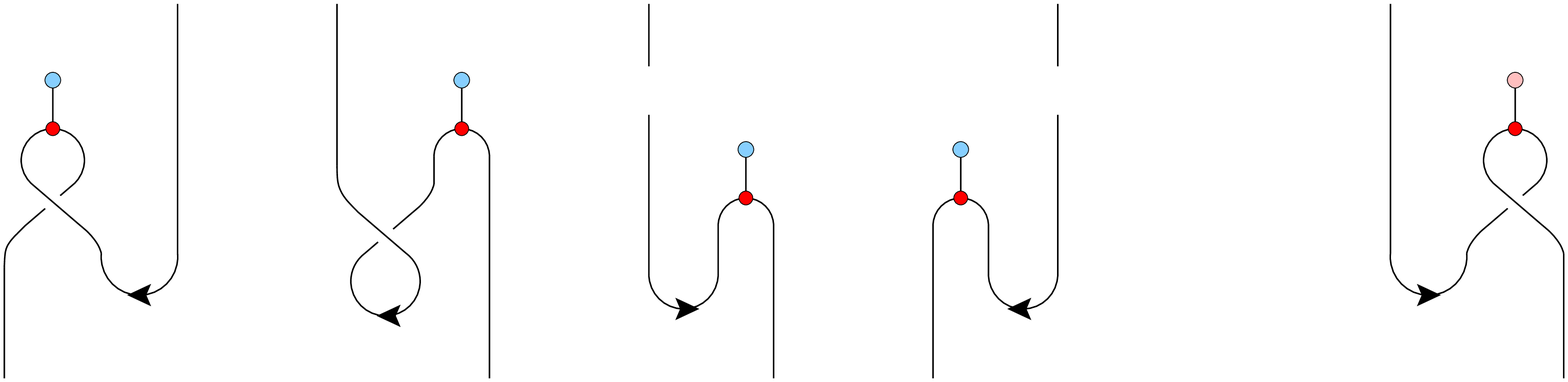}} \end{picture}}
  \put(58,42) {$=$}
  \put(135,42) {$=$}
  \put(209,42) {$=$}
  \put(277,42) {$= \ \cdots\  =$}
  \put(157,68) {$\theta$}
  \put(258,68) {$\theta$}
  \put(-2,-8) {\scriptsize$A$}
  \put(118,-8) {\scriptsize$A$}
  \put(188,-8) {\scriptsize$A$}
  \put(227,-8) {\scriptsize$A$}
  \put(383,-8) {\scriptsize$A$}
  \put(340,97) {\scriptsize$A^\vee$}
  \put(258,97) {\scriptsize$A^\vee$}
  \put(157,97) {\scriptsize$A^\vee$}
  \put(80,97) {\scriptsize$A^\vee$}
  \put(41,97) {\scriptsize$A^\vee$}
  \epicture21 \labl{pic17}
The first morphism is $\Phi_1$ for $A_{\rm op}$, expressed via the 
multiplication $m$ of $A$. Symmetry of $A$ enters in the third equality.
\qed

Note that even though $A_{\rm op}$ is equal to $A$ as an object in 
\calc, it has a different multiplication and is therefore, in general, not
isomorphic to $A$. In particular, this remark still applies when $A$ and
$A_{\rm op}$ are symmetric -- recall that the symmetry property, 
introduced in definition \ref{df:spec-etc}(ii), does not refer in any 
way to the braiding.
\vspace{-.8em}

\dt{Corollary}
(i)\hsp{.6}For any algebra $(A,m,\eta)$ and any $n\iN\zet$ also
$A^{(n)} \,{:=}\, (A,m\cir (c_{A,A})^n, \eta)$ is an algebra.
\\[.13em]
(ii) For any symmetric special Frobenius algebra $(A,m,\eta,\Delta,\eps)$ 
and any $n\iN\zet$ also
  \be A^{(n)} := (A,m\cir(c_{A,A})^n,\eta,(c_{A,A})^{-n}{\circ}\,\Delta,\eps) 
  \labl{Aon}
is a symmetric special Frobenius algebra.

\medskip\noindent
In particular we have $A\eq A^{(0)}$ and $A_{\rm op}\eq A^{(-1)}$.

\dt{Proposition}
(i)\hsp{.6}The twist $\theta_A$ is an intertwiner between the algebras 
$A^{(n)}$ and $A^{(n+2)}$, in the sense that
  \be
  \theta_A \circ \eta^{(n)} = \eta^{(n+2)}  
  \qquad{\rm and}\qquad
  \theta_A \circ m^{(n)} = m^{(n+2)} \circ (\theta_A \oti \theta_A) \,,
  \ee
with $m^{(n)}\eq m\cir (c_{A,A})^n$ and $\eta^{(n)}\eq\eta$ the product 
and unit of $A^{(n)}$. \\[.13em]
(ii) If $A^{(n)}$ is symmetric special Frobenius then in addition
  \be
  \eps^{(n)} = \eps^{(n+2)} \circ \theta_A
  \qquad{\rm and}\qquad
  (\theta_A \oti \theta_A) \circ \Delta^{(n)} = 
  \Delta^{(n+2)}\circ \theta_A \,.  \ee

\medskip\noindent
Proof:\\
By direct computation. One only has to use some of the defining properties 
of a \mtc, namely (see (\ref{ribax1}--\ref{ribax2})) that 
$\theta_V \cir f\eq f\cir\theta_U$, for any $f\iN\Hom(U,V)$ as well as 
$\theta_{U\otimes V}\eq c_{VU}^{} \cir c_{UV}^{} \cir (\theta_U\oti\theta_V)$.
For instance,
  \be
  \theta_A \circ m^{(n)} = m^{(n)} \circ \theta_{A\oti A}
  = m^{(n)} \circ (c_{A,A})^2 \circ (\theta_A \oti \theta_A)
  = m^{(n+2)} \circ (\theta_A \oti \theta_A) \,.  \ee 
\mbox{$ $}\\[-1.75em]\mbox{$ $}\qed

\medskip
By definition of the direct sum $A\opl B$ of two objects $A$ and $B$,
there exist morphisms
  \be  X_C \in \Hom(C, A\opl B) \qquad{\rm and}\qquad 
  Y_C \in \Hom(A\opl B, C) \ee
for $C\iN\{A,B\}$ such that 
  \be  Y_C \cir X_D = \delta_{C,D}\, \id_C \qquad {\rm and} \qquad
  X_A \cir Y_A + X_B \cir Y_B =\id_{A\oplus B} \,.  \labl{eq:XY-ortho}
for $C,D \iN \{A,B\}$. When $A$ and $B$ are algebras, then these 
morphisms can be used to endow $A\opl B$ with the structure of an \alg,
too:

\dtl{Proposition}{prop:alg-add} [{\em Direct sum of algebras\/}]
\\[.13em]
(i)\hsp{.6}The triple $(A \opl B,m_{A\oplus B},\eta_{A\oplus B})$ with
  \be  m_{A\oplus B}:= \sum_{C=A,B} X_C \cir m_C \cir (Y_C \oti Y_C) 
  \qquad{\rm and}\qquad
  \eta_{A\oplus B}:= \sum_{C=A,B} X_C \cir \eta_C  \labl{eq:A+B-alg}
furnishes an algebra structure on the object $A\opl B$. 
\\[.13em]
(ii) When $A,B$ are
symmetric special Frobenius and normalised such that
$\eps_A \cir \eta_A\eq\dim(A)$ and
$\eps_B \cir \eta_B\eq\dim(B)$, then $A \opl B$ is symmetric special 
Frobenius, with counit and coproduct given by
  \be  \eps_{A\oplus B} = \sum_{C=A,B} \eps_C \cir Y_C
  \qquad{\rm and}\qquad
  \Delta_{A\oplus B} = \sum_{C=A,B} (X_C \oti X_C) \cir 
  \Delta_C \cir Y_C \,.  \labl{eq:A+B-coalg}

\smallskip\noindent
Proof: \\
The proof proceeds again by direct computation. To check, 
for example, that $A \opl B$ is special, one computes
  \be\begin{array}{ll}
  m_{A\oplus B} \cir \Delta_{A\oplus B} \!\!
  &= \llb\! \dsty \sum_{C=A,B}  X_C \cir m_C \cir (Y_C \oti Y_C) \lrb
  \circ \llb\! \sum_{D=A,B} (X_D \oti X_D) \cir \Delta_D \circ Y_D \lrb
  \\{}\\[-.8em]
  &= \dsty\sum_C X_C \cir m_C \cir \Delta_C \cir Y_C
  = \sum_C X_C \cir \id_C \cir Y_C
  = \id_{A \oplus B}
  \end{array} \ee
In the second equality we insert the orthogonality relation in 
\erf{eq:XY-ortho}, while the third step
uses that $A$ and $B$ are special, with constants
$\beta_A$ and $\beta_B$ both equal to $1$. The last step is just
the completeness in \erf{eq:XY-ortho}.
\qed
\vspace{-.8em}

\dtl{Proposition}{prop:alg-mult} [{\em Tensor product of algebras\/}]
\\[.13em]
(i)\hsp{.6}For any two algebras $A$ and $B$
the triple $(A \oti B,m_{A\otimes B},\eta_{A\otimes B})$ with  
  \be
  m_{A\otimes B}:= (m_A \oti m_B) \circ (\id_A \oti (c_{A,B})^{-1}\oti\id_B)
  \qquad{\rm and}\qquad
  \eta_{A\otimes B}:= \eta_A \oti \eta_B \labl{eq:AxB-mult}
furnishes an algebra structure on the object $A \oti B$. 
\\[.13em]
(ii) If $A,B$ are symmetric special Frobenius, then also $A \oti B$ 
is symmetric special Frobenius, with counit and coproduct given by
  \be
  \eps_{A\oti B} = \eps_A \oti \eps_B \qquad {\rm and} \qquad
  \Delta_{A\oti B} = (\id_A \oti c_{A,B} \oti \id_B) \circ 
  (\Delta_A \oti \Delta_B) \,.  \ee

\smallskip\noindent
Proof:\\
In this case, too, one checks all the properties by direct computation. 
To give an example, the associativity of $A \oti B$ follows from
  \bea \begin{picture}(410,76)(0,7)
  \put(0,0)   {\begin{picture}(0,0)(0,0)
              \scalebox{.38}{\includegraphics{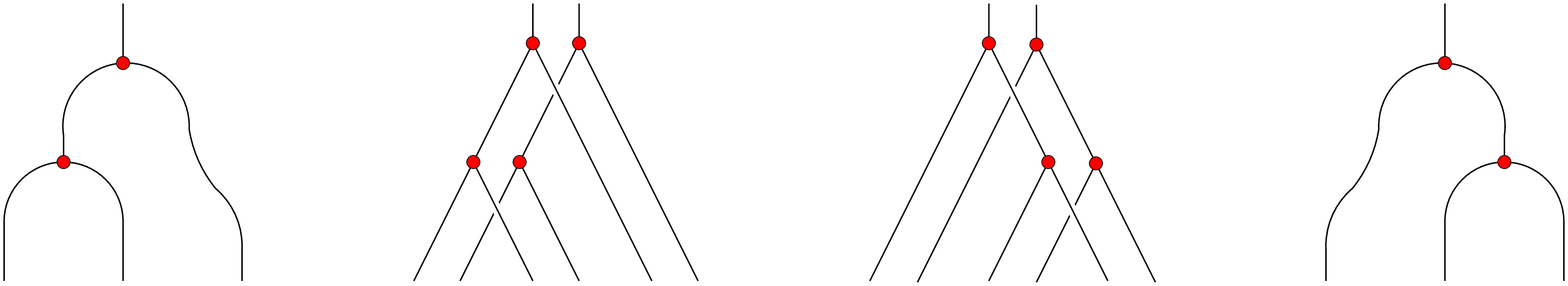}} \end{picture}}
   \put(81,34) {$=$}
   \put(197,34) {$=$}
   \put(312,34) {$=$}
   \put(-10,-8) {\scriptsize$A{\oti}B$}
   \put(21,-8) {\scriptsize$A{\oti}B$}
   \put(52,-8) {\scriptsize$A{\oti}B$}
   \put(21,78) {\scriptsize$A{\oti}B$}
   \put(103,-8) {\scriptsize$A$}
   \put(134,-8) {\scriptsize$A$}
   \put(134,78) {\scriptsize$A$}
   \put(165,-8) {\scriptsize$A$}
   \put(221,-8) {\scriptsize$A$}
   \put(252,-8) {\scriptsize$A$}
   \put(252,78) {\scriptsize$A$}
   \put(283,-8) {\scriptsize$A$}
   \put(116,-8) {\scriptsize$B$}
   \put(147,-8) {\scriptsize$B$}
   \put(147,78) {\scriptsize$B$}
   \put(178,-8) {\scriptsize$B$}
   \put(235,-8) {\scriptsize$B$}
   \put(266,-8) {\scriptsize$B$}
   \put(266,78) {\scriptsize$B$}
   \put(296,-8) {\scriptsize$B$}
   \put(332,-8) {\scriptsize$A{\oti}B$}
   \put(363,-8) {\scriptsize$A{\oti}B$}
   \put(394,-8) {\scriptsize$A{\oti}B$}
   \put(363,78) {\scriptsize$A{\oti}B$}
  \epicture-4 \labl{pic18}
\mbox{$\ $}\\[-1.86em]\mbox{$\ $}\qed 
\vspace{-.8em}

\dt{Remark} 
(i)\hsp{.6}In the definition \erf{eq:AxB-mult} one could have used the 
braiding $c_{B,A}$ instead of $(c_{A,B})^{-1}$. It is easy to check
that this alternative  multiplication on the object $A\oti B$ coincides 
with the multiplication on $(A_{\rm op} \oti B_{\rm op})^{(1)}$ given
above. We will not introduce a separate symbol for this way of 
defining a multiplication on $A\oti B$.  
\\[.13em]
(ii) For any three algebras $A,B,C$ there are two a priori distinct
ways to construct an algebra structure on the object $A\oti B\oti C$ by 
the tensor product of \alg s, namely $(A\oti B)\oti C$ and $A \oti 
(B\oti C)$. As $\calc$ is a strict tensor category, the two combinations
are identical as objects in \calc. One can quickly convince
oneself from the definitions in proposition \ref{prop:alg-mult}
that the unit and multiplication morphism (as well as the 
counit and comultiplication, if they exist) are identical in both
cases as well. Thus the tensor product of algebras defined in
proposition \ref{prop:alg-mult} is associative.
\\[.13em]
(iii) As will be discussed in a forthcoming paper, the operations of
taking sums, products and opposites of algebras are compatible with Morita 
equivalence. Denote by $[A]$ the Morita class of $A$, and let $A'$ be 
another representative of $[A]$, and similarly for $[B]$ and $B,B'$. Then 
  \be  [A \opl B] = [A' \opl B']\,, \qquad [A \oti B] = [A' \oti B']
  \qquad {\rm and}\qquad [A_{\rm op}] =[A'_{\rm op}] \,.  \ee
As a consequence, the operations on algebras can be lifted to the level 
of CFT. We will return to this issue in section \ref{sec:torus-pf}.

It is also worth mentioning that the resulting ring structure on Morita 
classes of symmetric special Frobenius algebras is reminiscent of 
Brauer groups for (finite-dimensional, central, semisimple) algebras 
over a field, compare e.g.\ chapter 4 of \cite{FAde}.
Indeed, consider the tensor category ${\rm Vect}(k)$ of vector spaces over 
some field $k$, which we do not assume to be algebraically closed. Then the
module categories over ${\rm Vect}(k)$ -- or, equivalently, the Morita 
classes of associative algebras in ${\rm Vect}(k)$ -- are classified 
\cite{ostr} by division algebras over $k$, which amounts to compute the 
Brauer group of all finite extensions of $k$. Quite generally, 
the problem of classifying full CFTs based on a modular tensor 
category $\calc$ can therefore be expected to amount to the following 
two tasks: Find all extensions of $\calc$, and compute the (suitably 
defined) Brauer group of each extended theory. Here we call a modular 
tensor category $\cald$ an {\em extension\/} of \calc\ iff there exists 
a haploid commutative symmetric special Frobenius algebra $A$ 
in \calc\ such that $\cald$ is equivalent to the modular tensor
category of local $A$-modules  (see section \ref{rep-and-mod} 
below).

\subsection{The case ${N_{ij}}^k \iN \{0,1\}$ 
            and $\dim\,\Hom(U_k,A) \iN \{0,1\}$} \label{sec:alg-N=1}

To continue our meta example, we will further limit ourselves to the case
where each simple subobject $a$ in the algebra $A$ occurs with
multiplicity one. In this case we can omit the index labelling the
basis of possible embeddings of $a$ in \erf{iaa,aai}, 
rendering the notation less heavy.

An algebra object is now described by a collection of 
pairwise non-isomorphic simple objects $\{\one,U_{a_1},...\,,U_{a_n}\}$. 
The multiplication on $A$ can be expressed by constants
${m_{ab}}^{\!c}$, as in formula \erf{eq:m-in-basis}, and a comultiplication 
similarly by constants $\Delta_c^{\,ab}$.
The associativity condition \erf{eq:general-assoc-m-basis} takes the form
  \be
  {m_{ab}}^{\!f}\, {m_{fc}}^{\!d}
  = \sum_{e\In A} {m_{bc}}^{\!e}\, {m_{ae}}^{\!d}\, \Fs abcdef \,,
  \labl{eq:easy-assoc}
Here we use the symbol ``$\In$'' to indicate the relation ``is a simple
subobject of''. Further, we choose the basis vector in $\Hom(\one,A)$ 
to be given by the unit $\eta$ of $A$; then the dual basis vector in 
$\Hom(A,\one)$ is fixed to be the multiple $\eps/\beta_\smallone$ of the
counit $\eps$. Thus in particular composing with the counit and
unit, \resp, gives
  \bea \begin{picture}(244,42)(0,24)
  \put(30,0)   {\begin{picture}(0,0)(0,0)
              \scalebox{.38}{\includegraphics{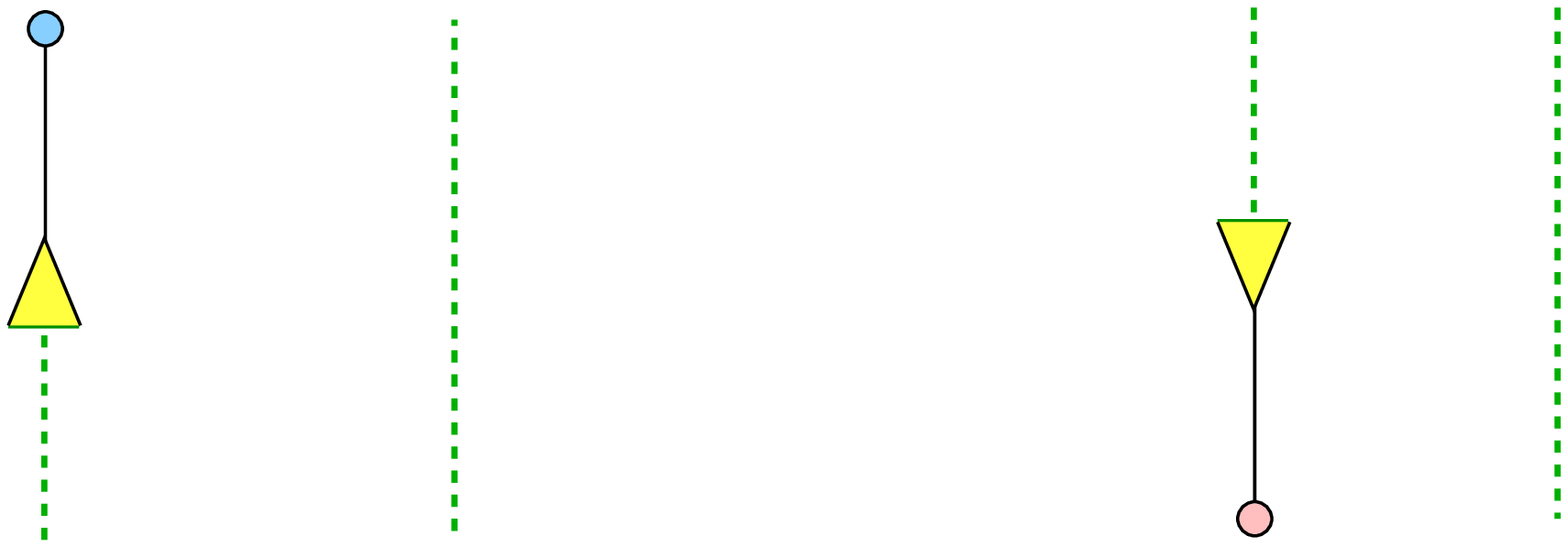}} \end{picture}}
  \put(50.1,30.5) {$=\;\beta_\smallone$} 
  \put(122.2,30.5) {and}
  \put(196.6,30.5) {$=$}
  \epicture08 \label{eq:N=1-unitnorm} \labl{unitnorm1} 

Now the multiplication already determines the comultiplication.
Let us make this relation explicit. The normalisation
\erf{eq:N=1-unitnorm} of the unit implies that 
${m_{a\One}}^{\!a}\eq1\eq{m_{\One a}}^{\!a}$. Since $A$ is haploid, the two
morphisms $\Phi_1$ and $\Phi_2$ in \erf{Phi12} are automatically equal. 
For them to be isomorphisms we need ${m_{a \bar a}}^{\!\One}\,{\ne}\,0$ for all 
$a\IN A$. To determine the inverse of $\Phi\,{\equiv}\,\Phi_1\eq\Phi_2$, we 
express it in a basis as 
  \bea \begin{picture}(380,74)(0,38)
  \put(0,0)   {\begin{picture}(0,0)(0,0)
              \scalebox{.38}{\includegraphics{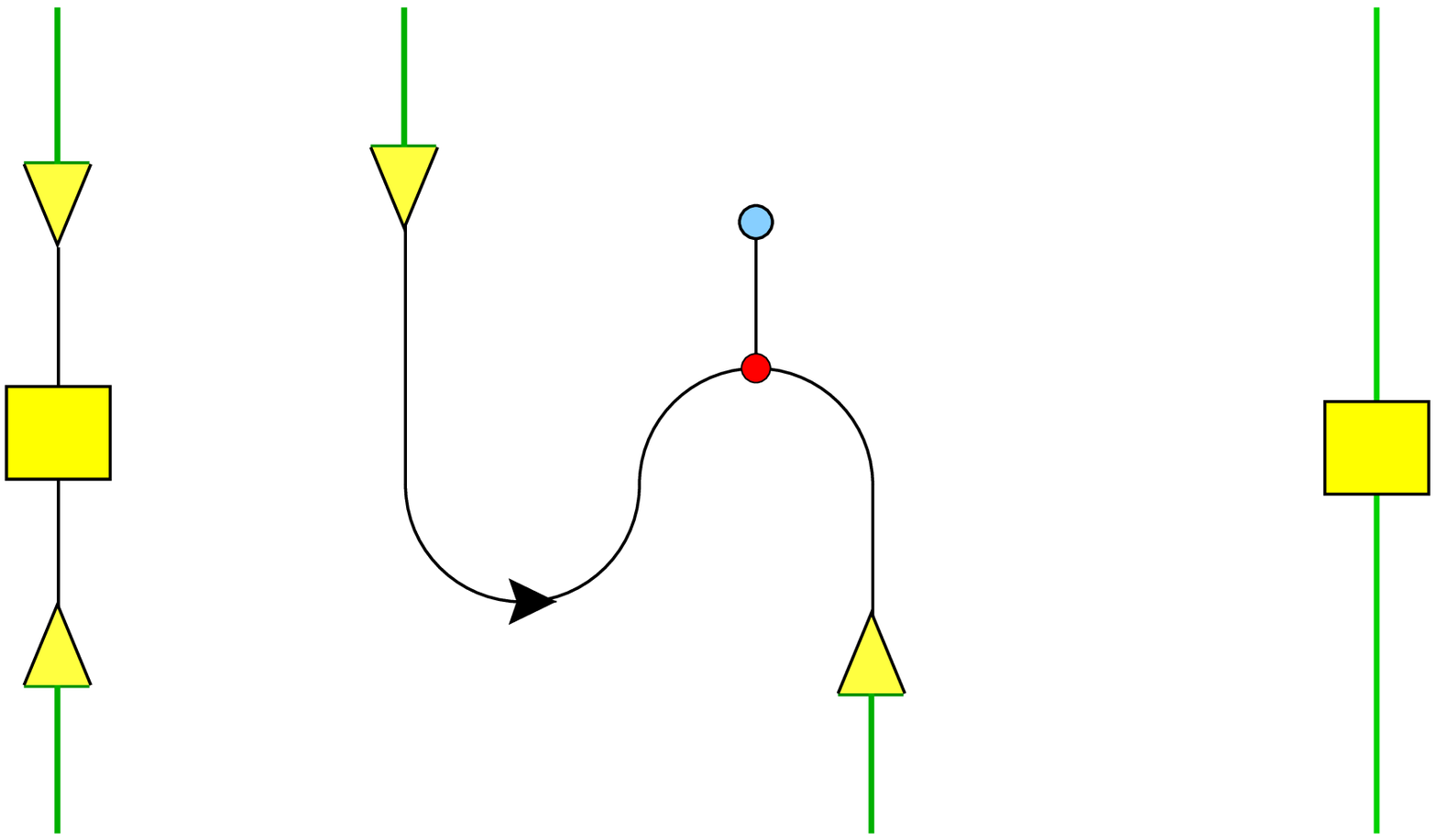}} \end{picture}}
  \put(30.9,48.5)   {$=$}
  \put(132.9,48.5)  {$=\;\phi_a$}
  \put(224.9,48.5)  {with $\qquad\phi_a = {m_{\bar a a}}^{\!\One}\,p_a\,
                    \lambda_a\,\dim(A)\,,$}
  \put(4.8,-6)      {\scriptsize$a$}
  \put(110.5,-6)    {\scriptsize$a$}
  \put(175.6,-6)    {\scriptsize$a$}
  \put(4.8,112)     {\scriptsize$\bar a^\vee$}
  \put(48.9,112)    {\scriptsize$\bar a^\vee$}
  \put(175.6,112)   {\scriptsize$\bar a^\vee$}
  \put(4.8,50.0)    {\scriptsize$\Phi$}
  \put(174.5,49.1)  {\scriptsize$\pi_a$}
  \epicture20 \labl{Phibasis} 
where $p_a$ and $\lambda_a$ are the numbers introduced in \erf{def-pi}
and \erf{xy-pi}, \resp.
It follows that the map $\Phi^{-1}$ is then just given by
  \bea \begin{picture}(170,77)(0,37)
  \put(0,0)   {\begin{picture}(0,0)(0,0)
              \scalebox{.38}{\includegraphics{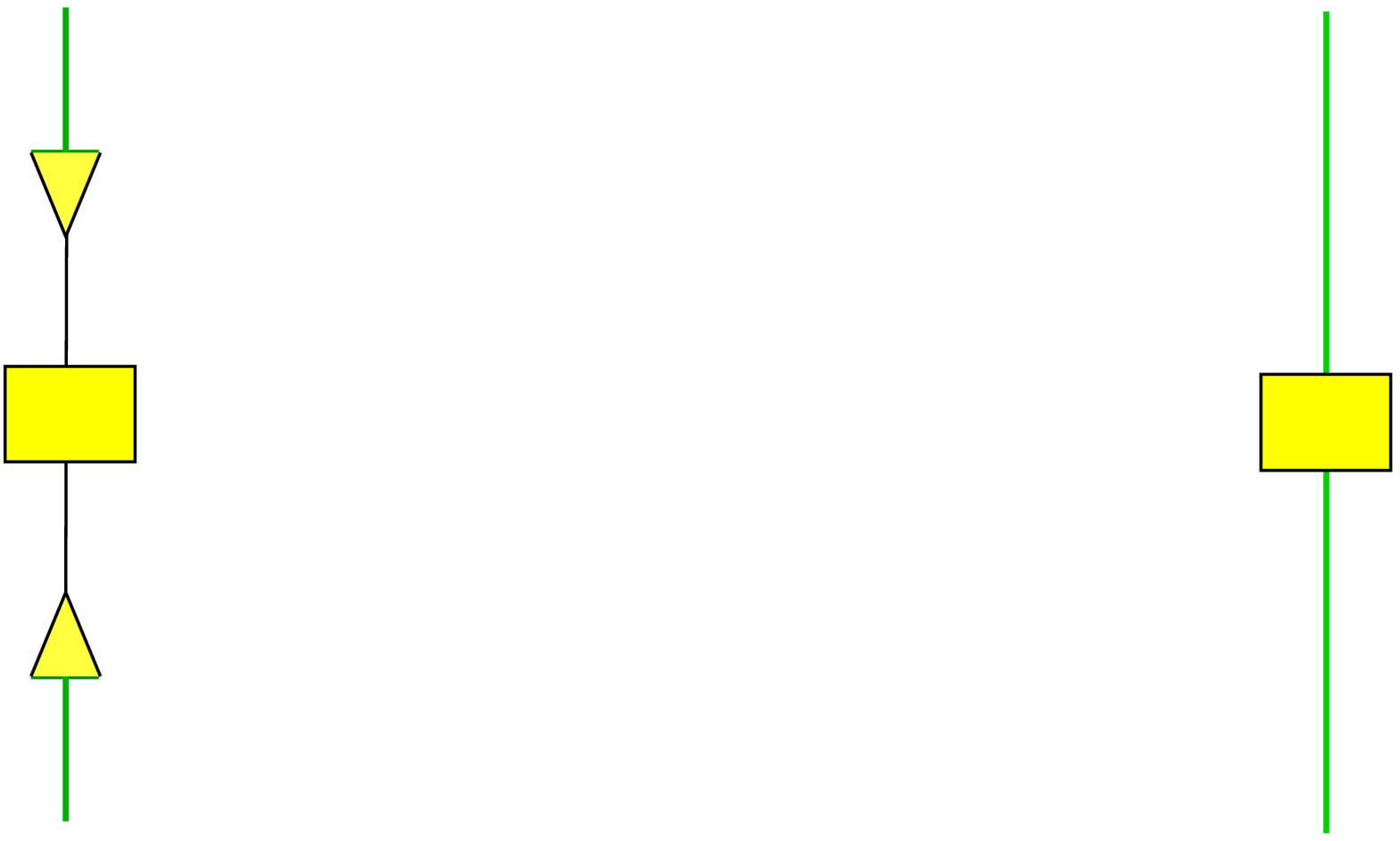}} \end{picture}}
  \put(35.9,51.2)   {$=\ \dsty\frac1{ {m_{\bar a a}}^{\!\One}\,p_a\,
                    \lambda_a\,\dim(A)}$}
  \put(6.5,109)    {\scriptsize$a$}
  \put(163.6,109)  {\scriptsize$a$}
  \put(4.8,-7)     {\scriptsize$\bar a^\vee$}
  \put(162,-7)     {\scriptsize$\bar a^\vee$}
  \put(1,50)       {\scriptsize$\Phi^{-1}$}
  \put(158,49.1)   {\scriptsize$\pi_a^{-1}$}
  \epicture23 \labl{Phi-basis} 
Using the expression \erf{Delta=} for the coproduct and the identities
\erf{eq:rel-3pt-cpl}, one finds
  \bea \begin{picture}(384,82)(0,44)
  \put(0,0)   {\begin{picture}(0,0)(0,0)
              \scalebox{.38}{\includegraphics{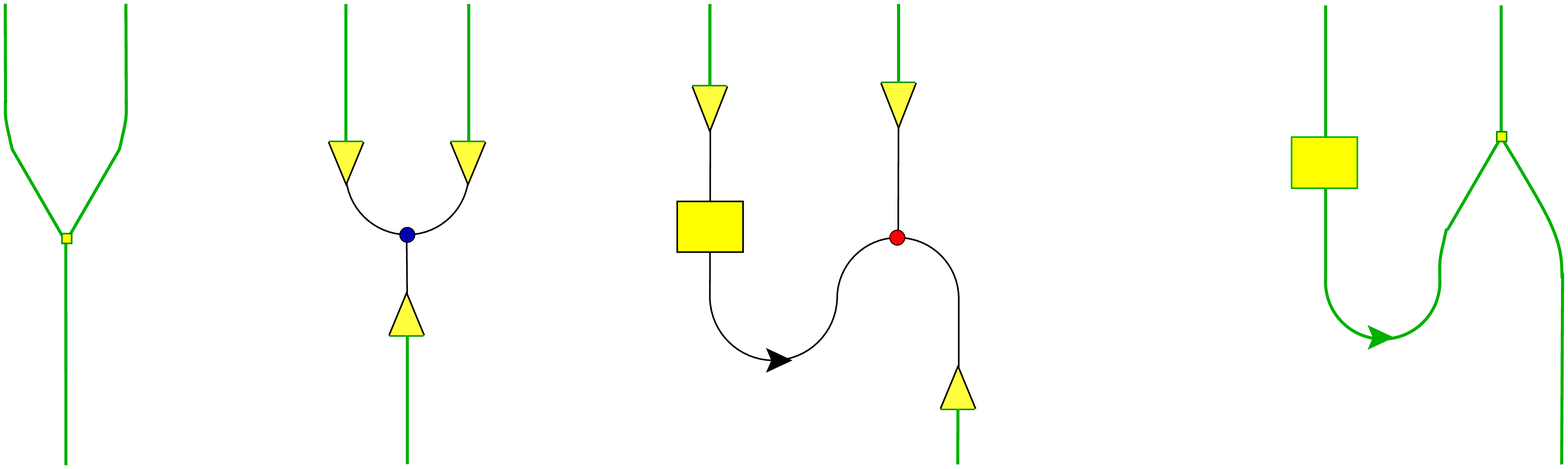}} \end{picture}}
   \put(-20,57) {$\Delta_{c}^{ab}$}
   \put(47,57) {$=$}
   \put(137,57) {$=$}
   \put(261,57) {$\dsty = \; \frac{{m_{\bar a c}}^{\!b}}{\phi_a} $}
   \put(167.5,57) {\scriptsize$\Phi^{-1}$}
   \put(319.6,73) {\scriptsize$\pi_a^{-1}$}
   \put(14,-6.5) {\scriptsize$c$}
   \put(98,-6.5) {\scriptsize$c$}
   \put(234,-6.5) {\scriptsize$c$}
   \put(382,-6.5) {\scriptsize$c$}
   \put(-1,118) {\scriptsize$a$}
   \put(29,118) {\scriptsize$b$}
   \put(82,118) {\scriptsize$a$}
   \put(113,118) {\scriptsize$b$}
   \put(173,118) {\scriptsize$a$}
   \put(220,118) {\scriptsize$b$}
   \put(325,118) {\scriptsize$a$}
   \put(368,118) {\scriptsize$b$}
   \put(354,65) {\scriptsize$\bar a$}
  \epicture29 \labl{pic5}
With the help of the relations \erf{xy-pi}, \erf{xy-pi-norm}, 
\erf{eq:dimF=pll} and \erf{eq:rel-3pt-cpl}, it then follows that
  \be
  \Delta_{c}^{ab} = \frac{1}{\dim(A)} \cdot
  \frac{ {m_{\bar a c}}^{\!b} }{ {m_{\bar a a}}^{\!\One} } \cdot
  \frac{\Fs a{\bar a}ccb\One}{\Fs a{\bar a}aa\One\One} \,.  
  \labl{eq:delta-by-mult}

\subsubsection{Example: Free boson}

Different algebra objects in the $\zet_{2N}$ free boson modular category 
correspond to different compactification radii. (This will be made more 
explicit in section \ref{sec:ex-free-boson-pf}
below, when we come to computing the torus partition function.) There is 
one algebra object associated to every subgroup of $\zet_{2N}$ that contains 
only `even elements', in the sense that there is one algebra object for every 
divisor $r$ of $N$. For such a subgroup, $2r$ with $0\,{\le}\,r\,{<}\,N$ 
is its `minimal' element. We denote as $A_{2r}$ the corresponding object
  \be  A_{2r} := \bigoplus_{n=0}^{N/r-1} [2nr] \,.
  \labl{eq:A2r-cont} 
The multiplication on $A_{2r}$ is given by
  \be
  {m_{[a][b]\phantom|}}^{\!\!\!\![a+b]} = 1 \qquad {\rm for\ all}\quad
  a,b,c\,{\In}\,A \,.  \labl{eq:A2r-mult}
One checks that all non-zero \FF-matrix elements \erf{eq:Z2N-F} with only 
even labels are equal to 1. As a consequence, with \erf{eq:A2r-mult} the 
associativity condition \erf{eq:easy-assoc} is satisfied trivially, as $1\eq1$.

The algebras presented above do not exhaust {\em all\/} algebra objects
of the free boson theory, but they do give all {\em haploid\/} ones. 
This follows from a general treatment of algebra objects containing only 
simple currents as simple subobjects, which will be presented in a separate 
paper. For now we only remark that simple current theory 
\cite{scya,intr,scya6,Sche2'} tells us that to obtain a modular invariant,
a simple current $J$ must be \cite{scya3} in the so-called 
{\em effective center\/}, i.e.\ the product of its order (the smallest 
natural number $\ell$ such that $J^{\otimes\ell}\,{\cong}\,\one$)
and its conformal weight must be an integer. 
In the case under study the simple current orbits are generated by objects
$r$ with $r$ a divisor of $2N$, i.e.\ by simple currents of order $2N/r$. 
Modulo integers, the conformal weight of $r$ is $r^2/4N$ and thus it 
is in the effective center iff $r\iN2{\mathbb Z}$.

\subsubsection{Example: \E modular invariant}\label{sec:E7-algebra-object}

According to \cite{boek2,ostr},
for the \E type modular invariant of the $\su(2)$ WZW model at level 16, 
the associated algebra object $A$ should be expected to have the form
  \be  A = (0)\,{\oplus}\,(8)\,{\oplus}\,(16) \,.  \labl{0816}
This object can indeed be turned into an algebra \cite{boek2,ostr}, and
we will show in the sequel how this can be established with our methods.
But of course, at this point we do not yet know whether the modular 
invariant of the CFT associated to the algebra \erf{0816} is of 
\E type -- this will be verified in section \ref{sec:E7-mod-inv}.

Note that $J\,{:=}\,(16)$ is a simple
current in $\su(2)_{16}$ and that $f\,{:=}\,(8)$ is its fixed point. 
Let us therefore investigate, more generally, algebra objects of the form
  \be  A = \one \oplus f \oplus J \,, \labl{eq:E7-alg-cont}
for which the fusion rules of the simple subobjects are 
  \be  J\oti J \cong \one\,, \qquad J\oti f \cong f \qquad{\rm and}\qquad
  f\oti f \cong \one\oplus f\oplus J\oplus \cdots \,.  \ee

Whereas in the free boson one would have easily guessed the
product structure on $A$, in the \E case at this point we do
have to find a solution to {\em polynomial\/} equations. However, we will 
see (though not in all detail in the present paper) that all the rest of
the calculations, down to the structure constants, then reduces to solving
systems of {\em linear\/} equations.
Let us write the relevant equations explicitly as relations that the
numbers ${m_{ab}}^{\!c}$ must satisfy in order to be the components of a
valid multiplication morphism. First, according to lemma 
\ref{thm:alg-from-bc}(ii) the Frobenius property imposes the restriction
${m_{aa}}^{\!\one}\,{\ne}\,0$; we can therefore choose the basis elements
in $\Hom(f,A)$ and $\Hom(J,A)$ such that
  \be
  {m_{ff}}^{\!\one} = {m_{JJ}}^{\!\one} = 1 \,.  \labl{eq:e7-alg-basis}
The system \erf{eq:easy-assoc} of polynomial equations encoding
associativity is finite, and it is not difficult to write out all equations
when $A$ is of the particular form \erf{eq:E7-alg-cont}. This system of
conditions is necessary and sufficient for the numbers ${m_{ab}}^{\!c}$ to 
provide an algebra structure. After some manipulations, one deduces that
these conditions imply
  \be\begin{array}{l}
  {m_{ff}}^J = {m_{fJ}}^f = {m_{Jf}}^f \,,\qquad\quad
  {({m_{ff}}^J)}^2 = \Fs JffJ\one f \,,
  \\{}\\[-.5em]
  {({m_{ff}}^f)}^2 = \llb \Fs fffff\one \lrb^{-1}\, \llb 1-\Fs ffff\one\one - 
  \Fs JffJ\one f \Fs ffffJ\one \lrb\,.
  \end{array}\labl{eq:E7-mult}
A priori this fixes the algebra structure up to sign choices only. But in fact
the convention \erf{eq:e7-alg-basis} still permits a sign flip both
in $\Hom(f,A)$ and in $\Hom(J,A)$. As a result, the choice of sign allowed by
\erf{eq:E7-mult} for the numbers ${m_{ff}}^{\!f}$ and ${m_{ff}}^{\!J}$ can 
be absorbed
into a change of basis, and hence \erf{eq:E7-mult} determines the algebra 
structure uniquely. Thus, if it exists, the algebra structure on $A$ is 
unique up to isomorphism.

Note that in \erf{eq:E7-mult} it is assumed that $\Fs fffff\one\,{\ne}\,0$
(which requires in particular that $f\IN f\Star f$). This holds true for 
$\su(2)_{16}$, but it need not hold in general. (If not, then the 
calculation will proceed differently from what is reported in the sequel; 
we do not discuss this case here.) In the same spirit we will also assume 
that ${m_{ff}}^f\,{\ne}\,0$. Again this may not be satisfied in some 
examples, but it does hold for $\su(2)_{16}$. With these two assumptions, 
the remaining associativity constraints are equivalent to
  \be\begin{array}{l}
  \nu_J = \nu_f = 1 \,,\qquad\quad
  \Fs JfJfff = \Fs fJfJff = 1\,,
  \\{}\\[-.5em]
  \Fs Jfffff = \Fs fJffff = \Fs ffJfff = \Fs fffJff = 1\,,
  \\{}\\[-.5em]
  \Fs JffJ\one f \Fs JJfff\one = 1 \,,\qquad\quad 
  \Fs JffJ\one f = \Fs fJJf\one f \,,\qquad\quad 
  \Fs JJfff\one = \Fs ffJJf\one \,,
  \\{}\\[-.5em]
  \Fs ffff\one k + ({m_{ff}}^J)^2\, \Fs ffffJk + ({m_{ff}}^f)^2\,
    \Fs fffffk =({m_{ff}}^k)^2 \quad\ {\rm for}\ k\iN\{J,f\}\,,
  \\{}\\[-.5em]
  \Fs ffff\one k + ({m_{ff}}^J)^2\, \Fs ffffJk + ({m_{ff}}^f)^2\,
    \Fs fffffk = 0 \quad\ {\rm for}\  k\iN\I\,{\setminus}\{\one,J,f\}
    \,\ {\rm and}\,\  \N ffk\eq1 \,.
  \end{array}\labl{eq:E7-assoc-F}
Owing to the pentagon identity fulfilled by the fusing matrices, not 
all of these requirements are
independent. At first sight it might look surprising that the pentagon 
identity for morphisms from $f{\oti}J{\oti}J{\oti}f$ to $f$ 
implies that $J$ has \fsi\ $\nu_J\eq1$ (recall formula~\erf{eq:nu=F}).
However, on general grounds \cite{fuRs3}, for self-dual $f$ and $J$ the
fact that ${N_{Jf}}^f$ is $1$ already implies that $\nu_J\nu_f\eq\nu_f$.

But not all of the relations \erf{eq:E7-assoc-F} follow from the pentagon, 
as one can easily convince oneself by finding explicit counter examples. 
So $A\eq\one{\oplus}f{\oplus}J$ can only be endowed 
with an algebra structure in rather special cases, as befits the fact
that it should describe an exceptional modular invariant.  
We verified numerically that the relations \erf{eq:E7-assoc-F} hold
true for $\su(2)_k$ with $k\eq16$. In fact they also hold for $k\eq8$ 
for which they yield the D-invariant (see section \ref{sec:E7-mod-inv} 
below), but not for various other $k$ we tested.  
For $k\eq4$ the object $\one\opl f\opl J$ possesses an \alg\ structure,
too (it yields just the A-invariant), but with ${m_{ff}}^f\eq0$ so that
the associativity constraints look different from \erf{eq:E7-assoc-F}.

%%%%%%%%%%%%%%%%%%%%%%%%%%%%%%%%%%%%%%%%%%%%%%%%%%%%%%%%%%%%%%%%%%%%%%%%
\newpage

\sect{Representation theory and boundary conditions}
\label{sec:representations} 

%%%%%%%%%%%%%%%%%%%%%%%%%%%%%%%%%%%%%%%%%%%%%%%%%%%%%%%%%%%%%%%%%%%%%%%%

\subsection{Representations and modules}\label{rep-and-mod}

When studying an ordinary \alg\ over the complex numbers (or any other 
number field), a key concept is that of a \rep\ of the \alg, together 
with the closely related notion of a module, i.e.\ the vector space
on which a \rep\ acts. Moreover, the modules of an 
\alg\ (or analogously, of a group, a Lie \alg, a quantum group 
or similar \alg ic structures) are the objects of a category, 
with morphisms given by \alg\ (or group, Lie \alg, ...) intertwiners.

It is not difficult to extend these notions in such a way that they apply
to \alg\ objects in arbitrary \tcs\ \calc.  The basic notion is that of 
an $A$-module:
\vspace{-.7em}

\dt{Definition}
For $A$ an \alg\ in a \tc\ \calc, a (left) {\em $A$-module\/} in \calc\ is 
a pair $N\eq(\n,\r)$ of two data: an object $\n$ of \calc\ and a morphism of 
\calc\ that specifies the action of $A$ on $\n$ -- the {\em \rep\ morphism\/}
$\r\,{\equiv}\,\r_N^{}\iN\Hom(A\Oti\n,\n)$. Further, $\r$ must
satisfy the {\em \rep\ properties\/}
  \be  \r\circ(m\oti\idN) = \r \circ (\id_A\oti\r)
  \qquad\mbox{and}\qquad \r\circ(\eta\oti\idN) = \idN \,.  \Labl1m

\noindent
Pictorially: 
  \bea \begin{picture}(408,54)(0,26)
  \put(0,0)    {\begin{picture}(0,0)(0,0)
               \scalebox{.38}{\includegraphics{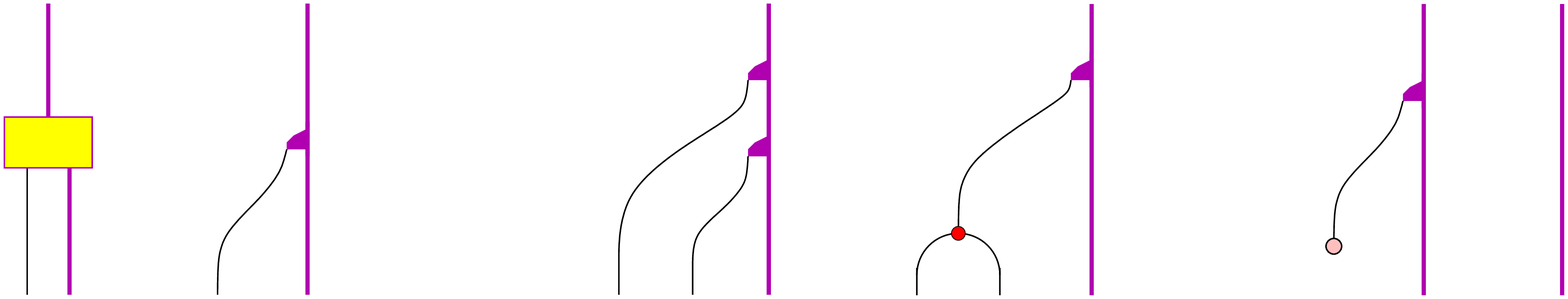}} \end{picture}}
  \put(1.8,-9.1)   {\scriptsize$A$}
  \put(8.6,38.3)   {\scriptsize$\r$}
  \put(8.1,76.3)   {\scriptsize$\n$}
  \put(12.8,-9.1)  {\scriptsize$\n$}
  \put(36,37)      {$=:$}
  \put(50.8,-9.1)  {\scriptsize$A$}
  \put(72.1,76.3)  {\scriptsize$\n$}
  \put(72.8,-9.1)  {\scriptsize$\n$}
  \put(149.3,-9.1) {\scriptsize$A$}
  \put(167.3,-9.1) {\scriptsize$A$}
  \put(185.3,-9.1) {\scriptsize$\n$}
  \put(185.9,76.3) {\scriptsize$\n$}
  \put(211,33)     {$=$}
  \put(223.3,-9.1) {\scriptsize$A$}
  \put(243.1,-9.1) {\scriptsize$A$}
  \put(266.1,-9.1) {\scriptsize$\n$}
  \put(267.2,76.3) {\scriptsize$\n$}
  \put(348.1,-9.1) {\scriptsize$\n$}
  \put(348.9,76.3) {\scriptsize$\n$}
  \put(365.9,33)   {$=$}
  \put(382.8,-9.1) {\scriptsize$\n$}
  \put(383.8,76.3) {\scriptsize$\n$}
  \epicture18 \labl{rho}
What is introduced here is called a {\em left\/} module because the action
of $A$ on $N$ is from the left. Analogously one can define right $A$-modules,
on which $A$ acts from the right, i.e.\ for which the \rep\ morphism is an 
element of $\Hom(\n\Oti A,\n)$. And there is also the notion of a (left or right)
{\em comodule\/} over a co-\alg, involving a morphism in $\Hom(\n,A\Oti\n)$ (or
$\Hom(\n,\n\Oti A$), \resp) that satisfies relations corresponding to
the pictures \erf{rho} turned upside-down, with coproduct and counit
in place of product and unit.  
Note that not every object $U\iN\objc$ needs to underlie some $A$-module, and 
that an object of $\calc$ can be an $A$-module in several inequivalent ways.

Among the morphisms $f$ between two objects $\n,\M\iN\objc$ that both carry 
the structure of an $A$-module, those that intertwine the action
of $A$ play a special role. Here the notion of intertwining the $A$-action is
analogous as for modules over ordinary \alg s:

\dt{Definition}
For (left) $A$-modules $N,M$, an $A$-{\em intertwiner\/} is a morphism $f$
between $\n$ and $\M$ satisfying $f \cir\r_N\eq\r_M\cir(\id_A\Oti f)$, i.e.
  \bea \begin{picture}(105,56)(0,28)
  \put(0,0)    {\begin{picture}(0,0)(0,0)
               \scalebox{.38}{\includegraphics{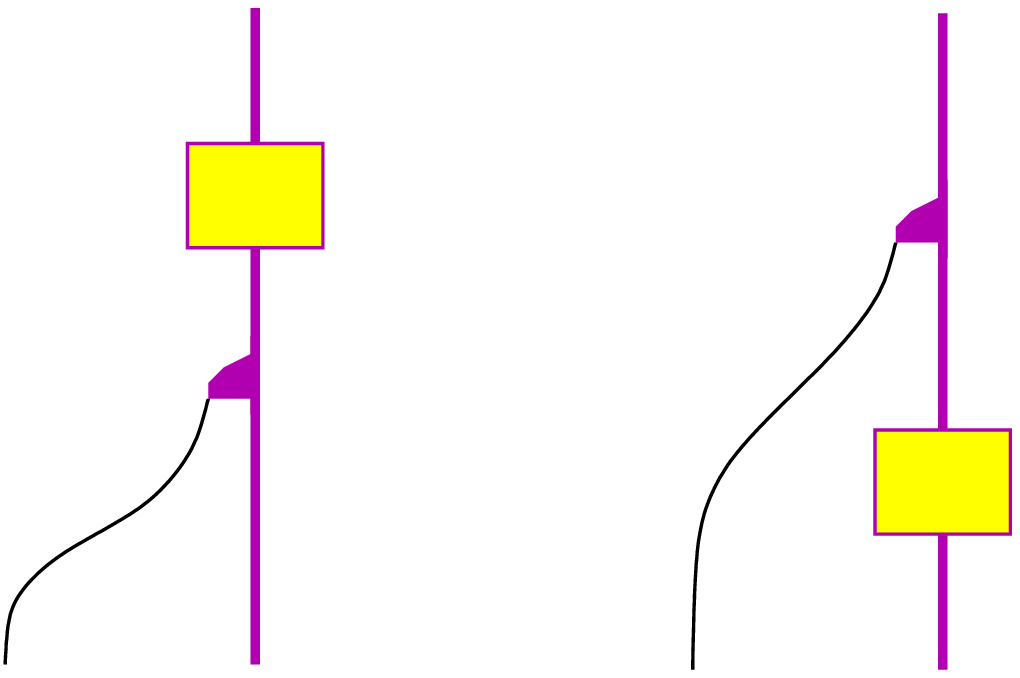}} \end{picture}}
  \put(-3.5,-9.1)  {\scriptsize$A$}
  \put(25.8,51.1)  {\scriptsize$f$}
  \put(23.6,-9.1)  {\scriptsize$\n$}
  \put(24.4,77.2)  {\scriptsize$\M$}
  \put(30.2,32.0)  {\scriptsize$\r_N^{}$}
  \put(57.5,37)    {$=$}
  \put(72.5,-9.1)  {\scriptsize$A$}
  \put(101.1,19.8) {\scriptsize$f$}
  \put(99.6,-9.1)  {\scriptsize$\n$}
  \put(99.6,77.2)  {\scriptsize$\M$}
  \put(106.3,48.9) {\scriptsize$\r_M^{}$}
  \epicture19 \labl{homA} 
Given an \alg\ $A$ in \calc, taking the $A$-modules in \calc\ as objects
and the subspaces
  \be  \HomA(N,M) := \{ f\iN \Hom(\n,\M) \,|\,
  f \cir\r_N\eq\r_M\cir(\id_A\Oti f) \}   \Labl ha
of the \calc-morphisms that intertwine the $A$-action
as morphisms results in another category, called 
the {\em category of\/} (left) $A$-{\em modules\/} and denoted by \calca.
Similarly as for morphisms in \calc\ (see formula \Erf lr), we will use 
the shorthand notation
  \be \dim\,\HomA(M,N) =: \lra MN  \labl{lra}
for any two $A$-modules $M,N\iN\objca$.

Typical representation theoretic tools, like induced modules and
reciprocity theorems, generalise to the category theoretic setting 
(see e.g.\ \cite{kios,kirI14,fuSc16,ostr}) and allow one to work out the 
representation theory in concrete examples. In particular, one shows that 
\calca\ inherits various properties of \calc. For instance, when \calc\ 
is modular and hence semisimple and when $A$ is special Frobenius,
then the category \calca\ is semisimple. 
On the other hand, that \calc\ is modular does {\em not\/} imply that
\calca\ is modular. In fact, it does not even imply that \calca\ is a \tc. 

However, a sufficient condition for \calca\ to be tensor
is then that $A$ is a commutative algebra and has
trivial twist, i.e.\ $\theta_{\!A}\eq\id_A$. The irreducible 
modules over such an algebra $A$ in a modular \tc\ fall into 
two different classes, the {\em local\/} and the {\em solitonic\/} modules. 
Local modules $M$ can be characterised \cite{kios} by the fact that their
twist is a morphism in $\HomA(M,M)$.
If $M$ is simple it follows that $\theta_M$ is a multiple of the identity, 
i.e.\ all irreducible subobjects of $M$
have the same conformal weight modulo integers. 
Let us denote by 
$(\calca)_{\rm loc}$ the full subcategory of \calca\ whose objects are the 
local $A$-modules. One can check that the twist and braiding on \calc\
induce a twist and braiding on $(\calca)_{\rm loc}$. Moreover, if \calc\ is 
a modular category, then $(\calca)_{\rm loc}$ is modular as well \cite{kios}.%
  \foodnode{The corresponding subcategory of \calc\ is not modular, but only
  `pre-modular'. It can be shown \cite{brug2,muge6} that such a category
  possesses a unique `modularisation'; this modularisation is precisely 
  $(\calca)_{\rm loc}$.}  
In CFT terms, $(\calca)_{\rm loc}$ is the modular category
for the chiral algebra \CA\ extended by the (primary) chiral fields that
correspond to the simple subobjects of the algebra $A$. Local modules 
correspond to boundary conditions that preserve the extended chiral algebra,
whereas solitonic modules break 
the extended symmetry (while still preserving \CA) \cite{fuSc11,bifs,fuSc12}.

A simple object in the category \calca\ is called a {\em simple $A$-module\/}.
Similarly as in section \ref{sec:mod-cat} (see before formula \erf{smat}), let 
us introduce a set of definite representatives for the isomorphism classes of 
simple $A$-modules; we denote them by $M_\kappa$ and the corresponding
label set by
  \be  \J = \{\kappa\} \,.  \labl J
As we will see later on, in the case of interest to us there are only finitely 
many isomorphism classes of simple $A$-modules, i.e.\ \J\ is a finite set.

\medskip

That the category \calca\ of $A$-modules is in general not a tensor category
even when the category \calc\ is modular, actually does not come as a 
surprise. The crucial property of \calca\ is that it is a so-called module 
category over \calc, and tensoriality is not a natural ingredient of that 
structure. A {\em module category\/} \cite{pare11,beRn,crfr,ostr} $\calm$ 
over a tensor category \calc\ is a category together with an exact bifunctor%
 \foodnode{It is conventional to denote this functor by the same symbol as
 the tensor product bifunctor of \calc.}
  \be \otimes: \quad \calm \oti \calc \to \calm  \labl{cs1}
that obeys generalised unit and associativity constraints: There are natural
isomorphisms from $(M\oti V)\oti W$ to $M\oti(V\oti W)$ and from
$M\oti\one$ to $M$, where $M$ is an object of $\calm$ while $V$ and
$W$ are objects of \calc.

In the case at hand, where $\calm\eq\calca$ is the category of left 
$A$-modules, the structure of a module 
category amounts to the observation that for any left $A$-module $M$ and
any object $V$ of \calc, also $M\oti V$ has a natural structure of
left $A$-module. Now recall that the tensor product bifunctor of \calc\ is 
exact, implying that the Grothendieck group $K_0(\calc)$ inherits from the 
tensor product on \calc\ the structure of a ring, the
fusion ring. It follows from the exactness of
the bifunctor \erf{cs1} that the Grothendieck group $K_0(\calm)$ carries
the structure of a right module over the ring $K_0(\calc)$. 
Thus the notion of a module category over a tensor category
is a categorification of the algebraic notion of a module over a ring,
in the same sense as the structure of a tensor category categorifies
the structure of a ring.

It can be shown \cite{ostr} that every module category over a tensor 
category $\calc$ is equivalent to the category of left modules for some 
associative algebra $A$ in \calc. Moreover, the analysis (see section 
\ref{sec:alg2bCFT}, and section \ref{bc-to-rep} below) of the OPE of boundary
fields, including those that change boundary conditions, shows that any 
unitary rational conformal field theory that possesses at least one boundary 
condition preserving the chiral algebra
gives rise to a module category over the category of 
Moore\hy Seiberg data. This provides another way to understand the emergence
of algebra objects which are central to our approach. (Indeed, the
reconstruction theorem of \cite{ostr} allows one to recover not just one
algebra in \calc, but rather a family of Morita equivalent algebras.)

We are now also in a position to explain why the algebras of open string 
states corresponding to different boundary conditions are Morita equivalent. 
The object $M$ of the module category $\calm$ that describes a boundary condition
is not necessarily simple,
i.e.\ we can allow for a superposition of elementary boundary conditions.
According to theorem 1 in \cite{ostr},
each such object $M$ gives rise to an algebra $A_M$ in the underlying tensor
category $\calc$, which is constructed using the so-called internal Hom:
  \be  A_M := \underline{\Hom} (M,M) \,.  \ee
In the same theorem it is shown that the category of $A_M$-modules is 
equivalent to the module category $\calm$. Thus in particular, any two objects 
$M_1$ and $M_2$ of the module category, i.e.\ any two boundary conditions, 
give rise to algebras $A_{M_1}$ and $A_{M_2}$ with equivalent categories
of left modules. By definition 10 of \cite{ostr}, these two algebras of
open string states are therefore Morita equivalent. In fact, the two
interpolating bimodules can be given explicitly in terms of internal
Hom's as well, namely as $\underline{\Hom}(M_1,M_2)$ and 
$\underline{\Hom}(M_2,M_1)$.

In any module category $\calm$, one can consider the morphism spaces
$\Hom_{\calm}(M\oti V, N)$ for any pair $M,N$ of objects of $\calm$ and 
any object $V$ of \calc. When $A$ is a special Frobenius algebra, as in 
the case of our interest, then semisimplicity of \calc\ implies that 
$\calm\eq\calca$ is semisimple as well. One can then generalise the arguments 
given in section \ref{sec:alg2bCFT} and choose definite bases of the spaces
$\Hom_{\calm}(M\oti V, N)$ with $M,N$ simple modules and $V$ a simple object,
and then write down the matrix elements of the isomorphism
  \be  \begin{array}l
  \dsty\bigoplus_{p\in\II} \Hom_{\calm}(M_i\oti U_p,M_l) \oti
  \Hom_{\mathcal C}(U_j\oti U_k, U_p) \\[-.1em]\hsp3
  \,\cong\, \Hom_{\calm} (M_i\oti U_j\oti U_k, M_l) 
  \,\cong\, \dsty\bigoplus_{q\in\JJ} (M_q\oti U_k,M_l)\oti\Hom(M_i\oti U_j,M_q) 
  \eear \labl{gz6j}
in these bases. Using the same graphs as in \erf{def-fmat}, but replacing
the simple objects $U_i$, $U_l$ and  $U_q$ of \calc\ by simple objects $M_i$,
$M_l$ and $M_q$ in the module category, one arrives at a generalisation of 
the 6j-symbols (\FF-matrices). These symbols, often denoted by
${}^{\sss(1)}\!{F}$, are labelled by three simple objects of $\calm$ and 
three simple objects of \calc\ (as well as one basis morphism of the type 
\erf{xijk,yijk} and three other basis morphisms involving also modules).
These mixed 6j-symbols express the associativity of the bifunctor \erf{cs1}.
They appear naturally in the theory of weak Hopf algebras \cite{bosz2}.
As already mentioned after formula \erf{eq:bope-sewing}, in CFT they arise
as structure constants for the operator products of boundary fields.

\medskip

Below it will be convenient to use the following construction of 
$A$-intertwiners:
\vspace{-.9em}

\dtl{Definition}{df:av}
For any $\Phi\iN\Hom(\dot M,\dot M')$, with $M,M'$ (left) $A$-modules, 
the $A$-{\em averaged morphism\/} $\overline\Phi$ is
  \be  \overline\Phi := \rho_{M'} \circ [\id_A\oti (\Phi\cir\rho_M)]
  \circ[(\Delta\cir\eta) \oti \idM] \,.  \ee
Pictorially:
  \bea \begin{picture}(98,96)(0,26)
  \put(0,47) {$\overline\Phi\ =$}
  \put(41,0) {\begin{picture}(0,0)(0,0)
             \scalebox{.38}{\includegraphics{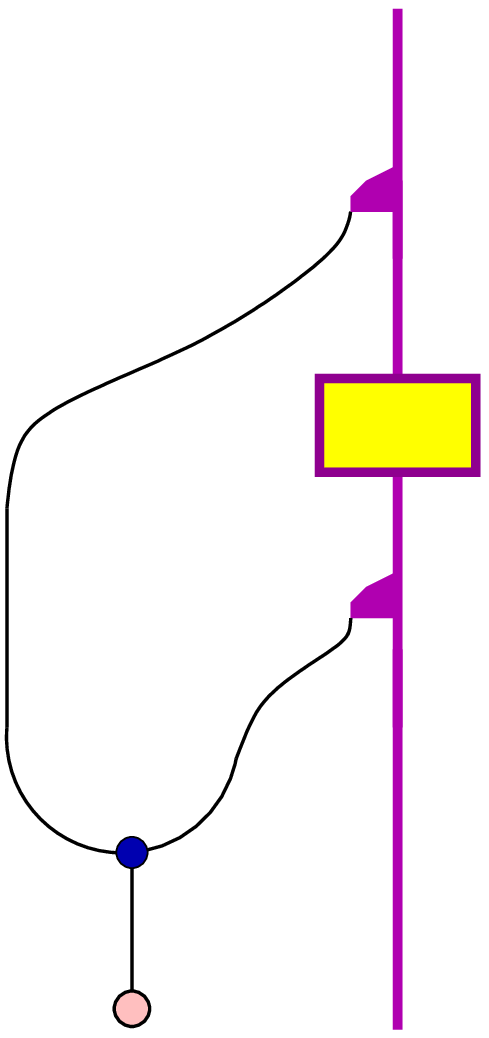}} \end{picture}}
  \put(79.9,-9.8)  {\scriptsize$\M$}
  \put(82.5,64.4)  {\scriptsize$\Phi$}
  \put(80.6,115.8) {\scriptsize$\M'$}
  \epicture18 \labl{average}

\dtl{Lemma}{le:av}
(i)\hsp{.77}When $A$ is a Frobenius \alg, then for every 
$\Phi\iN\Hom(\dot M,\dot M')$, where $M,M'$ are $A$-modules, the 
$A$-averaged morphism $\overline\Phi$ is an $A$-intertwiner,
  \be  \overline\Phi\in\HomA(M,M')\,.  \ee
(ii)\hsp{.51}When $A$ is special Frobenius and $\Phi \iN \Hom_A(\M,\M')$, then 
$\overline\Phi\eq\Phi$.
\\[.13em]
(iii) When $A$ is a symmetric special Frobenius \alg, then the trace of an
endomorphism does not change under averaging,
  \be \tr \overline\Phi = \tr\Phi \,.  \ee
Proof:\\
(i)\hsp{.72}The statement follows by a simple application of (both parts of) 
the Frobenius property of $A$.
\\[.13em]
(ii)\hsp{.51}Start by using the fact that $\Phi$ is an $A$-intertwiner, so as
to move the \rep\ morphism $\rho_M$ past $\Phi$ to get another $\rho_{M'}$. Next 
employ the representation property for $\rho_{M'}$. The $A$-lines can then be 
removed completely by first using specialness of $A$ (recall the convention
$\beta_{\!A}\eq1$) and the the unit property for $\rho_{M'}$.
\\[.13em]
(iii) Start with the left trace, as defined in \erf{trace} (recall that since 
$\calc$ is modular
and hence spherical, ${\rm tr}_{\rm r}\eq{\rm tr}_{\rm l} \,{=:}\,{\rm tr}$).
Owing to the cyclicity property of the trace we can pull the
upper $A$-line around until the two \rep\ morphisms are directly composed.
Now use symmetry of $A$ as in \erf{Phi12-} and the first representation
property \erf{1m}. Finally employ specialness \erf{spec2} of $A$
and the second representation property in \erf{1m}. Upon these 
manipulations the $A$-lines have completely disappeared.
\qed 

We will also need the notion of an \AB-bimodule. This is the
following structure:

\dtl{Definition}{def:bimod}
For $A$ and $B$ algebra objects in a tensor category \calc, an 
\AB-{\em bimodule\/} $M$ is a triple $(\M,\rho^{A},\tilde\rho^{B})$, 
consisting of an object $\M\iN\objc$ and two morphisms
$\rho^A \iN \Hom(A\oti \M,\M)$ and $\tilde\rho^{B} \iN \Hom(
\M$\linebreak[0]$\oti B,\M)$, such that $(\M,\rho^A)$ is a left 
$A$-module, $(\M,\tilde\rho^B)$ is a right $B$-module, and such 
that the actions $\rho^A$ and $\tilde \rho^B$ commute, i.e.\ 
  \be  \rho^A \circ (\id_A \oti \tilde \rho^B) 
  = \tilde\rho^B \circ (\rho^A \oti \id_B) \,. \ee

\medskip

When we want to make explicit the algebras over which $M$ is a bimodule,
we write it as ${}_{A\!}M_B$. 
If \calc\ is a braided tensor category, then bimodules can
equivalently be thought of as left $A\Oti B_{\rm op}$-modules,
see remark~12 in \cite{ostr}. 

This is an important point, because it implies that we do not need
to develop new techniques to find all bimodules, once we know how 
to deal with left modules. In particular the methods of induced modules
described in section \ref{sec:induced} below can be applied.

Let us formulate the correspondence between bimodules and left modules
more precisely.

\dtl{Proposition}{prop:bimod-leftmod}
For $A$, $B$ algebras in a braided tensor category, the mapping
(see picture \erf{eq:ff-bimod-iso} below)
  \be  f:\quad (\M, \rho^A, \tilde\rho^B ) \,\mapsto\,
  (\M, \rho^A \cir (\id_A \oti \tilde\rho^B) \cir (\id_A \oti c^{-1}_{\M,B}))
  \labl{eq:bimod-prop}
takes \AB-bimodules to left $A\Oti B_{\rm op}$-modules. $f$ is invertible, with
inverse $g$ given by
  \be g:\quad (\M, \rho^{A \otimes B_{\rm op}}) \,\mapsto\,
  (\M, \rho^{A\otimes B_{\rm op}} \cir (\id_A \oti \eta_{B_{\rm op}}
  \oti \id_{\M}), \rho^{A\otimes B_{\rm op}} \cir (\eta_A \oti c_{\M,B})) \,.
  \labl{eq:bimod-prop2}

\noindent
Proof:\\
To check that $g\cir f\eq\id$ and $f\cir g\eq\id$ is straightforward.
Now suppose we are given an \AB-bimodule $M$. 
Substituting the prescription \erf{eq:bimod-prop}, the representation
property for $M$ can be rewritten as in the first equality in
  \bea \begin{picture}(396,110)(0,37)
  \put(-20,0) {\begin{picture}(0,0)(0,0)
             \scalebox{.38}{\includegraphics{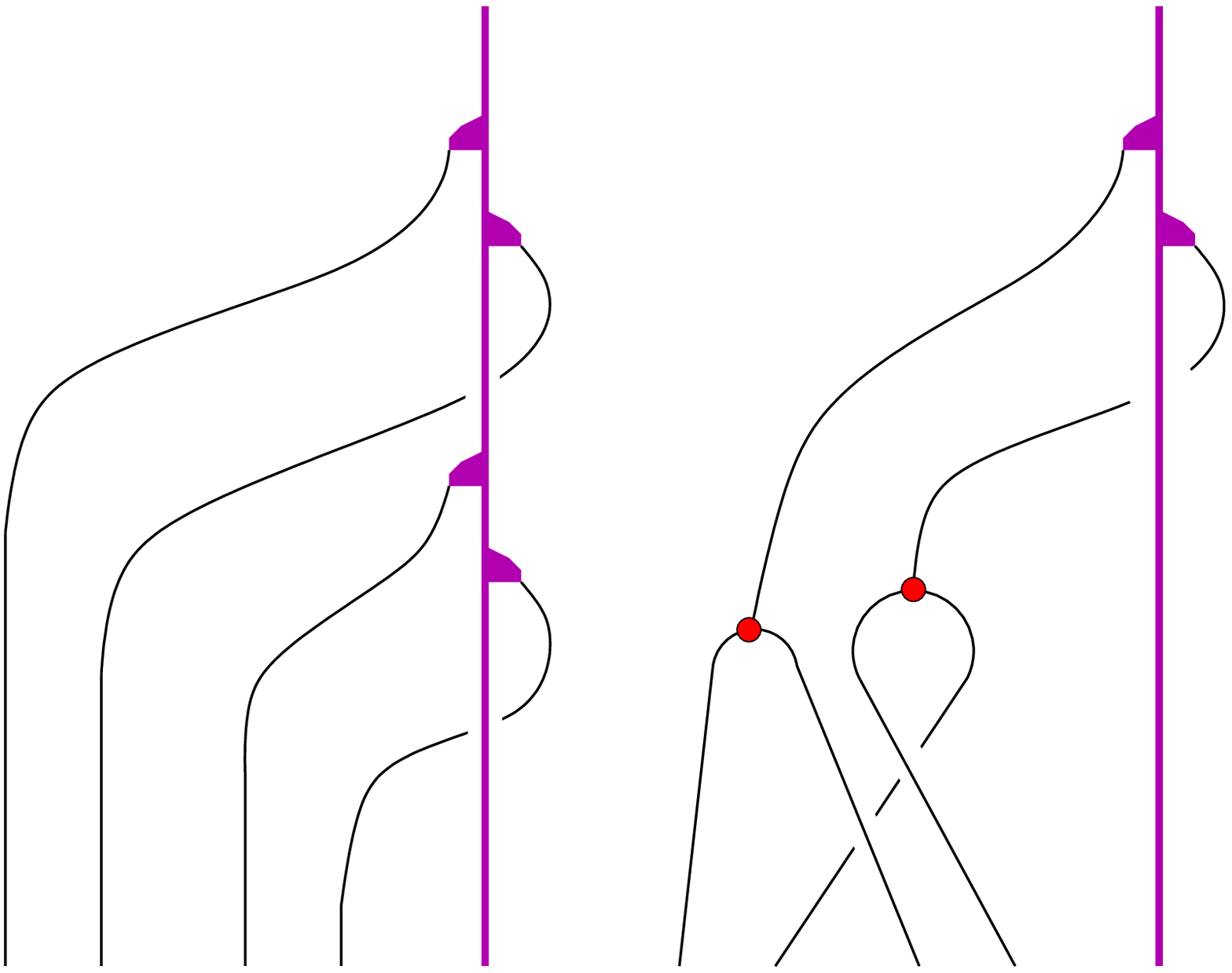}} \end{picture}
    \put(-2.8,-9)  {\scriptsize$A$} 
    \put(10.2,-9)   {\scriptsize$B$} 
    \put(29.8,-9)    {\scriptsize$A$} 
    \put(44.8,-9)   {\scriptsize$B$} 
    \put(64.8,-9)   {\scriptsize$\M$} 
    \put(64.8,141.3)  {\scriptsize$\M$} 
    \put(89,66)       {$=$}
    \put(94,-9) {\scriptsize$A$}
    \put(109,-9) {\scriptsize$B$}
    \put(128,-9) {\scriptsize$A$}
    \put(143,-9) {\scriptsize$B$}
    \put(161.8,-9)  {\scriptsize$\M$} 
    \put(161.8,141.3)  {\scriptsize$\M$} 
    }
  \put(177.4,66)      {$\Longrightarrow$}
  \put(220,0) {\begin{picture}(0,0)(0,0)
             \scalebox{.38}{\includegraphics{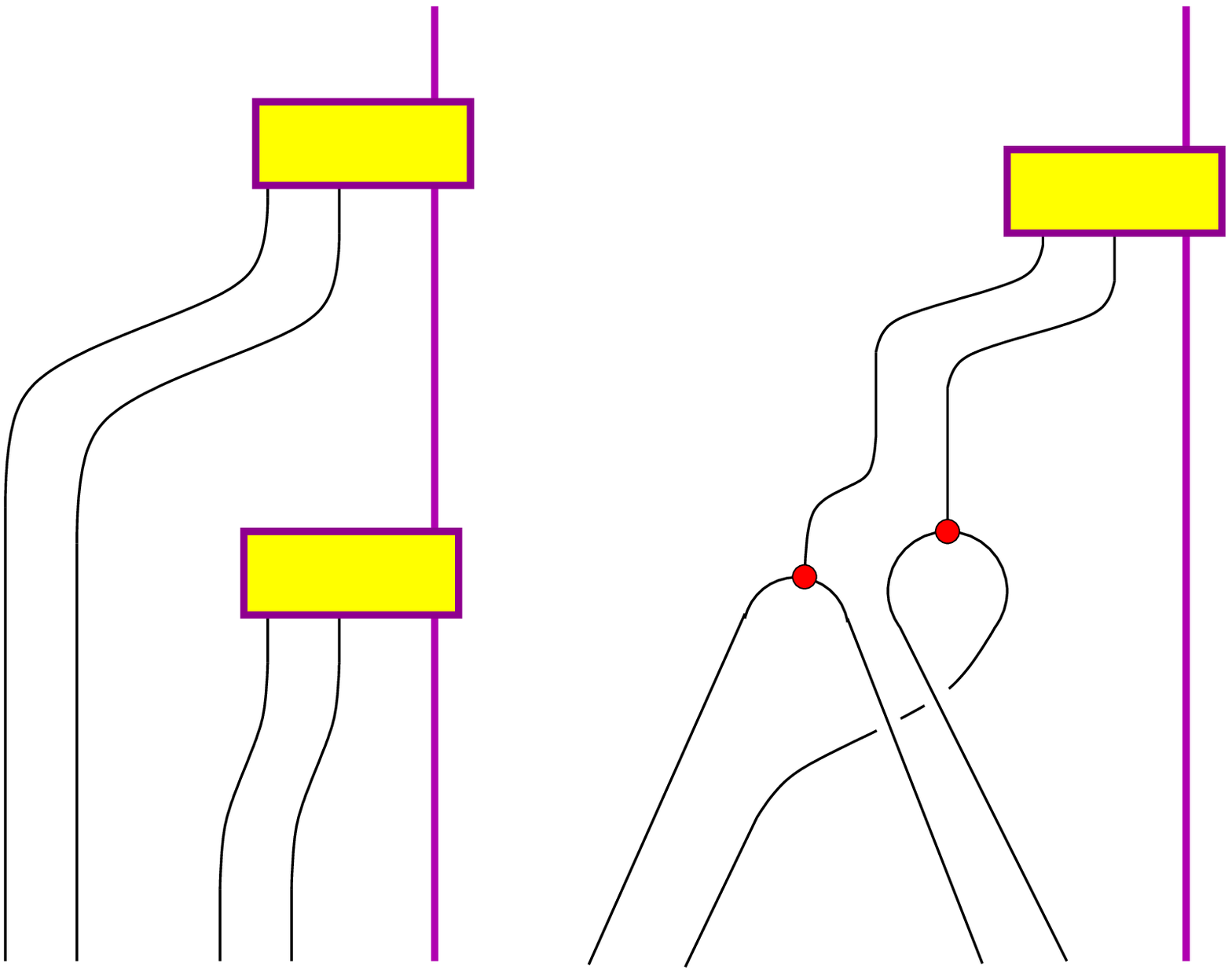}} \end{picture}
    \put(-3,-9) {\scriptsize$A$}
    \put(8,-9) {\scriptsize$B$}
    \put(29,-9) {\scriptsize$A$}
    \put(39,-9) {\scriptsize$B$}
    \put(79.8,-9) {\scriptsize$A$}
    \put(95,-9) {\scriptsize$B$}
    \put(136.8,-9) {\scriptsize$A$}
    \put(150,-9) {\scriptsize$B$}
    \put(57.8,-9)  {\scriptsize$\M$} 
    \put(57.8,142.3)  {\scriptsize$\M$} 
    \put(83,66)      {$=$}
    \put(166,-9)  {\scriptsize$\M$} 
    \put(166,142.3)  {\scriptsize$\M$} 
    \put(37,55) {\scriptsize$\rho^{A\!{\oti}\!B_{\rm op}}$}
    \put(37.5,117) {\scriptsize$\rho^{A\!{\oti}\!B_{\rm op}}$}
    \put(146,110) {\scriptsize$\rho^{A\!{\oti}\!B_{\rm op}}$}
             }
  \epicture24 \labl{eq:bimod1}
This equality, in turn, directly implies the second equality in
\erf{eq:bimod1}, which precisely spells out the representation property
for a left $A\Oti B_{\rm op}$-module. (Note that in the second equality, 
the multiplication morphisms are those of $A$ and $B$, not of $B_{\rm op}$.) 
\\
Conversely, let us suppose that $M$ is a left $A\Oti B_{\rm op}$-module, 
and verify that $g(M)$ provides an \AB-bimodule.  It is easily checked that 
$\rho^{A\otimes B_{\rm op}} \cir (\id_A \Oti \eta_{B_{\rm op}}\Oti\id_{\M})$
does provide a left representation of $A$. That $\rho^{A \otimes 
B_{\rm op}} \cir (\eta_A \Oti c_{\M,B})$ provides a right representation 
of $B$ and that the two actions commute amounts to verifying that
  \bea \begin{picture}(360,116)(0,29)
  \put(-20,0) {\begin{picture}(0,0)(0,0)
             \scalebox{.38}{\includegraphics{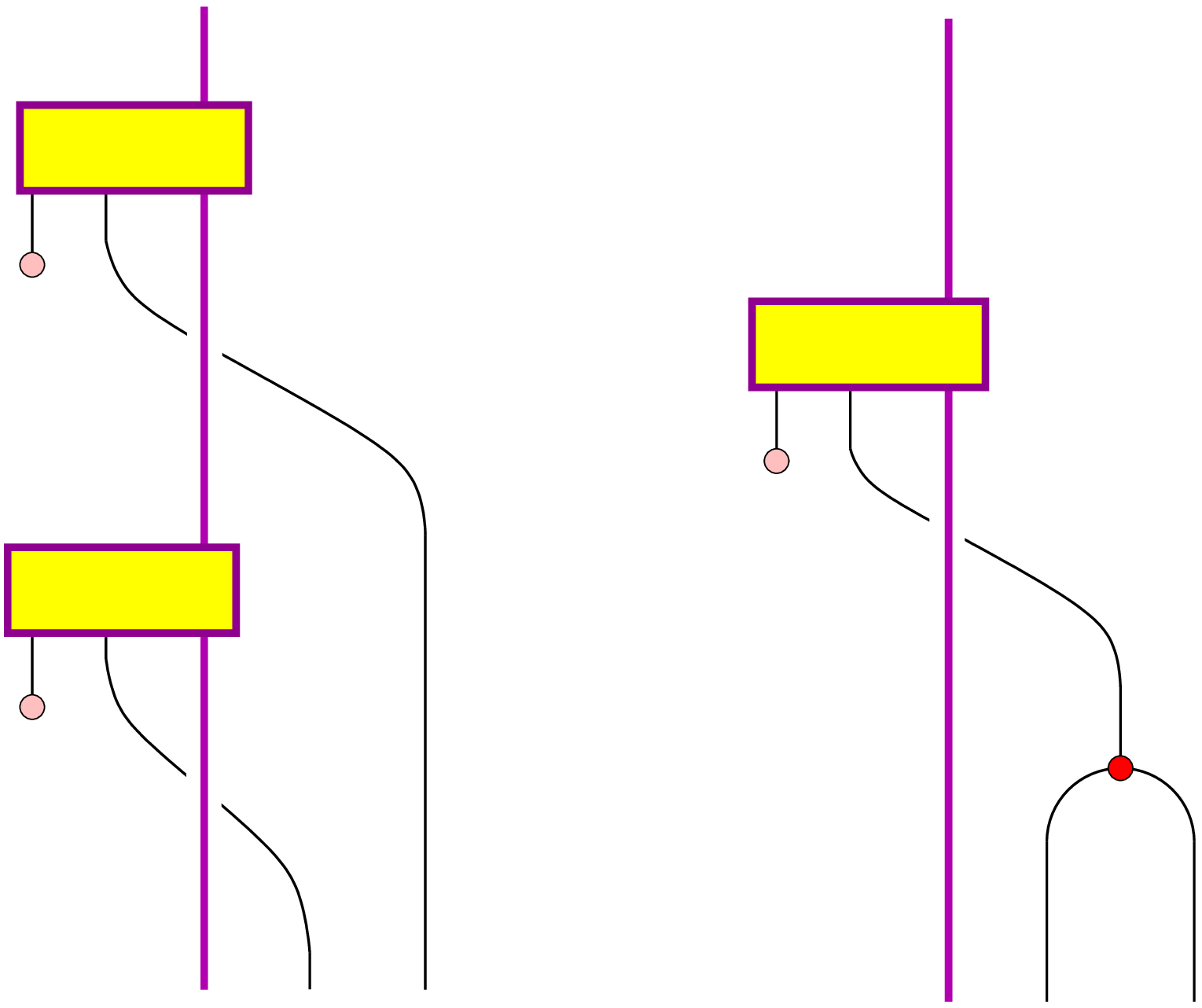}} \end{picture}
   \put(23,-8) {\scriptsize$\M$}
   \put(39.1,-8) {\scriptsize$B$}
   \put(55.7,-8) {\scriptsize$B$}
   \put(127,-9) {\scriptsize$\M$}
   \put(143,-9) {\scriptsize$B$}
   \put(163,-9) {\scriptsize$B$}
   \put(25,143) {\scriptsize$\M$}
   \put(129,141.6) {\scriptsize$\M$}
   \put(5,117) {\scriptsize$\rho^{A\!{\oti}\!B_{\rm op}}$}
   \put(3,55) {\scriptsize$\rho^{A\!{\oti}\!B_{\rm op}}$}
   \put(107,89) {\scriptsize$\rho^{A\!{\oti}\!B_{\rm op}}$}
  }
  \put(220,0) {\begin{picture}(0,0)(0,0)
             \scalebox{.38}{\includegraphics{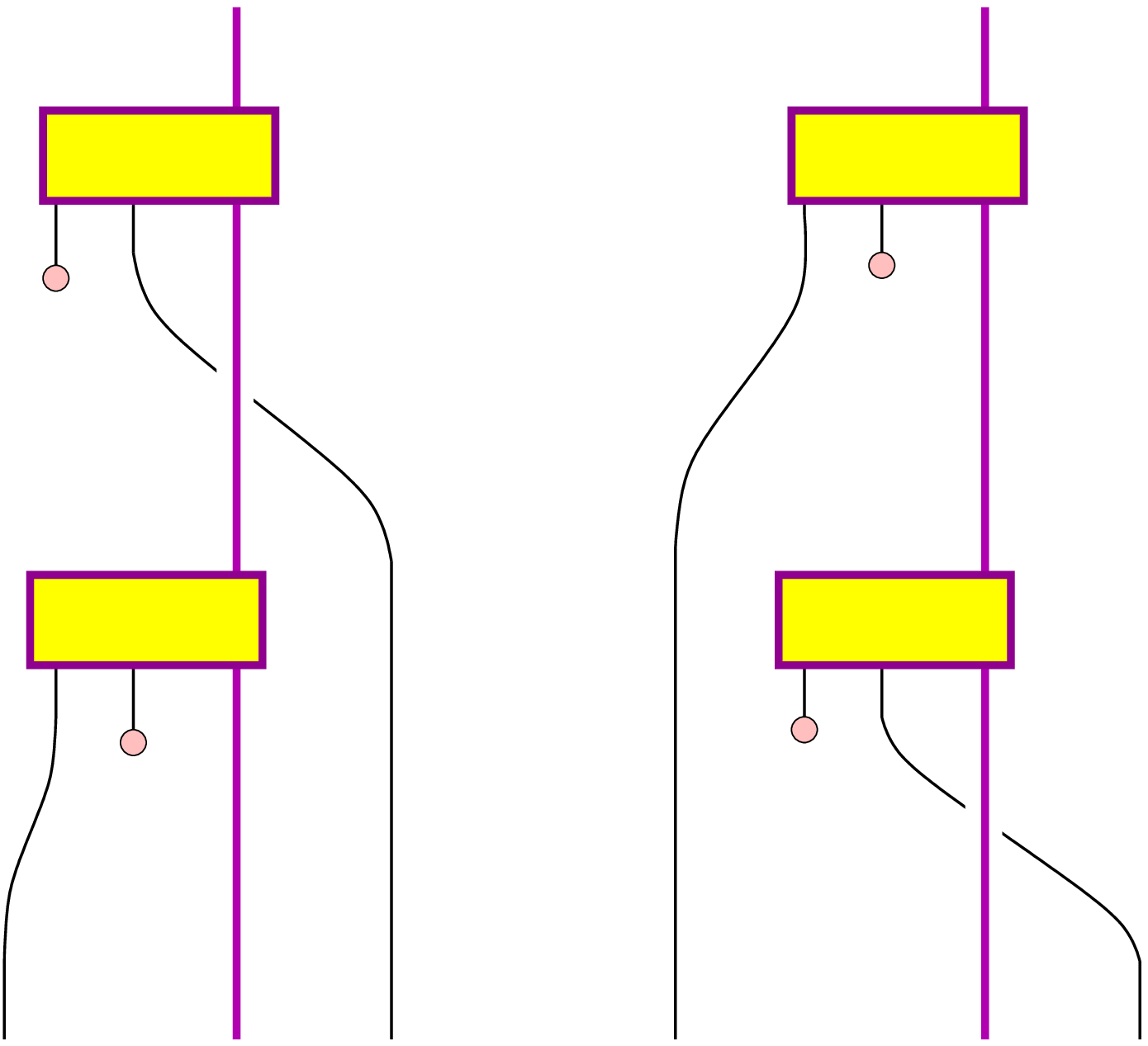}} \end{picture}
   \put(-3,-9) {\scriptsize$A$}
   \put(26,-9) {\scriptsize$\M$}
   \put(49,-9) {\scriptsize$B$}
   \put(86,-9) {\scriptsize$A$}
   \put(125,-9) {\scriptsize$\M$}
   \put(148,-9) {\scriptsize$B$}
   \put(28,141.6) {\scriptsize$\M$}
   \put(127,141.6) {\scriptsize$\M$}
   \put(6.5,115.5) {\scriptsize$\rho^{A\!{\oti}\!B_{\rm op}}$}
   \put(5,54.5) {\scriptsize$\rho^{A\!{\oti}\!B_{\rm op}}$}
   \put(105,54.5) {\scriptsize$\rho^{A\!{\oti}\!B_{\rm op}}$}
   \put(106,115.5) {\scriptsize$\rho^{A\!{\oti}\!B_{\rm op}}$}
  }
  \put(62,65)       {$=$}
  \put(170,65)      {and}
  \put(287,65)      {$=$}
  \epicture16 \ee
Both of these properties follow from the representation property of left 
$A\Oti B_{\rm op}$-modules as given in the second equality in \erf{eq:bimod1}.
\qed

\dt{Remark}
(i) Proposition \ref{prop:bimod-leftmod} provides an isomorphism
$f$ that takes \AB-bimodules to left $A\Oti B^{(-1)}$-modules. 
In an analogous manner one constructs an isomorphism $\tilde f$ that
takes \AB-bimodules to left $B^{(1)}\Oti A$-modules.
In pictures, the two isomorphisms are given by
  \bea \begin{picture}(295,44)(0,12)
  \put(0,0) {\begin{picture}(0,0)(0,0)
             \scalebox{.38}{\includegraphics{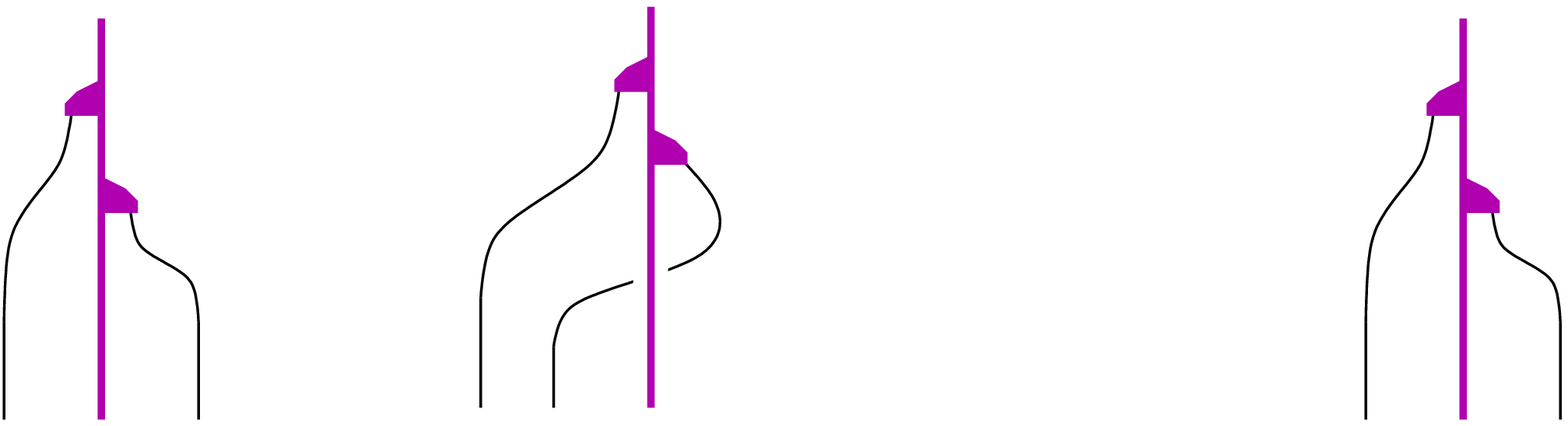}} \end{picture}
  \put(-25,22){$ f: $}
  \put(130,22){and}
  \put(170,22){$ \tilde f: $}
  \put(-3,-8) {\scriptsize$A$}
  \put(12,-8) {\scriptsize$\M$}
  \put(25,-8) {\scriptsize$B$}
  \put(65,-8) {\scriptsize$A$}
  \put(75,-8) {\scriptsize$B$}
  \put(89,-8) {\scriptsize$\M$}
  \put(189,-8) {\scriptsize$A$}
  \put(203,-8) {\scriptsize$\M$}
  \put(217,-8) {\scriptsize$B$}
  \put(258,-8) {\scriptsize$B$}
  \put(269,-8) {\scriptsize$A$}
  \put(281.2,-8) {\scriptsize$\M$}
  \put(40,22) {\scriptsize$\longmapsto$}
  \put(234,22) {\scriptsize$\longmapsto$}
  }
  \epicture02 \labl{eq:ff-bimod-iso}
(ii) It can be shown that for any Frobenius \alg\ $A$ in a \tc\ \calc,
the category \calcaa\ of \AA-bimodules is again a \tc. This is in 
remarkable contrast to the category \calca\ of left $A$-modules, which 
in general cannot can be equipped with a tensor product. Moreover, when 
\calc\ is sovereign and $A$ is symmetric, then \calcaa\ is equipped with 
a duality. However, in general the bimodule category does not have a 
braiding, and therefore it does not have a twist either.

\subsection{Representation functions}

For any given $A$-module $M$ we choose bases $b^M_{(i,\alpha)}$ for the 
morphism spaces%
  \foodnode{The label $i\iN\I$ refers to the chosen representatives $U_i$ 
  of the isomorphism classes of simple objects in \calc, see section 
  \ref{sec:mod-cat}.}
$\Hom(U_i,\M)$ and dual bases $b_{M}^{(j,\beta)}$ for $\Hom(\M,U_j)$ 
such that
  \be b_{M}^{(j,\beta)} \circ b^M_{(i,\alpha)}
  = \delta_i^j\, \delta_\alpha^\beta \,\id_{U_i}^{} \,.  \labl{b-dual}
Pictorially:
  \bea \begin{picture}(357,34)(0,17)
  \put(45,0) {\begin{picture}(0,0)(0,0)
             \scalebox{.38}{\includegraphics{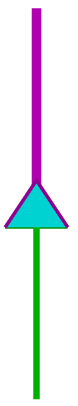}} \end{picture}}
  \put(0,20)      {$b^M_{(i,\alpha)}=:$}
  \put(47.1,-7.8) {\scriptsize$i$}
  \put(45.1,46.9) {\scriptsize$\M$}
  \put(53.1,21)   {\tiny$\alpha$}
  \put(153,0){\begin{picture}(0,0)(0,0)
             \scalebox{.38}{\includegraphics{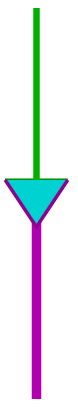}} \end{picture}}
  \put(77,20)     {$,\qquad b_M^{(j,\beta)}=:$}
  \put(151.9,-8.4){\scriptsize$\M$}
  \put(155.2,48.4){\scriptsize$j$}
  \put(161.3,21.2){\tiny$\bar\beta$}
  \put(265,-7) {\begin{picture}(0,0)(0,0)
             \scalebox{.38}{\includegraphics{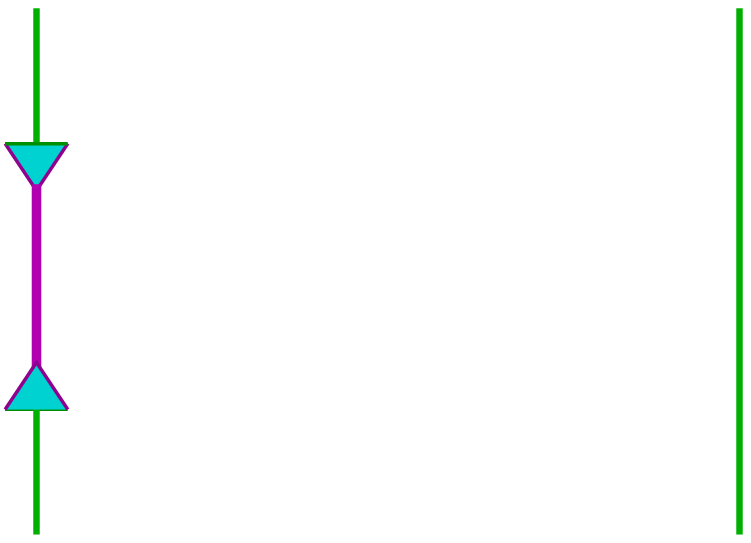}} \end{picture}}
  \put(206,20)    {and}
  \put(267.4,-14.8){\scriptsize$i$}
  \put(258.8,19.2){\scriptsize$\M$}
  \put(267.8,56.9){\scriptsize$j$}
  \put(274.1,7.2) {\tiny$\alpha$}
  \put(274.1,33.2){\tiny$\bar\beta$}
  \put(293,20)    {$=\ \delta_i^j\, \delta_\alpha^\beta$}
  \put(344.6,-14.8){\scriptsize$i$}
  \put(345.0,56.9){\scriptsize$i$}
  \epicture11 \labl{b-b}
Dominance in the category \calc\ implies the completeness property
  \be \sum_{i\in\II}\sum_\alpha
  b^M_{(i,\alpha)} \cir b_{M}^{(i,\alpha)} = \idM \,,  \labl{comp}
i.e.\
  \bea \begin{picture}(135,45)(0,22)
  \put(57,0)  {\begin{picture}(0,0)(0,0)
              \scalebox{.38}{\includegraphics{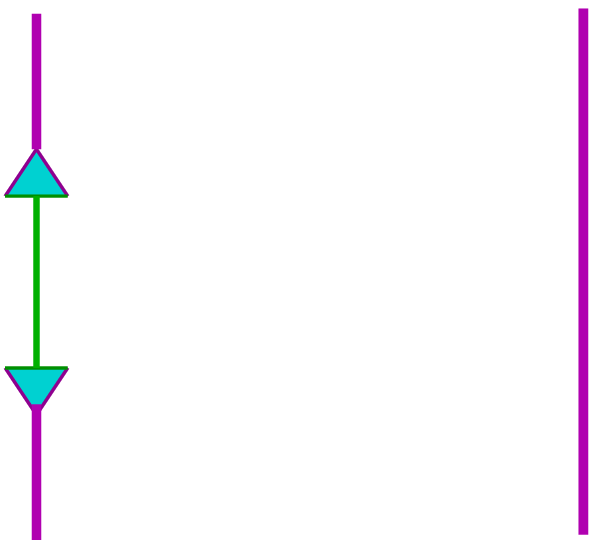}} \end{picture}}
  \put(5,27)      {$\dsty\sum_{i\in\II}\sum_\alpha$}
  \put(56.8,-8.8) {\scriptsize$\M$}
  \put(55.9,27.1) {\scriptsize$i$}
  \put(57.5,62.8) {\scriptsize$\M$}
  \put(64.6,14)   {\tiny$\bar\alpha$}
  \put(64.6,42.8) {\tiny$\alpha$}
  \put(88,27)     {$=$}
  \put(116.1,-8.8){\scriptsize$\M$}
  \put(116.5,62.8){\scriptsize$\M$}
  \epicture14 \labl{b-b-complete}

Next let us associate to each pair of morphisms $f\iN\Hom(X,\M)$ and 
$g\iN\Hom(\M,Y)$ the morphism
  \be \Psi(f,g) := g \circ \r_M^{} \circ (\id_A\oti f)
  \ \in \Hom(A\Oti X,Y) \,.  \labl{Psi}
We can then introduce the following notion:

\dt{Definition}
(i)\hsp{.6}For any $A$-module $M$ the morphism
  \bea \begin{picture}(181,47)(0,23)
  \put(152,0) {\begin{picture}(0,0)(0,0)
              \scalebox{.38}{\includegraphics{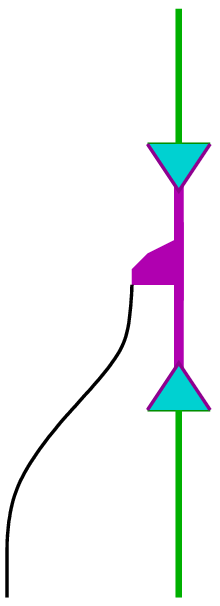}} \end{picture}}
  \put(0,30)  {$\r^{\ (j,\beta)}_{M(i,\alpha)}\,:=\,
                \Psi(b^M_{(i,\alpha)}{,}b_{M}^{(j,\beta)})\ =$}
  \put(148.8,-8.2){\scriptsize$A$}
  \put(169.8,-8.2){\scriptsize$i$}
  \put(169.8,69.7){\scriptsize$j$}
  \put(173.4,30.2)    {\scriptsize$\M$}
  \put(175.8,22.4){\tiny$\alpha$}
  \put(175.5,46.3){\tiny$\bar\beta$}
  \epicture17 \labl{def-rho}
(with $i,j\iN\I$ and $b^M_{(i,\alpha)},b_{M}^{(j,\beta)}$ basis elements
in $\Hom(U_i,\M)$ and $\Hom(\M,U_j)$ as described above)
is called a {\em \rep\ function\/} of $M$.
\\[.13em]
(ii) For every $j\iN\I$, the $j$th {\em character\/} of $M$ is the sum
  \be  \chii^j_M := \sum_\alpha \r_{M(j,\alpha)}^{\ (j,\alpha)}
  \in \Hom(A\oti U_j,U_j) \labl{chii}
of \rep\ functions, and the $j$th {\em dimension morphism\/} $\cald_j$
of $M$ is 
  \be  \cald_j(M) := \chii^j_M \cir (\eta\oti \id_{U_j})
  \in\End(U_j)\,.  \labl{dimi}

This diction is sensible because in the special case $\calc\eq\Vect(\complex)$
and $A$ the group algebra $\complex[{\rm G}]$ of a group $\rm G$, the 
quantities \erf{def-rho}
are indeed ordinary representation functions ${(\r_M(g))}_a^b$.
(Also, a module possesses then only a single character, since $\Vect(\complex)$
has only a single isomorphism class of simple objects.) 
Besides the representation property
  \be  \sum_b \rho_M(g)_a^b\, \rho_M(h)_b^c = \rho_M(gh)_a^c \,, \ee
ordinary representation functions for simple modules $M$, $M'$ also obey 
the orthogonality relation
  \be  \Frac1{|G|} \sum_{g\in G} (\rho_M(g)^*)_a^b\, \rho_{M'}(g)_c^d 
  = \Frac1{\dim(\M)}\, \delta_a^d\delta_c^b\, \delta_{M,M'} \,.  \ee 
In our context, the representation property amounts to the identity
  \be \rho_{M (i,\alpha)}^{\ (k,\gamma)} \cir (m\oti\id_{U_i})
  = \dsty\sum_{(j,\beta)} \rho_{M,(j,\beta)}^{\ (k,\gamma)}
  \circ (\id_A \oti \rho^{\ (j,\beta)}_{M,(i\alpha)} ) \ee
in $\Hom(A\Oti A \Oti U_i,U_k)$, i.e.\
  \bea \begin{picture}(170,73)(0,35)
  \put(0,0)  {\begin{picture}(0,0)(0,0)
             \scalebox{.38}{\includegraphics{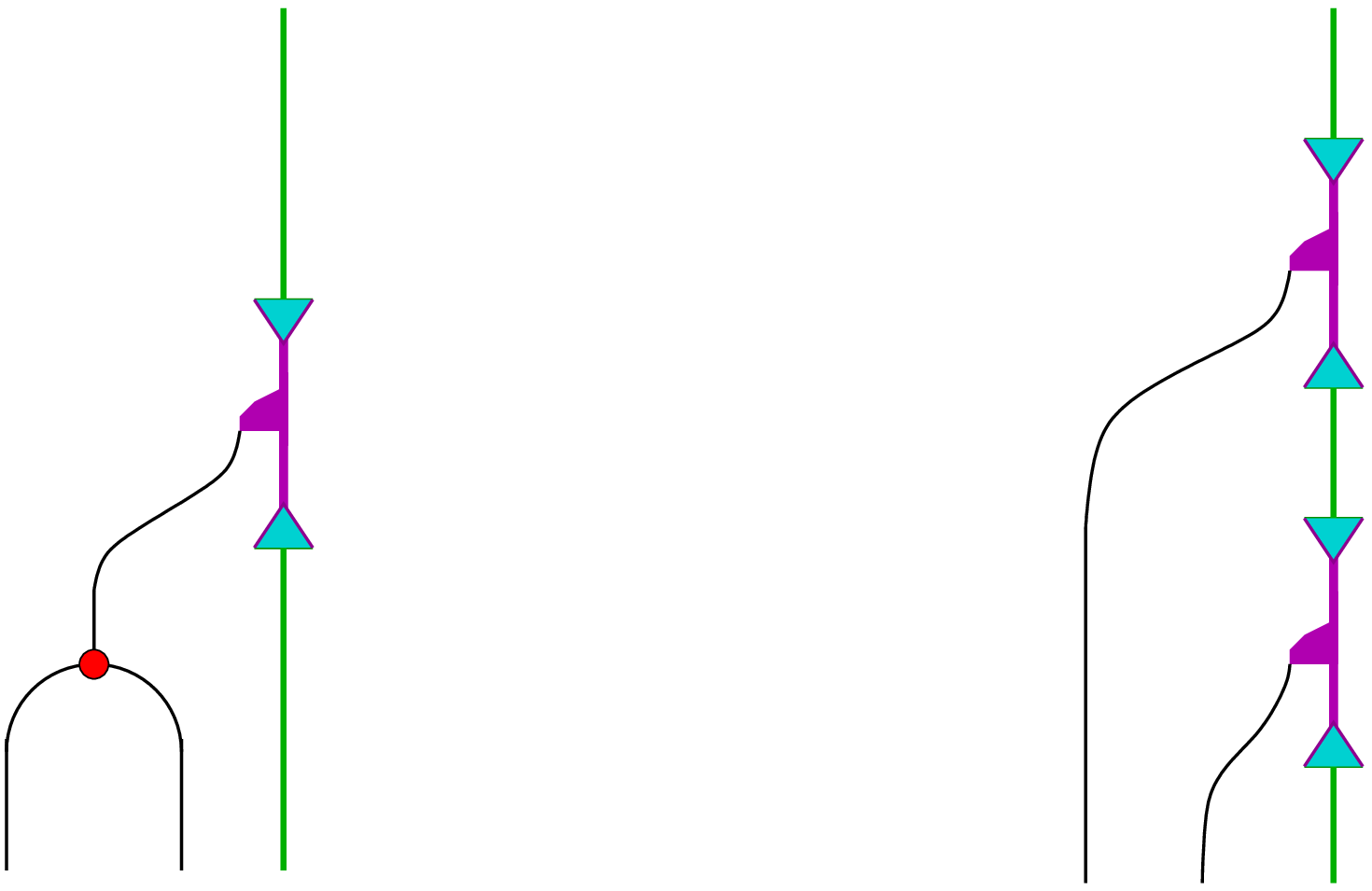}} \end{picture}}
  \put(-3.3,-8.2) {\scriptsize$A$}
  \put(15.5,-8.2) {\scriptsize$A$}
  \put(31.9,-8.2) {\scriptsize$i$}
  \put(31.1,108.3){\scriptsize$k$}
  \put(34.6,49.2) {\scriptsize$\M$}
  \put(37.1,37.1) {\tiny$\alpha$}
  \put(36.1,69.3) {\tiny$\bar\gamma$}
  \put(58.7,47.2) {$=\ \dsty\sum_{j\in\II}\sum_\beta$}
  \put(123.5,-8.2){\scriptsize$A$}
  \put(136.2,-8.2){\scriptsize$A$}
  \put(155.1,-8.2){\scriptsize$i$}
  \put(158.5,22.6){\scriptsize$\M$}
  \put(151.5,48.9){\scriptsize$j$}
  \put(158.5,69.3){\scriptsize$\M$}
  \put(154.9,108.3){\scriptsize$k$}
  \put(161.2,15.4){\tiny$\alpha$}
  \put(160.7,39.8){\tiny$\bar\beta$}
  \put(161.4,60.3){\tiny$\beta$}
  \put(161.9,86.2){\tiny$\bar\gamma$}
  \epicture23 \labl{rho-rep}
(Use completeness \erf{b-b-complete} and the representation property \erf{1m} 
to see this equality.)
The generalisation of orthogonality is less direct. We find:
\vspace{-.9em}

\dtl{Lemma}{le:rho-ortho}
The \rep\ functions \erf{def-rho} for simple modules $M_\kappa,\, M_{\kappa'}$
($\kappa,\kappa'\iN\J$) of a special Frobenius algebra $A$ satisfy
  \be  \rho_{M_{\kappa'}(j,\alpha)}^{\ (k,\delta)} \circ
  [\id_A \oti \rho_{M_\kappa(l,\gamma)}^{\ (i,\beta)}]
  \circ [(\Delta\cir\eta)\oti \id_{U_l} ]
  = \frac{\dim(U_i)}{\dim(\M_\kappa)}\, \delta_{\kappa,\kappa'} \,
  \delta^{(i,\alpha)}_{(j,\beta)} \delta^{(k,\delta)}_{(l,\gamma)}\, \id_{U_l}
  \,. \labl{eq:rho-ortho}
Pictorially:
  \bea \begin{picture}(315,85)(0,60)
  \put(40,0)  {\begin{picture}(0,0)(0,0)
              \scalebox{.38}{\includegraphics{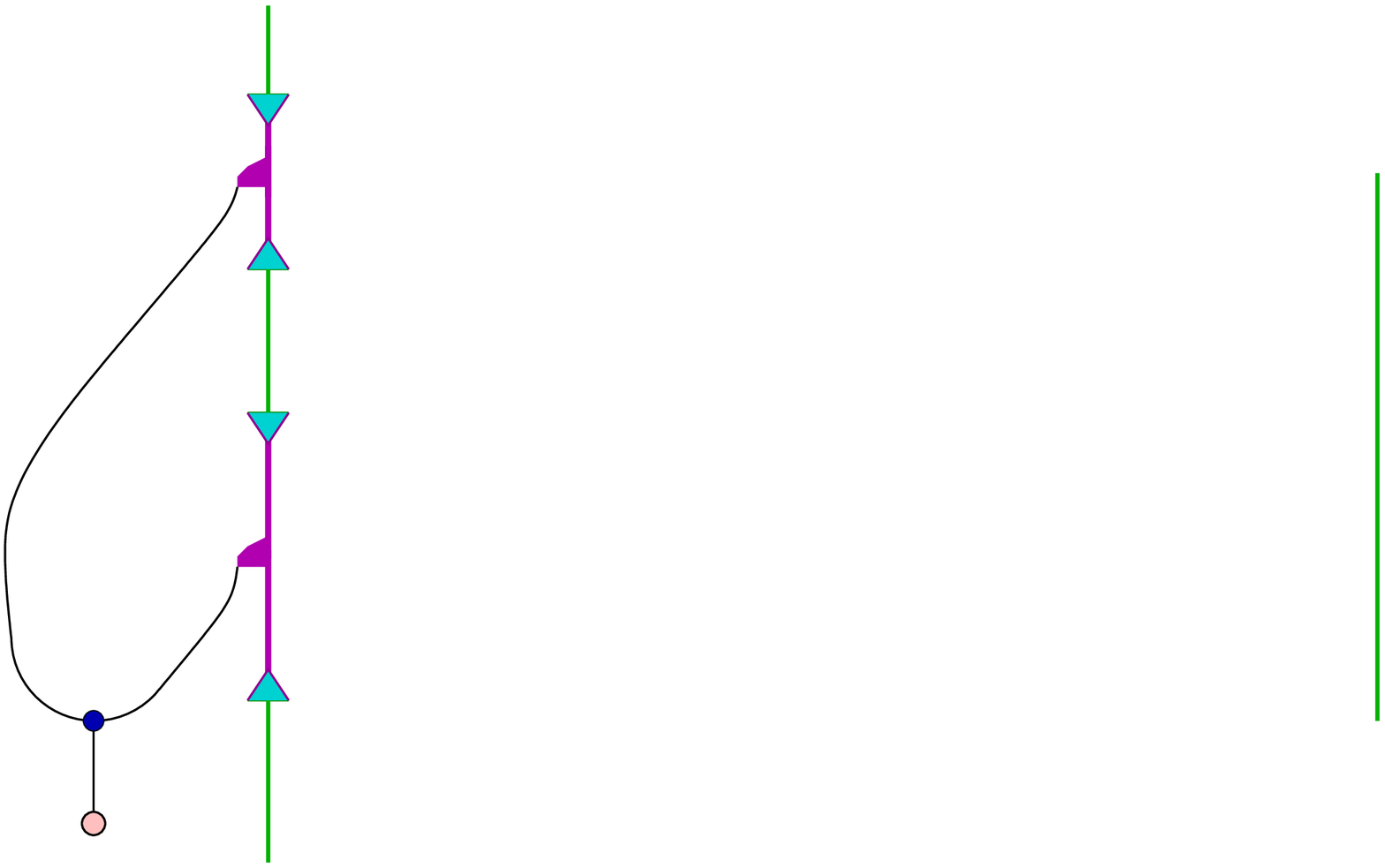}} \end{picture}}
  \put(83.3,-8.2) {\scriptsize$l$}
  \put(87,40)     {\scriptsize$\M_\kappa$}
  \put(79.9,80.2) {\scriptsize$j$}
  \put(85.9,92.8) {\scriptsize$i$}
  \put(87,110)    {\scriptsize$\M_{\kappa'}$}
  \put(82.8,148.1){\scriptsize$k$}
  \put(87.9,29.3) {\tiny$\gamma$}
  \put(87.9,71.6) {\tiny$\bar\beta$}
  \put(87.9,102.2){\tiny$\alpha$}
  \put(88.1,124.4){\tiny$\bar\delta$}
  \put(113.8,70)  {$=\ \dsty\frac{\dim(U_i)}{\dim(\M_\kappa)}\;\delta
                  _{\kappa,\kappa'}^{}\;\delta^{(i,\alpha)}_{(j,\beta)}\,
                  \delta^{(k,\delta)}_{(l,\gamma)}$}
  \put(268.9,14.2){\scriptsize$l$}
  \put(268.7,121.3){\scriptsize$l$}
  \epicture39 \labl{rho-ortho}
Proof:\\
Since $U_k$ and $U_l$ are simple objects (and we have restricted our attention
to only a single representative of each isomorphism class), the morphism 
on the \lhs\ can be non-zero only for $k\eq l$. For the same reason we also
need $i\eq j$, and we will tacitly assume this in the sequel. Similarly, 
noting that the \lhs\ involves an $A$-averaged morphism $\overline\Phi$
(see definition \ref{df:av}) -- namely the one associated to $\Phi\eq
b^{M_{\kappa'}}_{(i,\alpha)}\cir b_{M_\kappa}^{(j,\beta)}\iN\Hom(\M_\kappa,
\M_{\kappa'})$ -- and that by lemma \ref{le:av}(i) $\overline\Phi$ is 
an $A$-intertwiner, we can use the fact that $M_\kappa$ and $M_{\kappa'}$ are 
simple modules, and non-isomorphic for $\kappa\,{\ne}\,\kappa'$, to conclude 
that $M_\kappa\eq M_{\kappa'}\,{=:}\,M$. Moreover, $\overline\Phi$ 
must then be a multiple $\xi^i_{\alpha\beta}$ of $\idM$, 
and hence the morphism on the \lhs\ of \erf{eq:rho-ortho} is equal to
  \be  \varpi = \xi^i_{\alpha\beta}\, b_M^{(l,\delta)}\cir b^M_{(l,\gamma)}
%   = \xi^i_{\alpha\beta}\, \delta_\gamma^\delta\, \id_{U_l} 
  \,.  \Labl99 
To determine the coefficients $\xi^i_{\alpha\beta}$ we take the trace of 
\Erf99 and sum over $l\iN\I$ and over $\gamma=\delta$
(i.e.\ over dual bases in $\Hom(U_l,\M)$ and $\Hom(\M,U_l)$). 
Using the cyclicity of the trace and then the
completeness property \erf{b-b-complete} yields
  \be  \sumI_l \sum_\gamma \tr\varpi = \xi^i_{\alpha\beta}\, \dim(\M)
  \,.  \ee
On the other hand, when applying the trace and summation on the \lhs\ of
\erf{eq:rho-ortho}, and then again using the cyclicity and completeness,
we get
  \be  \sumI_l \sum_\gamma \tr\varpi = \tr\overline\Phi = \tr\Phi
  = \delta_\alpha^\beta\, \dim(U_i) \,,  \ee
where in the second step one invokes lemma \ref{le:av}(iii) and in the
last step once more cyclicity of the trace, as well as the orthogonality 
\erf{b-b} is used. Comparison of the two results yields 
$\xi^i_{\alpha\beta}\eq\delta_\alpha^\beta\,\dim(U_i)/\dim(\M)$ and thereby
establishes the validity of \erf{eq:rho-ortho}.
\qed

\dtl{Corollary}{co:cd}
The characters \erf{chii} of the simple $A$-modules $M_\kappa,M_{\kappa'}$ 
satisfy
  \be  \chii^i_{M_\kappa} \circ(\id_A \oti \chi^i_{M_{\kappa'}})
  \circ [(\Delta\cir\eta)\oti \id_{U_i} ]
  = \delta_{\kappa,\kappa'}\, \frac{\dim(i)}{\dim(\M_\kappa)}\,
  \lr{U_i}{\M_\kappa} \, \id_{U_i} \,,  \Labl98
and the dimension morphisms \erf{dimi} obey%
  \foodnode{Recall
 the shorthand $\lr XY\eq \dim\,\Hom(X,Y)$ introduced in \erf{lr}.}
  \be  \cald_i(M_\kappa) = \lr{U_i}{\M_\kappa}\, \id_{U_i} \,.  \Labl97 

\medskip\noindent
Proof:\\
Formula \Erf98 follows immediately from \erf{rho-ortho} by 
setting $j\eq k\eq l\eq i$, $\alpha\eq\delta$, $\beta\eq\gamma$ and summing 
over $\alpha$ and $\beta$. Equation \Erf97 is a direct consequence of the 
unit property of $\eta$ and the orthogonality \erf{b-b}.
\qed

Note that in the category of vector spaces, evaluating the character 
of a module at the unit element yields the dimension of the module.
In the situation studied
here, $\cald_i(M)$ tells us with which multiplicity the simple object $U_i$ 
``occurs in $M$", and we obtain the (quantum) dimension of $\M$ as
  \be  \dim(\M) = \sumI_i \tr\cald_i(M) = \sumI_i \lr{U_i}\M\,\dim(U_i)
  \,.  \ee

\subsection{Induced modules}\label{sec:induced}

Later on the $A$-modules will label the boundary conditions of the CFT
that $A$ defines. Given an algebra $A$ in \calc, we would therefore like 
to find all $A$-modules obtainable from objects in \calc.  
A powerful method to find $A$-modules is the one of induced 
modules, which we recall briefly in this section. The main result is that
if $A$ is special Frobenius, then every $A$-module is a submodule of an
induced module. The extension of the method of induced representations
to a general category theoretic setting was developed in \cite{kios,fuSc16}. 

\dt{Definition}
For $U\iN\obj(\calc)$, the {\em induced\/} (left) {\em module\/} $\inda(U)$ 
is the $A$-module that is equal to $A\Oti U$ as an object in \calc, 
and has representation morphism $m\oti\id_U$, i.e.
  \be \inda(U) := (\, A\Oti U \,,\,m\oti\id_U\,) \,. \labl{rho-ind}
Pictorially:
  \bea \begin{picture}(145,53)(0,24)
  \put(0,0)   {\begin{picture}(0,0)(0,0)
              \scalebox{.38}{\includegraphics{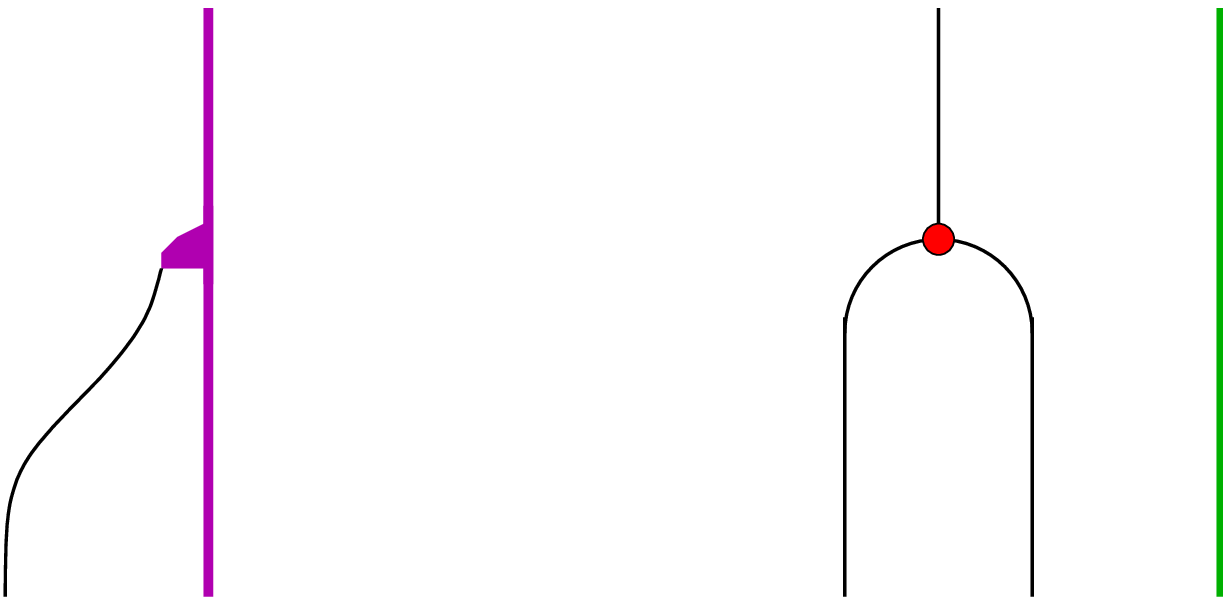}} \end{picture}}
  \put(-3.1,-8.2) {\scriptsize$A$}
  \put(12.1,-8.2) {\scriptsize$A\oti U$}
  \put(12.1,70.7) {\scriptsize$A\oti U$}
  \put(25.5,38.5) {\scriptsize$\rho_{\inda(U)}$}
  \put(68.8,30.4) {$=$}
  \put(88.1,-8.2) {\scriptsize$A$}
  \put(99.7,70.7) {\scriptsize$A$}
  \put(108.4,-8.2){\scriptsize$A$}
  \put(130.5,-8.2){\scriptsize$U$}
  \put(130.5,70.7){\scriptsize$U$}
  \epicture17 \labl{rho-inda}

That the morphism \erf{rho-ind} satisfies the \rep\ properties \Erf1m
follows directly from the associativity of $m$ and the unit property of
$\eta$. (Analogously, $\indar(U)\eq (U\Oti A,\id_U\Oti m)$
is a right $A$-module, called an induced right module. 
For the tensor unit $\one$ one obtains this way just the \alg\ itself,
$\inda(\one)\eq A\eq\indar(\one)$. Moreover, 
the product of $A$ is a morphism between $A\Oti A$ and $A$ as
\AA-bimodules, $m\iN\Hom_{A|A}(A\Oti A,A)$, while the coproduct 
$\Delta$ is in $\Hom_{A|A}(A,A\Oti A)$ if and only if $A$ is Frobenius. 

\dtl{Proposition}{prop:reci}
[{\em Reciprocity\/}]\, For $A$ a Frobenius algebra in a tensor category \calc,
$U\iN\obj(\calc)$ and $M$ an $A$-module, there are natural isomorphisms
  \be  
  \Hom(U,\M) \stackrel\cong\longrightarrow \HomA(\inda(U),M)
  \quad\; {\rm and} \quad\;
  \Hom(\M,U) \stackrel\cong\longrightarrow \HomA(M,\inda(U)) \,, \ee
given by $\varphi\,{\mapsto}\,\rho_M \cir (\id_A \oti \varphi)$
for $\varphi\iN\Hom(U,\M)$ and $\varphi\,{\mapsto}\,
(\id_A\oti(\varphi{\circ}\rho_M)) \cir ((\Delta{\circ}\eta)\oti\idM)$
for $\varphi\iN\Hom(\M,U)$. In particular,
  \be  \lr U\M = \lra{\inda(U)}M
  \qquad {\rm and} \qquad \lr \M U = \lra M{\inda(U)} \,.  \labl{recipro}

\medskip\noindent
For a proof see e.g.\ \cite{fuSc16}, propositions 4.7 and 4.11.%
  \foodnode{Via these maps one actually gets functors $\calc\,{\to}\,
  \calca$ (induction) and $\calca\,{\to}\,\calc$ (restriction). 
  Induction is a right-adjoint functor to restriction, and for
  modular \calc\ both functors are exact, i.e.\ carry exact sequences 
  of morphisms to exact sequences \cite{fuSc16}. Moreover, even when
  \calca\ is a \tc, the induction functor is not a tensor functor.
  In contrast, there are functors from \calc\ to the bimodule
  category \calcaa\ (so-called $\alpha$-induction, see section
  \ref{sec:alphaind} below) which are tensor functors.}

\dtl{Proposition}{modsubind}
If $A$ is a special Frobenius algebra in a semisimple tensor category 
\calc, then every $A$-module is a submodule of an induced module, and the
category \calca\ of $A$-modules is semisimple.

\medskip\noindent
For a proof see e.g.\ \cite{fuSc16}, lemma 5.24 and proposition 5.25.

\medskip
That \calca\ is semisimple means that every $A$-module can be
written as the direct sum of finitely many simple
$A$-modules. Furthermore, every simple module already
appears as a submodule of an induced module $\inda(U)$ with $U$ a simple
object of \calc. For \calc\ a modular tensor category, which is the case
under study, there are only finitely many isomorphism classes of simple
objects, and as a consequence there are then also only finitely many 
isomorphism classes of simple $A$-modules. Thus as announced,
the index set \J\ \erf J is finite.

One important aspect of proposition \ref{prop:reci} is that it
allows us to learn already quite a lot about the structure of simple
$A$-modules from the fusion ring of \calc\ alone. In simple cases -- like
for the free boson and for the algebra giving the \E modular invariant,
which are treated in section \ref{sec:N=1-repn} below -- the reciprocity 
relation already determines how many simple $A$-modules there are and 
into which simple objects of \calc\ they decompose as elements of $\objc$. 

Even when $U$ is a simple object of \calc, the induced module $\inda(U)$ 
will in general {\em not\/} be a simple $A$-module. Indeed, we obtain:
\vspace{-.9em}

\dtl{Corollary}{co:reci}
(i)\hsp{.6}The module $\inda(U_j)$ ($j\iN\II$) decomposes into simple 
modules $M_\kappa$ ($\kappa\iN\JJ$) according to
  \be  \inda(U_j) \,\cong\, \bigoplusJ_\kappa \lr{U_j}{\M_\kappa}\,
  M_\kappa \,.  \labl{reci}
(ii) For every simple object $U$ of \calc\ the sum rule
  \be  \dim(A) = \frac1{\dim(U)}\sumJ_\kappa \dim(\M_\kappa)\, \lr
  U{\M_\kappa} \labl{rule}
holds.

\medskip\noindent
Proof:\\
(i)\hsp{.6}Formula \erf{reci} follows immediately by applying the
reciprocity relation \erf{recipro} to the simple $A$-modules $M_\kappa$.
\\[.13em]
(ii) Since dimensions are constant on isomorphism classes of objects, we
can restrict our attention to $U\eq U_j$ for $j\iN\II$. For these objects
the result follows by taking the trace of both sides of the relation \erf{reci}.
\qed

In the vector space case the relation analogous to \erf{rule} reads
$\dim(A)\eq\sum_\kappa (\dim(V_\kappa))^2$. Thus when this formula is 
extended to general tensor categories, the two factors of $\dim(V_\kappa)$ 
get generalised in two distinct ways, one of them to
$\,\dim(\M_\kappa)/\dim(U_i)$, the other to $\lr{U_i}{\M_\kappa}$. Let us 
also mention that for simple current algebras, these two factors take the 
form $\,\dim(\M_\kappa)/\dim(U_i)$\linebreak[0]$ \eq |G|/\sqrt{s_iu_i}$ and
$\lr{U_i}{M_\kappa}\eq\sqrt{s_i/u_i}$, \resp, where $G$ is the simple current
group, $s_i$ is the order of the stabiliser of $U_i$ (the subgroup of $G$
of simple currents that leave $i$ fixed), and $u_i$ the order of the 
so-called untwisted stabiliser (for its definition see \cite{fusS6,bant7}).  

\medskip

We are now also in a position to make the following assertion that will 
be useful later on (see section \ref{sec:one-point-blocks}):
\vspace{-.6em}

\dtl{Theorem}{thm:mod-Aixj}
Let $X,Y$ be simple objects of \calc\ and $A$ a symmetric special 
Frobenius \alg\ in \calc. Then the prescription \erf{Psi} furnishes 
a natural isomorphism
  \be \Psi\,: \quad \bigoplusJ_\kappa \Hom(X,\M_\kappa) \oti
  \Hom(\M_\kappa,Y) \to \Hom(A\Oti X,Y) \,.  \labl{th}

\noindent
Proof:\\
Recall that the direct sum in \erf{th} is over one representative 
$\M_\kappa$ out of each 
isomorphism class of simple $A$-modules. In the following for simplicity we
will just write $\kappa$ in place of $\M_\kappa$. To show injectivity, we 
apply $\Psi$ to a basis of the space on the \lhs\ as in \erf{def-rho}. 
Suppose that 
  \be  \sumJ_\kappa \sumI_{i,j} \sum_{\alpha,\beta}
  \lambda_{\kappa(i,\alpha)}^{\ (j,\beta)}\, \rho_{\kappa(i,\alpha)}
  ^{\ (j,\beta)} = 0 \ee
for some complex numbers $\lambda_{\kappa(i\alpha)}^{\ (j\beta)}$.
Upon composition with another representation function
$\rho_{\kappa'(i',\alpha')}^{\ (j',\beta')}$ and averaging over $A$, the
\lhs\ yields $\lambda_{\kappa'(i',\alpha')}^{\ (j',\beta')}$ by lemma
\ref{le:rho-ortho}. Hence all the numbers 
$\lambda_{\kappa(i\alpha)}^{\ (j\beta)}$ are zero.
\\[.13em]
Surjectivity is checked as follows. Using the additivity of the $i$th 
dimension morphism \erf{dimi}, it follows from corollary \ref{co:reci} that
  \be  \bearll \cald_i(\inda(U_j)) \!\!
  &= \sumJ_\kappa \lr{U_j}{\M_\kappa}\, \cald_i(M_\kappa) 
   = \sumJ_\kappa \lr{U_j}{\M_\kappa}\, \lr{\M_\kappa}{U_i}\, \id_{U_i}
  \\{}\\[-.5em]
  &\equiv \dim\llb \bigoplusJ_\kappa \Hom(U_j,\M_\kappa) \oti
  \Hom(\M_\kappa,U_i) \lrb\, \id_{U_i} \,. \eear  \labl{s2}
On the other hand, using directly formula \Erf97 we get
  \be  \cald_i(\inda(U_j)) = \lr{U_i}{A\Oti U_j}\, \id_{U_i}
  \equiv \dim\llb \Hom(A\Oti U_j,U_i) \lrb\, \id_{U_i} \,.  \labl{s1}
Comparison of \erf{s1} and \erf{s2} establishes surjectivity of the map
\erf{th} for the case that $X\eq U_j$ and $Y\eq U_i$. By additivity this 
extends to the case of arbitrary objects $X$ and $Y$ (using also that \calc\
is semisimple).
\qed

\medskip

In the special case that $A$ is a group algebra in $\Vect(\complex)$, the
theorem amounts to the statement that for complex semisimple algebras the
representation functions span the dual space of the algebra.

\medskip

An interesting consequence of the relation \erf{rho-ortho} is that it allows us 
to invert the isomorphism \erf{th} explicitly. We find:

\dt{Lemma}
The coefficients $\lambda_{\kappa,\alpha}^{\ \beta}$ in the expansion
  \be  \phi = \sumJ_\kappa\sum_{\alpha,\beta} \lambda_{\kappa,\alpha}^{\ \beta}
  \,\rho_{M_\kappa(i,\alpha)}^{\ (j,\beta)}  \labl{Psi-1}
of a morphism $\phi\iN \Hom(A\oti U_i,U_j)$ \wrt the \rep\ functions
\erf{def-rho} read
  \be  \lambda_{\kappa,\alpha}^{\ \beta}\, \id_j = \frac{\dim(\M_\kappa)}
  {\dim(U_i)}\, \phi \circ [\id_A \oti (b_{M_\kappa}^{(i,\alpha)}\cir
  \rho_{M_\kappa})] \circ
  [(\Delta\cir\eta) \oti b^{M_\kappa}_{(j,\beta)} ] \,.  \labl{eq:lambda}

\medskip\noindent
Proof:\\
Since the morphisms $\rho_{M_\kappa(i,\alpha)}^{\ (j,\beta)}$ form a basis 
of $\Hom(A\oti U_i, U_j)$, there exists a unique set of numbers $\lambda
_{\kappa,\alpha}^{\ \beta}$ with property \erf{Psi-1}. Their values can 
be extracted by composing the \rhs\ of \erf{Psi-1} with a representation 
function and `integrating' over $A$. In this way one arrives at the expression
\erf{eq:lambda} for $\lambda_{\kappa,\alpha}^{\ \beta}$. Pictorially,
  \bea \begin{picture}(175,100)(0,31)
  \put(31,0)  {\begin{picture}(0,0)(0,0)
              \scalebox{.38}{\includegraphics{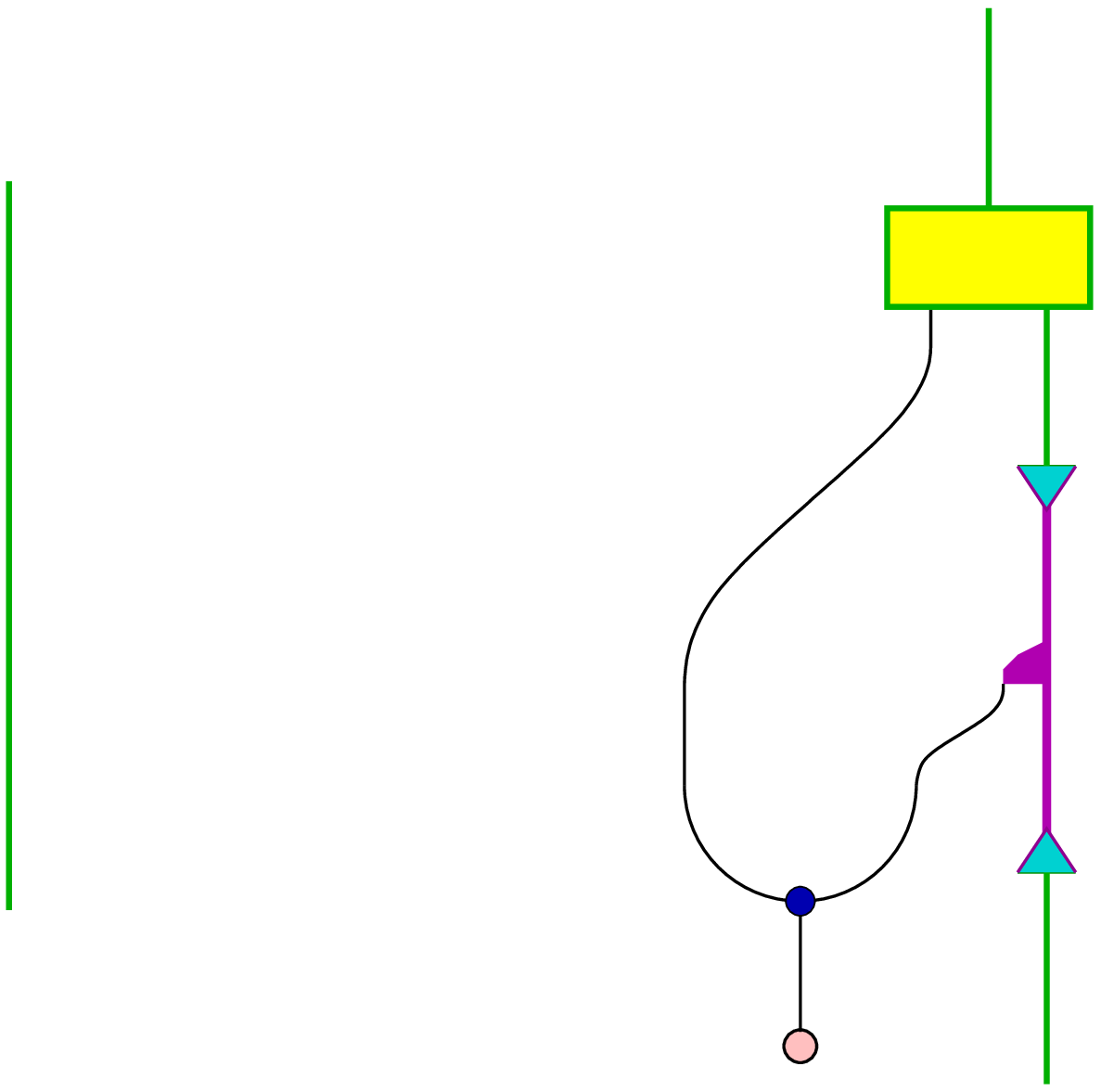}} \end{picture}}
  \put(0,62)     {$\lambda_{\kappa,\alpha}^\beta$}
  \put(29.2,13.5) {\scriptsize$j$}
  \put(29.8,112.5){\scriptsize$j$}
  \put(43.5,62)   {$=\;\dsty\frac{\dim(\M_\kappa)}{\dim(U_i)}$}
  \put(152.3,-7.4){\scriptsize$j$}
  \put(146.4,133.3){\scriptsize$j$}
  \put(113,40)    {\scriptsize$A$}
  \put(157,50)    {\scriptsize$\r_{M_\kappa}^{}$}
  \put(158.8,27)  {\tiny$\beta$} 
  \put(158.8,69)  {\tiny$\bar\alpha$}
  \put(156.5,81)  {\scriptsize$i$}
  \put(145,97)    {\scriptsize$\phi$}
  \epicture07 \labl{lambda}
\mbox{$ $}\\[-1.3em]\mbox{$ $}\qed

\subsection{From boundary conditions to representations}\label{bc-to-rep}

In section \ref{sec:alg2bCFT} we have seen how a boundary condition 
(preserving the chiral \alg\ \CA) in a rational CFT 
gives rise to a symmetric special Frobenius algebra $A$ in the \mtc\ of
the CFT (the \rep\ category of \CA). Let us henceforth also label the boundary 
condition from which this \alg\ arises by $A$. Recall that the OPE of two 
boundary fields living on the $A$-boundary furnishes the multiplication 
of the algebra $A$, and the sewing
constraint for four boundary fields on a disk assures its associativity.

The notion of an $A$-module has a direct interpretation in CFT
terms as well. Consider a stretch of boundary for which at some point $y$ 
the boundary condition changes from $A$ to another boundary condition, say 
$M$. Then at the point $y$ a `boundary changing field' $\Psi^{AM}(y)$ must be
inserted. The OPE of any field $\Phi(x)$ that lives on the boundary $A$ 
with the boundary changing field $\Psi^{AM}(y)$ is forced to be of the form
  \be 
  \Phi(x)\,\Psi^{AM}(y) = \sum_{k} 
  (x-y)^{\Delta_k-\Delta_{\Phi}-\Delta_{\Psi}} \Psi^{AM}_k(y) \,,  \ee
where $\Psi^{AM}_k$ is a suitable collection of boundary changing fields. The
boundary OPE thus defines a mapping
  \be  \left\{\!\!\begin{tabular}c fields\\living on $A$ \end{tabular}\!\!\right\}
  \;\times\; \left\{\!\!\begin{tabular}c
  boundary changing\\ fields $A\!\to\!M$ \end{tabular}\!\!\right\}
  \ \longrightarrow\ \left\{\!\!\begin{tabular}c
  boundary changing\\ fields $A\!\to\!M$ \end{tabular}\!\!\right\} \,. \ee
This is the analogue of the representation morphism $\r_M$
as defined in formula \Erf1m.

To make this qualitative description precise, let us write out the sewing
constraint and the representation property in a basis and check whether they
agree. We start with the general sewing constraint for
four boundary changing fields on a disk. For two (not necessarily distinct)
boundary conditions $M,N$, let $\Psi^{MN}_{a\alpha}(x)$ denote a boundary 
changing field\foodnode{%
  The boundary fields living on a given boundary $M$ are just those fields 
  that leave the boundary condition invariant. Thus in the present notation 
  they are written as $\Psi^{MM}_{a\alpha}(x)$.}
that is primary \wrt the chiral \alg\ $\CA$ and carries the representation
$a$ of $\CA$. The index $\alpha$ counts its multiplicity. For three 
boundary conditions $M,N,K$, let $\mCb a\alpha b\beta c\gamma\delta{MNK}$ 
be the boundary structure constants appearing in the OPE 
$\Psi^{MN}_{a\alpha}(x)\,\Psi^{NK}_{b\beta}(y)$, defined by a direct
generalisation of the OPE \erf{eq:bope-general}. Then the general sewing
constraint reads \cite{lewe3,bppz2,fffs}
  \be
  \sum_{\varphi=1}^{n_{fM}^{\;\;\; K}} \mCb a\alpha b\beta f\varphi\rho{MNK}\,
  \mCb f\varphi c\gamma d\delta\sigma{MKL}
  = \sum_{\varepsilon=1}^{n_{eN}^{\;\;\; L}}\sumI_e
    \sum_{\varrho=1}^{\N bce} \sum_{\tau=1}^{\N aed}
  \mCb b\beta c\gamma e\varepsilon\varrho{NKL}\, \mCb a\alpha e\varepsilon
  d\delta\tau{MNL}\, \F{a\U b}cdef\tau\varrho\rho\sigma \,.  
  \labl{eq:bope-sewing-ch}
Here $n_{fM}^{\quad K}$ gives the multiplicity of the boundary changing field
$\Psi^{MK}_{f\varphi}$, and similarly for $n_{eN}^{\;\;\; L}$. In fact, these 
are nothing but the annulus coefficients (see section \ref{sec:annulus-pf}),
$n_{iM}^{\;\;\; N}\eq\Ann {iM}N$.

Let us now write out the representation property \erf{1m} in a
basis. To do so define the numbers   
  \bea \begin{picture}(165,68)(0,34)
  \put(0,0)   {\begin{picture}(0,0)(0,0)
              \scalebox{.38}{\includegraphics{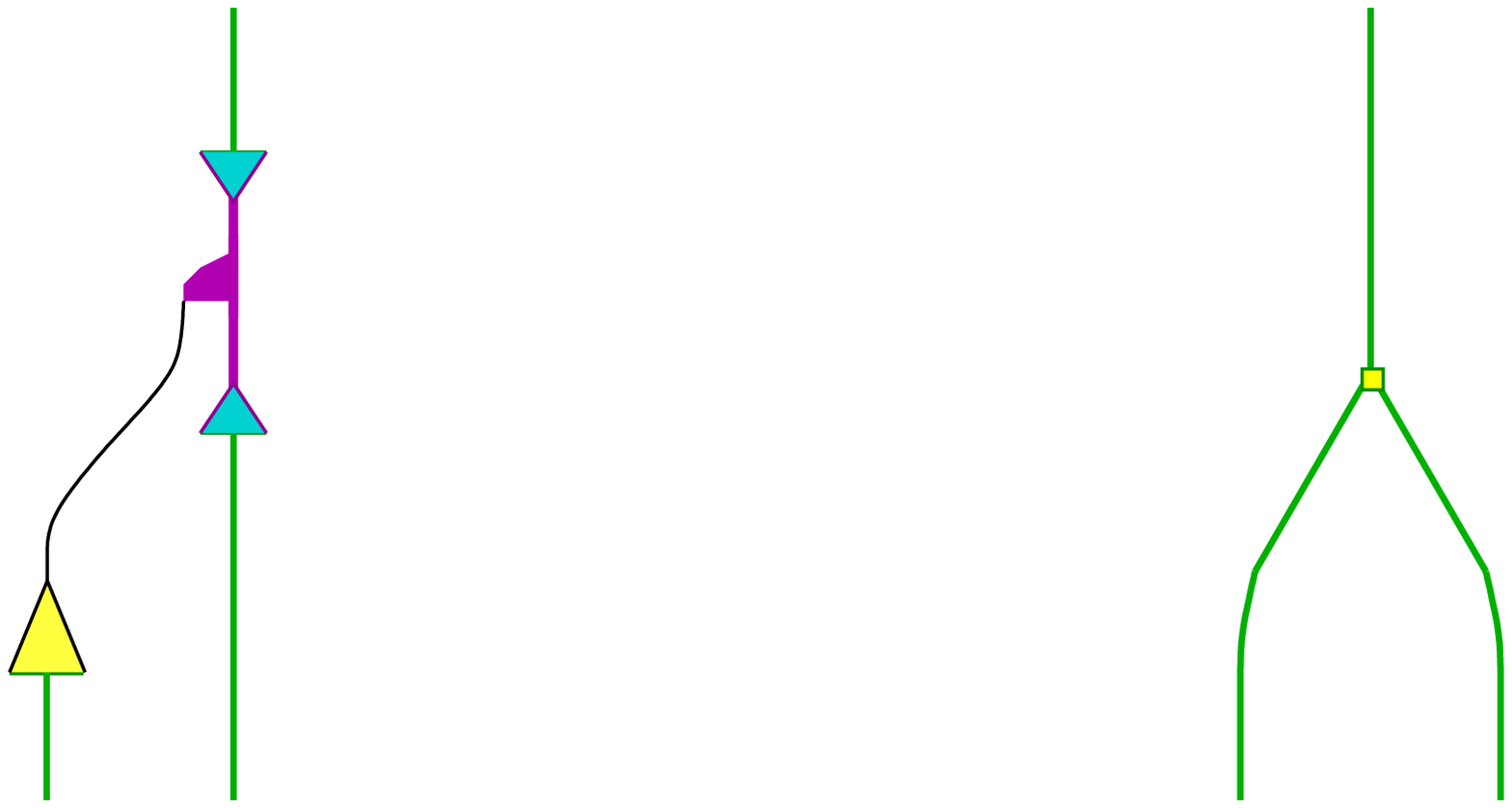}} \end{picture}}
  \put(3.5,17) {\tiny$\alpha$}
  \put(2.5,-7.5)  {\scriptsize$a$}
  \put(24,-7.5) {\scriptsize$i$}
  \put(24,96) {\scriptsize$j$}
  \put(28.5,59.2) {\scriptsize$\r_M$}
  \put(31.2,46) {\tiny$\beta$}
  \put(31.2,71) {\tiny$\bar\gamma$}
  \put(55,44.1)   {$=:\ \dsty \sum_{\delta=1}^{\N ai{\,j}} 
                  \rhobas a\alpha i\beta j\gamma\delta M$}
  \put(139,-7.5){\scriptsize$a$}
  \put(154,96){\scriptsize$j$}
  \put(150,48){\tiny$\delta$}
  \put(170,-7.5){\scriptsize$i$}
  \epicture23 \labl{rhobas}
In terms of these the representation property reads
  \be
  \sum_{\varphi=1}^{\lr f\M}  \m a\alpha b\beta f\varphi\rho\;
  \rhobas f\varphi c\gamma d\delta\sigma M
  \,=\, \sum_{e\in\II} \sum_{\varepsilon=1}^{\lr e\M}\sum_{\varrho=1}^{\N bce}
    \sum_{\tau=1}^{\N aed} \,
  \rhobas b\beta c\gamma e\varepsilon\varrho M\; \rhobas a\alpha e\varepsilon
  d\delta\tau M\; \F{a\U b}cdef\tau\varrho\rho\sigma \,.  
  \labl{eq:rep-in-bas-general}
Thus we see that upon setting
  \be
  \mCb a\alpha b\beta c\gamma\delta{AAA} = 
  \m a\alpha b\beta c\gamma\delta \qquad {\rm and } \qquad
  \mCb a\alpha i\beta j\gamma\delta{AAM} =
  \rhobas a\alpha i\beta j\gamma\delta M \,,
  \labl{eq:r=c}
formula \erf{eq:rep-in-bas-general} becomes the special case 
$M\eq N\eq K\eq A$, $L\eq M$ of \erf{eq:bope-sewing-ch}. As an object in
\calc, the module defined by the boundary condition $M$ is given by
  \be  \M \cong \bigoplus_{i\in\II} 
  \bigoplus_{\alpha=1}^{n_{iA}^{\;\;\;M}} \, U_i \,,\qquad
  {\rm i.e.}\ \ \,\lr i\M = n_{iA}^{\;\;\;M} \,. \ee
Thus the module $M$ consists of all 
representations of the chiral algebra $\CA$ that occur as
boundary changing fields $A\!\rightarrow\!M$, counted with multiplicities.

Together with the results of section \ref{sec:alg2bCFT} we have now learned
the following. Fix an arbitrary boundary condition $A$; this defines
an algebra object. Moreover, every boundary condition (including $A$ itself)
defines a module over this algebra. In other words, in our approach
{\em boundary conditions are labelled by modules of the algebra object\/}
$A$. It also turns out that elementary boundary conditions correspond to
simple $A$-modules, see section \ref{sec:annulus-pf} below. Direct sums of
simple modules, in contrast, correspond to Chan\hy Paton multiplicities,
which play an important role in type I string theories (for a recent review
see \cite{Ansa}). It will be shown in a forthcoming paper that with the help 
of the data provided
by the algebra $A$ one can indeed define structure constants that solve the
sewing constraint \erf{eq:bope-sewing-ch} in full generality
(not merely in the special cases \erf{eq:m=c} and \erf{eq:r=c}).  

There is no intrinsic meaning to the statement that \bc s in a CFT 
correspond to {\em left\/} rather than to {\em right\/} $A$-modules. This 
is just a convention we chose, and everything could as well be formulated 
in terms of right modules. Given a left $A$-module $M\eq(\M,\rho_M)$ we 
can construct a right $A$-module $(\M^\vee\!{,}\,f(\r_M))$, where 
$f(\rho_M)\iN\Hom(\M^\vee\Oti A,\M^\vee)$ is given by
  \be  f(\r_M) = (d_\M \oti \id_{\M^\vee}) \cir 
  (\id_{\M^\vee} \oti \r_M \oti \id_{\M^\vee}) \cir 
  (\id_{\M^\vee} \oti \id_A \oti b_\M) \,.  \ee
It is easy to check that the mapping 
$(\M,\r_M) \,{\mapsto}\,(\M^\vee,f(\r_M))$ is invertible and that
$f(\r_M)$ is a right representation morphism (as defined below \erf{rho}).
The right module $(\M^\vee\!{,}\,f(\r_M))$ describes the same boundary
condition as the left module $(\M,\r_M)$.

\medskip

The notion of an \AB-bimodule, given in definition \ref{def:bimod}, also 
has a CFT interpretation, which we mention briefly here (it will be studied 
in more detail elsewhere). For $A\eq B$, the bimodules correspond to
generalised disorder lines, a notion introduced in \cite{pezu5} 
and studied further in \cite{pezu7,coSc}. For $A\,{\ne}\,B$, bimodules 
describe tensionless interfaces between two CFTs. The CFTs on each 
side of the defect are the ones obtained from the algebra objects $A$ and 
$B$, respectively. Thus they contain the same chiral algebra \CA\ and 
in particular have the same Virasoro central charge. These defects are
special cases of those considered in \cite{bddo,qusc}, where they are treated
as boundary states in the product%
  \foodnode{This product must not be confused with the CFT associated
  to the tensor product of two algebras, see the discussion at
  the beginning of section \ref{sec:sum-prod-op}.}
of the two CFTs (this is known as the folding trick, see e.g.\ 
\cite{osaf2,lelu}).

\subsection{The case ${N_{ij}}^k \iN \{0,1\}$
            and $\dim\,\Hom(U_k,A) \iN \{0,1\}$} \label{sec:N=1-repn}

In this section we study the representation theory
of $A$ under the simplifying assumptions that ${N_{ij}}^k\iN\{0,1\}$
for all $i,j,k\iN\II$ and that $\,\dim\,\Hom(k,A)\iN\{0,1\}$ for all
$k\iN\II$. In contrast, we do {\em not\/} restrict our attention to such
modules which contain simple objects only with multiplicity zero or one.

The notation $\rhobas a\alpha i\beta j\gamma\delta M$ 
introduced in equation \erf{rhobas} to describe the representation
morphism in a basis now reduces to
$\rbas Ma{i\beta}{j\gamma}$, i.e.\
  \bea \begin{picture}(135,60)(0,35)
  \put(0,0)   {\begin{picture}(0,0)(0,0)
              \scalebox{.38}{\includegraphics{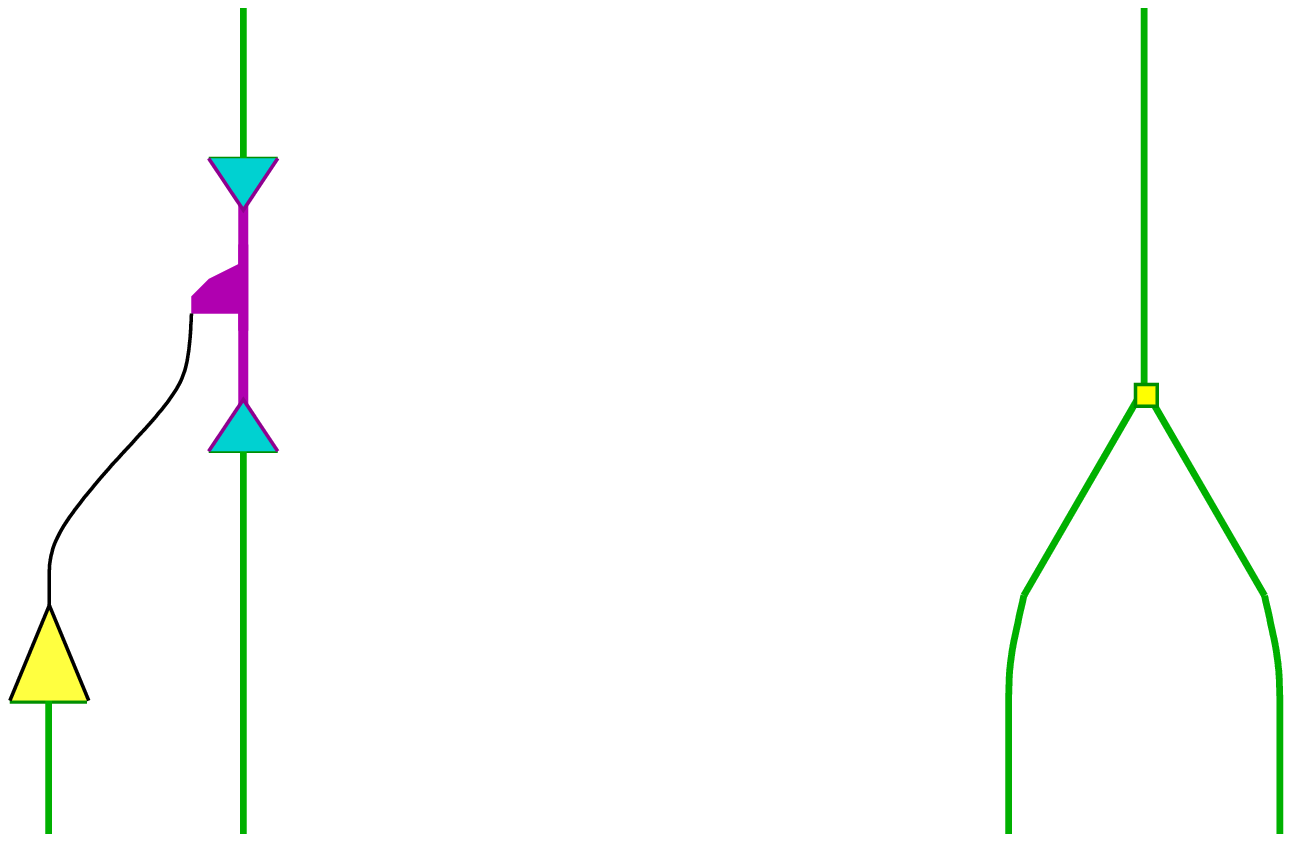}} \end{picture}}
  \put(2.5,-5.5)  {\scriptsize$a$}
  \put(24.8,-7.2) {\scriptsize$i$}
  \put(24.8,96.5) {\scriptsize$j$}
  \put(28.5,59.2) {\scriptsize$\r_M$}
  \put(31.2,44.6) {\tiny$\alpha$}
  \put(31.2,71) {\tiny$\bar\beta$}
  \put(57,44.1)   {$=\;\ \rbas Ma{i\alpha}{j\beta}$}
  \put(108,-5.5){\scriptsize$a$}
  \put(123.8,96.5){\scriptsize$j$}
  \put(139,-7.2){\scriptsize$i$}
  \epicture19 \labl{eq:rep-mat}
When expressed in terms of these numbers, the representation properties
\Erf1m also take an easier form than in formula \erf{eq:rep-in-bas-general}:
  \be
  \rbas M\One{i\alpha}{j\beta} = \delta_{i,j}\, \delta_{\alpha,\beta} 
  \qquad {\rm and} \qquad \sum_{(k\gamma)\In M}
  \rbas Ma{k\gamma}{j\beta}\, \rbas Mb{i\alpha}{k\gamma}\, \Fs abij kc
  = \delta_{c\In A} \, {m_{ab}}^{\!c} \, \rbas Mc{i\alpha}{j\beta}
  \,.  \labl{eq:rep-in-bas}
Solving this polynomial equation is not easy. But fortunately
we can invoke proposition \ref{modsubind}, which states
that every irreducible module is a submodule of an induced module. 
Therefore, instead of attacking \erf{eq:rep-in-bas} directly, we start 
by working out the matrix $\rbas{A{\otimes}i}a{k\alpha}{\ell\beta}$ of 
an induced module $\inda(U_i)$. We choose the bases
  \bea \begin{picture}(285,60)(0,30)
  \put(0,0)   {\begin{picture}(0,0)(0,0)
              \scalebox{.38}{\includegraphics{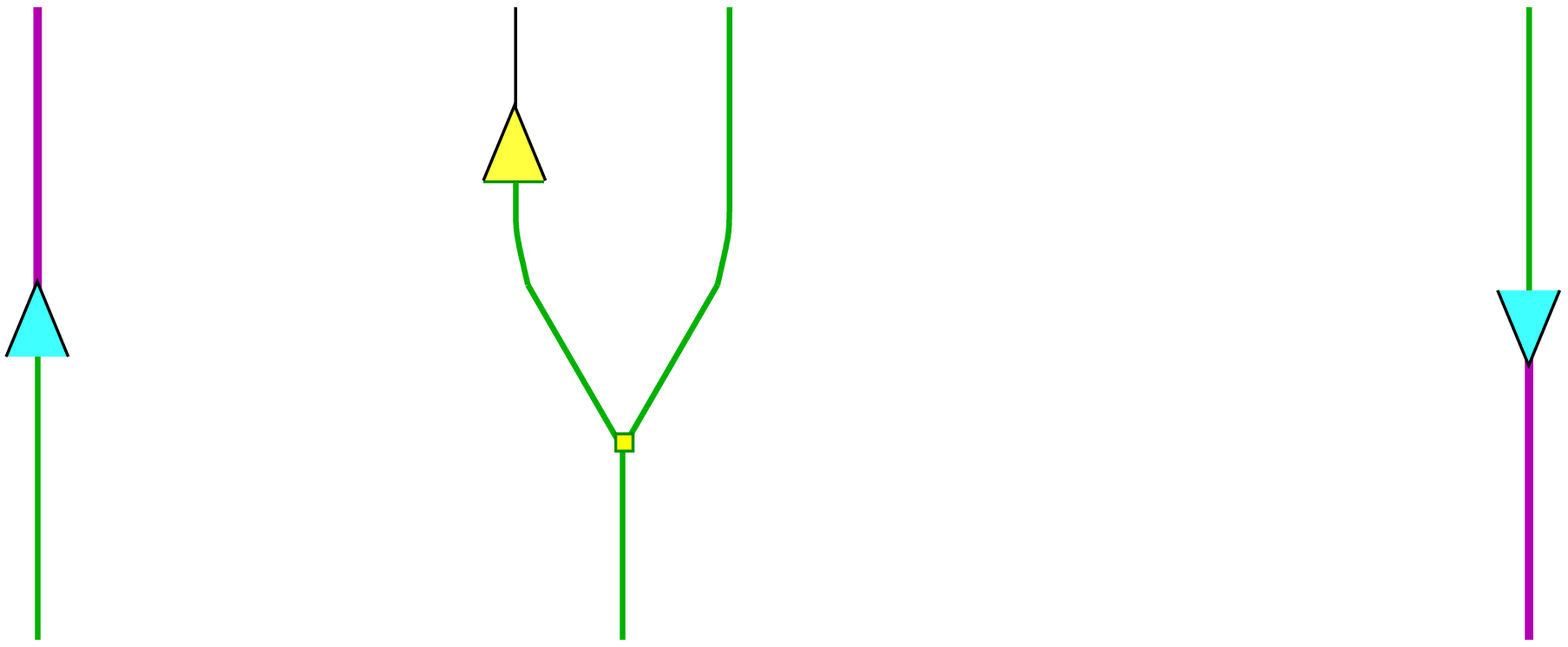}} \end{picture}}
  \put(-3,86) {\scriptsize$A\Oti i$}
  \put(2,-8)  {\scriptsize$k$}
  \put(3,38.2)  {\tiny$a$}
  \put(33.8,37.4) {$=$}
  \put(62,86) {\scriptsize$A$}
  \put(66.1,33.2) {\scriptsize$a$}
  \put(77.1,-8) {\scriptsize$k$}
  \put(91,86) {\scriptsize$i$}
  \put(128.8,37.4){and}
  \put(223.8,37.4){$=$}
  \put(184,-8)  {\scriptsize$A\Oti i$}
  \put(192,86)  {\scriptsize$k$}
  \put(192,41)  {\tiny$a$}
  \put(250,-6)  {\scriptsize$A$}
  \put(251,40)  {\scriptsize$a$}
  \put(266.1,86){\scriptsize$k$}
  \put(280,-6)  {\scriptsize$i$}
  \epicture17 \labl{hom-inda}
in $\Hom(U_k,A{\otimes}U_i)$ and $\Hom(A{\otimes}U_i,U_k)$;
they are dual to each other in the sense of \erf{b-b}. 
The basis vectors are labelled by those $a\IN A$ that occur 
in the fusion $U_k \oti U_{\bar\imath}$. The set is complete, since
$\dim\,\Hom(U_k,A{\otimes}U_i)\eq\dim\,\Hom(U_k{\otimes}U_{\bar\imath},A)$.

To find the representation matrix \erf{eq:rep-mat} in this basis,
one must evaluate
  \bea \begin{picture}(295,113)(0,52)
  \put(0,0)   {\begin{picture}(0,0)(0,0)
              \scalebox{.38}{\includegraphics{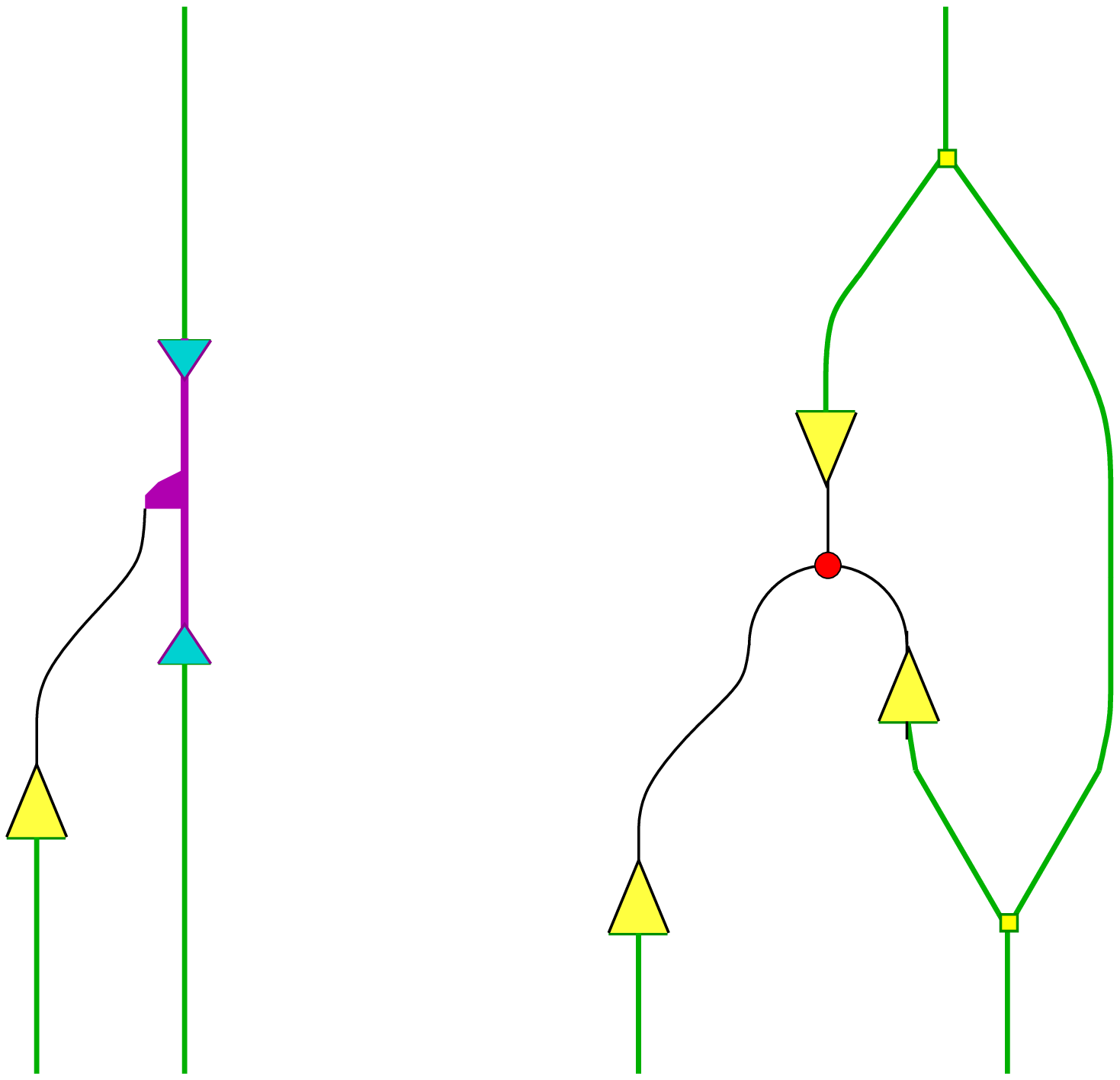}} \end{picture}}
  \put(3.3,-8.2)  {\scriptsize$a$}
  \put(24.1,-8.2) {\scriptsize$k$}
  \put(25,160)    {\scriptsize$l$}
  \put(28.1,84.2) {\scriptsize$\r_{\inda(U_i)}^{}$}
  \put(31.3,62)   {\tiny$b$}
  \put(31.3,103)  {\tiny$c$}
  \put(70.8,74.4) {$=$}
  \put(137,40)    {\scriptsize$b$}
  \put(124.4,112) {\scriptsize$c$}
  \put(90.1,-8.2) {\scriptsize$a$}
  \put(135,160){\scriptsize$l$}
  \put(144.1,-8.2){\scriptsize$k$}
  \put(162,78.2){\scriptsize$i$}
  \put(190.8,74.4){$=\;\ \rbas{A{\otimes}i}a{kb}{lc}$}
  \put(271.1,160){\scriptsize$l$}
  \put(255,-8.2)  {\scriptsize$a$}
  \put(285,-8.2) {\scriptsize$k$}
  \epicture35 \labl{rhoindbas}
Using the definition of $\GG$ \erf{gmat}, this leads to the simple formula
  \be  \rbas{A{\otimes}i}a{kb}{lc} 
  = {m_{ab}}^c \, \Gs abilck \,.  \labl{eq:ind-repn}
In case the induced module $\inda(U_i)$ is a simple module, then we are
already done. If it is not, then we must find a projector decomposition of 
the space $\HomA(\inda(U_i),\inda(U_i))$ to extract the 
simple submodules. This may require some work, but it amounts to
solving only {\em linear\/} equations only.

Suppose now that a simple module $M$
that is not itself an induced module occurs as a submodule in 
  \be  \inda(U_p) \,\cong\, \inda(U_{i_1}) \oplus \cdots \oplus 
  \inda(U_{i_n}) \oplus M \,, \labl{eq:easy-M}
with all induced modules $\inda(U_{i_m})$ being simple.%
  \foodnode{This is of course not the generic situation. In particular, it
  may well happen that two
  unknown simple modules $M$ and $N$ always occur as a pair $M\opl N$, in 
  which case one cannot escape the full projector calculation. But the
  special situation studied here is e.g.\ realised in the \E example that
  we will investigate below.} 
Given \erf{eq:easy-M}, we can extract the representation matrix 
$\rho^M$ with a little extra work from our knowledge of the induced 
representations. To this end, we choose a special basis in the spaces
$\Hom(U_k,A{\otimes}U_p)$ together with a dual basis in 
$\Hom(A{\otimes}U_p,U_k)$. We do so in two steps. First, define the morphisms
  \bea \begin{picture}(400,88)(0,45)
  \put(0,0)   {\begin{picture}(0,0)(0,0)
              \scalebox{.38}{\includegraphics{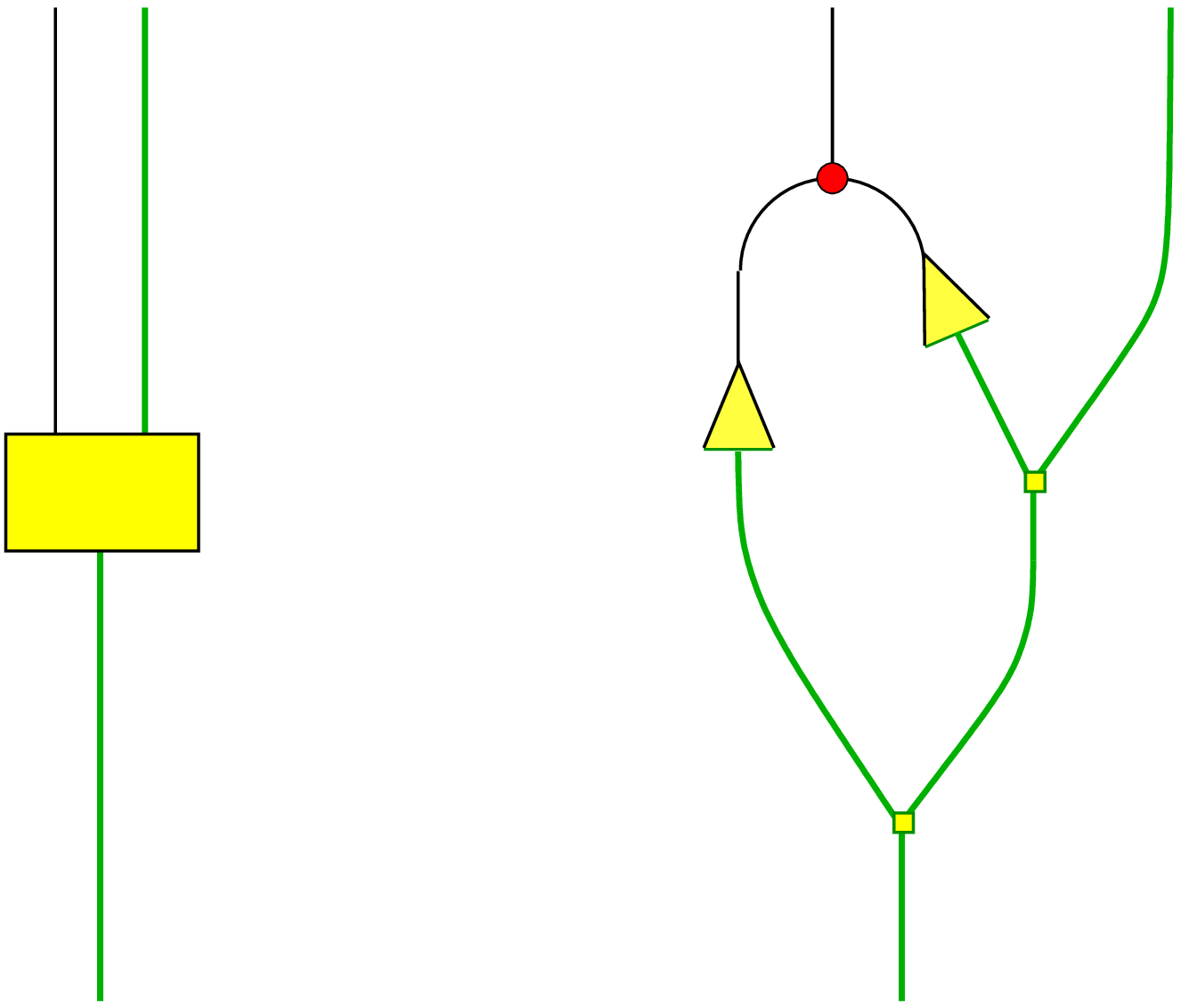}} \end{picture}}
  \put(235,0) {\begin{picture}(0,0)(0,0)
              \scalebox{.38}{\includegraphics{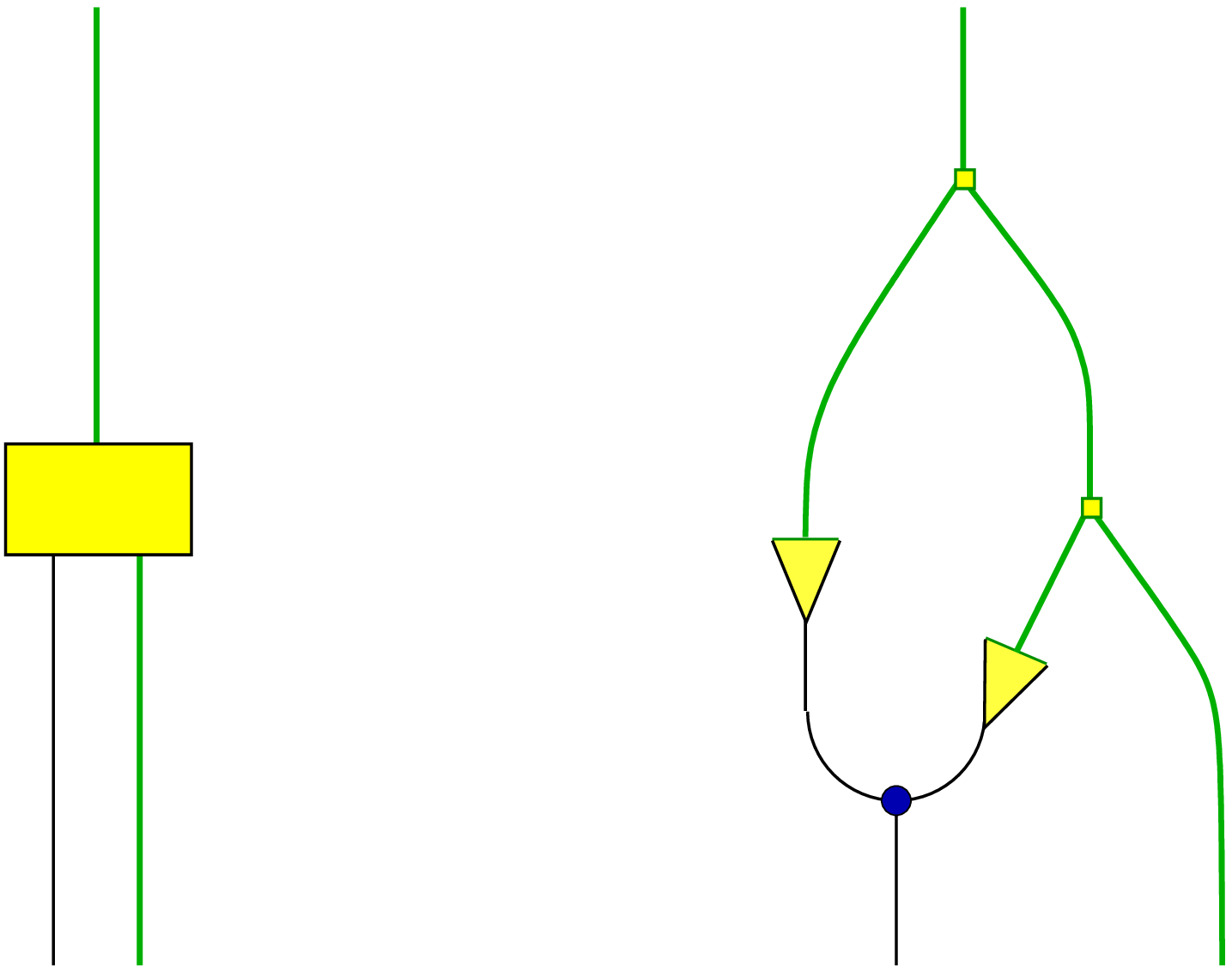}} \end{picture}}
  \put(1.5,128) {\scriptsize$A$}
  \put(2,61) {\scriptsize$X^{A\otimes i}_{k,ab}$}
  \put(10,-8) {\scriptsize$k$}
  \put(16.5,128){\scriptsize$p$}
  \put(49.8,59) {$=$}
  \put(93,38.2) {\scriptsize$a$}
  \put(99.1,128){\scriptsize$A$}
  \put(107.9,-8){\scriptsize$k$}
  \put(126.5,38){\scriptsize$i$}
  \put(117,71){\scriptsize$b$}
  \put(141.9,128){\scriptsize$p$}
  \put(176.8,59){and}
  \put(237,-8) {\scriptsize$A$}
  \put(237,58) {\scriptsize$Y^{A\otimes i}_{k,ab}$}
  \put(246,128) {\scriptsize$k$}
  \put(251.2,-8){\scriptsize$p$}
  \put(270.8,59) {$=\;\dimA$}
  \put(339,82) {\scriptsize$a$}
  \put(345.7,-8){\scriptsize$A$}
  \put(355.9,128){\scriptsize$k$}
  \put(373,82){\scriptsize$i$}
  \put(363,49){\scriptsize$b$}
  \put(389,-8){\scriptsize$p$}
  \epicture31 \label{eq:ind-embed} \labl{Xpk}
These are indeed dual to each other, 
  \be
  Y^{A\otimes j}_{k,cd} \circ X^{A\otimes i}_{k,ab}
  = \delta_{i,j}\, \delta_{a,c}\, \delta_{b,d}\, \id_{U_k} \,.  \ee
This follows from the fact that $\dim\,\HomA(\inda(U_i),\inda(U_j))
\eq\delta_{i,j}$ when $\inda(U_i)$ and $\inda(U_j)$ are simple 
$A$-modules. In particular, the morphism on the \lhs\ of
  \bea \begin{picture}(140,92)(0,46)
  \put(0,0)   {\begin{picture}(0,0)(0,0)
              \scalebox{.38}{\includegraphics{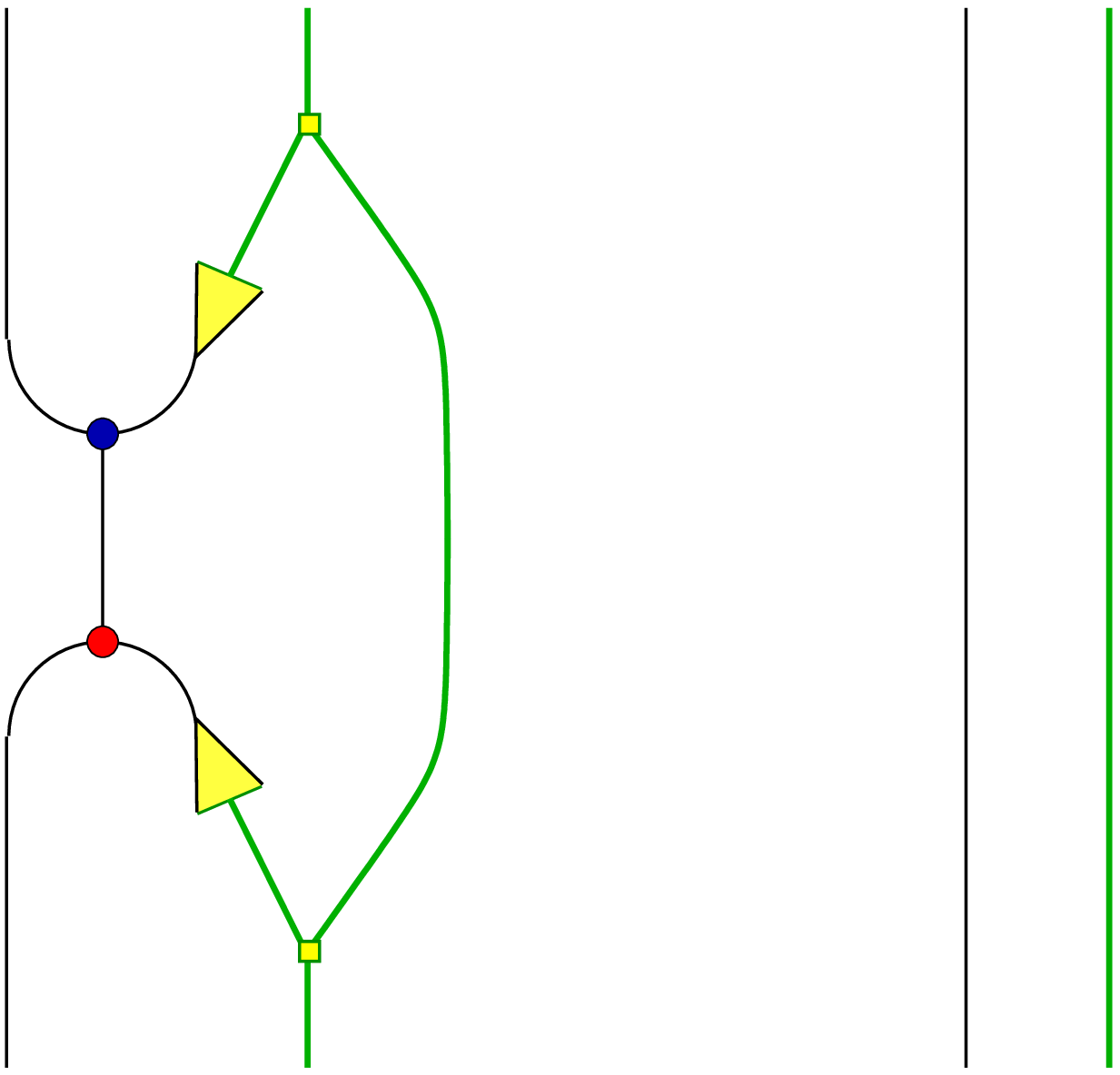}} \end{picture}}
  \put(-2,133.2){\scriptsize$A$}
  \put(-2,-8.2) {\scriptsize$A$}
  \put(36,133.2){\scriptsize$i$}
  \put(36,-8.2) {\scriptsize$i$}
  \put(27,22)   {\scriptsize$b$}
  \put(27,105)  {\scriptsize$c$}
  \put(48,68)   {\scriptsize$p$}
  \put(72.2,63.4) {$=\ \xi_{bc}$}
  \put(113,-8.2) {\scriptsize$A$}
  \put(113,133.2) {\scriptsize$A$}
  \put(133,-8.2) {\scriptsize$i$}
  \put(133,133.2) {\scriptsize$i$}
  \epicture29 \labl{xiidai}
must be a scalar multiple of $\id_{A{\otimes}i}$. Composing both sides of 
this relation with unit and counit, one determines $\xi_{bc}\eq\delta_{b,c}
/\dim(A)$. Another important property of the morphisms $X,Y$ is that they 
allow us to build elements in $\HomA(\inda(U_p),\inda(U_p))$:
  \begin{eqnarray} \begin{picture}(400,172)(17,0)
  \put(135,0) {\begin{picture}(0,0)(0,0)
              \scalebox{.38}{\includegraphics{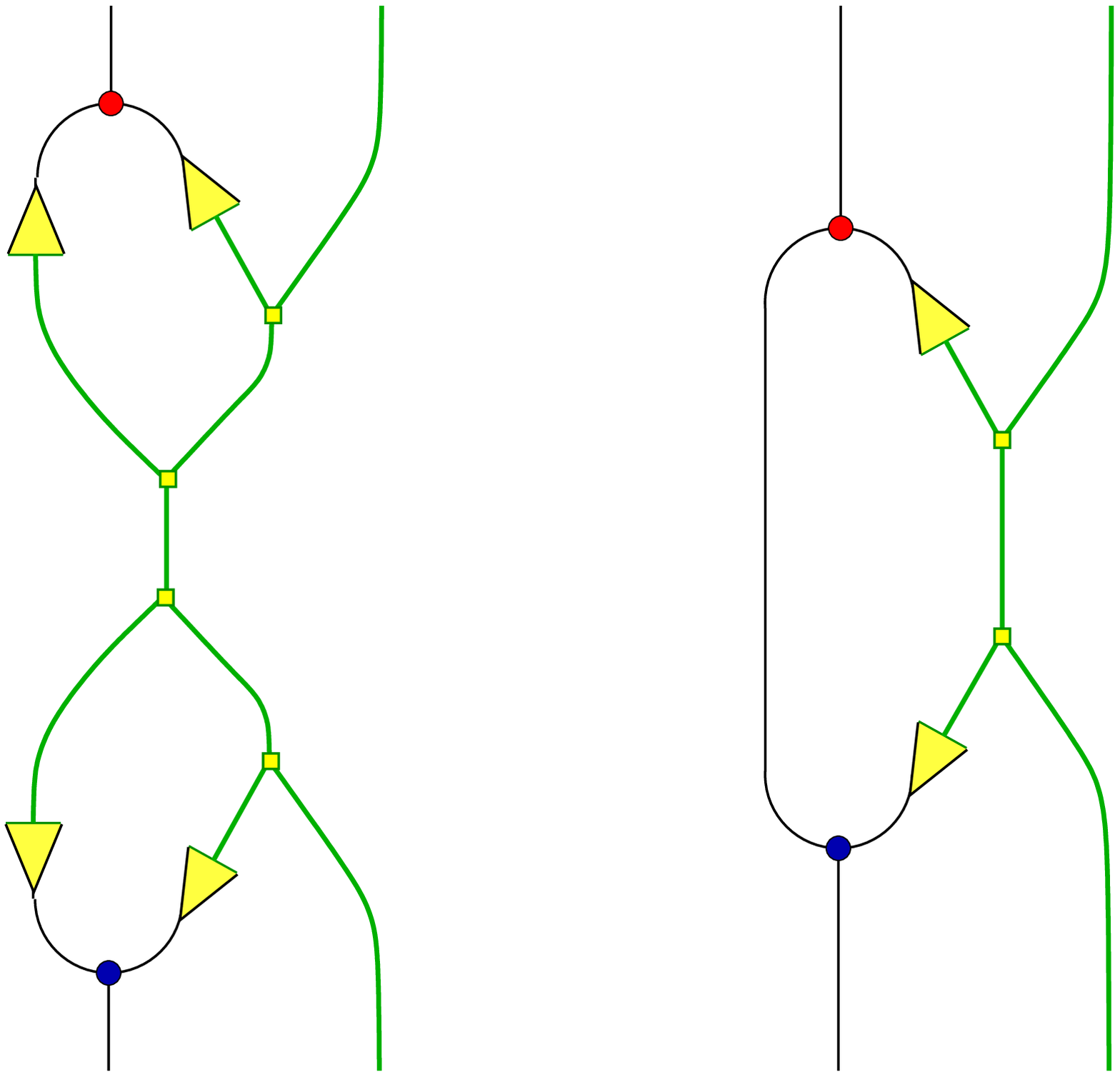}} \end{picture}
    \put(14,169){\scriptsize$A$}
    \put(56,169){\scriptsize$p$}
    \put(12.5,-8){\scriptsize$A$}
    \put(56,-8){\scriptsize$p$}
    \put(32.4,121) {\scriptsize$b$}
    \put(10.4,97) {\scriptsize$a$}
    \put(32,104) {\scriptsize$i$}
    \put(27,78) {\scriptsize$k$}
    \put(7,60) {\scriptsize$a$}
    \put(38,61) {\scriptsize$i$}
    \put(34,42) {\scriptsize$b$}
    % 2nd part
    \put(127,169) {\scriptsize$A$}
    \put(169,169) {\scriptsize$p$}
    \put(125.5,-8) {\scriptsize$A$}
    \put(169,-8) {\scriptsize$p$}
    \put(145,101) {\scriptsize$b$}
    \put(156.5,79.7) {\scriptsize$i$}
    \put(145,59) {\scriptsize$b$}
  }
  \put(-6,80.4)    {$\dsty\sumI_{k,a}X^{A\otimes i}_{k,ab}\circ
                   Y^{A\otimes i}_{k,ab} \ = \;\sumI_{k,a}$}
  \put(191.2,80.4){$=\ \dimA$}
  \put(313.2,80.4){$\in \HomA(\inda(U_p),\inda(U_p))$}
  \end{picture} \nonumber \\[-4.1em]{} \label{sumXY} \\[1.2em] \nonumber
  \end{eqnarray}
The morphisms $X$ are linearly independent, but do not yet form a basis
for all of $\Hom(U_k,A{\otimes}U_p)$, just because the module
$M$ is still missing. So we must find elements
  \be
  X^M_{k\alpha} \in \Hom(U_k,A{\otimes}U_p)
  \qquad {\rm and} \qquad
  Y^M_{k\alpha} \in \Hom(A{\otimes}U_p,U_k) \ee
such that we have a basis in each of these spaces, and such that the $Y$s 
are still dual to the $X$s.  Using this basis, we can apply dominance in 
the form
  \be
  \id_{A{\otimes}U_p}
  = \sumI_{i,k} \sum_{a,b\In A}
  X^{A\otimes i}_{k,ab} \circ Y^{A\otimes i}_{k,ab}
  + \sumI_{k}\sum_{\alpha}
  X^{M}_{k\alpha} \circ Y^{M}_{k\alpha} \,.  \ee
Now since both the \lhs\ and the first term of the \rhs\ of this formula 
are in $\HomA$, the last term is in $\HomA$ as well. To finally obtain the 
representation matrix $\rho^M$, express $X^M$ and $Y^M$ in a basis as
  \bea \begin{picture}(390,60)(0,31)
  \put(0,0)   {\begin{picture}(0,0)(0,0)
              \scalebox{.38}{\includegraphics{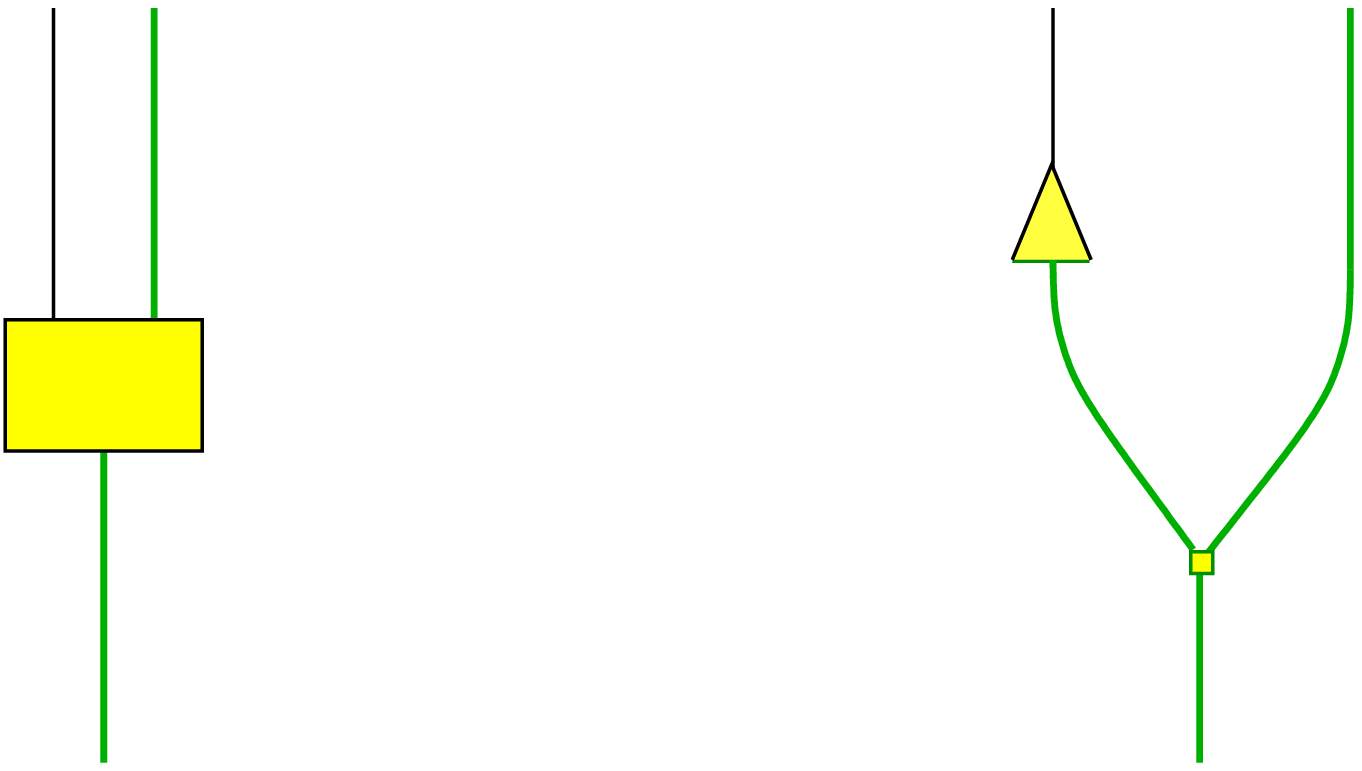}} \end{picture}
    \put(4,87) {\scriptsize$A$}
    \put(15,87) {\scriptsize$p$}
    \put(4,40) {\scriptsize$X^M_{k\alpha}$}
    \put(10,-8.6) {\scriptsize$k$}
    % 2nd part
    \put(113,87) {\scriptsize$A$}
    \put(146,87) {\scriptsize$p$}
    \put(122,38) {\scriptsize$b$}
    \put(129,-8.6) {\scriptsize$k$}
  }
  \put(240,0) {\begin{picture}(0,0)(0,0)
              \scalebox{.38}{\includegraphics{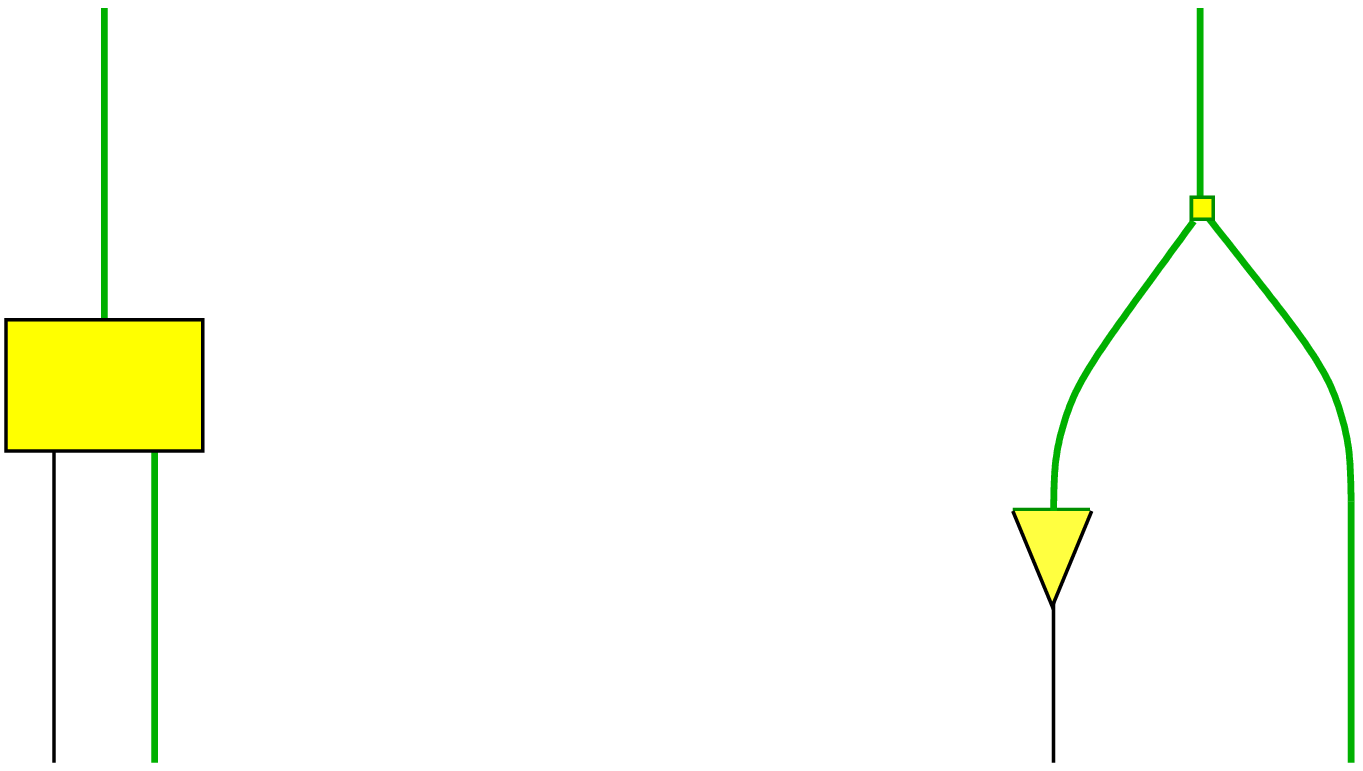}} \end{picture}
    \put(1.4,-8.6) {\scriptsize$A$}
    \put(15,-8.6) {\scriptsize$p$}
    \put(4,40) {\scriptsize$Y^M_{k\alpha}$}
    \put(10,87) {\scriptsize$k$}
    % 2nd part
    \put(112,-8.6) {\scriptsize$A$}
    \put(146,-8.6) {\scriptsize$p$}
    \put(122,38) {\scriptsize$b$}
    \put(129,87) {\scriptsize$k$}
  }
  \put(36.2,40.4){$=\ \dsty\sum_{b\In A}\big(X^M_{k\alpha}\big)_b$}
  \put(185.2,40.4){and}
  \put(276.2,40.4){$=\ \dsty\sum_{b\In A}\big(Y^M_{k\alpha}\big)_b$}
  \epicture20 \labl{eq:XY-in-bas}
The relation
  \bea \begin{picture}(360,94)(0,38)
  \put(37,0)  {\begin{picture}(0,0)(0,0)
              \scalebox{.38}{\includegraphics{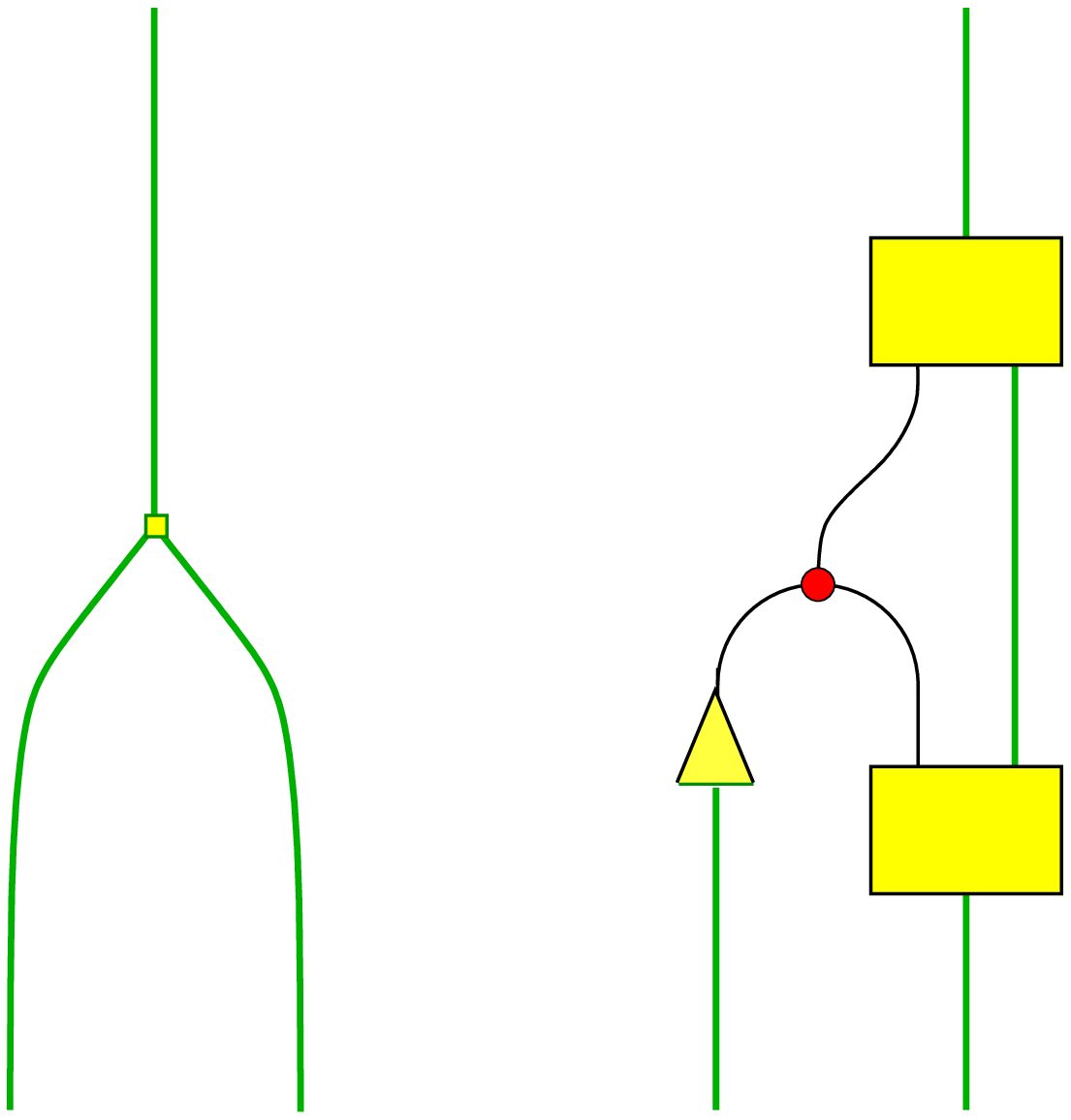}} \end{picture}
   \put(16,131) {\scriptsize$l$}
   \put(0,-8) {\scriptsize$a$}
   \put(32,-8) {\scriptsize$k$}
   %2nd part
   \put(108,131) {\scriptsize$l$}
   \put(102,91) {\scriptsize$Y^M_{\ell\beta}$}
   \put(117,59) {\scriptsize$p$}
   \put(102,31) {\scriptsize$X^M_{k\alpha}$}
   \put(79,-8) {\scriptsize$a$}
   \put(108,-8) {\scriptsize$k$}
   %3rd part
   \put(283,131) {\scriptsize$l$}
   \put(266,101) {\scriptsize$c$}
   \put(310,60) {\scriptsize$p$}
   \put(282,20) {\scriptsize$b$}
   \put(254,-8) {\scriptsize$a$}
   \put(292,-8) {\scriptsize$k$}
  }
  \put(2,55.4)   {$\rbas Ma{k\alpha}{\ell\beta}$}
  \put(88.2,55.4){$=$}
  \put(177.2,55.4){$=\ \dsty\sum_{b,c\In A}\big(X^M_{k\alpha}\big)_b 
                   \big(Y^M_{\ell\beta}\big)_c$}
  \epicture25 \labl{rhoXY}
then determines $\rho^M$ to be
  \be  \rbas Ma{k\alpha}{\ell\beta} =
  \sum_{b,c} \big(X^M_{k\alpha}\big)_b \,
  \big(Y^M_{\ell\beta}\big)_c\, {m_{ab}}^c \,
  \Gs abp\ell ck \,.  \labl{eq:sub-rep-mat}
(Recall that the label $p$ refers to the induced module $\inda(U_p)$ form 
which the simple submodule $M$ has been extracted.) 

\subsubsection{Example: Free boson}

For the free boson, all induced modules turn out to be simple.
(Thus the second half of the arguments above is not needed). To see this,
recall the definition of the \alg\ object $A_{2r}$ in formula
\erf{eq:A2r-cont}. The dimensions of the morphism spaces $\HomA$ are
then given by
  \be
  \dim\,\HomA(\indaz([n]),\indaz([m]))
  = \dim\,\Hom([n],A_{2r}{\otimes}[m])
  = \Big\{ \begin{tabular}{ll}
  $\!1\!$ & if $n{-}m = 0\;{\rm mod}\;2r\,,$\\ $\!0\!$ & otherwise\,.
  \end{tabular} \ee
In particular, indeed every induced module is simple. But not all
of them are distinct. Rather, we have
  \be  \indaz([n])\cong \indaz([n{+}2r])  \ee
for any $n$ (and no further isomorphisms). A set of representatives for
each equivalence class of simple modules is thus given by
  \be
  M_k = \indaz([k]) \qquad {\rm for} \quad
  k\eq0,1,...\,,2r{-}1 \,.  \ee
The multiplication can be read off the formula for induced
modules \erf{eq:ind-repn}:
  \be  \rbas{M_n}{[a]}{[k]}{[k{+}a]} =
  {m_{[a][k{-}n]}}^{[a{+}k{-}n]}\,
  \Gs{[a]}{[k{-}n]}{[n]}{[k{+}a]}{[a{+}k{-}n]}{[k]} = 1 
  \labl{eq:Z2N-repmat}
(and 0 else). In deriving this formula one uses that $a\IN A$ is even 
and that $k{-}n\,{\ge}\, 0$ is a multiple of $2r$ and thus is even, too. 
Note that the 1$\times$1-matrix \FF\ in \erf{eq:Z2N-F} is just a
sign and hence equals its own inverse, $\GG\eq\FF$.  

\subsubsection{Example: \E modular invariant} \label{sec:ex-E7-repn}

For the \E modular invariant we need to determine the representation
theory of the algebra object $A\eq(0)\opl(8)\opl(16)$ in the modular 
category formed by the integrable highest weight representations of 
$\su(2)_{16}$. 

As a first step we work out the embedding structure of the induced
modules. The following table gives $\dim\,\HomA(\inda(i),\inda(j))$ 
for $i,j\eq0,1,...\,,8$. 
(For easier reading, in the table zero entries are indicated by a dot.)
  \be
\begin{tabular}{c|ccccccccc}
   & 0 & 1 & 2 & 3 & 4 & 5 & 6 & 7 & 8 \\ \hline &\\[-.95em]
 0 & 1 & $\cdot$ & $\cdot$ & $\cdot$ & $\cdot$ & $\cdot$ & $\cdot$ & $\cdot$ & 1 \\
 1 & $\cdot$ & 1 & $\cdot$ & $\cdot$ & $\cdot$ & $\cdot$ & $\cdot$ & 1 & $\cdot$ \\
 2 & $\cdot$ & $\cdot$ & 1 & $\cdot$ & $\cdot$ & $\cdot$ & 1 & $\cdot$ & 1 \\
 3 & $\cdot$ & $\cdot$ & $\cdot$ & 1 & $\cdot$ & 1 & $\cdot$ & 1 & $\cdot$ \\
 4 & $\cdot$ & $\cdot$ & $\cdot$ & $\cdot$ & 2 & $\cdot$ & 1 & $\cdot$ & 1 \\
 5 & $\cdot$ & $\cdot$ & $\cdot$ & 1 & $\cdot$ & 2 & $\cdot$ & 1 & $\cdot$ \\
 6 & $\cdot$ & $\cdot$ & 1 & $\cdot$ & 1 & $\cdot$ & 2 & $\cdot$ & 1 \\
 7 & $\cdot$ & 1 & $\cdot$ & 1 & $\cdot$ & 1 & $\cdot$ & 2 & $\cdot$ \\
 8 & 1 & $\cdot$ & 1 & $\cdot$ & 1 & $\cdot$ & 1 & $\cdot$ & 3 
\end{tabular}
  \labl{123}
All other cases follow from the table via $\inda(r)\,{\cong}\,\inda(16{-}r)$,
an isomorphism due to the simple current $(16)$. 
{}From the diagonal entries of \erf{123} we see that the four 
induced modules
  \be \inda(0) \, , \quad \inda(1) \, , \quad 
  \inda(2) \, , \quad \inda(3) \, .  \ee
are simple, while the other induced modules decompose as
  \be\begin{array}{l}
  \inda(4)\cong P\opl Q \, , \qquad 
  \inda(5) \cong \inda(3) \opl R \, , \qquad 
  \inda(6) \cong \inda(2) \opl Q \, , \qquad \\{}\\[-.6em]
  \inda(7)\cong \inda(1) \opl \inda(3) \, , \qquad \quad
  \inda(8) \cong \inda(0) \opl  \inda(2) \opl P \, ,
  \end{array}\ee
with simple $A$-modules $P,Q,R$ that are not isomorphic to any induced module. 
As objects in \calc, these simple modules decompose into simple objects as
  \be
  \dot P \cong (4) \opl (8) \opl (12) \, , \quad\;\
  \dot Q \cong (4) \opl (6) \opl (10) \opl (12) \, , \quad\;\
  \dot R \cong (5) \opl (11) \, .  \ee

In agreement with the discussion above we can extract $P$ from $\inda(8)$, 
$Q$ from $\inda(6)$ and $R$ from $\inda(5)$. Let us start with $\inda(8)$.
Its decomposition into simple objects of \calc\ reads
  \be  \begin{array}{rccccccccccccccr} A{\otimes}(8) \;\cong \
  & & (0)\!\! & & & & & \Opl & (8) & \Opl & & & & & \!\!\!\!(16) & 
      \ \ [\,{\rm from}\; A{\otimes}(0)\,]\\[.3em]
  & \Opl & & \!\!(2)\!\! & & \Opl & \!(6) & \Opl & (8) & \Opl & (10)\! &
       \Opl & &\!\!\!\! (14)& &    [\,{\rm from}\; A{\otimes}(2)\,]\\[.3em]
  & \Opl & & & \!\!(4)\!\! & & & \Opl & (8) & \Opl & & & \!(12) & & &    
      [\,{\rm from}\; P\,]
  \end{array}\ee
One verifies that to complete the bases in the spaces
$\Hom((4),A{\otimes}(8))$, $\Hom((8),A{\otimes}(8))$ and
$\Hom((12),A{\otimes}(8))$ one may choose the elements
  \bea \begin{picture}(375,69)(0,43)
  \put(37,0)  {\begin{picture}(0,0)(0,0)
              \scalebox{.38}{\includegraphics{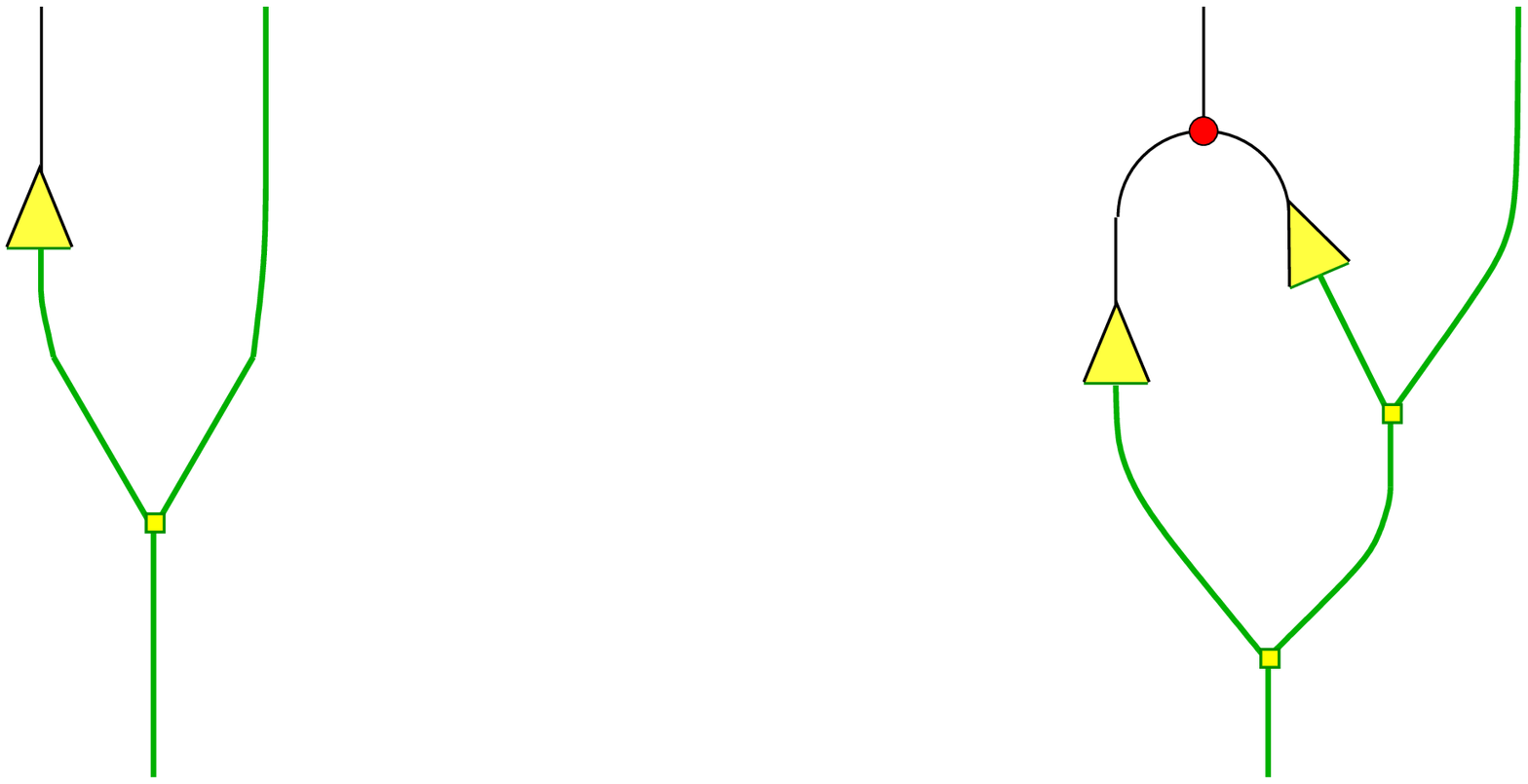}} \end{picture}
   \put(3,108) {\scriptsize$A$}
   \put(0,44) {\scriptsize$(8)$}
   \put(28,108) {\scriptsize$(8)$}
   \put(17,-8) {\scriptsize$(4)$}
   %2nd part
   \put(156,108) {\scriptsize$A$}
   \put(195,108) {\scriptsize$(8)$}
   \put(169,56) {\scriptsize$(8)$}
   \put(147,23) {\scriptsize$(8)$}
   \put(182,25) {\scriptsize$(4)$}
   \put(164,-8) {\scriptsize$(8)$}
   %3rd part
   \put(297,108) {\scriptsize$A$}
   \put(323,108) {\scriptsize$(8)$}
   \put(295,44) {\scriptsize$(8)$}
   \put(306,-8) {\scriptsize$(12)$}
  }
  \put(-12.2,55.4){$X^P_{(4)}\ =$}
  \put(111.2,55.4){$X^P_{(8)}\ =\ \xi^{-1}$}
  \put(281.2,55.4){$X^P_{(12)}\ =$}
  \epicture26 \labl{XP4812}
where the normalisation constant $\xi$ is defined
implicitly by the condition $Y^P_{(8)} \cir X^P_{(8)} \eq \id_{(8)}$,
and the dual basis elements are
  \bea \begin{picture}(370,75)(0,40)
  \put(37,0)  {\begin{picture}(0,0)(0,0)
              \scalebox{.38}{\includegraphics{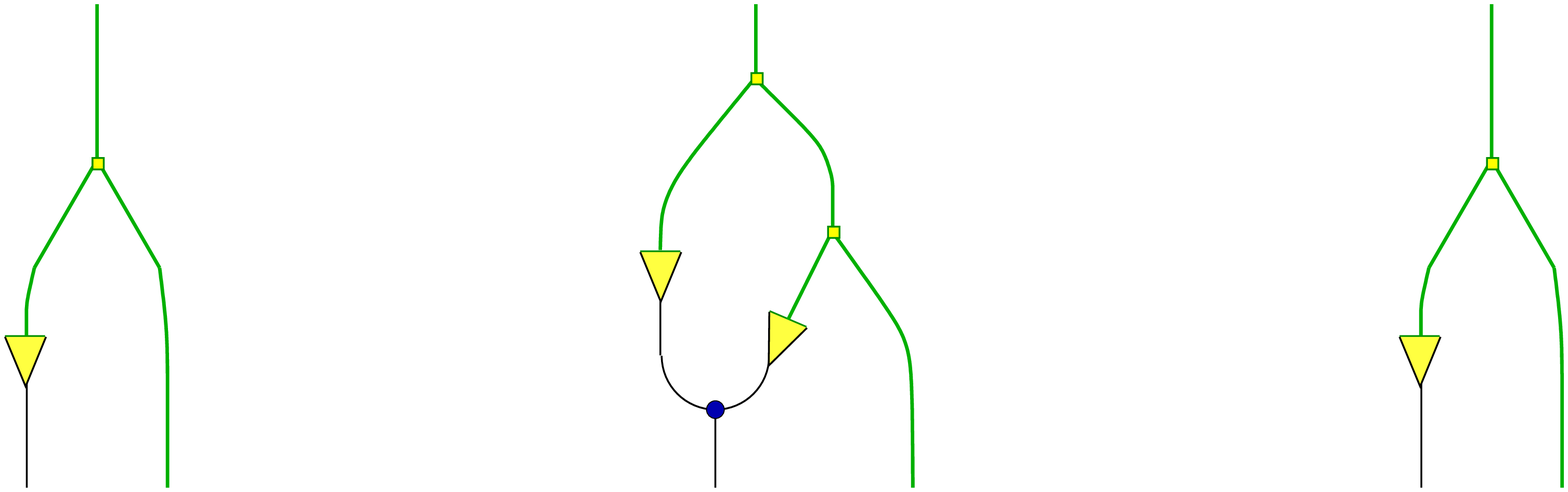}} \end{picture}
   \put(3,-8) {\scriptsize$A$}
   \put(0,57) {\scriptsize$(8)$}
   \put(28,-8) {\scriptsize$(8)$}
   \put(17,108) {\scriptsize$(4)$}
   %2nd part
   \put(146,-8) {\scriptsize$A$}
   \put(185,-8) {\scriptsize$(8)$}
   \put(159,46) {\scriptsize$(8)$}
   \put(137,78) {\scriptsize$(8)$}
   \put(172,76) {\scriptsize$(4)$}
   \put(154,108) {\scriptsize$(8)$}
   %3rd part
   \put(297,-8) {\scriptsize$A$}
   \put(323,-8) {\scriptsize$(8)$}
   \put(295,57) {\scriptsize$(8)$}
   \put(306,108) {\scriptsize$(12)$}
  }
  \put(-6.2,50.4) {$Y^P_{(4)}\ =$}
  \put(126.2,50.4){$Y^P_{(8)}\ =$}
  \put(284.2,50.4){$Y^P_{(12)}\ =$}
  \epicture26 \labl{eq:mod-X-duals}
The only non-trivial case is $X^P_{(8)}$. The calculation that this indeed extends 
$X^{A{\otimes}(0)}_{(8)}$ and $X^{A{\otimes}(2)}_{(8)}$
(in the conventions defined in \erf{eq:ind-embed}) to a basis
of $\Hom((8),A{\otimes}(8))$ with dual basis given by
$Y^{A{\otimes}(0)}_{(8)}$, $Y^{A{\otimes}(2)}_{(8)}$ and $Y^P_{(8)}$
amounts to the observation that 
$\dim\,\HomA(\inda(4),\inda(r))\eq0$ for $r\eq0,2$.

The same procedure works to extract $Q$ from $\inda(6)$, 
giving the additional basis elements
  \begin{eqnarray} \begin{picture}(435,112)(0,0)
  \put(37,0)  {\begin{picture}(0,0)(0,0)
              \scalebox{.38}{\includegraphics{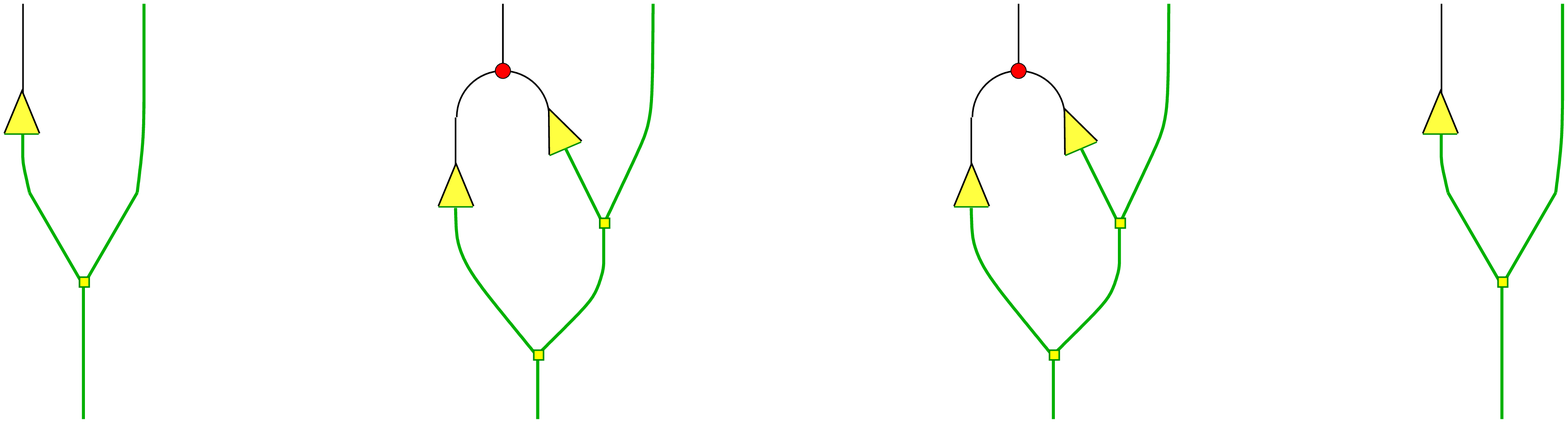}} \end{picture}
  %1st part
   \put(3,108) {\scriptsize$A$}
   \put(31,108) {\scriptsize$(6)$}
   \put(3,42) {\scriptsize$(8)$}
   \put(16,-8) {\scriptsize$(4)$}
  %2nd part
   \put(120,108) {\scriptsize$A$}
   \put(154,108) {\scriptsize$(6)$}
   \put(133,54) {\scriptsize$(8)$}
   \put(111,24) {\scriptsize$(8)$}
   \put(143,23) {\scriptsize$(4)$}
   \put(127,-8) {\scriptsize$(6)$}
  %3rd part
   \put(347,0){
     \put(3,108) {\scriptsize$A$}
     \put(31,108) {\scriptsize$(6)$}
     \put(3,42) {\scriptsize$(8)$}
     \put(16,-8) {\scriptsize$(12)$}
   }
  %4th part
   \put(127,0){
     \put(120,108) {\scriptsize$A$}
     \put(154,108) {\scriptsize$(6)$}
     \put(133,54) {\scriptsize$(8)$}
     \put(111,24) {\scriptsize$(8)$}
     \put(143,23) {\scriptsize$(4)$}
     \put(127,-8) {\scriptsize$(10)$}
    }
  }
  \put(-1.4,55.4) {$X^Q_{(4)}\,=$}
  \put(91.2,55.4) {$\xi_1\,X^Q_{(6)}\,=$}
  \put(215.2,55.4){$\xi_2\,X^Q_{(10)}\,=$}
  \put(344.2,55.4){$X^Q_{(12)}\,=$}
  \end{picture} \nonumber \\[-2.9em]{} \label{XQ461012} \\[.5em] \nonumber 
  \end{eqnarray}
with duals defined in the same manner as in \erf{eq:mod-X-duals} and
$\xi_1, \xi_2$ defined implicitly by the conditions $Y^Q_{(6)}\cir X^Q_{(6)}
 \eq \id_{(6)}$ and $Y^Q_{(10)} \cir X^Q_{(10)} \eq \id_{(10)}$.

For the module $R$ this short-cut does not work, and we need to solve a 
linear system to complete the basis and its dual. But since both
morphism spaces (embedding $(5)$ or $(11)$ into $A{\otimes}(5)$) are 
two-di\-mensional, we can easily write down (one choice for) the 
missing vectors:
  \bea \begin{picture}(425,64)(0,32)
  \put(57,0)  {\begin{picture}(0,0)(0,0)
              \scalebox{.38}{\includegraphics{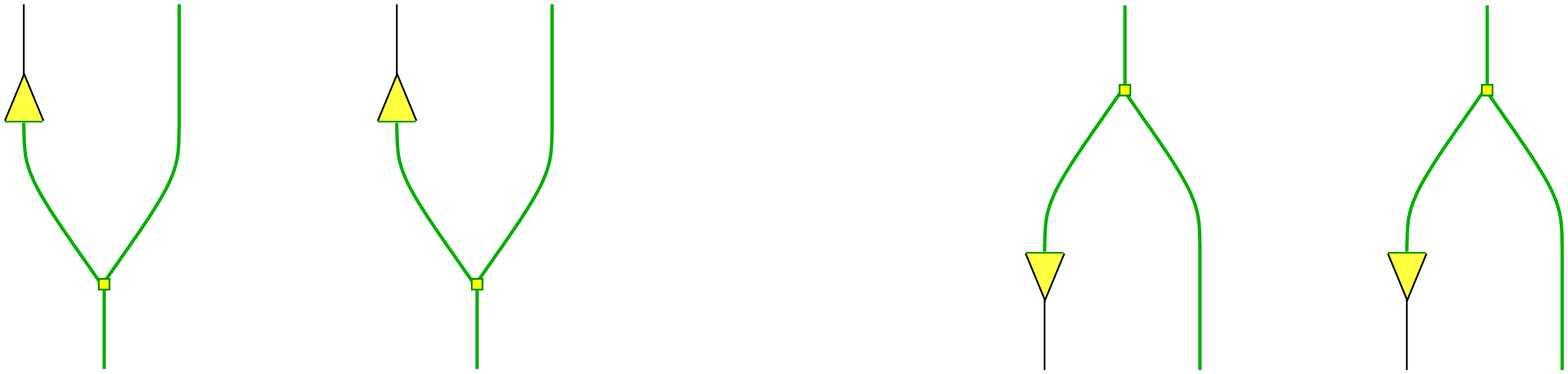}} \end{picture}
   \put(2,89) {\scriptsize$A$}
   \put(34,89) {\scriptsize$(5)$}
   \put(2,28) {\scriptsize$(0)$}
   \put(18,-8) {\scriptsize$(5)$}
   % part 2
   \put(86,89) {\scriptsize$A$}
   \put(118,89) {\scriptsize$(5)$}
   \put(88,28) {\scriptsize$(8)$}
   \put(100,-8) {\scriptsize$(5)$}
   % part 3
   \put(245,89) {\scriptsize$(5)$}
   \put(228.9,50) {\scriptsize$(0)$}
   \put(231,-8) {\scriptsize$A$}
   \put(262,-8) {\scriptsize$(5)$}
   % part 4
   \put(327,89) {\scriptsize$(5)$}
   \put(311,50) {\scriptsize$(8)$}
   \put(312,-8) {\scriptsize$A$}
   \put(344,-8) {\scriptsize$(5)$}
}
  \put(-1.4,42.4) {$X^R_{(5)}\;=\;u$}
  \put(113.2,42.4){$+\;\ v$}
  \put(231.2,42.4){$Y^R_{(5)}\;=\;x$}
  \put(339.8,42.4){$+\;\ y$}
  \epicture18 \labl{XYR55}
where, in the notation \erf{eq:XY-in-bas} from above,
  \be
  u = (Y^{A{\otimes}(3)}_{(5)})_{(8)} \,,\quad\
  v = - (Y^{A{\otimes}(3)}_{(5)})_{(0)} \,,\quad\
  x = (X^{A{\otimes}(3)}_{(5)})_{(8)} \,,\quad\
  y = - (X^{A{\otimes}(3)}_{(5)})_{(0)} \,.  \ee
The morphisms $X^R_{(11)}$ and $Y^R_{(11)}$ 
can be determined in the same way.

We have now gathered the necessary information to give the 
representation matrix of the simple $A$-modules $P,Q,R$ via equation 
\erf{eq:sub-rep-mat}.  As an additional check of the result, we 
verified numerically for a random sample of cases that these 
representation matrices indeed solve the non-linear constraint 
\erf{eq:rep-in-bas} for $\su(2)_{16}$.

%%%%%%%%%%%%%%%%%%%%%%%%%%%%%%%%%%%%%%%%%%%%%%%%%%%%%%%%%%%%%%%%%%%%%%%%
\newpage

\sect{Partition functions} 

%%%%%%%%%%%%%%%%%%%%%%%%%%%%%%%%%%%%%%%%%%%%%%%%%%%%%%%%%%%%%%%%%%%%%%%%

\label{sec:parfu-gen}

Before proceeding, let us recall a few aspects of our general philosophy.
The construction we want to implement can be summarised as
  \be
  \fbox{\rm chiral~data} \,\ + \,\ \fbox{\rm symm.~special~Frobenius~algebra}
  \quad \longrightarrow \quad  \fbox{\rm full~CFT} \;\ \labl{gen-const}
By chiral data we mean the representation theory of the chiral algebra and 
the conformal blocks of a rational \cft. The corresponding modular category 
contains strictly less information than the chiral data. 
Roughly speaking, the category encodes only the monodromies of
the blocks, but not their functional dependence on the insertion points and
the moduli of the world sheet or the information which state of a given
representation of the chiral \alg\ is inserted.%
  \foodnode{More precisely, the modular tensor category encodes the
  information about the conformal blocks as abstract vector bundles $\calv$
  over the moduli space (which for $m$ insertion points labelled by
  $\vec\imath\,{:=}\,(i_1,i_2,...\,,i_m)\iN\I^m$ is 
  the moduli space $\calm_{g,m}$ of complex curves
  of genus $g$ with $m$ marked points). There are actually two different
  formulations of TFT: The topological modular functor works with vector
  spaces and provides a representation of the mapping class group. The complex
  modular functor gives a local system over the moduli space $\calm_{m,g}$.
  The two descriptions are related by a Riemann\hy Hilbert problem, so that
  they contain equivalent information \cite{BAki}.
  The full chiral data provide much more: They specify
  an embedding of $\calv_{\vec\imath}$ in the trivial bundle over
  $\calm_{g,m}$ whose fiber is the algebraic dual of the tensor product
  $\calh_{i_1}\oti\calh_{i_2}\,\Oti\cdots\Oti\,\calh_{i_m}$
  of the vector spaces underlying the irreducible representations of
  the chiral algebra. This information allows to obtain the values of
  conformal blocks on all chiral states.}
But it turns out that much important information about a CFT, like its
field content, its boundary conditions and defect lines, the OPE structure 
constants, and the consistency of these data with factorisation, can be 
discussed entirely at the level of the modular category. It is a strength 
of the present approach that, in a sense, in dealing with the modular 
category \calc\ one `forgets' exactly the right amount
of complexity of the chiral data to render those problems tractable.  

In order to select which full CFT is to be constructed from a given
set of chiral data, additional input is required. In our approach this 
input is a symmetric special Frobenius algebra object $A$ in \calc.
The full CFT that one obtains this way -- the one indicated on the \rhs\ of 
\erf{gen-const} -- will be specified in terms of its correlation functions. 
To this end, the correlators on an arbitrary world sheet $\rmX$ are
expressed as specific elements in the vector spaces of conformal blocks on
the complex double $\hat\rmX$ of $\rmX$. Such an element is described by
a ribbon graph in a three-manifold $M_\rmX$, called the {\em connecting
three-manifold\/} whose boundary $\partial M_\rmX$
is equal to the double $\hat \rmX$. Which three-manifold $M_\rmX$  
is to be used, and which ribbon graph in $M_\rmX$, was established for the 
Cardy case in \cite{fffs2,fffs3} (for all possible world sheets), and
described in the general case in \cite{fuRs} (with restriction to
orientable world sheets).

The idea to study a chiral CFT as the edge system of a (topological) theory
on a three-dimensional manifold is quite natural and has been put forward
in various guises, see for instance \cite{mose4,hora7,feko}. It finds
a physical realisation in the CFT description of quantum Hall fluids in 
the scaling limit (see \cite{fpsw} for a review). By exploiting chiral CFT
on the complex double $\hat\rmX$, this relationship can be used to study
{\em full\/} CFT on the world sheet $\rmX$.

In the following we present this construction in the form that it takes for
orientable world sheets without field insertions. Afterwards we further
specialise to the cases where the world sheet is a torus or an annulus.

\subsection{The connecting manifold and the ribbon graph}
\label{sec:con-mf-rib}

Having fixed a modular category \calc\ and a symmetric special Frobenius 
algebra $A$ in \calc, to construct the correlator on a world sheet without 
field insertions we must in addition provide the following data:
\\[.3em]
\nxt An orientable world sheet $\rmX$, possibly with boundaries;
\\[.3em] 
\nxt an orientation on $\rmX$; 
\\[.3em] 
\nxt for each component of the boundary $\partial \rmX$ an $A$-module 
that specifies the boundary condition.
\vskip.4em

Here an important point is that orientability of the world sheet
is not sufficient. Rather, to determine the correlator uniquely we 
need to select an actual {\em orientation\/}.
For the correlator to be defined unambiguously without specifying
the orientation, more structure than just a symmetric special
Frobenius algebra is needed. This is linked to the fact that there exist
CFTs which cannot be defined on non-orientable world sheets while still
preserving the chiral \alg\ \CA, most notably those whose torus 
partition function is not symmetric, $Z_{ij}\,{\ne}\,Z_{ji}$.
Since also such modular invariants can be described in terms of a
Frobenius algebra, some {\em extra structure\/} is needed to
ensure that the theory is consistent also on non-orientable world
sheets. It turns out that the relevant extra structure is given by
a {\em conjugation\/} on $A$, which furnishes an isomorphism between
$A$ and $A_{\rm op}$ and may be thought of as providing a square
root of the twist. Not every symmetric special Frobenius algebra 
possesses such an extra structure and for those who do, the conjugation 
need not be unique. Moreover, the number of conjugations is {\em not\/} 
Morita invariant.  We will return to this issue in a future publication. 

The second important aspect of the prescription is that boundary 
conditions are labelled by $A$-modules. We will prove below that 
{\em simple\/} $A$-modules provide a complete set of {\em elementary\/} 
boundary conditions. This implies in particular that
the number of elementary boundary conditions is the same as the
number of Ishibashi states in the (dual of the) bulk state space. 
A general $A$-module is a direct sum of simple modules and corresponds
to a superposition of elementary boundary conditions.

All these boundary conditions preserve the chiral algebra \CA\ whose \rep\ 
category is the modular category \calc. Recall that \CA\ is not necessarily 
the maximally extended chiral \alg\ -- in the most extreme case, it is just 
the Virasoro algebra. As a consequence, our formalism covers symmetry 
breaking boundary conditions as well, as long as the 
subalgebra of the chiral algebra that is preserved by 
all boundary conditions under study is still rational.

\medskip

The complex double $\hat \rmX$ of an orientable world sheet $\rmX$ can be
obtained by taking two disjoint copies of $\rmX$, with opposite orientation,
and gluing them together along the boundary:
  \be  \hat \rmX := \big( \rmX \times \{-1,1\} ) /_{\dsty\sim}
  \qquad {\rm with} \quad
  (x,1) \,{\sim}\, (x,-1) \quad {\rm iff} \quad x\iN\partial \rmX \,.  \ee
The connecting three-manifold $M_\rmX$ is then 
the following natural interval bundle over $\rmX$ \cite{fffs2}:
  \be  M_\rmX := \big( \rmX \,{\times}\, [-1,1] \big) /_{\dsty\sim}\,; \ee
the equivalence relation $\sim$ now identifies the values $t$ and 
$-t$ in all intervals $\{x\}\,{\times}\,[-1,1]$ that lie above
a boundary point of $\rmX$, i.e.\ 
  \be
  (x,t) \sim (x,-t) \quad {\rm for~all} \quad
  x\iN\partial \rmX \quad{\rm and~all} \quad t\iN [-1,1] \,.  \labl{xtxt}
One quickly checks that in this manner one indeed achieves
$\partial M_\rmX\eq\hat \rmX$. Moreover,
the world sheet $\rmX$ is naturally embedded in $M_\rmX$, via the map
  \be \iota :\quad \rmX \to M_\rmX \,, \quad x \mapsto (x,0) \,.  \labl{iot}
Thereby $\rmX$ is a retract of $M_\rmX$, and conversely the connecting 
manifold $M_\rmX$ can be regarded as a fattening of the world sheet (and 
hence in particular its construction does not introduce any additional 
homotopical information).  In the sequel we will always think of the 
world sheet $\rmX$ as being embedded in $M_\rmX$ in this fashion. Via 
the embedding \erf{iot} each boundary component of $\rmX$ coincides with
a circular line of singular points of $M_\rmX$ that results from the 
fixed points under the identification \erf{xtxt}.

We now describe how the ribbon graph in $M_\rmX$ is constructed.
This construction involves several choices. As we will show afterwards,
the invariant associated to the graph is independent of all these choices.
\\[-1.4em]{}
\noindent\begin{itemize} 
\item[\Nxt]
First choose a triangulation of the world sheet $\rmX$ -- choice~\#1. 
To be precise, the faces of the `triangulation'
are allowed to possess arbitrarily many edges, while all the vertices 
are required to be trivalent (so strictly speaking, this is the dual 
of a genuine triangulation).%
  \foodnode{It is in fact sufficient to place an edge along each
  boundary component and along a basis of non-contractible cycles of
  the two-manifold (resolving any four-point vertices that result from
  this prescription into two three-point vertices). This amounts to
  the rule that the `triangulation' needs to be just fine enough such
  that any further edge that is added can be removed by repeated use
  of the fusion and bubble moves given in \erf{bubble2} below.} 
\\[-1.7em]{}
\item[\Nxt]
All ribbons lie in the two-dimensional submanifold $\rmX$ of $M_\rmX$. 
Furthermore, all ribbons (regarded as embedded two-manifolds) are to be 
oriented in such a way that their
orientation coincides with the one induced from $\rmX$ -- in short, the
``white'' side of each ribbon faces ``up''. The boundary of $\rmX$ is
taken to be oriented such that upon an orientation preserving
map of a patch of $\rmX$ that contains a boundary segment
to the upper half plane the orientation of the image of the segment 
agrees with that of the real axis. 
\\[-1.7em]{}
\item[\Nxt]
For every boundary component of $\rmX$, labelled by an $A$-module $M$, 
place an annular $M$-ribbon in $\rmX\,{\subset}\,M_\rmX$, along the circular 
line of singular points of $M_\rmX$ that corresponds to the boundary component.
The orientation of the core of the ribbon must be taken to agree with 
the orientation of the boundary component.
\\[-1.7em]{}
\item[\Nxt]
On those edges of the triangulation that are not part of the boundary 
of $\rmX$, place $A$-ribbons that are directed away from the vertices. 
In the middle of each edge these are to be joined by the 
morphism $\Phi_2\iN\Hom(A,A^\vee)$ that we defined in \erf{eq:Phi-def}
and display once again on the \lhs\ of figure \erf{Phi1} below.
As indicated on the \rhs\ of \erf{Phi1}, this can be done in two distinct ways; 
pick one of them on each edge -- this is choice~\#2.
  \bea \begin{picture}(270,91)(0,40)
  \put(0,4)   {\begin{picture}(0,0)(0,0)
              \scalebox{.38}{\includegraphics{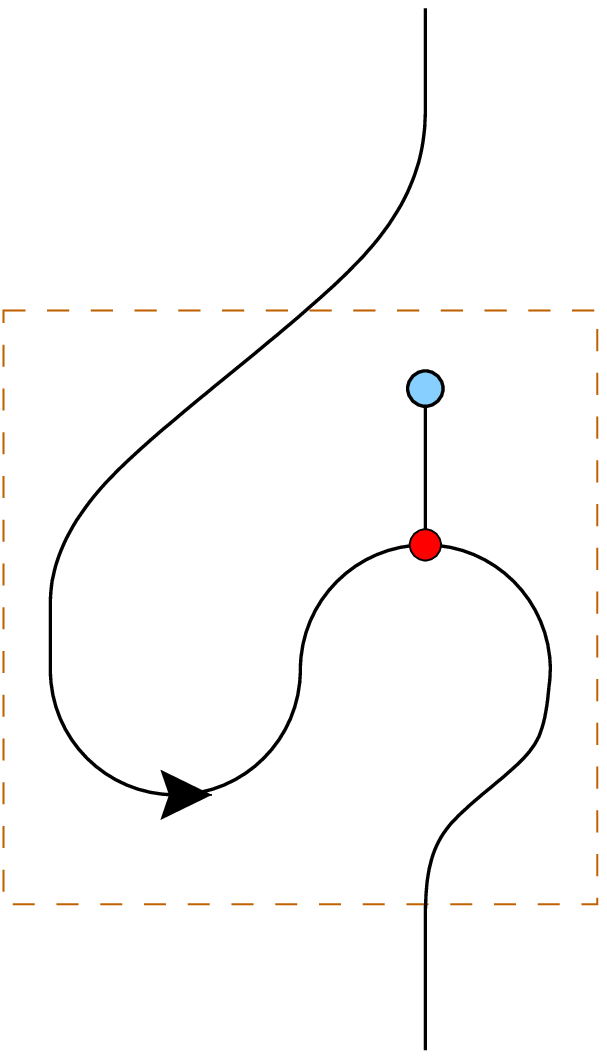}} \end{picture}}
  \put(-39,55)    {$\Phi_2\;=$}
  \put(42.9,-2.8) {\scriptsize$A$}
  \put(41.5,122.9){\scriptsize$A\Vee$}
  \put(150,0) {\begin{picture}(0,0)(0,0)
              \scalebox{.38}{\includegraphics{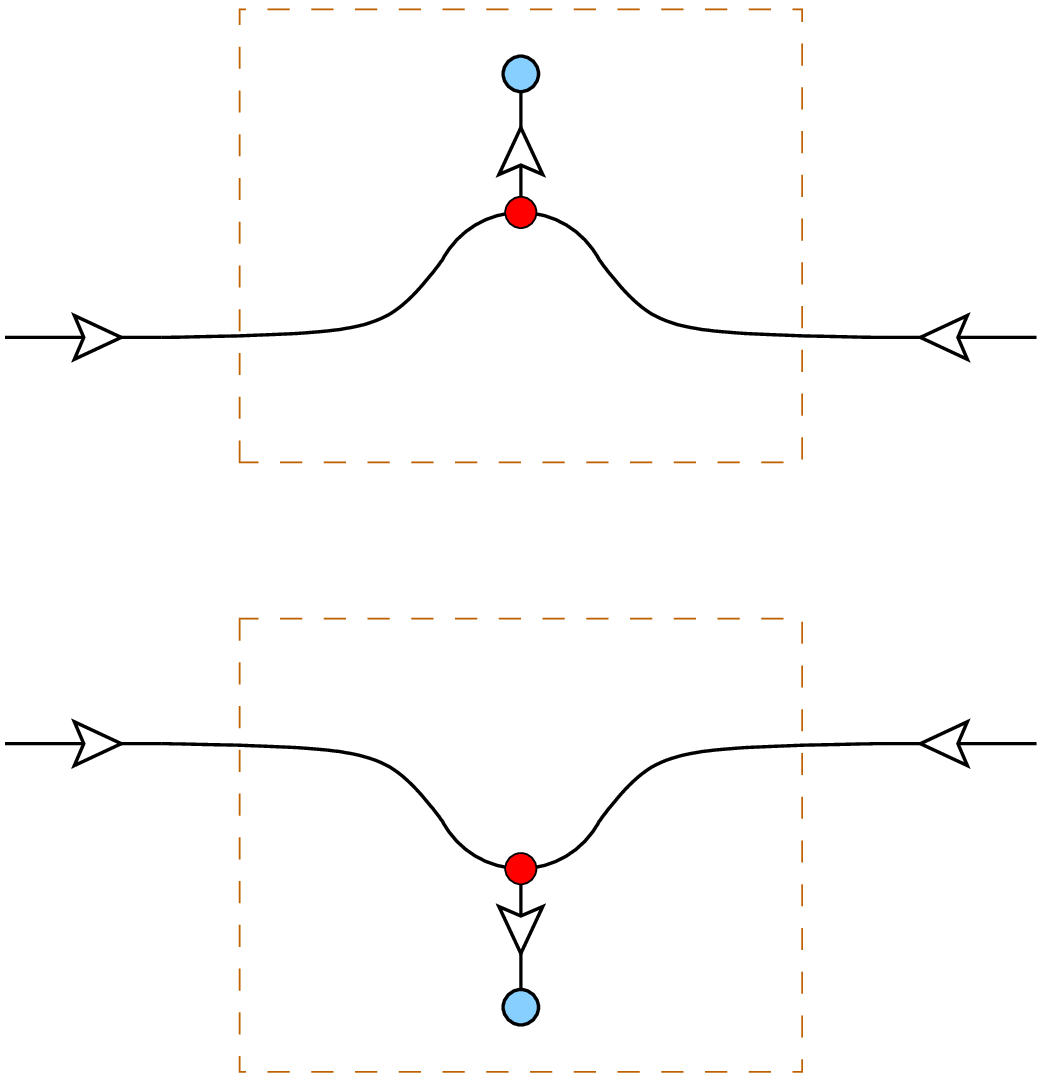}} \end{picture}
    \put(33,86) {\scriptsize$A$}
    \put(76,86) {\scriptsize$A$}
    \put(33,38) {\scriptsize$A$}
    \put(76,38) {\scriptsize$A$}
  }
  \epicture23 \labl{Phi1}
\item[\Nxt]
At every trivalent vertex in the interior of $\rmX$, join the
three outgoing $A$-lines with the morphism 
$\Delta \cir\Phi_1^{-1} \iN \Hom(A^\vee, A{\otimes}A)$,
see the left part of figure \erf{AAA}.
This morphism can be inserted in three different ways -- choice~\#3.
\\[-1.7em]{}
\item[\Nxt]
At every trivalent vertex on a boundary component of $\rmX$ that is labelled
by $M$, put the morphism $(\id_A\oti\r_M) \cir (\Delta \oti\idM) 
\cir (\eta\oti\idM) \iN \Hom(\M,A{\otimes}\M)$, with $\r_M$ the 
representation morphism of $M$, as in the right part of figure \erf{AAA}.
\end{itemize}
  \bea \begin{picture}(285,85)(0,44)
  \put(0,0)   {\begin{picture}(0,0)(0,0)
              \scalebox{.38}{\includegraphics{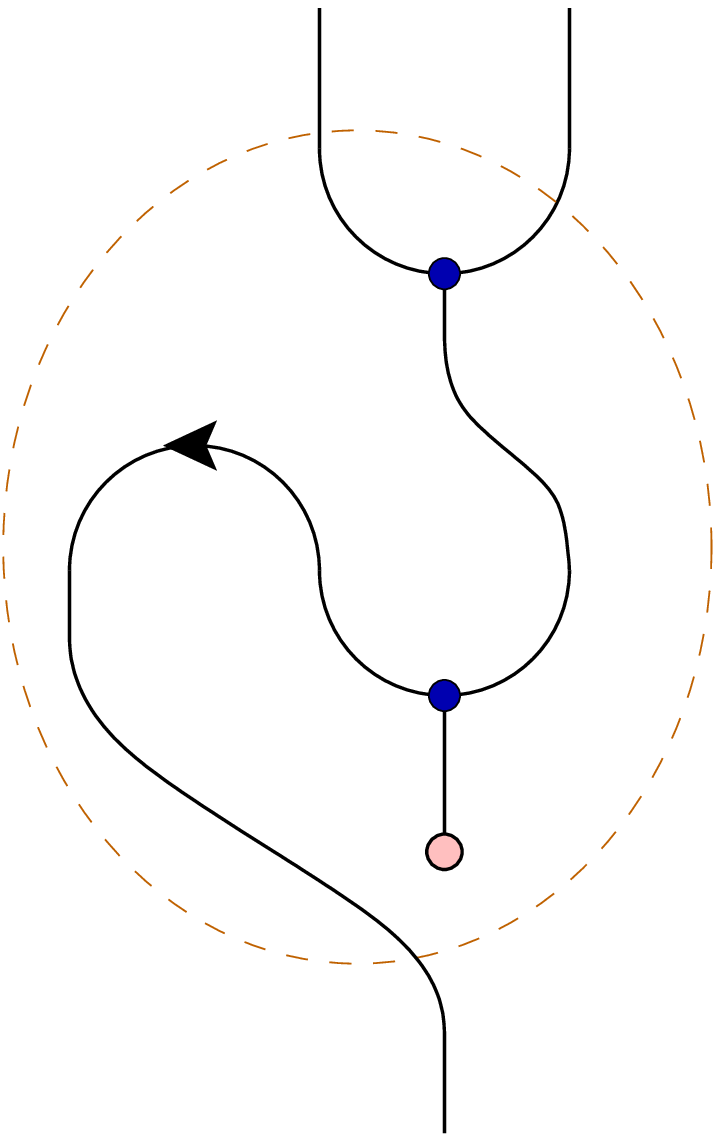}} \end{picture}}
  \put(31.5,127.7){\scriptsize$A$}
  \put(43.5,-8.4) {\scriptsize$A^\vee$}
  \put(59.5,127.7){\scriptsize$A$}
  \put(190,8) {\begin{picture}(0,0)(0,0)
              \scalebox{.38}{\includegraphics{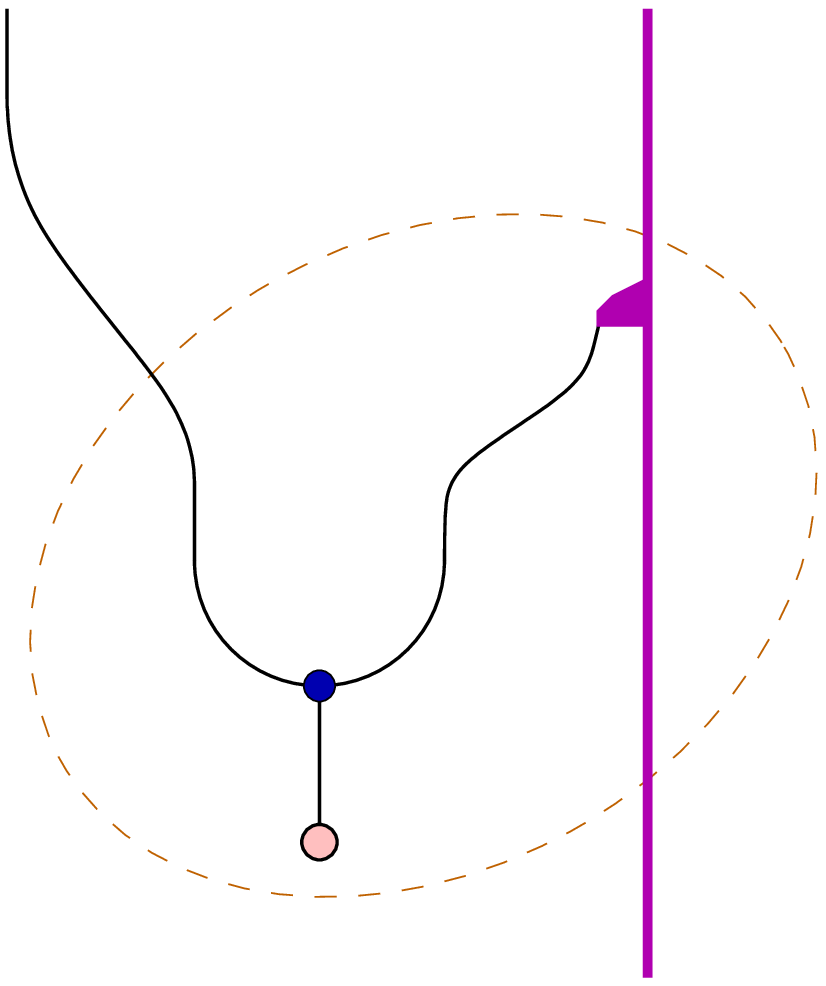}} \end{picture}}
  \put(188.5,119.1){\scriptsize$A$}
  \put(256.1,-2.4){\scriptsize$\M$}
  \put(256.9,119.1){\scriptsize$\M$}
  \epicture31 \labl{AAA}
This construction will be illustrated for the torus and the annulus
in sections \ref{sec:torus-pf} and \ref{sec:annulus-pf}, \resp.

\medskip 

According to the discussion in section \ref{sec:TFT}, the invariant of 
the ribbon graph in $M_\rmX$ constructed as above is a vector in the 
space of conformal zero-point blocks on $\partial M_\rmX\eq\hat \rmX$, 
which is the value of the correlator on $\rmX$. We proceed to show 
that this element is independent of the three choices made in the 
construction. We will call two ribbon graphs in a three-manifold 
{\em \equi\/} if they possess the same invariant for any modular 
category \calc. In this terminology, what we are going to
show is that different choices always lead to \equi\ ribbon graphs.

\begin{itemize}
\item[\Nxt] 
Choice~\#2 -- two ways to insert $\Phi_2$: The two fragments of the
ribbon graph can be transformed into each other as follows 
(all ribbons are $A$-ribbons):
  \bea \begin{picture}(350,87)(0,29)
  \put(0,0)   {\begin{picture}(0,0)(0,0)
              \scalebox{.38}{\includegraphics{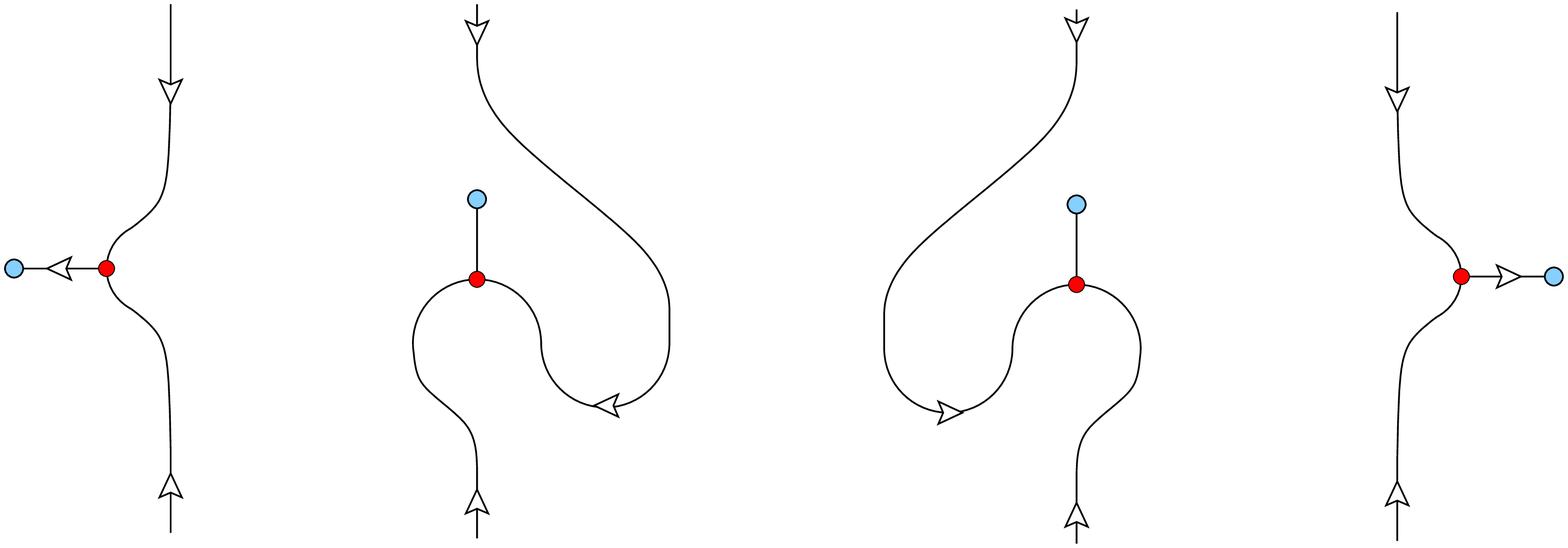}} \end{picture}}
  \put(53,52) {$=$}
  \put(157,52){$=$}
  \put(262,52){$=$}
  \epicture09 \labl{Phi123}
The first equality is just a deformation of the ribbon graph, while
the second equality amounts to the fact that the \alg\ $A$ is symmetric, 
as in \erf{Phi12}.
\item[\Nxt] 
Choice~\#3 -- coupling three $A$-ribbons: It is sufficient to show
that a 120$^\circ$ clockwise rotation of the encircled vertex that 
appears on the \lhs\ of \erf{AAA} does produce \equi\
ribbon graphs, i.e.\ that (all ribbons in the figure are $A$-ribbons)
  \bea \begin{picture}(215,85)(0,43)
  \put(0,0)   {\begin{picture}(0,0)(0,0)
              \scalebox{.38}{\includegraphics{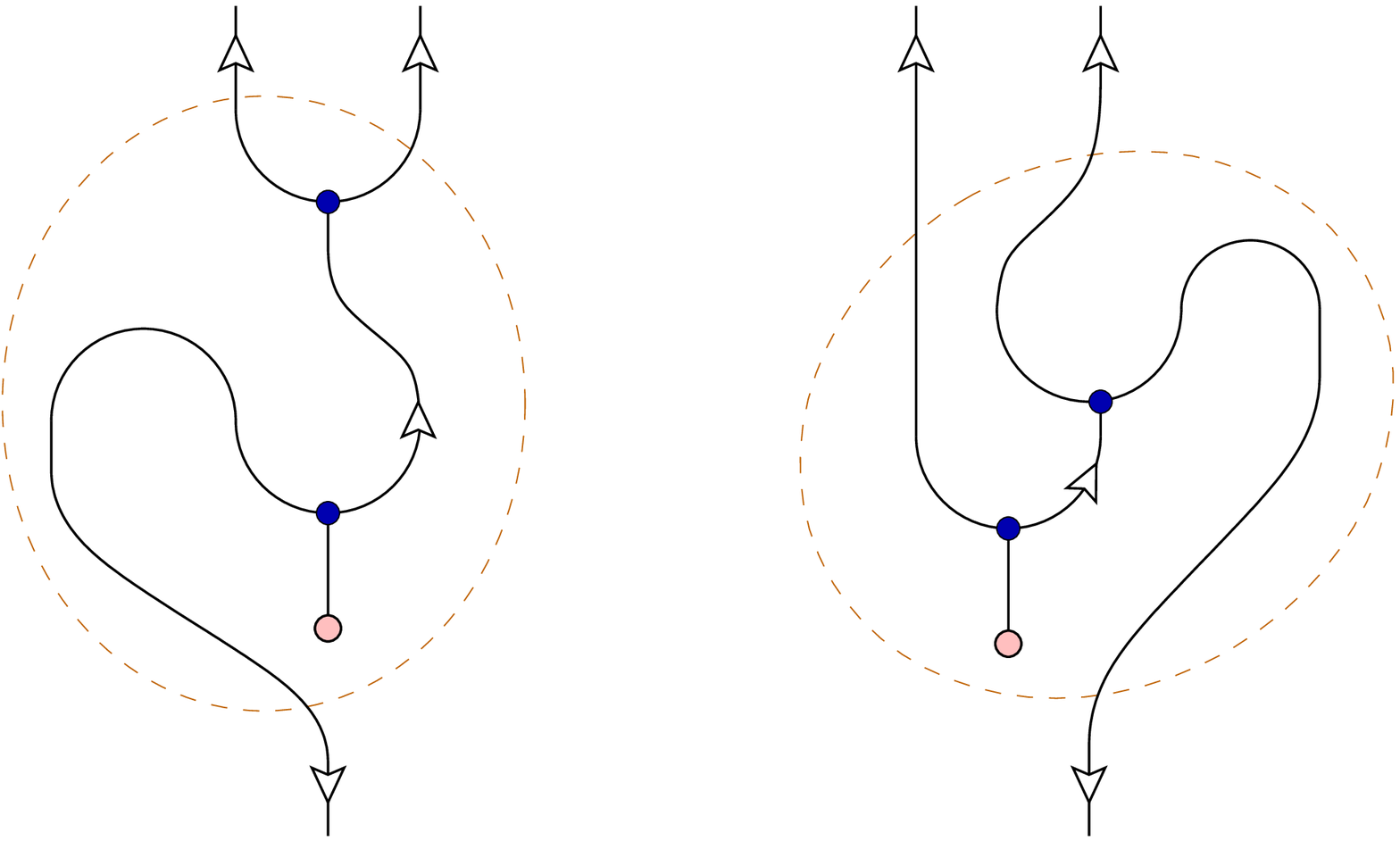}} \end{picture}}
  \put(95.6,62.5){$=$}
  \epicture22 \labl{AAA120}
The \lhs\ of this equation can be transformed into the \rhs\
by first using that $A$ is symmetric, i.e.\ that $\Phi_1^{-1}\eq\Phi_2^{-1}$
for the morphisms \erf{Phi12-}, on the lower coproduct, and then 
coassociativity as in \erf{Delta-eps}.
\item[\Nxt]
Choice~\#1 -- triangulation of the world sheet: Any two triangulations of a 
Riemann surface can be transformed into each other via the so-called 
{\em fusion\/} and {\em bubble\/} moves (see e.g.\ \cite{fuhk,chfs,kamo}), 
which look as follows:
  \bea \begin{picture}(368,61)(0,28)
  \put(0,0)   {\begin{picture}(0,0)(0,0)
              \scalebox{.38}{\includegraphics{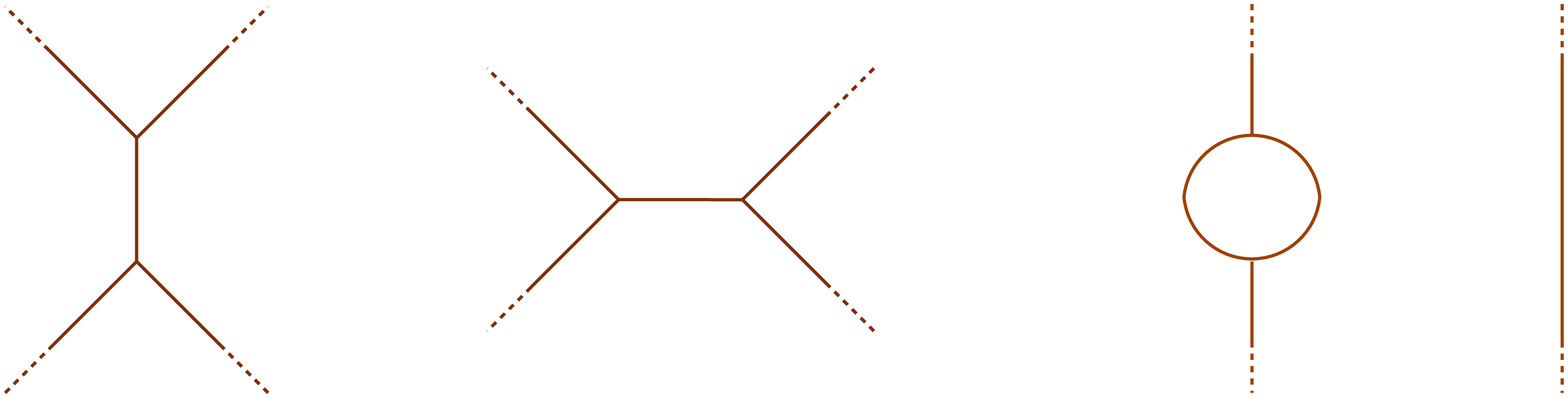}} \end{picture}}
  \put(66.3,50.6) {\small fusion}
  \put(69.6,39.1) {$\longleftrightarrow$}
  \put(303,50.6)  {\small bubble}
  \put(308,39.1)  {$\longleftrightarrow$}
  \epicture11 \labl{bubble2}
Two triangulations related by one of these moves give rise to equivalent 
ribbon graphs if the following transformations can be performed
with the help of the properties of the algebra $A$ and the module $M$: 
In the interior,
  \bea \begin{picture}(357,67)(0,34)
  \put(0,0)   {\begin{picture}(0,0)(0,0)
              \scalebox{.38}{\includegraphics{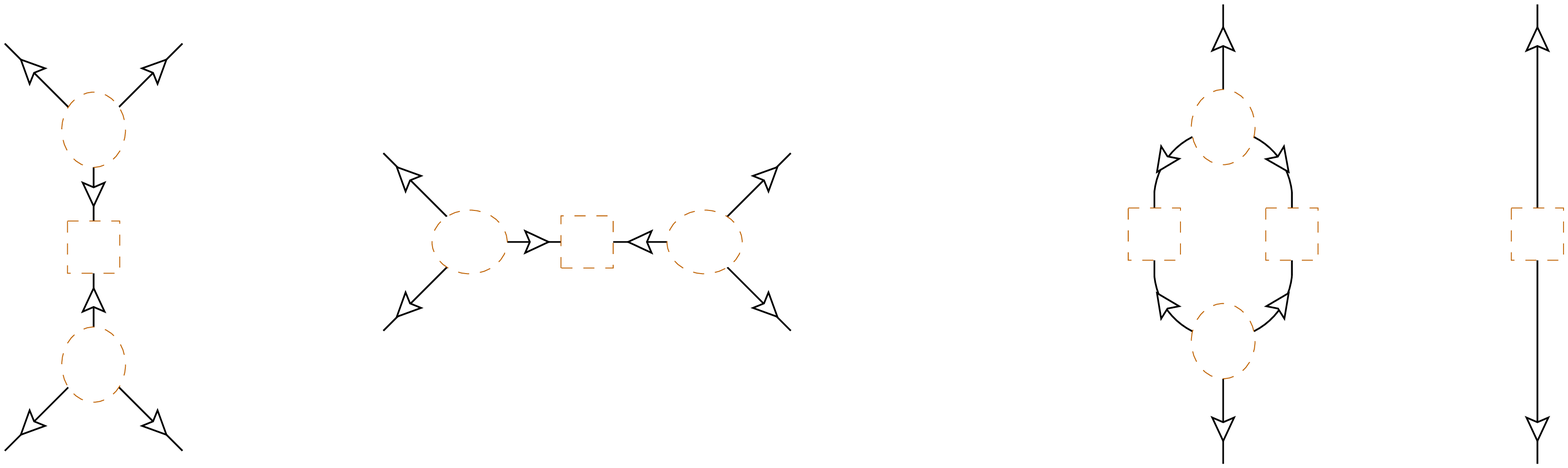}} \end{picture}}
  \put(5.3,2.1)    {\scriptsize$A$}
  \put(-3.5,84.1)  {\scriptsize$A$}
  \put(26.4,2.1)   {\scriptsize$A$}
  \put(35.5,84.1)  {\scriptsize$A$}
  \put(48.6,47.1)  {$=$}
  \put(80.1,59.5)  {\scriptsize$A$}
  \put(90.1,29.4)  {\scriptsize$A$}
  \put(158.5,29.4) {\scriptsize$A$}
  \put(170.1,59.5) {\scriptsize$A$}
  \put(196.6,47.1) {and}
  \put(258.5,3.1)  {\scriptsize$A$}
  \put(257.5,94.1) {\scriptsize$A$}
  \put(303.6,47.1) {$=$}
  \put(338.1,3.1)  {\scriptsize$A$}
  \put(338.9,94.1) {\scriptsize$A$}
  \epicture15 \labl{bubbleA}
and on the boundary,
  \bea \begin{picture}(369,77)(0,36)
  \put(0,0)   {\begin{picture}(0,0)(0,0)
              \scalebox{.38}{\includegraphics{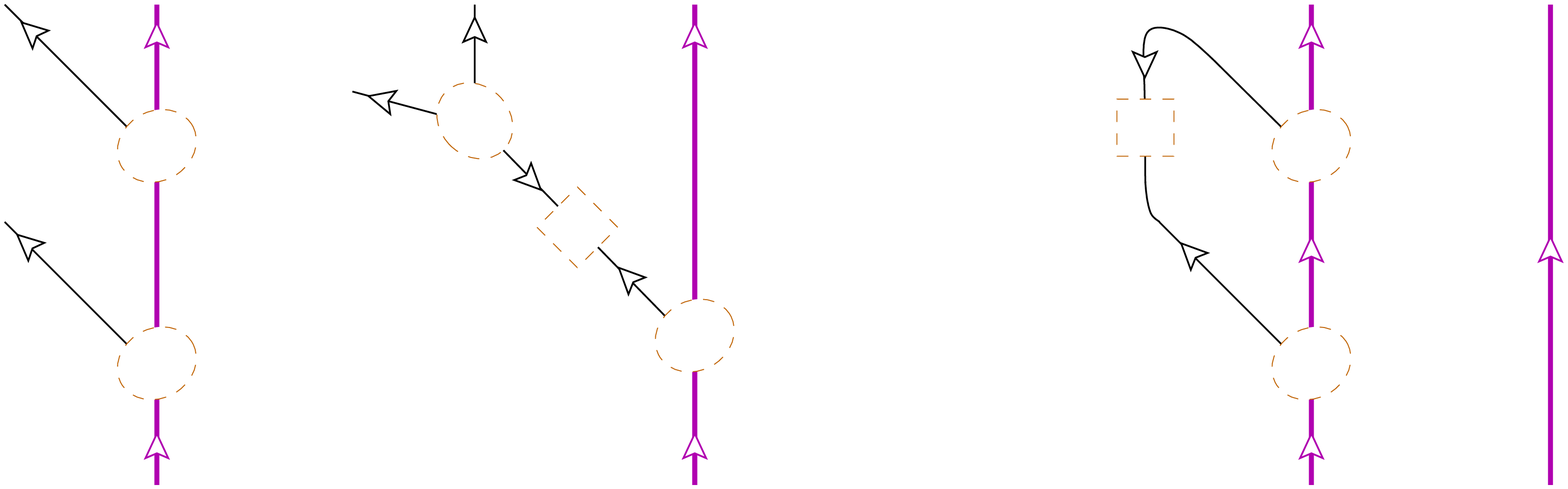}} \end{picture}}
  \put(-1.5,47.1)  {\scriptsize$A$}
  \put(-1.5,95.1)  {\scriptsize$A$}
  \put(37.5,3.1)   {\scriptsize$\M$}
  \put(37.5,97.1)  {\scriptsize$\M$}
  \put(61.6,54.1)  {$=$}
  \put(76.3,79.1)  {\scriptsize$A$}
  \put(104.3,102.1){\scriptsize$A$}
  \put(156.5,3.1)  {\scriptsize$\M$}
  \put(156.5,99.1) {\scriptsize$\M$}
  \put(194.6,54.1) {and}
  \put(256.3,93.7) {\scriptsize$A$}
  \put(254.3,47.1) {\scriptsize$A$}
  \put(292.9,3.1)  {\scriptsize$\M$}
  \put(292.9,99.1) {\scriptsize$\M$}
  \put(313.6,54.1) {$=$}
  \put(345.2,53.1) {\scriptsize$\M$}
  \epicture19 \labl{bubbleM}
(Here the parts enclosed by dashed lines stand for the corresponding parts 
of figures \erf{Phi1} and \erf{AAA}.) A slightly lengthy but straightforward 
calculation, using only the various properties of the $A$ and $M$, i.e.\ the 
defining axioms of a symmetric special Frobenius algebra and its \rep s, shows
that this is indeed the case and thus proves independence of the triangulation.
\end{itemize}

\medskip

In the remainder of this section we will mainly deal with the following 
three-manifolds: $S^2\,{\times}\,S^1$, $D\,{\times}\,S^1$ and $S^3$, 
with $D$ denoting a disk. The pictorial representation that we will use 
for ribbon graphs in these manifolds is illustrated in the following figure,
for the example of the Hopf link:
  \bea \begin{picture}(375,72)(0,10)
  \put(-20,-10)   {\begin{picture}(0,0)(0,0)
              \scalebox{.38}{\includegraphics{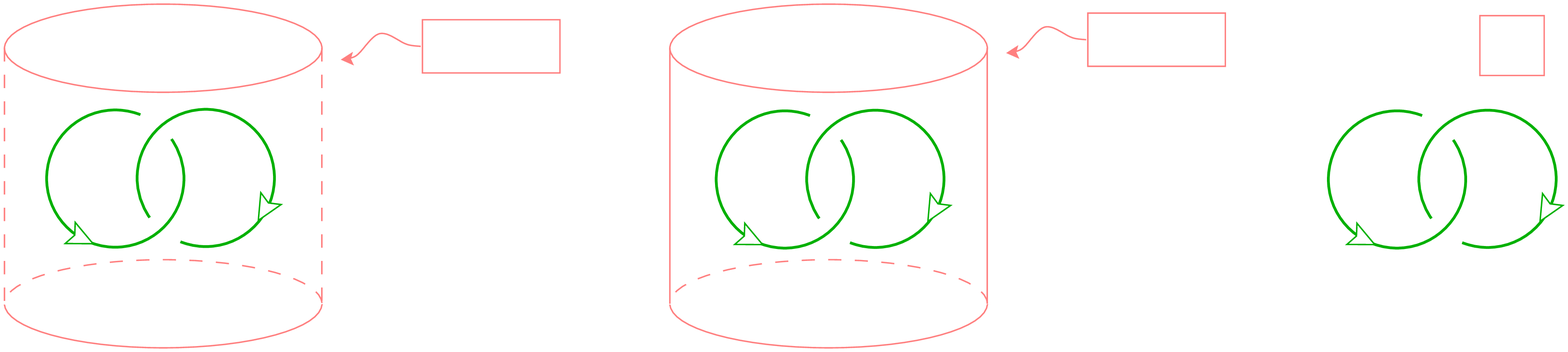}} \end{picture}
   \put(111,74) {$S^2{\times}S^1$}
   \put(284,76) {$D{\times}S^1$}
   \put(385,74) {$S^3$}
  }
  \epicture02 \labl{inv-conv}
In the first two pictures, the vertical direction corresponds to the
$S^1$ factor, and accordingly top and bottom are to be identified.
The first picture stands for the complex number given by the invariant of 
the relevant ribbon graph (here the Hopf link) in $S^2\,{\times}\,S^1$; 
the second picture denotes an element in the vector space
$\calh(\emptyset,\torus)$ where $\torus$ is a two-torus. The third picture 
again stands for the complex number that is given by the invariant
associated to the relevant ribbon graph, but now {\em multiplied
by the factor\/} $1/S_{0,0}$. 

The latter convention avoids a proliferation of factors $S_{0,0}$.
Indeed, recall (see formula \erf{norm-Z} and the subsequent text) that the 
invariant of a ribbon graph embedded in $S^3$ is given by $S_{0,0}$ times 
the complex number obtained from translating the ribbon graph into a 
morphism in $\Hom(\one,\one)$. The convention allows us to
replace ribbon graphs in $S^3$ by morphisms without the need to 
introduce explicit factors of $S_{0,0}$.

\medskip
 
Let us also remark that on orientable world sheets, independence from the
triangulation is precisely
what is needed for topological invariance in two-dimensional lattice
topological theory. Indeed, triangulation independence of our results
follows by the same arguments \cite{fuhk} as in two-dimensional lattice TFT.
Furthermore, the construction in \cite{fuhk} shows that two-dimensional
lattice TFTs are in one-to-one correspondence with symmetric special
Frobenius algebras in the category of complex vector spaces. Our
construction of conformal field theory amplitudes can therefore be understood
\cite{fuRs} as a natural generalisation of lattice TFTs from the category of
finite-dimensional complex vector spaces to more general modular tensor
categories.
 
It is also worth mentioning that \cite{peSc2} what is needed in
order for triangulations with the same number of faces to yield
the same results, are only the properties of $A$ to be a symmetric Frobenius
algebra. The property of $A$ to be also special, on the other hand, allows one
to reduce the number of faces in a triangulation by the bubble move
\erf{bubble2}. The latter issue can be expected
to be much more subtle for irrational conformal field theories.

\subsection{Zero-point blocks on the torus} \label{sec:toruszero}

The torus and annulus partition functions can both be expressed in terms of
conformal blocks on the torus. Let us therefore have a closer
look at the space of torus blocks from the TFT point of view.

Let $\torus$ be the extended surface given by the oriented torus
$S^1{\times}S^1$, without any embedded arcs (and with a choice of 
Lagrangian subspace, to be detailed below). As discussed in \ref{sec:TFT},
the TFT assigns to $\torus$ a vector space $\calh(\emptyset;\torus )$ (the 
symbol $\emptyset$ makes explicit that the extended surface $\torus$ does
not carry any arc). The space $\calh(\emptyset;\torus)$ has dimension $|\II|$. 
A distinguished basis can be obtained in the following way: Let $M_1$ 
be a solid torus with a $k$-ribbon running along its non-contractible cycle. 
Choose $\partial_+ M_1\eq\torus $ and $\partial_- M_1\eq\emptyset$, and
take the Lagrangian subspace in the first homology of $\torus$ to be 
spanned by the cycle in $\torus$ that is contractible within $M_1$.
The basis vectors $\tbas{k}{}$ are then obtained by applying the map 
$Z(M_1,\emptyset,{\rm T})$ to the number $1\iN\complex\eq\calh(\emptyset)$:
  \be
  \tbas{k}{} = Z(M_1,\emptyset,{\rm T}) \, 1 \ \ \in \calhT \,.  \ee
In pictures,
  \bea \begin{picture}(110,84)(0,51)
  \put(0,0)   {\begin{picture}(0,0)(0,0)
              \scalebox{.38}{\includegraphics{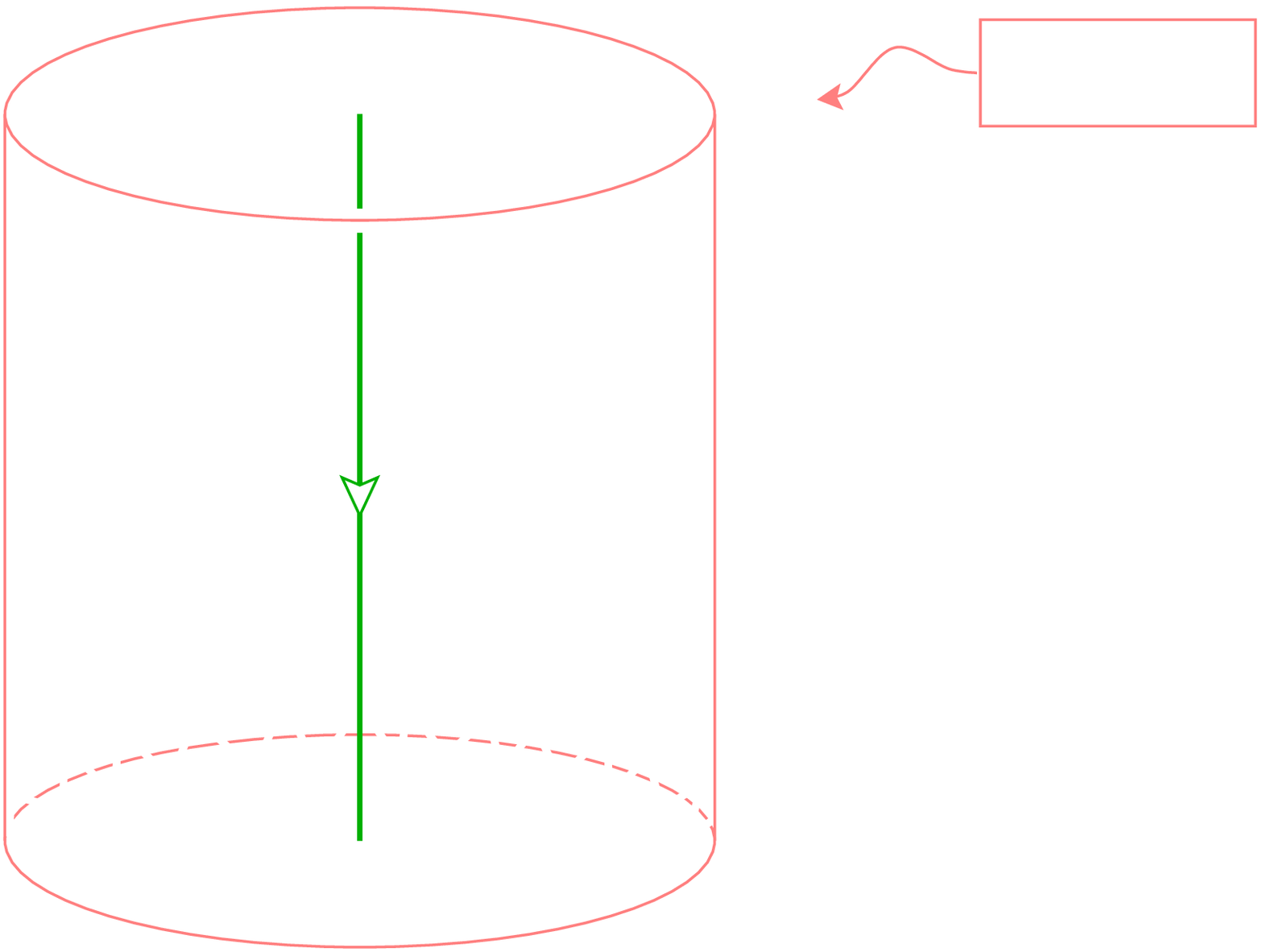}} \end{picture}}
  \put(-63,66.1){$\tbas{k}{}\; :=$}
  \put(53.5,59.9) {\scriptsize$k$}
  \put(122.6,66.1){$\in\, \calhT$}
  \put(141.2,120.5){$D\,{\times}\,S^1$}
  \epicture29 \labl{eq:tor-bas} 

Interpreting $\calh(\emptyset;{\rm T})$ as the space of conformal blocks on 
the torus, these basis elements correspond to the characters $\chii_k$ of 
the irreducible highest weight modules of the chiral algebra \CA.  
Note that one should actually think of the character as a {\em one\/}-point 
block on the torus, with insertion of the vacuum representation. The character 
thus depends on the modulus $\tau$ as well as on a field insertion
$\varphi(v;z)$ with $v\iN\CA$ (owing to translation invariance there is no 
dependence on $z$). The Virasoro specialised characters are obtained
when choosing $v$ to be the vacuum vector, corresponding to $\varphi\eq\one$,
upon which only the $\tau$-dependence is left. 
When \CA\ is larger than the VOA associated to the Virasoro algebra, 
the specialised characters need not be linearly independent. For
example, $\chii_k(\tau)\eq\chii_{\bar k}(\tau)$ even when $k\,{\ne}\,\bar k$. 

Reversing the orientation of $\torus$ takes the modulus $\tau$ to $-\tau^*$. 
In terms of specialised characters the correspondence is
  \be
  \tbas{k}{}  \,\leadsto\, \chi_k(\tau) \qquad {\rm and} \qquad
  \tbas{k}{-} \,\leadsto\, \chi_k(-\tau^*) \,.  \labl{eq:vect-to-char}

In order to expand a general element 
$|\psi;{\rm T}\rangle \iN \calhT$ in terms of this basis, we also need
the dual basis vectors $\tdbas{k}{}$ in $\calhT^*$. Then we can write 
$|\psi;{\rm T}\rangle\eq\sumI_k \tdbas{k}{}\psi;{\rm T}\rangle \, \tbas{k}{}$.
Let $M_2$ be a solid torus with a $k$-ribbon running along its non-contractible
cycle. Take its orientation such that $\partial M_2\eq{-}\torus$. We choose
$\partial_+ M_2\eq\emptyset$ and $\partial_- M_2\eq\torus$ (recall from 
section \ref{sec:TFT} that $\partial_-M_2$ is defined to have reversed 
orientation). Then $Z(M_2,\torus,\emptyset)$ is a linear function
from $\calhT$ to $\complex$, and we have
  \be
  \tdbas{k}{} = Z(M_2,\torus,\emptyset) \ \ \in \calhT^* \,. \ee
In pictures,
  \bea \begin{picture}(115,89)(0,45)
  \put(0,0)   {\begin{picture}(0,0)(0,0)
              \scalebox{.38}{\includegraphics{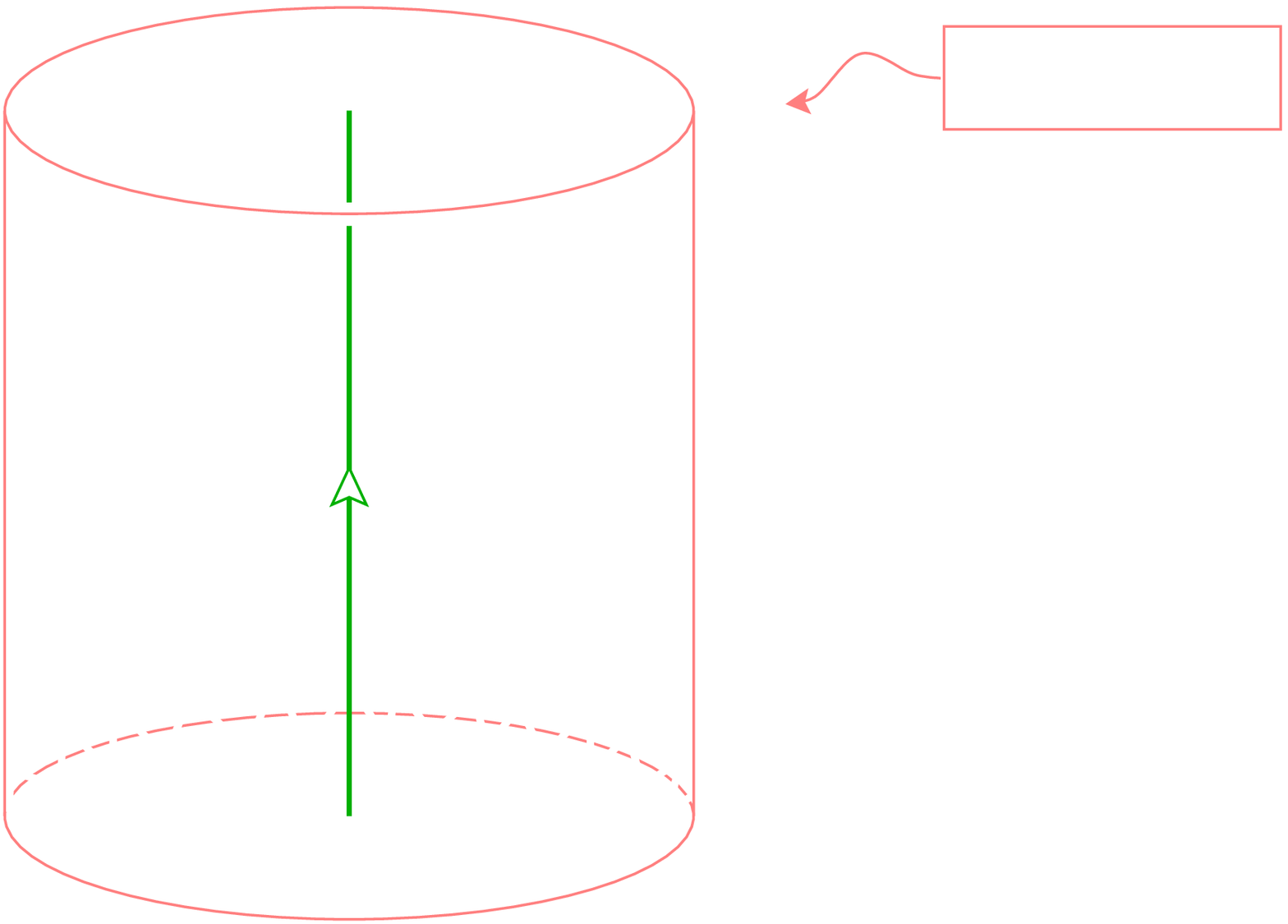}} \end{picture}}
  \put(-63,63.1){$\tdbas{k}{}\; =$}
  \put(53.5,64.9) {\scriptsize$k$}
  \put(122.6,63.1){$\in\, \calhT^*$}
  \put(141.2,120.5){$-D\,{\times}\,S^1$}
  \epicture28 \labl{eq:tor-dual}
To verify that this is indeed dual to \erf{eq:tor-bas}, use the identity map 
to glue $\partial M_1$ to $\partial M_2$. Then functoriality 
\erf{eq:Zfunct} implies
$\langle\chi_k;\torus\,|\,\chi_\ell;\torus\rangle\eq Z(M_2,\torus,\emptyset) 
\cir Z(M_1,\emptyset,\torus)\eq Z(M,\emptyset,\emptyset)$, where
$M$ is the three-manifold $S^2{\times}S^1$, with two ribbons as in the
following figure:
  \bea \begin{picture}(115,89)(0,45)
  \put(0,0)   {\begin{picture}(0,0)(0,0)
              \scalebox{.38}{\includegraphics{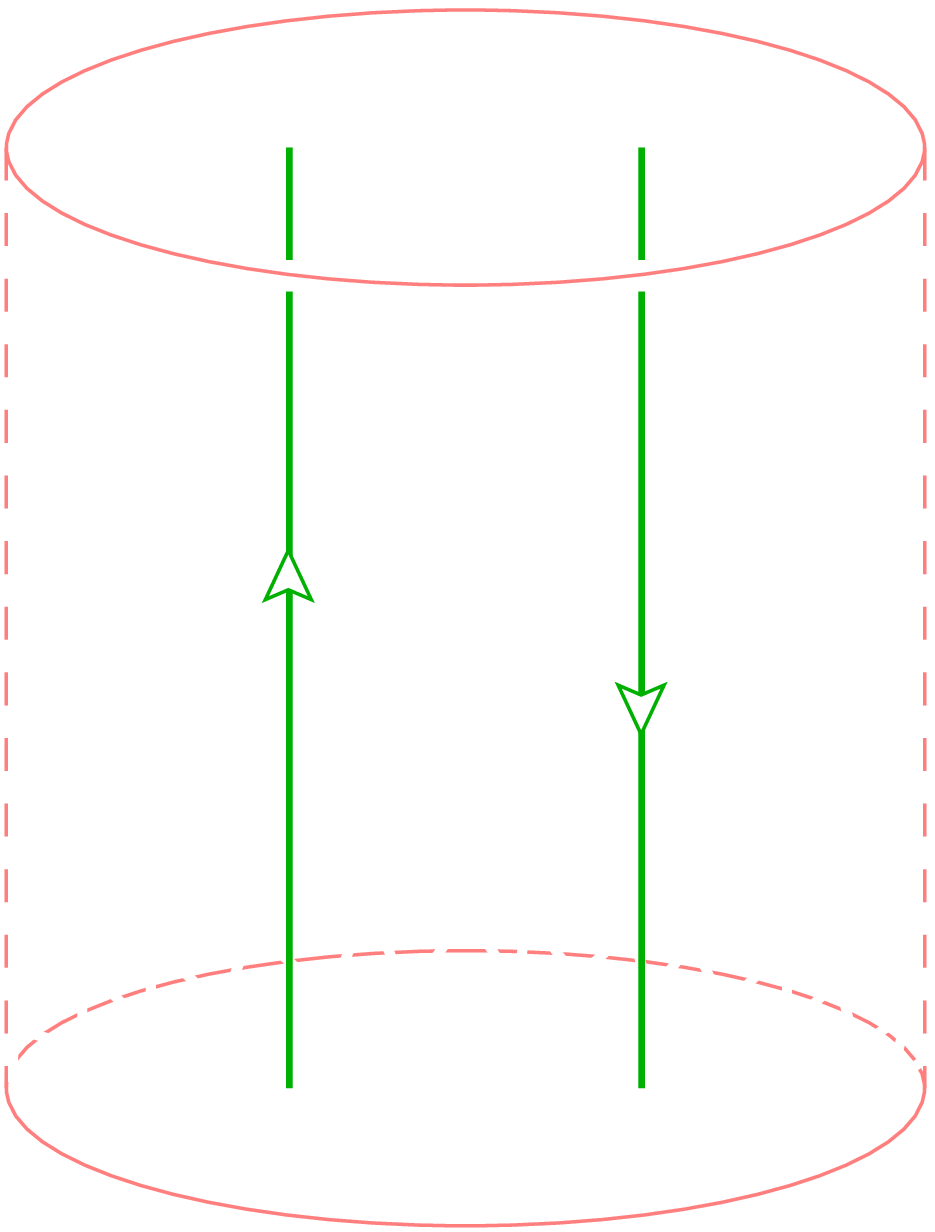}} \end{picture}}
  \put(-69.6,63.1){$\langle\chii_k|\chii_\ell\rangle\; =$}
  \put(25.3,58.3) {\scriptsize$k$}
  \put(72.5,64.7) {\scriptsize$\ell$}
  \put(122.6,63.1){$=\;\delta_{k,\ell}$}
  \epicture26 \labl{tor-prod}
To see how the delta symbol arises, use relation \erf{eq:TFT-Ztrace} to 
rewrite the number $Z(M,\emptyset,\emptyset)$ as a trace over 
$\calh(k,(\ell,-);S^2)$. Property \erf{eq:TFT-Zid} implies that the trace 
is taken over the identity operator. 
This leads to $\langle\chi_k;\torus\,|\,\chi_\ell;\torus\rangle\eq
\dim\,\calh(k,(\ell,-);S^2)$. This dimension, in turn, equals 
$\delta_{k,\ell}$, as discussed in section \ref{sec:TFT}.

By using surgery on the relation \erf{tor-prod} one can relate the Hopf 
link \erf{smat} to the matrix $(S_{i,j})$ of modular transformations of
characters. To see this first recall \cite{witt27} that any
link in $S^2\,{\times}\,S^1$ is related to an \equi\
link in $S^3$ via surgery; in the present situation,
  \bea \begin{picture}(350,86)(0,49)
  \put(0,0)   {\begin{picture}(0,0)(0,0)
              \scalebox{.38}{\includegraphics{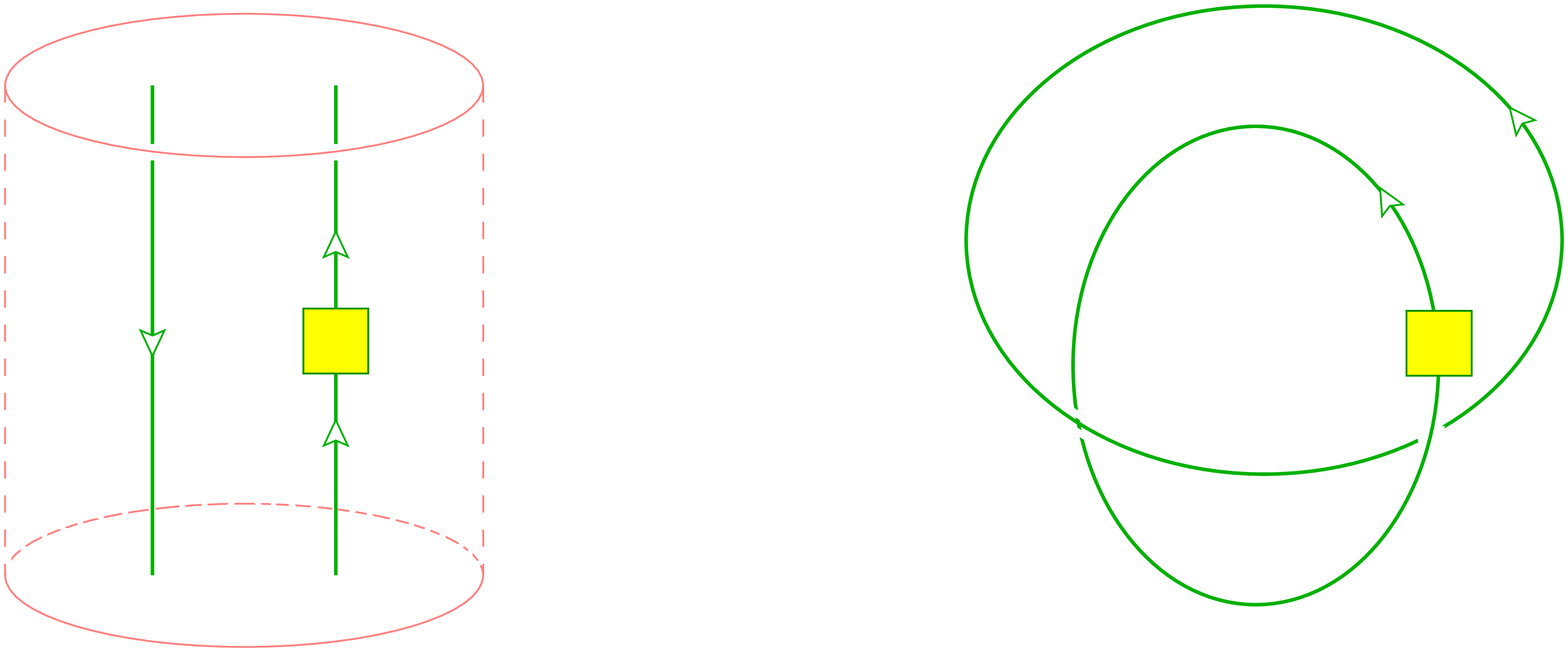}} \end{picture}}
  \put(25.8,60.1) {\scriptsize$i$}
  \put(73.0,89.9) {\scriptsize$X$}
  \put(67.3,62.2) {\scriptsize$\Phi$}
  \put(73.0,34.9) {\scriptsize$X$}
  \put(122.6,63.1){$=\;S_{0,0} \dsty\sumI_j S_{i,j}$}
  \put(282.0,91.1){\scriptsize$X$}
  \put(300.3,62.2){\scriptsize$\Phi$}
  \put(325,112){\scriptsize$j$}
  \epicture28 \labl{S-xfer}
Here $X$ is an arbitrary (not necessarily simple) object and 
$\Phi$ a morphism in $\Hom(X,X)$. Concretely, this result is obtained by
cutting out a small solid torus containing the $i$-ribbon on the \lhs\
of \erf{S-xfer} and gluing it back after an $S$-transformation,
yielding the three-manifold $S^3$ with ribbon graph as on the \rhs.

Upon setting $i\eq\ell$, $X\eq U_k$ and $\Phi\eq\id_{U_k}$, \erf{S-xfer} 
turns into 
  \be \delta_{k,\ell} = S_{0,0} \sum_j S_{\ell ,j}\, s_{j,\bar k} \,,
  \labl{eq:SxS=C}
which by $(S^2)_{i,j}\eq C_{i,j}\,{\equiv}\,\delta_{i,\bar\jmath}$ 
implies that
  \be  s_{i,j} = S_{i,j}/S_{0,0} \,.  \ee

%%%%%%%%%%%%%%%%%%%%%%%%%%%%%%%%%%%%%%%%%%%%%%%%%%%%%%%%%%%%%%%%%%%%%%%%

\subsection{The torus partition function} \label{sec:torus-pf}   

Applied to the torus partition function, the general construction in
section \ref{sec:con-mf-rib} proceeds as follows. The world sheet $\rmX$
is a torus $\torus$, thus the double $\hat\rmX$ is the disconnected sum
$\torus{\sqcup}({-}\torus)$ of two copies of the torus, and
the connecting manifold is $M_\torus\eq\torus\,{\times}\,[-1,1]$.
Next we pick a triangulation of the world sheet, and then convert
this triangulation to a ribbon graph embedded in $M_\torus$. 
The \lhs\ of the following figure displays the triangulation we choose.
The ribbon graph in $M_T$ obtained by inserting the 
elements \erf{Phi1} and \erf{AAA} is displayed on the \rhs\
(the extension in the direction of the interval $[-1,1]$ is suppressed,
all ribbons are labelled by $A$):
  \bea \begin{picture}(340,57)(0,29)
  \put(0,0)   {\begin{picture}(0,0)(0,0)
              \scalebox{.38}{\includegraphics{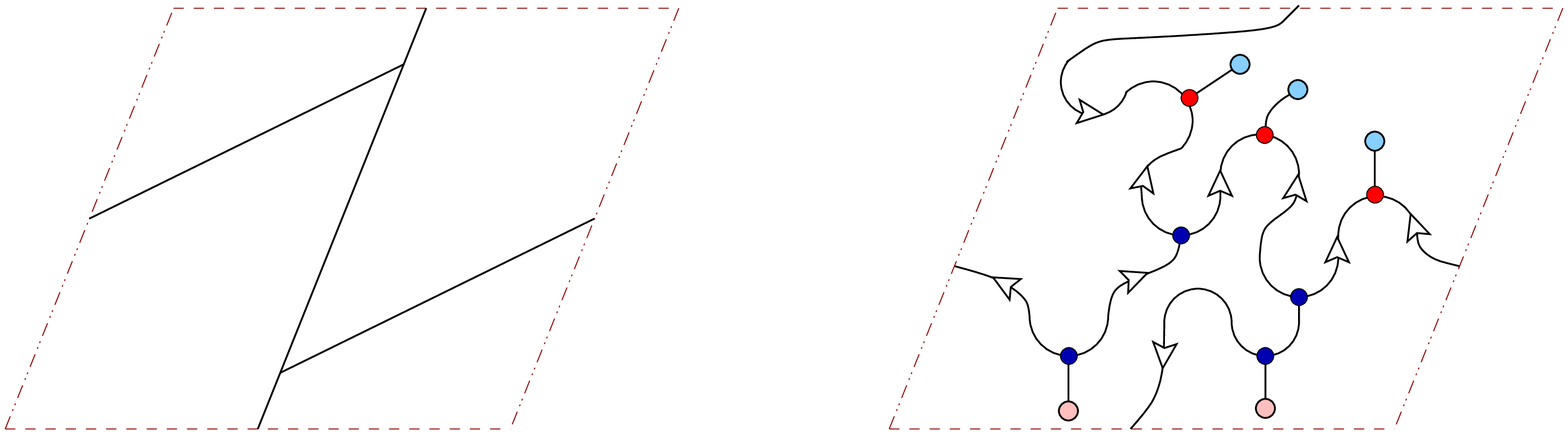}} \end{picture}}
  \put(144.6,42.1) {$\longrightarrow$}
  \epicture12 \labl{tor}
This graph can be simplified a lot using the relations for multiplication 
and comultiplication of $A$; this way we arrive at
  \bea \begin{picture}(175,67)(0,35)
  \put(0,0)   {\begin{picture}(0,0)(0,0)
              \scalebox{.38}{\includegraphics{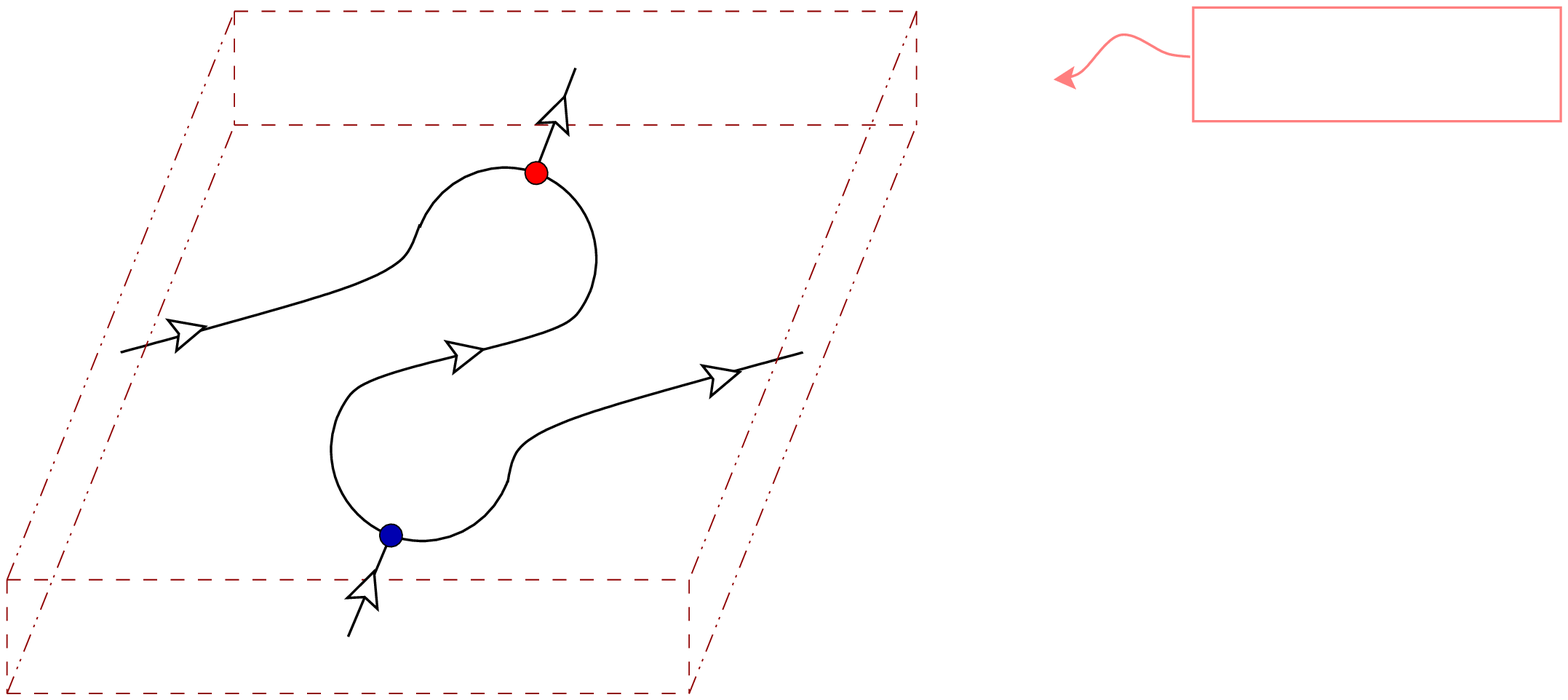}} \end{picture}}
  \put(41,63) {\scriptsize$A$}
  \put(91,61) {\scriptsize$A$}
  \put(93,48) {\scriptsize$A$}
  \put(45,8) {\scriptsize$A$}
  \put(182,92.5) {$T {\times} [-1,1]$}
  \put(-28.6,55.1) {$Z\;=$}
  \epicture11 \label{eq:tor-Z}\labl{tor-Z}
It is this ribbon graph which we will work with in the rest of this section.

\medskip

Often we will need a move that reverses the orientation of (the core of)
an $A$-ribbon to see that two given ribbon graphs are \equi.
The following equality on segments of a ribbon graph is a direct consequence
of the Frobenius property of $A$: 
  \bea \begin{picture}(300,45)(0,10)
  \put(0,0)   {\begin{picture}(0,0)(0,0)
              \scalebox{.38}{\includegraphics{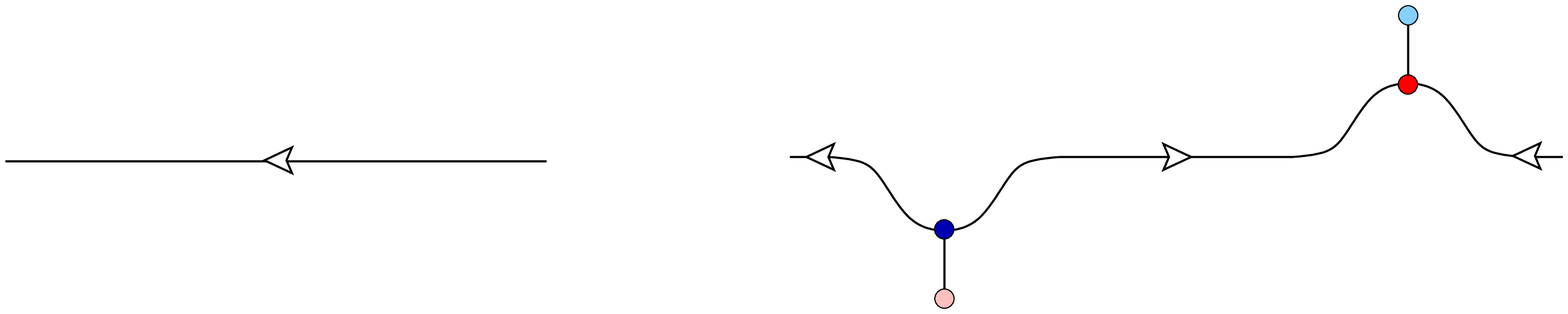}} \end{picture}}
  \put(15,28.7) {\scriptsize$A$}
  \put(140,30.5) {\scriptsize$A$}
  \put(191,28.8) {\scriptsize$A$}
  \put(274,31.5) {\scriptsize$A$}
  \put(115,22) {$=$}
  \epicture-4 \labl{eq:reverse-A}
Using this identity, one can reverse the orientation of the core of an 
$A$-ribbon. This is achieved as follows: First replace a stretch of $A$-ribbon 
with the \rhs\ of \erf{eq:reverse-A}; then move the multiplication and 
comultiplication appearing there to the respective ends of the $A$-ribbon. 
Whenever both ends of ribbon end on (the coupon for) a multiplication or
comultiplication morphism, one can then use the Frobenius or
associativity properties of $A$ to remove the unit and counit.

\medskip

It was already shown in section \ref{sec:con-mf-rib}
that the invariant \erf{eq:tor-Z} is
independent of the triangulation we start from.
Further, invariance under the action of the relative
modular group amounts to the usual
modular invariance of the torus partition function. In more detail,
consider the two transformations
  \be
  {\mathcal U}:\quad \tau \,\mapsto\,  \tau / (\tau\,{+}\,1)
  \qquad{\rm and}\qquad
  {\mathcal T}:\quad \tau \,\mapsto\,  \tau\,{+}\,1 \ee
of the complex upper half-plane.
The map $\mathcal U$ corresponds to the change of basis in the lattice
that takes $\{\tau,1\}$ to $\{\tau,1{+}\tau\}$. It is expressible in terms
of $\mathcal T$ and ${\mathcal S}{:}\ \tau\,{\mapsto}\,{-}1/\tau$ as 
${\mathcal U}\eq\mathcal{TST}$.
These changes of the fundamental region of the torus
modify the graph \erf{eq:tor-Z} as follows:
  \begin{eqnarray} \begin{picture}(400,84)(27,0)
  \put(0,0)   {\begin{picture}(0,0)(0,0)
              \scalebox{.38}{\includegraphics{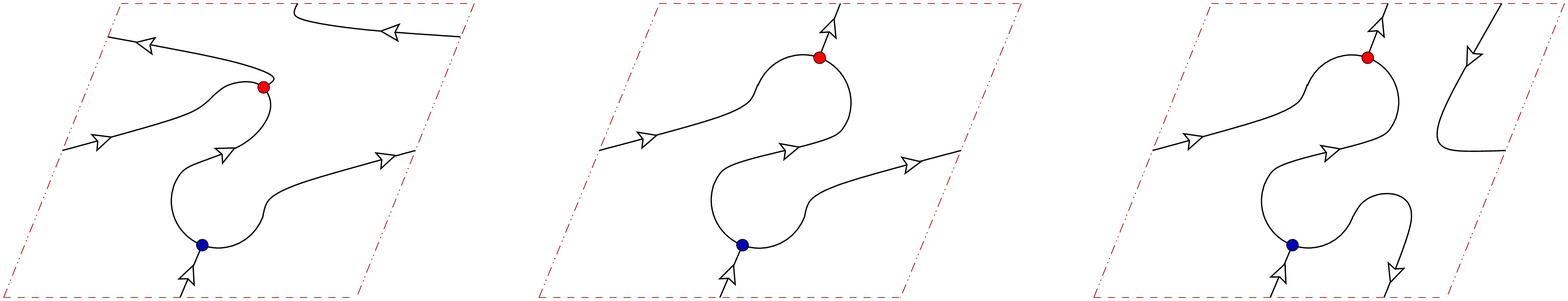}} \end{picture}}
  \put(138.6,40.1) {$\stackrel{\ \dsty\mathcal T}{\longleftarrow}$}
  \put(299.6,40.1) {$\stackrel{\dsty\mathcal U}{\longrightarrow}$}
  \end{picture} \nonumber \\[-1.64em]{} \label{tor-UT}
  \end{eqnarray}
In agreement with our general arguments about triangulation independence,
both resulting graphs can be transformed back to \erf{eq:tor-Z} by using 
properties of $A$. Explicitly, for the change of fundamental domain 
induced by $\mathcal T$ we have:
  \begin{eqnarray} \begin{picture}(400,86)(28,0)
  \put(0,0)   {\begin{picture}(0,0)(0,0)
              \scalebox{.38}{\includegraphics{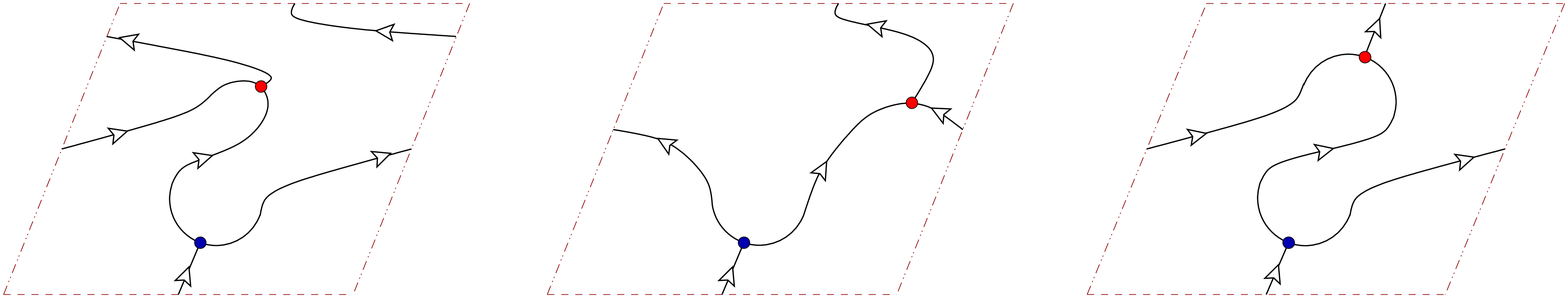}} \end{picture}}
  \put(143.6,40.1) {$=$}
  \put(302.6,40.1) {$=$}
  \end{picture} \nonumber \\[-1.61em]{} \label{tor-T}
  \end{eqnarray}

The ribbon graph \erf{eq:tor-Z} for $Z$ describes an element in
$\calh(\emptyset;\torus)\,{\times}\,\calh(\emptyset;-\torus)$. It can be 
expanded in a standard basis of conformal blocks (i.e.\ characters) as
  \be  Z 
  = \sumI_{i,j} Z_{ij}\, |\chii_i,\torus\rangle \otimes |\chii_j,
  -\torus\rangle \,.  \labl{eq:tor-pf}
Via the correspondence \erf{eq:vect-to-char}, this tells us that in terms 
of specialised characters of the chiral algebra the CFT partition
function reads $Z\eq\sumI_{i,j} Z_{ij}\,\chii_i(\tau)\,\chii_j{(\tau)}^*$.

To obtain the coefficients $Z_{ij}$ we glue the dual basis
elements for $|\chii_i,\torus\rangle$ and $|\chii_j,-\torus\rangle$ to 
the two boundaries of $Z$. This yields the ribbon graph
  \bea \begin{picture}(85,121)(0,38)
  \put(0,0)   {\begin{picture}(0,0)(0,0)
              \scalebox{.38}{\includegraphics{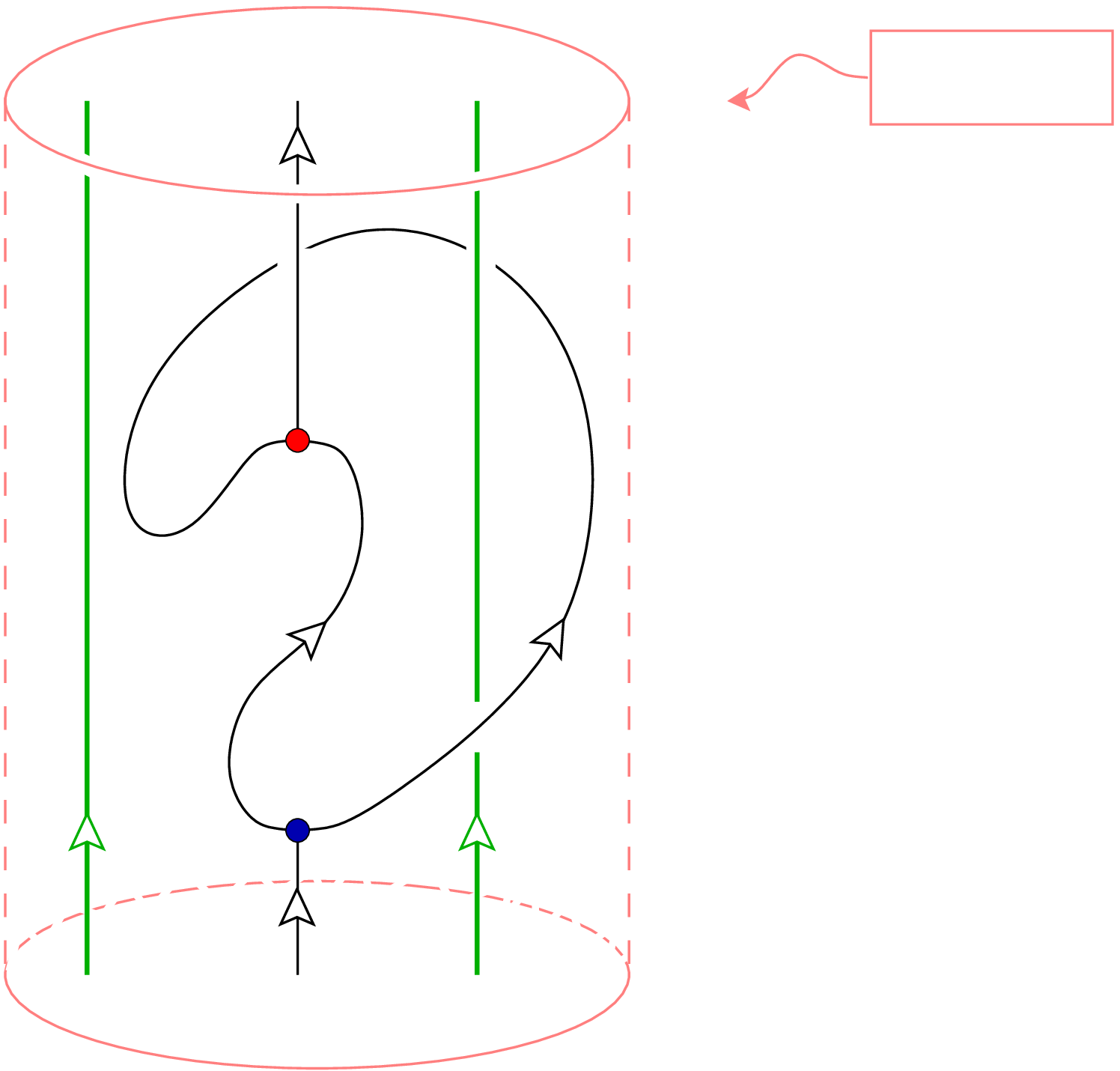}} \end{picture}}
  \put(-54.6,77.1) {$Z_{ij}\;:=$}
  \put(15.5,14.9) {\scriptsize$i$}
  \put(44.5,14.9) {\scriptsize$A$}
  \put(71.5,14.9) {\scriptsize$j$}
  \put(30,56) {\scriptsize$A$}
  \put(79,83) {\scriptsize$A$}
  \put(45,104) {\scriptsize$A$}
  \put(128.5,141.6) {$S^2{\times}S^1$}
  \epicture11 \labl{Zij2}
in $S^2\,{ \times}\, S^1$.

\dtl{Theorem}{ZZZ}
The numbers $Z_{ij}$ given by the invariant
of the ribbon graph \erf{Zij2} enjoy the following properties:
\begin{eqnarray}
{\rm(i)}\;\ && [Z,S] = 0 \quad{\rm and}\quad [Z,T]=0 \,. \\ [.4em]
{\rm(ii)}\  && Z_{ij} \in \zet_{\ge 0} \,.      \label{ZZZii}\\[.4em]
{\rm(iii)}  && Z_{00} = \dim \, \centreA(\Atop) \label{ZZZiii} \,.
\end{eqnarray}

\medskip\noindent
(As usual, we denote the matrices that implement the modular transformations
$\mathcal S$ and $\mathcal T$ on the space of characters by $S$ and $T$,
\resp.) Before proving these claims, it is useful to introduce, for any 
object $X$, a specific morphism $P_X\iN\Hom(A{\otimes}X,A{\otimes}X)$:
  \bea  \begin{picture}(25,91)(0,37)
  \put(0,0)   {\begin{picture}(0,0)(0,0)
              \scalebox{.38}{\includegraphics{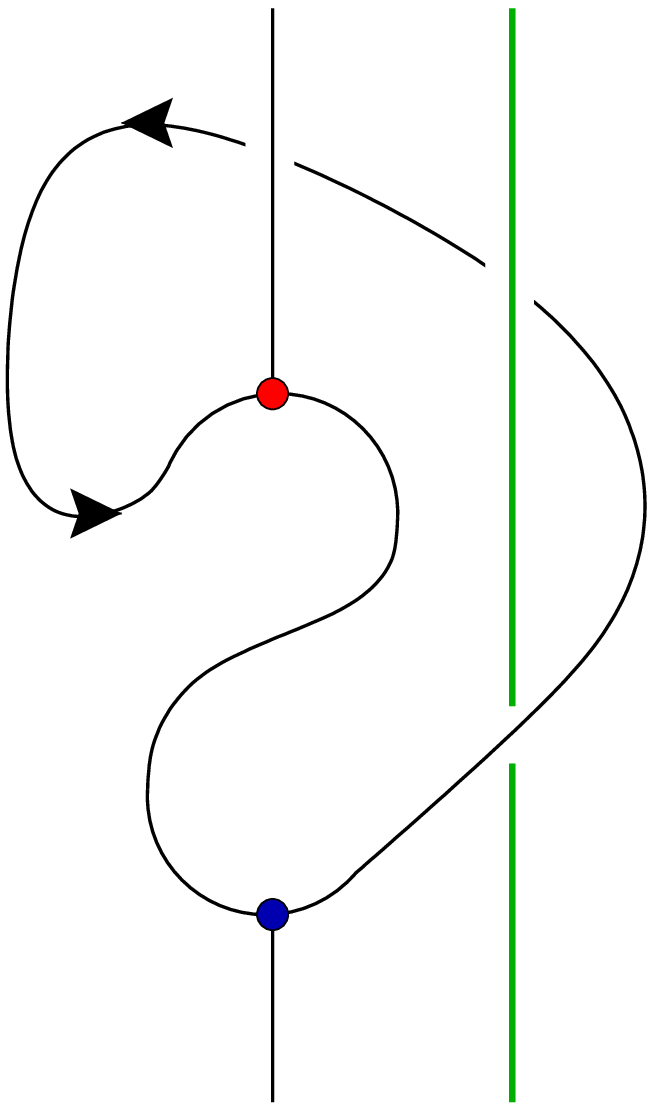}} \end{picture}}
  \put(-48.6,57.1) {$P_X\;:=$}
  \put(25.5,-7.9) {\scriptsize$A$}
  \put(52.5,-7.9) {\scriptsize$X$}
  \put(25.5,125) {\scriptsize$A$}
  \put(53.3,125) {\scriptsize$X$}
   \put(72,67) {\scriptsize$A$}
   \put(10,32) {\scriptsize$A$}
  \epicture15\label{eq:PX-def} \labl{PX-def}

\dt{Lemma} $P_X$ is a projector:
  \be  P_X \circ P_X = P_X \,.  \ee

%\smallskip
   \newpage
   %% pagebreak enforced
\noindent
Proof:\\[.16em]
It is straightforward to establish the sequence 
  \bea \begin{picture}(395,117)(0,53)
  \put(0,0)   {\begin{picture}(0,0)(0,0)
              \scalebox{.38}{\includegraphics{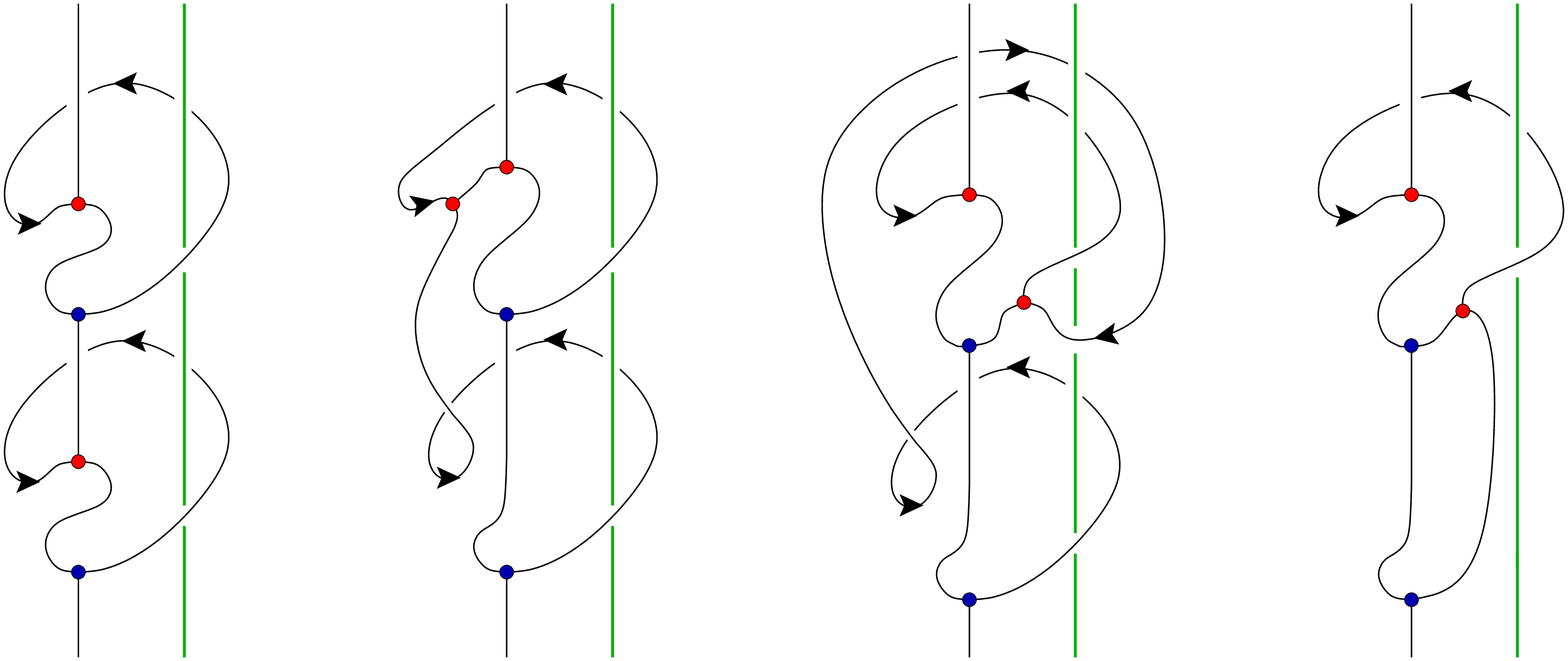}} \end{picture}}
  \put(14.5,-7.9) {\scriptsize$A$}
  \put(41.5,-7.9) {\scriptsize$X$}
  \put(14.5,167) {\scriptsize$A$}
  \put(41.5,167) {\scriptsize$X$}
  \put(122,-7.9) {\scriptsize$A$}
  \put(148,-7.9) {\scriptsize$X$}
  \put(122,167) {\scriptsize$A$}
  \put(148,167) {\scriptsize$X$}
  \put(237,-7.9) {\scriptsize$A$}
  \put(263,-7.9) {\scriptsize$X$}
  \put(237,167) {\scriptsize$A$}
  \put(263,167) {\scriptsize$X$}
  \put(346,-7.9) {\scriptsize$A$}
  \put(372,-7.9) {\scriptsize$X$}
  \put(346,167) {\scriptsize$A$}
  \put(372,167) {\scriptsize$X$}
  \put(72.6,80.1)  {$=$}
  \put(182.6,80.1) {$=$}
  \put(305.6,80.1) {$=$}
  \epicture36 \labl{PXPX}
of equalities for $P_X\cir P_X$.
By first using the Frobenius property and then the specialness of $A$
one concludes that the morphism on the \rhs\ is equal to $P_X$.
\qed

\medskip\noindent
Proof of Proposition \ref{ZZZ}:\\
Property (i) has already been derived: $Z$ is invariant under
$T$ and $U$, and hence under $T$ and $S$.
\\[.13em]
To obtain (ii), cut the picture \erf{Zij2} along an $S^2$ to arrive at
a ribbon graph in $S^2\,{\times}\,[0,1]$, which defines a linear map
  \be
  P_{ij}:\quad \calh(i,A,j;S^2) \,\to\, \calh(i,A,j;S^2) \,.  \ee
The coefficients $Z_{ij}$ are then recovered as
  \be  Z_{ij} = {\rm tr}_{\calH(i,A,j;S^2)}\, P_{ij} \,.  \ee
The morphism described by $P_{ij}$ is nothing but
$\id_{U_i}\oti P_j$ with $P_j\,{\equiv}\,P_{U_j}$ as defined in
\erf{eq:PX-def}. Since $P_j$ is a projector, it follows in particular
that $P_{ij}$ is a projector as well, 
  \be  P_{ij}\cir P_{ij} = P_{ij} \,.  \ee
Now the trace of a projector equals the dimension of its image. Hence
  \be  {\rm tr}_{\calH(i,A,j;S^2)}\,P_{ij}\in\zet_{\ge 0} \,, \ee
which establishes \erf{ZZZii}.
\\[.13em]
To show (iii) we write $Z_{00}\eq{\rm tr}_\calH P_{00}
\eq{\rm tr}_\calH P_0$ with $\calh\,{\equiv}\,\calh(A;S^2)$.
Then we use dominance to insert a basis in the $A$-line, leading to
$Z_{00}\eq\sum_{k,\alpha}\! {\rm tr}_\calH(P_0 \cir
\iaa k\alpha \cir \aai k\alpha)$ (the basis morphisms are those
defined in \erf{iaa,aai}). The space $\calh(k;S^2)$ of blocks is
non-zero only for $k\eq0$; thus we obtain
  \bea \begin{picture}(211,93)(30,59)
  \put(0,0)   {\begin{picture}(0,0)(0,0)
              \scalebox{.38}{\includegraphics{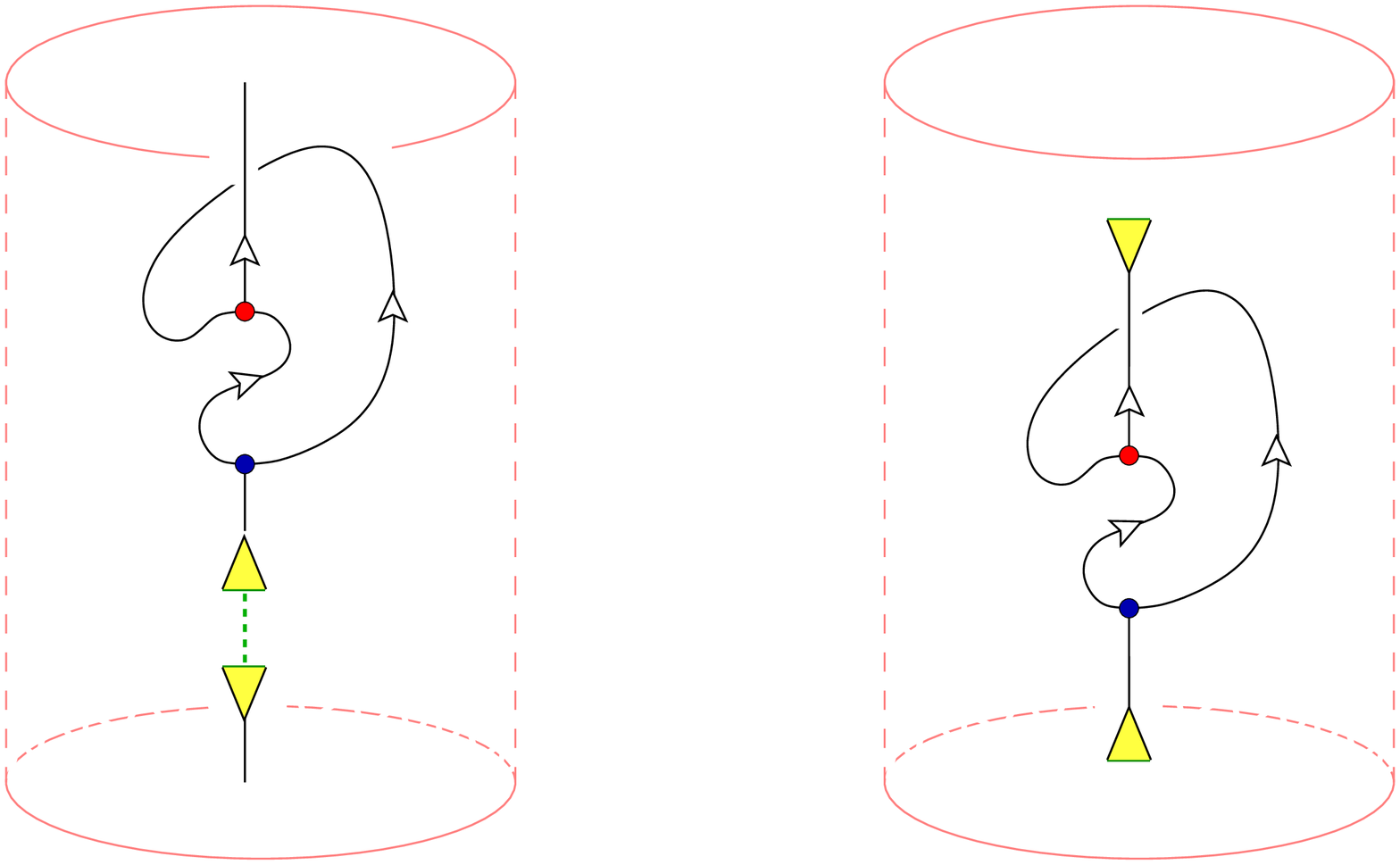}} \end{picture}}
  \put(47,53) {\scriptsize$\alpha$}
  \put(48,29) {\scriptsize$\overline\alpha$}
  \put(45,112) {\scriptsize$A$}
  \put(206,21) {\scriptsize$\alpha$}
  \put(192,107) {\scriptsize$\overline\alpha$}
  \put(204,86) {\scriptsize$A$}
  \put(-84.6,75.1){${\rm tr}_\calH P_0\;=\;\dsty\sum_\alpha$}
  \put(34.5,12.9) {\scriptsize$A$}
  \put(108.5,75.1){$=\,\ \dsty\sum_\alpha$}
  \put(260,75.1){$=\;{\rm tr}_{\!\ATop}P\top$}
  \epicture31 \labl{traceP0}
with the linear map
  \be  \begin{array}{rl} P\top:
  \quad \Atop \!\!& \to\, \Atop \\[.42em]
  \alpha \!\!& \mapsto\, P_0 \cir\alpha \,.  \eear \ee
The second equality of \erf{traceP0} is just a translation
in $S^2\,{\times}\,S^1$, while in the last step
the summation over $\alpha$ is recognised as a trace in \Atop.
We conclude that $Z_{00}\eq\dim\,{\rm Im}(P\top)$. Noting the identity
  \bea \begin{picture}(150,69)(0,33)
  \put(0,0)   {\begin{picture}(0,0)(0,0)
              \scalebox{.38}{\includegraphics{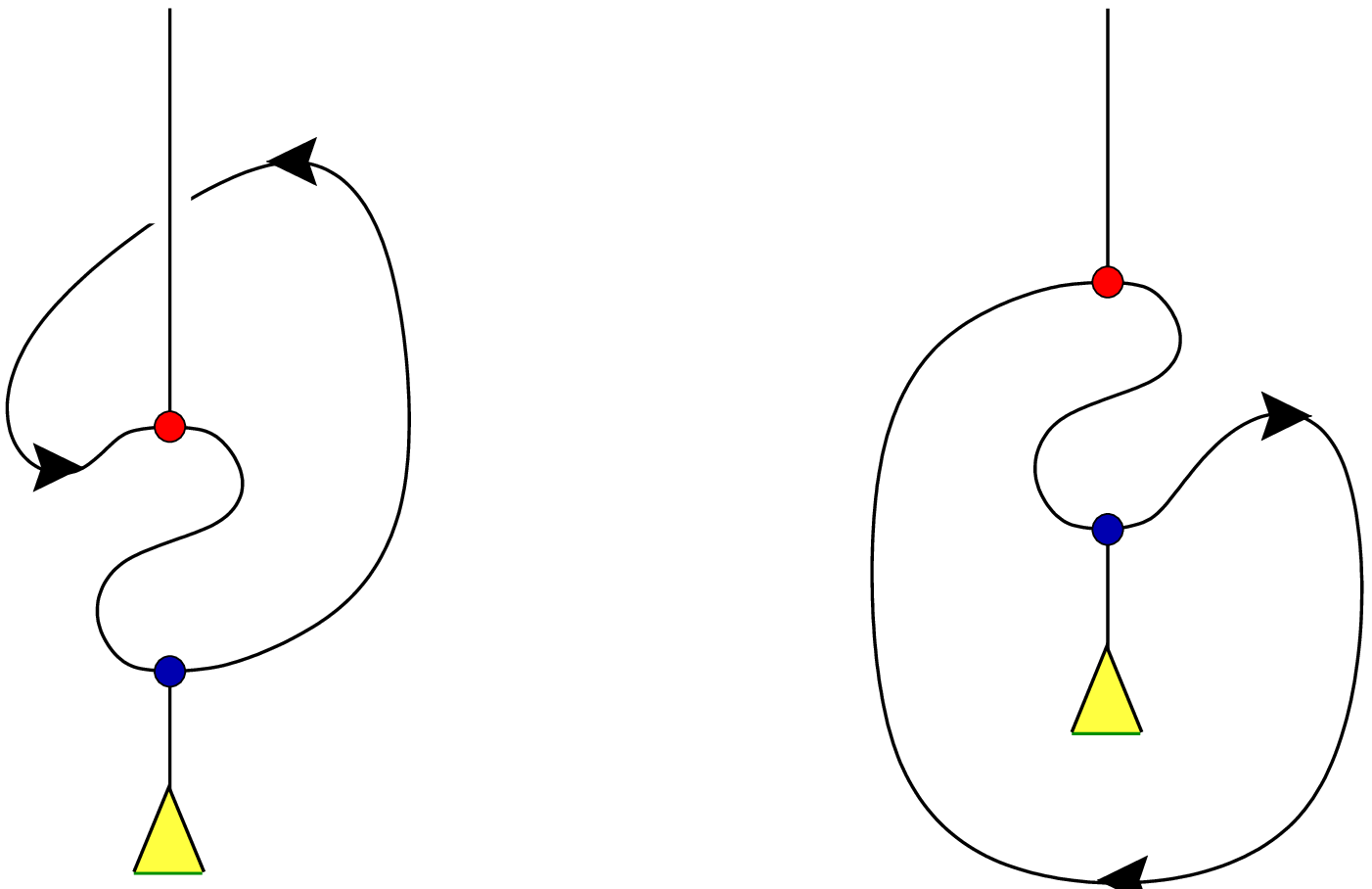}} \end{picture}}
   \put(22,7) {\scriptsize$\alpha$}
   \put(127,22) {\scriptsize$\alpha$}
   \put(16,103) {\scriptsize$A$}
   \put(121,103) {\scriptsize$A$}
  \put(69.6,50.1){$=$}
  \epicture14 \labl{P-top}
we can now use lemma \ref{lem:alpha-cent} to conclude that $P\top\alpha
\eq\alpha$ is equivalent to $\alpha \iN\centreA(\Atop)$. It follows that 
$\dim\,{\rm Im}(P\top)\eq\dim\,\centreA(\Atop)$, thus proving formula 
\erf{ZZZiii}.
\qed

\medskip

Let us denote the torus partition function \erf{Zij2}
by $Z(A)_{ij}$ so as to make its dependence on the algebra
object $A$ explicit. Denote further by $C$ the charge conjugation matrix
$C_{k\ell}\eq\delta_{\bar k,\ell}$ and set
  \be  \tilde Z(A)_{ij} := \sumI_k C_{ik}\, Z(A)_{kj} \,.  \ee
Also recall from section \ref{sec:sum-prod-op} the definition
of the opposite \alg\ $A_{\rm op}$ and of the direct sum
$A\,{\oplus}\,B$ and product $A\oti B$ of \alg s.

\dtl{Proposition}{prop:Ztilde}
The following relations are valid as matrix equations:
  \be  \bearll
  {\rm(i)}  & \tilde Z(A{\oplus}B) = \tilde Z(A) + \tilde Z(B) \,,
  \\{}\\[-.6em]
  {\rm(ii)} & \tilde Z(A \Oti B) = \tilde Z(A)\, \tilde Z(B) \,,
  \\{}\\[-.6em]
  {\rm(iii)}& \tilde Z(A_{\rm op}) = \tilde Z(A)\oT_{} \,,\qquad
  \mbox{or equivalently,}\qquad Z(A_{\rm op}) = Z(A)\oT_{} \,.
  \eear  \labl{ZABC}

%\medskip
\noindent
Proof:\\
(i) The definition of $Z(A)$ gives immediately 
  \bea \begin{picture}(325,93)(0,38)
  \put(0,0)   {\begin{picture}(0,0)(0,0)
              \scalebox{.38}{\includegraphics{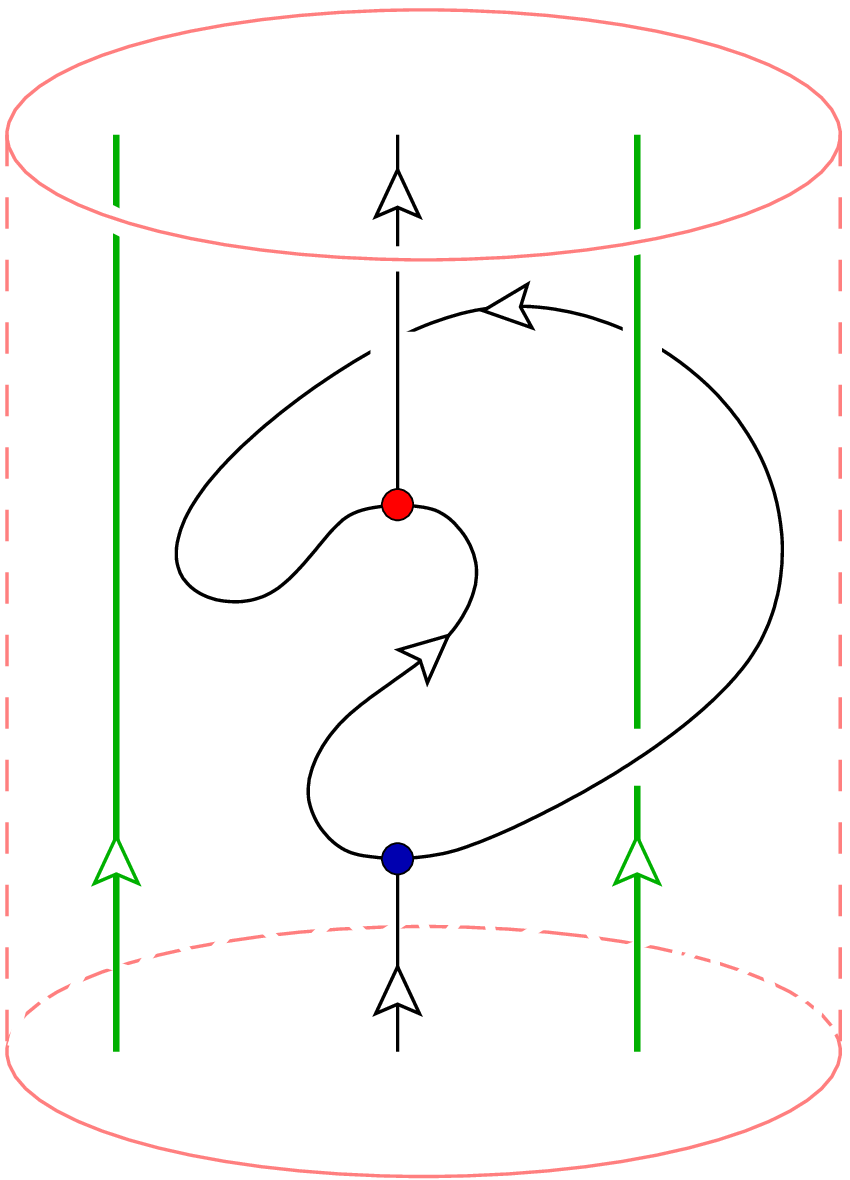}} \end{picture}}
  \put(-75.7,63.1) {$Z(A{\oplus}B)_{ij}\,=$}
  \put(15.5,14.9)  {\scriptsize$i$}
  \put(44.5,14.9)  {\scriptsize$A{\oplus}B$}
  \put(71.5,14.9)  {\scriptsize$j$}
  \put(105.7,63.1) {$=\,\dsty\sum_{X=A,B}$}
  \put(160,0)   {\begin{picture}(0,0)(0,0)
              \scalebox{.38}{\includegraphics{Zij2s.eps}} \end{picture}}
  \put(206.5,14.9) {\scriptsize$X$}
  \put(263.3,63.1) {$=\,Z(A)_{ij}+Z(B)_{ij}$}
  \epicture16 \labl{Zij2s}
in $S^2\,{ \times}\, S^1$.
Multiplying by $C$ from the left yields relation (\ref{ZABC}(i)).
\\[.13em]
(ii) For ribbon graphs in a solid torus $D \,{\times}\, S^1$ the relation
  \bea \begin{picture}(305,88)(0,38)
  \put(0,0)   {\begin{picture}(0,0)(0,0)
              \scalebox{.38}{\includegraphics{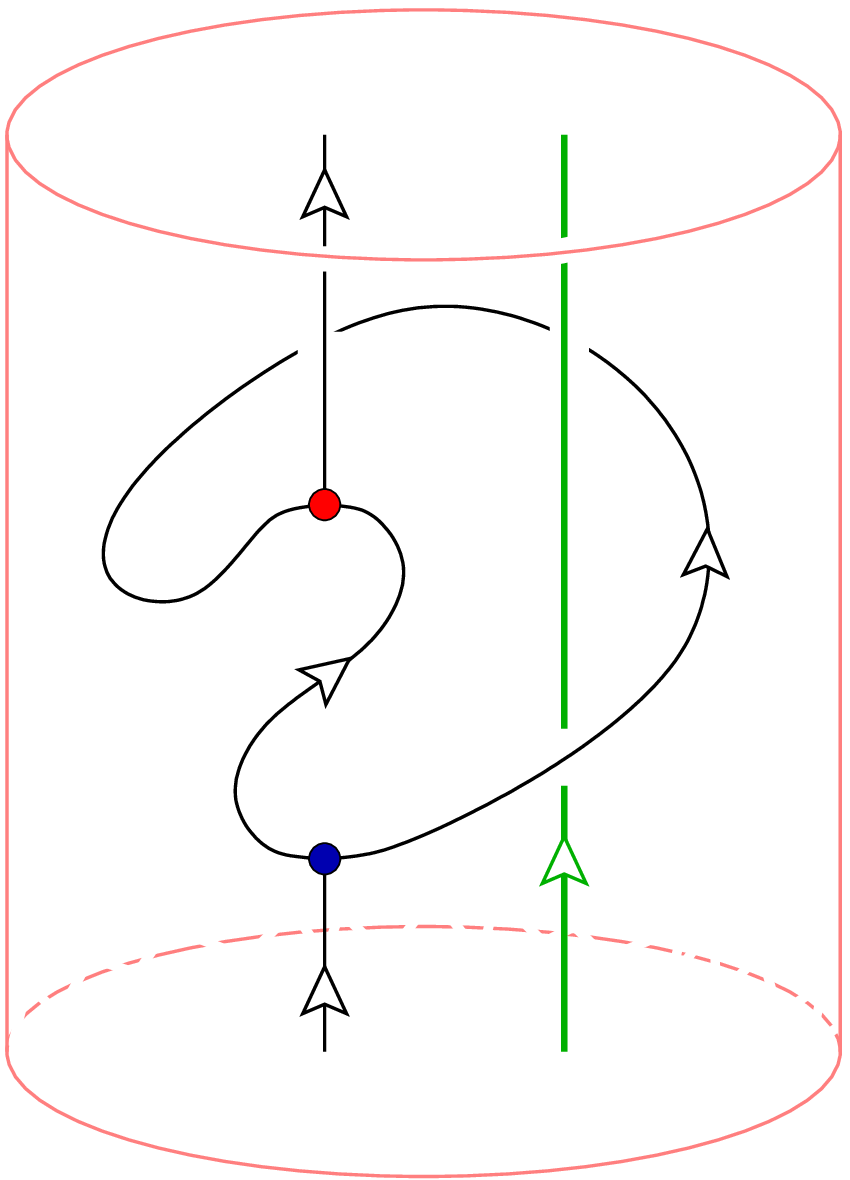}} \end{picture}}
  \put(25.8,14.9)  {\scriptsize$A$}
  \put(63.9,14.9)  {\scriptsize$j$}
  \put(113.7,63.1) {$=\;\ \dsty\sumI_k Z(A)_{kj}$}
  \put(199,0)   {\begin{picture}(0,0)(0,0)
              \scalebox{.38}{\includegraphics{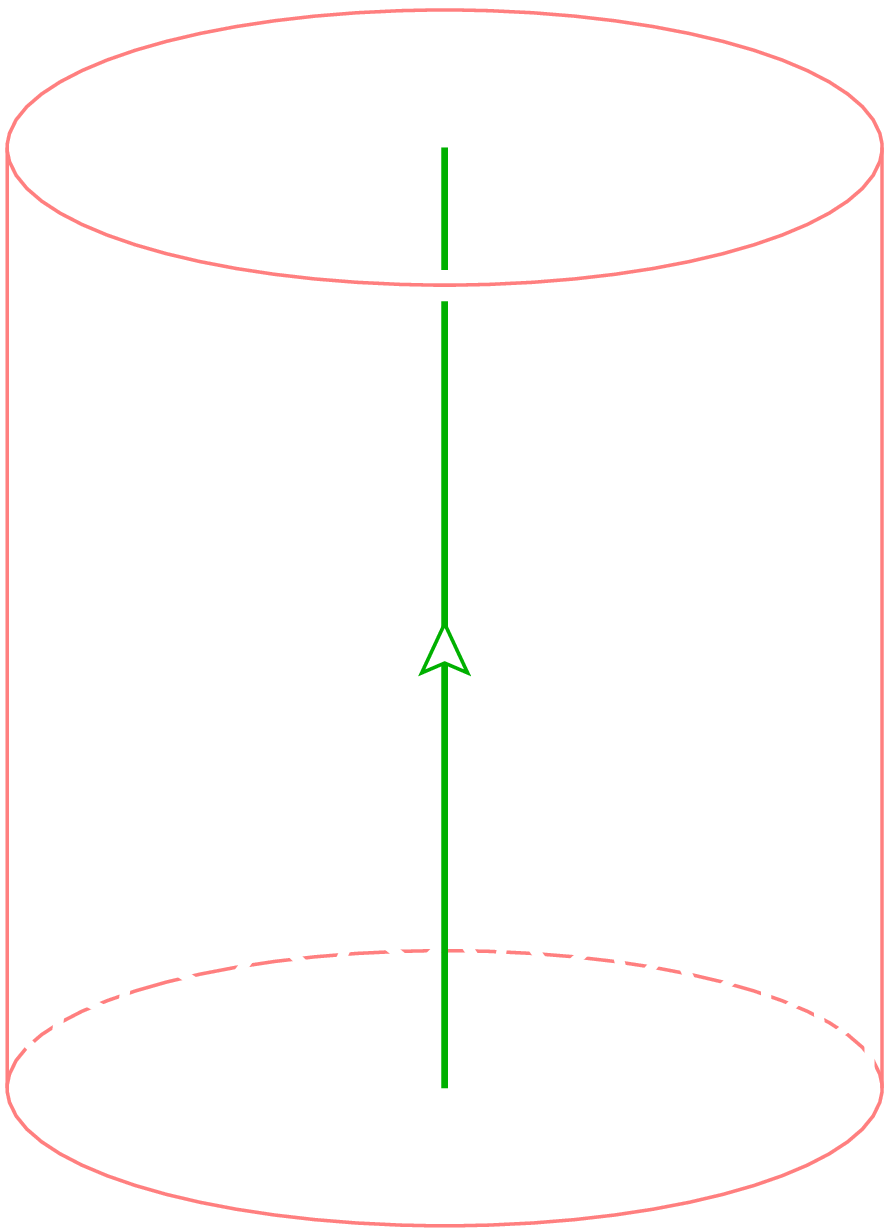}} \end{picture}}
  \put(249.8,14.9) {\scriptsize$\bar k$}
  \epicture16 \labl{eq:Z-in-torus}
holds. Here the \rhs\ is expanded in a basis of zero-point blocks on the
torus, as discussed in section \ref{sec:toruszero}. That the
coefficients are precisely given by $Z(A)_{kj}$ can be seen by gluing a 
$D\,{\times}\,S^1$ containing a single $i$-ribbon to the left and \rhs s 
of \erf{eq:Z-in-torus}. On the \lhs\ one thereby obtains the graph \erf{Zij2} 
for the torus partition function, while the \rhs\ gives $\delta_{ik}$.
%\\
Next note that the ribbon graph for $Z(A\Oti B)_{ij}$ can be deformed as 
  \bea \begin{picture}(270,97)(0,47)
  \put(0,0)   {\begin{picture}(0,0)(0,0)
              \scalebox{.38}{\includegraphics{Zij2s.eps}} \end{picture}}
  \put(-84.7,63.1) {$Z(A{\Oti}B)_{ij}\ =$}
  \put(15.5,14.9)  {\scriptsize$i$}
  \put(44.5,11.9)  {\scriptsize$A{\Oti}B$}
  \put(71.5,14.9)  {\scriptsize$j$}
  \put(113.7,63.1) {$=$}
  \put(144,-20) {\begin{picture}(0,0)(0,0)
              \scalebox{.38}{\includegraphics{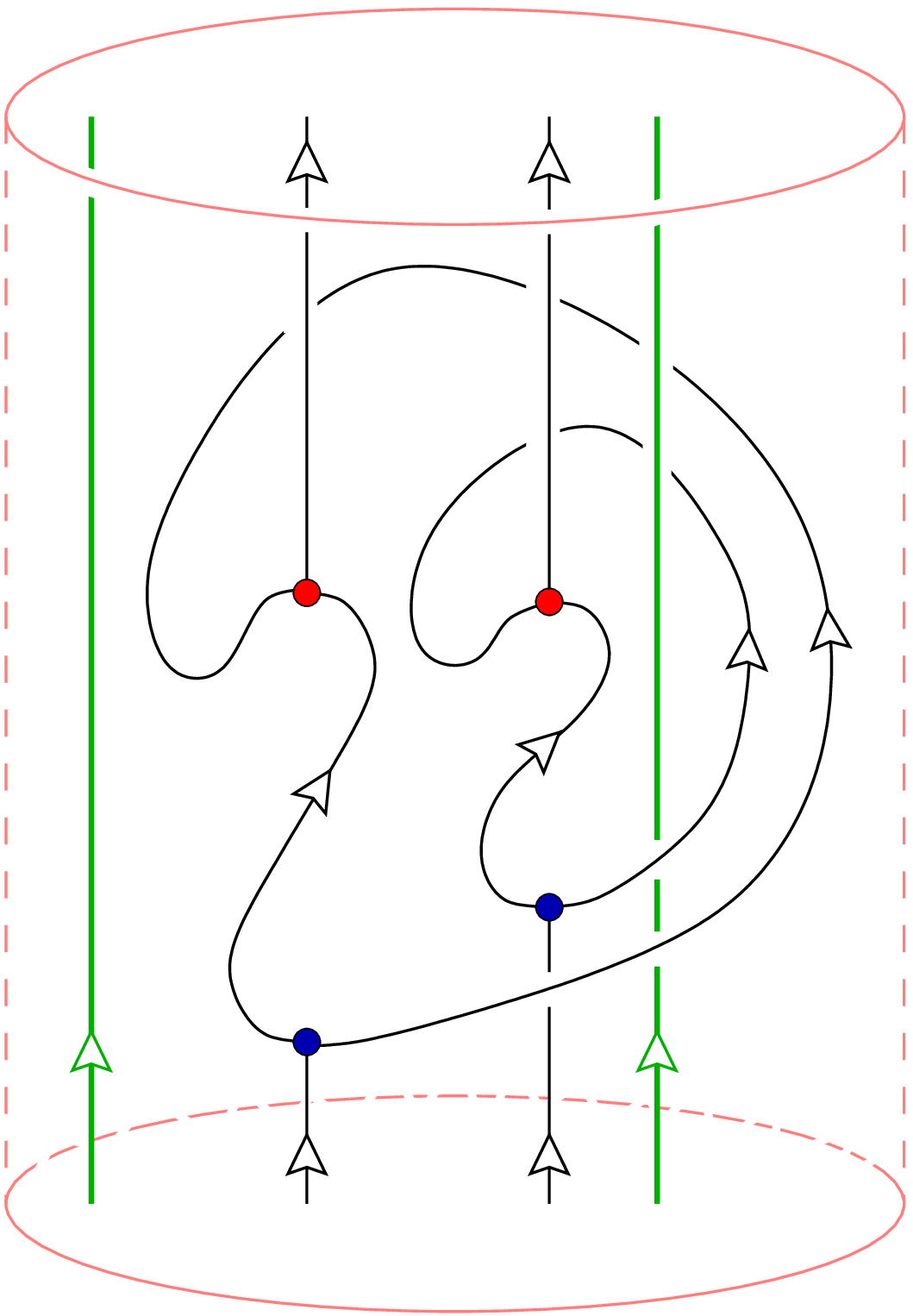}} \end{picture}}
  \put(157.5,-7.9) {\scriptsize$i$}
  \put(182.5,-7.9) {\scriptsize$A$}
  \put(205.5,-7.9) {\scriptsize$B$}
  \put(229.5,-7.9) {\scriptsize$j$}
  \epicture41 \labl{ZiABj}
Now we can apply surgery to cut out a solid torus containing the
$B$- and $j$-ribbons, and then use \erf{eq:Z-in-torus} to obtain
  \begin{eqnarray} && \begin{picture}(156,122)(-68,0)
  \put(0,0)   {\begin{picture}(0,0)(0,0)
              \scalebox{.38}{\includegraphics{Zij2s.eps}} \end{picture}}
  \put(-138.7,63.1){$Z(A{\Oti}B)_{ij}\,=\,\dsty\sumI_k Z(B)_{kj}$}
  \put(15.5,14.9)  {\scriptsize$i$}
  \put(44.5,14.9)  {\scriptsize$A$}
  \put(71.5,14.9)  {\scriptsize$\bar k$}
  \put(109.7,63.1) {$=\; \dsty\sumI_k Z(B)_{kj}\,Z(A)_{i\bar k}$}
% \put(166.7,38.8) {$=\, (Z(A) C Z(B))_{ij} \,.$}
  \end{picture} \nonumber\\[-4.5em]
  &&\hsp{20.5} =\, (Z(A)\, C\, Z(B))_{ij}^{} \,. 
  \\[1.2em]\nonumber \end{eqnarray} 
This implies (ii). 
\\[.18em]
(iii) The following relations are valid:
  \begin{eqnarray} && \begin{picture}(326,128)(0,0)
  \put(0,0)   {\begin{picture}(0,0)(0,0)
              \scalebox{.38}{\includegraphics{Zij2s.eps}} \end{picture}}
  \put(-67.7,63.1) {$Z(A_{\rm op})_{ij}\,=$}
  \put(15.5,14.9)  {\scriptsize$i$}
  \put(44.5,14.9)  {\scriptsize$A_{\rm op}$}
  \put(71.5,14.9)  {\scriptsize$j$}
   \put(45,76) {\scriptsize$m_{\rm op}$}
  \put(106.2,63.1) {$=$}
  \put(127,0) {\begin{picture}(0,0)(0,0)
              \scalebox{.38}{\includegraphics{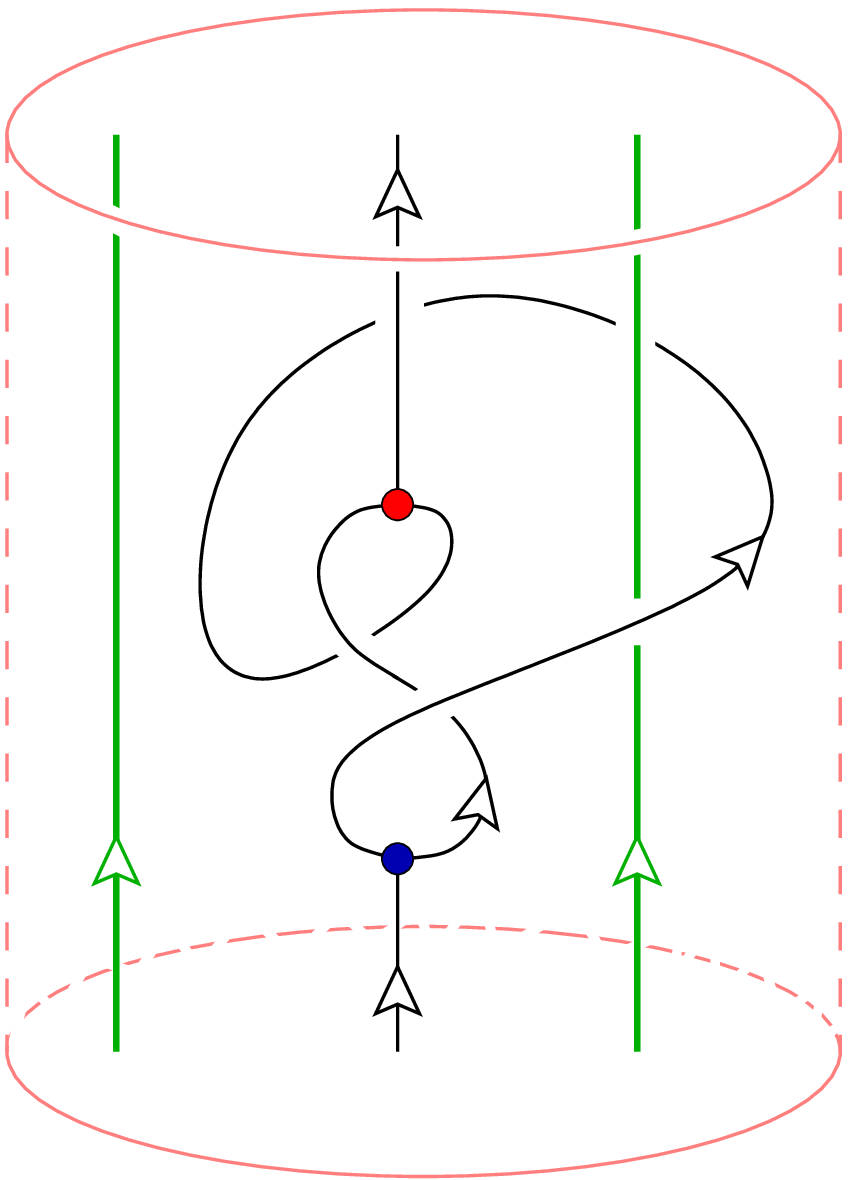}} \end{picture}
   \put(46,76) {\scriptsize$m$}
   \put(14,16) {\scriptsize$i$}
   \put(71,16) {\scriptsize$j$}
   \put(47,16) {\scriptsize$A$}
  }
  \put(230.2,63.1) {$=$}
  \put(251,0)   {\begin{picture}(0,0)(0,0)
              \scalebox{.38}{\includegraphics{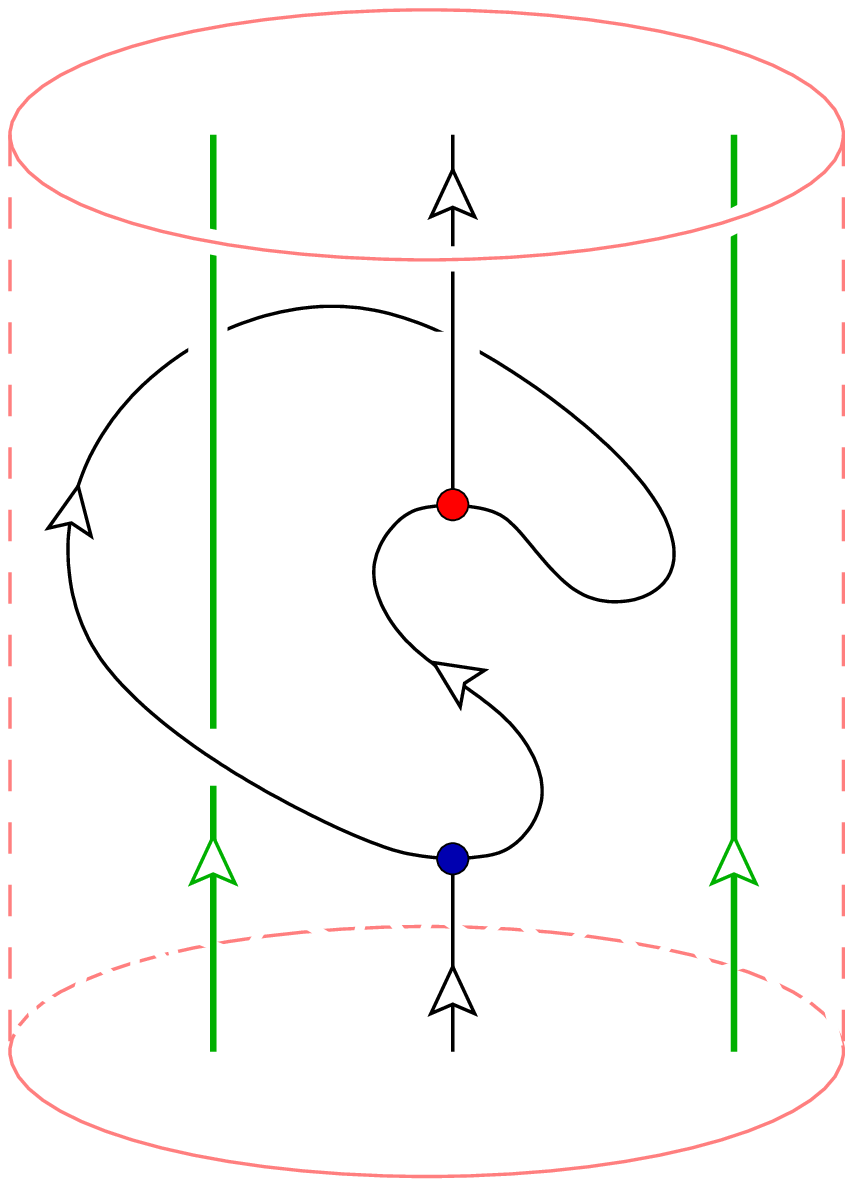}} \end{picture}
   \put(27,15) {\scriptsize$i$}
   \put(84,15) {\scriptsize$j$}
   \put(56,15) {\scriptsize$A$}
   \put(53,76) {\scriptsize$m$}
  }
% \put(300.3,-27.1){$=\,Z(A)_{ji}\,.$}
  \end{picture} \nonumber\\[.13em]
  && \hsp{25.9} =\,Z(A)_{ji}^{} \,. \label{ZAop}
  \\[-.9em]\nonumber \end{eqnarray} 
Here in the first step the definition of the multiplication on $A_{\rm op}$ 
in terms of $A$ is inserted. The second step uses the fact that in the 
`horizontal' direction the manifold has the topology of a two-sphere $S^2$. 
The $A$-ribbon is deformed around the two-sphere to wrap around the 
$i$-ribbon (this step is best checked by visualising it with actual ribbons,
rather than lines, which facilitates keeping track of the twists). Finally 
we move the $j$-ribbon to the left side of the graph and use Frobenius
and associativity properties to change the $A$-ribbon so as to match
the one appearing in the defining relation \erf{Zij2}.
\\
Thus we have $Z(A_{\rm op})\eq Z(A)\oT_{}$. This, in turn, is equivalent 
to $\tilde Z(A_{\rm op})\eq C\,\tilde Z(A)\oT_{}\,C$ and hence (because
of $[Z,C]\eq0$) to $\tilde Z(A_{\rm op})\eq\tilde Z(A)\oT_{}$.
\qed

\dt{Remark}
(i) Suppose $A$ and $B$ are algebras such that
$Z(A)_{00}\eq Z(B)_{00}\eq1$. Then $Z(A{\oplus} B)_{00}\eq2$. Accordingly
 the CFT should be interpreted as a superposition of two CFTs with
$Z_{00}\eq1$, see the brief discussion at the end of 
section~\ref{sec:alg2bCFT}. Unsurprisingly, it is a superposition 
of the CFTs associated to $A$ and $B$, as is confirmed on the
level of partition functions by proposition~\erf{prop:Ztilde}(i). 

\medskip\noindent
(ii) It is obvious that for two matrices $Z_1$ and $Z_2$ that commute
with $S$ and $T$, the product $Z_1 C Z_2$ also commutes with $S$ and $T$.
It is less obvious that this matrix appears as the torus partition 
function of any consistent CFT.
However, when both $Z_1$ and $Z_2$ are obtained from algebra objects,
then according to the result above $Z_1 C Z_2$ is obtained from an algebra 
object as well, and is therefore
indeed realised as the torus partition function of a CFT.
(If the tensor product $A\oti B$ is isomorphic to a direct sum of
algebras, the resulting theory can be interpreted as a superposition of 
several CFTs with $Z_{00}\eq1$, as discussed in point (i).)

\medskip\noindent
(iii) In terms of Morita classes of algebras, the tensor product of CFTs
amounts to a product given by $[A]\,{\times}\,[B]\,{:=}\,[A\Oti B]$. As an 
example, consider the WZW theories for $\su(2)_k$ and $\mathfrak u(1)_{2N}$.  
For $\su(2)_k$, denote by $[{\rm A}]$ the Morita class of all algebras 
giving rise to an A-series modular invariants, and similarly for $[{\rm D}]$
and for $[{\rm E}6]$, $[{\rm E}7]$ and $[{\rm E}8]$. The A-invariant occurs at
all levels $k$; its product with itself is $[{\rm A}]\,{\times}\,[{\rm A}]
\eq [{\rm A}]$. In all other cases, too, the multiplication is commutative 
and $[{\rm A}]$ acts as unit. Further, one computes
  \be\begin{array}{lll}
  k \iN 4\zet    : & \ [{\rm D}] \times [{\rm D}] = 2\, [{\rm D}] \,,
  \\[.45em]
  k \iN 4\zet{+}2: & \ [{\rm D}] \times [{\rm D}] = [{\rm A}] \,,
  \\[.45em]
  k=10           : & \ [{\rm D}] \times [{\rm E}_6] = [{\rm E}_6] \,,
                   & \ [{\rm E}_6] \times [{\rm E}_6] = 2\, [{\rm E}_6]\,,
  \\[.45em]
  k=16           : & \ [{\rm D}] \times [{\rm E}_7] = 2\, [{\rm E}_7]\,,
                   & \ [{\rm E}_7] \times [{\rm E_7}] = [{\rm D}] + [{\rm E}_7]\,,
  \\[.45em]
  k=28           : & \ [{\rm D}] \times [{\rm E}_8] = 2\, [{\rm E}_8]\,,
                   & \ [{\rm E}_8] \times [{\rm E}_8] = 4\, [{\rm E}_8]\,.
  \end{array}\ee
For $\mathfrak u(1)_{2N}$ the situation is more involved. We find
  \be
  [A_{2r}] \times [A_{2s}] = n \, [A_{2t}]  \ee
with integers $n$ and $t$ determined as follows: For $x\eq r,s$ set 
$g_x\,{:=}\,{\rm gcd}(x,N/x)$ and $\alpha_x\,{:=}\,N/(x g_x)$. Let $a_x$ 
be any integer obeying $a_x\,\alpha_x \,{\equiv}\,1\,{\rm mod}\,(x/g_x)$ 
(the result does not depend on the particular choice of $a_x$). Set 
further $G\,{:=}\,{\rm gcd}(N/g_r,N/g_s)$. Then $n$ and $t$ are given by
  \be
  n={\rm gcd}(g_r,g_s) \,, \qquad
  t=\frac{ n N }{G\, g_r g_s} \, 
  {\rm gcd}(g_r g_s (a_r\alpha_r{-}a_s\alpha_s)/n,G) \,.  \ee
One can verify that, as expected, $[A_{2N}]$ is the identity element,
and further $[A_{2r}] \,{\times}\, [A_2] \eq [A_{2N/r}]$ as well as
$[A_{2r}] \,{\times}\, [A_{2r}] \eq {\rm gcd}(r,N/r)\,
[A_{2\,{\rm lcm}(r,N/r)}]$. 

Note that the matrices $Z(A_{2r})$ are simple current modular invariants.
Products of simple current modular invariants 
have been considered in \cite{krSc}.

\medskip

As will be explained in more detail elsewhere, on orientable world sheets
bulk fields of the CFT determined by $A$ are triples $(i,j,\alpha)$, 
where $U_i$ and $U_j$ are simple objects and $\alpha$ is an element of
the space $\Hom(A{\otimes}U_j,U_i^\vee)$ of three-point couplings.
However, it turns out that not all triples $(i,j,\alpha)$ are allowed. 
Instead $\alpha$ has to be {\em local\/}, in the sense defined below.  

\dtl{Definition}{def:localHom}
Let $A$ be a symmetric special Frobenius algebra and $X,\,Y$ objects in 
\calc. A morphism $\varphi$ in the space $\Hom(A{\otimes}X,Y)$ is called 
{\em local\/} iff $\varphi \cir P_X\eq\varphi$. Local morphisms 
in $\Hom(A{\otimes}X,Y)$ are denoted by
  \be
  \Loc(A{\otimes}X,Y) := \{ \varphi \iN \Hom(A{\otimes}X,Y) \,|\,
  \varphi \cir P_X\eq\varphi \} \,.  \ee
(The morphism $P_X$ is defined in \erf{eq:PX-def}.)

\medskip

Consider a morphism $\alpha\iN\Hom(A{\otimes}U_j, U_{\bar\imath})$. 
Via $\alpha \,{\mapsto}\, \alpha \cir P_{U_j}$, the morphism $P_{U_j}$ 
induces a projector $p$ on $\alpha\iN\Hom(A{\otimes}U_j,U_{\bar\imath})$.
This follows immediately from \erf{PXPX}. Denote by 
$\{ \mu^{ij}_\alpha \}$ an eigenbasis of $p$, i.e.
  \be
  \{ \mu^{ij}_\alpha \} \; {\rm basis~of }\;
  \Hom(A{\otimes}U_j, U_{\bar\imath}) \quad {\rm s.t.} \quad
  \mu^{ij}_\alpha \circ P_{U_j} =  \eps \mu^{ij}_\alpha 
  \quad {\rm with}\;\eps \in \{0,1\} \,.  \labl{eq:mubasis}
Denote by $\{ \bar\mu^{ij}_\alpha \} \,{\subset}\, \Hom(U_{\bar\imath},
A{\otimes}U_j)$ a basis dual to $\{ \mu^{ij}_\alpha \}$, that is (using 
also dominance)
  \be
  \mu^{ij}_\alpha \cir \bar\mu^{ij}_\beta = \delta_{\alpha,\beta}\,
  \id_{{\bar\imath}} \qquad {\rm and} \qquad
  \sumI_{i} \sum_{\alpha} \bar\mu^{ij}_\alpha \cir \mu^{ij}_\alpha
  = \id_{A\otimes U_j} \,.  \labl{eq:mubasis-dual}
One verifies that $P_{U_j} \cir \bar\mu^{ij}_\alpha\eq\bar\mu^{ij}_\alpha$ 
iff $\mu^{ij}_\alpha\cir P_{U_j}\eq\mu^{ij}_\alpha$, and that it is
zero otherwise.

We also fix the bases
$\{ \varphi^{ij}_\alpha \}$ of $\Hom(A{\otimes}U_j, U_i^\vee)$ and
$\{ \bar \varphi^{ij}_\alpha \}$ of $\Hom(U_i^\vee, A{\otimes}U_j)$ via
  \be
  \varphi^{ij}_\alpha := \pi_{\bar\imath} \cir \mu^{ij}_\alpha
  \quad {\rm and} \quad
  \bar \varphi^{ij}_\alpha := \bar\mu^{ij}_\alpha\cir\pi_{\bar\imath}^{-1}\,.
  \labl{eq:phibasis}
One quickly checks that the bases $\{ \varphi^{ij}_\alpha \}$ and
$\{ \bar \varphi^{ij}_\alpha \}$ are dual to each other, similar to
\erf{eq:mubasis-dual}, and that
  \be
  \varphi^{ij}_\alpha \;{\rm local} \;\ \Leftrightarrow\;\ 
  \mu^{ij}_\alpha \;{\rm local} \,.  \ee

\medskip

Unless mentioned otherwise, from here on, whenever summing over a basis 
of morphisms in $\Hom(A{\otimes}U_i,U_j^\vee)$, $\Hom(A{\otimes}U_i,U_j)$, 
$\Hom(U_j^\vee,A{\otimes}U_i)$ or $\Hom(U_j,A{\otimes}U_i)$,
it is understood that the basis is chosen in the manner described above.

\medskip

It was stated above that bulk fields are labelled by elements in 
$\Loc(A{\otimes}U_j,U_i^\vee)$. This is consistent with the following 
observation.
\vspace{-.7em}

\dtl{Lemma}{lem:dim-loc-Hom}
The dimension of the subspace $\Loc(A{\otimes}U_j,U_i^\vee)$ of local 
morphisms in $\Hom(A{\otimes}U_j,U_i^\vee)$ is equal to $Z_{ij}$:
  \be  \dim\llb\Loc(A{\otimes}U_j,U_i^\vee)\lrb = Z_{ij} \,.  \ee 

\smallskip\noindent
Proof:\\
Using dominance and the fact that $\calh(i,(m,-){;}\,S^2)$ is
non-zero only for $i\eq m$, we can rewrite $Z_{ij}$ from \erf{Zij2} as
  \bea \begin{picture}(250,92)(0,30)
  \put(20,0)   {\begin{picture}(0,0)(0,0)
              \scalebox{.38}{\includegraphics{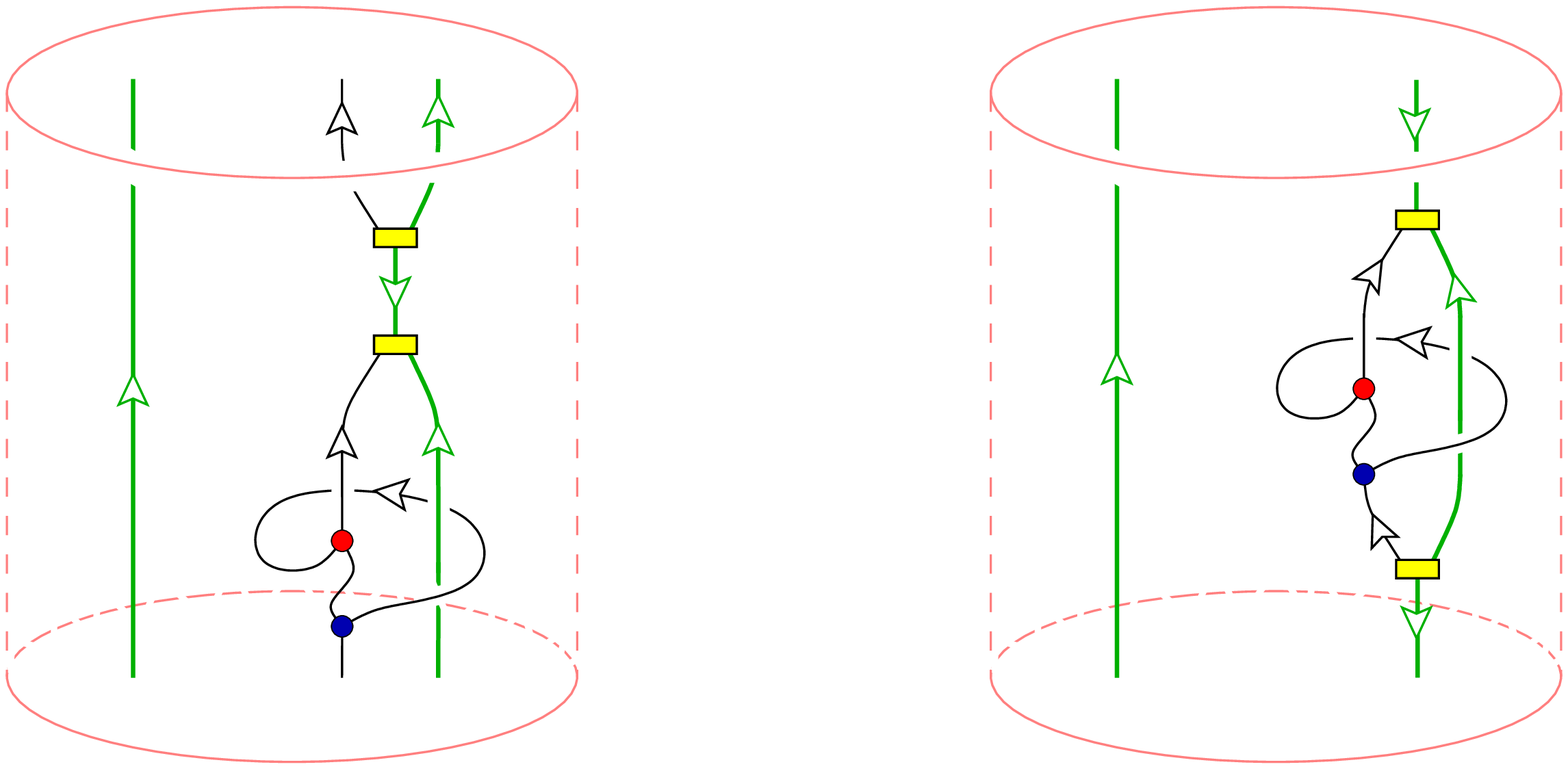}} \end{picture}
   \put(46,16) {\scriptsize$A$}
   \put(46,106) {\scriptsize$A$}
   \put(53,81.9) {\tiny$\overline\alpha$}
   \put(53.4,66.3) {\tiny$\alpha$}
   \put(66,75) {\scriptsize$m$}
   \put(22,17) {\scriptsize$i$}
   \put(72,17) {\scriptsize$j$}
  \put(-70,58){$Z_{ij} \;=\; \dsty \sum_{m,\alpha}$}
  \put(110,58){$=\; \dsty \sum_\alpha$}
   \put(180,17) {\scriptsize$i$}
   \put(229,17) {\scriptsize$i$}
   \put(217.4,86) {\tiny$\alpha$}
   \put(217,28.9) {\tiny$\overline\alpha$}
   \put(208,46) {\scriptsize$A$}
   \put(235,40) {\scriptsize$j$}
  }
  \epicture13  \labl{tor-pic8}
Here the sum over $\alpha$ runs over the basis introduced in \erf{eq:phibasis}.
The last ribbon graph in \erf{tor-pic8} can be seen to contain the
element $\varphi^{ij}_\alpha \circ P_{U_j}$, thus the sum can be restricted to
a basis of $\Loc(A{\otimes}U_j, U_i^\vee)$. By definition of the basis
$\varphi^{ij}_\alpha$ we can replace $\varphi^{ij}_\alpha \circ P_{U_j}$
by $\varphi^{ij}_\alpha$ in the last graph of \erf{tor-pic8}. But now 
$\varphi^{ij}_\alpha$ and its dual cancel to $\id_{U_i^\vee}$ and the
resulting ribbon graph takes the constant value 1. The sum is thus equal to
the number of local basis elements in $\Hom(A{\otimes}U_j, U_i^\vee)$.
\qed

\subsection{Bulk fields and $\alpha$-induced bimodules}\label{sec:alphaind}

We now further investigate the space of bulk fields. This allow us in
particular to show that our prescription for the modular invariant
torus \parfu\ coincides with the one obtained by different methods
in \cite{boek2}, and at the same time give a deeper understanding of the 
space $\Loc(A{\otimes}X,Y)$ of local morphisms. Bimodules over $A$ are a
crucial ingredient in this analysis, so
we start with a few comments on the category \calcaa\ of \AA-bimodules.
In contrast to the category of left $A$-modules, \calcaa\ is 
naturally endowed with the structure of a tensor category, with tensor 
product $M\,{\otimes_A}\, N$ defined to be the tensor product over $A$. 
The tensor unit of \calcaa\ is $A$ itself. We denote the morphism 
spaces of \calcaa\ by $\Hom_{A|A}(\cdot\,,\cdot)$.

The following prescription defines two tensor functors $\alpha^{(\pm)}$ 
from \calc\ to \calcaa. For every object $V$ of \calc\ the induced left 
module $\inda(V)\eq(A\Oti V,m\Oti\id_V)$ can be endowed with
two different structures of a right $A$-module: In the first case the
right action $\r_\rmr\iN\Hom(A\oti V\oti A,A)$ is
  \be  \r_\rmr^{(+)} := (m\oti\id_V) \cir (\id_A\oti c_{V,A}) \,,  \ee
and in the second case,
  \be  \r_\rmr^{(-)} := (m\oti\id_V) \cir (\id_A\oti(c_{A,V})^{-1})\,.
  \ee 
Since $A$ is associative, both of these right actions commute with the 
left action of $A$ on $A\oti V$. We denote the two induced bimodules 
obtained this way by $\alpha^{(\pm)}(V)$. (In the introduction we have 
used the notation $(A{\otimes}V)^\pm$ instead, because it adapts easier 
to the general situation, where one has $(B{\otimes}V)^\pm$ with $V$ 
an object of \calc\ and $B$ an arbitrary \AA-bimodule.) Both functors 
$\alpha^{(\pm)}$ act on morphisms $f\iN\Hom(V,W)$ in the same way as 
the induction functor of left modules:
  \be \alpha^{(\pm)}(f) := \id_A \oti f \,\in\Hom(A\oti V,A\oti W)\,.
  \ee

The two tensor functors $\alpha^{(\pm)}$ have first appeared in the 
theory of subfactors \cite{lore}, where they play a crucial role (see 
e.g.\ \cite{xu3,boev,boek,boev5,boek3}) and have been termed 
$\alpha$-{\em induction\/}. The category-theoretic reformulation 
presented here was obtained in \cite{ostr}. For those cases where 
$A$ describes the embedding of a subfactor into a factor, 
it has been shown \cite{boek2} that the matrix with entries
  \be  Z^{(\rm BEK)}_{k,l} := \dim\llb\Hom_{A|A}
  (\alpha^{(-)}(U_k),\alpha^{(+)}(U_l))\lrb  \labl{zbek}
($k,l\iN\II$) commutes with the matrices that describe the action of 
the modular group and hence provides a modular invariant. 

We now show that the matrix \erf{zbek} is related to the modular invariant 
partition function discussed in section \ref{sec:torus-pf} above by
  \be  Z^{(\rm BEK)}_{k,l} = Z_{\bar l,k} \,.  \labl{Zbek}
Along with the construction of general correlators outlined in 
\cite{fuRs}, this shows in particular that $Z^{(\rm BEK)}$ is realised 
as the torus partition function of a physical conformal field theory,  
a property that does not follow from the subfactor considerations.

To derive the relation \erf{Zbek} we make use of lemma 
\ref{lem:dim-loc-Hom}, in view of which
we need to relate the space $\Loc(A{\otimes}X,Y)$ of local morphisms 
to morphisms of $\alpha$-induced bimodules. This is done as follows.
\vspace{-.9em}

\dt{Proposition}
For any symmetric special Frobenius algebra $A$ in a braided tensor
category the linear map
  \be  \begin{array}{lc}
  f : \ & \Loc(A{\otimes}X,Y) \longrightarrow 
  \Hom_{A|A}(\alpha^-(X),\alpha^+(Y)) \\{}\\[-.88em]
  & \hsp{4.8} \beta \,\longmapsto\, (\id_A\Oti\beta)\cir(\Delta\Oti\id_X)
  \eear  \ee
is an isomorphism. \\
Its (left and right) inverse is given by $g{:}\ \gamma \,{\mapsto}\,(\eps
\Oti\id_Y)\cir\gamma$ for $\gamma \iN \Hom_{A|A}(\alpha^-(X), \alpha^+(Y))$.

\medskip\noindent
Proof:
\\[.16em]
We first show that $f$ maps into the correct space.
That $f$ is a linear map from $\Loc(A{\otimes}X,Y)$ to
$\Hom(\alpha^-(X),\alpha^+(Y))$ is obvious. What we still must check 
is that $f(\beta)$ is in fact even in $\Hom_{A|A}$, i.e.\ a morphism of
\AA-bimodules; this is done as follows.
\\
It is a direct consequence of the Frobenius property of $A$ that 
for every $\beta\iN\Loc(A{\otimes}X,Y)$ the image $f(\beta)$ is a
morphism of left $A$-modules. To see that it is a morphism of right 
$A$-modules as well, we consider the equalities
  \bea \begin{picture}(400,155)(0,12)
  \put(0,0)   {\begin{picture}(0,0)(0,0)
              \scalebox{.38}{\includegraphics{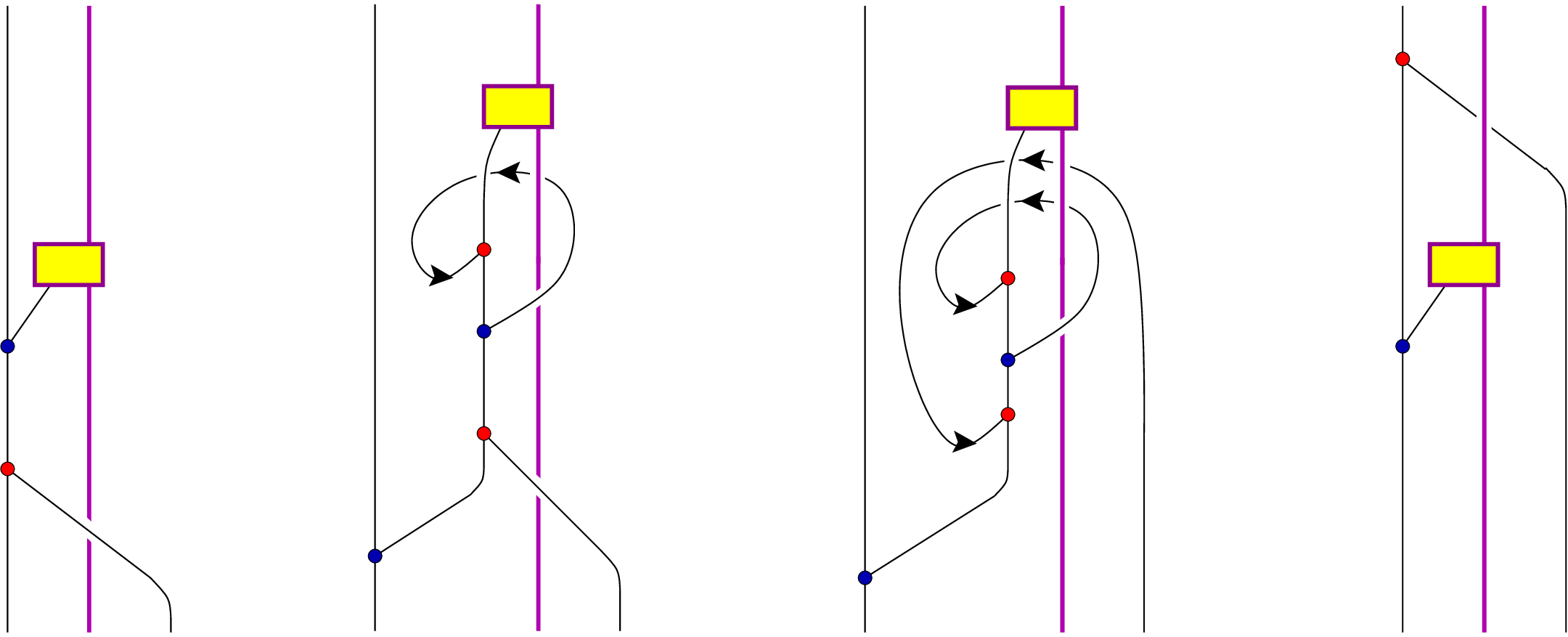}} \end{picture}
   \put(-2,-8) {\scriptsize$A$}
   \put(-2,164) {\scriptsize$A$}
   \put(39,-8) {\scriptsize$A$}
   \put(91,-8) {\scriptsize$A$}
   \put(91,164) {\scriptsize$A$}
   \put(152,-8) {\scriptsize$A$}
   \put(214,-8) {\scriptsize$A$}
   \put(214,164) {\scriptsize$A$}
   \put(152,-8) {\scriptsize$A$}
   \put(214,-8) {\scriptsize$A$}
   \put(214,164) {\scriptsize$A$}
   \put(284,-8) {\scriptsize$A$}
   \put(349,-8) {\scriptsize$A$}
   \put(349,164) {\scriptsize$A$}
   \put(390,-8) {\scriptsize$A$}
   \put(19,-8) {\scriptsize$X$}
   \put(132,-8) {\scriptsize$X$}
   \put(264,-8) {\scriptsize$X$}
   \put(370,-8) {\scriptsize$X$}
   \put(19,164) {\scriptsize$Y$}
   \put(132,164) {\scriptsize$Y$}
   \put(264,164) {\scriptsize$Y$}
   \put(370,164) {\scriptsize$Y$}
   \put(15,91.3) {\scriptsize$\beta$}
   \put(128,132) {\scriptsize$\beta$}
   \put(260,132) {\scriptsize$\beta$}
   \put(367,91.3){\scriptsize$\beta$}
   \put(53,76) {$=$}
   \put(170,76) {$=$}
   \put(313,76) {$=$}
  }
  \epicture02
  \ee
In the first equality one uses the Frobenius property of $A$ and the
locality property of $\beta$ to insert a projector $P_X$ (as defined in 
formula \erf{PX-def}). The second step summarises a number of moves that 
take the $A$-ribbon around the projector; these employ the Frobenius and 
associativity properties of $A$. The third step uses again that $\beta$ 
is local, which allows one to leave out the projector, and that $A$ is 
symmetric Frobenius to take the $A$-ribbon past the comultiplication.
Notice that on $A\oti X$ the action of $A$ from the right is defined
via the braiding, while on $A\oti Y$ it is defined using the inverse
of the braiding.
\\[.16em]
Next we establish that $g$ maps into the correct space, too.
By definition, $g$ is a linear map from $\Hom_{A|A}(\alpha^-(X),\alpha^+(Y))$
to $\Hom(A{\otimes}X,Y)$. It remains to show that $g$ is local.
\\
For every $\gamma\iN\Hom_{A|A}(\alpha^-(X),\alpha^+(Y))$ we have the 
equalities 
  \bea \begin{picture}(400,112)(0,15)
  \put(0,0)   {\begin{picture}(0,0)(0,0)
              \scalebox{.38}{\includegraphics{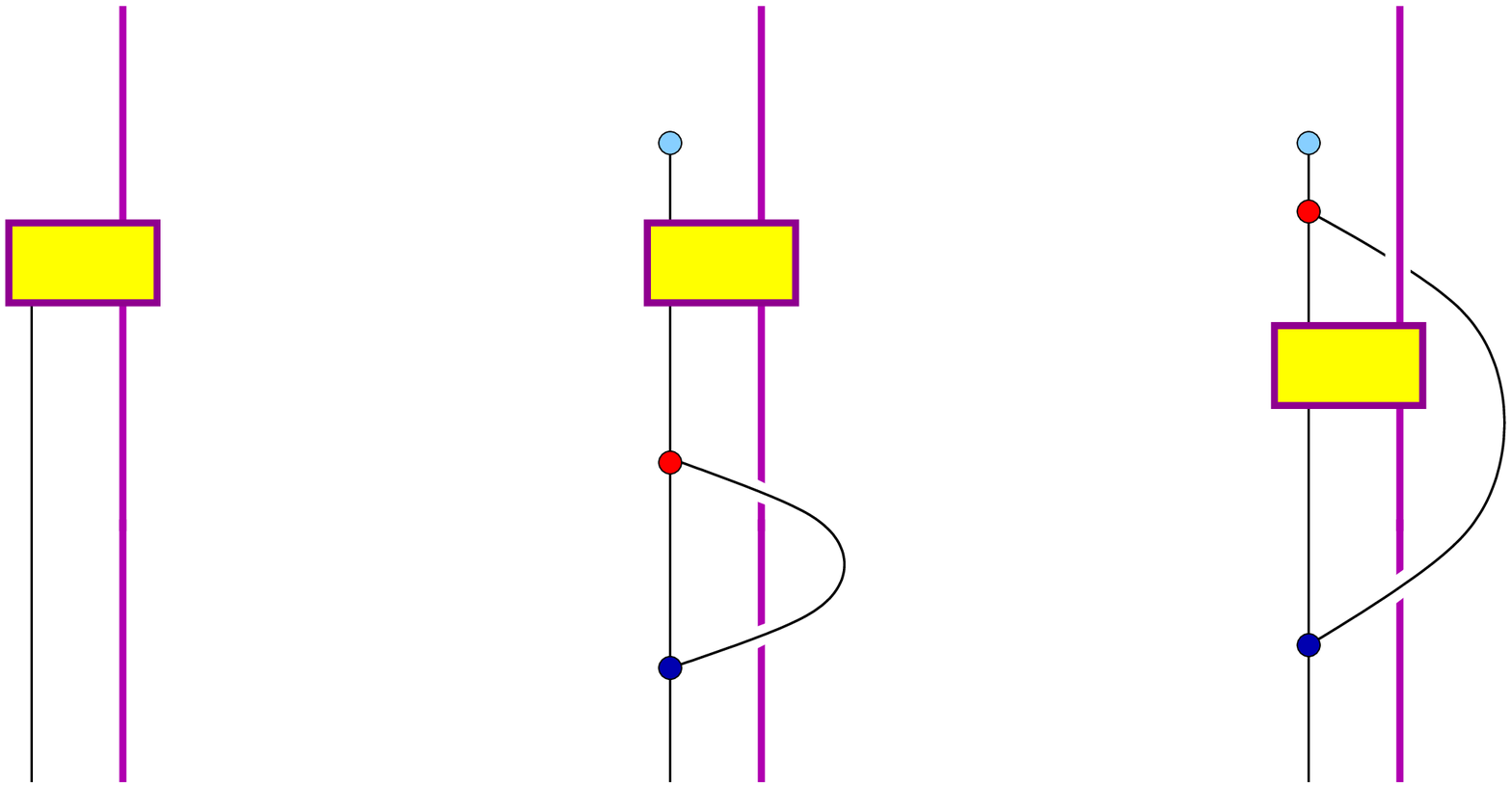}} \end{picture}
   \put(51,56) {$=$}
   \put(140,56) {$=$}
   \put(249,56) {$=$}
   \put(341,56) {$=$}
   \put(2,-7) {\scriptsize$A$}
   \put(98,-7) {\scriptsize$A$}
   \put(193,-7) {\scriptsize$A$}
   \put(289,-7) {\scriptsize$A$}
   \put(385,-7) {\scriptsize$A$}
   \put(15,-7) {\scriptsize$X$}
   \put(111,-7) {\scriptsize$X$}
   \put(208,-7) {\scriptsize$X$}
   \put(303,-7) {\scriptsize$X$}
   \put(399,-7) {\scriptsize$X$}
   \put(15,122) {\scriptsize$Y$}
   \put(111,122) {\scriptsize$Y$}
   \put(208,122) {\scriptsize$Y$}
   \put(303,122) {\scriptsize$Y$}
   \put(399,122) {\scriptsize$Y$}
   \put(4.9,77.4) {\scriptsize$g(\gamma)$}
   \put(105.4,78.7) {\scriptsize$\gamma$}
   \put(200.3,62.4) {\scriptsize$\gamma$}
   \put(296.7,59.4) {\scriptsize$\gamma$}
   \put(393,78.4) {\scriptsize$\gamma$}
  }
  \epicture01
  \labl{gbeta}
In the first step the definition of $g(\gamma)$ is inserted, and it is 
used that $A$ is special. The second step follows because $\gamma$ is 
a morphism of right $A$-modules. The third step employs symmetry of $A$,
and the last one that $\gamma$ is a left $A$-morphism. Comparing the 
left and right sides of \erf{gbeta}, we see that $g(\gamma)\eq 
g(\gamma)\cir P_X$, which means that indeed $g(\gamma)\iN\Loc(A\Oti X,Y)$.
\\[.16em]
Finally we show that $f$ and $g$ are each others' inverses. That
$g\cir f\eq\id$ on $\Loc(A{\otimes}X,Y)$ is an immediate consequence of
the defining property of the counit. To see that also
$f(g(\gamma))\eq\gamma$ for all $\gamma\iN\Hom_{A|A}(\alpha^-(X),\alpha^+(Y))$ 
one uses that $A$ is Frobenius and that $\gamma$ is in particular a left 
$A$-morphism. These properties imply that $f$ is injective and surjective.
\qed

\subsection{The extended left and right chiral algebras} 
\label{sec:left-right-chiral}

Let us recall that the chiral algebra $\CA$, which corresponds to a 
subsector of the holomorphic fields of a CFT, is the vertex operator \alg\
whose representation theory gives rise to the modular category \calc. 
The full, maximally extended, chiral algebra of the CFT can be
larger than $\CA$. It can also be different for the left and right
(holomorphic and anti-holomorphic) parts of the CFT.
Let us denote these maximally extended left and right chiral algebras
by $A_\ell$ and $A_r$, \resp. Their $\CA$-representation content can be read 
off the torus partition function $Z$ as
  \be
  A_\ell \,\cong\, \bigoplus_{j\in\II} \, Z_{j0}\, U_j \qquad {\rm and} \qquad
  A_r \,\cong\, \bigoplus_{i\in\II} \, Z_{0i}\, U_i \,.  \labl{eq:lr-chiral-Z}
These left and right chiral algebras also turn out to possess a nice 
interpretation in terms of the
algebra object $A$. To this end let us introduce the notion of the left and
right center of an algebra \cite{ostr}.

\dtl{Definition}{def:lr-centre}
The {\em left center\/} $\centlA$ of an algebra $A$ is the maximal subobject
of $A$ such that
  \be  m \circ c_{A,A} \circ (\beta_\ell \oti \id_A) =
  m \circ (\beta_\ell \oti \id_A) \,,  \ee
with $\beta_\ell \iN \Hom(\centlA,A)$ is the embedding morphism of $\centlA$ 
into $A$.\\[.13em]
The {\em right center\/} $\centrA$ of an algebra $A$ is the maximal subobject
of $A$ such that the equality $m \cir c_{A,A} \cir (\id_A \oti \beta_r)
\eq m \cir (\id_A \oti \beta_r)$ holds.\\[.13em]
In pictures,
  \bea \begin{picture}(235,58)(0,28)
  \put(0,0)   {\begin{picture}(0,0)(0,0)
              \scalebox{.38}{\includegraphics{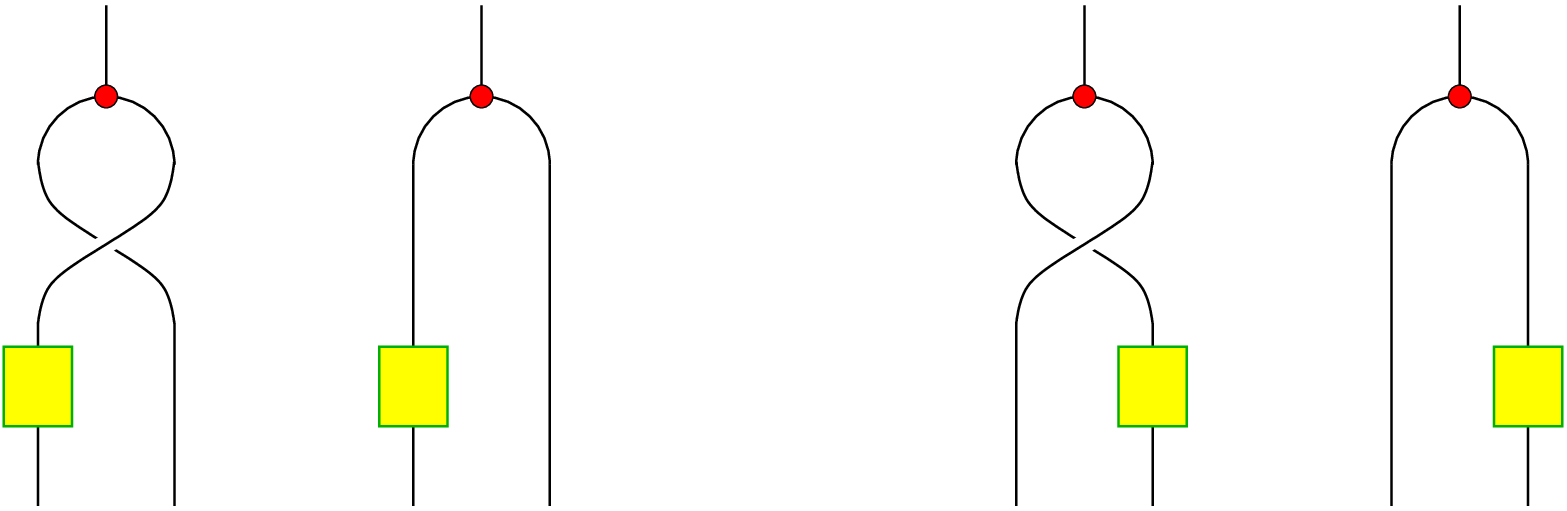}} \end{picture}}
  \put(1.8,16.4)   {\scriptsize$\beta_\ell$}  
  \put(-6,-9.1)   {\scriptsize$\centlA$}  
  \put(12.2,79.5)  {\scriptsize$A$}  
  \put(22.4,-9.1)  {\scriptsize$A$}  
  \put(40.0,32)    {$=$}
  \put(58.3,16.4)  {\scriptsize$\beta_\ell$}  
  \put(51,-9.1)  {\scriptsize$\centlA$}  
  \put(68.9,79.5)  {\scriptsize$A$}  
  \put(78.4,-9.1)  {\scriptsize$A$}  
  \put(109.5,32)   {and}
  \put(149.8,-9.1) {\scriptsize$A$}  
  \put(160.8,79.5) {\scriptsize$A$}  
  \put(169.6,16.4) {\scriptsize$\beta_r$}  
  \put(166,-9.1) {\scriptsize$\centrA$}  
  \put(189.0,34)   {$=$}
  \put(205.1,-9.1) {\scriptsize$A$}  
  \put(214.8,79.5) {\scriptsize$A$}  
  \put(226,16.4) {\scriptsize$\beta_r$}  
  \put(220,-9.1) {\scriptsize$\centrA$}  
  \epicture21 \ee

As already mentioned, \twodim\ conformal field theories that are even
topological are included in our construction by specialising to the
tensor category of complex vector spaces. In this specific context,
the idea has been put forward \cite{moor10,moor12,laza} that the boundary
fields yield a non-commutative Frobenius algebra -- the \alg\ of open string
states -- and that its center describes an \alg\ of bulk fields (closed 
string states). We will now see how the situation looks like for general
conformal field theories.

\medskip

An important property \cite{ostr} of the left and right centers of an
algebra $A$ is that they are invariant under Morita equivalence. This is 
consistent with the fact that the CFT only depends on the Morita class of 
the algebra $A$, so that physical quantities extracted from the algebra $A$
must be invariant under Morita equivalence, too.
Furthermore, the left and right chiral algebras are precisely given by the
left and right centers of $A$ (see claim~5 in \cite{ostr}), i.e.:
\vspace{-.9em}

\dtl{Proposition}{prop:A=B}
As objects in \calc, 
  \be  A_\ell \,\cong\, \centlA \qquad {\rm and} \qquad A_r \,\cong\,
  \centrA \,.  \ee

\medskip\noindent
The proof of proposition \ref{prop:A=B} will fill the remainder of this 
section. Let us start by giving an alternative definition of the left 
and right centers:
\vspace{-.9em}

\dt{Definition}
The {\em left center\/} ${\rm cent}_{A,{\rm left}}$ of an algebra $A$ is 
the operation that assigns to any object
$X$ of $\calc$ the subspace of all elements in $\Hom(X,A)$ such that the
multiplication becomes commutative (\wrt the braiding $c_{A,A}$), i.e.
  \be
  {\rm cent}_{A,{\rm left}}(X) :=
  \big\{ \,\alpha\iN\Hom(X,A) \;\big|\; m \cir c_{A,A} \cir (\alpha \oti \id_A)
  = m \cir (\alpha \oti \id_A) \,\big\} \,.  \ee
Analogously, the {\em right center\/} ${\rm cent}_{A,{\rm right}}$ of an algebra 
$A$ is defined by
  \be
  {\rm cent}_{A,{\rm right}}(X) :=
  \big\{\, \alpha\iN\Hom(X,A) \;\big|\; m \cir c_{A,A} \cir (\id_A \oti \alpha)
  = m \cir (\id_A \oti \alpha)\, \big\} \,.  \ee

\medskip
The relation to the previous definition is
  \be
  \centlA \,\cong\, \bigoplus_{i\in\II}
  \dim\,\llb{\rm cent}_{A,{\rm left}}(U_i)\lrb \, U_i \,, \ee
and similarly for $\centrA$. Thus in this alternative formulation the relevant
subobject of $A$ is characterised via the components of its embedding morphism. 
There is also a close relationship with the relative center of $\Atop$ as 
introduced in definition \ref{def:cent-Atop}:
  \be
  \centreA(\Atop) = {\rm cent}_{A,{\rm left}}(\one) =
  {\rm cent}_{A,{\rm right}}(\one) \,.  \ee

\dtl{Lemma}{lem:centre-proj}
The morphism
$\alpha\iN\Hom(U_i,A)$ is in ${\rm cent}_{A,{\rm right}}(U_i)$ iff
  \bea \begin{picture}(94,74)(0,18)
  \put(0,0)   {\begin{picture}(0,0)(0,0)
              \scalebox{.38}{\includegraphics{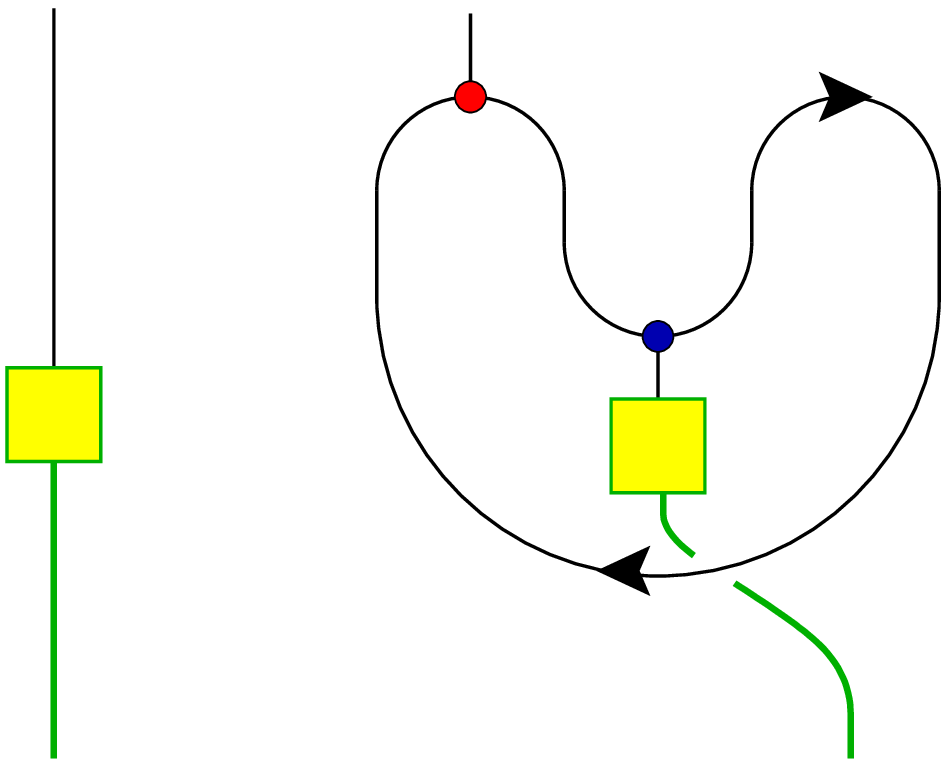}} \end{picture}}
  \put(21.5,37) {$=$}
  \put(3,87) {\scriptsize$A$} 
  \put(48,87) {\scriptsize$A$}
  \put(4.3,-8.6) {\scriptsize$i$}
  \put(91.9,-8.6) {\scriptsize$i$}
  \put(4,37) {\scriptsize$\alpha$}
  \put(70,33) {\scriptsize$\alpha$}
  \epicture09 \labl{eq:lem-cent}
An analogous statement holds for the left center when the inverse braiding 
is used in \erf{eq:lem-cent}.

\medskip\noindent
Proof:\\
To see the relation for the right center, consider the equivalence
  \bea \begin{picture}(400,80)(0,21)
  \put(0,0)   {\begin{picture}(0,0)(0,0)
              \scalebox{.38}{\includegraphics{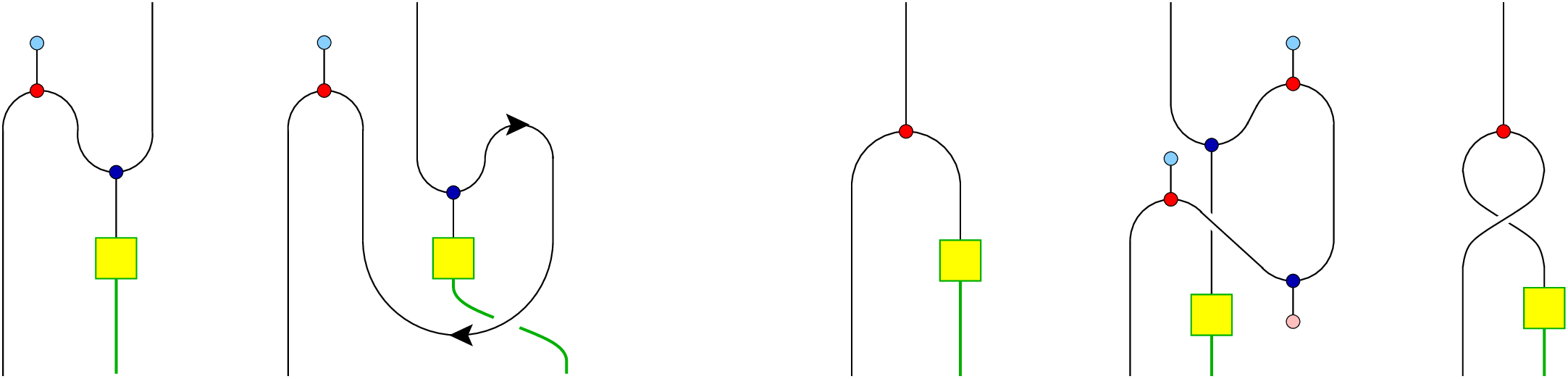}} \end{picture}}
   \put(-3,-8) {\scriptsize$A$}
   \put(25,-8) {\scriptsize$X$}
   \put(49,41){$=$}
   \put(261,41){$=$}
   \put(349,41){$=$}
   \put(168.3,41){$\Longleftrightarrow$}
   \put(70,-8) {\scriptsize$A$}
   \put(138.4,-8) {\scriptsize$X$}
   \put(212,-8) {\scriptsize$A$}
   \put(240,-8) {\scriptsize$X$}
   \put(283,-8) {\scriptsize$A$}
   \put(303,-8) {\scriptsize$X$}
   \put(366,-8) {\scriptsize$A$}
   \put(386,-8) {\scriptsize$X$}
   \put(36,99) {\scriptsize$A$}
   \put(102,99) {\scriptsize$A$}
   \put(225,99) {\scriptsize$A$}
   \put(292,99) {\scriptsize$A$}
   \put(376,99) {\scriptsize$A$}
   \put(27,29) {\scriptsize$\alpha$}
   \put(112,29) {\scriptsize$\alpha$}
   \put(241,29) {\scriptsize$\alpha$}
   \put(304,15) {\scriptsize$\alpha$}
   \put(388,16) {\scriptsize$\alpha$} 
  \epicture10 \labl{tor-pic26}
The first of these equalities is equivalent to the left equality in 
\erf{eq:cent-lem};  indeed, it can be obtained by composing 
$\id\oti$\erf{eq:cent-lem} with $(\eps{\circ}m)\oti\id_A$. Further,
lemma \ref{lem:alphaX} implies that this first equality is, in turn,
equivalent to the relation \erf{eq:lem-cent}. To get the 
equivalence asserted in \erf{tor-pic26}, one uses the Frobenius property
several times, on both sides of the first equality, as well as the fact 
that $A$ is symmetric, i.e.\ relation \erf{Phi1-Phi2inv}.
\\
For the left center the proof proceeds analogously.
\qed

To proceed, we introduce, for all $i,j\iN\I$, the isomorphism
$L_{ij}{:}\ \Hom(A{\oti}U_i,U_j^\vee)\,{\to}\,\Hom(U_i{\oti}U_j, A)$ via
  \bea \begin{picture}(120,55)(0,21)
  \put(0,0)   {\begin{picture}(0,0)(0,0)
              \scalebox{.38}{\includegraphics{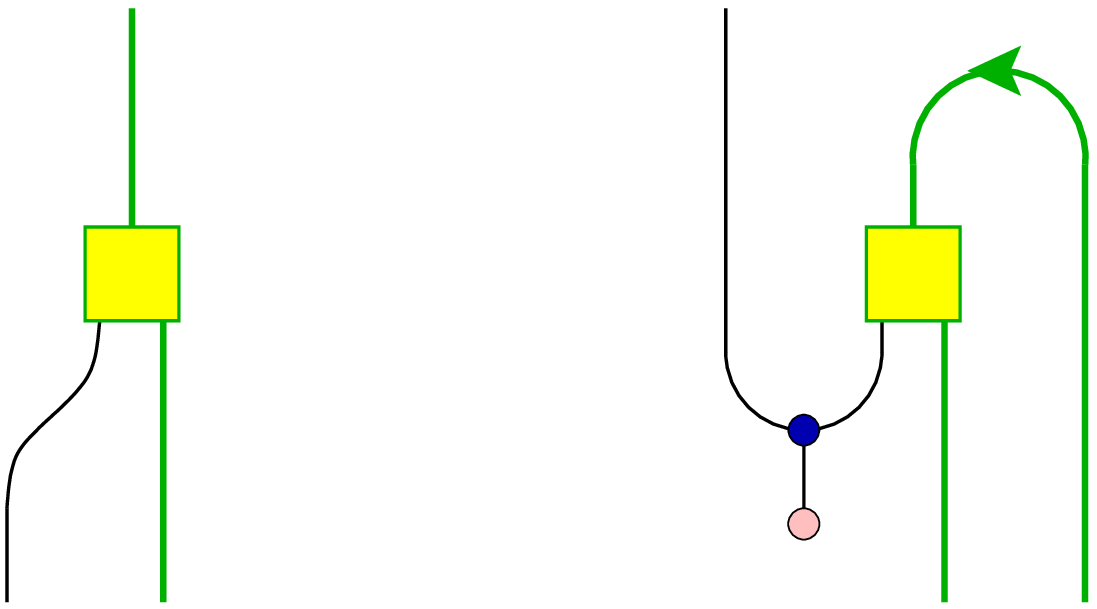}} \end{picture}}
   \put(-3,-8) {\scriptsize$A$}
   \put(16,-8) {\scriptsize$i$}
   \put(102,-8) {\scriptsize$i$}
   \put(117,-8) {\scriptsize$j$}
   \put(13,70) {\scriptsize$j^\vee$}
   \put(77,70) {\scriptsize$A$}
   \put(39.9,33.4) {$\longmapsto$}
   \put(12,35.5) {\scriptsize$\varphi$}
   \put(98,35.5) {\scriptsize$\varphi$}
  \epicture09 \labl{eq:L-def}

\dtl{Lemma}{lem:L-right-centre}
  \mbox{$\ $}\\[-.95em]
For each $i\iN\I$, the map $L_{i0}$ defines an isomorphism between 
${\rm cent}_{A,{\rm right}}(U_i)$ and $\Loc(A{\oti}U_i,\one)$.

\medskip\noindent
Proof:\\
Consider the moves
  \bea \begin{picture}(360,99)(0,32)
  \put(0,0)   {\begin{picture}(0,0)(0,0)
              \scalebox{.38}{\includegraphics{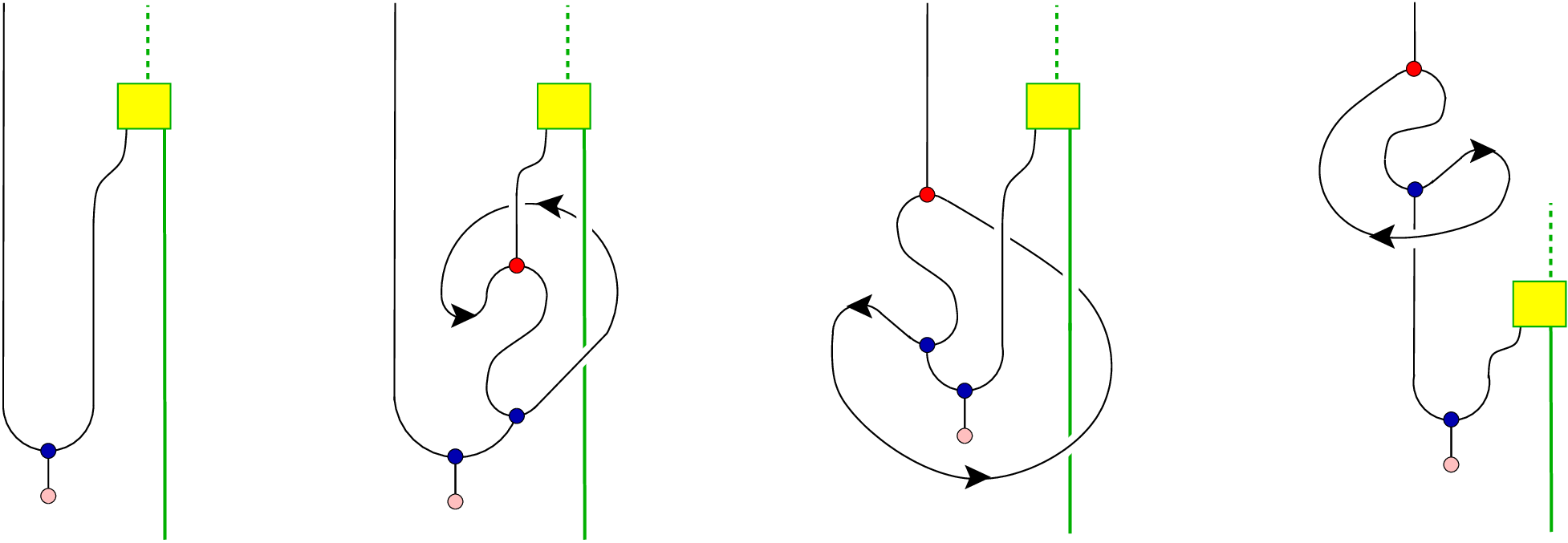}} \end{picture}}
   \put(-2.3,126.2) {\scriptsize$A$}
   \put(36.1,-8.7)  {\scriptsize$i$}
   \put(58.5,59)    {$=$}
   \put(87.3,126.2) {\scriptsize$A$}
   \put(131.1,-8.7) {\scriptsize$i$}
   \put(162.5,59)   {$=$}
   \put(208.3,126.2){\scriptsize$A$}
   \put(242.1,-7.7) {\scriptsize$i$}
   \put(273.5,59)   {$=$}
   \put(319.3,126.2){\scriptsize$A$}
   \put(352.1,-7.7) {\scriptsize$i$}
   \put(30,99) {\scriptsize$\varphi$}
   \put(126,99) {\scriptsize$\varphi$}
   \put(237.5,99) {\scriptsize$\varphi$}
   \put(348,54.5) {\scriptsize$\varphi$}
  \epicture17 \ee
The first equality is the definition of $\varphi$ being local.
In the equivalence we composed both sides with
$(\Delta\cir\eps)\oti\id_{U_i}$. The \rhs\ of the resulting equation can 
be rewritten as sketched above by using that $A$ is symmetric and Frobenius.
The resulting equality is equivalent, by lemma
\ref{lem:centre-proj} to $L_{i0}(\varphi)$ being in the right center.
\qed

\dtl{Lemma}{lem:L-left-centre}
For each $j\iN\I$, the map $L_{0j}$ defines an isomorphism between 
${\rm cent}_{A,{\rm left}}(U_j,A)$ and the subspace of local elements
in $\Hom(A{\oti}\one,U_j^\vee)$.

\medskip\noindent
Proof:\\
This is shown similarly to the previous lemma. The manipulations 
look as follows:
  \bea \begin{picture}(398,97)(0,31)
  \put(0,0)   {\begin{picture}(0,0)(0,0)
              \scalebox{.38}{\includegraphics{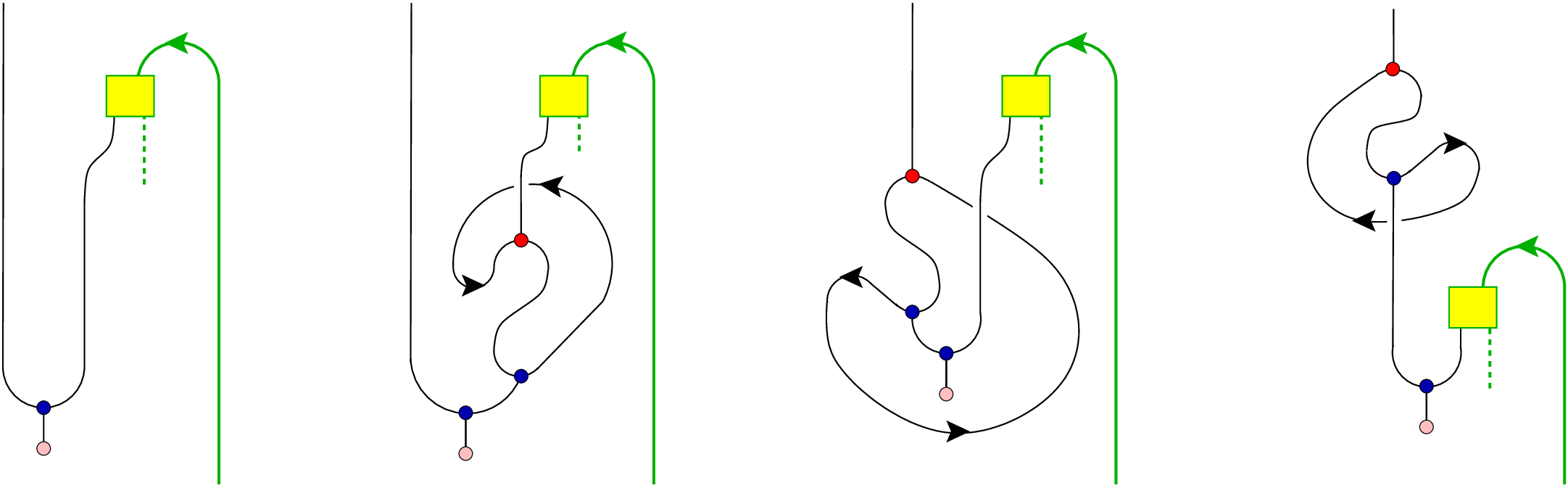}} \end{picture}}
   \put(-2.3,126.2) {\scriptsize$A$}
   \put(54.1,-8.7)  {\scriptsize$j$}
   \put(74.5,59)    {$=$}
   \put(100.3,126.2){\scriptsize$A$}
   \put(163.1,-8.7) {\scriptsize$j$}
   \put(182.5,59)   {$=$}
   \put(227.5,126.2){\scriptsize$A$}
   \put(280.1,-8.7) {\scriptsize$j$}
   \put(306.5,59)   {$=$}
   \put(349.4,126.2){\scriptsize$A$}
   \put(393.4,-8.7) {\scriptsize$j$}
   \put(30,97.5) {\scriptsize$\varphi$}
   \put(139,97.5) {\scriptsize$\varphi$}
   \put(256,97.5) {\scriptsize$\varphi$}
   \put(369.5,44.5) {\scriptsize$\varphi$}
  \epicture17 \ee
The left equality is again the definition of $\varphi$ being local, while
by lemma \ref{lem:centre-proj} the right equality is equivalent to
$\varphi$ being in the left center.
\qed

\medskip\noindent
Proof of proposition \ref{prop:A=B}:\\[.13em]
According to lemma \ref{lem:dim-loc-Hom}, the dimension of the subspace 
of local elements in $\Hom(A\oti U_j,U_i^\vee)$ equals $Z_{ij}$. Thus
by the lemmata \ref{lem:L-right-centre} and \ref{lem:L-left-centre}
it follows that
  \be
  Z_{0i} = \dim\llb{\rm cent}_{A,{\rm right}}(U_i)\lrb
  \qquad {\rm and} \qquad
  Z_{j0} = \dim\llb{\rm cent}_{A,{\rm left}}(U_j)\lrb \,.  \ee
As a consequence, the definition of $A_\ell$ and $A_r$ in 
equation \erf{eq:lr-chiral-Z} does coincide with
the definition \ref{def:lr-centre} of $\centlA$ and $\centrA$.
\qed

\subsection{The case ${N_{ij}}^k \iN \{0,1\}$
            and $\dim\,\Hom(U_k,A) \iN \{0,1\}$}

In this section we express the ribbon graph \erf{Zij2} giving
the torus partition function in terms of the Moore\hy Seiberg data of the
modular category \calc. For notational simplicity we concentrate on the
situation that both ${N_{ij}}^k \iN \{0,1\}$ for all $i,j,k\iN\I$ and
$\dim\,\Hom(U_k,A) \iN \{0,1\}$ for all $k\iN\I$.

The ribbon graphs defined by
  \begin{eqnarray} \begin{picture}(447,153)(0,0)
  \put(40,0)  {\begin{picture}(0,0)(0,0)
              \scalebox{.38}{\includegraphics{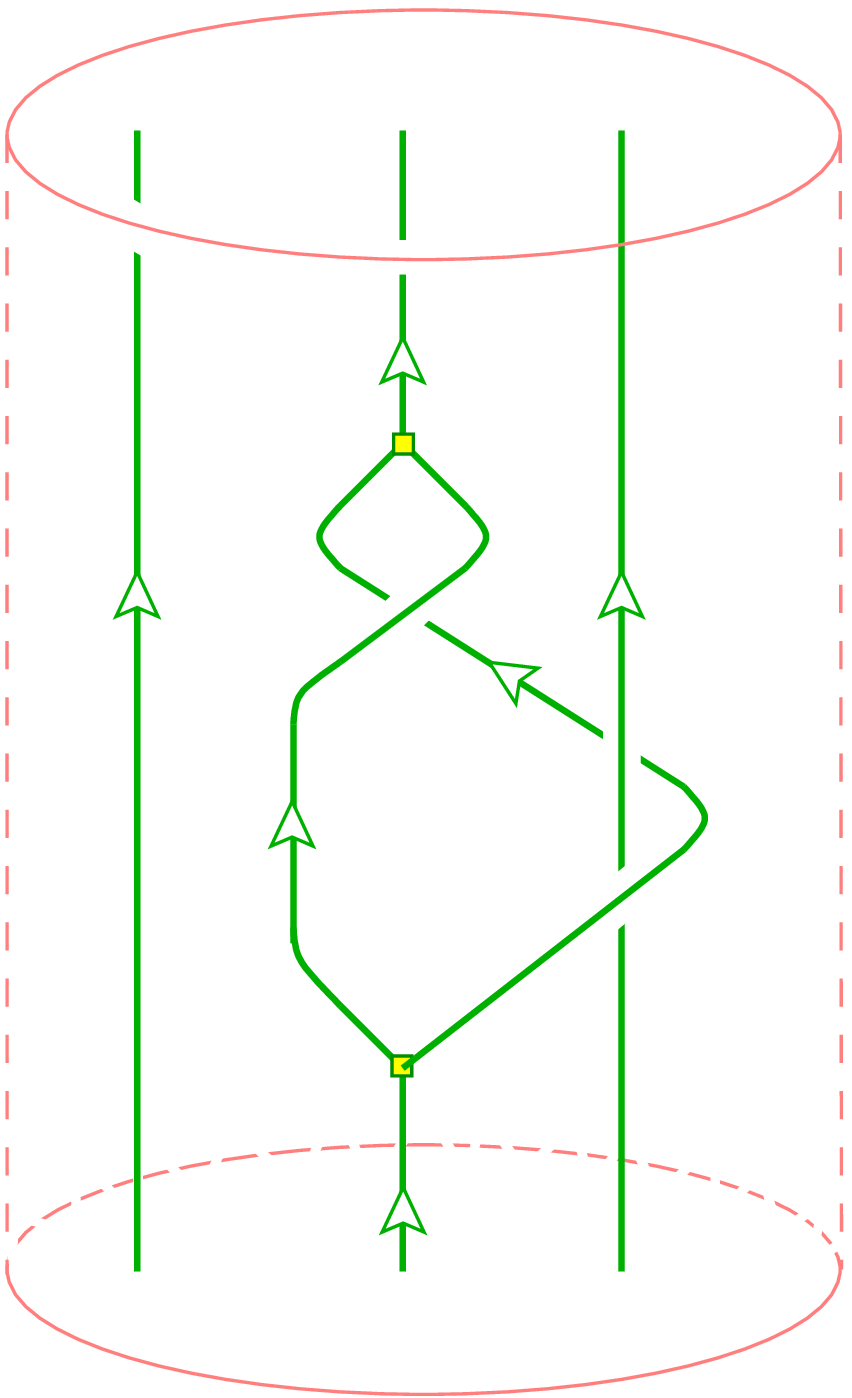}} \end{picture}
   \put(48,16) {\scriptsize$a$}
   \put(46,135) {\scriptsize$a$}
   \put(33,52) {\scriptsize$c$}
   \put(51,46) {\scriptsize$b$}
   \put(10,38) {\scriptsize$i$}
   \put(70,38) {\scriptsize$j$}
   }
  \put(197,0) {\begin{picture}(0,0)(0,0)
              \scalebox{.38}{\includegraphics{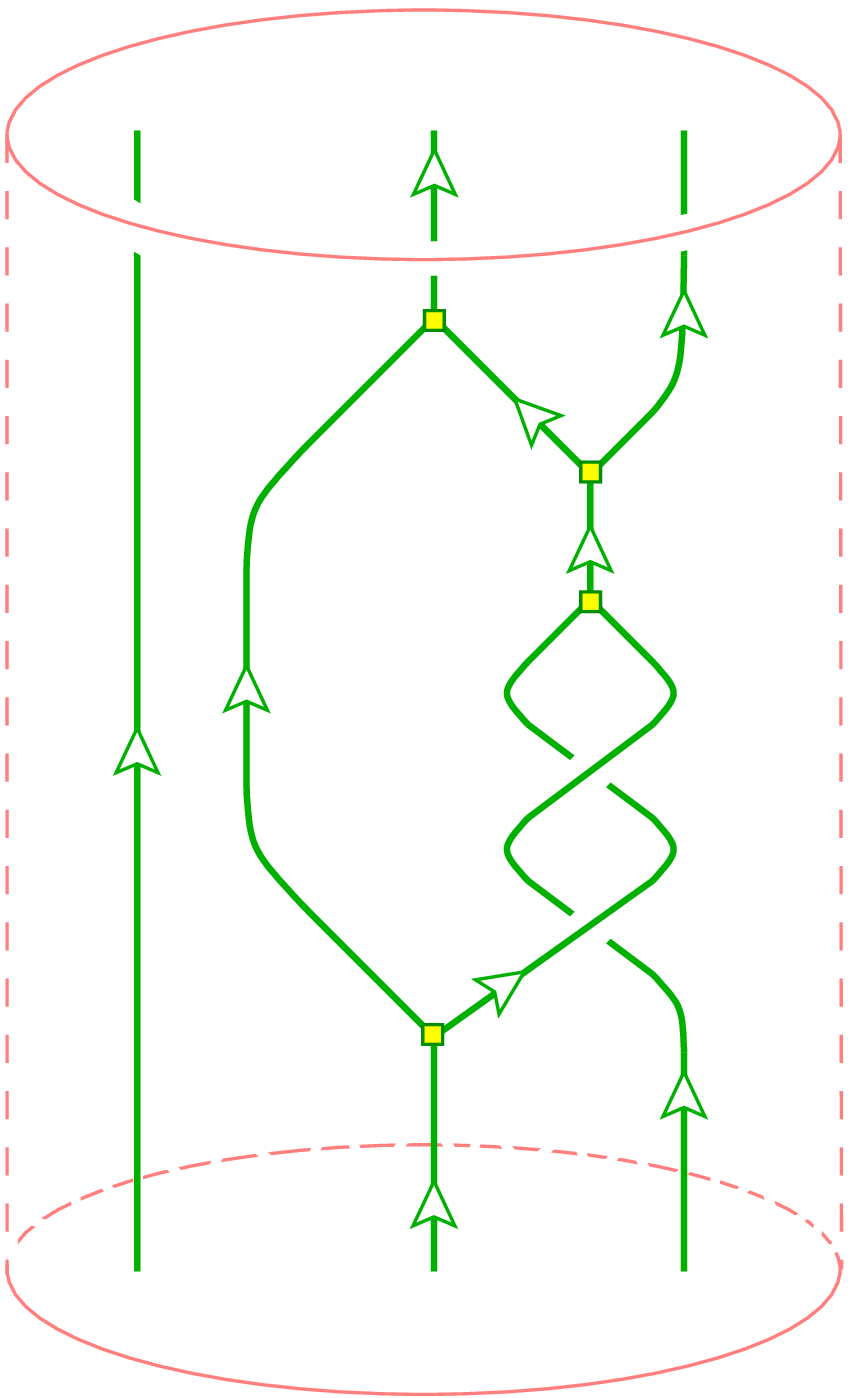}} \end{picture}
   \put(10,38) {\scriptsize$i$}
   \put(77,17) {\scriptsize$j$}
   \put(77,137) {\scriptsize$j$}
   \put(41,135) {\scriptsize$a$}
   \put(41,14) {\scriptsize$a$}
   \put(29,65) {\scriptsize$c$}
   \put(75,57) {\scriptsize$b$}
   \put(56,111) {\scriptsize$b$}
   \put(67,92) {\scriptsize$k$}
   }
  \put(354,0) {\begin{picture}(0,0)(0,0)
              \scalebox{.38}{\includegraphics{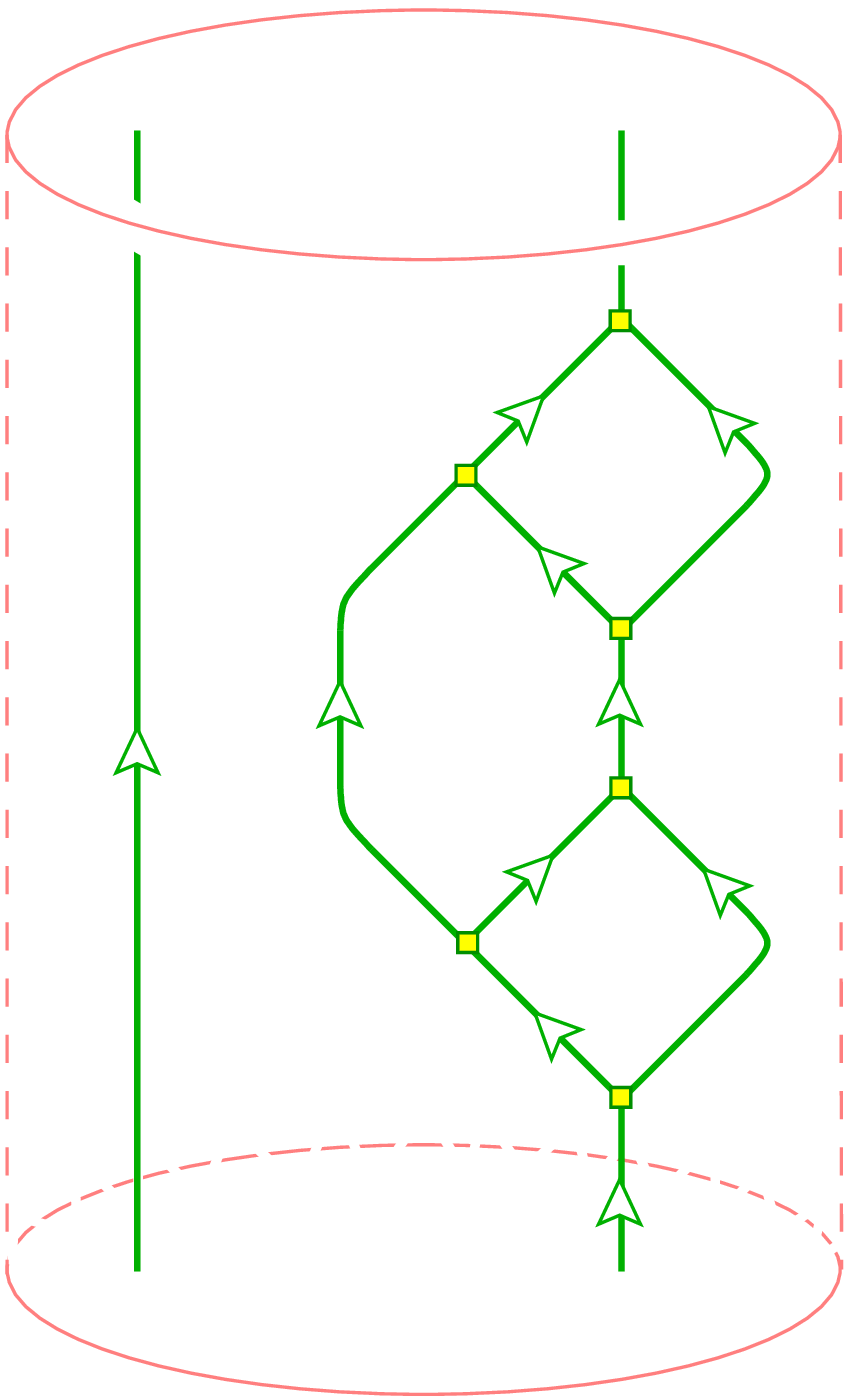}} \end{picture}
   \put(10,38) {\scriptsize$i$}
   \put(39,66) {\scriptsize$c$}
   \put(53,38) {\scriptsize$a$}
   \put(56,112) {\scriptsize$a$}
   \put(85,98) {\scriptsize$j$}
   \put(85,47) {\scriptsize$j$}
   \put(71,72) {\scriptsize$k$}
   \put(60,95) {\scriptsize$b$}
   \put(55,60) {\scriptsize$b$}
   \put(72,16) {\scriptsize$\overline\imath$}
   \put(70,134) {\scriptsize$\overline\imath$}
   }
  \put(0,73)       {$\Gamma_1\;:=$}
  \put(157,73)     {$\Gamma_2\;:=$}
  \put(314,73)     {$\Gamma_3\;:=$}
  \end{picture}\nonumber \\[-.5em]{} \label{tor-pic20abc} \end{eqnarray} 
can be related to each other by
inserting bases as described in sections \ref{sec:alg-obj} and 
\ref{sec:alg-N=1}, using the compatibility constraint between braiding and
twist, and accounting for $\dim\,\calh(i,j;S^2)\eq\delta_{j,\bar\imath}$.
This way we simplify the expression for the ribbon graph \erf{Zij2} to
  \be \bearll
  Z_{ij} \!\!
  &= \dsty\sum_{a,b,c\In A}\! {m_{bc}}^{\!a} \Delta_a^{\,cb}\, \theta_b\, \Gamma_1
  = \sum_{a,b,c\In A}\sumI_{k} {m_{bc}}^{\!a} \Delta_a^{\,cb}\, \theta_b\,
  \Rs {}cba \, \Gamma_2
  \\{}\\[-.6em]
  &= \dsty\sum_{a,b,c\In A}\sumI_{k} {m_{bc}}^{\!a} \Delta_a^{\,cb}\,
  \Frac{\theta_k}{\theta_j}\, \Rs {}cba \, \Gamma_3 \,.
  \eear \ee
Via a fusion and an inverse fusion operation one then arrives at
a graph that can be further simplified by using the definition of the dual
coupling. The final result is
  \be  Z_{ij} = \sum_{a,b,c\In A}\! {m_{bc}}^{\!a} \Delta_a^{\,cb}
  \sum_{k\in\II} \Gs cbj{\bar\imath}ak\, \Rs {}cba\,
  \Frac{\theta_k}{\theta_j}\, \Fs cbj{\bar\imath}ka \,.  \labl{eq:tor-MS}
(Recall that the coproduct can also be expressed in terms of the product
using \erf{eq:delta-by-mult}.)
Note that in this formulation, it is not obvious why
$Z_{ij}$ should be real, let alone a non-negative integer. This
illustrates the power of the graphical approach, which readily supplies a
proof of non-negativity and integrality.

\subsubsection{Example: Free boson} \label{sec:ex-free-boson-pf}

The expression \erf{eq:tor-MS} can now be used to compute
the \tparfu\ associated to the algebra $A_{2r}$ as defined
in \erf{eq:A2r-cont} and \erf{eq:A2r-mult}. We also need the
$\zet_{2N}$ free boson modular data given in \erf{eq:Z2N-F}.
Substituting these into formula \erf{eq:tor-MS}, one notes that the
simple fusion rules make three of the four sums disappear,
and we end up with
  \be  
  Z_{[x][y]}(A_{2r}) = \delta_{[x{+}y]\In A} \cdot \Frac{r}{N}
  \sum_{a\In A} \exp( 2\pi\ii\,\textstyle{\frac{x-y}{2}\,\frac{a}{2N}} )
  = \delta_{x{+}y,0 \;{\rm mod}\, 2r} \,
  \delta_{x{-}y,0 \;{\rm mod}\, 2N/r} \,.  \ee
Let us have a look at the two extremal cases. For $r\eq N$ we have
$A_{2r}\eq A_{2N}\eq [0]$. This leads to
  \be  Z_{[x][y]}(A_{2N}) = \delta_{x+y,2N} \,.  \labl{c=1ZC}
(Recall that we chose representatives of
simple objects such that $0\,{\le}\,x,y\,{<}\, 2N$.)
The \parfu \erf{c=1ZC} is just the charge
conjugation modular invariant. On the other extreme, for $r\eq 1$ we deal
with $A_2\eq[0]\,{\oplus}\,[2]\,{\oplus}\cdots{\oplus}\,[2N{-}2]$ and find
  \be  Z_{[x][y]}(A_{2}) = \delta_{x,y} \,,  \ee
which pairs \CA-representations of the same (minimal) $\mathfrak u(1)$
charge; this is precisely the T-dual of the \parfu\ \erf{c=1ZC}.

More generally, one easily verifies that
  \be  Z_{[x][2N{-}y]}(A_{2r}) = Z_{[x][y]}(A_{2N/r}) \,,  \ee
which means that the two
algebras $A_{2r}$ and $A_{2N/r}$ produce T-dual theories.

\subsubsection{Example: \E modular invariant} \label{sec:E7-mod-inv}

It is only now that we can verify that the title of this example indeed 
deserves its name. We combine the definition of the modular category for 
$\su(2)_k$ in section \ref{sec:modcat-su2} at level $k\eq16$, the explicit 
form \erf{eq:E7-mult} for the product of $A$, and the formula \erf{eq:tor-MS}
for the coefficients $Z_{ij}$ in $Z\eq\sum_{i,j\in\II}Z_{ij}\,\chii_i^{}\,
\chii_j^*$. We can then check numerically that we get precisely the \E 
modular invariant:
  \be  Z =
  |\chii_0{+}\chii_{16}{|}^2 + |\chii_4{+}\chii_{12}{|}^2 +
  |\chii_6{+}\chii_{10}{|}^2 + |\chii_8{|}^2
  + \chii_8\, (\chii_2^*\,{+}\,\chii_{14}^*)
  + (\chii_2\,{+}\,\chii_{14})\, \chii_8^* \,.  \ee 
Incidentally, one also checks that $A$ of the form
  \be  A = (0) \oplus (k/2) \oplus (k) \,,  \ee
i.e.\ precisely containing the unit, the simple current and its fixed point
(see \erf{eq:E7-mult}),
is an algebra also at the levels $k\eq4$ and $k\eq8$. In the first case
it is Morita equivalent to $\one$, since
(recall from section \ref{sec:E7-algebra-object} that the multiplication on 
$\one{\oplus}f{\oplus}J$ is unique)
  \be  (0) \oplus (2) \oplus (4) \cong (2) \otimes (2) \,,  \ee
and accordingly
gives rise to the $A$-series modular invariant, while in the second case
one obtains the $D$-series modular invariant.

\subsection{One-point blocks on the torus} \label{sec:one-point-blocks}

Similarly as for the zero-point blocks on the torus, we may study the 
behavior of torus blocks with one $A$-insertion, i.e.\ of elements of 
$\calh(A;\torus)$, under an $\mathcal S$-transformation. In analogy 
with \erf{S-xfer} we fix our conventions by defining the matrix
$L^A\eq(L^A_{i\alpha,j\beta})$ via
  \bea \begin{picture}(350,94)(0,39)
  \put(0,0)   {\begin{picture}(0,0)(0,0)
              \scalebox{.38}{\includegraphics{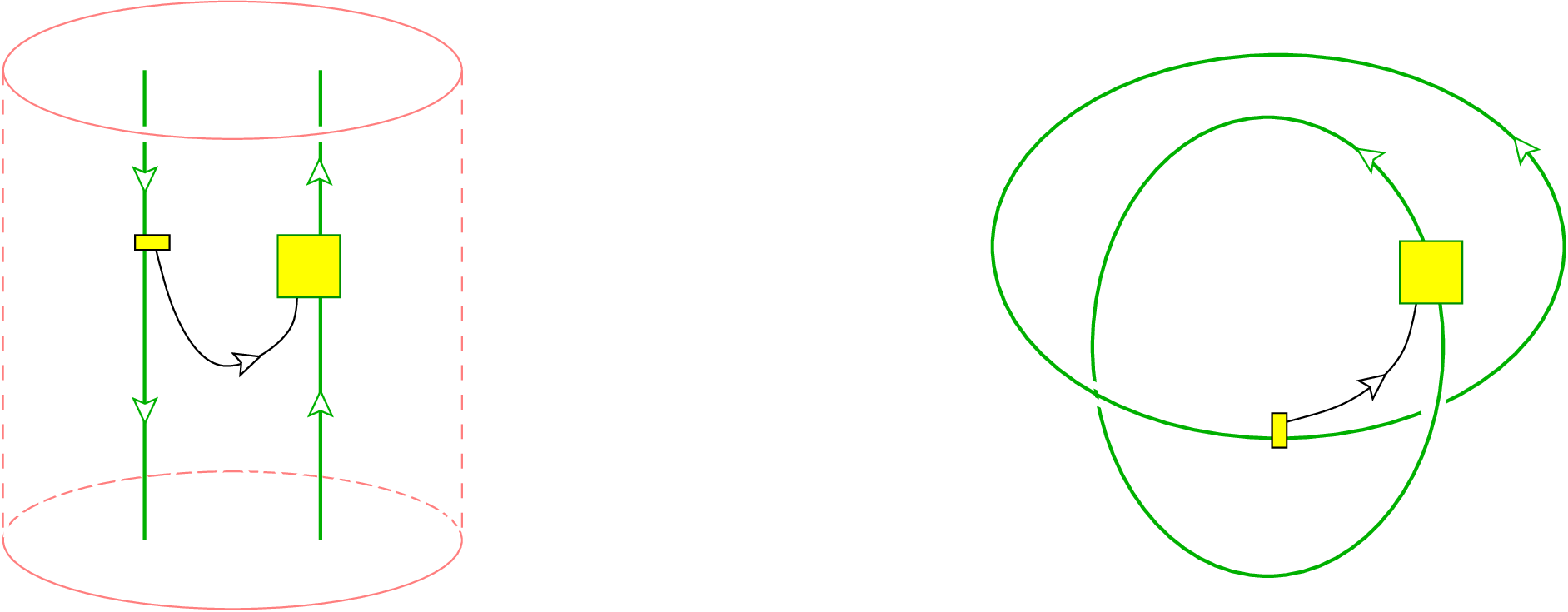}} \end{picture}}
  \put(26.5,60.9) {\scriptsize$i$}
  \put(73.0,88.9) {\scriptsize$X$}
  \put(64.7,73.3) {\scriptsize$\Phi$}
  \put(72.2,34.1) {\scriptsize$X$}
  \put(22,79) {\scriptsize$\overline\alpha$}
  \put(26.5,113) {\scriptsize$i$}
  \put(46,59) {\scriptsize$A$}
  \put(122.6,63.1){$=\;S_{0,0} \dsty\sum_{j,\beta} L^A_{i\alpha,j\beta}$}
  \put(291.0,94)  {\scriptsize$X$}
  \put(311.7,72.2){\scriptsize$\Phi$}
  \put(339.8,100.5){\scriptsize$j$}
  \put(278.7,27.1) {\scriptsize$\overline\beta$}
  \put(296,54) {\scriptsize$A$}
 \epicture18 \labl{eq:U-def}
Here $X$ is an arbitrary (not necessarily simple) object,
$\Phi \iN \Hom(A{\otimes}X,X)$, $U_i,U_j$ are simple objects,
$\bar\alpha$ denotes a basis vector in $\Hom(U_i,A{\otimes}U_i)$,
and $\bar\beta$ runs over the basis in $\Hom(U_j,A{\otimes}U_j)$.
That the ribbon graphs on the two sides are indeed linearly related
-- and thus that the matrix $L^A$ exists --
follows from the general result that the TFT provides us with a
projective \rep\ of the mapping class group.

Let us examine some properties of $L^A$.

\dtl{Lemma}{U-Ut}
The matrix $L^A$ is invertible, with inverse $\tilde L^A$ given by
  \bea \begin{picture}(5,60)(-5,31)
  \put(0,0)   {\begin{picture}(0,0)(0,0)
              \scalebox{.38}{\includegraphics{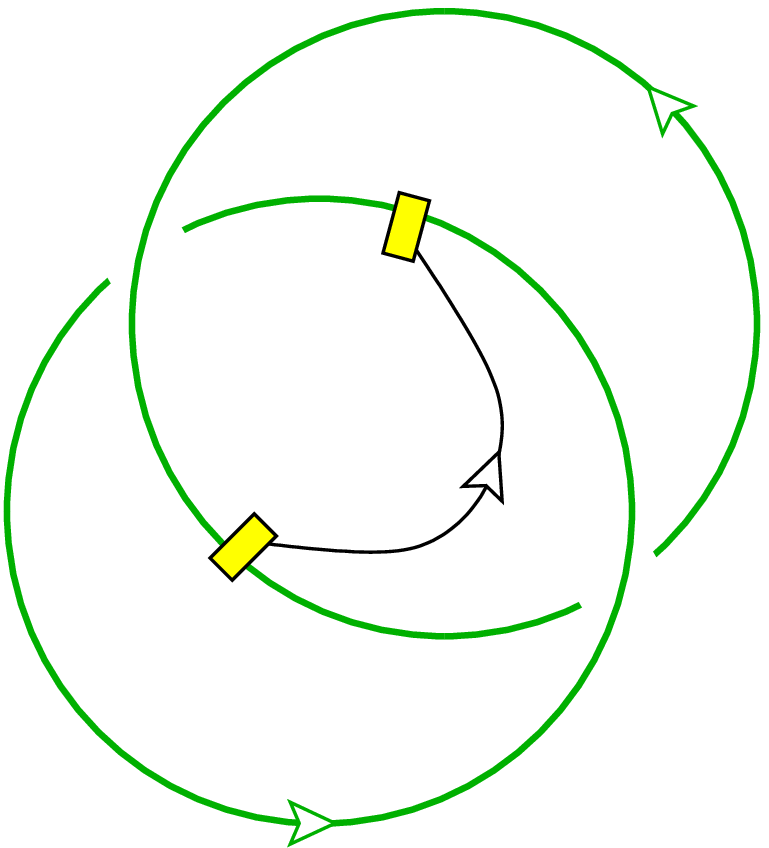}} \end{picture}}
   \put(-2,19) {\scriptsize$k$}
   \put(17.9,25.3) {\scriptsize$\overline\beta$}
   \put(43,74) {\scriptsize$\alpha$}
   \put(44,39) {\scriptsize$A$}
   \put(77,61) {\scriptsize$j$}
  \put(-80,48) {$\tilde L^A_{j\beta,k\gamma} \; =\;  S_{0,0}$}
  \epicture09 \labl{eq:Ut-def}
Proof:\\
For $X\eq U_k$ and $\Phi\eq\gamma$, the \lhs\ of \erf{eq:U-def} reads
  \bea \begin{picture}(333,104)(0,34)
  \put(0,0)   {\begin{picture}(0,0)(0,0)
              \scalebox{.38}{\includegraphics{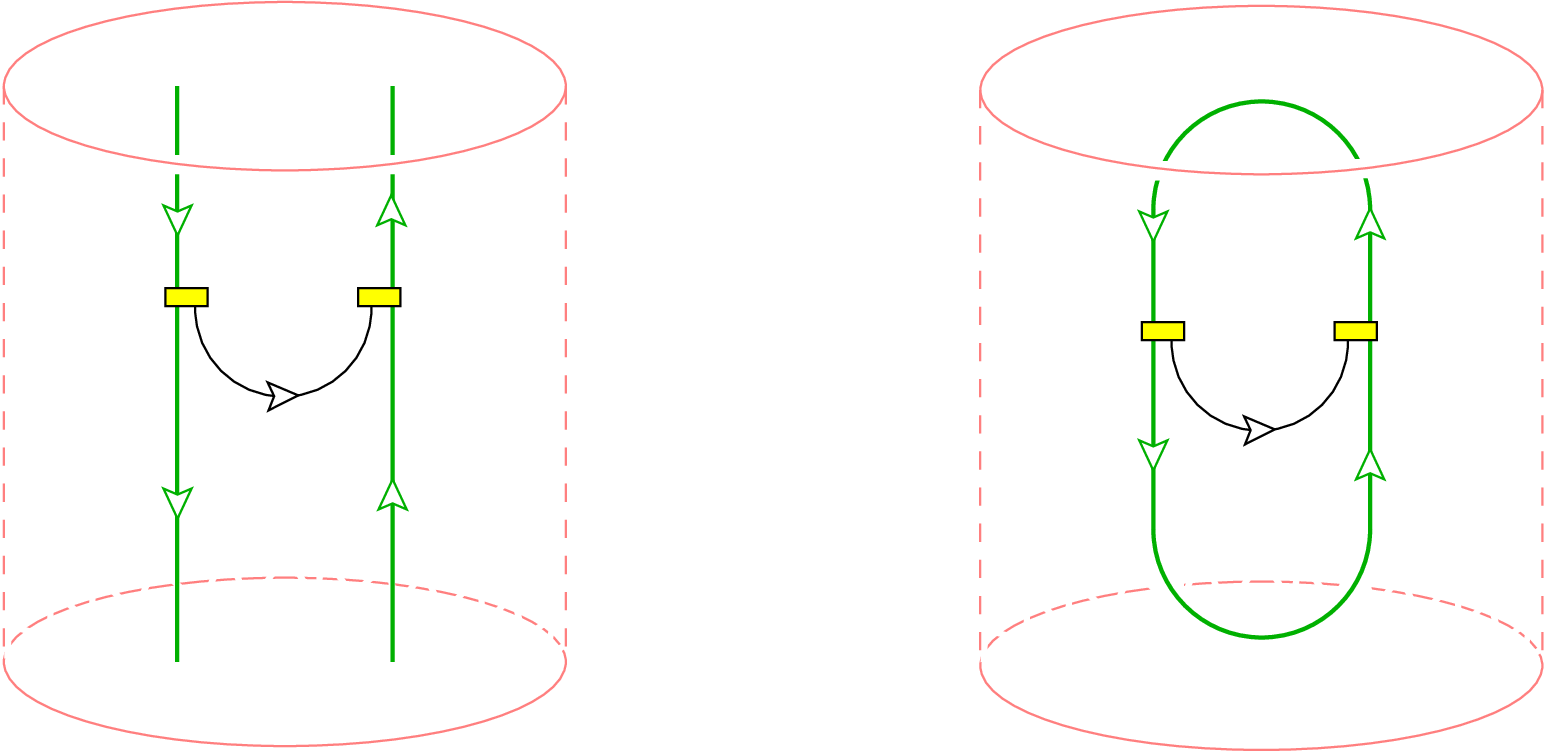}} \end{picture}}
   \put(26.5,60.9) {\scriptsize$i$}
   \put(26.5,113) {\scriptsize$i$}
   \put(72,60) {\scriptsize$k$}
   \put(72,113) {\scriptsize$k$}
   \put(22,79) {\scriptsize$\overline\alpha$}
   \put(73,79) {\scriptsize$\gamma$}
   \put(46,55) {\scriptsize$A$}
   \put(211,18) {\scriptsize$i$}
   \put(237.6,115) {\scriptsize$i$}
   \put(197,73) {\scriptsize$\overline\alpha$}
   \put(248.5,73) {\scriptsize$\gamma$}
   \put(222,48) {\scriptsize$A$}
  \put(-49,65) {${\rm l.h.s.} \; = \;$}
  \put(110,65) {$\dsty = \, \frac{\delta_{ik}}{\dim(U_i)}$}
  \put(290,65) {$\; = \; \delta_{i,k}\, \delta_{\alpha,\gamma}\,,$}
  \epicture15 \ee
while the ribbon graph on the \rhs\ of \erf{eq:U-def} reduces to
$\tilde L^A_{j\beta,k\gamma}/S_{0,0}$, so that \erf{eq:U-def} becomes
  \be
  \delta_{i,k}\, \delta_{\alpha,\gamma} = \sumI_{j}\sum_{\beta}
  L^A_{i\alpha,j\beta}\, \tilde L^A_{j\beta,k\gamma} \,.
  \labl{eq:UA-UtA-sum}
Thus $L^A$ has $\tilde L^A$ as its right-inverse. Since $L^A$ is a square
matrix, this is also a left-inverse.
\qed

Note that the relation \erf{eq:SxS=C} for the zero-point blocks
is recovered as a special case of \erf{eq:UA-UtA-sum}, by
setting $A\eq\one$ and $\alpha\eq\gamma\eq\id$. In this case
$\tilde L^A$ is just the inverse of the $S$-matrix,
i.e.\ $\tilde L^\one_{j,k}\eq S_{0,0}\, s_{j,\bar k}$ with $s$
as defined in \erf{smat}.

In the next section we will heavily use matrices $S^A\eq
(S^A_{\kappa,i\alpha})$ and $\tilde S^A\eq(\tilde S^A_{i\alpha,\kappa})$, 
where $\kappa\iN\J$ (corresponding to a simple $A$-module $M_\kappa$) and
$i\iN\I$ (corresponding to a simple object $U_i$), and
$\alpha \iN \Hom(A{\otimes}U_i,U_i)$ is local. We start by defining
  \bea \begin{picture}(5,59)(-4,34)
  \put(0,0)   {\begin{picture}(0,0)(0,0)
              \scalebox{.38}{\includegraphics{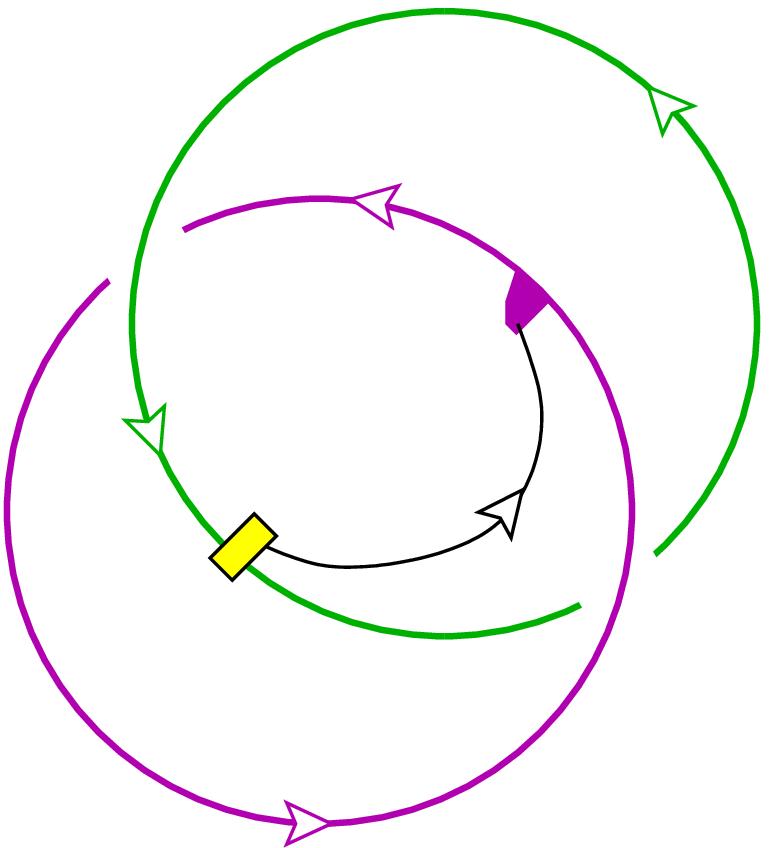}} \end{picture}}
  \put(2,5) {\scriptsize$\M_\kappa$}
  \put(60,65) {\scriptsize$\rho_\kappa$}
  \put(20,23) {\scriptsize$\overline\alpha$}
  \put(78.2,57) {\scriptsize$i$}
  \put(49.6,39.7) {\scriptsize$A$}
  \put(-80,46) {$\tilde S^A_{i\alpha,\kappa} \; :=\; S_{0,0}$}
  \epicture13 \labl{eq:SAt-def}
Note that the row and column labels of $\tilde S^A$
take their values in two rather different index sets.
(Thus, while being similar to the $S$-matrix
for the category of local $A$-modules that is used in \cite{kios},
it is not quite the same matrix. The notion of a local
module as defined in \cite{kios} refers to \alg s $A$ that are commutative 
and have trivial twist, neither of which properties is imposed here.)

The matrix $S^A$ is defined to implement an $\mathcal S$-transformation on 
the torus one-point blocks, analogously to $L^A$ in \erf{eq:U-def}:
  \bea \begin{picture}(350,96)(0,38)
  \put(0,0)   {\begin{picture}(0,0)(0,0)
              \scalebox{.38}{\includegraphics{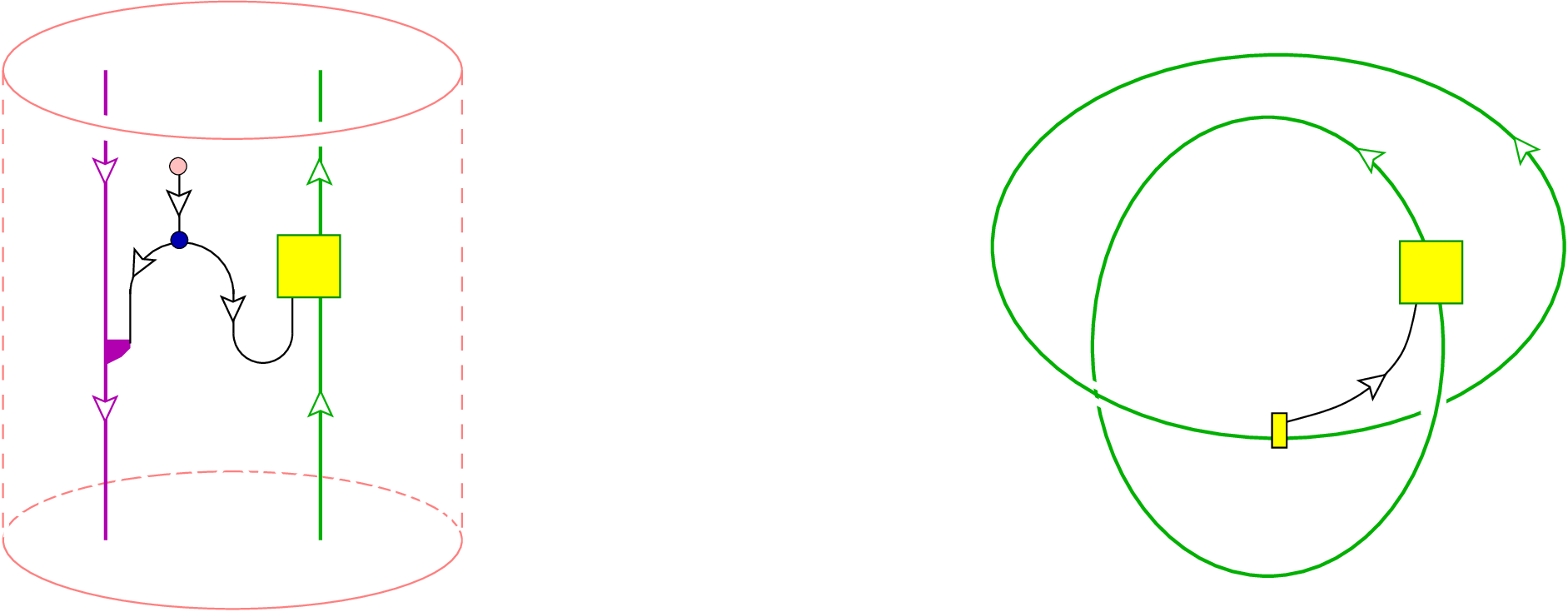}} \end{picture}}
  \put(113,62){$=\;S_{0,0}\dsty\sumI_i\sum_\alpha S^A_{\kappa,i\alpha}$}
   \put(24.6,35) {\scriptsize$\M_\kappa$}
   \put(73.0,88.9) {\scriptsize$X$}
   \put(64.7,73.3) {\scriptsize$\Phi$}
   \put(72.2,34.1) {\scriptsize$X$}
   \put(13,58) {\scriptsize$\rho_\kappa$}
   \put(43,59) {\scriptsize$A$}
   \put(291.0,94)  {\scriptsize$X$}
   \put(311.7,72.2){\scriptsize$\Phi$}
   \put(339.5,100.5)  {\scriptsize$i$}
   \put(277,27) {\scriptsize$\overline\alpha$}
   \put(292,54) {\scriptsize$A$}
  \epicture17 \labl{eq:SA-def}
Here $M_\kappa$ is a simple $A$-module, and the second sum extends over
the dual basis $\{\bar\alpha\}$ of the morphism space
$\Hom(U_i,A{\otimes}U_i)$. When $X\eq\M_\kappa$ and $\Phi\eq\r_\kappa$ is a \rep\
morphism, then the ribbon graph on the \rhs\ of \erf{eq:SA-def} looks 
like \erf{eq:SAt-def}, except for
the fact that $\alpha$ in \erf{eq:SAt-def} is restricted to be local.
This constraint does, however, not pose any problem:

\dtl{Lemma}{lem:only-local}
The relation \erf{eq:SA-def}, with $X\eq\M_\kappa$ and $\Phi\eq\r_\kappa$,
remains true when the sum
on the \rhs\ is restricted to local couplings $\alpha$.

\medskip\noindent
Proof:\\
We have the following identities of ribbon graphs in $S^3$:
  \bea \begin{picture}(353,85)(0,20)
  \put(0,0)   {\begin{picture}(0,0)(0,0)
              \scalebox{.38}{\includegraphics{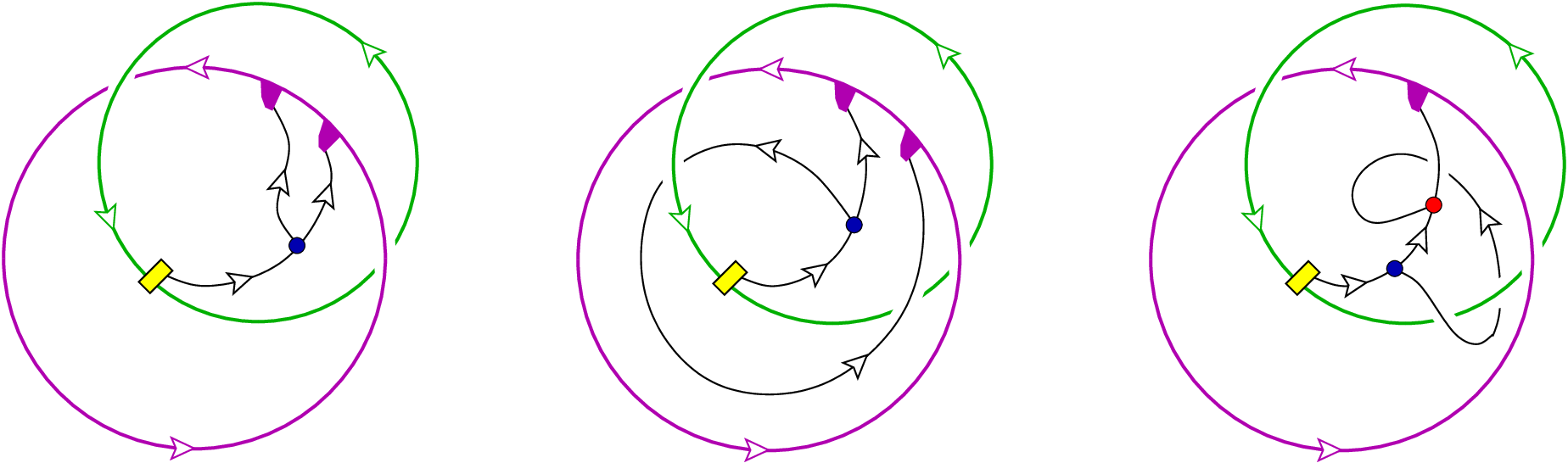}} \end{picture}}
   \put(5,6) {\scriptsize$\M_\kappa$}
   \put(128,6) {\scriptsize$\M_\kappa$}
   \put(251,6) {\scriptsize$\M_\kappa$}
   \put(25,30) {\scriptsize$\overline\alpha$}
   \put(149,30) {\scriptsize$\overline\alpha$}
   \put(272,30) {\scriptsize$\overline\alpha$}
   \put(51,49) {\scriptsize$A$}
   \put(172,50) {\scriptsize$A$}
   \put(284,59) {\scriptsize$A$}
   \put(84,86.8) {\scriptsize$i$}
   \put(208,86.8) {\scriptsize$i$}
   \put(331,86.8) {\scriptsize$i$}
   \put(101,46) {$=$}
   \put(226,46) {$=$}
  \epicture05 \labl{eq:only-local}
This means that one may always dress the coupling $\alpha$ on the \rhs\ of 
\erf{eq:SA-def} by the projector $P_i$. This, in turn, implies that 
the sum does not change its value when restricted to local couplings.
\qed

The following two results imply that, when restricted to a basis of 
local couplings in $\Hom(A{\otimes}U_i,U_i)$, $S^A$ is invertible.
\vspace{-.9em}

\dtl{Proposition}{pr:XMN}
The matrix $S^A$ is a left-inverse
for $\tilde S^A$. That is, for two simple $A$-modules $M_{\kappa},
M_{\kappa'}$ one has
  \be
  \sum_{i\in\II} \sum_{\alpha\;{\rm local}} S^A_{\kappa,i\alpha} \,
  \tilde S^A_{i\alpha,\kappa'} = \delta_{\kappa,\kappa'} \,.
  \labl{eq:SA-leftinv}

\smallskip\noindent
Proof:\\
Consider the defining relation \erf{eq:SA-def} for $X\eq\M_{\kappa'}$, 
where $M_{\kappa'}$ is a simple $A$-module and $\Phi\eq\r_{\kappa'}$ the 
\rep\ morphism. By lemma \ref{lem:only-local}, the range of summation 
on the \rhs\ of \erf{eq:SA-def} can be restricted to local $\alpha$, thus 
resulting in the \lhs\ of \erf{eq:SA-leftinv}. On the other hand, defining 
the graphs
  \begin{eqnarray} \begin{picture}(400,173)(29,0)
  \put(0,0)   {\begin{picture}(0,0)(0,0)
              \scalebox{.38}{\includegraphics{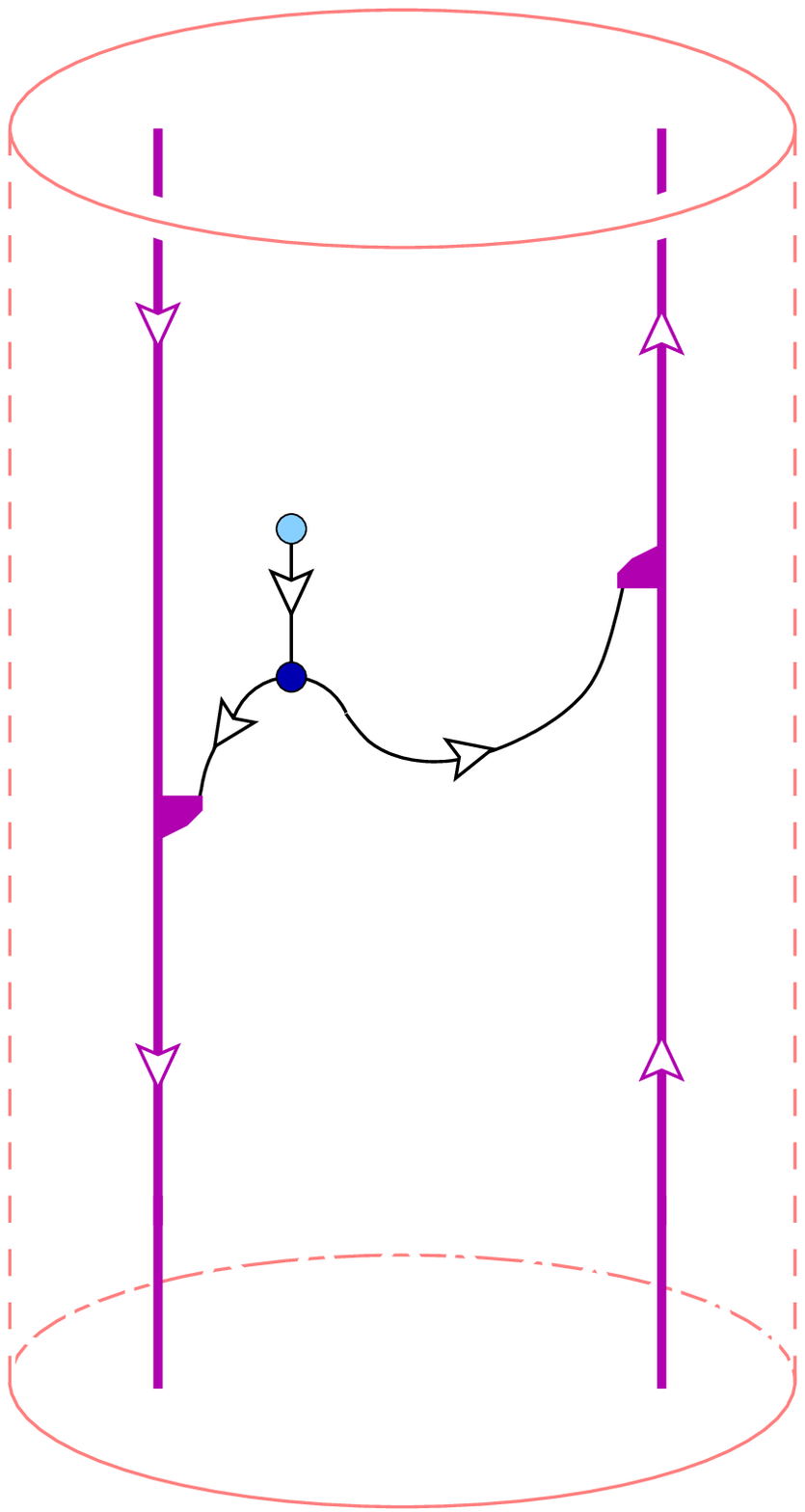}} \end{picture}
   \put(21,157) {\scriptsize$\M_\kappa$}
   \put(61,157) {\scriptsize$\M_{\kappa'}$}
   \put(8,78) {\scriptsize$\rho_\kappa$}
   \put(79,108) {\scriptsize$\rho_{\kappa'}$}
   \put(49,91) {\scriptsize$A$}
  \put(5,-4){$\dsty\underbrace{\hsp{7}}_{\ \ \ \dsty=:\,\Gamma_1}$}
   }
  \put(128,0)   {\begin{picture}(0,0)(0,0)
              \scalebox{.38}{\includegraphics{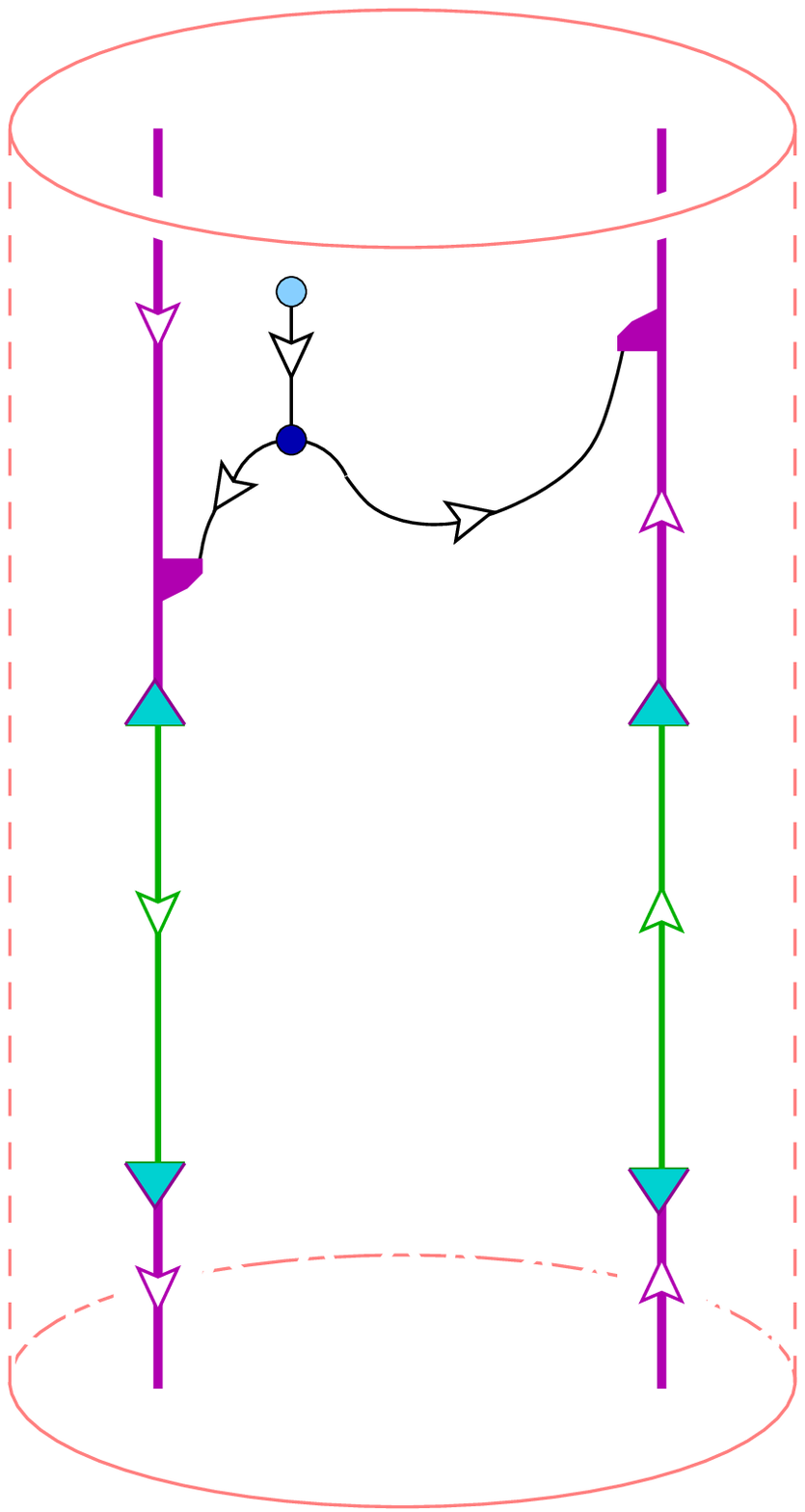}} \end{picture}
   \put(21,157) {\scriptsize$\M_\kappa$}
   \put(61,157) {\scriptsize$\M_{\kappa'}$}
   \put(49,119) {\scriptsize$A$}
   \put(8,106) {\scriptsize$\rho_\kappa$}
   \put(79,135) {\scriptsize$\rho_{\kappa'}$}
   \put(23,91) {\scriptsize$\overline\alpha$}
   \put(23,37) {\scriptsize$\alpha$}
   \put(65,37) {\scriptsize$\overline\beta$}
   \put(65,91) {\scriptsize$\beta$}
   \put(20,57) {\scriptsize$a$}
   \put(70,57) {\scriptsize$b$}
  \put(5,-4){$\dsty\underbrace{\hsp{7}}_{\ \ \ \dsty=:\,\Gamma_2}$}
   }
  \put(256,0)   {\begin{picture}(0,0)(0,0)
              \scalebox{.38}{\includegraphics{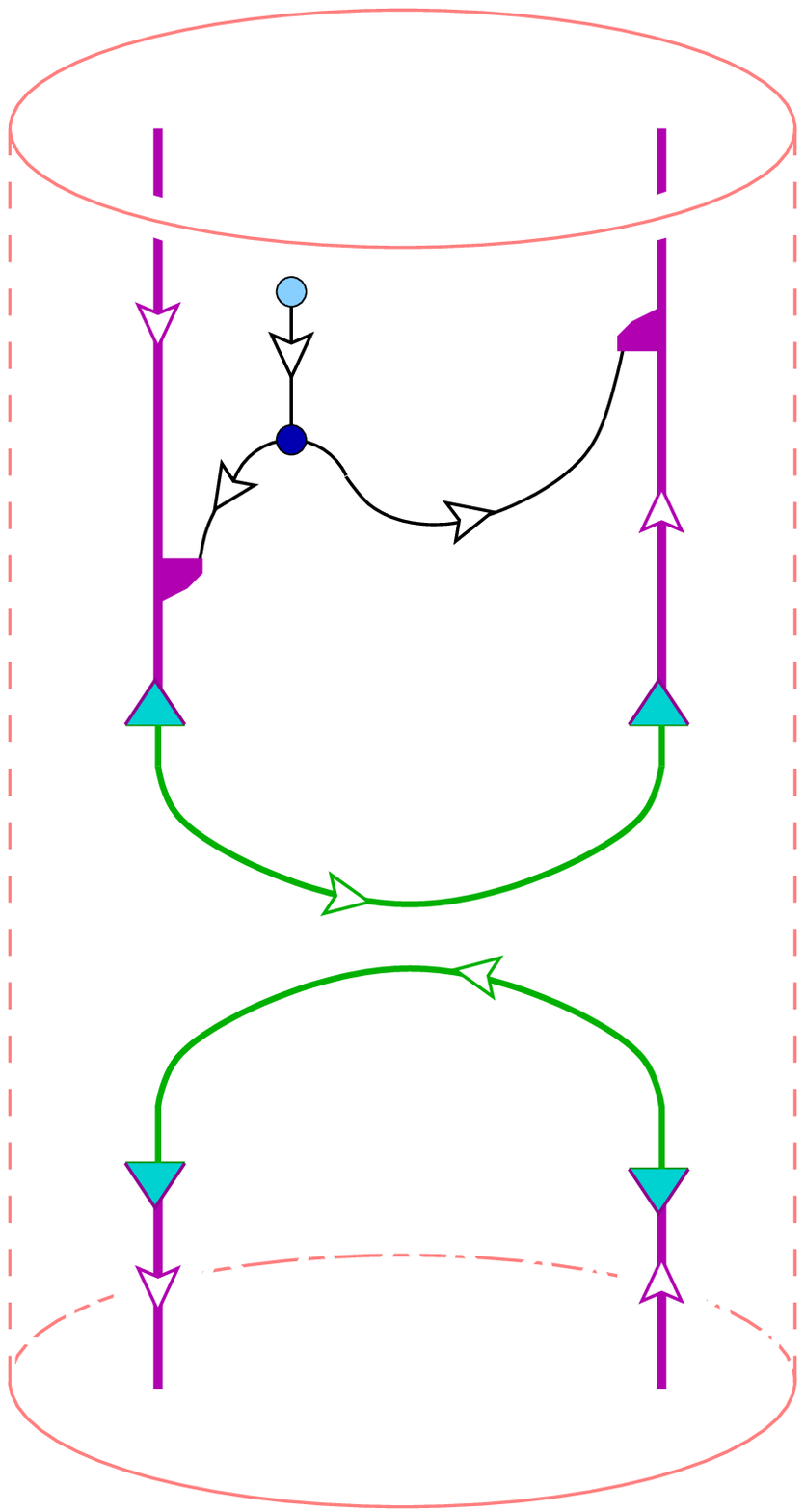}} \end{picture}
   \put(21,157) {\scriptsize$\M_\kappa$}
   \put(61,157) {\scriptsize$\M_{\kappa'}$}
   \put(49,119) {\scriptsize$A$}
   \put(8,106) {\scriptsize$\rho_\kappa$}
   \put(79,135) {\scriptsize$\rho_{\kappa'}$}
   \put(23,91) {\scriptsize$\overline\alpha$}
   \put(23,37) {\scriptsize$\alpha$}
   \put(65,37) {\scriptsize$\overline\beta$}
   \put(65,91) {\scriptsize$\beta$}
   \put(43,73) {\scriptsize$a$}
   \put(39,57) {\scriptsize$a$}
  \put(5,-4){$\dsty\underbrace{\hsp{7}}_{\ \ \ \dsty=:\,\Gamma_3}$}
   }
  \put(384,0)   {\begin{picture}(0,0)(0,0)
              \scalebox{.38}{\includegraphics{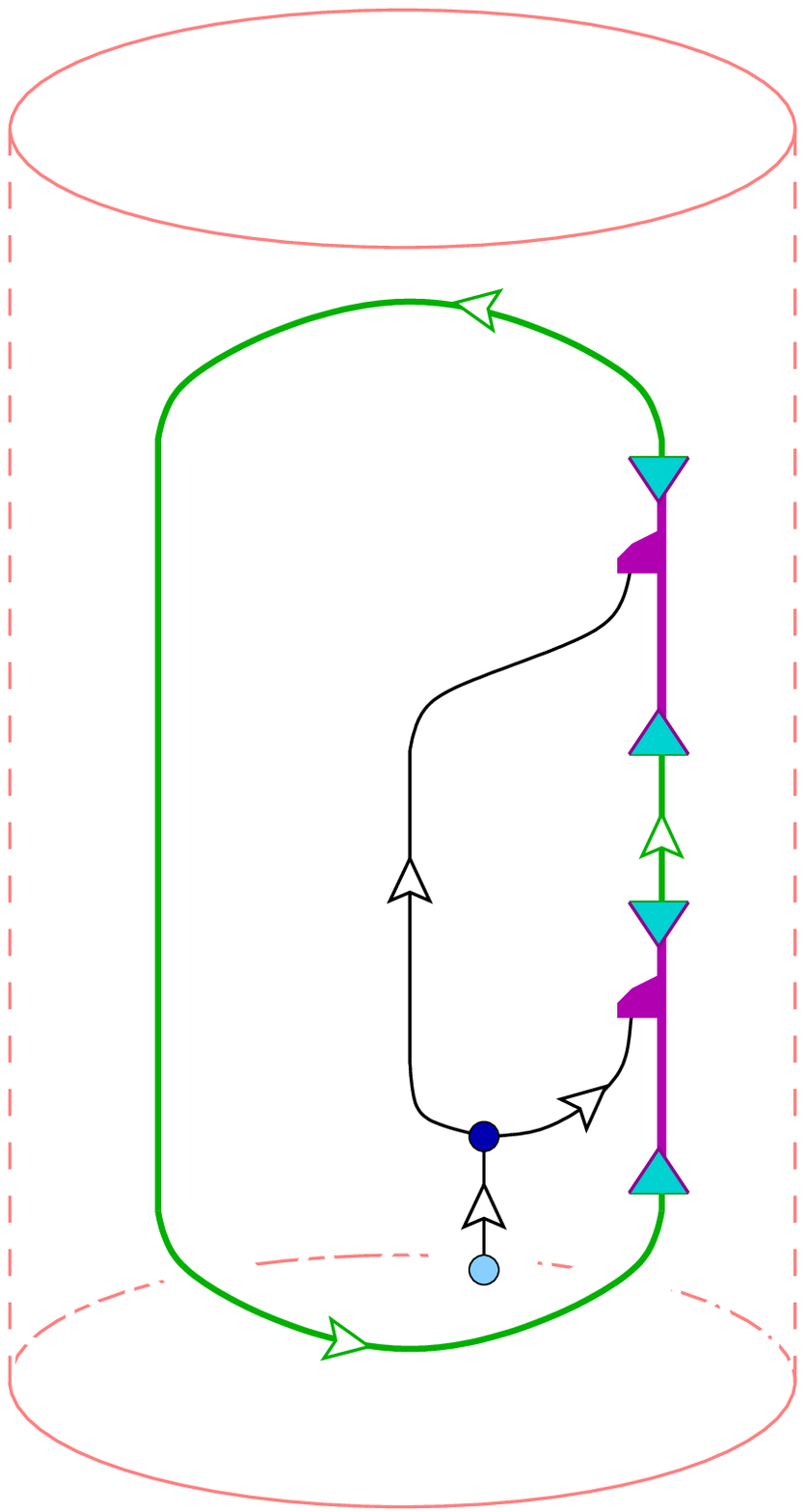}} \end{picture}
   \put(40,78) {\scriptsize$A$}
   \put(11,80) {\scriptsize$a$}
   \put(80.5,75) {\scriptsize$a$}
   \put(66,118) {\scriptsize$\overline\beta$}
   \put(66,87) {\scriptsize$\beta$}
   \put(66,66) {\scriptsize$\overline\alpha$}
   \put(66,36) {\scriptsize$\alpha$}
   \put(78,106) {\scriptsize$\rho_{\kappa'}$}
   \put(78,57) {\scriptsize$\rho_\kappa$}
  \put(5,-4){$\dsty\underbrace{\hsp{7}}_{\ \ \ \dsty=:\,\Gamma_4}$}
   }
  \end{picture} \nonumber \\[1.95em]{} \end{eqnarray} \label{93}
the \lhs\ of \erf{eq:SA-def} becomes
  \be
  D_{\kappa,\kappa'} := \Gamma_1 = \sumI_{a,b} \sum_{\alpha,\beta} \Gamma_2
  = \sumI_a \sum_{\alpha,\beta} \frac{1}{\dim(U_a)}\,
  \Gamma_3 = \sumI_a \sum_{\alpha,\beta} 
  \frac{1}{\dim(U_a)}\, \Gamma_4 \,. \labl{eq:XMN-def}
Here in the first step bases of the type \erf{b-b} for the morphisms between
modules and their subobjects are inserted. The second step uses dominance 
and the fact that $\dim(\calh(m;S^2))\eq\delta_{m,0}$ for the one-point 
blocks on the sphere, implying that only the tensor unit $\one$ survives 
in the intermediate channel. In the third step the ribbon graph is shifted 
along the $S^1$ direction and is shown from a slightly different angle.
\\
One can now use relation \erf{rho-ortho}, which applies because 
$M_{\kappa}$ and $M_{\kappa'}$ are simple $A$-modules. It follows that
  \bea \begin{picture}(315,36)(0,29)
  \put(200,0)   {\begin{picture}(0,0)(0,0)
              \scalebox{.38}{\includegraphics{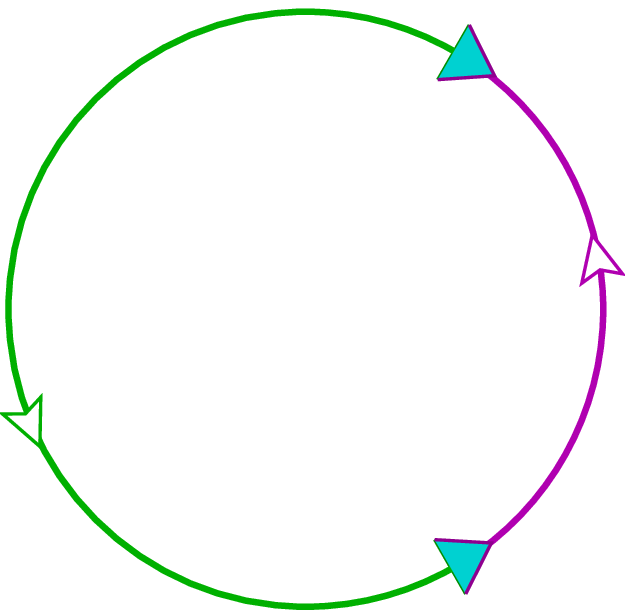}} \end{picture}
  \put(11,13) {\scriptsize$a$}
  \put(51,25) {\scriptsize$\M_\kappa$}
  \put(54,1) {\scriptsize$\alpha$}
  \put(56,61) {\scriptsize$\overline\beta$}   
  \put(-220,32.2) {$D_{\kappa,\kappa'} \; = \;
      \dsty\sumI_a \sum_{\alpha,\beta} \Frac{1}{\dim(U_a)} \,
      \Frac{\dim(U_a)}{\dim(\M_{\kappa})}\, \delta_{\kappa,\kappa'}\,
      \delta_{\alpha,\beta}$}
  \put(78,35) {$=\, \delta_{\kappa,\kappa'} \,, $}
  }
  \epicture05 \labl{eq:XMN-dMN}
where in the last equality it is used that by the completeness 
property \erf{comp} the remaining sum over $a$ and
$\alpha$ simply amounts to a factor of $\dim(\M_{\kappa})$.
\qed

\dt{Proposition}
The matrix $\tilde S^A$ has a right-inverse. That is, there exists a matrix
$\Sigma\eq(\Sigma_{\kappa,j\beta})$, with $\kappa\iN\J$, $j\iN\I$ 
  %% enforce page break:
\\[.2em]
and $\beta$ local in $\Hom(A{\otimes}U_j,U_j)$, such that
  \be  \sumJ_{\kappa} \tilde S^A_{i\alpha,\kappa} \,
  \Sigma_{\kappa,j\beta} = \delta_{i,j}\, \delta_{\alpha,\beta} \,.  \ee

%\medskip
\noindent
Proof:\\
According to lemma \ref{U-Ut}
the matrices $L^A$ and $\tilde L^A$ are each others' inverses:
  \be
  \sumI_{j} \sum_\beta \tilde L^A_{i\alpha,j\beta}\,
  L^A_{j\beta, k\gamma} = \delta_{i,k} \delta_{\alpha,\gamma}
  \,.  \labl{eq:SA-rightinv-a}
Note that the $\beta$-summation runs over a basis of the {\em full\/} 
coupling space $\Hom(A{\otimes}U_j,U_j)$. Let us switch from
this basis $\{\beta\}$, which consists of eigenvectors of $P_j$,
to the basis \erf{def-rho} of $\Hom(A{\otimes}U_j,U_j)$ that is labelled 
by $(\kappa\,\rho\sigma)$ (recall also theorem \ref{thm:mod-Aixj}). 
Let $D^j_{\beta,\kappa\,\rho\sigma}$ denote the (invertible) 
matrix that implements this change of basis, and define
  \be
  K_{(k,\kappa\,\rho\sigma),k\gamma}
  := \sum_{\beta}  (D^j)^{-1}_{\kappa\,\rho\sigma,\beta}\, L^A_{j\beta,k\gamma}
  \qquad{\rm and}\qquad
  \tilde K_{i\alpha,(j,\kappa\,\rho\sigma)}
  := \sum_\beta\tilde L^A_{i\alpha,j\beta}\,D^j_{\beta,\kappa\,\rho\sigma}
  \,. \ee
The matrices $K$ and $\tilde K$ are inverse to each other, too:
  \be
  \sumI_{j} \sumJ_\kappa \sum_{\rho,\sigma}
  \tilde K_{i\alpha,(j,\kappa\,\rho\sigma)}\, K_{(j,\kappa\,\rho\sigma),k\gamma}
  = \delta_{i,k}\, \delta_{\alpha,\gamma} \,.  \labl{eq:SA-rightinv-c}
The pictorial representation of $\tilde K$ is
  \bea \begin{picture}(5,62)(-4,31)
  \put(0,0)   {\begin{picture}(0,0)(0,0)
              \scalebox{.38}{\includegraphics{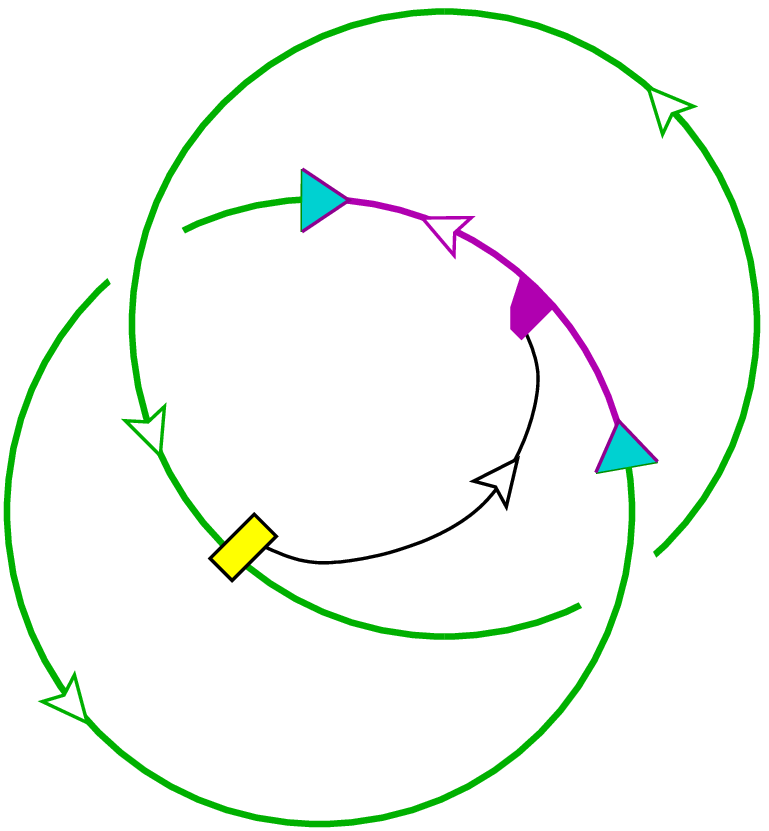}} \end{picture}
   \put(48,69) {\scriptsize$\M_\kappa$}
   \put(49,52) {\scriptsize$\rho_\kappa$}
   \put(20,22) {\scriptsize$\overline\alpha$}
   \put(78,57) {\scriptsize$i$}
   \put(45,37) {\scriptsize$A$}
   \put(33,60) {\scriptsize$\overline\sigma$}
   \put(70,45) {\scriptsize$\rho$}
   \put(15,10) {\scriptsize$j$}
   }
  \put(-80,40) {$\tilde K_{i\alpha,(j,\kappa\,\rho\sigma)} \; = \; $}
  \epicture15 \labl{eq:SA-rightinv-b}
We now impose the condition that $\alpha$ is local.
We are then allowed to insert a projector $P_i$ in the graph
\erf{eq:SA-rightinv-b}. This leads to the following identities:
  \bea \begin{picture}(395,98)(5,19)
  \put(50,0)   {\begin{picture}(0,0)(0,0)
              \scalebox{.38}{\includegraphics{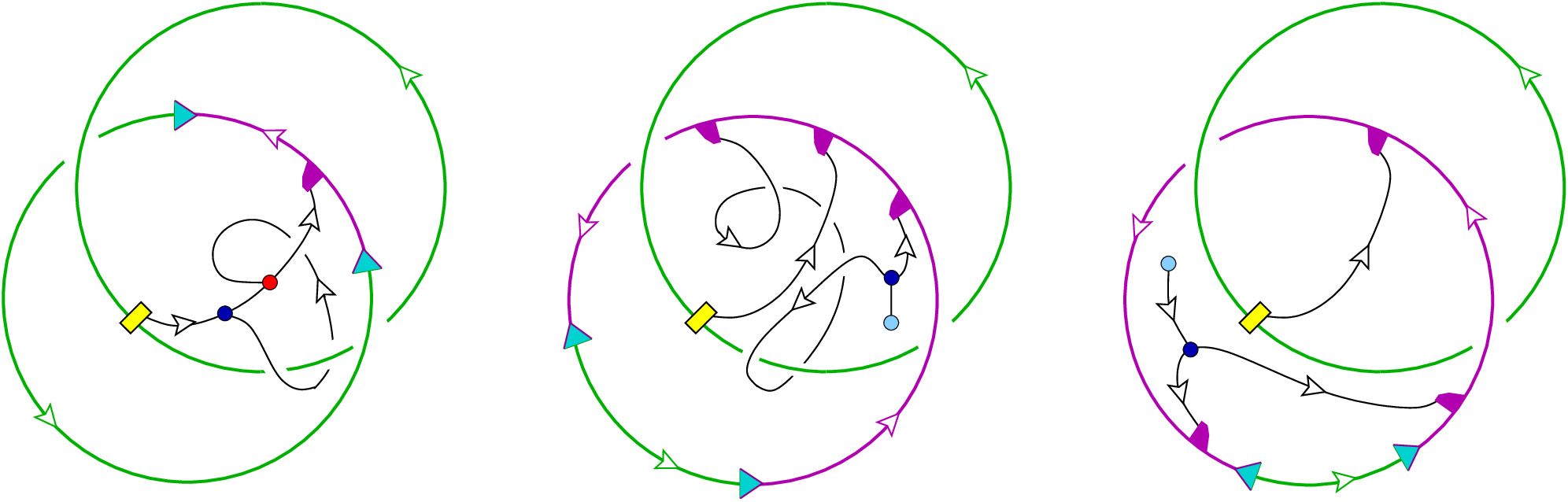}} \end{picture}
   \put(68,85) {\scriptsize$\M_\kappa$}
   \put(199,85) {\scriptsize$\M_\kappa$}
   \put(327,85) {\scriptsize$\M_\kappa$}
   \put(42,58) {\scriptsize$A$}
   \put(182,59) {\scriptsize$A$}
   \put(311,61) {\scriptsize$A$}
   \put(282,27) {\scriptsize$A$}
   \put(24,33) {\scriptsize$\overline\alpha$}
   \put(155,33) {\scriptsize$\overline\alpha$}
   \put(289,48) {\scriptsize$\overline\alpha$}
   \put(97,70) {\scriptsize$i$}
   \put(229,70) {\scriptsize$i$}
   \put(357,70) {\scriptsize$i$}
   \put(10,10) {\scriptsize$j$}
   \put(140,11) {\scriptsize$j$}
   \put(303,8) {\scriptsize$j$}
   \put(88,58) {\scriptsize$\rho$}
   \put(176,8) {\scriptsize$\rho$}
   \put(330,7) {\scriptsize$\rho$}
   \put(44,93) {\scriptsize$\overline\sigma$}
   \put(124,36) {\scriptsize$\overline\sigma$}
   \put(282,0) {\scriptsize$\overline\sigma$}
   \put(-70,55) {$\tilde K_{i\alpha,(j,\kappa\,\rho\sigma)}\, = $}
   \put(112,55) {$=$}
   \put(243,55) {$=$}
    }
  \epicture02 \ee
In the last picture we can use the orthogonality relation \erf{rho-ortho} of
representation functions (in \erf{rho-ortho}, take $k\eq l$, take the trace on
both sides and sum over $\gamma\eq\delta$).
As a consequence, for local $\alpha$ we have the equality
  \be
  \tilde K_{i\alpha,(j,\kappa\,\rho\sigma)} = \delta_{\rho,\sigma}\,
  \Frac{\dim(U_j)}{\dim(\M_\kappa)} \,\tilde S^A_{i\alpha,\kappa\,} \,.
  \ee
Using this relation, \erf{eq:SA-rightinv-c} takes the form
  \be
  \sumI_j \sumJ_\kappa \sum_{\rho,\sigma} \delta_{\rho,\sigma}\,
  \Frac{\dim(U_j)}{\dim(\M_\kappa)} \, \tilde S^A_{i\alpha,\kappa}
  K_{(j,\kappa\,\rho\sigma),k\gamma}
  = \sumJ_\kappa \tilde S^A_{i\alpha,\kappa} \, \Sigma_{\kappa,k\gamma}
  = \delta_{i,k}\, \delta_{\alpha,\gamma} \ee
with
  \be
  \Sigma_{\kappa,k\gamma} := \sumI_j \sum_\rho 
  \Frac{\dim(U_j)}{\dim(\M_\kappa)} \, K_{(j,\kappa\,\rho\rho),k\gamma}
  \,.  \ee
Thus the matrix $\Sigma$ defined this way, with $\gamma$ restricted to be
local, is right-inverse to $\tilde S^A$.
\qed

We conclude that $\tilde S^A$ has both a left- and a right-inverse. Thus
the two inverses coincide, i.e.\ $S^A$ and $\tilde S^A$ are square 
matrices and inverse to each other.  This provides us with a key 
information for the next section: For $M_{\kappa},M_{\kappa'}$ simple 
$A$-modules, $U_i,U_j$ simple objects and $\alpha,\beta$
local basis vectors in the corresponding coupling spaces, we have
  \be
  \sumI_i \sum_{\alpha\;{\rm local}} S^A_{\kappa,i\alpha}\,
  \tilde S^A_{i\alpha,\kappa'} = \delta_{\kappa,\kappa'}
  \qquad{\rm and} \qquad
  \sumJ_\kappa \tilde S^A_{i\alpha,\kappa} \, S^A_{\kappa,j\beta} =
  \delta_{i,j} \, \delta_{\alpha,\beta} \,.  \ee
For now let us just note one astonishing consequence of this result.

\dtl{Theorem}{thm:simp=trace}
Let $A$ be a symmetric special Frobenius algebra in the modular tensor
category $\calc$. Then
  \be
  |\,{\rm isom.\,classes~of~simple~left~}A\mbox{-modules}\,|
  = \tr\llb C\, Z(A)\lrb \,.  \ee
Proof:\\
Consider the space of local couplings in $\Hom(A\oti U_k,U_k)$, or
equivalently, in $\Hom(A\oti U_k,U_{\bar k}^\vee)$. By lemma
\ref{lem:dim-loc-Hom}, its dimension is equal to $Z_{k,\bar k}$. 
Thus a basis $B$ for all local couplings in $\bigoplus_{k\in\II} 
\Hom(A\oti U_k,U_k)$ has $\sumI_k Z_{k,\bar k}$ elements. 
Moreover, since $S^A$ is a square matrix, the number
of isomorphism classes of simple modules must be the same as
the number of basis vectors in $B$.
\qed

\dtl{Remark}{samenumber}
(i)\hsp{.6}In CFT terms, this result tells us that for the CFT based
on $A$ the number of elementary boundary conditions (corresponding 
to simple $A$-modules) is the same as the number of Ishibashi 
states, whose total number is $\sumI_k Z_{k\bar k}$). This is the 
completeness condition for boundary conditions \cite{prss3,bppz2}.  
The completeness condition enters in the proof \cite{bppz2} that 
the annulus coefficients of a CFT furnish a NIM-rep of the fusion 
rules. Theorem \ref{thm:simp=trace} implies that CFTs constructed
from algebra objects fulfill this condition, and hence their annulus 
coefficients must provide a NIM-rep. In theorem \ref{thm:annuluscoeff} 
below we will see explicitly that this is indeed the case. 
\\[.13em]
(ii) By combining theorem \ref{thm:simp=trace} with 
propositions \ref{prop:bimod-leftmod} and \ref{prop:Ztilde} we 
arrive at the following relation for the category of \AB-bimodules:
  \be \bearll
  |\,{\rm isom.\,classes~of~simple~}A\mbox{-}B\mbox{-bimodules}\,| \!\!
  &= \tr\llb C Z(A \Oti B_{\rm op})\lrb
  \\[.49em]
  &= \tr \llb C Z(A) C Z(B)\oT \lrb\qquad
  \\[.49em]
  &= \tr \llb Z(A) Z(B)\oT\lrb \,,
  \eear \labl{eq:number-bimod}
where in the last step we used that the charge conjugation
matrix commutes with a modular invariant matrix $Z$.

Recall the comment in the end of section \ref{bc-to-rep} that 
\AB-bimodules describe tensionless interfaces between the CFTs 
associated to the algebras $A$ and $B$. In the special case $A\eq B$ 
equation \erf{eq:number-bimod} reproduces the relation for the 
number of generalised defect lines found in \cite{pezu5}.

%%%%%%%%%%%%%%%%%%%%%%%%%%%%%%%%%%%%%%%%%%%%%%%%%%%%%%%%%%%%%%%%%%%%%%%%

\subsection{Annulus partition functions} \label{sec:annulus-pf} 

Boundary conditions correspond to modules of a symmetric special 
Frobenius algebra $A$, and elementary boundary conditions are simple 
$A$-modules.  In this section we work out the partition function 
$\Ann MN$ of an annulus with boundary conditions $M$ and $N$.
Let us for now take $M$ and $N$ to be simple $A$-modules. 
The world sheet $\rmX$ for the annulus amplitude is an annulus.
Its complex double $\hat \rmX$ is a torus $\torus\eq S^1\,{\times}\, S^1$, 
and the connecting three-manifold $M_\rmX$ is a
full torus $D\,{\times}\, S^1$, where $D$ is a disk.

According to the prescription in section \ref{sec:con-mf-rib}, the first 
step in the construction of the \parfu\ is to specify a triangulation of
the world sheet $\rmX$; let us choose the one given in the picture on the 
\lhs\ of 
  \bea  \begin{picture}(335,70)(0,20)
  \put(0,0)   {\begin{picture}(0,0)(0,0)
              \scalebox{.38}{\includegraphics{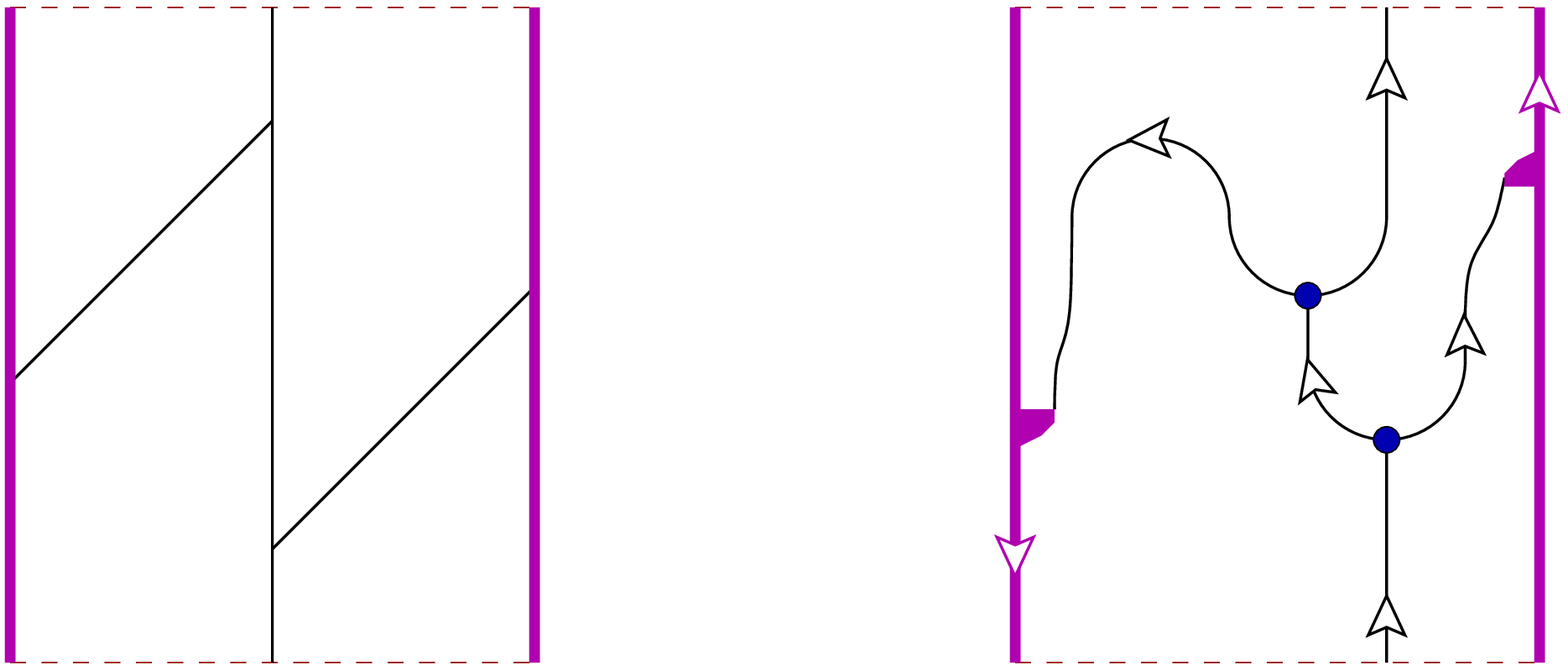}} \end{picture}}
  \put(123.2,2.9)  {\scriptsize$\n$}
  \put(204.0,79.6) {\scriptsize$\M$}
  \put(224.6,43.1) {$=$}
  \put(245.8,2.9)  {\scriptsize$\n$}
  \put(329.0,79.6) {\scriptsize$\M$}
  \put(175,15) {\scriptsize$A$}
  \put(162,57) {\scriptsize$A$}
  \put(190,31) {\scriptsize$A$}
  \put(286,69) {\scriptsize$A$}
  \put(299,65) {\scriptsize$A$}
  \put(307,44) {\scriptsize$A$}
  \epicture06 \labl{ann}
(Here top and bottom are to be identified, which is indicated by the
dashed lines.) Then we substitute the elements \erf{Phi1}, \erf{AAA} 
to convert the triangulation into a
ribbon graph; simplifying the result, we arrive at the graph in the middle
of \erf{ann}. Using the Frobenius property of $A$ and the representation 
property of $\r_M$, this graph can in turn be rewritten as on the \rhs.
It follows that the $A$-ribbon that runs around the annulus can be removed. 
Then the ribbon graph describing the annulus amplitude finally becomes
  \bea  \begin{picture}(100,106)(0,16)
  \put(0,0)   {\begin{picture}(0,0)(0,0)
              \scalebox{.38}{\includegraphics{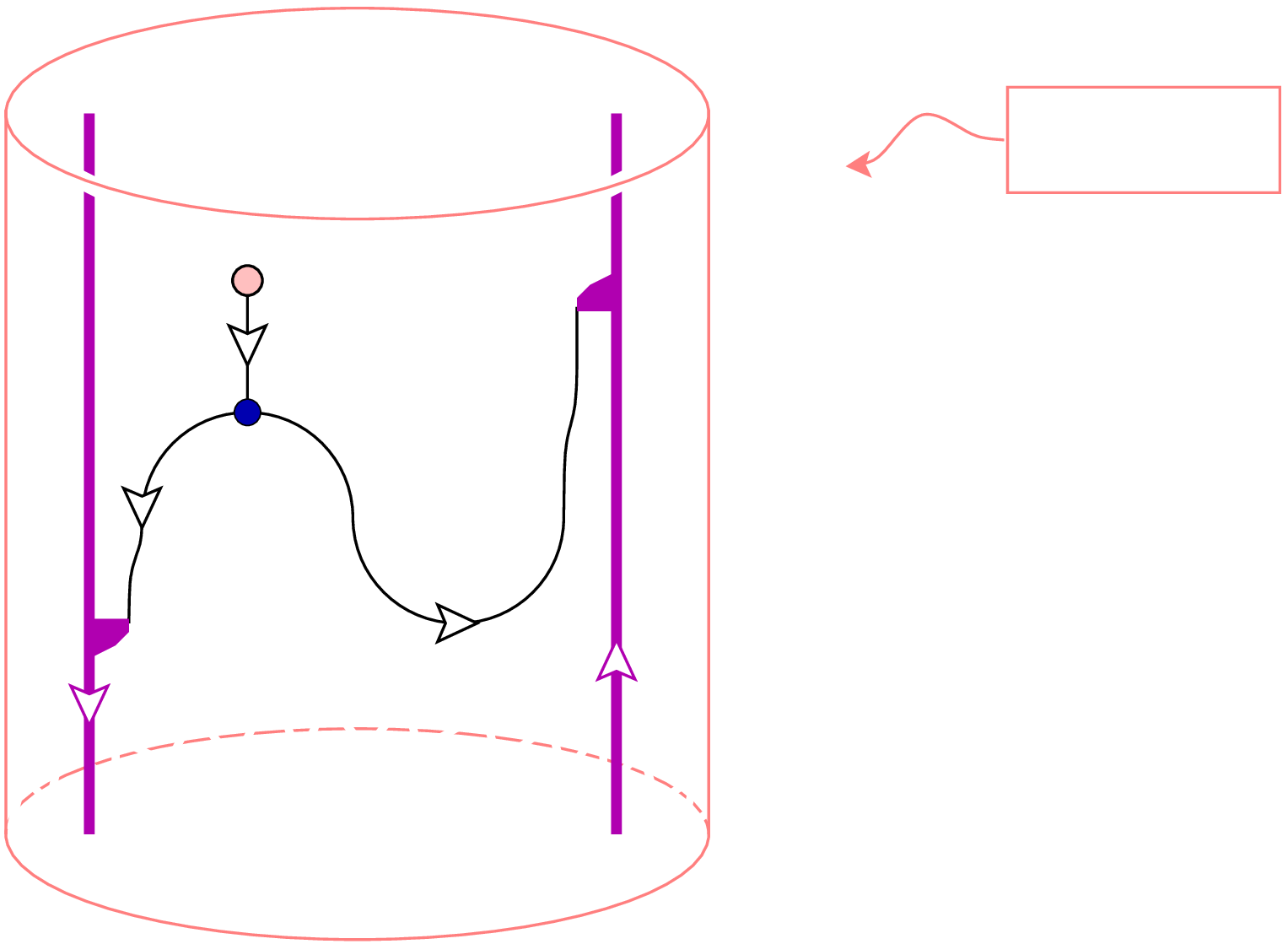}} \end{picture}}
  \put(-50.5,55.9){$ \Ann MN\;=$}
  \put(14,14.9) {\scriptsize$\n$}
  \put(70,14.9) {\scriptsize$\M$}
  \put(133.5,101.2) {$D{\times}S^1$}
  \put(16,35) {\scriptsize$\rho_N$}
  \put(64,86) {\scriptsize$\rho_M$}
  \put(21,50) {\scriptsize$A$}
  \put(47,53) {\scriptsize$A$}
  \epicture-1 \label{eq:AMN-ribbon} \labl{Amn2}
This graph determines an element $\Ann MN$ in the space $\calhT$ of 
zero-point conformal blocks on the torus.
To obtain the annulus coefficients we must expand $\Ann MN$ in a basis 
of blocks on the torus, i.e.\ in terms of characters \erf{eq:tor-bas}:
  \be  \Ann MN = \sumI_k \Ann {kM}N\, \tbas k{} \,.  \ee
Just like for the torus partition function, we can isolate the
coefficient $\Ann {kM}N$ by gluing the basis dual to $\tbas k{}$
to the boundary of the three-manifold on both sides of the equation. 
We are then left with
  \bea  \begin{picture}(90,98)(0,26)
  \put(0,0)   {\begin{picture}(0,0)(0,0)
              \scalebox{.38}{\includegraphics{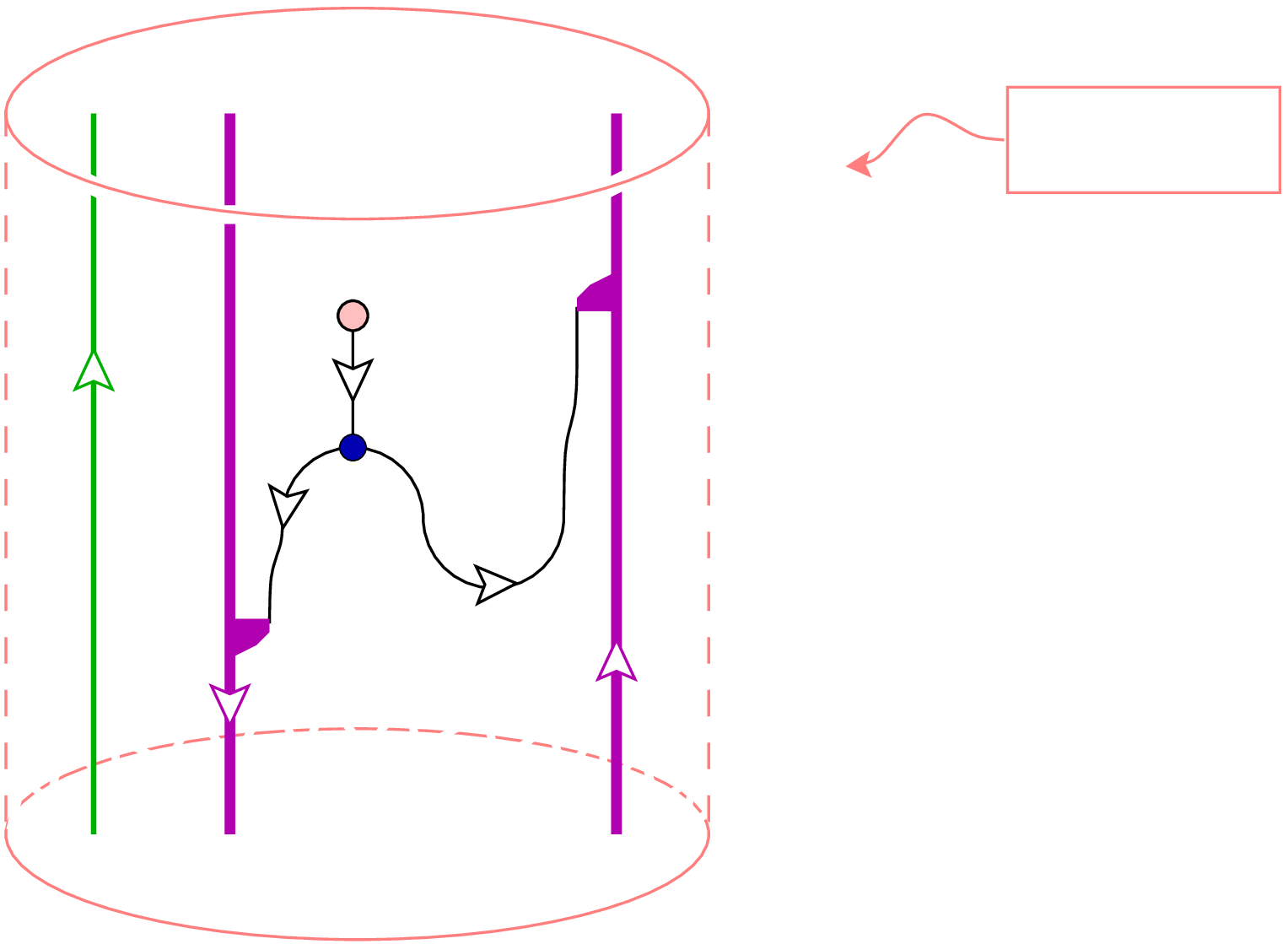}} \end{picture}}
  \put(-54.5,55.9){$ \Ann{kM}N\;=$}
  \put(13.5,14.9) {\scriptsize$k$}
  \put(31.1,14.9) {\scriptsize$\n$}
  \put(71.5,14.9) {\scriptsize$\M$}
  \put(132,100.6) {$S^2{\times}S^1$}
  \put(34,35) {\scriptsize$\rho_N$}
  \put(64,86) {\scriptsize$\rho_M$}
  \put(38,50) {\scriptsize$A$}
  \put(57,53) {\scriptsize$A$}
  \epicture10 \label{eq:AkMN-graph} \labl{Akmn}

We now establish that the quantities defined by formula \erf{Amn2} indeed
possess the properties that befit a partition function on an annulus.

\dtl{Theorem}{thm:annuluscoeff} 
Let $A$ be a symmetric special Frobenius algebra, $M\,{\equiv}\,
M_{\kappa}$ and $N\,{\equiv}\,M_{\kappa'}$ ($\kappa,\kappa'\iN\J$)
simple $A$-modules, and $U_k$ ($k\iN\I$) a
simple object. Then the numbers $\Ann {kM}N$ given by the invariant
of the ribbon graph \erf{eq:AkMN-graph} have the following properties:
\\[.08em]
(i)\hsp{.9}Non-negativity: 
  \be  \Ann {kM}N \in \zet_{\ge 0} \,.  \ee
(ii)\hsp{.6}Uniqueness of the vacuum:  
  \be  \Ann {0M}N = \delta_{M,N} \,.  \labl{Ann0MN}
(iii) Exchange of boundary conditions:
  \be \Ann {k\,M}N = \Ann {{\overline k}\,N}M \;. \labl{eq:AnnTranspose}
(iv) Consistency with the bulk spectrum:
  \be  \Ann {kM}N = \sumI_p \sum_{\alpha\; {\rm local}}
  S^A_{N,p\alpha}\, \frac{S_{kp}}{S_{0p}}\,\tilde  S^A_{p\alpha,M} \,.
  \labl{eq:Ann-closed}
(v) \,NIM-rep of the fusion rules:
  \be \Ann i{}\;\Ann j{} = \sumI_k {N_{ij}}^{\!k}\, \Ann k{} \;,
  \labl{eq:Ann-NIMrep}
where the $\Ann k{\,}$ are the matrices with entries $\Ann {kM}N$.

\medskip\noindent
Proof:\\
To show (i) one cuts the three-manifold \erf{eq:AkMN-graph} along
an $S^2$ to obtain a ribbon graph in $S^2\,{\times}\,[-1,1]$. This 
defines a linear map
  \be Q_k :\quad
  \calh(k,(\n,-),\M ; S^2) \rightarrow \calh(k,(\n,-),\M ; S^2) \,,
  \ee
from which the annulus coefficients are recovered via
  \be  \Ann {kM}N = {\rm tr}_{\calH(k,(\n,-),\M ; S^2)} Q_k \,. 
  \labl{eq:ann-Q-trace}
Next one considers the identities
  \bea  \begin{picture}(370,129)(0,23)
  \put(0,0)   {\begin{picture}(0,0)(0,0)
              \scalebox{.38}{\includegraphics{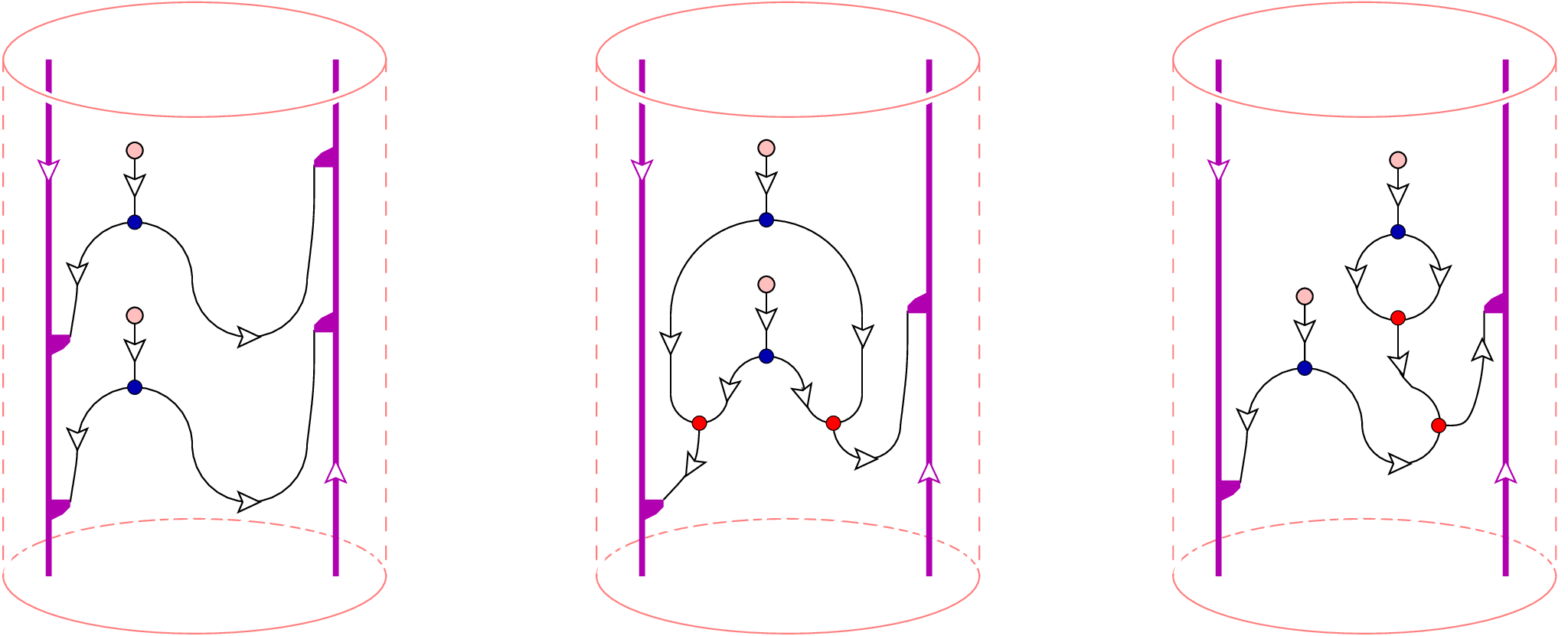}} \end{picture}}
  \put(13,14.9) {\scriptsize$\n$}
  \put(71,14.9) {\scriptsize$\M$}
  \put(155,14.9) {\scriptsize$\n$}
  \put(213,14.9) {\scriptsize$\M$}
  \put(292,14.9) {\scriptsize$\n$}
  \put(350,14.9) {\scriptsize$\M$}
  \put(114.6,71.1){$=$}
  \put(251.6,71.1){$=$}
  \put(23,100) {\scriptsize$A$}
  \put(48,83) {\scriptsize$A$}
  \put(21,41) {\scriptsize$A$}
  \put(48,43) {\scriptsize$A$}
  \put(162,74) {\scriptsize$A$}
  \put(199,76) {\scriptsize$A$}
  \put(169,41) {\scriptsize$A$}
  \put(196,37) {\scriptsize$A$}
  \put(318,90) {\scriptsize$A$}
  \put(318,49) {\scriptsize$A$}
  \put(343,56) {\scriptsize$A$}
  \put(301,48) {\scriptsize$A$}
  \epicture04 \labl{Q2Q}
Here the first step uses the representation property, while the second
step results from repeated application of the associativity and 
Frobenius property of $A$. Using also that $A$ is special, as well as the
defining property of the unit, one concludes that the map $Q_k$ is a 
projector, 
  \be  Q_k \circ Q_k = Q_k \,, \ee
and therefore
  \be  {\rm tr}_{\calH(k,(\n,-),\M ; S^2)} \in \zet_{\ge 0} \,.  \ee
(ii): For $k\eq\One$, $\Ann {kM}N$ is given by the graph $\Gamma_{\!1}$ 
in picture \Erf93. The statement of (ii) is therefore equivalent to
the result \erf{eq:XMN-dMN} that was obtained in the course 
of proving proposition \ref{pr:XMN}.
\\[.16em]
To see (iii) it is sufficient to `rotate' the ribbon graph \erf{Akmn}
according to
  \bea  \begin{picture}(195,96)(0,23)
  \put(0,0)   {\begin{picture}(0,0)(0,0)
              \scalebox{.38}{\includegraphics{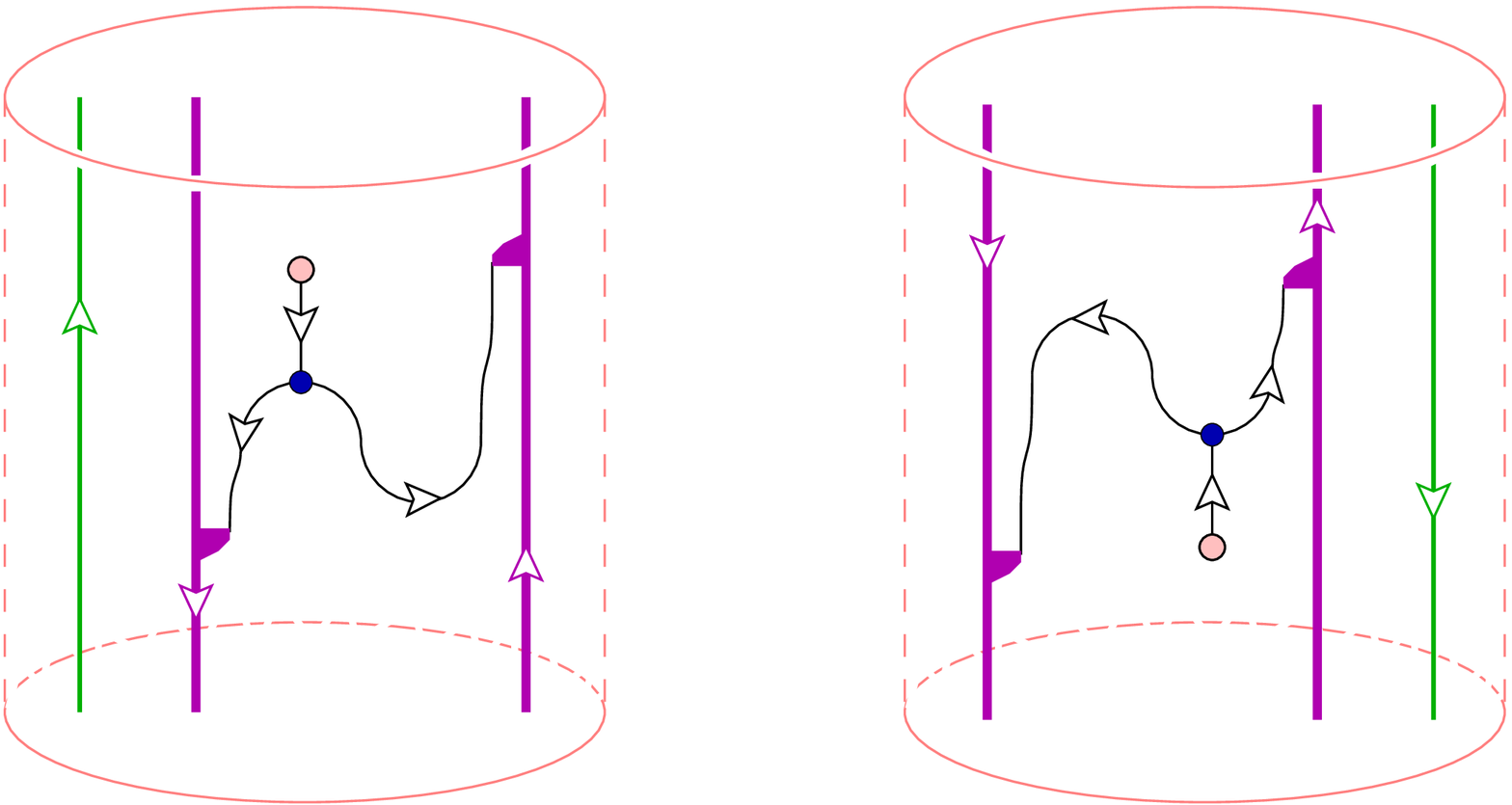}} \end{picture}}
  \put(-59.5,55.9){$\Ann{{\overline k}\,M}N\,=$}
  \put(13.8,14.9) {\scriptsize$\overline k$}
  \put(31.6,14.9) {\scriptsize$\n$}
  \put(70.2,14.9) {\scriptsize$\M$}
  \put(34.9,36.5) {\scriptsize$\rho_{\!N}^{}$}
  \put(64,86) {\scriptsize$\rho_{\!M}^{}$}
  \put(37,50) {\scriptsize$A$}
  \put(57,53) {\scriptsize$A$}
  \put(110,55.9){$=$}
  \put(152.2,13.9) {\scriptsize$\M$}
  \put(156.2,34.9) {\scriptsize$\rho_{\!M}^{}$}
  \put(157.1,50.5) {\scriptsize$A$}
  \put(169.1,61.5) {\scriptsize$A$}
  \put(185.9,80.5) {\scriptsize$\rho_{\!N}^{}$}
  \put(191.8,13.9) {\scriptsize$\n$}
  \put(212.6,13.9) {\scriptsize$k$}
  \epicture04 \labl{Akmn-rot}
The second equality holds because the two three-manifolds with embedded
ribbon graphs are related by an (invertible) orientation preserving
homeomorphism and thus have the same topological invariant. 
The second ribbon graph in \erf{Akmn-rot} equals $\Ann{k\,N}M$, as can 
be seen by using that $A$ is symmetric and by inserting a pair 
$\pi_k \cir \pi_k^{-1}$ to replace the $\bar k$-ribbon by
a $k$-ribbon with reversed orientation of the core.
\\[.13em]
For the proof of (iv) we use the surgery relation \erf{S-xfer} with
$i\eq k$ and $X\eq \n^\vee{\otimes}\M$, and with $\Phi$ corresponding to
an $A$-ribbon suspended between $N^\vee$ and $M$ as in \erf{eq:AkMN-graph}. 
This yields
  \bea  \begin{picture}(390,325)(12,7)
              \put(0,210){\begin{picture}(0,0)(0,0)
  \put(0,0)     {\begin{picture}(0,0)(0,0)
                \scalebox{.38}{\includegraphics{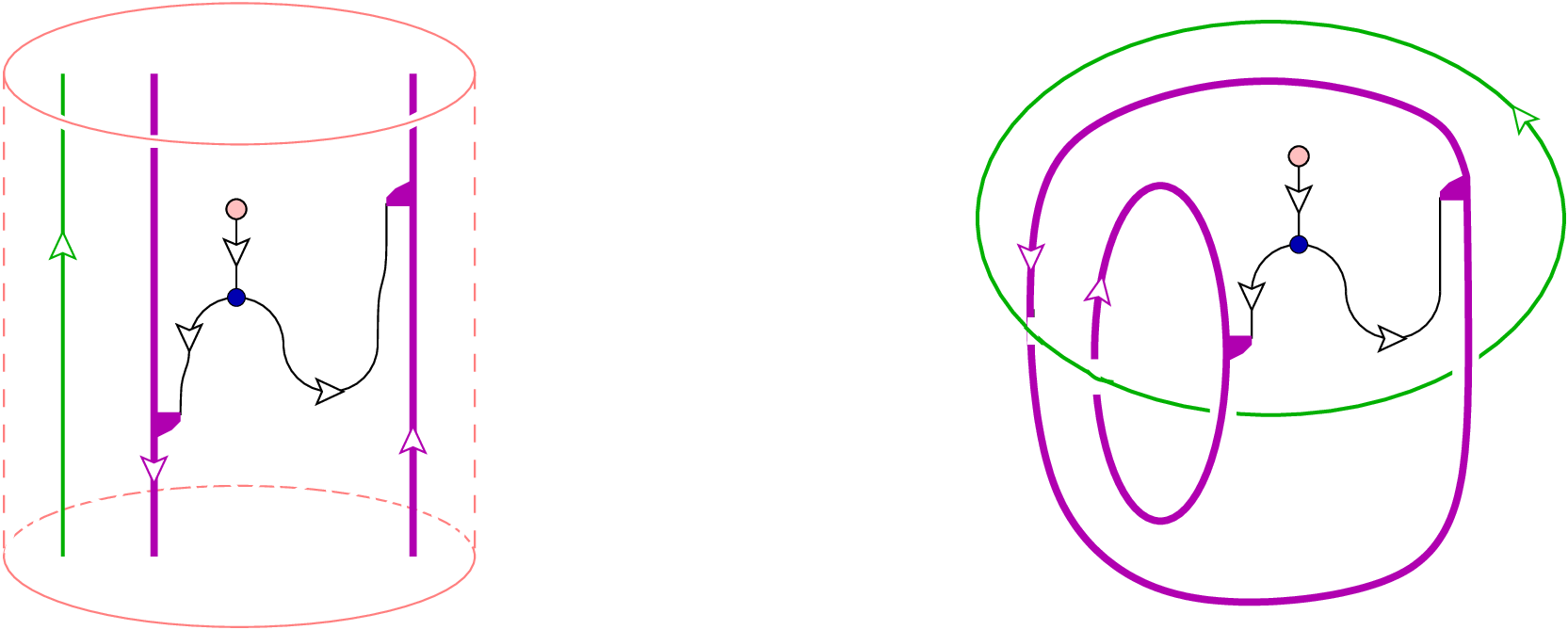}} \end{picture}
  \put(13.5,14.9) {\scriptsize$k$}
  \put(30.5,14.9) {\scriptsize$\n$}
  \put(71.5,14.9) {\scriptsize$\M$}
   \put(56,54) {\scriptsize$A$}
   \put(262,65) {\scriptsize$A$}
   \put(184,76) {\scriptsize$p$}
   \put(237,27) {\scriptsize$\n$}
   \put(194,22) {\scriptsize$\M$}
  }
  \put(110.5,54.9){$=\;S_{0,0} \dsty\sumI_p S_{k,p}$}
              \end{picture}}
              \put(0,105){\begin{picture}(0,0)(0,0)
  \put(160,0)   {\begin{picture}(0,0)(0,0)
                \scalebox{.38}{\includegraphics{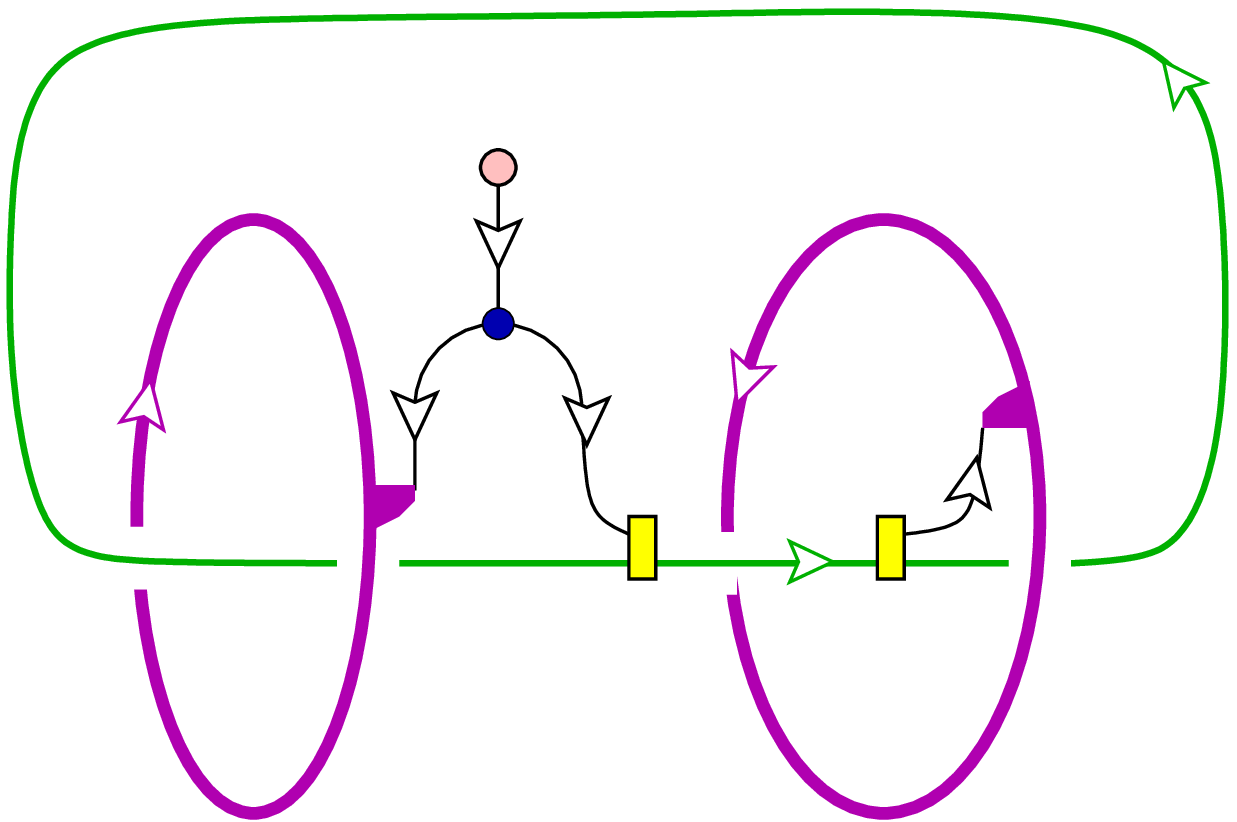}} \end{picture}
   \put(137,56) {\scriptsize$p$}
   \put(9.1,5) {\scriptsize$\n$}
   \put(111,5) {\scriptsize$\M$}
   \put(62,52) {\scriptsize$A$}
   \put(97,35) {\scriptsize$A$}
   \put(88,23) {\scriptsize$p$}
   \put(67,22) {\scriptsize$\alpha$}
   \put(95,20) {\scriptsize$\overline\alpha$}
  }
  \put(56,54.9) {$=\;S_{0,0} \dsty\sumI_p S_{k,p}\sum_\alpha$}
              \end{picture}}
              \put(0,0){\begin{picture}(0,0)(0,0)
  \put(198,0)   {\begin{picture}(0,0)(0,0)
                \scalebox{.38}{\includegraphics{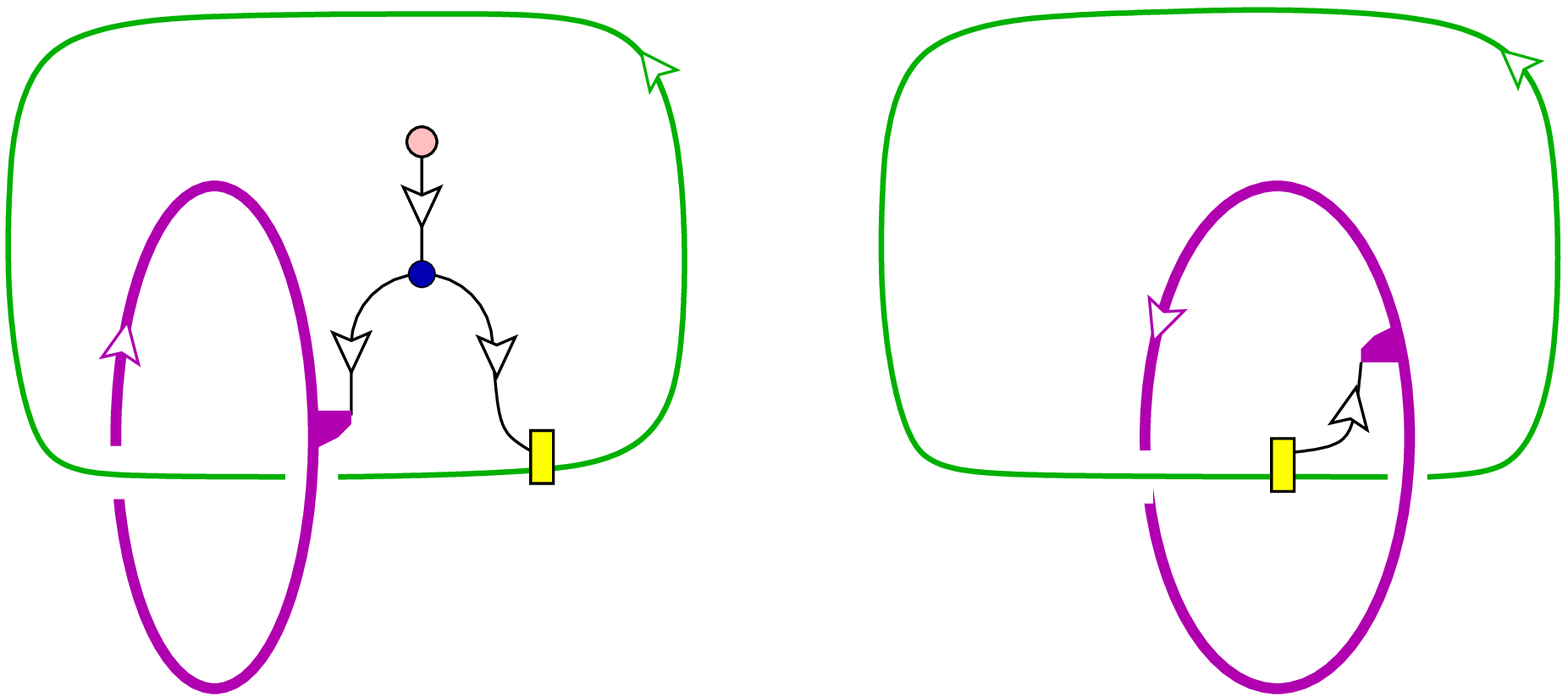}} \end{picture}
   \put(90,58) {\scriptsize$p$}
   \put(204,58) {\scriptsize$p$}
   \put(9.1,5) {\scriptsize$\n$}
   \put(179.3,5) {\scriptsize$\M$}
   \put(62,51) {\scriptsize$A$}
   \put(166,35) {\scriptsize$A$}
   \put(68,22) {\scriptsize$\alpha$}
   \put(163,20) {\scriptsize$\overline\alpha$}
  }
  \put(56,54.9) {$=\;S_{0,0} \dsty\sumI_p \frac{S_{k,p}}{\dim(U_p)}
                 \sum_{\alpha\; {\rm local}}$}
              \end{picture}}
  \epicture-6 \label{eq:AMN-prop-a} \labl{Akmn-S}
Here the first step is the relation \erf{S-xfer}. 
In the second step dominance is used to insert a basis
for $\Hom(A{\otimes}U_p,U_p)$ (in the summation over intermediate objects
$U_m$, only $U_p$ survives, as is seen by reading the picture from left 
to right, which involves a morphism from $\one$ to $U_m{\otimes}U_p^\vee$ 
to $\one$, implying that $m\eq p$). The index $\alpha$ runs over the 
eigenbasis in $\Hom(A{\otimes}U_p,U_p)$, and $\bar\alpha$ denotes the dual 
basis. In the last step, dominance is used once more, and again only a 
single intermediate channel contributes, this time the tensor unit $\one$. 
Moreover, the sum over $\alpha$ can be restricted to only local basis 
vectors of $\Hom(A{\otimes}U_p,U_p)$, by the same reasoning that was used in
the proof of lemma \ref{lem:only-local}, i.e.\ by similar moves as in
\erf{eq:only-local}.
\\
The second ribbon graph in the last line of \erf{eq:AMN-prop-a} is nothing
but the quotient $\tilde S^A_{p\alpha,M} / S_{0,0}$, as defined in 
\erf{eq:SAt-def}. Consider now the result \erf{eq:AMN-prop-a} for the 
special case $k\eq\One$. Then according to \erf{Ann0MN} we must get
$\delta_{M,N}$. Using also that $S_{0,p}/\dim(U_p)\eq S_{0,0}$,
it follows that the first ribbon graph in the last line describes the
matrix that is inverse to $\tilde S^A$. Thus it equals
$S^A_{N,p\alpha}/S_{0,0}$, with $S^A$ as defined in \erf{eq:SA-def}. 
In pictures,
  \bea  \begin{picture}(172,55)(0,29)
  \put(74,0)  {\begin{picture}(0,0)(0,0)
              \scalebox{.38}{\includegraphics{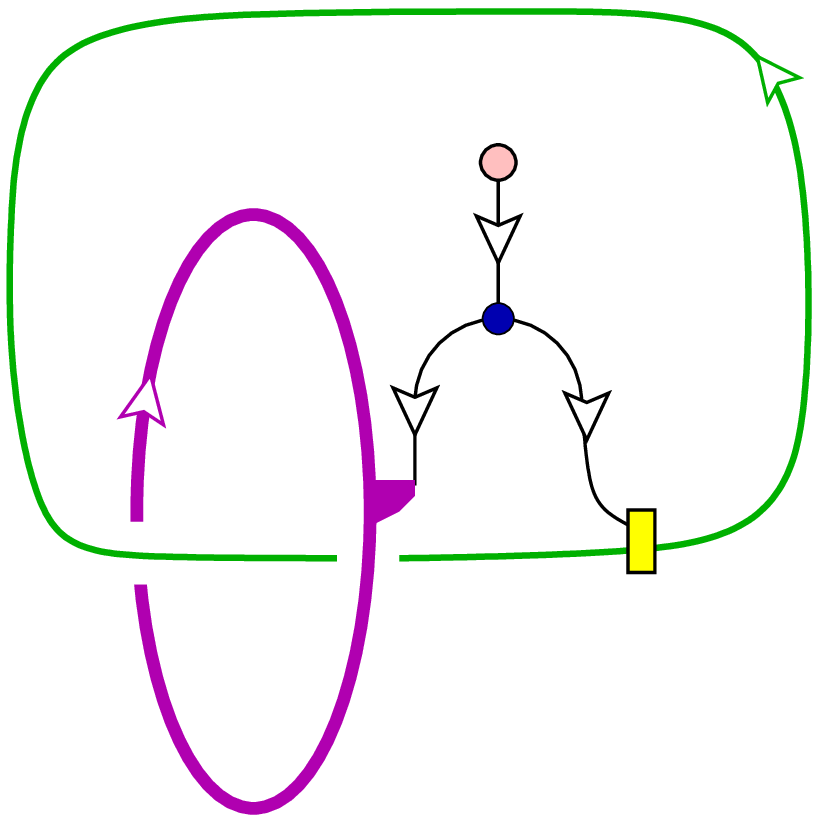}} \end{picture}
   \put(9.5,3.5) {\scriptsize$\n$}
   \put(90.7,58) {\scriptsize$p$}
   \put(63.3,51) {\scriptsize$A$}
   \put(67,21) {\scriptsize$\alpha$}
   \put(28.3,35) {\scriptsize$\r_N^{}$}
  }
  \put(-2,44.9)    {$S^A_{N,p\alpha} \;=\; S_{0,0}$}
  \epicture05 \labl{SAmpa}
This completes the proof of part (iv).
\\[.13em]
Finally, property (v) is a direct consequence of (iv). Indeed, substituting 
\erf{eq:Ann-closed} into the \rhs\ of \erf{eq:Ann-NIMrep}, we get
  \be\begin{array}{ll} 
  \dsty\sumJ_R \Ann {iM}R \Ann {jR}N \!\!
  &= \dsty\sumJ_R \sum_{p,\alpha} 
  S^A_{R,p\alpha} \,\frac{S_{p,i}}{S_{p,0}}\, \tilde S^A_{p\alpha,M}
  \sum_{q,\beta}
  S^A_{N,q\beta} \,\frac{S_{q,j}}{S_{q,0}}\, \tilde S^A_{q\beta,R}
  \\{}\\[-.87em]
  &= \dsty\sum_{p,\alpha} \sum_{q,\beta}
  S^A_{N,q\beta} \,\frac{S_{p,i}}{S_{p,0}}\, \delta_{p,q}\,
  \delta_{\alpha,\beta} \,\frac{S_{q,j}}{S_{q,0}}\, \tilde S^A_{p\alpha,M} 
  \\{}\\[-.87em]
  &= \dsty\sum_{p,\alpha} \sumI_k \frac{S_{p,i}S_{p,j}S^{-1}_{p,k}}{S_{p,0}}
  \,S^A_{N,p\alpha} \,\frac{S_{p,k}}{S_{p,0}}\, \tilde S^A_{p\alpha,M} 
  \, = \,\dsty\sumI_k {N_{ij}}^k\, \Ann {kM}N \,,
  \end{array}\ee
which establishes that we are indeed dealing with a NIM-rep.
\qed

%\medskip \noindent
\dt{Remark}
(i)\hsp{.9}For $A\eq\one$, i.e.\ for the charge conjugation modular 
invariant $Z_{ij}\eq\delta_{i,\bar \jmath}$, one recovers \cite{fffs3} 
the original result \cite{card9} that in this situation the annulus 
coefficients coincide with the fusion rules.
\\[.13em]
(ii)\hsp{.6}The matrices $\Ann{k}{}\,$\ commute and are normal (since
${\Ann{k}{}}\oT\eq\Ann{\bar k}{}\;$, see formula \erf{eq:AnnTranspose})
and hence can be simultaneously diagonalised, with eigenvalues
$S_{k,a}/S_{0,a}$.
Requiring consistency between the open and closed string channel in 
the annulus amplitude leads to the condition that the multiplicity
$m(a)$ of the set $\{S_{k,a}/S_{0,a}\,|\,k\iN\II\}$ of eigenvalues must be
equal to $Z_{a\bar a}$. This condition has been thoroughly investigated 
in \cite{gann17}, where many examples of modular invariants without 
associated NIM-rep as well as NIM-reps without associated modular invariant
were found.
\\
Since the number of basis elements in the space
$\Loc(A\Oti U_p,U_p)$ of local couplings is $Z_{p \bar p}$ (see lemma 
\ref{lem:dim-loc-Hom}), part (iv) of theorem \ref{thm:annuluscoeff} 
implies that for any CFT obtained from an algebra object, the
annulus coefficients $\Ann{kM}N$ are consistent with the modular 
invariant $Z_{ij}$ in the sense of definition 4 of \cite{gann17}.
\\[.13em]
(iii) In \cite{scst}, polynomial equations and trace identities for the
(integral) coefficients appearing in the torus, cylinder, M\"obius and Klein
bottle amplitudes have been presented. Those identities of \cite{scst} that
involve $Z_{ij}$ and $\Ann{kM}N$ follow from completeness of the boundary 
conditions, and are thus automatically satisfied by CFTs constructed from 
algebra objects. For example, one derives the trace identity \cite{scst}
  \be \bearll  \dsty\sum_M \Ann{kM}M \!\!
  &= \dsty\sum_M \sumI_p \sum_{\alpha\; {\rm local}}
  S^A_{M,p\alpha}\, \frac{S_{k,p}}{S_{0,p}}\,\tilde  S^A_{p\alpha,M}
  = \sumI_p \sum_{\alpha\; {\rm local}} \frac{S_{k,p}}{S_{0,p}} 
  \\{}\\[-.7em] & \dsty
  = \sumI_p Z_{p \bar p}\, \frac{S_{k,p}}{S_{0,p}}
  = \sumI_{i,j,p} S_{p,i}\,Z_{ij}\,S^{-1}_{\bar p,j}\,\frac{S_{k,p}}{S_{0,p}}
  = \sumI_{i,j,p} \N ik{\bar\jmath}\, Z_{ij} \,. \eear\ee
Here we used the facts that $S^A$ and $\tilde S^A$ are inverse to each 
other and that $Z$ commutes with $S$.

\medskip

Boundary fields will be labelled by $(M,N,k,\alpha)$, where
$M,N$ are $A$-modules, $U_k$ is a simple object and $\alpha$ an 
element of $\Hom_A(M{\otimes}U_k,N)$. That this prescription
is consistent with the annulus amplitude $\Ann {kM}N$ follows from
\vspace{-1.2em}

\dtl{Proposition}{prop:bndfield}
Let $U_k$ ($k\iN\I$) be a simple object and $M,N$ two (not necessarily 
simple) $A$-modules of a symmetric special Frobenius algebra $A$. Then
  \be
  \Ann{kM}N = \dim\llb\Hom_A(M{\otimes}U_k,N)\lrb \,, \ee
where $\Ann{kM}N$ is the invariant associated to the ribbon graph 
\erf{Akmn}.

\medskip

Before proving the proposition, it is helpful to introduce some 
additional notation. Define the linear map 
$Q_{kM}^N$ acting on $\Hom(\M{\otimes}U_k,\n)$ via
$Q_{kM}^N(\Phi) \,{:=}\,\overline \Phi$, where $\overline\Phi$ denotes
the $A$-averaged morphism as introduced in formula \erf{average}.
By combining parts (i) and (ii) of lemma \ref{le:av} it follows that 
$Q_{kM}^N$ is a projector and that 
  \be
  {\rm Im}\, Q_{kM}^N = \Hom_A(M{\otimes}U_k,N)
  \subseteq \Hom(\M{\otimes}U_k,\n) \,.  \labl{eq:ImQkMN}
Let us choose an eigenbasis $\{ \psi^{kMN}_\alpha \}$ of $Q_{kM}^N$, i.e.
  \be
  \{\psi^{kMN}_\alpha\} \subset \Hom(\M{\otimes}U_k, \n) \quad
  {\rm such\;that} \quad
  Q_{kM}^N \psi^{kMN}_\alpha =  \eps_\alpha^{} \psi^{kMN}_\alpha 
  \quad {\rm with}\;\eps_\alpha \iN \{0,1\} \,.  \labl{eq:psibasis}
Further, we fix a basis $\{\bar\psi^{kMN}_\alpha\}$ in $\Hom(\n,\M\Oti U_k)$ 
that is dual to $\{ \psi^{kMN}_\alpha \}$ in the sense that
  \be
  \tr \llb \psi^{kMN}_\alpha{\circ}\, \bar\psi^{kMN}_\beta \lrb
  = \delta_{\alpha,\beta} \,.  \ee
Then we can write
  \bea \begin{picture}(35,84)(0,30)
  \put(100,0) {\begin{picture}(0,0)(0,0)
             \scalebox{.38}{\includegraphics{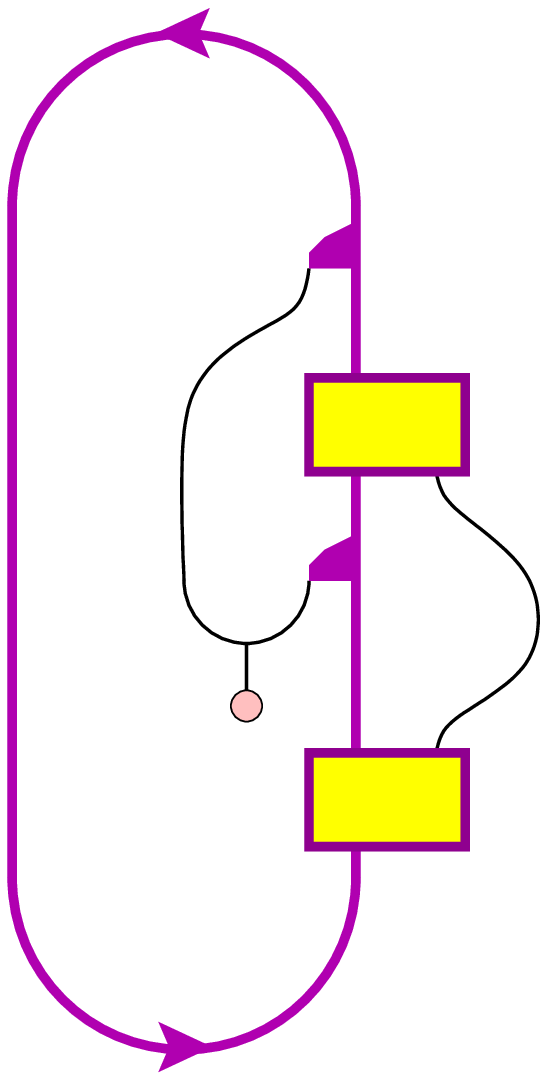}} \end{picture}
   \put(39.8,69.9) {\scriptsize$\alpha$}
   \put(39.8,28) {\scriptsize$\overline\alpha$}
   \put(40.1,45) {\scriptsize$\M$}
   \put(40.1,13) {\scriptsize$\n$}
   \put(40.1,86) {\scriptsize$\n$}
   \put(61,48) {\scriptsize$k$}
   \put(12,59) {\scriptsize$A$}
    \put(-242,50){$\dsty \tr Q_{kM}^N \,=\, \sum_{\alpha} 
        \tr \llb Q_{kM}^N(\psi^{kMN}_\alpha) \cir \bar\psi^{kMN}_\alpha\lrb
        \,=\, \sum_\alpha$} 
  }
  \epicture11 \labl{eq:trQkMN}
where the first trace is a trace over the vector space $\Hom(\M\Oti U_k,\n)$,
while the second trace is a trace in the category theoretic sense of 
\erf{trace}.

\medskip

\noindent
Proof of proposition \ref{prop:bndfield}:\\[.16em]
Consider the two ribbon graphs 
  \bea \begin{picture}(310,80)(0,0)
  \put(0,0) {\begin{picture}(0,0)(0,0)
             \scalebox{.38}{\includegraphics{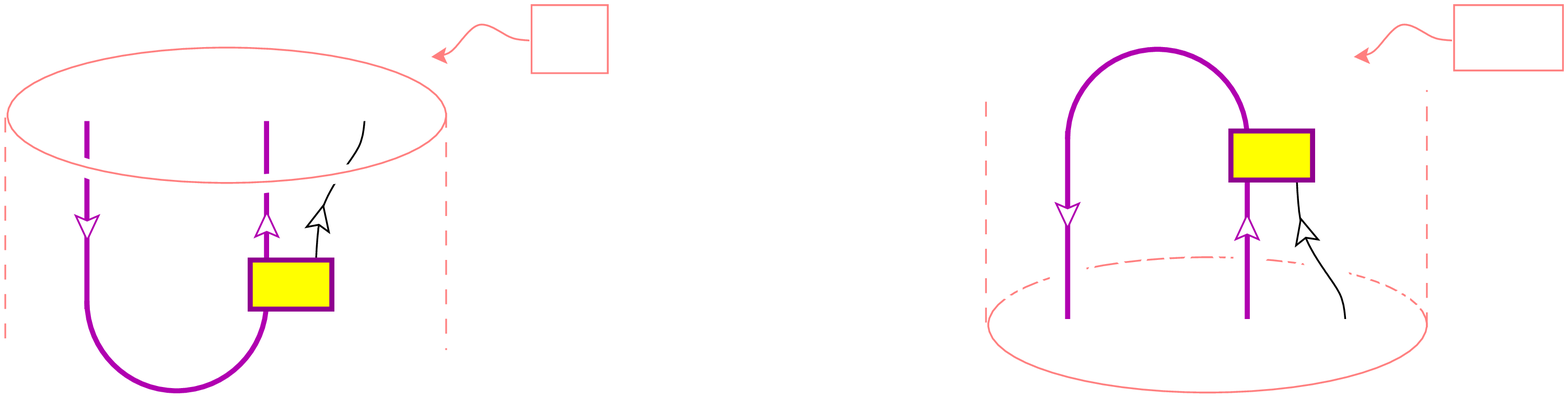}} \end{picture}
  \put( 9,21) {\scriptsize$\n$}
  \put(45,49) {\scriptsize$\M$}
  \put(58,21) {\scriptsize$\overline\alpha$}
  \put(77,53) {\scriptsize$k$}
  \put(114,71) {$B$}
  \put(225.3,16.8) {\scriptsize$\n$}
  \put(251.1,16.8) {\scriptsize$\M$}
  \put(283.9,16.6) {\scriptsize$k$}
  \put(264.3,48.5) {\scriptsize$\alpha$}
  \put(307,71) {$-B$}
  \put(-45,30) {$\Gamma_1 \; :=$}
  \put(160,30) {$\Gamma_2 \; :=$}
  }
  \epicture-8 \ee
The three-manifolds in which these ribbon graphs are embedded are both 
solid three-balls, but with opposite orientation, $\pm B$.
In the drawing, the disk bounded by the circle indicates the
boundary $S^2$ of $\pm B$. Define elements 
  \be\bearll
  v_\alpha:= Z(\Gamma_1,\emptyset,S^2)\,1 & 
      \in\, \calh((\n,-),\M,k;S^2) \,, \\[5pt]
  \bar v_\alpha:= (S_{0,0}\, \dim(\n))^{-1}_{} \, 
      Z(\Gamma_2,S^2,\emptyset)
  \,& \in\, {(\calh((\n,-),\M,k;S^2))}^* \,.  \eear\ee
That indeed $\bar v_\alpha(v_\beta)\eq\delta_{\alpha,\beta}$ can be
seen as follows:
  \bea \begin{picture}(315,60)(0,33)
  \put(11,46) {$\dsty \bar v_\alpha(v_\beta)\;=\;(S_{0,0}\,\dim(\n))^{-1}
              \,S_{0,0}$}
  \put(258,46) {$\dsty =\; \delta_{\alpha,\beta}$}
  \put(175,0) {\begin{picture}(0,0)(0,0)
             \scalebox{.38}{\includegraphics{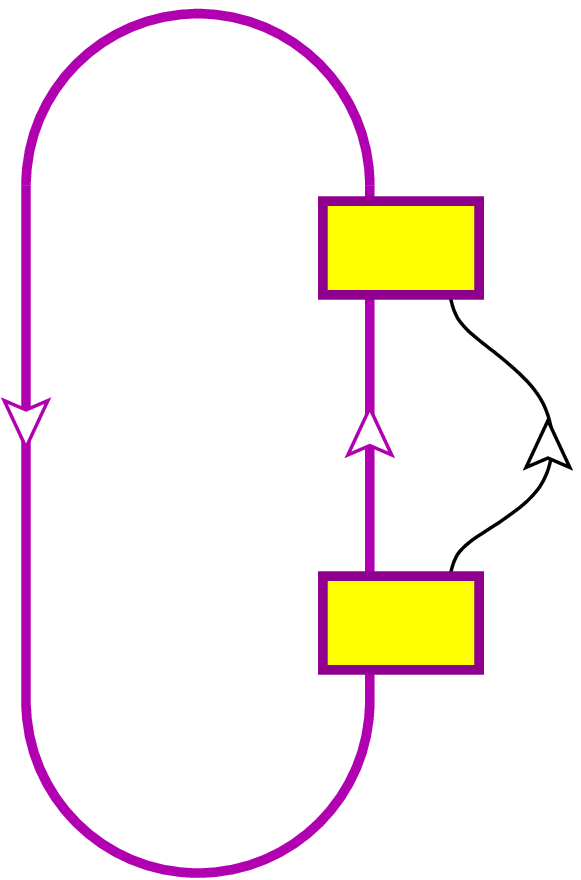}} \end{picture}
   \put(4.3,42) {\scriptsize$\n$}
   \put(27.8,42) {\scriptsize$\M$}
   \put(63.1,42) {\scriptsize$A$}
   \put(41,26) {\scriptsize$\overline\alpha$}
   \put(41,68) {\scriptsize$\alpha$}
  }
  \epicture10 \ee
In the first equality the concatenation of $\bar v$ and $v$ is expressed as
a ribbon graph in $S^3$ obtained by glueing the two three-balls along their
boundary. Recall the convention described below \erf{inv-conv},
introducing a factor $S_{0,0}$ for the
invariant associated to a ribbon graph in $S^3$. 

Having established that the vectors $v_\alpha$ and $\bar v_\beta$ constitute 
dual bases of $\calh((\n,-),\M,k;S^2)$ and $\calh((\n,-),\M,k;S^2{)}^*$, we 
can rewrite the invariant \erf{Akmn} as in \erf{eq:ann-Q-trace} and express 
the trace as (we can move the $k$-ribbon in \erf{Akmn}
from the left side of the module ribbons to the right side)
  \begin{eqnarray} \begin{picture}(400,105)(23,0)
  \put(0,56.4){$\dsty \Ann{kM}N = \Tr_{\calH((\n,-),\M,k;S^2)} Q_k
    = \sum_{\alpha} \bar v_\alpha( Q_k v_\alpha )
    = \frac{S_{0,0}}{S_{0,0} \, \dim(\n)} \sum_\alpha$}
  \put(408,56.4) {$\dsty = \tr Q_{kM}^N$}
  \put(320,0) {\begin{picture}(0,0)(0,0)
             \scalebox{.38}{\includegraphics{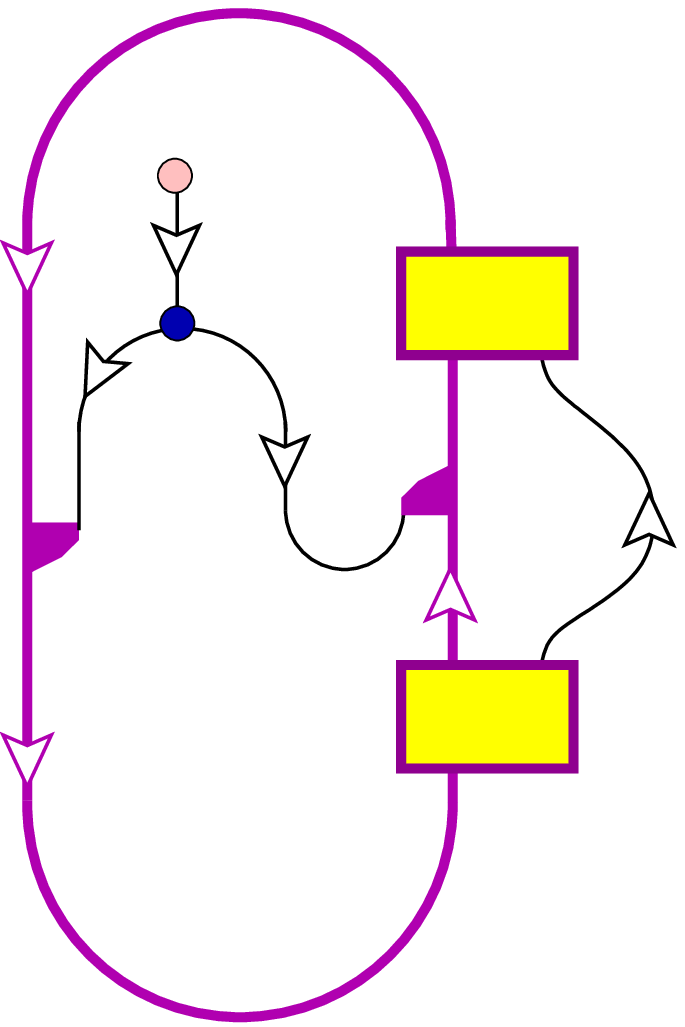}} \end{picture}
   \put(51.6,53) {\scriptsize$\M$}
   \put(51,32) {\scriptsize$\overline\alpha$}
   \put(51,77) {\scriptsize$\alpha$}
   \put(-5.8,37) {\scriptsize$\n$}
   \put(22,62) {\scriptsize$A$}
   \put(71.9,59) {\scriptsize$k$}
  }
  \end{picture} \nonumber \\[-3.1em]{} \\[-.1em] \nonumber
  \end{eqnarray}  
To see the last equality, we note that the ribbon graph above can be 
transformed into the one shown in figure \erf{eq:trQkMN} (using that the 
\alg\ $A$ is symmetric). Equation \erf{eq:ImQkMN} now implies the proposition.
\qed

\subsection{The case ${N_{ij}}^k \iN \{0,1\}$
            and $\dim\,\Hom(U_k,A) \iN \{0,1\}$}

This is the last part of our meta example (as far as the present paper
is concerned). We will
illustrate how to compute the invariant associated to the annulus
partition function \erf{Akmn}. This can be done by the following
series of transformations:
  \bea  \begin{picture}(274,317)(0,40)
              \put(0,212){\begin{picture}(0,0)(0,0)
  \put(0,0)     {\begin{picture}(0,0)(0,0)
                \scalebox{.38}{\includegraphics{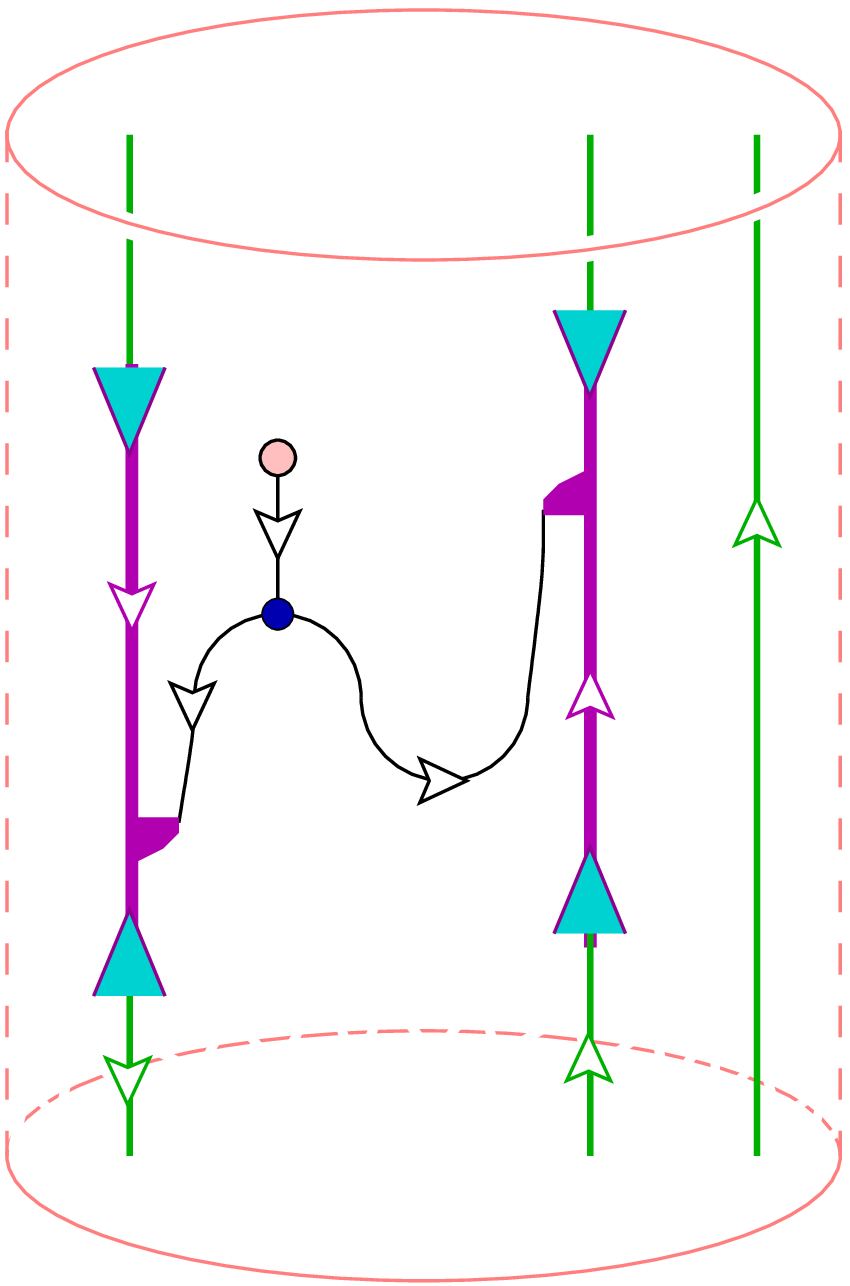}} \end{picture}
   \put(5,93) {\scriptsize$\alpha$}
   \put(5,35) {\scriptsize$\overline\alpha$}
   \put(70,37) {\scriptsize$\beta$}
   \put(70,99) {\scriptsize$\overline\beta$}
   \put(5,63) {\scriptsize$\n$}
   \put(20,46) {\scriptsize$\rho_N$}
   \put(68,66) {\scriptsize$\M$}
   \put(47,88) {\scriptsize$\rho_M$}
   \put(40,65) {\scriptsize$A$}
   \put(17,19) {\scriptsize$i$}
   \put(68,19) {\scriptsize$j$}
   \put(84,49) {\scriptsize$k$}
    }
  \put(-100.1,67.5){$\Ann {kM}N\,=\;\dsty\sumI_{i,j}\sum_{\alpha,\beta}$}
  \put(107.1,67.5){$=\;\dsty\sumI_{i,j}\sum_{\alpha,\beta}\frac1{\dim(U_i)}$}
              \end{picture}}
              \put(221,205){\begin{picture}(0,0)(0,0)
  \put(0,0)   {\begin{picture}(0,0)(0,0)
                \scalebox{.38}{\includegraphics{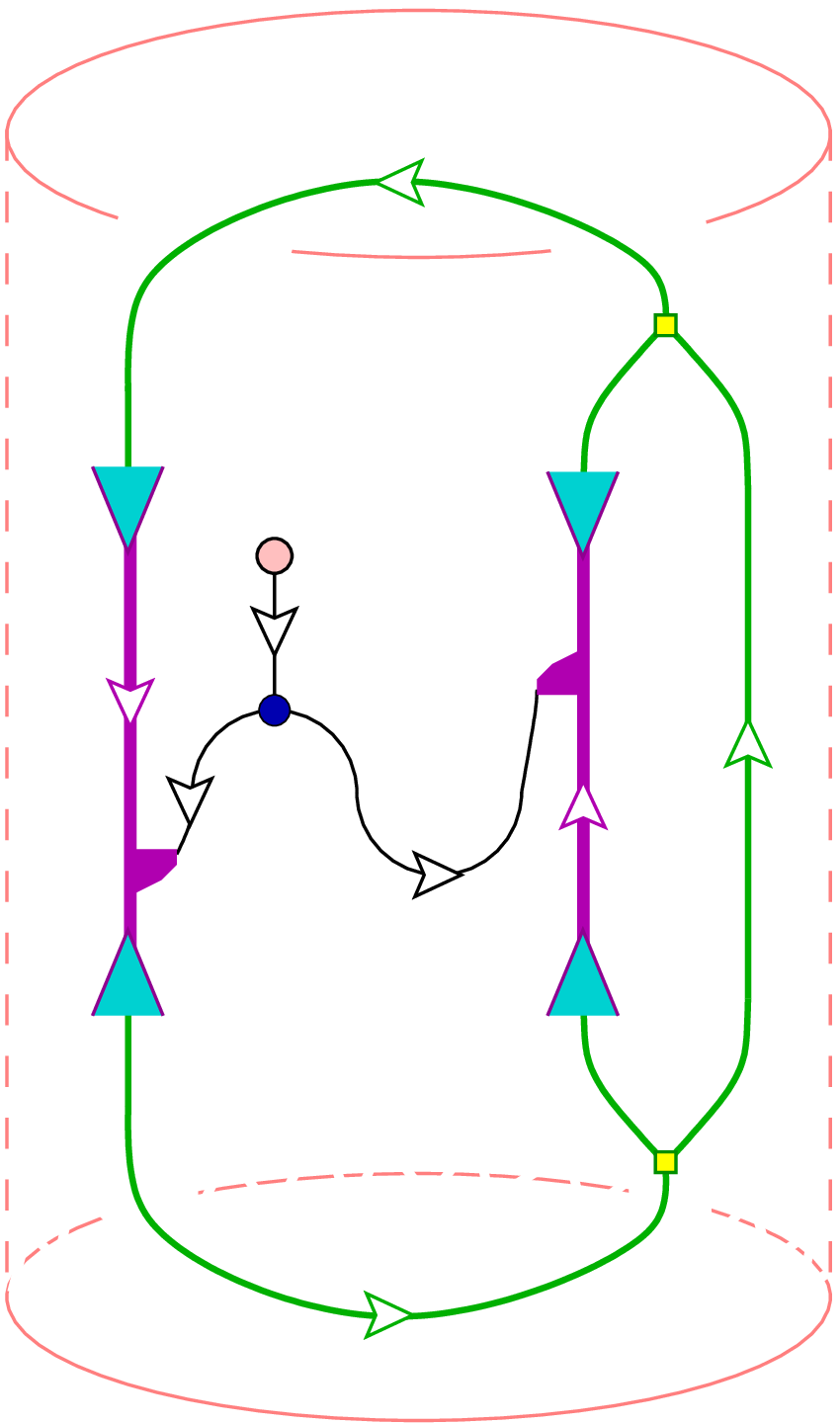}} \end{picture}
   \put(5,98) {\scriptsize$\alpha$}
   \put(5,49) {\scriptsize$\overline\alpha$}
   \put(68,51) {\scriptsize$\beta$}
   \put(68,97) {\scriptsize$\overline\beta$}
   \put(6,63) {\scriptsize$\n$}
   \put(68,61) {\scriptsize$\M$}
   \put(40,71) {\scriptsize$A$}
   \put(56,16) {\scriptsize$i$}
   \put(40,140) {\scriptsize$i$}
   \put(62,32) {\scriptsize$j$}
   \put(62,114) {\scriptsize$j$}
   \put(84.5,55) {\scriptsize$k$}
    }
              \end{picture}}
              \put(0,0){\begin{picture}(0,0)(0,0)
  \put(0,0)    {\begin{picture}(0,0)(0,0)
                \scalebox{.38}{\includegraphics{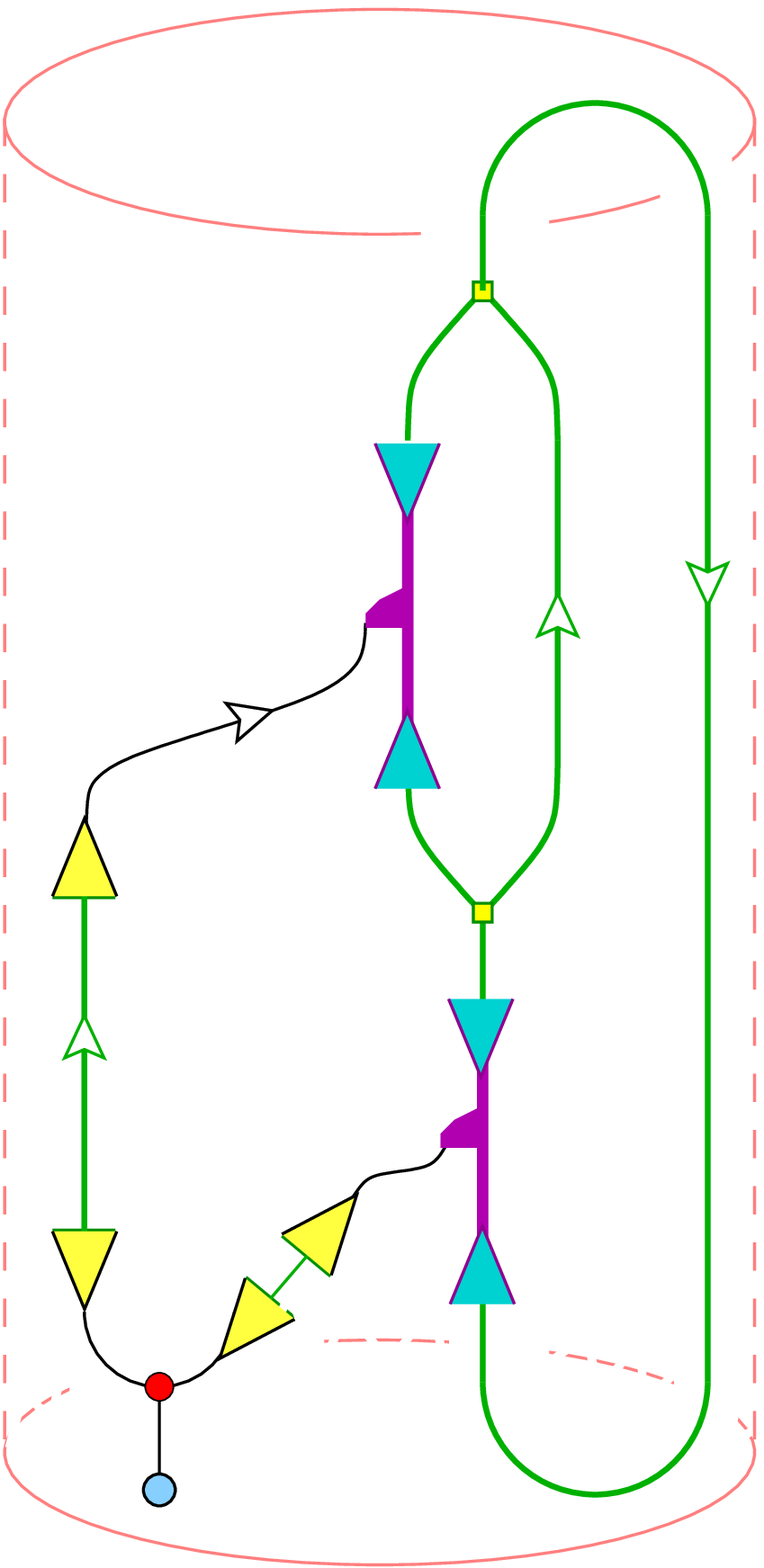}} \end{picture}
   \put(40,130) {\scriptsize$\overline\beta$}
   \put(39,95) {\scriptsize$\beta$}
   \put(49,61) {\scriptsize$\overline\alpha$}
   \put(50,35) {\scriptsize$\alpha$}
   \put(13,47) {\scriptsize$\overline a$}
   \put(30,36.5) {\scriptsize$a$}
   \put(60,71) {\scriptsize$i$}
   \put(81,71) {\scriptsize$i$}
   \put(47,84) {\scriptsize$j$}
   \put(47,148) {\scriptsize$j$}
   \put(69,125) {\scriptsize$k$}
   \put(51,110) {\scriptsize$\M$}
   \put(61,46) {\scriptsize$\n$}
   \put(15,103) {\scriptsize$A$}
    }
  \put(-98,94.9)  {$=\,\dsty\sum_{\scs i,j,a\in\II\atop\scs\alpha,\beta}
                  \frac1{\dim(U_i)}$}
  \put(103.5,94.9){$=\,\dsty\sum_{\scs i,j,a\atop\scs\alpha,\beta}\frac{\Delta
                  _0^{\bar a a}}{\dim(U_i)}\,\rbas M{\bar a}{j\beta}{j\beta}\,
                  \rbas Na{i\alpha}{i\alpha}$}
              \end{picture}}
              \put(247,0){\begin{picture}(0,0)(0,0)
  \put(0,34)   {\begin{picture}(0,0)(0,0)
                \scalebox{.38}{\includegraphics{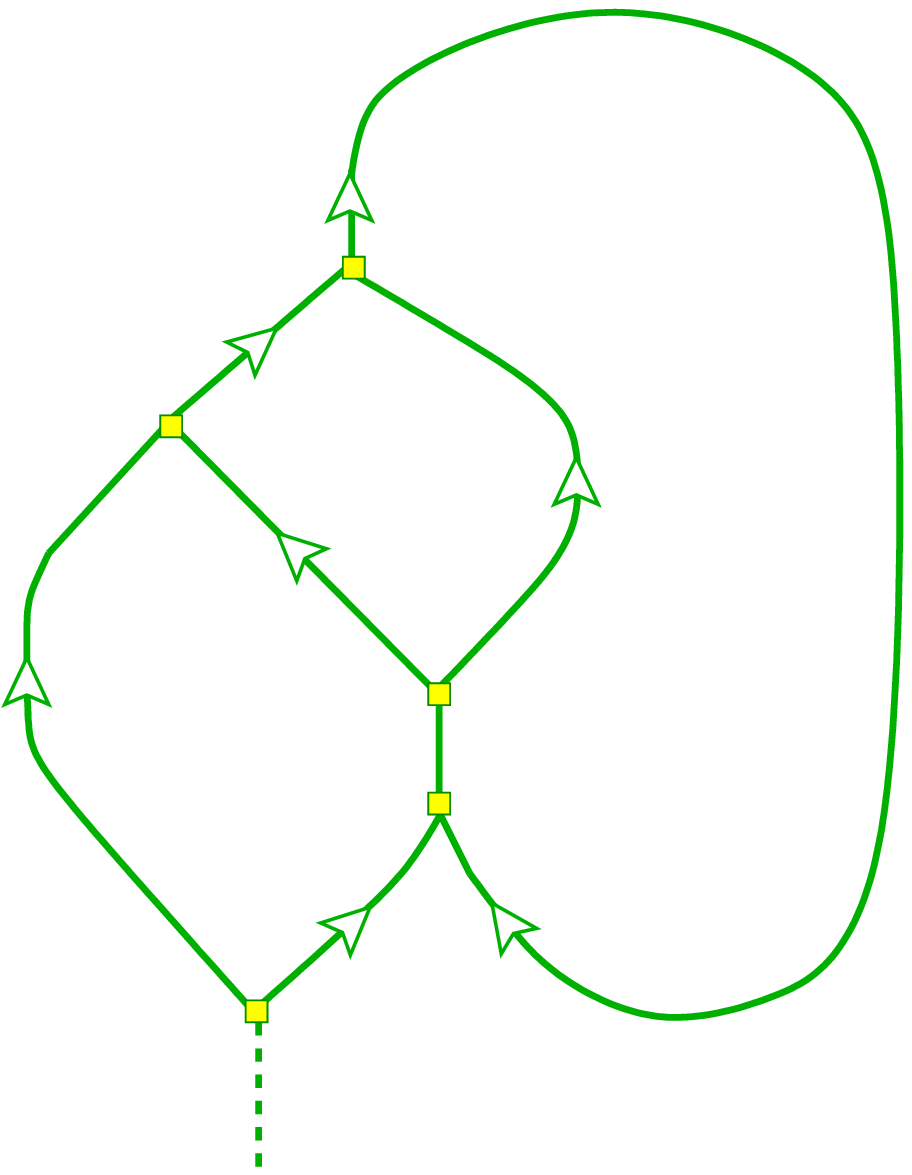}} \end{picture}
   \put(10,41) {\scriptsize$\overline a$}
   \put(39,22) {\scriptsize$a$}
   \put(49,44) {\scriptsize$i$}
   \put(62,24) {\scriptsize$i$}
   \put(42,114) {\scriptsize$i$}
   \put(38,68) {\scriptsize$j$}
   \put(31,89) {\scriptsize$j$}
   \put(62,65) {\scriptsize$k$}
}
              \end{picture}}
  \epicture16 \labl{Akmnij}
Here the usual calculational devices are used: In the first step bases for 
the morphisms involving the simple $A$-modules $M$ and $N$ are inserted. 
The second step uses dominance twice, together with 
$\dim(\calh((i,-),m;S^2))\eq\delta_{m,i}$ and
$\dim(\calh(m;S^2))\eq\delta_{m,0}$, which imply that that the sums over 
intermediate simple objects each reduce to a single term. The third step 
consists again in the insertion of a basis, and in the fourth step one
substitutes the definition \erf{eq:rep-mat} for $\rho$. 

To arrive at our final formula, we also use an inverse fusion move on the 
last graph above, together with relation \erf{eq:rel-3pt-cpl}. The result is
  \be
  \Ann {kM}N = \sum_{a\In A} \sumI_{i,j} \sum_{\alpha,\beta}
  \rbas Na{i\alpha}{i\alpha} \rbas M{\bar a}{j\beta}{j\beta}\,
  \Delta_0^{\bar a a} \cdot
  \Gs{\bar a}jkiji\, \Fs{\bar a}aiii\One \,,
  \labl{eq:ann-MS}
where $\alpha$ labels a basis of
$\Hom(U_i,\M)$ and $\beta$ a basis of $\Hom(U_j,\n)$. 
Here the coproduct $\Delta_0^{\bar a a}$ can also be
expressed through the multiplication using \erf{eq:delta-by-mult}.

\subsubsection{Example: Free boson}

Substituting the free boson modular data \erf{eq:Z2N-F},
the $A_{2r}$-algebra \erf{eq:A2r-cont}, \erf{eq:A2r-mult} and
the expressions \erf{eq:Z2N-repmat} for the representation matrices
into the general formula \erf{eq:ann-MS}, we obtain
  \be
  A_{[k]\,M_m}^{\quad\; M_n} = \frac{r}{N} \sumI_{j}
  \delta_{[j{+}k{-}m]\In A} \, \delta_{[j{-}n]\In A}
  = \delta_{n+k,m \;{\rm mod}\,2r} \,.  \ee
Note that when the two boundary conditions are equal, then the result 
no longer depends on $m\eq n$. Thus each of the $2r$ distinct 
elementary boundary conditions has the same field content.

\subsubsection{Example: \E modular invariant}

As seen in section \ref{sec:ex-E7-repn},
the algebra object $A$ that gives the \E modular invariant of 
the $\su(2)_{16}$ WZW model has seven isomorphism classes of
simple modules. We label representatives for these classes as
  \be  \begin{array}{ll}
  M_1 = A\ \,,\ &
  M_2 = \inda(1)\,,\quad\
  M_3 = \inda(2)\,,\quad\
  M_4 = \inda(3)\,, \\{}\\[-.6em]
  M_5 = P \,, &
  M_6 = Q \,,\quad\
  M_7 = R \,.  \eear \labl{eq:E7-bcs}
(Also recall from  section \ref{sec:ex-E7-repn} that the latter three 
are not induced modules.) We can now numerically evaluate formula
\erf{eq:ann-MS} for the annulus coefficients ${{\rm A}_{kM}}^{\!N}$.
To obtain the representation matrices we proceed as 
described in section \ref{sec:ex-E7-repn} and use 
formula \erf{eq:sub-rep-mat}.  
One then directly verifies that the numbers ${{\rm A}_{kM}}^{\!N}$ are
non-negative integers, satisfy ${{\rm A}_{\One M}}^{\!N}\eq\delta_{M,N}$,
and furnish a NIM-rep of the fusion rules.

To make contact with the classification of boundary
conditions in \cite{bePz,bppz} we also present the matrix
${\rm A}_{(1)}$:
  \be {\rm A}_{(1)} = \pmatrix{
  \cdot & 1 & \cdot & \cdot & \cdot & \cdot & \cdot \cr 
  1 & \cdot & 1 & \cdot & \cdot & \cdot & \cdot \cr 
  \cdot & 1 & \cdot & 1 & \cdot & \cdot & \cdot \cr 
  \cdot & \cdot & 1 & \cdot & 1 & 1 & \cdot \cr 
  \cdot & \cdot & \cdot & 1 & \cdot & \cdot & \cdot \cr 
  \cdot & \cdot & \cdot & 1 & \cdot & \cdot & 1 \cr 
  \cdot & \cdot & \cdot & \cdot & \cdot & 1 & \cdot } \ee
(Rows and columns are ordered according to the labelling \erf{eq:E7-bcs},
and zero entries have been replaced by dots to improve readability.)
In agreement with the results of \cite{bePz,bppz}, this is indeed 
the adjacency matrix of the Dynkin diagram of the Lie algebra $E_7$.

\subsection{Defect lines and double fusion algebra}

In this section we consider the partition function of a torus with two 
defect lines inserted -- a setup studied in \cite{pezu5}.
In particular we recover the property that the coefficients 
of such partition function furnish a NIM-rep of the `double fusion
algebra', a structure also seen in \cite{ocne9,boek}.

As already mentioned at the end of section \ref{bc-to-rep} and in 
remark \ref{samenumber}, in our framework defect lines are described 
(and labelled) by \AA-bimodules. Consider the situation where the 
world sheet is a two-torus $\torus$ without field insertions,
but with two defect lines $X,\,Y$ running parallel and with opposite
orientation. The setup is shown in the following picture, together with
the triangulation we will use (compare also \erf{tor}):
  \bea \begin{picture}(335,58)(0,29)
  \put(100,0)   {\begin{picture}(0,0)(0,0)
              \scalebox{.38}{\includegraphics{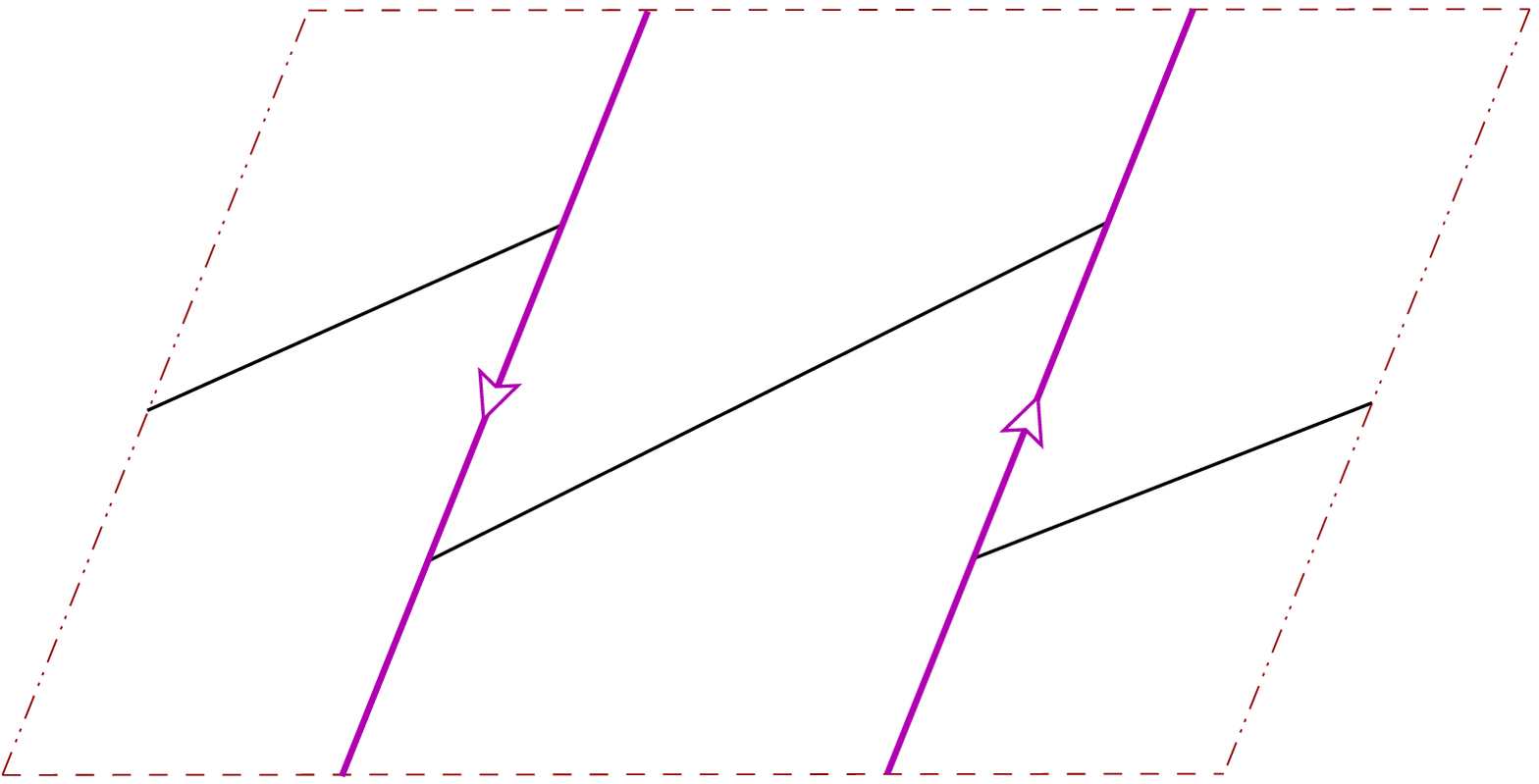}} \end{picture}
   \put(34.1,7) {\scriptsize$Y$}
   \put(93.8,7) {\scriptsize$X$}
  }
  \epicture08 \labl{defecttor}
Proceeding similarly to section \ref{sec:torus-pf} we can express
the coefficients $\Zdfct XYkl$ in the twisted partition function
  \be
  \Zdfct XY{}{} = \sum_{k,l \in \II} \Zdfct XYkl
  \tbas k{} \otimes \tbas l{-} \ee
as the invariant of a ribbon graph. We obtain
  \bea \begin{picture}(62,77)(0,37)
  \put(0,0)   {\begin{picture}(0,0)(0,0)
              \scalebox{.38}{\includegraphics{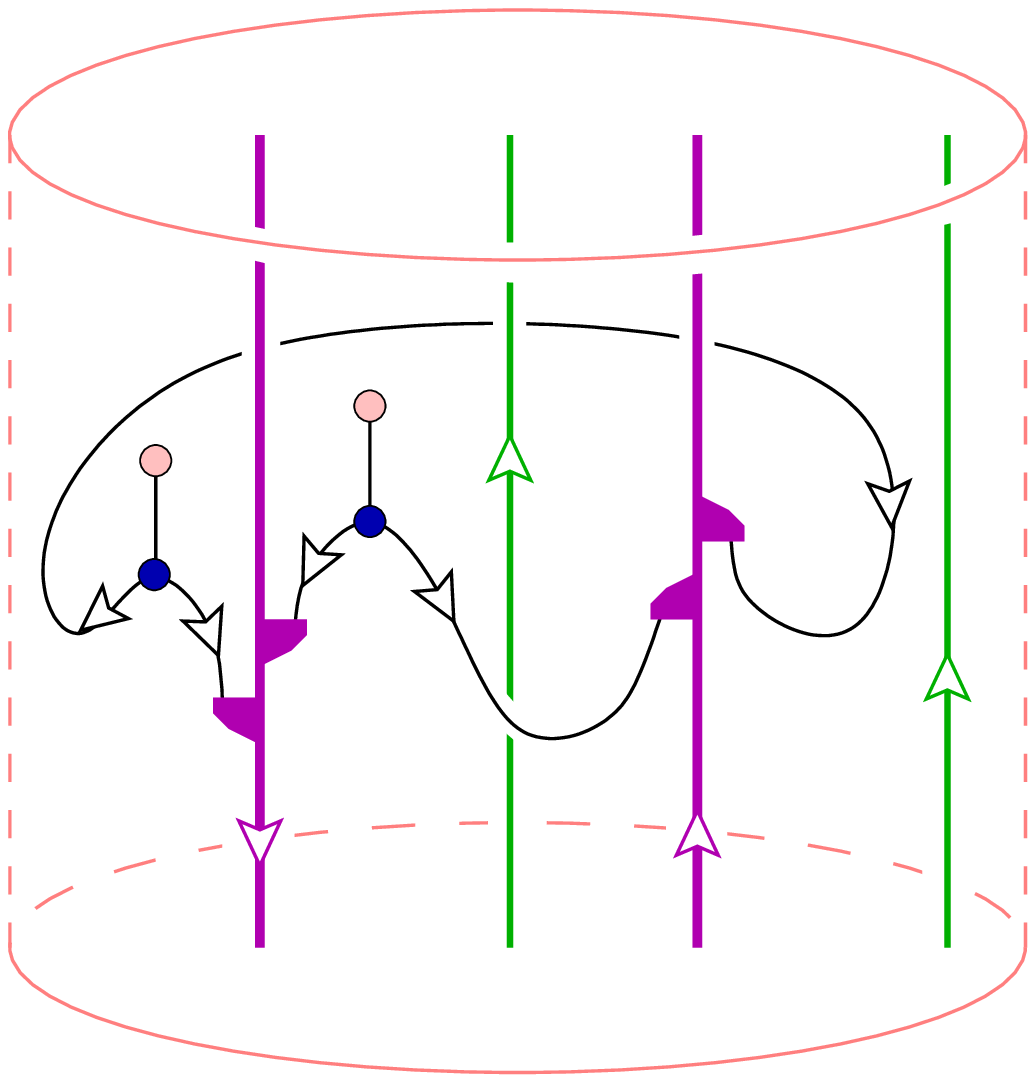}} \end{picture}
   \put(-55,55) {$\Zdfct XYkl \;=$}
   \put(20.8,15) {\scriptsize$Y$}
   \put(68,15) {\scriptsize$X$}
   \put(51.7,15) {\scriptsize$l$}
   \put(98,15) {\scriptsize$k$}
   \put(89,66) {\scriptsize$A$}
   \put(44,44) {\scriptsize$A$}
  }
  \epicture13 \labl{ZXYkl}
Before proceeding to prove some properties of these numbers 
it is useful to slightly change the notation for annulus coefficients, so
as to make the dependence
on the algebra explicit. Thus for a symmetric special Frobenius algebra $B$
and for left $B$-modules $M,\,N$, let $\AnnB{kM}N$ denote the number 
defined in \erf{Akmn}, but with all algebra ribbons labelled by $B$.

\dt{Theorem}
Let $A$ be a symmetric special Frobenius algebra,
let $U_k$, $U_l$ be simple objects and $X,Y$ be \AA-bimodules.
Abbreviate $B\eq A{\otimes}A^{(-1)}$ and $\tilde B\eq A^{(1)}{\otimes}A$.
The numbers $\Zdfct XYkl$ have the following properties:
  \begin{eqnarray}
{\rm(i)}\hsp{.9}  && \Zdfct XYkl \in \zet_{\ge 0}\,. \\ [.4em]
{\rm(ii)}\hsp{.6} && \Zdfct AAkl = Z(A)_{kl}   \,.\\[.4em]
{\rm(iii)}\hsp{.3}&& \Zdfct XY{\bar k}{\bar l} = \Zdfct YXkl \,.\\[.4em]
{\rm(iv)}\hsp{.32}&& \Zdfct XYk0 = \AnnB{k\,f(X)}{f(Y)}\quad\ {\rm and}\quad\
                     \Zdfct XY0l = \AnntB{l\,\tilde f(X)}{\tilde f(Y)} \,.
                     \label{eq:ZXY-AXY} \\{}\nonumber\\[-.6em]
{\rm(v)}\hsp{.55} && \Zdfct XYkl = \sum_R \Zdfct XRk0 \Zdfct RY0l
                     \quad {\rm or,~as~matrix~equation,} \quad
                     Z_{kl} = Z_{k0}\, Z_{0l} \,. \hsp{3.5}{}\\[.1em]
{\rm(vi)}\hsp{.3} && {\rm As~matrix~equations,} \ \ 
          [Z_{l0},Z_{k0}] = [Z_{0l},Z_{0k}] = [Z_{0l},Z_{k0}] = 0 \,.\\[.4em]
{\rm(vii)}        && {\rm As~matrix~equation,} \;\ \
          Z_{ij}\, Z_{kl} = \sumI_{r,s} \N ikr\, \N jls\, Z_{rs} \,.
\end{eqnarray}
In (ii), $Z(A)_{kl}$ denotes the coefficients of the untwisted
torus partition function \erf{Zij2}. In (v)--(vii) 
$Z_{kl}$ is understood as a matrix with entries
$(Z_{kl})_{X,Y}\eq\Zdfct XYkl$. The sum in (v) is over
(representatives of isomorphism classes of) simple \AA-bimodules
$R$. The notation $A^{(n)}$ was defined in \erf{Aon}.
In (iv), $f$ and $\tilde f$ are the isomorphisms 
defined in \erf{eq:ff-bimod-iso} taking 
\AA-bimodules to left $A{\otimes}A^{(-1)}$- and 
$A^{(1)}{\otimes}A$-modules, respectively.

\medskip\noindent
Proof:\\[.13em]
Property (i) follows along the same lines as the proof of
theorem \ref{ZZZ}(ii). That is, the invariant \erf{ZXYkl} can
be rewritten as the trace of a projector. The projector is again
obtained by cutting the three-manifold in \erf{ZXYkl} along a
`horizontal' $S^2$; the projector property follows from the
representation property of the bimodules $X,Y$ and the properties of $A$.
\\[.16em]
To see (ii), note that the ribbon graph resulting from \erf{ZXYkl}
when replacing $X$ and $Y$ by $A$ is almost identical to the graph
\erf{Zij2}, except for an additional $A$-ribbon running vertically. 
The latter can be removed by transformations similar to
\erf{ann} with $M,N$ set to $A$ (that the geometry in that situation
is actually a cylinder, rather than a torus, does not play a role in
the calculation).
\\[.16em]
The proof of (iii) uses the same argument as the proof of
theorem \ref{thm:annuluscoeff}(iii): The ribbon graph for
$\Zdfct XY{\bar k}{\bar l}$ is that of $\Zdfct YXkl$ turned
upside down (this is a rotation, preserving the orientation of
the three-manifold) and with the $k$- and $l$-ribbons replaced by
$\bar k$- and $\bar l$-ribbons with opposite orientation of their cores.
\\[.16em]
To show (iv) we draw the ribbon graphs for the annulus
coefficients appearing in \erf{eq:ZXY-AXY}, inserting the relation 
\erf{eq:ff-bimod-iso} between \AA-bimodules and $A{\otimes}A^{(-1)}$-
(respectively left $A^{(1)}{\otimes}A$-) modules and the definition of 
the algebras $A^{(\pm 1)}$. The resulting graphs are:
  \bea \begin{picture}(320,109)(0,33)
  \put(0,0)   {\begin{picture}(0,0)(0,0)
              \scalebox{.38}{\includegraphics{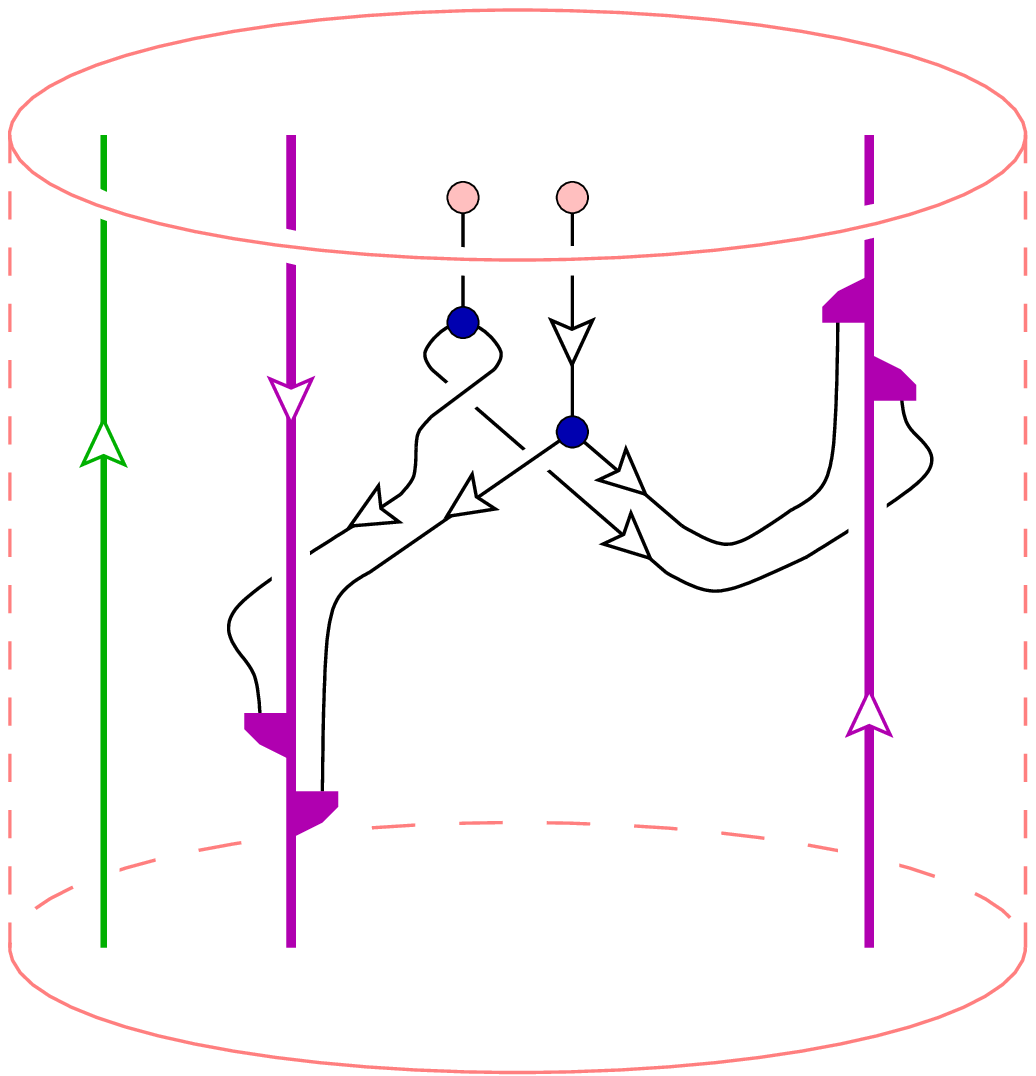}} \end{picture}
  \put(-69,128) {$\AnnA{A{\otimes}A^{(-1)}}{k\,f(X)}{f(Y)}$}
  \put(-20,59) {$=$}
   \put(23.9,15) {\scriptsize$Y$}
   \put(87.1,15) {\scriptsize$X$}
   \put(6,39) {\scriptsize$k$}
   \put(46,53) {\scriptsize$A$}
   \put(73,65) {\scriptsize$A$}
  }
  \put(220,0) {\begin{picture}(0,0)(0,0)
              \scalebox{.38}{\includegraphics{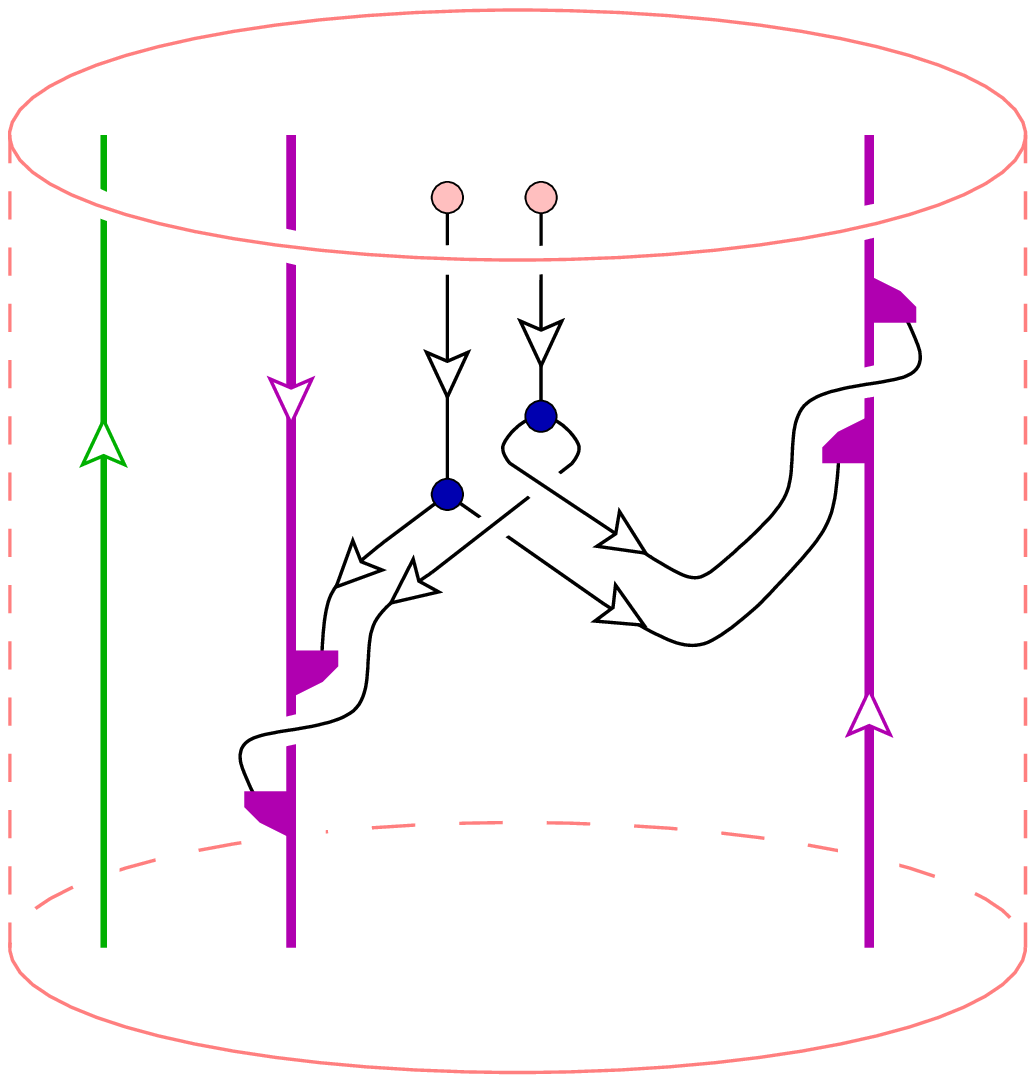}} \end{picture}
  \put(-69,128) {$\AnnA{A^{(1)}{\otimes}A}{l\,\tilde f(X)}{\tilde f(Y)}$}
  \put(-20,59) {$=$}
   \put(23.9,15) {\scriptsize$Y$}
   \put(87.1,15) {\scriptsize$X$}
   \put(6,39) {\scriptsize$l$}
   \put(43,45) {\scriptsize$A$}
   \put(73,58) {\scriptsize$A$}
  }
  \epicture19 \ee
(Here all \rep\ morphisms are those of \AA-bimodules, and all 
comultiplications are given by the coproduct of $A$.)
In the case of $\AnnA{A{\otimes}A^{(-1)}}{f(X)}{f(Y)}$, the 
ribbon graph is obviously equal to the graph \erf{ZXYkl} with $l\eq0$.
For $\AnnA{A^{(1)}{\otimes}A}{\tilde f(X)}{\tilde f(Y)}$,
the required moves are slightly more complicated; they
are again best visualised by using actual ribbons. The main step is to 
recall that the `horizontal' direction is an $S^2$, which allows us to
move one of the `horizontal' $A$-ribbons around the $S^2$ in such a way
that in the pictorial representation the $A$-ribbons now seem to wrap 
around the $l$-ribbon.  Once this step is performed, equality with 
$\Zdfct XY0l$ is easily established.
\\[.16em]
Proof of (v): Let $B\eq A{\otimes}A^{(-1)}$. Given a left simple object
$U_k$ and a left $B$-module $M$, the tensor product 
$M{\otimes}U_k$ is again a left $B$-module (compare
the discussion of module categories in section \ref{rep-and-mod}). 
By proposition \ref{modsubind} the category of left $B$-modules is 
semisimple, and hence we can decompose $M{\otimes}U_k$ in
terms of simple left $B$-modules as 
  \be
  M \oti U_k \, \cong \,
  \bigoplus_{i\in\II_B} {\lr{M{\otimes}U_k}{S_i}}_{\!B} \, S_i \,,
  \labl{eq:B-mod-decomp}
as an isomorphism of left $B$-modules. Here the sum is over 
representatives $S_i$ of isomorphism classes of simple $B$-modules. 
The multiplicities ${\lr{M{\otimes}U_k}S}_{\!B}$ have been computed 
in proposition \ref{prop:bndfield} to be given by annulus coefficients:
  \be
  {\lr{M{\otimes}U_k}S}_{\!B} = \AnnB{kM}{\ S} \,.  \labl{eq:BHom-Ann}
We can apply this relation to \erf{ZXYkl} by understanding the 
\AA-bimodule $Y$ as a left $B$-module $f(Y)$. This results in
  \bea \begin{picture}(100,83)(0,35)
  \put(0,0)   {\begin{picture}(0,0)(0,0)
              \scalebox{.38}{\includegraphics{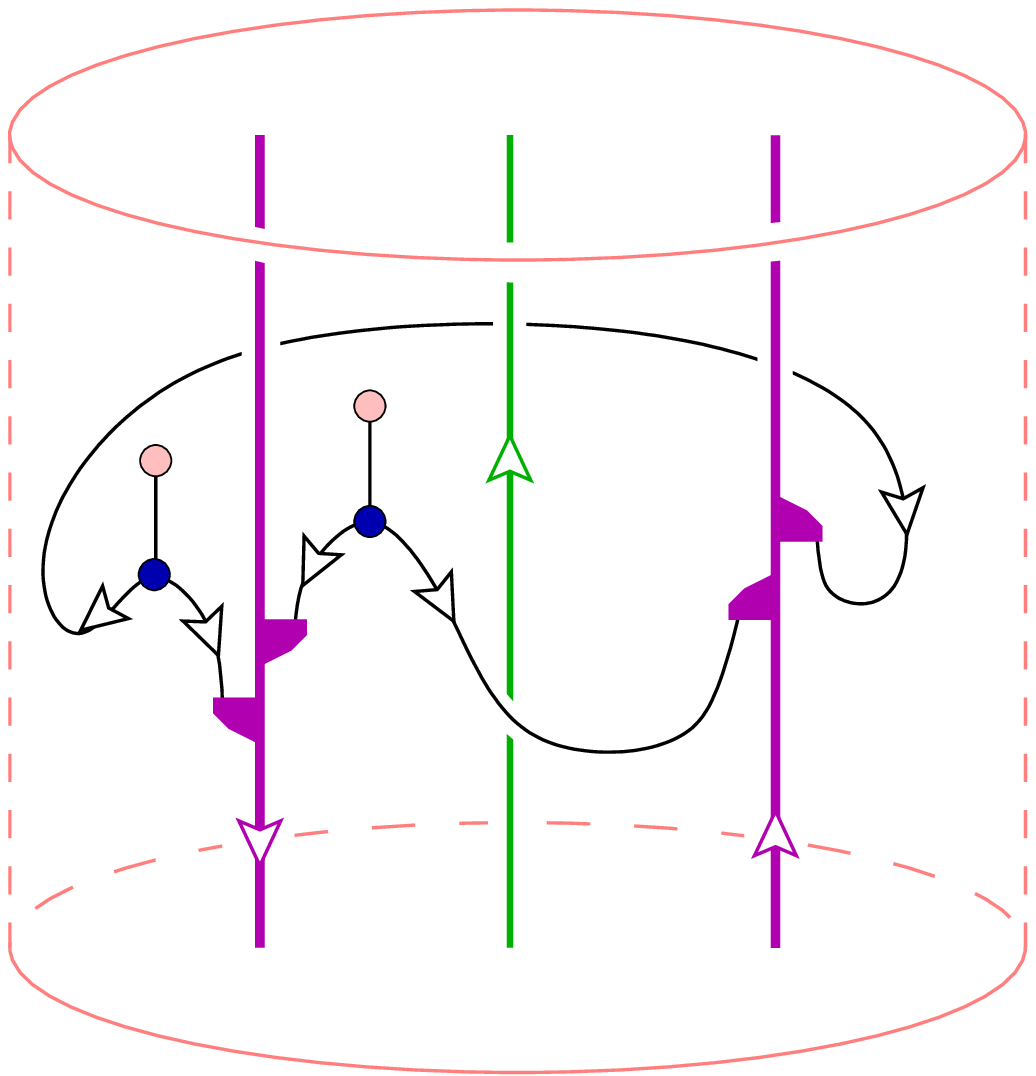}} \end{picture}
   \put(-150,58) {$\dsty \Zdfct XYkl \;=\ \sum_R\AnnB{k\,f(X)}{f(R)}$}
   \put(128,58)  {$\dsty =\; \sum_R \,\Zdfct XRk0 \,\Zdfct RY0l$}
   \put(20.6,15) {\scriptsize$Y$}
   \put(77.1,15) {\scriptsize$R$}
   \put(51.8,15) {\scriptsize$l$}
   \put(101,66) {\scriptsize$A$}
   \put(44,44) {\scriptsize$A$}
  }
  \epicture15 \ee
The first step makes use of relations \erf{eq:B-mod-decomp} and 
\erf{eq:BHom-Ann}; the sum is over representatives of isomorphism 
classes of simple \AA-bimodules, which are taken to simple left $B$-left 
modules via $f$. The second step inserts definition \erf{ZXYkl} for 
the ribbon graph and the previous result \erf{eq:ZXY-AXY}.
\\[.16em]
That the first two commutators in part (vi) vanish follows directly from 
(iv) together with the NIM-rep property \erf{eq:Ann-NIMrep}. The last
equality is obtained by combining (iii) and (v). Indeed, by (v) we have
  \be
  \Zdfct XYkl = \sum_R \Zdfct XRk0 \Zdfct RY0l \labl{eq:arglarg1}
as well as, using (iii) and $\bar0\eq0$,
  \be
  \Zdfct YX{\bar k}{\bar l} = \sum_R \Zdfct YR{\bar k}0 \Zdfct RX0{\bar l}
  = \sum_R \Zdfct XR0l \Zdfct RYk0 .  \labl{eq:arglarg2}
Employing (iii) once more we see that \erf{eq:arglarg1} and 
\erf{eq:arglarg2} are equal, thereby establishing (vi).
\\[.16em]
Finally, (vii) follows by a short calculation from (iv), (v) and (vi). 
With $B\eq A{\otimes}A^{(-1)}$ and $\tilde B\eq A^{(1)}{\otimes}A$ we have
  \be\bearll
  \dsty (Z_{ij} Z_{kl})^{X|Y} \!\!\!
  &=(Z_{i0} Z_{0j} Z_{k0} Z_{0l})^{X|Y}
  = (Z_{i0} Z_{k0} Z_{0j} Z_{0l})^{X|Y} 
  = \dsty\sum_{R,S,T} \Zdfct XRi0 \Zdfct RSk0 \Zdfct ST0j \Zdfct TY0l
  \\{}\\[-.93em]
  &= \dsty\sum_{R,S,T} \AnnB{i\,f(X)}{f(R)}\, \AnnB{k\,f(R)}{f(S)}\,
    \AnntB{j\,\tilde f(S)}{\tilde f(T)}\,
    \AnntB{l\,\tilde f(T)}{\tilde f(Y)}\,
  \\{}\\[-.93em]
  &= \dsty\sumI_{r,s} \sum_S \N ikr\, \AnnB{r\,f(X)}{f(S)}\,
    \N jls\, \AnntB{s\,\tilde f(S)}{\tilde f(Y)}
  \\{}\\[-.9em]
  &= \dsty\sumI_{r,s}  \N ikr \N jls \sum_S \Zdfct XSr0\, \Zdfct SY0s \,.
  \eear \ee
In the next-to-last step it is used that the annulus coefficients
furnish a NIM-rep, see theorem \ref{thm:annuluscoeff}.
\qed

\dt{Remark}
(i) In \cite{pezu5}, the double NIM-rep property of the twisted 
partition functions was proven under the assumption that there is a 
complete set of defect lines.  Now according to formula 
\erf{eq:number-bimod}, the number of (elementary) defect lines 
present in a CFT constructed from an algebra object $A$ is given 
by $\tr[Z(A) Z(A)\oT]$. This is precisely the number needed for 
completeness in \cite{pezu5}; thus their arguments apply, in 
agreement with point (vii) of the theorem.
\\[.13em]
(ii) The properties of $\Zdfct XYkl$ derived above can already be 
found explicitly or implicitly in \cite{pezu5,pezu7}, with the 
exception of (iv). This point, together with (v), has a curious 
interpretation: The twisted torus partition functions of the CFT 
associated to an algebra $A$ can be expressed in terms of the annulus 
coefficients of (in general) {\em different\/} CFTs possessing the
same chiral data -- the full CFTs associated to the tensor product
algebras $A{\otimes}A^{(-1)}$ and $A^{(1)}{\otimes}A$. 
\\
That there exists a relation between defect lines in a CFT and 
boundary conditions in a product CFT bears some similarity with the
`folding trick' mentioned at the end of section
\ref{bc-to-rep}. However, the folding trick uses boundary
states in a CFT of {\em twice\/} the central charge of the CFT whose 
defects are described, whereas the annulus coefficients in point (iv) are
those of a CFT with the {\em same\/} central charge.
The physical interpretation of this observation remains to be clarified.
\\[.13em]
(iii) The structure of a double NIM-rep can in fact be generalised 
further. So far we have established that annulus coefficients associate 
a single NIM-rep to left $A$-modules and that defect lines associate
a double NIM-rep to left $A_1{\otimes}A_2$-modules, with suitable
\alg s $A_1$ and $A_2$. We may view the corresponding coefficients 
schematically as follows:
  \bea \begin{picture}(227,33)(0,30)
  \put(0,0)   {\begin{picture}(0,0)(0,0)
              \scalebox{.38}{\includegraphics{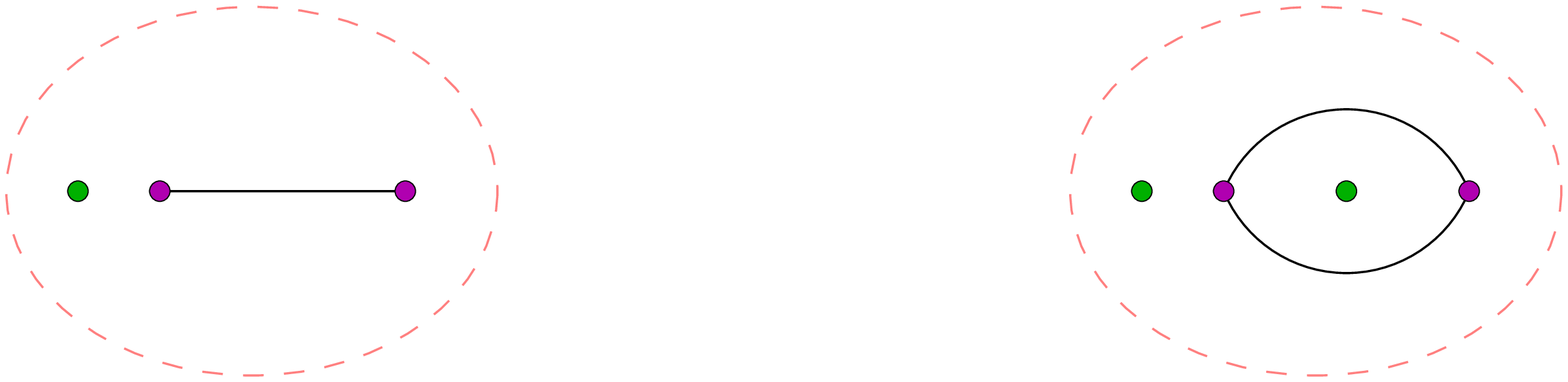}} \end{picture}
  \put(-52,30){$\Ann{kX}{Y}\;=$}
  \put(126,30){$\Zdfct XYkl \,=$}
  \put(4,30) {\scriptsize$k$}
  \put(182,30) {\scriptsize$k$}
  \put(219,30) {\scriptsize$l$}
  \put(23.2,35) {\scriptsize$Y$}
  \put(64.4,35) {\scriptsize$X$}
  \put(200.7,35) {\scriptsize$Y$}
  \put(245.5,35) {\scriptsize$X$}
  \put(42,34) {\scriptsize$A$}
  \put(211,12.3) {\scriptsize$A_1$}
  \put(212,48) {\scriptsize$A_2$}
  }
  \epicture09 \ee
What is shown are horizontal sections of ribbon graphs in $S^2{\times}S^1$. 
More specifically, the figures are cross sections of the graphs \erf{Akmn} 
and \erf{ZXYkl}, \resp; vertical ribbons for simple objects and left modules 
are indicated by filled circles, while the lines symbolise $A$-ribbons. 
\\
Analogously, for any pair of left $A_1{\otimes}\cdots{\otimes}A_n$-modules 
$X,\,Y$, the numbers $Z^{X|Y}_{i_1 , \dots , i_n}$ defined by
  \bea \begin{picture}(78,87)(0,21)
  \put(0,0)   {\begin{picture}(0,0)(0,0)
              \scalebox{.38}{\includegraphics{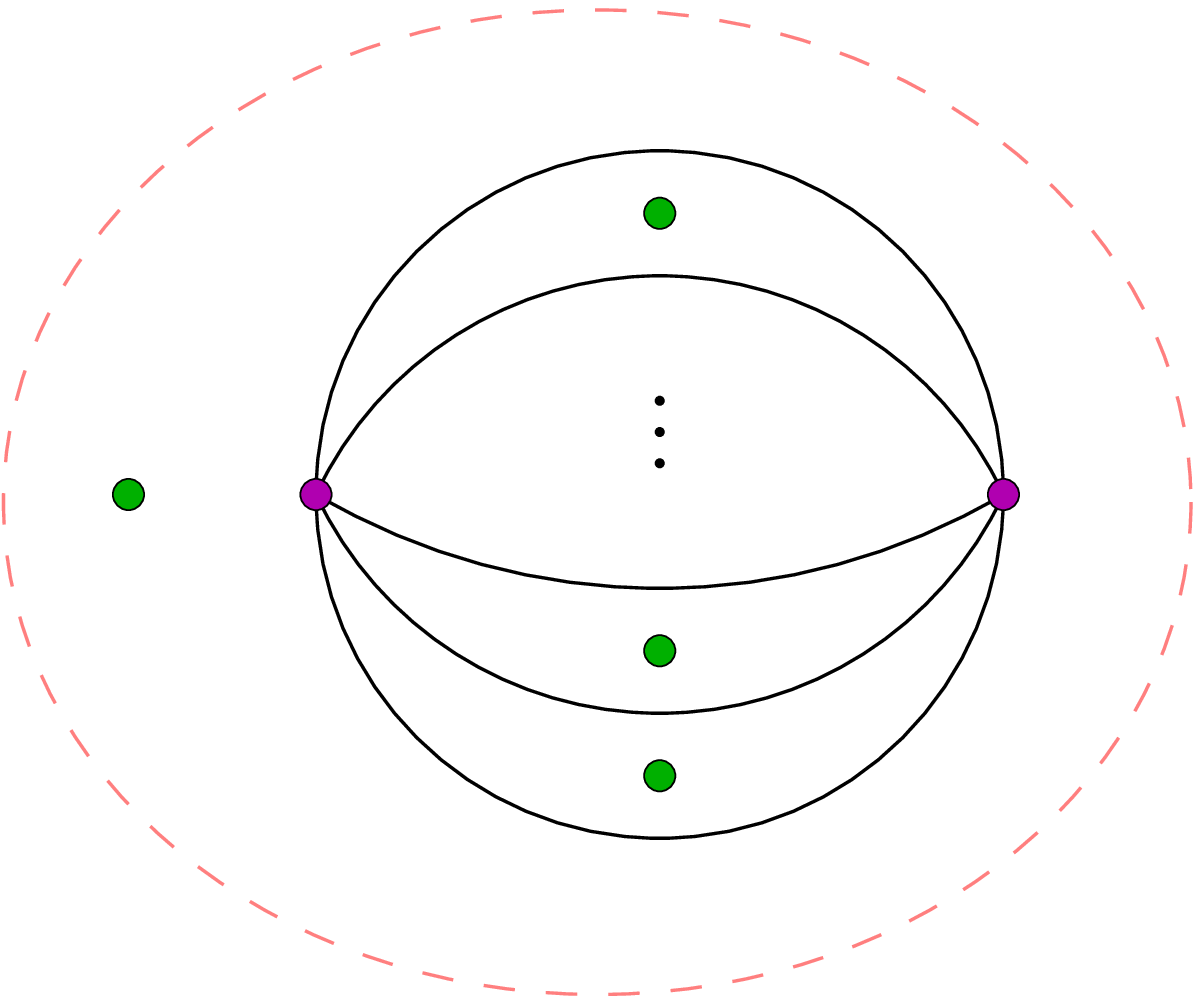}} \end{picture}
  \put(-65,53){$Z^{X|Y}_{i_1 , \dots , i_n} \; :=$}
   \put(25.6,52.6) {\scriptsize$Y$}
   \put(112.3,52.8) {\scriptsize$X$}
   \put(5,54) {\scriptsize$i_1$}
   \put(75,22) {\scriptsize$i_2$}
   \put(75,36) {\scriptsize$i_3$}
   \put(75,84) {\scriptsize$i_n$}
   \put(43,19) {\scriptsize$A_1^{}$}
   \put(49,29) {\scriptsize$A_2^{}$}
   \put(51,50.7) {\scriptsize$A_3^{}$}
   \put(45,65) {\scriptsize$A_{n-1}$}
   \put(36,86) {\scriptsize$A_n$}
  }
  \epicture02 \ee
furnish an {\em $n$-fold NIM-rep\/} of the fusion rules, i.e.\
  \be
  \sum_R Z^{X|R}_{i_1,..., i_n} \, Z^{R|Y}_{j_1, \dots ,j_n}
  = \sumI_{k_1, \cdots ,k_n} \N{i_1}{j_1}{k_1} \cdots
  \N{i_n}{j_n}{k_n} \, Z^{X|Y}_{k_1, \dots ,k_n} \,.  \ee

Applications of this structure in string theory remain to be clarified.
It is, however, tempting to conjecture that they appear in situations
where the world sheet is no longer a smooth manifold. Such world sheets 
play an important role in the description of string junctions and, 
more generally, in string networks (see e.g.\ \cite{sen14}).

\newpage
%%%%%%%%%%%%%%%%%%%%%%%%%%%%%%%%%%%%%%%%%%%%%%%%%%%%%%%%%%%%%%%%%%%%%%%%

\section{Epilogue:\\ Non-commutative geometry in tensor categories}

Our results relate rational conformal field theory to the theory 
of non-commutative algebras and their representations in modular tensor 
categories. Now a convenient way to think about non-commutative algebras 
in the tensor category of vector spaces is non-commutative geometry. 
It is therefore tempting to relate conformal field theory to a version 
of non-commutative geometry over a modular tensor category.

Consider a compact topological manifold $M$ with measure $\mu$. On the
commutative algebra $A\eq C^0(M)$ of continuous functions a counit 
$\eps{:}\ A\,{\to}\,\complex$ is provided by the integral, $\eps(f)\eq
\int\!{\rm d}\mu\, f$. This way, $A$ becomes a symmetric Frobenius 
algebra (though not necessarily a special one). In this sense, the 
Frobenius algebra $A$ in \calc\ that we used to describe conformal 
field theory can be regarded as non-commutative measure theory in 
the tensor category \calc.

The \alg\ $A$ itself provides all information needed to analyze \corfu s
on arbitrary closed oriented world sheets. The study of boundary 
conditions and defect lines requires in addition the study of
$A$-modules and \AA-bimodules. They should be thought of as non-commutative
vector bundles. As will be discussed in a future publication, our formalism 
can be extended to unorientable world sheets as well. This requires
the choice of a `conjugation' on $A$, i.e.\ the category-theoretic
analogue of a conjugation or, in other words, of a *-structure. This can 
be interpreted as the choice of a real structure in the non-commutative 
geometry over the tensor category $\calc$.

We summarise these ideas in the following table:

\begin{center}
\begin{tabular}{lllll}
\multicolumn1c{world sheet} &\hsp{.7} & \multicolumn1c{algebraic data}
              &\hsp{.7} & \multicolumn1c{NC geometry over \calc} \\[3pt]
\cline{1-1}\cline{3-3}\cline{5-5}\\[-7pt]
closed oriented && symm.\,special Frobenius object $A$ 
                      && NC measure theory \\[1pt]
boundaries   && $A$-modules && NC vector bundles \\[1pt]
defect lines && \AA-bimodules &&  \\[1pt]
orientifolds && generalised *-structure on $A$ && real NC geometry \\[1pt]
\multicolumn1c\ldots && \multicolumn1c\ldots && \multicolumn1c\ldots  
\end{tabular}
\end{center}

\noindent
Obviously, in all three columns more entries must be added to arrive at
a complete picture. For example, cyclic
cohomology, the non-commutative analogue of de Rham cohomology, should
play an important role for describing deformations of conformal field 
theories. Indeed, in the case of complex algebras it is known that the
second degree of cyclic cohomology controls deformations of associative
algebras with bilinear invariant form (see e.g.\ \cite{peSc2}). In any case,
already on the basis of the presently available evidence it is reasonable
to expect that viewing conformal field theory as non-commutative
geometry over a tensor category can serve as a fruitful guiding 
principle in the future.

The emergence of non-commutative structures in conformal field theory
does not come as a surprise. There exist {\em families\/} of CFTs which 
are known to yield, in a certain limit, non-commutative field theories 
\cite{sewi8}. The present results are, however, much closer to a different
proposal \cite{frgA} according to which {\em every single\/}
conformal field theory gives rise to a non-commutative geometry.
These ideas have been formulated within the category of vector spaces.
Our results indicate that we can realise them by lifting all relevant
structures, algebraic and geometric, to more general tensor categories.

\newpage
%%%%%%%%%%%%%%%%%%%%%%%%%%%%%%%%%%%%%%%%%%%%%%%%%%%%%%%%%%%%%%%%%%%%%%%%

%\vskip4em
% \vfill
\noindent
{\sc Acknowledgements:}\\[.3em]
We are grateful to J\"urg Fr\"ohlich and Albert Schwarz for discussions
and encouragement, and to Kevin Graham, Viktor Ostrik and Bodo Pareigis
for correspondence.
J\"urgen Fuchs thanks the LPTHE for hospitality.
Ingo Runkel is supported by the EU grant HPMF-CT-2000-00747. 
\vskip4em
% \vfill
%\newpage

%\end{document}
%%%%%%%%%%%%%%%%%%%%%%%%%%%%%%%%%%%%%%%%%%%%%%%%%%%%%%%%%%%%%%%%%%%%%%%%

\newcommand\wb{\,\linebreak[0]} \def\wB {$\,$\wb}
 \newcommand\Bi[1]    {\bibitem{#1}}
 \renewcommand\J[5]     {{\sl #5\/}, {#1} {#2} ({#3}) {#4} }
 \newcommand\K[6]     {{\sl #6\/}, {#1} {#2} ({#3}) {#4}}
 \newcommand\PhD[2]   {{\sl #2\/}, Ph.D.\ thesis (#1)}
 \newcommand\Prep[2]  {{\sl #2\/}, pre\-print {#1}}
 \newcommand\BOOK[4]  {{\sl #1\/} ({#2}, {#3} {#4})}
 \newcommand\inBO[7]  {{\sl #7\/}, in:\ {\sl #1}, {#2}\ ({#3}, {#4} {#5}), p.\ {#6}}
 \newcommand\webb[2]  {{\sl #2\/}, available at http:/$\!$/#1}
 \newcommand\webp[2]  {{\sl #2\/}, available at\\ http:/$\!$/#1}
 \def\jf    {J.\ Fuchs}
 \def\dim   {dimension}  
 \def\adma  {Adv.\wb Math.}
 \def\atmp  {Adv.\wb Theor.\wb Math.\wb Phys.} 
 \def\coia  {Com\-mun.\wB in\wB Algebra} 
 \def\coma  {Con\-temp.\wb Math.}
 \def\comp  {Com\-mun.\wb Math.\wb Phys.}
 \def\cpma  {Com\-pos.\wb Math.}
 \def\duke  {Duke\wB Math.\wb J.}
 \def\fiic  {Fields\wB Institute\wB Commun.}
 \def\foph  {Fortschritte\wB d.\wb Phys.}
 \def\gafa  {Geom.\wB and\wB Funct.\wb Anal.}  
 \def\ijmp  {Int.\wb J.\wb Mod.\wb Phys.\ A}
 \newcommand\ilag[2] {\inBO{\Infdim Lie Algebras and Groups}
            {V.G.\ Kac, ed.} \WS\Si{1989} {{#1}}{{#2}}}
 \def\jams  {J.\wb Amer.\wb Math.\wb Soc.}
 \def\jgap  {J.\wb Geom.\wB and\wB Phys.}
 \def\jhep  {J.\wb High\wB Energy\wB Phys.} 
 \def\jktr  {J.\wB Knot\wB Theory\wB and\wB its\wB Ramif.}
 \def\joal  {J.\wB Al\-ge\-bra}
 \def\jomp  {J.\wb Math.\wb Phys.}
 \def\jopa  {J.\wb Phys.\ A} 
 \def\josp  {J.\wb Stat.\wb Phys.} 
 \def\jpaa  {J.\wB Pure\wB Appl.\wb Alg.}      
 \def\maan  {Math.\wb Annal.}
 \def\mams  {Memoirs\wB Amer.\wb Math.\wb Soc.}
 \newcommand\nqft[2] {\inBO{%{\rm 1987 Carg\`ese Lectures on}
            Nonperturbative \QFT} {G.\ 't Hooft, A.\ Jaffe, G.\ Mack, P.K.\
            Mitter, and R.\ Stora, eds.} \PL\NY{1988} {{#1}}{{#2}} }
 \def\nupb  {Nucl.\wb Phys.\ B}
 \newcommand\phgt[2] {\inBO{Physics, Geometry, and Topology}
            {H.C.\ Lee, ed.} \PL\NY{1990} {{#1}}{{#2}} }
 \def\phep  {Proc.\wb HEP$\!$}
 \def\phlb  {Phys.\wb Lett.\ B}
 \def\phrl  {Phys.\wb Rev.\wb Lett.}
 \def\phrp  {Phys.\wb Rep.}           
 \def\rims  {Publ.\wB RIMS}  
 \def\rvmp  {Rev.\wb Math.\wb Phys.}
 \def\slnm  {Sprin\-ger\wB Lecture\wB Notes\wB in\wB Ma\-the\-matics }
 \def\tams  {Trans.\wb Amer.\wb Math.\wb Soc.}  
 \def\thmp  {Theor.\wb Math.\wb Phys.} 
 \def\topo  {Topology} 
 \def\AMS    {{American Mathematical Society}}
 \def\AP     {{Academic Press}}
 \def\IPC    {{International Press Company}}
 \def\PL     {{Plenum Press}}
 \def\SV     {{Sprin\-ger Ver\-lag}}
 \def\WI     {{Wiley Interscience}}
 \def\WS     {{World Scientific}} 
 \def\Be     {{Berlin}}
 \def\Ca     {{Cambridge}}
 \def\PR     {{Providence}}
 \def\Si     {{Singapore}} 
 \def\NY     {{New York}}

\end{document}